\documentclass[11pt,a4paper,twoside]{book}
\pdfoutput=1
\usepackage{graphicx}
\usepackage{amssymb,amsmath}
\usepackage{cite,color,url}
 \usepackage{fancyhdr}   

\usepackage{epsfig,psfrag,rotating,soul}
\usepackage{rotfloat}

\usepackage[linktocpage=true
,pdfproducer=medialab
,colorlinks=true
,citecolor=green
,linkcolor=blue
,menucolor=blue
,pdfa=true
]{hyperref}
\usepackage{amsfonts,feynarts}

\input paperdef

    \makeatletter
    \def\cleardoublepage{\clearpage\if@twoside \ifodd\c@page\else%
        \hbox{}%
        \thispagestyle{empty}
        \newpage%
        \if@twocolumn\hbox{}\newpage\fi\fi\fi}
    \makeatother

\allowdisplaybreaks

\def\signofmetric{1}

\if0\signofmetric

\fi

\if1\signofmetric

\fi

\if2\signofmetric

\fi

\if3\signofmetric

\fi

\def\perspectives{in {\it Perspectives on Supersymmetry}, ed.
G.L.~Kane (World Scientific, 1998)}

\def\beq{\begin{eqnarray}}
\def\eeq{\end{eqnarray}}
\def\bea{\begin{eqnarray*}}
\def\eea{\end{eqnarray*}}
\def\Baryon{{\rm B}}
\def\Lepton{{\rm L}}
\def\sbar{\overline}
\def\stilde{\widetilde}

\def\lagr{{\cal L}}
\def\drbar{\overline{\rm DR}}

\def\conj{{{\rm c.c.}}}

\def\cbeta{c_{\beta}}
\def\sbeta{s_{\beta}}

\def\half{{1\over 2}}

\def\centeron#1#2{{\setbox0=\hbox{#1}\setbox1=\hbox{#2}\ifdim
\wd1>\wd0\kern.5\wd1\kern-.5\wd0\fi
\copy0\kern-.5\wd0\kern-.5\wd1\copy1\ifdim\wd0>\wd1
\kern.5\wd0\kern-.5\wd1\fi}}
\def\ltap{\;\centeron{\raise.35ex\hbox{$<$}}{\lower.65ex\hbox{$\sim$}}\;}
\def\gtap{\;\centeron{\raise.35ex\hbox{$>$}}{\lower.65ex\hbox{$\sim$}}\;}
\def\gsim{\mathrel{\gtap}}
\def\lsim{\mathrel{\ltap}}

\def\slashchar#1{\setbox0=\hbox{$#1$}           
   \dimen0=\wd0                                 
   \setbox1=\hbox{/} \dimen1=\wd1               
   \ifdim\dimen0>\dimen1                        
      \rlap{\hbox to \dimen0{\hfil/\hfil}}      
      #1                                        
   \else                                        
      \rlap{\hbox to \dimen1{\hfil$#1$\hfil}}   
      /                                         
   \fi}                                        %

\begin{document}
\thispagestyle{empty}


\begin{titlepage}
\thispagestyle{empty}
\begin{center}
\vspace*{-5em}

\begin{tabular}{cc}
\includegraphics[width=60mm]{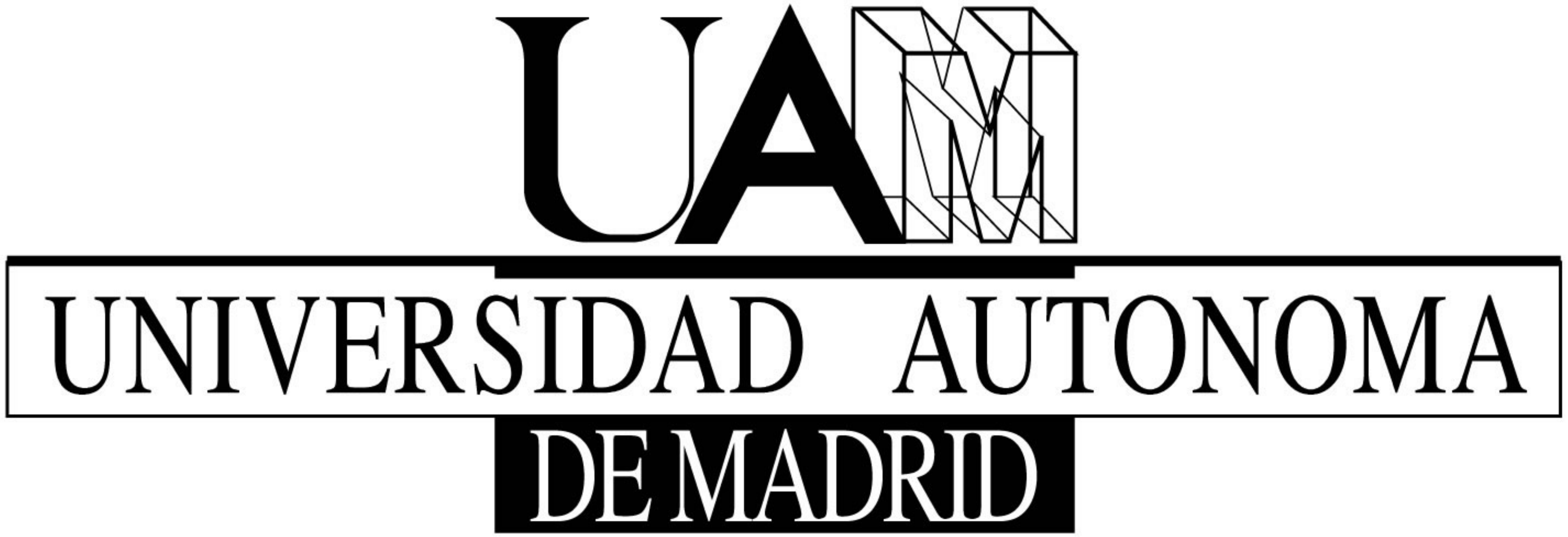} &
\includegraphics[width=55mm]{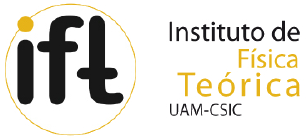} \\
\end{tabular}

\vspace*{1em}
{\Large \bf Universidad Aut\'onoma de Madrid}\\[.2cm]
{\Large \bf Facultad de Ciencias}\\[.2cm]
{\Large \bf Departamento de F\'{\i}sica Te\'orica}\\[7em]
{\Huge\bfseries The flavour of supersymmetry:\\[1ex]} 
{\Huge\bfseries Phenomenological implications\\ of sfermion mixing}\\[8em]
Memoria de Tesis Doctoral realizada por\\
\vspace*{0.5em}
{\LARGE\bfseries Miguel Arana Catania}\\
\vspace*{0.5em}
presentada ante el Departamento de F\'{\i}sica Te\'orica\\
de la Universidad Aut\'onoma de Madrid\\
%
\vfill
Trabajo dirigido por la\\
{\large\bfseries Dra. Mar\'{\i}a Jos\'e Herrero Solans},\\
Catedr\'atica del Departamento de F\'{\i}sica Te\'{o}rica\\
y miembro del Instituto de F\'{\i}sica Te\'orica, IFT-UAM/CSIC\\
y por el\\
{\large\bfseries Dr. Sven Heinemeyer},\\
Investigador cient\'{\i}fico del Instituto de F\'{\i}sica de Cantabria IFCA (CSIC-UC)\\[1em]

{\Large Madrid, diciembre de 2013\\
}
\end{center}
\end{titlepage}


\newpage
\thispagestyle{empty}
\mbox{}

\clearpage
\thispagestyle{empty}

$\phantom{x}$

\vspace{4 cm}

\begin{minipage}{0.25\textwidth}
$\phantom{x}$
\end{minipage}
\begin{minipage}{0.7\textwidth}
[...] o simplemente agarrando una tacita de caf\'e y mir\'andola por todos lados, no ya como una taza sino como un testimonio de la inmensa burrada en que estamos metidos todos, creer que ese objeto es nada m\'as que una tacita de caf\'e cuando el m\'as idiota de los periodistas encargados de resumirnos los quanta, Planck y Heisenberg, se mata explic\'andonos a tres columnas que todo vibra y tiembla y est\'a como un gato a la espera de dar el enorme salto de hidr\'ogeno o de cobalto que nos va a dejar a todos con las patas para arriba.

\bigskip
Julio Cort\'azar. {\it Rayuela. Cap\'\i{}tulo 71, Morelliana.}

\end{minipage}
\begin{minipage}{0.05\textwidth}
$\phantom{x}$
\end{minipage}

\tableofcontents


\newpage
\thispagestyle{empty}


\chapter*{Agradecimientos}
\addcontentsline{toc}{chapter}{Agradecimientos}

\parskip 0.5mm

Este cap\'\i{}tulo es uno de los m\'as importantes de esta tesis. No trata sobre ciencia, sino sobre el mundo que hay ah\'\i{} fuera mientras hacemos ciencia. Ese mundo deber\'\i{}a saltar por los aires en un mill\'on de pedazos a cada segundo, y sin embargo hay gente, hay formas de vivirlo, que permiten que siga girando, que mantenga el calor, que al acabar un d\'\i{}a empiece otro. Ante eso no hay nada que decir, s\'olo inclinarse y agradecer, profundamente:

\bigskip
\parskip 2.0mm
 a mi madre, mi padre, mi hermano y mis hermanas, sus parejas, y mis sobrinas, por ese amor que me ha arropado siempre, con un calor total e impenetrable, y que ha estado por encima de todo; por ese hogar tan \'unico, tan privilegiado, desde el que salir al mundo, por hacer de lo elevado la normalidad, por esa visi\'on profunda y completa del mundo con la que poder orientarme, por comprender siempre tan claramente lo que importa, por los pasos que seguir e imitar con profundo orgullo, desde aqu\'\i{} o desde el recuerdo de otros, por hacerlo todo f\'acil, y ense\~narme que no tiene por qu\'e ser de otra manera.

 a Adri\'an, \'Angel, Antonio, Mar, Ana, Victor, Mar\'\i{}a, Sandra, Sergio y \'Alvaro, por haber estado ah\'\i{} desde siempre, tan mezclados y diferentes, por ser un refugio frente a lo de fuera, por las costumbres, los roles, lo eternamente repetido que hace el hogar.\parskip 0.5mm

 a Germ\'an, Irene, Victor, Arianna, Edoardo y otros compa\~neros de laboratorio que hemos vivido juntos esa nueva extra\~na madurez de pasar a ser profesores. Me alegro mucho de que haya sido con vosotros ese momento \'unico.

 a Mar\'\i{}a, Juan y Sof\'\i{}a, por haberme abierto el camino a, y compartido, esa locura, esa obsesi\'on; porque encontrarse bajo los focos es encontrarse en una parte muy \'unica del mundo, eso ya no se nos va a pasar; por el cari\~no, por las risas, por las millones de horas unos con otros, muy cerca.

 a Mar\'\i{}a Jos\'e, por estos a\~nos compartiendo la investigaci\'on, por ense\~narme todas estas cosas, por darte tanto en lo que haces, por la cercan\'\i{}a, por crear todos esos momentos buenos y divertidos, que han sido muchos. Sobre los malos momentos, espero que te libres de esa incapacidad para permitir las formas de hacer las cosas de los otros, vas a hacer que la gente sea m\'as feliz a tu alrededor.

 a Sven, por hacer ciencia conmigo, por ense\~narme a hacerla, por la sonrisa infatigable (en un sentido totalmente literal), por no poner barreras, por estar siempre preparado para soltar una carcajada, por las noches de cervezas, por cada momento divertido y cercano, tambi\'en excepcionalmente numerosos.

 a Juanjo, por pensar a lo grande siempre, por pensar diferente, y pensar conmigo; por ense\~narme, por esas noches tremendas, por las que vendr\'an.

 a todos los profesores que he tenido desde el colegio hasta la universidad que han querido ense\~nar, especialmente a los que han sabido encontrar el verdadero conocimiento oculto entre monta\~nas de ense\~nanzas prescindibles; a los que han sido capaces de volver a las aulas d\'\i{}a tras d\'\i{}a, a\~no tras a\~no, sin perder la felicidad y la fuerza.

 a Andrea, mi bichillo, por haberme querido tanto, por querer compartir toda tu vida, esa tan \'unica, por haber hecho mi vida mejor, m\'as importante, por haberme dado tantas risas, por esa energ\'\i{}a inagotable que lo cambia todo.

 a Ernesto, por confiar, por dejarme jugar, por dejarme crecer. Y por bella persona. Muy bella.

 a todos los que contribuyen de alguna manera a que exista la universidad y la investigaci\'on p\'ublica. Que nos permiten que podamos dedicar los d\'\i{}as de nuestra vida a esto tan fascinante. A los que entienden la importancia absoluta de la tarea.

 a Javi por esa mirada siempre limpia, siempre abierta sobre el mundo, tan poco usual; por llevarla con la total certeza de que as\'\i{} el mundo es mejor, y a pesar de todos esos hombres grises obsesionados porque seas uno de ellos; por salir victorioso tantas veces de la lucha contra tu parte oscura, por todos los grandes juegos que hemos vivido juntos, en la f\'\i{}sica, en la pol\'\i{}tica, en el mundo,... qu\'e divertido.

 a los que no se resignan a un mundo gris y mediocre donde la vida no importe; a los que exigen belleza, a los que exigen felicidad.

 a mis compas de la casa de las flores y de tantas vueltas por el mundo (tambi\'en Jules, Etor y Susana!), por hacer all\'\i{} y aqu\'\i{} f\'aciles y divertidos los miles de minutos que tiene cada d\'\i{}a, que es tan fundamental. Those are for you.

 a \'Alvaro, por todo, aunque a veces se haga complicado.

 a Jorge Alem\'an, por ayudarme con esa otra investigaci\'on tan complicada, con tanta energ\'\i{}a y dedicaci\'on, poniendo en ella toda esa cabeza tan genial.

 a Ana, porque hemos sobrevivido a la gran guerra, y ahora desde la residencia para veteranos tullidos y borderline podemos contar batallitas; por la absurdidad, por esa risa ultradecib\'elica, por eso que hace que todos se arremolinen siempre a tu alrededor (incluso aunque intentemos escapar). Hubiera sido tan aburrido sin ti!

 a los que buscan trascender esta realidad.

 a Alice y Daniel, por hacerme querer una y otra vez parar todo para simplemente estar con vosotros, porque en esa casa no se paren de escuchar risas.

 a toda la gente que sali\'o a las plazas, que con naturalidad fue a encontrarse y a iniciar una revoluci\'on; que le dieron la espalda al supuesto poder, para hablar y pensar juntos durante 1001 noches lo que deb\'\i{}a ser el mundo; que callaron a tertulianos, analistas, falsos, manipuladores, demagogos y cobardes, y demostraron qui\'enes somos realmente

 a Helen, por ese tiempo de cuento.

 a Siannah y Arian, por ese car\'acter siempre tranquilo, esa disposici\'on amable, que hace tan agradable investigar y estar con vosotros.

 a los que comparten cultura, software y conocimiento en internet, y ayudan a acabar con el concepto de ``propiedad intelectual", porque mi vida ha sido tremendamente m\'as rica gracias a ello.

 a todos mis compa\~neros de despacho por convertir un espacio de trabajo muerto en un lugar de convivencia donde querer volver. Nos pasamos la mayor parte de nuestra vida ah\'\i{}, no es algo menor, es muy importante.

 a Javier Men\'endez (und \'Africa nat\"urlich) por dejarme, divertido y tranquilo como siempre eres, tomar el despacho y armar ruido y radionovelas; porque para un doctorando novato es fundamental tener modelos por encima de uno de los que aprender, y yo he tenido mucha suerte en esto.

 a Steve Blanchet, que trajo la belleza y la elegancia al despacho, por hacernos un hueco tan c\'alido y cercano en tu vida, por querer compartir con nosotros tantas risas explosivas (sobre todo cuando podr\'\i{}as estar machacando a Federer).

 a Liliana, que calladita calladita nos fuiste conquistando a todos. Y a Luis, por ser un show con patas (aunque no te dejes querer).

 a Josu y Xabi, porque es un orgullo tener unos hermanos peque\~nos como vosotros, por aceptarme como hermano mayor, por traer mucho ruido y buen humor al despacho, por ser el par m\'as \'unico de todo el instituto.

 a Sylwia, por la dulzura, por hacer el mundo f\'acil y feliz, y bonito, siempre, en cualquier lugar, meine sch\"one kleine Pferd.

 a los amigos del 15M (Clara, St\'ephane, Pablo, Aya, Juanlu, los compas de wet, Eric, Gregorio, Eduardo,... y cientos otros) con los que hemos vivido con pasi\'on c\'omo la historia tomaba forma delante de nuestros ojos, c\'omo nuestras fantas\'\i{}as se hac\'\i{}an realidad; que hemos pasado las noches y los d\'\i{}as, y cada minuto juntos, pensando mundos, trazando revoluciones, hasta que ca\'\i{}amos rendidos unos sobre otros.

 a mi familia argentina: los Larriera, Silvia, Mem\'e, Ren\'ee, Miguel, Luly, Dar\'\i{}o, Amanda, Maruja, Rodo, Ana, Justo, Mennella, los Regueiro (que ya est\'ais bastante argentinizados), Alejandro, Nory, Liliana, Vilma, Diana y Cacho, por el calor familiar que no ha faltado nunca, por la mitad de mi educaci\'on que han sido esas comidas familiares en casa, por crecer rodeado de la risa y la palabra.

 a todos los que emplean el tiempo de sus vidas en cambiar el mundo, en que los dem\'as vivan mejor. A los que para hacerlo tienen que buscar grietas en sus vidas y trabajos por donde escaparse.

 a \'Angel, nuestro fil\'osofo, por el humor siempre presente a pesar de la conciencia de la tragedia, por la entrega incondicional a tus amigos, por guiar, por ense\~nar, por ser un referente, por no querer nunca ser m\'as que el otro, sea quien sea, por mantenerte siempre elevado, con total ligereza.

 a los hyp\'aticos, por haber convertido ese cuartucho de la facultad en el rinc\'on m\'as interesante y excitante de toda la universidad, por la maravillosa mezcla, por la brillantez, por la educaci\'on recibida, por los caminos que seguir, por jugarlo todo siempre en los l\'\i{}mites, por buscar el d\'\i{}a siguiente.

 al No-grupo por habernos enamorado tan intensamente, por esas cabecitas tan geniales, por las noches de conspiraci\'on para cambiar el mundo, por cuidarnos unos a otros.

 a los que fallan, a los que no llegan, a los que no son capaces.

 a Mario Sanz, por el cari\~no, la cercan\'\i{}a, la sensibilidad, por compartir tanto, tan personal, tan especial.

 a todos los cient\'\i{}ficos que ofrecen sus vidas para poder colaborar en la construcci\'on del conocimiento humano; millones de personas que mano con mano empujan haci\'endonos avanzar a todos; que llegan cada d\'\i{}a a sus despachos y laboratorios sabiendo que la sociedad no les va a recompensar, ni agradecer; que pasar\'an desapercibidos, que s\'olo habr\'a silencio. Y en ese silencio se dedican a trabajar contentos.

 a \'Alvaro pensamiento, por ser una persona de una calidad humana excepcional.

 a Nadir, por sobrevivir hombro con hombro (y gracias a eso precisamente), al fr\'\i{}o de Berl\'\i{}n; por lanzarte conmigo a abrir ruta tras ruta, y estar ah\'\i{} siempre que se te necesita.

 a Carlos, por las infinitas noches pensando universos, planeando revoluciones, disfrutando los infinitos otros mundos que vamos encontrando, pensando la estructura escondida de este; por pensar libre, sin l\'\i{}mites ni esquemas, y lanzarte a darle vida a cualquier idea genial; por la sabidur\'\i{}a, por la incorruptibilidad, por la ausencia de ego, por la constante tranquilidad.

$\phantom{x}$

\bigskip
A todos ellos va dedicado este trabajo.

\parskip 1.5mm

\chapter{Introducci\'on}
\label{introduccion}

El 4 de julio de 2012 fue anunciado internacionalmente en un evento en vivo que el bos\'on de Higgs hab\'\i{}a sido descubierto \cite{higgsdiscoveryannouncement,Aad:2012tfa,Chatrchyan:2012ufa}. En los pocos segundos que tard\'o la se\~nal transmitida en viajar alrededor del mundo, la humanidad dio un enorme salto hacia adelante en nuestro conocimiento del universo. No es muy com\'un que el avance cient\'\i{}fico cristalice en un evento tan espec\'\i{}fico, y de tal magnitud, as\'\i{} que los que vivimos ese momento debemos atesorarlo. ?`Qu\'e ocurri\'o ese d\'\i{}a? ?`qu\'e clase de cambio produjo en el mundo? y despu\'es del salto ?`d\'onde estamos ahora?

El descubrimiento es un movimiento dial\'ectico de apertura y cierre. El cierre se produce con respecto al Modelo Est\'andar (SM) de las interacciones fundamentales, y produjo en muchos f\'\i{}sicos un sentimiento de alivio despu\'es del anuncio. El bos\'on de Higgs es la \'ultima pieza que faltaba del SM, nuestro modelo actual de f\'\i{}sica de part\'\i{}culas desarrollado durante m\'as de 40 a\~nos \cite{Glashow:1961tr,GellMann:1964nj,Zweig:1964jf,Weinberg:1967tq,Salam:1968rm}. El SM es una teor\'\i{}a cu\'antica de campos renormalizable en 4 dimensiones con una simetr\'\i{}a gauge $SU(3)_C\times SU(2)_L\times U(1)_Y$, invariante bajo transformaciones de Poincar\'e y CPT. El bos\'on de Higgs es la part\'\i{}cula que aparece como consecuencia del mecanismo de Brout-Englert-Higgs \cite{Higgs:1964ia,Higgs:1964pj,Higgs:1966ev,Englert:1964et,Guralnik:1964eu}, necesario en el SM para dar masa a las part\'\i{}culas elementales. Al mismo tiempo que el SM, extendido con masas de neutrinos, ha demostrado experimento tras experimento ser capaz de coincidir en sus predicciones con cada medida que se ha podido imaginar jam\'as en el campo de la f\'\i{}sica de part\'\i{}culas de altas energ\'\i{}as, el bos\'on de Higgs se ha mantenido escondido. El extraordinario \'exito del modelo entraba en conflicto con la elusividad de su \'ultima predicci\'on. Pero finalmente el descubrimiento del bos\'on de Higgs se ha producido en el laboratorio CERN, por los dos experimentos  ATLAS~\cite{Aad:2012tfa} y CMS~\cite{Chatrchyan:2012ufa}. Estos experimentos han medido una part\'\i{}cula tipo Higgs con spin 0, sin cargas el\'ectrica o de color y paridad positiva (aunque esto \'ultimo est\'a todav\'\i{}a estudi\'andose), como predec\'\i{}a el SM, y con una masa entre 125 y 126 GeV  \cite{Aad:2013wqa,Chatrchyan:2013lba}  compatible con las restricciones del polo de Landau y de la estabilidad del vac\'\i{}o del SM hasta energ\'\i{}as muy altas, cercanas a la masa de Planck $M_P=10^{19}$GeV (el SM no predice un valor concreto de la masa del bos\'on de Higgs, pero pone l\'\i{}mites superiores e inferiores al mismo; en concreto dentro del SM puro parece que vivimos en un vac\'\i{}o metaestable de vida larga \cite{Buttazzo:2013uya,Alekhin:2012py}). Con este descubrimiento se cierra un cap\'\i{}tulo importante en la historia de la f\'\i{}sica de part\'\i{}culas. Probablemente ninguna otra construcci\'on humana pueda ser comparada ahora en complejidad y solidez al SM. El nivel de concordancia entre las predicciones del modelo y los resultados de los experimentos alcanza por ejemplo el orden de $10^{-12}$ en la medida del momento magn\'etico del electr\'on \cite{Hanneke:2008tm,Hanneke:2010au}.

Sin embargo, el SM no es una teor\'\i{}a final en f\'\i{}sica de part\'\i{}culas. Y aqu\'\i{} viene la apertura. Hay muchos fen\'omenos en el universo que el SM no es capaz de explicar, as\'\i{} que est\'a claro que no es un modelo completo de la f\'\i{}sica de nuestro universo, y por lo tanto debe ser ampliado o introducido en una teor\'\i{}a m\'as extensa. Por ejemplo, el SM no incluye ni la masa ni las oscilaciones de los neutrinos, a pesar de que experimentos como Super-Kamiokande hayan probado ambos hechos \cite{Fukuda:1998mi}. Tampoco incluye un candidato como part\'\i{}cula de materia oscura. De hecho, el bos\'on de Higgs est\'a en conexi\'on directa con el principal fen\'omeno que escapa a nuestro control: la masa de las part\'\i{}culas elementales. Aparte de esto y desde un punto de vista m\'as general en f\'\i{}sica, con respecto a la masa gravitacional, otra de las principales preguntas abiertas ser\'\i{}a c\'omo unir los niveles gravitacional y cu\'antico en una teor\'\i{}a unificada que describa todas las fuerzas en un marco com\'un. El objetivo ser\'\i{}a cuantizar la gravedad, si miramos el problema desde el punto de vista de la teor\'\i{}a cu\'antica de campos; una cuesti\'on que evidentemente no est\'a resuelta en el SM ya que la gravitaci\'on es la \'unica interacci\'on que no est\'a inclu\'\i{}da en el modelo. Sin embargo, desde la perspectiva que ser\'a m\'as relevante para nuestro trabajo, donde nos mantendremos al nivel de la teor\'\i{}a cu\'antica de campos y las fuerzas gravitacionales ser\'an despreciables, el problema de la masa puede ser reformulado de otra manera en lo que se conoce como el problema del sabor.

El problema del sabor, formulado a grandes rasgos, podr\'\i{}a ser entendido como la ausencia de una teor\'\i{}a capaz de explicar el papel que juega la masa en nuestro modelo de f\'\i{}sica de part\'\i{}culas. Esto se traduce en un conjunto de caracter\'\i{}sticas o fen\'omenos observados que no somos capaces de explicar. Por ejemplo, las part\'\i{}culas elementales aparecen como `familias' de part\'\i{}culas, donde cada familia es una copia de la anterior en los n\'umeros cu\'anticos, pero difiere en las masas. La existencia de estas familias y su n\'umero (3 en el SM, siendo esto totalmente compatible con los actuales l\'\i{}mites experimentales, v\'ease por ejemplo \cite{limits4thgenerationfermions}) no tienen explicaci\'on en el SM. El patr\'on de las masas de las part\'\i{}culas tambi\'en parece arbitrario, con vastas regiones vac\'\i{}as en la escala de masas como el espacio entre las masas de los neutrinos y del electr\'on, desde aproximadamente  $10^{-9}$ GeV a $0.51\times 10^{-3}$ GeV, o los dos \'ordenes de magnitud entre la masa del quark bottom con 4.18 GeV y la masa del quark top con 173.2 GeV. Los valores de los elementos de las matrices de mezcla entre familias de fermiones, CKM  \cite{Cabibbo:1963yz,Kobayashi:1973fv} y PMNS \cite{Pontecorvo:1957cp,Pontecorvo:1967fh,Maki:1962mu}, tampoco son predichos por el SM. Estas y otras preguntas sobre el sabor abren una nueva ventana a la fenomenolog\'\i{}a m\'as all\'a del SM que consideraremos a lo largo de este trabajo. En este sentido, entender las caracter\'\i{}sticas del bos\'on de Higgs, siendo este el campo responsable de las masas de las dem\'as part\'\i{}culas a trav\'es del mecanismo de Brout-Englert-Higgs, que rompe espont\'aneamente la simetr\'\i{}a electrod\'ebil $SU(2)_L \times U(1)_Y$ en la simetr\'\i{}a electromagn\'etica $U(1)_{em}$, puede darnos pistas para entender el problema del sabor y para buscar se\~nales de nueva f\'\i{}sica a trav\'es de la fenomenolog\'\i{}a del bos\'on de Higgs y del sabor. La \'ultima pieza del `viejo' modelo puede convertirse as\'\i{} en la primera piedra del `nuevo'.

El descubrimiento del bos\'on de Higgs abre situaciones muy prometedoras al mismo tiempo que va cerrando los cap\'\i{}tulos del SM. Este movimiento en zigzag entre el pasado y el futuro puede verse en diferentes aspectos del descubrimiento: las interacciones entre el bos\'on de Higgs y el resto de part\'\i{}culas est\'an siendo medidas en este momento en el LHC, y por el momento son compatibles con los valores del SM  \cite{Aad:2012mea,TheATLAScollaboration:2013lia,ATLAS:2013qma,Aad:2013wqa,higgscouplingsatlas,Chatrchyan:2013lba,higgscouplingscms}. Sin embargo, en la pr\'oxima fase del LHC la precisi\'on en estas medidas ser\'a ampliamente mejorada  \cite{higgscouplingsfuture,higgscouplingsfuturecms,higgscouplingsfuturecms2,higgscouplingsfutureatlas} abriendo una puerta a la medida de efectos de f\'\i{}sica m\'as all\'a del SM en el valor de estos acoplamientos. Por otro lado, aunque la naturaleza fundamental versus compuesta del bos\'on de Higgs no ha sido todav\'\i{}a desentra\~nada por los experimentos, si la hip\'otesis del SM de ser una part\'\i{}cula escalar fundamental es finalmente confirmada, esto supondr\'a un nuevo hito. De esta forma se dar\'a un paso adelante en el entendimiento de teor\'\i{}as con escalares fundamentales en ellas, como diferentes modelos de f\'\i{}sica m\'as all\'a del SM, y en particular el que consideraremos en este trabajo. Como otro movimiento, la medida de la masa del bos\'on de Higgs cierra la casilla vac\'\i{}a de la tabla de propiedades de las part\'\i{}culas del SM, pero abre nuevas posibilidades dado que esta masa puede estar relacionada con la escala de la nueva f\'\i{}sica, o la masa de sus nuevas part\'\i{}culas, todav\'\i{}a desconocidas, como veremos a lo largo de este trabajo. De nuevo una medida de mayor precisi\'on, en este caso de la masa del bos\'on de Higgs, nos dar\'a informaci\'on sobre nueva f\'\i{}sica, ya que esta estar\'a tambi\'en relacionada con las caracter\'\i{}sticas de los nuevos modelos de f\'\i{}sica m\'as all\'a del SM.

Respecto a la nueva f\'\i{}sica, supersimetr\'\i{}a (SUSY) \cite{Golfand:1971iw,Volkov:1973ix,Wess:1974tw} ser\'a nuestra elecci\'on como nueva simetr\'\i{}a subyacente a la f\'\i{}sica m\'as all\'a del SM, y en particular nos centraremos en el Modelo Est\'andar Supersim\'etrico M\'\i{}nimo (MSSM) \cite{Haber:1984rc,Gunion:1984yn,Gunion:1986nh}. La principal idea en la base de los modelos supersim\'etricos es a\~nadir una nueva simetr\'\i{}a que relaciona bosones y fermiones, como aspectos parciales de un elemento m\'as fundamental en la construcci\'on del universo llamado supercampo. De nuevo se repite la idea que ha conducido gran parte de la historia de la f\'\i{}sica de encontrar constituyentes m\'as simples que dan lugar a las m\'ultiples entidades que nos rodean, y de nuevo la idea de un universo m\'as sim\'etrico bajo la apariencia del que vemos. De hecho, del teorema de Haag-Lopuszanski-Sohnius \cite{HLS} sabemos que supersimetr\'\i{}a es la \'unica extensi\'on no trivial del grupo de Poincar\'e de teor\'\i{}as cu\'anticas relativistas en 3+1 dimensiones. Esta sencilla idea de una simetr\'\i{}a entre fermiones y bosones se convierte en un poderoso motor cuyo desarrollo ha llevado a la construcci\'on de teor\'\i{}as supersim\'etricas en las que se resuelven diferentes problemas o aspectos poco atractivos del SM: supersimetr\'\i{}a incluye muchas caracter\'\i{}sticas atractivas como la unificaci\'on de las interacciones del SM, la posibilidad de resolver el problema de la materia oscura, la naturalidad de la ruptura de la simetr\'\i{}a electrod\'ebil o la conexi\'on de supersimetr\'\i{}a con una versi\'on gauge local de la gravedad o con interesantes teor\'\i{}as de altas energ\'\i{}as como teor\'\i{}a de cuerdas.

El MSSM es la versi\'on supersim\'etrica m\'\i{}nima del Modelo Est\'andar. Supone la introducci\'on de una nueva part\'\i{}cula supersim\'etrica compa\~nera para cada grado de libertad del SM, con los mismos n\'umeros cu\'anticos bajo el grupo gauge $SU(3)_C\times SU(2)_L\times U(1)_Y$, pero con un spin que difiere en 1/2. Por cada fermi\'on se introducen dos sfermiones, con spin 0: dos squarks, ${\stilde q_{L}}$ y ${\stilde q_{R}}$, para cada quark, y dos sleptones, ${\stilde l_{L}}$ y ${\stilde l_{R}}$, para cada lept\'on. Por cada bos\'on gauge con spin 1 se introduce un gaugino con spin 1/2: gluinos $\stilde g$, winos $\stilde W^\pm$ y binos $\stilde B^0$. El sector de Higgs del MSSM es diferente al del SM, con dos dobletes de Higgs en lugar de uno, y los correspondientes compa\~neros de spin 1/2, los Higgsinos. Esto producir\'a 5 bosones de Higgs f\'\i{}sicos: dos bosones neutros $h$ y $H$ con  ${\cal CP} = +1$ (siendo el primero el m\'as ligero), un bos\'on neutro pseudoescalar $A$ con  ${\cal CP} = -1$, y dos bosones cargados  $H^+$ y $H^-$. El bos\'on de Higgs observado en el LHC corresponder\'\i{}a, en la versi\'on m\'as plausible del MSSM, al bos\'on de Higgs neutro m\'as ligero del MSSM $h$, y esta ser\'a la hip\'otesis que consideremos a lo largo de este trabajo. En la actualidad todas las observaciones son compatibles con un bos\'on de Higgs de tipo SM, pero un bos\'on de Higgs supersim\'etrico podr\'\i{}a imitar el comportamiento del bos\'on del SM, as\'\i{} que se necesitar\'an medidas de alta precisi\'on de sus propiedades para concluir finalmente si la part\'\i{}cula observada es supersim\'etrica o no. Despu\'es de la ruptura de simetr\'\i{}a electrod\'ebil los Higgsinos se mezclan con los winos y binos, produciendo los charginos ${\tilde \chi_{1,2}}^\pm$ (con carga el\'ectrica), y los neutralinos  ${\tilde \chi_{1...4}}^0$ (sin carga el\'ectrica). El neutralino m\'as ligero es generalmente el candidato preferido como part\'\i{}cula de materia oscura \cite{Ellis:1983ew,Goldberg:1983nd}.

Un modelo con supersimetr\'ia exacta implicar\'\i{}a que las nuevas part\'\i{}culas supersim\'etricas tendr\'\i{}an las mismas masas que sus compa\~neras conocidas del SM, y dado que no han sido vistas todav\'\i{}a en los experimentos, un modelo consistente necesita implementar una ruptura suave de supersimetr\'\i{}a capaz de elevar el valor de las masas de estas spart\'\i{}culas por encima del alcance de los experimentos. Los l\'\i{}mites inferiores de exclusi\'on actuales provenientes de los datos del LHC respecto a las masas de las spart\'\i{}culas est\'an situados aproximadamente a un nivel de ${\cal O} \textrm{(1 TeV)}$ \cite{atlassusysearches,cmssusysearches}, siendo los m\'as restrictivos aquellos para las spart\'\i{}culas con interacci\'on fuerte: los gluinos y los squarks, que se predice que ser\'an producidos m\'as abundantemente que el resto de part\'\i{}culas supersim\'etricas. Una SUSY natural se esperaba por debajo o alrededor de la escala del TeV, siendo esta la soluci\'on del llamado problema de la jerarqu\'\i{}a. Este problema de la jerarqu\'\i{}a consiste en el hecho de que la masa del bos\'on de Higgs obtiene correcciones radiativas que crecen cuadr\'aticamente con la escala cut-off de la teor\'\i{}a de alta energ\'\i{}a, generando correcciones gigantes por ejemplo de 15 \'ordenes de magnitud m\'as grandes que la masa inicial a nivel arbol para un cut-off de la escala de Planck de  $10^{19}$ GeV. Por lo tanto los contrat\'erminos deben ser ajustados con una precisi\'on extrema para cancelar estas correcciones y producir la masa del bos\'on de Higgs observada. SUSY se propon\'\i{}a como soluci\'on al problema de la jerarqu\'\i{}a; sin embargo la capacidad de SUSY para resolver esta cuesti\'on se vuelve m\'as d\'ebil seg\'un SUSY se va volviendo m\'as pesada, y por lo tanto m\'as rota.

La ausencia de part\'\i{}culas supersim\'etricas en el LHC vuelve crucial el estudio de sus consecuencias fenomenol\'ogicas de otras formas diferentes a las b\'usquedas directas. Los efectos indirectos de SUSY en los observables de precisi\'on y en la f\'\i{}sica del sabor son formas \'unicas de encontrar se\~nales indirectas de SUSY. Estas b\'usquedas indirectas de nuevas part\'\i{}culas se han usado con anterioridad en el pasado para construir el propio SM y para comprobar la existencia de quarks pesados del SM antes de su descubrimiento. Por ejemplo, una b\'usqueda indirecta del quark top en LEP a trav\'es de observables de precisi\'on hizo posible constre\~nir el valor de su masa a un intervalo peque\~no \cite{Alexander:1991vi,lepewwg}, que fue luego utilizado en Tevatron para descubrirlo en una b\'usqueda directa  \cite{Abe:1995hr,Abachi:1995iq}.  Lo mismo ocurri\'o con el bos\'on de Higgs \cite{lepewwg,ALEPH:2010aa}. Los efectos de los quarks charm y bottom aparecieron indirectamente en la fenomenolog\'\i{}a de mesones en procesos con cambio de sabor antes de que estos quarks fueran descubiertos \cite{Aubert:1974js,Augustin:1974xw,Herb:1977ek}. La ventana que se abre ahora para probar la f\'\i{}sica m\'as all\'a del SM a trav\'es de procesos con cambio de sabor es incomparable debido a la gran supresi\'on de procesos con cambio de sabor en el SM. Por ejemplo, en el sector lept\'onico, incluso introduciendo masas y mezclas de neutrinos, las tasas de procesos de Violaci\'on de Sabor Lept\'onico (LFV) est\'an enormemente suprimidas porque est\'an determinadas por los min\'usculos acoplamientos de Yukawa de los leptones. Estos generan por ejemplo valores alrededor de $10^{-60}$ en el cociente de ramificaci\'on de las desintegraciones LFV del Higgs con neutrinos masivos de Dirac o alrededor de $10^{-35}-10^{-42}$ si las masas de los neutrinos est\'an generadas a trav\'es de un mecanismo seesaw I con masas pesadas de Majorana del orden de  $10^{3}-10^{6}$ GeV \cite{Arganda:2004bz}. En principio, en la f\'\i{}sica m\'as all\'a del SM los procesos con cambio de sabor podr\'\i{}an no estar tan suprimidos como en el SM, y por lo tanto ser detectables en los experimentos en estos espacios vac\'\i{}os dejados por el SM. En concreto, en los modelos SUSY las tasas predichas de procesos con cambio de sabor son en general de hecho demasiado grandes y est\'an en contradicci\'on con los experimentos. Por ello se proponen generalmente nuevas simetr\'\i{}as de sabor que incorporan la hip\'otesis de Violaci\'on de Sabor M\'\i{}nima (MFV)  \cite{D'Ambrosio:2002ex}, donde los acoplamientos de Yukawa son los \'unicos generadores de los procesos con cambio de sabor. Dado que el tama\~no de estos acoplamientos de Yukawa es peque\~no en general, las tasas de procesos neutros de cambio de sabor son tambi\'en muy peque\~nas. La \'unica excepci\'on es el acoplamiento de Yukawa del top que es en general el que domina en estas tasas.

La b\'usqueda de nueva f\'\i{}sica m\'as all\'a del SM a trav\'es de procesos con cambio de sabor ha sido ya hasta cierto punto exitosa en la f\'\i{}sica de part\'\i{}culas: el descubrimiento de las oscilaciones de neutrinos en 1998  \cite{Fukuda:1998mi}, se\~nalando cambios de sabor en el sector de los neutrinos y como consecuencia el hecho de que los neutrinos tienen masa, no estaba contemplado en el SM, y por lo tanto es una se\~nal de f\'\i{}sica m\'as all\'a del SM. El sector cargado de los leptones no ha mostrado todav\'\i{}a procesos con violaci\'on de sabor pero hay experimentos desarroll\'andose en este momento con esperanza de detectarlos \cite{webmeg,webbelle}. La violaci\'on de sabor en el sector de los quarks s\'\i{} que est\'a incorporada en el SM a trav\'es de la matriz CKM, y este fen\'omeno ha sido observado en diferentes observables. Es m\'as, los experimentos de muy alta precisi\'on como las factor\'\i{}as de B \cite{webbelle,webbabar} o el LHCb \cite{weblhcb}, los convierten en candidatos \'unicos para distinguir posibles efectos de nueva f\'\i{}sica. Incluso cuando esta nueva f\'\i{}sica no se detecte directamente, sus masas y par\'ametros quedar\'an m\'as restringidos por estos tipos de medidas indirectas. Las instalaciones futuras como las super factor\'\i{}as de B  \cite{webbelle2,Abe:2010sj,lhcbupgrade}, COMET \cite{lettercomet}, Mu2e \cite{webmu2e,Abrams:2012er} o PRISM \cite{Barlow:2011zza} mejorar\'an significativamente las medidas de los procesos con cambio de sabor, haciendo muy prometedor el futuro de las b\'usquedas de violaci\'on de sabor.

Nuestro objetivo en esta tesis es describir los procesos con cambio de sabor en teor\'\i{}as supersim\'etricas con un enfoque lo m\'as general que sea posible. Por ello utilizaremos una parametrizaci\'on gen\'erica de la mezcla de sabor en el sector sfermi\'onico a trav\'es de un conjunto de par\'ametros sin dimensi\'on $\deXYij$ (con $X,Y=L,R$ haciendo referencia a las quiralidades de los compa\~neros fermi\'onicos; e $i$ y $j$ siendo las generaciones implicadas en la mezcla) y estudiaremos sus implicaciones fenomenol\'ogicas. Por lo tanto no nos limitaremos a la hip\'otesis de MFV sino que estudiaremos las implicaciones fenomenol\'ogicas del caso m\'as general de Violaci\'on de Sabor No M\'\i{}nima (NMFV). A diferencia de otros trabajos en este tema, nuestro estudio no depender\'a de aproximaciones para introducir la mezcla de sabor sfermi\'onico, como la Aproximaci\'on de Inserci\'on de Masa (MIA)~\cite{Hall:1985dx}, sino que realizaremos una diagonalizaci\'on completa de las matrices de masa de los sfermiones con t\'erminos generales de mezcla de sabor. Este estudio nos permitir\'a entender en detalle las consecuencias fenomenol\'ogicas de la mezcla general de sabor sfermi\'onico incluso antes de descubrir los sfermiones en los experimentos.

A trav\'es de este trabajo estudiaremos diferentes observables con cambio de sabor y exploraremos sus predicciones en SUSY intentando entender sus comportamientos y consecuencias al variar los diferentes par\'ametros de mezcla sfermi\'onicos $\deXYij$, as\'\i{} como los par\'ametros del MSSM. Esto nos permitir\'a trazar un mapa del espacio de par\'ametros de mezcla de sabor sfermi\'onico de SUSY, conociendo qu\'e regiones est\'an permitidas y cu\'ales no, y tambi\'en nos permitir\'a encontrar las mejores ventanas experimentales a trav\'es de la cuales buscar efectos indirectos de SUSY.

Comenzaremos nuestro estudio de la mezcla de sfermiones centr\'andonos en diferentes observables con cambio de sabor en el sector fermi\'onico. En primer lugar en el sector de los quarks, estudiaremos el cociente de ramificaci\'on de la desintegraci\'on radiativa de $B$,  \bsg, el cociente de ramificaci\'on de la desintegraci\'on mu\'onica de $B_s$, \bmm, y la diferencia de masas  $B_s-{\bar B_s}$, \dmbs. En segundo lugar en el sector lept\'onico nos centraremos en desintegraciones como las desintegraciones radiativas $l_j \to l_i \gamma$,  las desintegraciones lept\'onicas $l_j \to 3 l_i$ y  las desintegraciones semilept\'onicas $\tau \to \mu \eta$ y $\tau \to e \eta$. Tambi\'en consideraremos las tasas de conversi\'on LFV de muones en electrones en n\'ucleos pesados. Una vez que esto est\'e hecho, y conociendo las \'areas permitidas para los par\'ametros de sabor de los sfermiones, esta tarea se completar\'a centr\'andonos en los observables del bos\'on de Higgs, con un estudio detallado de las correcciones de masa del bos\'on de Higgs inducidas por mezcla de sabor de squarks. Finalmente estudiaremos  las desintegraciones con violaci\'on de sabor lept\'onico del Higgs  $\phi \to \tau \mu$ y $\phi \to \tau e$ (donde $\phi=h,H,A$) inducidos por mezclas de sabor slept\'onico, dentro de las regiones permitidas por las b\'usquedas de LFV m\'as restrictivas en este momento.

Durante el estudio de  las desintegraciones del bos\'on de Higgs prestaremos especial atenci\'on al comportamiento de estos observables cuando la escala SUSY se vuelve muy pesada. La ausencia de efectos de SUSY en los experimentos puede ser entendida como consecuencia de una SUSY de una escala muy alta, donde generalmente los efectos de la nueva f\'\i{}sica desacoplan y son dif\'\i{}ciles de ser encontrados en las observaciones. Pero como demostraremos en esta tesis, algunos observables como  las desintegraciones LFV del Higgs tienen un comportamiento no desacoplante con SUSY pesada y pueden manifestarse en se\~nales indirectas de SUSY incluso si esta SUSY pesada no aparece directamente. Tener observables no desacoplantes puede ser vital a la hora de probar SUSY, as\'\i{} que parte de nuestra investigaci\'on se centrar\'a en este importante tema.

Como vemos, el descubrimiento del bos\'on de Higgs no se\~nala ning\'un punto final. Con el cierre del SM nos embarcamos en la b\'usqueda abierta de supersimetr\'\i{}a y las respuestas al problema del sabor, ya que nuestra sed de entendimiento de la realidad parece no estar satisfecha nunca. Ser\'a un nuevo viaje emocionante.

\bigskip
Esta tesis est\'a estructurada como sigue: el Cap\'\i{}tulo  \ref{supersymmetry} est\'a dedicado a resumir los aspectos m\'as relevantes de supersimetr\'\i{}a. Despu\'es se introduce el modelo con el que trabajaremos, el MSSM, detallando los diferentes sectores de spart\'\i{}culas. El Cap\'\i{}tulo  \ref{flavour} resume los aspectos principales del sabor.

En el Cap\'\i{}tulo \ref{paramnmfvscen} presentamos la parametrizaci\'on general de sabor sfermi\'onico que ser\'a utilizada a lo largo del resto del trabajo, para describir la fenomenolog\'\i{}a relevante de los procesos con cambio de sabor m\'as all\'a de la hip\'otesis de MFV. Se muestra c\'omo estos escenarios nos permitir\'an tratar con la mezcla de sabor en supersimetr\'\i{}a de una manera general. Despu\'es de esto, presentamos los diferentes conjuntos de escenarios con los que trabajaremos aqu\'\i{}, con diferentes tipos de hip\'otesis y libertad en la elecci\'on de par\'ametros. Estos escenarios van desde escenarios motivados desde alta energ\'\i{}a, definidos en el marco de supergravedad m\'\i{}nima, hasta escenarios de baja energ\'\i{}a m\'as enfocados en los actuales experimentos como los escenarios del MSSM fenomenol\'ogico.

Los siguientes cuatro cap\'\i{}tulos recogen el n\'ucleo de nuestra investigaci\'on y los resultados principales de esta tesis. El Cap\'\i{}tulo  \ref{sec:Bphysics} cubre el estudio de los efectos de mezcla de squarks en los observables de f\'\i{}sica de $B$ mencionados anteriormente. Se realiza un estudio num\'erico en estos observables de $B$, para obtener l\'\i{}mites en los par\'ametros de sabor de los squarks.

Despu\'es de conocer las \'areas permitidas en el espacio de mezcla de sabor de los squarks, entramos en el estudio de las correcciones a la masa del bos\'on de Higgs por mezcla de sabor de squarks en el Cap\'\i{}tulo  \ref{higgsmasssquark}. Primero realizamos un estudio anal\'\i{}tico de las correcciones de masa y derivamos las f\'ormulas a un loop necesarias. Despu\'es de esto, usamos esas f\'ormulas para estudiar los efectos en las masas de los bosones de Higgs de los par\'ametros permitidos de mezcla de sabor de los squarks.

En el Cap\'\i{}tulo \ref{phenoflavourslep} estudiamos las consecuencias de la mezcla de sfermiones en el sector slept\'onico. En este cap\'itulo estudiamos los procesos m\'as relevantes con cambio de sabor en el sector lept\'onico incluyendo desintegraciones radiativas $l_j \to l_i \gamma$, desintegraciones lept\'onicas $l_j \to 3 l_i$, las desintegraciones semilept\'onicas $\tau \to \mu \eta$ y $\tau \to e \eta$ y la conversi\'on $\mu - e$ en n\'ucleos, y extraemos de ellos los observables m\'as restrictivos que imponen los principales l\'\i{}mites a los par\'ametros de mezcla de sabor slept\'onico.

El Cap\'\i{}tulo \ref{lfvhiggsdecaysslepton} concluye el estudio de las consecuencias fenomenol\'ogicas de la mezcla de sleptones con un estudio de las desintegraciones LFV de bosones de Higgs. Estas desintegraciones se estudian compar\'andolas con los observables LFV m\'as restrictivos estudiados anteriormente, para encontrar las ventanas experimentales m\'as interesantes para probar indirectamente SUSY. Se presta especial atenci\'on al comportamiento desacoplante/no desacoplante con SUSY pesada de estos observables, que ser\'a de gran inter\'es en un futuro cercano ya que el no tener por el momento ninguna se\~nal de SUSY sit\'ua las part\'\i{}culas supersim\'etricas en la regi\'on pesada por encima de la escala del TeV.

Esta tesis se concluye con el Cap\'\i{}tulo \ref{conclusions}, donde se recogen las conclusiones generales derivadas de este trabajo.

Los resultados presentados en esta tesis, compendiados a lo largo de los Cap\'\i{}tulos \ref{paramnmfvscen}, \ref{sec:Bphysics}, \ref{higgsmasssquark}, \ref{phenoflavourslep} y \ref{lfvhiggsdecaysslepton}, las conclusiones y los ap\'endices, son trabajos originales que han sido publicados en los siguientes documentos: \cite{AranaCatania:2011ak}, \cite{AranaCatania:2012sn}, 
\cite{Arana-Catania:2013nha}, \cite{Arana-Catania:2013xma} y \cite{Arana-Catania:updatedsquark}.


\chapter{Introduction}
\label{introduction}

On the 4th of July of 2012 it was announced worldwide on a live event that the Higgs boson had been discovered  \cite{higgsdiscoveryannouncement,Aad:2012tfa,Chatrchyan:2012ufa}. In the few seconds that took the broadcast signal to travel around the world, the mankind took a huge leap forward in our understanding of the universe. It is not very common that the scientific advance crystallises in such a specific event, and of such magnitude, so the ones who lived that moment should treasure it. What happened that day? what kind of change was brought to the world? and after the leap, where are we now?

The discovery is a dialectical movement of opening and closure. The closure is produced with respect to the Standard Model (SM) of fundamental interactions, and brought to many physicists a feeling of relief after the announcement. The Higgs boson is the last missing piece of the SM, our current model of particle physics developed during more than 40 years \cite{Glashow:1961tr,GellMann:1964nj,Zweig:1964jf,Weinberg:1967tq,Salam:1968rm}. The SM is a renormalizable quantum field theory in 4 dimensions with a $SU(3)_C\times SU(2)_L\times U(1)_Y$ gauge symmetry, and invariant under Poincar\'e and CPT transformations. The Higgs boson is the particle that appears as a consequence of the Brout-Englert-Higgs mechanism \cite{Higgs:1964ia,Higgs:1964pj,Higgs:1966ev,Englert:1964et,Guralnik:1964eu}, needed in the SM to give mass to the elementary particles. At the same time the SM, extended with neutrino masses, has demonstrated experiment after experiment being able to match its predictions with every imaginable measurement ever done in the field of high energy particle physics, the Higgs boson kept hiding. The extraordinary success of the model was in conflict with the elusiveness of its last prediction. But finally the discovery of a Higgs boson has occurred in the CERN laboratory by the two experiments ATLAS~\cite{Aad:2012tfa} and CMS~\cite{Chatrchyan:2012ufa}. These experiments have measured a Higgs-like particle with spin 0, no electric or colour charges and positive parity (although the latter is still under investigation), as predicted by the SM, and with a mass between 125 and 126 GeV \cite{Aad:2013wqa,Chatrchyan:2013lba}  compatible with the constraints of the Landau pole and the vacuum stability of the SM up to very high energies, close to the Planck mass $M_P=10^{19}$GeV (the SM does not predict a specific value for the Higgs boson mass, but sets some upper and lower bounds; in particular within the pure SM we  seem to live in a long-lived metastable vacuum \cite{Buttazzo:2013uya,Alekhin:2012py}). With this discovery an important chapter in the history of particle physics is closed. Probably no other human construction could be compared now in complexity and strength to the SM. The level of agreement between the predictions of the model and the outcome of the experiments reaches for example an order of $10^{-12}$ in the measurement of the electron magnetic moment \cite{Hanneke:2008tm,Hanneke:2010au}. 

However, 
the SM is not a final theory in particle physics. And here comes the opening. There are many phenomena in the universe that the SM has is not able to explain, so it is clear that it is not a complete model of the physics of our universe, and therefore should be enlarged or introduced in a wider theory. For example, the SM does not include neither the mass nor the oscillations of the neutrinos, in spite that the experiments as the Super-Kamiokande proved both facts \cite{Fukuda:1998mi}. It also does not include a candidate for a dark matter particle. In fact, the Higgs boson is in direct connection with the main phenomenon that escapes to our control: the mass of the elementary particles. Apart from this, and from a more general point of view in physics, regarding the gravitational mass, another of the main opened questions would be how to merge the quantum and gravitational levels in a unified theory that describes all the forces in one common framework. This leads to the aim of quantizing gravity, if we watch the problem from the point of view of the quantum field theory; an issue which is obviously not solved within the SM since gravitation is the only interaction not included in the model. However, from the perspective that will be more relevant to our work here, where we stay at the quantum field theory level and the gravitational forces become negligible, the problem of the mass can be reformulated differently in what is called the flavour problem. 

The flavour problem, stated in a broad way, could be understood as the absence of a theory able to explain the role that the mass plays in our model of particle physics. This is translated in a set of observed features or phenomena that we are not able to explain. For example, we found that the elementary particles appear as `families' of particles, where each family is a copy of the previous one in the quantum numbers, but differs in the masses. The existence of these families, and also the number of families (3 in the SM, totally compatible with the current experimental bounds, see for instance \cite{limits4thgenerationfermions}) have no explanation within the SM. The pattern of masses of the particles seems also arbitrary with vast regions in the mass scale empty as the space between the masses of the neutrinos and the electron, ranging from approximately $10^{-9}$ GeV to $0.51\times 10^{-3}$ GeV, or the two orders of magnitude between the mass of the bottom quark of 4.18 GeV and the top quark mass of 173.2 GeV. The values of the elements of the mixing matrices, CKM \cite{Cabibbo:1963yz,Kobayashi:1973fv} and PMNS \cite{Pontecorvo:1957cp,Pontecorvo:1967fh,Maki:1962mu}, between families of fermions are not predicted either in the SM. These and other questions on flavour open a new window to phenomenology beyond SM that we will consider through this work. In that sense, the understanding of the Higgs features, being this field the responsible of the masses of the rest of the particles through the Brout-Englert-Higgs mechanism, which breaks spontaneously the electroweak symmetry $SU(2)_L \times U(1)_Y$ into the electromagnetic symmetry $U(1)_{em}$, could give us some clues to understand the flavour problem and also to look for new physics signals through the Higgs and flavour phenomenology. The last piece of the `old' model could turn into the first stone of the `new' one. 

The discovery of the Higgs boson opens very promising situations at the same time the chapters of the SM are being closed. This zigzag movement between the past and the future can be seen in different aspects of the discovery: The interactions between the Higgs boson and the rest of the particles are being currently measured at the LHC, and for the moment they are compatible with the SM values \cite{Aad:2012mea,TheATLAScollaboration:2013lia,ATLAS:2013qma,Aad:2013wqa,higgscouplingsatlas,Chatrchyan:2013lba,higgscouplingscms}. However, in the next phase of the LHC the precision in these measurements will be largely upgraded \cite{higgscouplingsfuture,higgscouplingsfuturecms,higgscouplingsfuturecms2,higgscouplingsfutureatlas} and a door will be opened to measure the effects of physics beyond the SM on the values of these couplings. In addition to this, although the fundamental versus composite nature of the Higgs boson has not been disentangled yet in the experiments, if the SM hypothesis of a fundamental scalar particle is finally confirmed, another milestone will be reached. Thus we will take one step forward in the understanding of theories that have fundamental scalars on them, as many models of physics beyond the SM, and in particular the one we will consider in this work. As another movement, the measurement of the mass of the Higgs boson closes the empty cell in the table of properties of the SM particles, but opens new possibilities since this mass could be related with the scale of the new physics, or the masses of its new particles, still unknowns, as we will see through this work. Again a higher precision measurement, in this case of the Higgs boson mass, will give us insights into new physics, since it will also be related with the characteristics of the new models of physics beyond the SM.

Respect to the new physics, Supersymmetry (SUSY) \cite{Golfand:1971iw,Volkov:1973ix,Wess:1974tw} will be our choice of the new symmetry underlying the physics beyond the SM, and in particular we will focus on the Minimal Supersymmetric Standard Model (MSSM) \cite{Haber:1984rc,Gunion:1984yn,Gunion:1986nh}. The main idea in the basis of the supersymmetric models is to add a new symmetry relating bosons and fermions, as partial aspects of a more general building block of the universe called superfield. Again it is repeated the leading idea that drove much of the history of physics of finding simpler constituents who lead to the multiple entities around us, and again the idea of a more symmetrical universe under the appearance we see. In fact, from the Haag-Lopuszanski-Sohnius Theorem \cite{HLS} we know that supersymmetry is the unique non-trivial extension of the Poincar\'e group of relativistic quantum field theories in 3+1 dimensions. This simple idea of a symmetry between fermions and bosons turns into a powerful motor whose development has led to the building of supersymmetric theories where many different problems or unappealing aspects of the SM are solved: supersymmetry includes many attractive features as the unification of SM interactions, the possibility of solving the dark matter problem, the naturalness of the breaking of the electroweak symmetry or the connection of supersymmetry with a local gauge version of gravity or with interesting high energy theories as string theory.

The MSSM is the minimal supersymmetric version of the Standard Model. It implies introducing a new supersymmetric partner particle for each degree of freedom of the SM, with the same quantum numbers under the gauge group $SU(3)_C\times SU(2)_L\times U(1)_Y$, but with a spin differing in 1/2. For each fermion there are introduced two sfermions, with spin 0: two squarks, ${\stilde q_{L}}$ and ${\stilde q_{R}}$, for each quark, and two sleptons, ${\stilde l_{L}}$ and ${\stilde l_{R}}$, for each lepton. For each gauge boson with spin 1 a gaugino with spin 1/2 is introduced: gluinos $\stilde g$, winos $\stilde W^\pm$ and binos $\stilde B^0$. The Higgs sector of the MSSM is different to the one of the SM, having two Higgs doublets instead of one, and the corresponding spin 1/2 partners, the Higgsinos. This will produce 5 physical Higgs bosons: two neutral bosons $h$ and $H$ with ${\cal CP} = +1$ (being the first one the lightest one), one neutral pseudoscalar boson $A$ with ${\cal CP} = -1$, and two charged bosons $H^+$ and $H^-$. The observed Higgs boson in the LHC could correspond, in the most plausible version of the MSSM, to the lightest neutral MSSM Higgs boson $h$, and this hypothesis will be considered through this work. At present all the observations are compatible with a SM-like Higgs boson, but a supersymmetric Higgs boson could mimic the behaviour of the SM one, so high precision measurements on its properties are needed to finally conclude if the observed Higgs particle is supersymmetric or not. After the electroweak symmetry breaking the Higgsinos mix with the winos and binos, producing the charginos ${\tilde \chi_{1,2}}^\pm$ (with electrical charge), and the neutralinos ${\tilde \chi_{1...4}}^0$ (without electrical charge). The lightest neutralino is usually the preferred supersymmetric candidate for dark matter particle \cite{Ellis:1983ew,Goldberg:1983nd}.

A model with exact supersymmetry model would imply that the new supersymmetric particles would have the same masses as their known SM partners, and since they have not been seen yet in the experiments, a consistent model needs to implement a soft supersymmetry breaking able to lift the value of the masses of these sparticles above the reach of the experiments. The present exclusion lower bounds from LHC data on the masses of the sparticles are roughly at the ${\cal O} \textrm{(1 TeV)}$  level \cite{atlassusysearches,cmssusysearches}, with the most restrictive ones being those for the strongly interacting sparticles: the gluinos and the squarks, which are predicted to be more abundantly produced than the rest of SUSY particles. A natural SUSY was expected to appear below or around the TeV scale, being SUSY a solution of the so-called hierarchy problem. This hierarchy problem consists in the fact that the Higgs boson mass has loop corrections that grow quadraticaly with the cut-off scale of the high energy theory, generating huge corrections for example of 15 orders of magnitude larger than the starting tree level mass for a cut-off at the Planck scale of $10^{19}$ GeV. Thus the counterterms should be extremely fine-tuned for cancelling these corrections and produce the observed Higgs mass. SUSY was claimed to solve this problem; however the ability of SUSY to solve this issue becomes weaker as SUSY gets more heavy, and thus the supersymmetry gets more broken. 

The absence of supersymmetric particles in the LHC makes crucial the study of its phenomenological consequences in other ways than direct searches. The indirect effects of SUSY in the precision observables and in the flavour physics are unique ways of finding indirect signs of SUSY. These indirect searches of new particles have already being used in the past for the building of the SM itself and to test the existence of the heavy SM quarks before their discovery. For instance, an indirect search of the top quark at LEP through precision observables made possible to constrain the value of its mass to a narrow gap  \cite{Alexander:1991vi,lepewwg}, that then was used at Tevatron to discover it in a direct search \cite{Abe:1995hr,Abachi:1995iq}. The same happened with the Higgs boson \cite{lepewwg,ALEPH:2010aa}. The effects of the charm and bottom quarks appeared indirectly in the meson phenomenology in flavour changing processes before these quarks were discovered \cite{Aubert:1974js,Augustin:1974xw,Herb:1977ek}. The window that is open now to probe the physics beyond the SM through flavour changing processes is incomparable due to the large suppression of flavour changing processes in the SM. For example, in the lepton sector, even introducing neutrino masses and mixing, the rates of Lepton Flavour Violating (LFV) processes are hugely suppressed because they are driven by the tiny lepton Yukawa couplings. This generates for example values around $10^{-60}$ in the branching ratio of the LFV Higgs decays with massive Dirac neutrinos or around $10^{-35}-10^{-42}$ if the neutrino masses are generated via a seesaw I mechanism with heavy Majorana masses of order $10^{3}-10^{6}$ GeV \cite{Arganda:2004bz}. In principle, in the physics beyond the SM the flavour changing processes could not be so suppressed as in the SM, and therefore be detectable in the experiments in these void spaces left by the SM. In particular, in SUSY models the predicted rates of flavour changing processes are in general actually too large and in contradiction with the experiments. Thus new flavour symmetries are usually proposed incorporating the Minimal Flavour Violation (MFV) hypothesis \cite{D'Ambrosio:2002ex}, where the Yukawa couplings are the unique generators of flavour changing processes. Since the size of these Yukawa couplings is small in general, the rates of the flavour changing neutral processes are also very tiny. The only exception being the top Yukawa coupling which is generally governing these rates.

The search of physics beyond the SM through flavour changing processes has been already to some extent successful in the history of particle physics: the discovery of neutrino oscillations in 1998 \cite{Fukuda:1998mi}, signalling flavour changings in the neutrino sector and therefore the fact that neutrinos have masses, were not contemplated in the SM, and therefore they are a sign of physics beyond the SM. The charged lepton sector has not shown yet flavour violating processes but there are experiments ongoing hoping to detect them \cite{webmeg,webbelle}. The flavour violation in the quark sector, is indeed incorporated in the SM via the CKM matrix, and this phenomenon has been observed in different observables. Furthermore, the very high precision experiments as the B-factories \cite{webbelle,webbabar} or the LHCb \cite{weblhcb}, make them unique candidates to distinguish possible new physics effects. Even when not detecting directly yet the new physics, its masses and parameters would get more constrained by these types of indirect measurements. The future facilities as the super-$B$ factories \cite{webbelle2,Abe:2010sj,lhcbupgrade}, COMET \cite{lettercomet}, Mu2e \cite{webmu2e,Abrams:2012er} or PRISM \cite{Barlow:2011zza} will improve significantly the measurements of flavour changing processes, making the future of flavour violating searches very promising.

Our goal in this thesis work is to describe the flavour changing processes in supersymmetric theories with an approach as general as possible. Therefore we will use a general parametrization of flavour mixing in the sfermion sector through a set of dimensionless parameters $\deXYij$ (with $X,Y=L,R$ referring to the chiralities of the fermionic partners; and $i$ and $j$ being the generations involved in the mixing) and study its phenomenological implications. We will therefore not stick here to the MFV hypothesis but study the phenomenological implications of the most generic Non Minimal Flavour Violation (NMFV) case. Differently to other works on this subject, our study will not rely on approximations to introduce the sfermion flavour mixing as the Mass Insertion Approximation (MIA)~\cite{Hall:1985dx}, but we will perform a full diagonalization of the sfermion mass matrices with general flavour mixing terms. This study will allow us to understand the phenomenological consequences of general sfermion flavour mixing in detail even before discovering the sfermions at the experiments.

Through this work we will study different flavour changing observables and explore their predictions in SUSY trying to understand their behaviours and consequences when varying the different $\deXYij$ sfermion mixing parameters as well as the MSSM parameters. This will allow us to build a map on the sfermion flavour mixing parameter space of SUSY, knowing which regions are allowed and which ones are not, and also to find the best experimental windows where to look for indirect effects of SUSY. 

We will begin our study of sfermion mixing by focusing on different flavour changing observables on the fermionic sector. First on the quark sector, we will study the branching ratio of the radiative $B$ decay \bsg, the branching ratio of the $B_s$ muonic decay \bmm, and the $B_s-{\bar B_s}$ mass difference \dmbs. Second on the lepton sector we will focus on decays like the radiative $l_j \to l_i \gamma$ decays, the leptonic $l_j \to 3 l_i$ decays and the semileptonic $\tau \to \mu \eta$ and $\tau \to e \eta$ decays. Also we will consider the LFV conversion rate of muon to electron in heavy nuclei. Once this is done, and knowing the allowed areas for the flavour parameters of the sfermions, this task will be completed focusing on the Higgs boson observables, with a detailed study on the Higgs boson mass corrections induced from squark flavour mixing. Finally we will study the lepton flavour violating Higgs decays $\phi \to \tau \mu$ and $\phi \to \tau e$ (where $\phi=h,H,A$) induced from slepton flavour mixing, within the allowed regions by the presently most restrictive LFV searches.

During the study of the Higgs boson decays we will pay special attention to the behaviour of these observables when the SUSY scale becomes very heavy. The absence of SUSY effects in the experiments could be understood as having a high scale SUSY, where usually the effects of the new physics decouple and are hard to be found in the observations. But as we will demonstrate in this thesis, some observables like the LFV Higgs decays have a non-decoupling behaviour with heavy SUSY and could be able to manifest in indirect SUSY signals even if this heavy SUSY is not directly produced. Having non-decoupling observables could be vital to probe SUSY, so part of our research will be focused on this important issue.

As we see, the discovery of the Higgs boson marked no end point. With the closure of the SM we get on board in the open quest for supersymmetry and the answers to the flavour problem, for our thirst of understanding the reality seems never satisfied. It will be an exciting new trip.

\bigskip
This thesis is structured as follows: Chapter \ref{supersymmetry} is devoted to summarise the most relevant aspects of supersymmetry. Then it is introduced the model we will work with, the MSSM, detailing the different sectors of particles. Chapter \ref{flavour} summarises the main aspects of flavour.

In Chapter \ref{paramnmfvscen} we present the general sfermion flavour parametrization that will be used along the rest of the work, to describe the relevant phenomenology of flavour changing processes beyond the MFV hypothesis. It is shown how these scenarios will allow us to deal with the flavour mixing issue in supersymmetry in a general way. After it, we present the different sets of scenarios that we will work with here, with different kinds of hypothesis and freedom in the election of parameters. The scenarios go from high energy motivated scenarios, defined in the minimal supergravity framework, to purely low energy scenarios more focused on the current experiments as the phenomenological MSSM scenarios.

The following four chapters collect the core of our investigation and the main results of this thesis. Chapter \ref{sec:Bphysics} covers the study of the squark mixing effects on the $B$ physics observables mentioned above. A numerical study of these $B$ observables is performed on them, to obtain bounds on the squark flavour parameters.

After knowing the allowed areas in the squark flavour mixing space, we enter into the study of the Higgs boson mass corrections from squark flavour mixing in Chapter \ref{higgsmasssquark}. First we perform an analytical study on the mass corrections and derive the needed one-loop formulas. After it, we use these formulas to study the effects on the masses of the Higgs bosons from the allowed squark flavour mixing parameters.

In Chapter \ref{phenoflavourslep} we study the consequences on sfermion mixing in the slepton sector. In this chapter we study the most relevant flavour changing processes in the leptonic sector including LFV radiative lepton decays $l_j \to l_i \gamma$, leptonic decays $l_j \to 3 l_i$, some semileptonic decays $\tau \to \mu \eta$ and $\tau \to e \eta$ and $\mu - e$ conversion in nuclei, and extract from them the most constraining LFV observables that set the main bounds on the slepton flavour mixing parameters.

Chapter \ref{lfvhiggsdecaysslepton} conclude the study of the phenomenological consequences from slepton mixing by a research on the LFV Higgs decays. These decays are studied comparing them with the most restrictive LFV observables studied before, to find the most interesting experimental windows to probe indirectly SUSY. A special attention is paid to the decoupling/non-decoupling behaviour with heavy SUSY of these observables, that will be of great interest at the near future since having no experimental sign of SUSY so far is driving the SUSY particles to the heavy energy region beyond the TeV scale.

This thesis come to an end with Chapter \ref{conclusions}, where the general conclusions derived from this work are collected.

The results presented in this thesis, summarized along the Chapters \ref{paramnmfvscen}, \ref{sec:Bphysics}, \ref{higgsmasssquark}, \ref{phenoflavourslep} and \ref{lfvhiggsdecaysslepton}, the conclusions and the appendices, are original works that have been published in the following papers: \cite{AranaCatania:2011ak}, \cite{AranaCatania:2012sn}, 
\cite{Arana-Catania:2013nha}, \cite{Arana-Catania:2013xma} and \cite{Arana-Catania:updatedsquark}.

\chapter{Supersymmetry and the MSSM}
\label{supersymmetry}

The SM is not a completely satisfactory model of particle physics, as we explained in the introduction. Supersymmetry, and in particular the Minimal Supersymmetric Standard Model is one of the possible next steps beyond the SM that has been extensively explored by the physics community during the last decades, and is the one we will explore in this work. 

\section{Motivations for SUSY}
\label{motiv-susy}

Usually supersymmetry is introduced as a solution to some problems of the SM, being the most common one the hierarchy problem. As we will see next this is not a real problem, but more a kind of ugliness of our theory, and thus a poor justification of the theory from the utilitarian point of view. However it touches the main concept that, from our point of view, drove so much attention over this theory, being it the beauty. The SM is constructed using as building blocks two objects with very different behaviours, fermions and bosons, in a not very satisfactory aesthetic way. Supersymmetry relates both objects as parts of a more abstract and only entity called the superfield, and thus takes us one step forward in our quest for simplification. All the equations are rewritten in a much more elegant way in terms of these superfields, by imposing a new symmetry. And since our current physics models are defined through the symmetries of the Universe, from which we derive our conservation laws, forces and matter fields, an extra symmetry is not perceived as an ad hoc ingredient, but as a missing part of the recipe. 

The generators of the supersymmetry transformations act upon fermions and bosons transforming ones into the others:

\beq
\label{susytransfs}
Q |{\rm Boson}\rangle = |{\rm Fermion }\rangle, \qquad\qquad
Q |{\rm Fermion}\rangle = |{\rm Boson }\rangle .
\eeq

From a mathematical point of view, it is a very beautiful symmetry, since from the Haag-Lopuszanski-Sohnius Theorem \cite{HLS} (being it an extension of the Coleman-Mandula Theorem \cite{ColemanMandula}) we know that supersymmetry is the unique non-trivial extension of the Poincar\'e group of relativistic quantum field theories in 3+1 dimensions. It is the only extension that relates the internal symmetries of the theory with the space-time symmetry. The generators of supersymmetry $Q$ and $Q^\dagger$ obey the following algebra:

\beq
&&\{ Q, Q^\dagger \} = -2\sigma^\mu P_\mu , \label{susyalgone}
\\
&&\{ Q,Q \} = \{ Q^\dagger , Q^\dagger \} = 0 , \label{susyalgtwo}
\\
&&[ P^\mu , Q  ] = [P^\mu, Q^\dagger ] = 0 ,\label{susyalgthree}
\eeq
where $P_\mu$ is the four-momentum (and generator of space-time translations), $\sigma^\mu$ are $2\times 2$ Pauli matrices as defined in Eq. 2.7 of \cite{Martin:1997ns} and the spinor indices are not shown. 

The superfields are the irreducible representations of the supersymmetry algebra, where each superfield contains bosons and fermions. Since the supersymmetry generators commute with $P^2$ both elements of the superfield should have the same $P^2$ eigenvalue, that is the same mass. They also commute with the generators of the gauge groups, so the fields in the superfield should have the same charges with respect to these groups. Therefore a very interesting consequence of the new symmetry arises: the spectrum of particles should be doubled, introducing a new set of particles with the same mass and quantum numbers as the ones we know from the SM, but with a different spin. These particles are called sparticles (for supersymmetric-particles). The full spectrum of particles for the minimal model that can be constructed, named Minimal Supersymmetric Standard Model, can be found in tables \ref{tab:chiral} and \ref{tab:gauge}. In fact it can be seen that the spectrum is a bit more than doubled, since the MSSM needs two Higgs doublets for the Brout-Englert-Higgs mechanism. We will study this in more detail in the following sections.

\renewcommand{\arraystretch}{1.4}
\begin{table}[tb]
\begin{center}
\begin{tabular}{|c|c|c|c|c|}
\hline
Particles & Superfields & spin 0 & spin 1/2 & $(SU(3)_C ,\, SU(2)_L ,\, U(1)_Y)$
\\  
\hline
squarks, quarks & $\hat Q$ & $({\stilde u}_L\>\>\>{\stilde d}_L )$&
 $(u_L\>\>\>d_L)$ & $(\>{\bf 3},\>{\bf 2}\>,\>{1\over 6})$
\\
($u=u,c,t$ ; $d=d,s,b$) & $\hat U$
&${\stilde u}^*_R$ & $u^\dagger_R$ & 
$(\>{\bf \overline 3},\> {\bf 1},\> -{2\over 3})$
\\ & $\hat D$ &${\stilde d}^*_R$ & $d^\dagger_R$ & 
$(\>{\bf \overline 3},\> {\bf 1},\> {1\over 3})$
\\  \hline
sleptons, leptons & $\hat L$ &$({\stilde \nu}\>\>{\stilde e}_L )$&
 $(\nu\>\>\>e_L)$ & $(\>{\bf 1},\>{\bf 2}\>,\>-{1\over 2})$
\\
($e=e,\mu,\tau$ ; $\nu=\nu_{e},\nu_{\mu},\nu_{\tau}$) & $\hat E$
&${\stilde e}^*_R$ & $e^\dagger_R$ & $(\>{\bf 1},\> {\bf 1},\>1)$
\\  \hline
Higgs, higgsinos &$\hat H_2$ &$({\cal H}_2^+\>\>\>{\cal H}_2^0 )$&
$(\stilde {\cal H}_2^+ \>\>\> \stilde {\cal H}_2^0)$& 
$(\>{\bf 1},\>{\bf 2}\>,\>+{1\over 2})$
\\ &$\hat H_1$ & $({\cal H}_1^0 \>\>\> {\cal H}_1^-)$ & $(\stilde {\cal H}_1^0 \>\>\> \stilde {\cal H}_1^-)$& 
$(\>{\bf 1},\>{\bf 2}\>,\>-{1\over 2})$
\\  \hline
\end{tabular}
\caption{Chiral supermultiplets in the MSSM, their field content, and their representations in the gauge groups.\label{tab:chiral}}
\vspace{-0.6cm}
\end{center}
\end{table}

\renewcommand{\arraystretch}{1.55}
\begin{table}[t]
\begin{center}
\vspace{1 cm}
\begin{tabular}{|c|c|c|c|c|}
\hline
Particles & Superfields & spin 1/2 & spin 1 & $(SU(3)_C, \> SU(2)_L,\> U(1)_Y)$\\
\hline
gluino, gluon & $\hat G^a$ &$ \stilde g$& $g$ & $(\>{\bf 8},\>{\bf 1}\>,\> 0)$
\\
\hline
winos, W bosons & $\hat W^i$ & $ \stilde W^\pm\>\>\> \stilde W^0 $&
 $W^\pm\>\>\> W^0$ & $(\>{\bf 1},\>{\bf 3}\>,\> 0)$
\\
\hline
bino, B boson & $\hat B$ & $\stilde B^0$&
 $B^0$ & $(\>{\bf 1},\>{\bf 1}\>,\> 0)$
\\
\hline
\end{tabular}
\caption{Gauge supermultiplets in
the MSSM, their field content, and their representations in the gauge groups.\label{tab:gauge}}
\vspace{-0.45cm}
\end{center}
\end{table}

\begin{figure}[ht!]
\begin{center}
\begin{tabular}{cc}
\includegraphics[width=80mm]{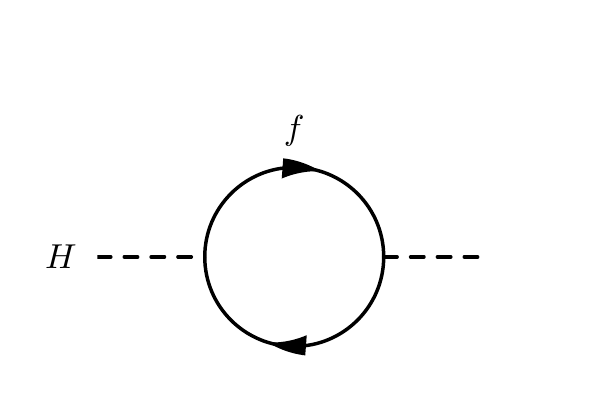} &
\includegraphics[width=80mm]{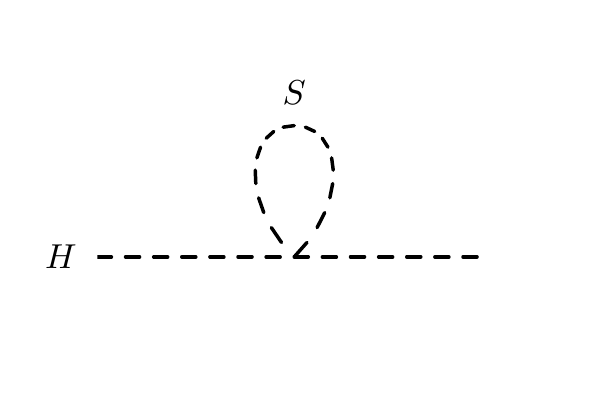} \\
\end{tabular}
\caption{Loop corrections to Higgs boson mass by a fermion (left) and a scalar particle (right).}
\label{higgs-loops-fermions-bosons}
\end{center}
\end{figure}

\bigskip
Following our search for beauty, let us remind next on how supersymmetry is able to solve the so-called hierarchy problem. Since the SM is not a complete model for particle physics, it should be understood as an effective theory, valid just in some energy domain. In particular, the energy of the processes involved when using the effective theory should be below a value called the cut-off of the theory. The so-called hierarchy problem arises when one computes the loop corrections of fermions and bosons to the Higgs boson mass taking in account this cut-off, that we will name $\Lambda_{\rm UV}$. From the one loop calculation of the diagrams of Figure \ref{higgs-loops-fermions-bosons} one obtains the following value:

\beq
\Delta m_H^2 \>=\>  
-{|\lambda_f|^2\over 8 \pi^2} \Lambda_{\rm UV}^2 
+{\lambda_S\over 16 \pi^2}
\left [\Lambda_{\rm UV}^2 - 2 m_S^2
\> {\rm ln}(\Lambda_{\rm UV}/m_S)
\right ]  + \ldots .
\label{quaddivsum}
\eeq
where $\lambda_f$ and $\lambda_S$ are the couplings of the Higgs boson to the fermion and scalar respectively, $m_S$ is the mass of the scalar and $\Lambda_{\rm UV}$ is the ultraviolet cut-off.

The above logarithmic corrections appear also in other observables, but here we obtain as a novelty quadratic corrections with the cut-off scale. In the higher-order corrections to the masses of other non-scalar particles it is common to find also logarithmic corrections, but multiplied by the mass squared of the particle involved instead of $\Lambda_{\rm UV}^2$. This happens because the mass terms in the Lagrangian for these other particles are not allowed by some symmetry (e.g. chiral and gauge, for electrons and photons) and then the corrections should also vanish when the mass is set to zero and the symmetry holds (what is called corrections ``protected'' by a symmetry). For the Higgs boson, if we set for instance the Planck mass as the value of the cut-off, where the gravitational effects are important and the SM is no longer valid, we obtain corrections 30 orders of magnitude larger than the value of the Higgs boson mass, around 125 GeV. This could be understood as not being a problem, since we could think in other corrections that could appear at this high energy scale and cancel these large corrections, but this would imply that Nature chose different numbers for the corrections that are exactly equal for the first 30 significant figures and then differ from each other, so their difference gives us this small Higgs boson mass (and different appropriate cancellations happen for each order of perturbation theory). This is what is usually called a fine-tuning problem, in this case of the radiative Higgs boson mass corrections, and although it is not a real mathematical problem, it leaves us with the suspicion that some other mechanism should be responsible of evading these quadratic corrections. 

Here is where supersymmetry plays its role. The fermions and the bosons are just part of common supermultiplets that we can use to construct our Lagrangian, and therefore all the fermionic and bosonic terms of the Lagrangian, and thus the couplings, are related. In particular, the couplings are related through $\lambda_S = |\lambda_f|^2$. If we apply this relation to the previous corrections, considering our new doubled supersymmetric spectrum we see how the quadratic divergences are cancelled exactly. And this happens to all orders of perturbation theory, therefore no fine-tuning is required.

\bigskip
Another motivation for SUSY is the unification of interactions. If we look for simplification as a mean of better understanding our universe, supersymmetry is an answer not only for the relation of fermions and bosons, but for the forces between themselves. In Quantum Field Theory the couplings that set the strength of the interactions are not fixed values but vary as we change the energy of the experiments. If we calculate the running of the SM gauge couplings $g_i (i=1,2,3)$ we obtain the behaviour of $\alpha_i^{-1}$, with $\alpha_i=g^2_i/(4\pi)$, with respect to $\mu$, the scale of the renormalization, given in figure \ref{coupling-unification-sm}:

\begin{figure}[ht!]
\begin{center}
\includegraphics[width=100mm]{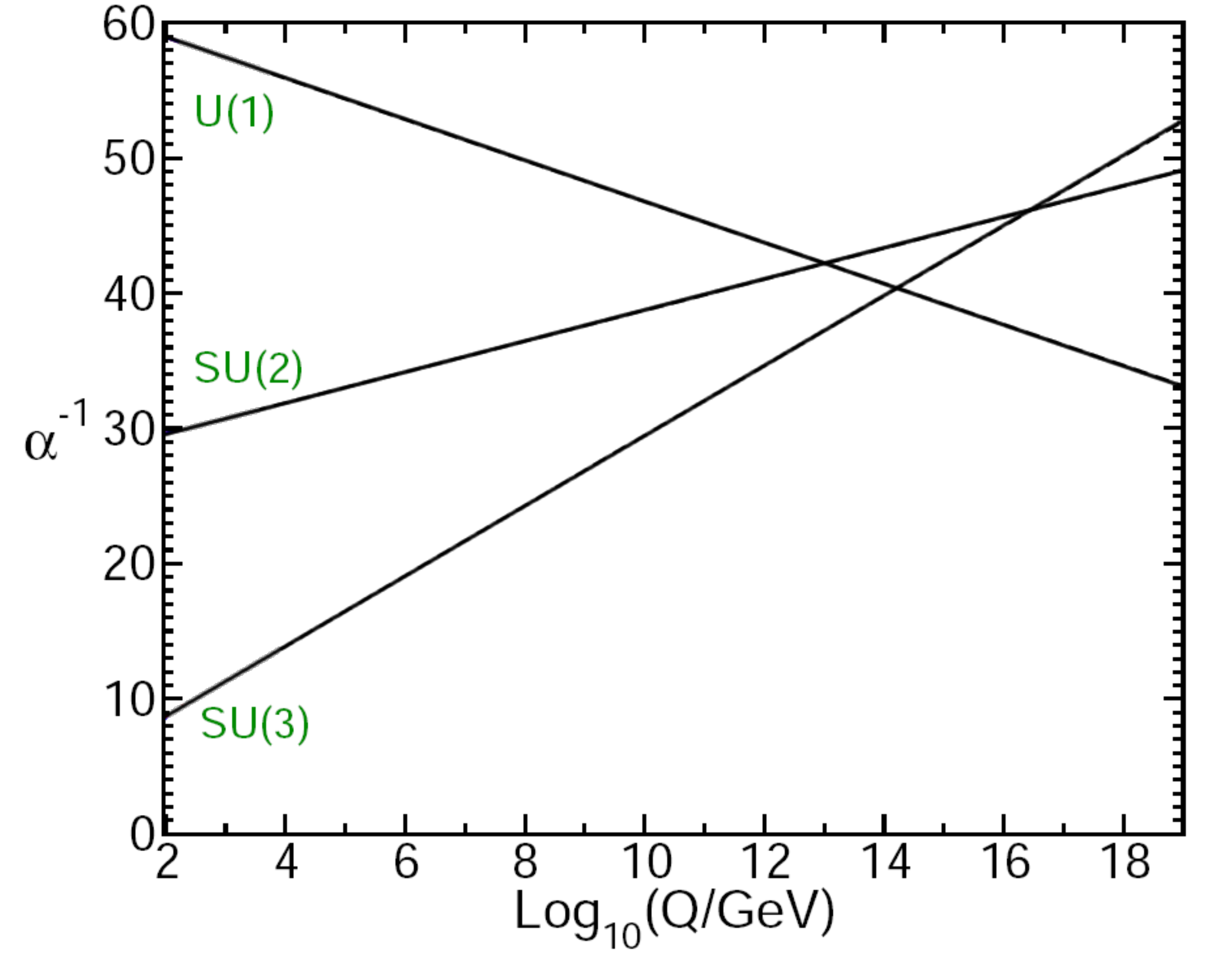}
\caption{Two-loop renormalization group evolution of the inverse of the gauge couplings $\alpha_i (i=1,2,3)$ in the SM, as shown in \cite{Martin:1997ns}.}
\label{coupling-unification-sm}
\end{center}
\end{figure}

\begin{figure}[ht!]
\begin{center}
\includegraphics[width=100mm]{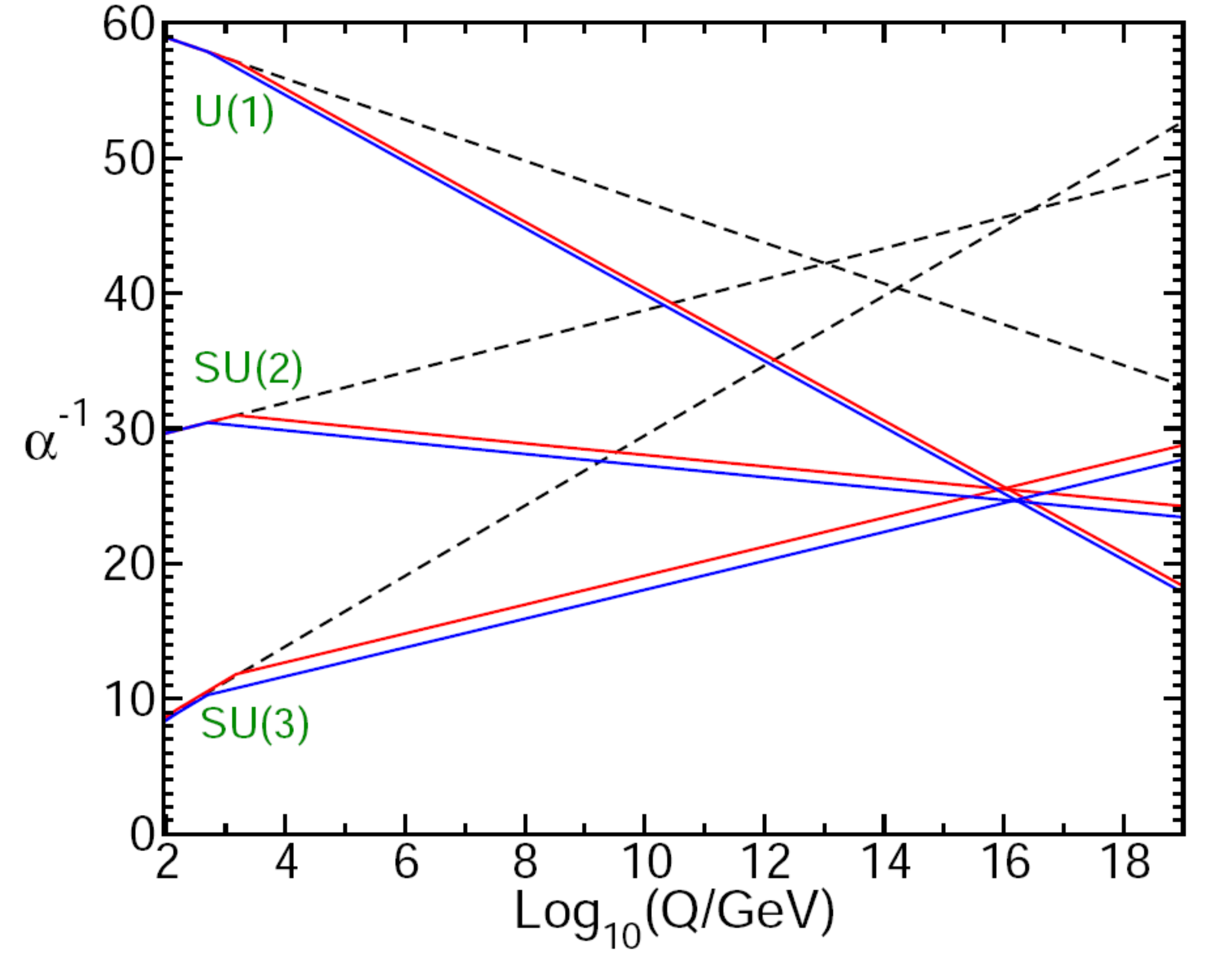}
\caption{Two-loop renormalization group evolution of the inverse of the gauge couplings $\alpha_i (i=1,2,3)$ in the MSSM (solid lines) and in the SM (dashed lines), as shown in \cite{Martin:1997ns}. For the MSSM the sparticles masses are varied between 500 GeV and 1.5 TeV and $\alpha_3(M_Z)$ is varied between 0.117 and 0.121.}
\label{coupling-unification-mssm}
\end{center}
\end{figure}

We see three lines that almost cross in a small region of the plot, and again suggest that could be something new beyond the SM that make them exactly cross, therefore giving rise to the unification of interactions. And again the beauty is recovered as we introduce supersymmetry, obtaining the result of Fig. \ref{coupling-unification-mssm}, where we see how now the three lines bend around $m_{SUSY}\sim{\cal O} \textrm{(}1 \textrm{ TeV)}$ and then cross at some high scale $m_{GUT}\sim{\cal O} \textrm{(}10^{16} \textrm{ TeV)}$, and therefore the unification of the three forces related to the gauge groups of the SM is produced.

Another appealing aspect of supersymmetry is that it is a needed ingredient for String theory, being this one of the few theories that seem good candidates to solve the unification between the quantum and the relativistic world including gravity. The introduction of supersymmetry in String theory led to a series of discoveries \cite{superstring} in the eighties called the First Superstring Revolution, that draw a lot of attention to this theory pointing it as a possible correct approach to a final unification of all the interactions in physics under a unique framework.

\bigskip
Another interesting feature of supersymmetry is related to the dark matter problem and its connection to the conserved quantum numbers. In the SM, the baryon ($B$) and lepton ($L$) numbers are accidental conserved quantities, due to the fact that is not possible to introduce renormalizable terms in the Lagrangian able to break them. To do it we should introduce non-renormalizable 5 and 6 dimension operators as $LLHH/\Lambda_L$ or $QQQL/\Lambda_B^2$, as shown in \cite{Weinberg:1979sa}. Operators like these could be introduced however in a non-perturbative way by effects like instantons \cite{Espinosa:1989qn}, but there would be highly suppressed. These are good news for us, since not conserving these numbers would have consequences as the proton decay (see \cite{Nath:2006ut} for a review). Luckily the proton is known to have a half-life larger than $2.1\times10^{29}$ years \cite{Ahmed:2003sy}. Other measurements set also important constraints to the violation of any of these numbers (see \cite{rparityconstraints,RPVreviews,Barbier:2004ez}). However in supersymmetry is possible to introduce terms in the Lagrangian that would violate the conservation of these numbers. These terms, that will be introduced later in Eqs. \ref{WLviol} and \ref{WBviol}, imply proton decay for example through the following diagram:

\begin{figure}[ht!]
\begin{center}
\includegraphics[width=80mm, trim=0 5mm 0 0]{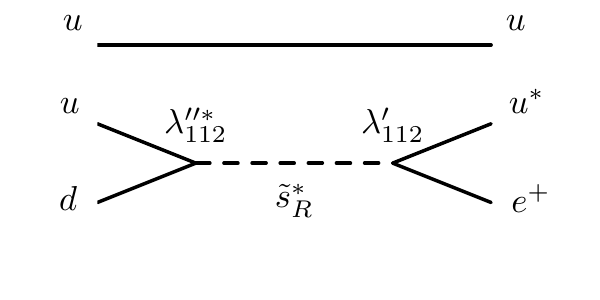}
\caption{Proton decay to a pion and a positron, mediated by a squark. Diagram produced through terms in the supersymmetric Lagrangian violating lepton and baryon number.}
\label{proton-decay-diag}
\end{center}
\end{figure}
leading to a partial decay width of:

\beq
\Gamma_{p \rightarrow e^+ \pi^0} \>\sim\> m_{{\rm proton}}^5 \sum_{i=2,3}  
|\lambda^{\prime 11i}\lambda^{\prime\prime 11i}|^2/m_{\stilde d_i}^4,
\eeq
that means a decay of seconds for couplings $\lambda$ of order 1 and squarks masses around the TeV. The solution to this unobserved behaviour of the proton could be to ask for a new symmetry as the R-parity \cite{Rparity}. This is defined as the conservation of the following number: 

\beq
P_R = (-1)^{3(\Baryon-\Lepton) + 2 s}
\label{defRparity}
\eeq
where $s$ is the spin of each particle.

With this new symmetry, all particles would have a defined R-parity ($+1$ for the known particles and the particles of the two Higgs doublet, and $-1$ for the supersymmetric partners), and each term of the Lagrangian should have a positive R-parity, and hence no process will change its value between the initial and final state. The consequence of this would be the absence of the worrying terms in the Lagrangian. Moreover, any interaction between particles and sparticles would always involve an even number of sparticles. This means that any decay of any sparticle would have at least one sparticle on the final state, and therefore the lightest supersymmetric particle (LSP) would be stable, being unable of decaying in other particles/sparticles. If this sparticle is neutral it could be the responsible for the unknown dark matter of the Universe, one of the main problems of the cosmology. The existence of charged LSP has very strict bounds, since the ones produced in the Big Bang would have produced bound states with some nuclei, leading to exotic isotopes whose abundance is very constrained (see e.g. \cite{Wolfram:1978gp}). The favourite candidate to dark matter within the MSSM is the lightest neutralino  \cite{Ellis:1983ew,Goldberg:1983nd} which as we will see next is a mixture between the SUSY partner of the photon and the SUSY partner of the Higgs bosons. The fact that SUSY provides good candidates for dark matter is a very convincing argument in favour of SUSY.

\section{The Minimal Supersymmetric Standard Model}
\label{sec:mssm}
The Minimal Supersymmetric Standard Model is the simplest model that can incorporate supersymmetry and the Standard Model. 
The particle content of the model is shown in tables~\ref{tab:chiral} and~\ref{tab:gauge}, where we see how it doubles the content of the SM (and also modifies the Higgs sector as explained below). In addition to the known SM particles it contains their supersymmetric partners, denoted with a tilde. The SUSY partners of the fermions are the sfermions (that are scalar bosons), and the partners of the gauge bosons are the gauginos (that are fermions). Specifically, the partners of the quarks $q$ and leptons $l$ are the squarks $\stilde q$ and sleptons $\stilde l$ respectively, that are spin 0 particles. The sub-indices $L$ and $R$ in the sparticles refer to the chiralities of the corresponding fermionic particles. The squarks and sleptons are scalars, and thus have no chirality. The SUSY partners of the SM gauge bosons are the gluinos $\stilde g$, winos $\stilde W$ and binos $\stilde B$, all of them spin 1/2 particles. The last two combine after electroweak symmetry breaking to form the photinos. The Higgs sector is a bit more complex than the one of the SM. It is composed of two $SU(2)_L$ Higgs doublets $H_{1,2}$ instead of one, and their SUSY partners, the Higgsinos $\stilde H_{1,2}$, have spin 1/2. We will explain them in detail in its corresponding section.
 
The gauge symmetries and a superpotential function should be defined to set the interactions of the model. The gauge groups are the ones of the SM, $SU(3)_C \times SU(2)_L \times U(1)_Y$, and the superpotential is:
\beq
W_{\rm MSSM} =
\hat U Y^u \hat Q \hat H_2 -
\hat D Y^d \hat Q \hat H_1 -
\hat E Y^e \hat L \hat H_1 +
\mu \hat H_1 \hat H_2 \> .
\label{MSSMsuperpot}
\eeq
where the fields are the scalars of each specific superfield appearing in table \ref{tab:chiral} (usually this superfield notation is used when specifying the superpotential). The flavour and gauge indices are not shown, a proper expansion of them would mean terms like  $\hat U^{ia}\, {(Y^u)_i}^j\, \hat Q_{j\alpha a}\, (\hat H_2)_\beta \epsilon^{\alpha \beta}$ and $\mu (\hat H_1)_\alpha (\hat H_2)_\beta \epsilon^{\alpha\beta}$. The Yukawa couplings $Y^{u,d,e}$ are $3\times3$ matrices in flavour space.

Other terms, changing in one unit the lepton and baryon number respectively, could be added to the superpotential compatible with the gauge groups and renormalizability as the following:

\beq
W_{\Delta {\rm L} =1} &=&
{1\over 2} \lambda^{ijk} \hat L_i \hat L_j \hat E_k
+ \lambda^{\prime ijk} \hat L_i \hat Q_j \hat D_k
+ \mu^{\prime i} \hat L_i \hat H_2 ,
\label{WLviol} \\
W_{\Delta {\rm B}= 1} &=& {1\over 2} \lambda^{\prime\prime ijk}
\hat U_i \hat D_j \hat D_k ,
\label{WBviol}
\eeq
where the $\lambda$ terms could be matrices in flavour space, but they would violate lepton or baryon number respectively, triggering processes as the proton decay we commented at the end of Section \ref{supersymmetry}. We will work here with a conserving R-parity MSSM so these terms will be absent (for R-parity violating SUSY see for instance \cite{Barbier:2004ez} and \cite{Hirsch:2004he}, for bounds on these terms see also \cite{Cheng:2013qpa,Dreiner:2012mx,atlassusysearches,cmssusysearches}). 

In the superpotential~\ref{MSSMsuperpot} we can recognise the Yukawa terms and a bilinear Higgs boson term, with a slight difference respect to the SM ones since here it is needed two different Higgs fields to give masses to the up and down-type particles. Therefore, the original SM spectrum will be a bit more than doubled. This requirement of having two fields comes from the fact that the superpotential must be holomorphic in the scalar fields (i.e. it should be a complex analytic function on them), and not in the complex conjugated fields. Thus, there are no terms like $\hat U Y^u \hat Q \hat H_1^*$ as we would naively had expected looking at the SM. Instead, $\hat U Y^u \hat Q \hat H_2$, is the one allowed here. This second Higgs doublet is also needed to cancel the triangle $SU(2)_L$ and $U(1)_Y$ anomalies produced by the fermionic partners of the Higgs bosons, i.e. 
\beq
Tr(Y^3)=Tr(T_{3}^2Y)=0,
\eeq 
that would not be cancelled with only one Higgs doublet, and therefore one fermionic superpartner to this Higgs doublet with $Y=1/2$ or $Y=-1/2$ spoiling this last relation (tables \ref{tab:chiral} and \ref{tab:gauge} can be used to check this cancellation, $T_{3}$ is the third component of weak isospin and $Y$ is the weak hypercharge).

The last ingredient we need for the model is the soft SUSY breaking Lagrangian:
\beq
\lagr_{\rm soft}^{\rm MSSM} &=& -\half\left ( M_3 \stilde g\stilde g
+ M_2 \stilde W \stilde W + M_1 \stilde B\stilde B 
+\conj \right )
\nonumber
\\
&&
-\left ( \tilde {\cal Q}_i {\bar {\cal  A}}^{u}_{ij} \tilde {\cal U}^{*}_j {\cal H}_2
- \tilde {\cal Q}_i {\bar {\cal A}}^{d}_{ij} \tilde {\cal D}^{*}_j {\cal H}_1  
- \tilde {\cal L}_i {\bar {\cal A}}^{e}_{ij} \tilde {\cal E}^{*}_j {\cal H}_1  
+ \conj \right ) 
\nonumber
\\
&&
- \tilde {\cal Q}^{\dagger}_i m_{\tilde Q_{ij}}^2 \tilde {\cal Q}_j 
- \tilde {\cal L}^{\dagger}_i m_{\tilde L_{ij}}^2 \tilde {\cal L}_j 
- \tilde {\cal U}^{*}_i m_{\tilde U_{ij}}^2 \, \tilde {\cal U}_j
- \tilde {\cal D}^{*}_i m_{\tilde D_{ij}}^2 \, \tilde {\cal D}_j
- \tilde {\cal E}^{*}_i m_{\tilde E_{ij}}^2 \, \tilde {\cal E}_j
\nonumber \\
&&
- \, m_{H_1}^2 {\cal H}_1^* {\cal H}_1 - m_{H_2}^2 {\cal H}_2^* {\cal H}_2
- \left ( b {\cal H}_2 {\cal H}_1 + \conj \right ) .
\label{MSSMsoft}
\eeq
where we have used calligraphic capital letters for the sfermion fields in the interaction basis with generation indices,
\beq
\tilde {\cal U}_{1,2,3}=\tilde u_R,\tilde c_R,\tilde t_R ;\quad 
\tilde {\cal D}_{1,2,3}=\tilde d_R,\tilde s_R,\tilde b_R &;& \quad 
\tilde {\cal Q}_{1,2,3}=(\tilde u_L \, \tilde d_L)^T, (\tilde c_L\, \tilde s_L)^T, (\tilde t_L \, \tilde b_L)^T  \\
\tilde {\cal E}_{1,2,3}=\tilde e_R,\tilde \mu_R,\tilde \tau_R &;& \quad  
\tilde {\cal L}_{1,2,3}=(\tilde \nu_{eL} \, \tilde e_L)^T, (\tilde \nu_{\mu L}\, \tilde \mu_L)^T, (\tilde \nu_{\tau L} \, \tilde \tau_L)^T
\eeq
and all the gauge indices have been omitted. All the trilinear couplings ${\bar {\cal  A}}_{ij}$ and the soft squared masses $m_{ij}^2$ are $3\times 3$ matrices in flavour space. 

Although we do not assume here any particular origin of SUSY breaking, it is reasonable to expect that all the terms with mass dimensions ($M_3, {\bar {\cal  A}}^{u}_{ij}, m_{\tilde Q_{ij}},...$) are of the same order as the scale of SUSY breaking, generically called here and from now on $m_{SUSY}$.

\bigskip


\subsection{The particles of the MSSM}
\label{particles-mssm}

In this section we summarise the different particle sectors of the model, and the mass spectrum of the particles according to the MSSM parameters. To celebrate the recent discovery \cite{Aad:2012tfa,Chatrchyan:2012ufa,Chatrchyan:2012jja,Chatrchyan:2013lba,Aad:2013xqa,Aad:2013wqa} of the Higgs boson,
 we start with the Higgs sector.

\subsubsection{The Higgs sector}
\label{particles-mssm-higgs}

The Higgs sector of the MSSM is composed of two Higgs doublets:
\BEA
\cHe &=& \VL \cHe^0 \\[0.5ex] \cHe^- \VR \; = \; \VL v_1 
        + \ed{\wz}(\phi_1^0 - i\chi_1^0) \\[0.5ex] -\phi_1^- \VR~,  
        \non \\
\cHz &=& \VL \cHz^+ \\[0.5ex] \cHz^0 \VR \; = \; \VL \phi_2^+ \\[0.5ex] 
        v_2 + \ed{\wz}(\phi_2^0 + i\chi_2^0) \VR~.
\label{higgsfeldunrot}
\EEA
where $v_1$ and $v_2$ are the vacuum expectation values (VEV) for the fields and the ratio between the two is defined as $\Tb = v_2/v_1$.

These 8 degrees of freedom, once the mass matrices are diagonalized, will produce a spectrum with the following particles:

\BEA
h,H &:& \mbox{2 neutral bosons ($h$ will be the notation for the lightest one) with $\cp$} = +1 \non \\
A &:& \mbox{1 neutral boson with $\cp$ = -1} \mbox{ (pseudoscalar)}  \non \\
H^+, H^- &:& \mbox{2 charged bosons}                   \non \\
 G, G^+, G^- &:& \mbox{3 unphysical Goldstone bosons (which will give mass to $Z$ and $W^{\pm}$)}       .
\EEA

The Higgs bosons in the MSSM are needed as in the SM to break the electroweak symmetry $SU(2)_L \times U(1)_Y$ to $U(1)_{em}$ and give masses to the $Z$ and $W^{\pm}$, to break the chiral symmetry ($\psi_{L,R}\rightarrow e^{i\theta_{L,R}}\psi_{L,R}$) and give masses to the fermions (otherwise impossible since by gauge invariance is impossible to introduce mass terms for them in the Lagrangian), and are also responsible of solving the unitarity problem of the $WW$ scattering. This last problem appears when we calculate the scattering between longitudinal modes of the $W$. For instance in the SM there are contributions to this scattering from the following diagrams in Fig. \ref{ww-scattering-1}, whose amplitude grows with the energy as:

\begin{figure}[ht!]
\begin{center}
\begin{tabular}{ccc}
\includegraphics[width=45mm]{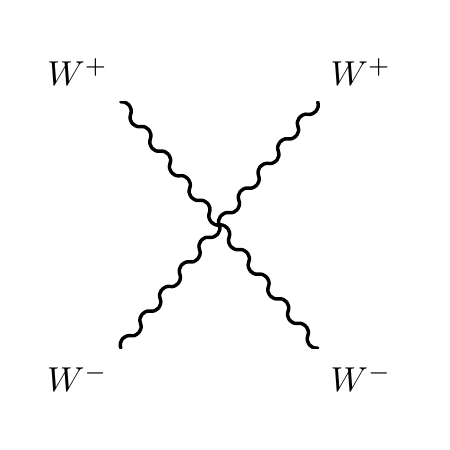} &
\includegraphics[width=45mm]{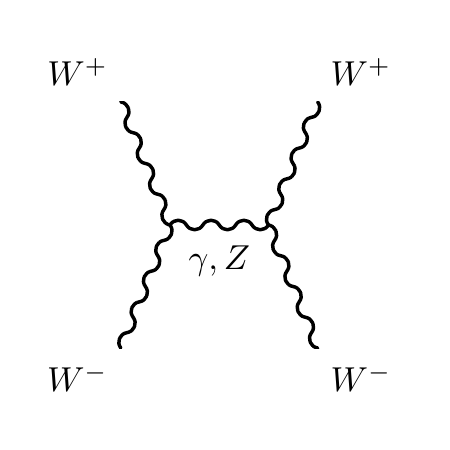} &
\includegraphics[width=45mm]{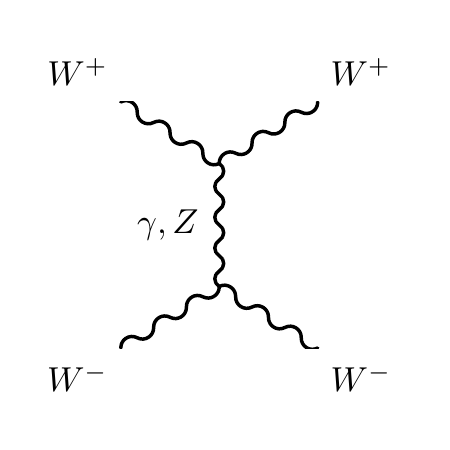} \\
\end{tabular}
\caption{$WW$ scattering without Higgs boson}
\label{ww-scattering-1}
\end{center}
\end{figure}

\beq
M_V &=& -g^2\frac{E^2}{M_W^2}+\ldots ,
\eeq
where $g=e/\sin \theta_W$ and $E$ is the energy of the process. This amplitude violates unitarity at high energies. When the Higgs diagrams in Fig. \ref{ww-scattering-2} are added,

\begin{figure}[ht!]
\begin{center}
\begin{tabular}{cc}
\includegraphics[width=45mm]{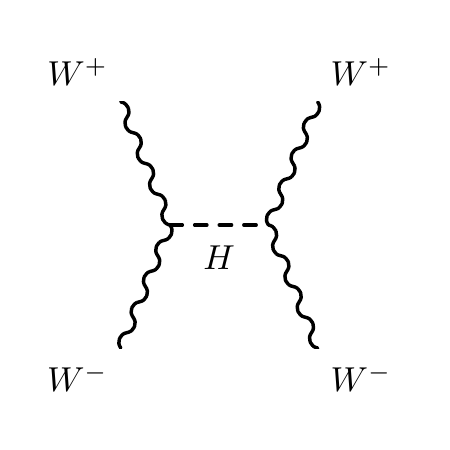} &
\includegraphics[width=45mm]{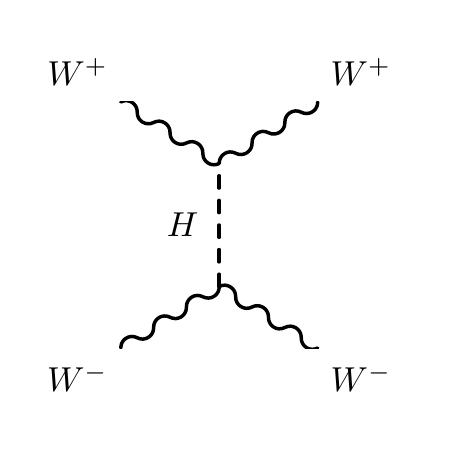} \\
\end{tabular}
\caption{$WW$ scattering mediated by a Higgs boson}
\label{ww-scattering-2}
\end{center}
\end{figure}
then one gets:
\beq
M_{TOT} &=& M_V+M_S = \frac{E^2}{M_W^4}(g_{WWH}^2-g^2M_W^2)+\ldots ,
\eeq
When the relation between the couplings is the right one, as it happens for the SM Higgs boson $g_{WWH}=g M_W$, then the previously commented dangerous growing of the amplitude with energy cancels exactly and unitarity is preserved. Similar arguments apply for the MSSM Higgs sector. 

\bigskip
To build up the MSSM Higgs sector, one starts with the Higgs potential~\cite{hhg} given by:

\beq
V\! &=&\!
(|\mu|^2 + m^2_{H_2}) (|{\cal H}_2^0|^2 + |{\cal H}_2^+|^2)
+ (|\mu|^2 + m^2_{H_1}) (|{\cal H}_1^0|^2 + |{\cal H}_1^-|^2)
\nonumber \\ &&
+\, [b\, ({\cal H}_2^+ {\cal H}_1^- - {\cal H}_2^0 {\cal H}_1^0) + \conj]
\nonumber \\ &&
+ {1\over 8} (g^2 + g^{\prime 2})
(|{\cal H}_2^0|^2 + |{\cal H}_2^+|^2 - |{\cal H}_1^0|^2 - |{\cal H}_1^-|^2 )^2
+ \half g^2 |{\cal H}_2^+ {\cal H}_1^{0*} + {\cal H}_2^0 {\cal H}_1^{-*}|^2. 
\phantom{xxx}
\label{bighiggsv}
\eeq
The $\mu$ terms come from the F-term, the terms with $g$ and $g^{\prime}$ from the D-term, and the rest from the soft SUSY breaking terms.

In this potential we have 5 independent combinations of parameters (besides the known $g$ and $g^{\prime}$): $(|\mu|^2 + m^2_{H_1})$, $(|\mu|^2 + m^2_{H_2})$, $b$, $v_2\equiv\langle {\cal H}_2^0\rangle$ and $v_1\equiv \langle {\cal H}_1^0\rangle$. But they can be reduced to two with the following relations, which are obtained from the minimization of the potential and the value of the gauge boson masses. The minimum of the potential satisfies $\partial V / \partial {\cal H}_1^0=\partial V / \partial {\cal H}_2^0=0$. When applied to the previous potential translates into:
\beq
&&m_{H_2}^2 + |\mu |^2 -b \cot\beta - (M_Z^2/2) \cos (2\beta) 
\>=\> 0 ,
\label{mubsub2}
\\
&&m_{H_1}^2 + |\mu |^2 -b \tan\beta + (M_Z^2/2) \cos (2\beta) \>=\> 0,
\label{mubsub1}
\eeq
giving us two of the commented relations. On the other hand, the gauge boson masses are:
\BE
M_W^2 = \frac{1}{2} g^2 (v_1^2+v_2^2) \qquad;\qquad
M_Z^2 = \frac{1}{2}(g'^2+g^2)(v_1^2+v_2^2) \qquad;\qquad M_\ga=0.
\end{equation}
where $g=e/\sin \theta_W$ and $g'=e/\cos \theta_W$.

As the masses are known, they fix the combination $(v_1^2+v_2^2)$ and therefore we have the third relation needed to reduce the unknown parameters of our potential to two, that usually are taken as: 
\BE
\Tb = v_2/v_1\quad;\quad M_A^2 = -{1\over 8} (g^2 + g^{\prime 2})(\Tb+\CTb),
\end{equation}
where $M_A$ is the mass of the $\cp$-odd Higgs boson~$A$. The rest of the Higgs masses are fixed in terms of these parameters at tree level.

\bigskip
If the equations \ref{mubsub2} and \ref{mubsub1} are used to work out the value of $M_Z$ we obtain:

\beq
M_Z^2 &=& \frac{|m^2_{H_1} - m^2_{H_2}|}{\sqrt{1 - \sin^2(2\beta)}}
- m^2_{H_2} - m^2_{H_1} -2|\mu|^2
.
\label{eq:solvemzsq}
\eeq
If we look at the size of the parameters in this last equation we see that a problem appears. The parameters $m_{H_1}$ and $m_{H_2}$ have to do with the soft SUSY breaking scale, while $\mu$ has nothing to do with this breaking, it is a SUSY preserving parameter, and therefore one would expect it to come from a much higher scale. But if that was the situation, being of such a different orders of magnitude it would be impossible to get a cancellation between them to obtain the small value of $M_Z$.  
 A popular model that solves this issue is the Next-to-Minimal Supersymmetric Standard Model (NMSSM) (for a review, see for instance \cite{Ellwanger:2009dp}) which introduces a new scalar singlet that gets a VEV generating the $\mu$ term, now of the proper size to get the above commented cancellation. But even when getting the right size, we see that the cancellation should be very precise to get the right value of $M_Z$ in the electroweak scale. Again the need of some fine-tuning, that was precisely one of the motivations to go when going from the SM to the MSSM, reappears in our model. 
Anyway, as we commented when talking about the hierarchy problem, this could be understood as not being a real problem, but a hint of possible more complex models beyond the MSSM, as the one commented before.

With respect to the potential, two important inequalities should be satisfied if we want our model to work properly:
First, we want our potential to be bounded from below, so the vacuum is stable. 
 This will be satisfied thanks to the positive quartic interactions of the potential that dominate over the others terms at large value of the fields, but for the directions where the quartic terms are zero, the following condition should be satisfied in order to have a bounded potential from the interaction between the others positive and negative terms:
\beq
2 b < 2 |\mu |^2 + m^2_{H_1} + m^2_{H_2}.
\label{eq:boundedfrombelow}
\eeq
Second, we want to break the electroweak symmetry, because electromagnetism is the symmetry we see in our universe. Some linear combination of ${\cal H}^0_1$ and ${\cal H}^0_2$ will have negative squared mass for ${\cal H}^0_1={\cal H}^0_2=0$ (and thus this point will be unstable, and the breaking will take place) if this relation is true:
\beq
b^2 > (|\mu|^2 + m^2_{H_1} )(|\mu|^2 + m^2_{H_2}).
\label{eq:destabilizeorigin}
\eeq
If we take a look to the Renormalization Group Equations (RGE), we can see that the large Yukawa coupling for the top drives $m^2_{H_2}$ to negative values and therefore helps to satisfy the previous equation. This is another success of SUSY, since in the SM the equivalent $\mu$ term is just a free parameter and there is no theoretical reason for $\mu^2$ to be negative and break the electroweak symmetry. Furthermore, if we evolve the top mass from high scale down to low energy, the RGEs tell us that at some moment the evolution will stop (what is called a fixed point) \cite{Pendleton:1980as} giving us a top Yukawa around 1, almost independently of the high scale chosen. Hence, this requirement of having a large Yukawa is also understood in SUSY, and the success is double. 

\bigskip
Once we get the potential, we can write the mass matrix of the bosons. Since there are terms that mix the scalar bosons, the corresponding mass matrix will not be diagonal, so we have to rotate it to find the mass basis. This is performed via the orthogonal transformations: 
\BEA
\label{hHdiag}
\VL H \\[0.5ex] h \VR &=& \ML \Ca & \Sa \\[0.5ex] -\Sa & \Ca \MR 
\VL \phi_1^0 \\[0.5ex] \phi_2^0 \VR~,  \\
\label{AGdiag}
\VL G \\[0.5ex] A \VR &=& \ML \Cb & \Sbe \\[0.5ex] -\Sbe & \Cb \MR 
\VL \chi_1^0 \\[0.5ex] \chi_2^0 \VR~,  \\
\label{Hpmdiag}
\VL G^{\pm} \\[0.5ex] H^{\pm} \VR &=& \ML \Cb & \Sbe \\[0.5ex] -\Sbe & 
\Cb \MR \VL \phi_1^{\pm} \\[0.5ex] \phi_2^{\pm} \VR~.
\EEA

The mixing angle $\al$ is determined through
\BE
\al = {\rm arctan}\KKL 
  \frac{-(\MA^2 + \MZ^2) \Sbe \Cb}
       {\MZ^2 \CQb + \MA^2 \SQb - m^2_{h,{\rm tree}}} \KKR~, ~~
 -\frac{\pi}{2} < \al < 0~.
\label{alphaborn}
\end{equation}

At tree level the mass matrix of the neutral $\cp$-even Higgs bosons
is given in the $\Pe$-$\Pz$-basis 
in terms of $\MZ$, $\MA$, and $\Tb$ by
\BEA
M_{\rm Higgs}^{2, {\rm tree}} &=& \ML \mpe^2 & \mpez^2 \\ 
                           \mpez^2 & \mpz^2 \MR \non\\
&=& \ML \MA^2 \SQb + \MZ^2 \CQb & -(\MA^2 + \MZ^2) \Sbe \Cb \\
    -(\MA^2 + \MZ^2) \Sbe \Cb & \MA^2 \CQb + \MZ^2 \SQb \MR,
\label{higgsmassmatrixtree}
\EEA
which by diagonalization according to \refeq{hHdiag} yields the
tree-level Higgs boson masses
\BE
M_{\rm Higgs}^{2, {\rm tree}} 
   \stackrel{\al}{\longrightarrow}
   \ML m_{H,{\rm tree}}^2 & 0 \\ 0 &  m_{h,{\rm tree}}^2 \MR~,
\end{equation}
where
\BE
\label{rMSSM:mtree}
(m_{H,h}^2)_{\rm tree}=
\edz\left[\MA^2+\MZ^2 \pm\sqrt{(\MA^2+\MZ^2)^2-
4\MZ^2\MA^2\cos^2 2\be}\right] ~.
 \end{equation}

At tree level the light Higgs boson is too light, having as upper bound to its mass the value of $\MZ$. 

The charged Higgs boson mass is given by
\BE
\label{rMSSM:mHp}
m^{2}_{H^{\pm},{\rm tree}} = \MA^2 + \MW^2~.
\end{equation}

Once we have all the Higgs boson masses it is interesting to give a fast look to the couplings or at least to some features of them. Nowadays these couplings are being measured for the discovered Higgs boson, and it is important to know the differences between a SM and a MSSM Higgs boson. At tree level, the MSSM couplings can be expressed in terms of the corresponding SM couplings and some extra factors. See next some of these relations:

\beq
g_{hVV}=g_{HVV}^{SM} \sin(\beta-\alpha)\quad;\quad g_{HVV}=g_{HVV}^{SM} \cos(\beta-\alpha)\quad;\quad g_{AVV}=0
\eeq
\beq
g_{hAZ}=\frac{g'}{2c_W}\cos(\beta-\alpha)\quad&;&\quad g_{HAZ}=\frac{g'}{2c_W}\sin(\beta-\alpha)\\
g_{hb\sbar b,h\tau^+\tau^-}=g_{Hb\sbar b,H\tau^+\tau^-}^{SM}(-\frac{\sin \alpha}{\cos \beta})\quad&;&\quad g_{ht\sbar t}=g_{Ht\sbar t}^{SM}\frac{\cos \alpha}{\sin \beta}\\
g_{Hb\sbar b,H\tau^+\tau^-}=g_{Hb\sbar b,H\tau^+\tau^-}^{SM}\frac{\cos \alpha}{\cos \beta}\quad&;&\quad g_{Ht\sbar t}=g_{Ht\sbar t}^{SM}\frac{\sin \alpha}{\sin \beta}\\
g_{Ab\sbar b,A\tau^+\tau^-}=g_{Hb\sbar b,H\tau^+\tau^-}^{SM}\gamma_5\tan \beta\quad&;&\quad g_{At\sbar t}=g_{Ht\sbar t}^{SM}\gamma_5\cot \beta
\eeq
where $V=Z,W^{\pm}$, and the results are the same for the other two generations of fermions.

\begin{figure}[ht!]
\begin{center}
\includegraphics[width=90mm]{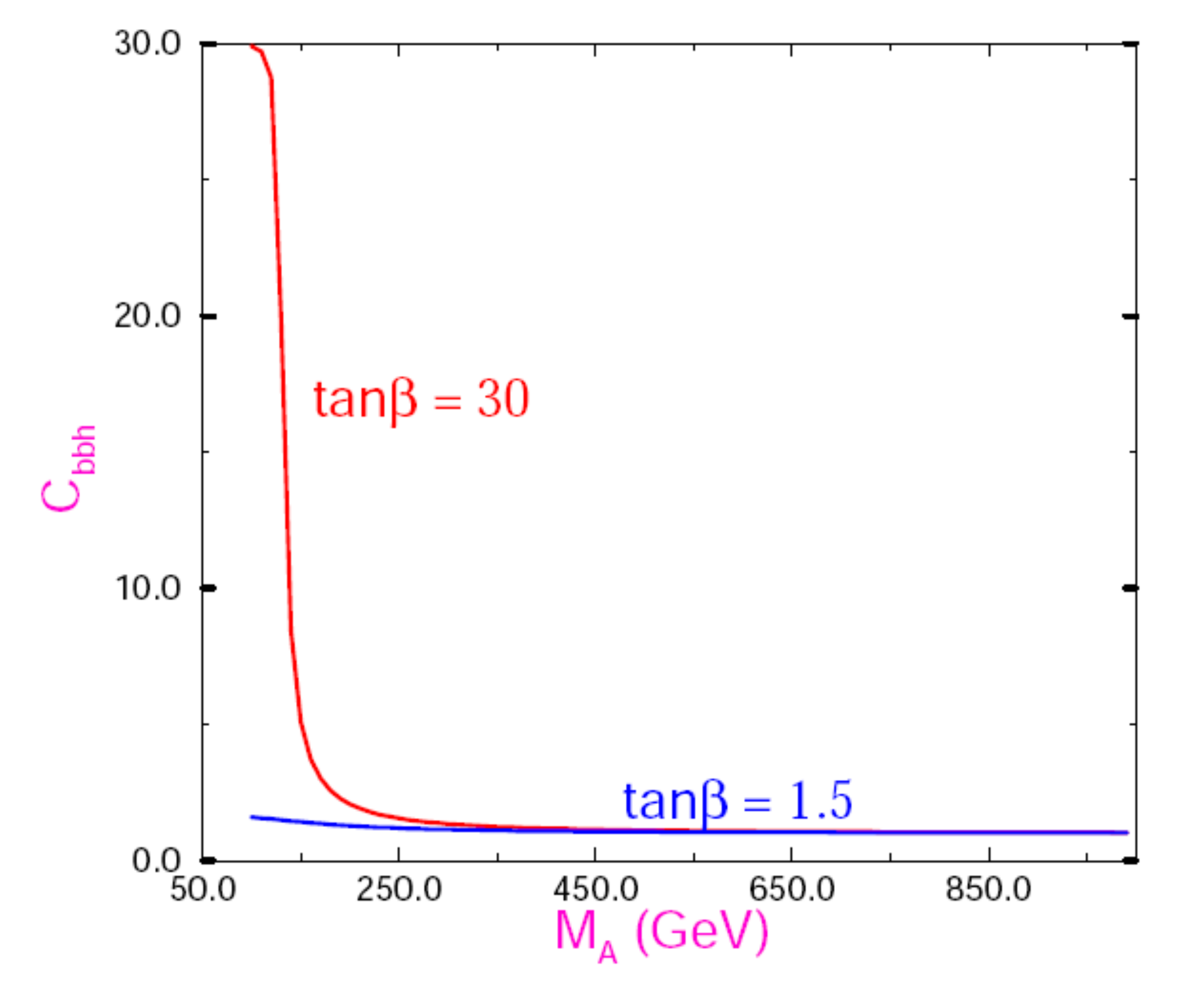}
\caption{Ratio between the tree-level couplings of the MSSM and the SM between a down type fermion and the Higgs boson (light Higgs boson and the SM Higgs boson respectively). Plot obtained from \cite{Dawson:1997tz}.}
\label{fig:hbbma}
\end{center}
\end{figure}

\begin{figure}[ht!]
\begin{center}
\includegraphics[width=90mm]{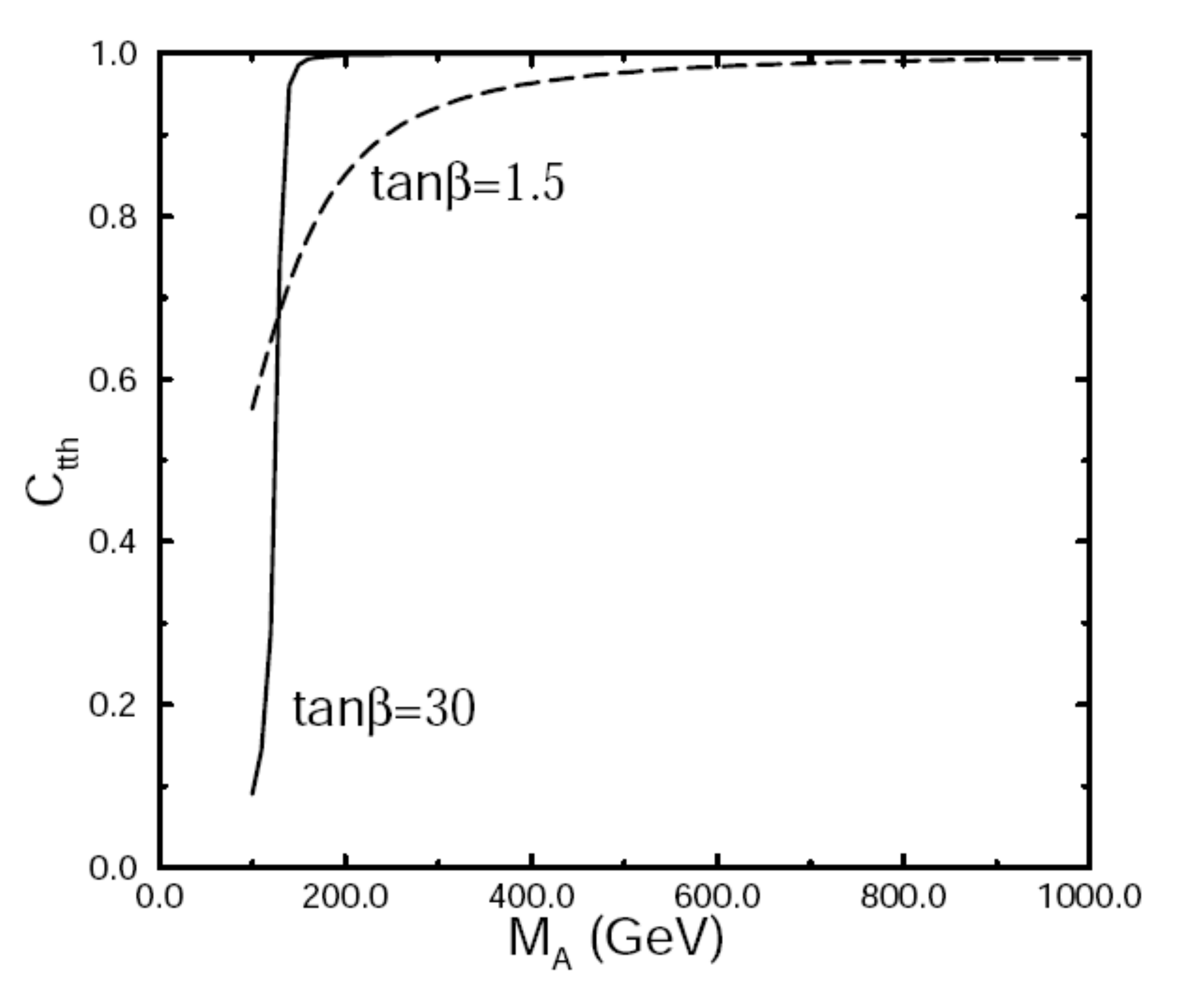}
\caption{Ratio between the tree-level couplings of the MSSM and the SM between an up type fermion and the Higgs boson (light Higgs boson and the SM Higgs boson respectively). Plot obtained from \cite{Dawson:1997tz}.}
\label{fig:httma}
\end{center}
\end{figure}

We can see how depending on the $\alpha$ and $\beta$ angles, we can obtain very different situations to the SM. For example, the coupling of the light Higgs boson with the pseudoscalar $A$ and the gauge boson $Z$ that can give rise to new phenomenology, or the huge increase we can find in the coupling with the bottom-type fermions for the large $\tan \beta$ case, that could play a very important role in the MSSM phenomenology (see e.g. \cite{Isidori:2006pk,Babu:1998er,Boos:2002ze,Carena:2000uj,Carena:1999py}). Another interesting situation is the limit:
\beq
-\frac{\sin \alpha}{\cos \beta}\to 1\qquad;\qquad \beta-\alpha \to \pi /2
\eeq
where the light Higgs boson will behave as the SM Higgs boson, being almost impossible to distinguish between the two. This situation is realised in the so called decoupling limit, for $M_A\gg 150$ GeV, when $h$ behaves SM-like and $M_A\simeq M_H\simeq M_H^{\pm}$, with the heavy Higgs bosons decoupling from the low energy electroweak physics. As an example of this limit, see the dependence of two of the tree level couplings of the light MSSM Higgs boson to the $b$ and $t$ quarks with respect to $M_A$ in Figs. \ref{fig:hbbma} and \ref{fig:httma}. The values represented $C_{ffh}$ are the ratios between the MSSM value of the coupling divided to the corresponding SM Higgs coupling. That is, their values are 1 when the MSSM coupling coincides with the SM coupling. 

\bigskip
Since finally we have measurements for the Higgs mass, and these will improve a lot in a close future, is important to have precision predictions for it, and for the rest of the Higgs boson parameters. 
In the next section we will study the one loop corrections to the Higgs boson masses, and we will see how these corrections are very important.


\subsubsection{The Higgs sector at one-loop}
\label{sec:renrMSSM}

In the Feynman diagrammatic (FD) approach that we are following here, the higher-order corrected 
$\cp$-even Higgs boson masses are derived by finding the
poles of the $(h,H)$-propagator 
matrix. The inverse of this matrix is given by
\BE
\left(\Delta_{\rm Higgs}\right)^{-1}
= - i \ML p^2 -  \mHtree^2 + \hSi_{HH}(p^2) &  \hSi_{hH}(p^2) \\
     \hSi_{hH}(p^2) & p^2 -  \mhtree^2 + \hSi_{hh}(p^2) \MR~.
\label{higgsmassmatrixnondiag}
\end{equation}
Determining the poles of the matrix $\Delta_{\rm Higgs}$ in
\refeq{higgsmassmatrixnondiag} is equivalent to solving
the equation
\begin{equation}
\left[p^2 - \mhtree^2 + \hSi_{hh}(p^2) \right]
\left[p^2 - \mHtree^2 + \hSi_{HH}(p^2) \right] -
\left[\hSi_{hH}(p^2)\right]^2 = 0\,.
\label{eq:proppole}
\end{equation}
Similarly, in the case of the charged Higgs sector, the corrected Higgs mass is derived by the position of the pole in the charged Higgs propagator, which is defined by: 
\noindent \begin{equation}
p^{2}-m^{2}_{H^{\pm},{\rm tree}} +
\hat{\Sigma}_{H^{-}H^{+}}\left(p^{2}\right)=0.
\label{eq:proppolech}
\end{equation}

To solve the previous equations, the renormalized Higgs boson
self-energies are needed. Here we follow the procedure used in
\citeres{mhiggsf1lC,mhcMSSMlong} (and references therein) and review it for
completeness. 
The parameters appearing in the Higgs
potential, see \refeq{bighiggsv}, are renormalized as follows:
\begin{align}
\label{rMSSM:PhysParamRenorm}
  \MZ^2 &\to \MZ^2 + \dMZsq,  & \tadh &\to \tadh +
  \dtadh, \\ 
  \MW^2 &\to \MW^2 + \dMWsq,  & \tadH &\to \tadH +
  \dtadH, \notag \\ 
  M_{\rm Higgs}^2 &\to M_{\rm Higgs}^2 + \de M_{\rm Higgs}^2, & 
  \tanb &\to \tanb (1+\dtanb). \notag 
\end{align}
$M_{\rm Higgs}^2$ denotes the tree-level Higgs boson mass matrix given
in \refeq{higgsmassmatrixtree}. $\tadh$ and $\tadH$ [are the tree-level
tadpoles (i.e.\ the terms linear in $h$ and $H$ in the Higgs potential).

The field renormalization matrices of both Higgs multiplets
can be set up symmetrically, 
\begin{align}
\label{rMSSM:higgsfeldren}
  \begin{pmatrix} h \\[.5em] H \end{pmatrix} \to
  \begin{pmatrix}
    1+\tfrac{1}{2} \dZ{hh} & \tfrac{1}{2} \dZ{hH} \\[.5em]
    \tfrac{1}{2} \dZ{hH} & 1+\tfrac{1}{2} \dZ{HH} 
  \end{pmatrix} \cdot
  \begin{pmatrix} h \\[.5em] H \end{pmatrix}~.
\end{align}

\noindent
For the mass counter term matrices we use the definitions:
\begin{align}
  \delta M_{\rm Higgs}^2 =
  \begin{pmatrix}
    \dmhsq  & \dmhHsq \\[.5em]
    \dmhHsq & \dmHsq  
  \end{pmatrix}~.
\end{align}
The renormalized self-energies, $\hSi(p^2)$, can now be expressed
through the unrenormalized self-energies, $\Si(p^2)$, the field
renormalization constants and the mass counter terms.
This reads for the $\cp$-even part,
\begin{subequations}
\label{rMSSM:renses_higgssector}
\begin{align}
\ser{hh}(p^2)  &= \se{hh}(p^2) + \dZ{hh} (p^2-\mhtree^2) - \dmhsq, \\
\ser{hH}(p^2)  &= \se{hH}(p^2) + \dZ{hH}
(p^2-\tfrac{1}{2}(\mhtree^2+\mHtree^2)) - \dmhHsq, \\ 
\ser{HH}(p^2)  &= \se{HH}(p^2) + \dZ{HH} (p^2-\mHtree^2) - \dmHsq~.
\end{align}
\end{subequations}

Inserting the renormalization transformation into the Higgs mass terms
leads to expressions for their counter terms which consequently depend
on the other counter terms introduced in~(\ref{rMSSM:PhysParamRenorm}). 

For the $\cp$-even part of the Higgs sectors, these counter terms are:
\begin{subequations}
\label{rMSSM:HiggsMassenCTs}
\begin{align}
\dmhsq &= \de\MA^2 \cos^2(\alpha-\beta) + \delta \MZ^2 \sin^2(\alpha+\beta) \\
&\quad + \tfrac{e}{2 \MZ \sw \cw} (\dtadH \cos(\alpha-\beta)
\sin^2(\alpha-\beta) + \dtadh \sin(\alpha-\beta)
(1+\cos^2(\alpha-\beta))) \notag \\ 
&\quad + \dtanb \sinb \cosb (\MA^2 \sin 2 (\alpha-\beta) + \MZ^2 \sin
2 (\alpha+\beta)), \notag \\ 
\dmhHsq &= \tfrac{1}{2} (\de\MA^2 \sin 2(\alpha-\beta) - \dMZsq \sin
2(\alpha+\beta)) \\ 
&\quad + \tfrac{e}{2 \MZ \sw \cw} (\dtadH \sin^3(\alpha-\beta) -
\dtadh \cos^3(\alpha-\beta)) \notag \\ 
&\quad - \dtanb \sinb \cosb (\MA^2 \cos 2 (\alpha-\beta) + \MZ^2 \cos
2 (\alpha+\beta)), \notag \\ 
\dmHsq &= \de\MA^2 \sin^2(\alpha-\beta) + \dMZsq \cos^2(\alpha+\beta) \\
&\quad - \tfrac{e}{2 \MZ \sw \cw} (\dtadH \cos(\alpha-\beta)
(1+\sin^2(\alpha-\beta)) + \dtadh \sin(\alpha-\beta)
\cos^2(\alpha-\beta)) \notag \\ 
&\quad - \dtanb \sinb \cosb (\MA^2 \sin 2 (\alpha-\beta) + \MZ^2 \sin
2 (\alpha+\beta))~. \notag 
\end{align}
\end{subequations}

\bigskip
For the field renormalization we chose to give each Higgs doublet one
renormalization constant,
\begin{align}
\label{rMSSM:HiggsDublettFeldren}
  \cHe \to (1 + \tfrac{1}{2} \dZ{\cHe}) \cHe, \quad
  \cHz \to (1 + \tfrac{1}{2} \dZ{\cHz}) \cHz~.
\end{align}
This leads to the following expressions for the various field
renormalization constants in \refeq{rMSSM:higgsfeldren}:
\begin{subequations}
\label{rMSSM:FeldrenI_H1H2}
\begin{align}
  \dZ{hh} &= \sinasq \dZ{\cHe} + \cosasq \dZ{\cHz}, \\[.2em]
  \dZ{hH} &= \sina \cosa (\dZ{\cHz} - \dZ{\cHe}), \\[.2em]
  \dZ{HH} &= \cosasq \dZ{\cHe} + \sinasq \dZ{\cHz}~.
\end{align}
\end{subequations}
The counter term for $\tb$ can be expressed in terms of the vacuum
expectation values as
\begin{equation}
\de\tb = \frac{1}{2} \KL \dZ{\cHz} - \dZ{\cHe} \KR +
\frac{\de v_2}{v_2} - \frac{\de v_1}{v_1}~,
\end{equation}
where the $\de v_i$ are the renormalization constants of the $v_i$:
\begin{equation}
v_1 \to \KL 1 + \dZ{\cHe} \KR \KL v_1 + \de v_1 \KR, \quad
v_2 \to \KL 1 + \dZ{\cHz} \KR \KL v_2 + \de v_2 \KR~.
\end{equation}
Similarly for the charged Higgs sector, the renormalized self-energy is expressed in terms of the unrenormalized one and the corresponding counter-terms as:
\noindent \begin{equation}
\hat{\Sigma}_{H^{-}H^{+}}\left(p^{2}\right)=\Sigma_{H^{-}H^{+}}\left(p^{2}\right)+\delta Z_{H^{-}H^{+}}\left(p^{2}-m^{2}_{H^{\pm},{\rm tree}} \right)-\delta m_{H^{\pm}}^{2},\end{equation}
where, 
\noindent \begin{equation}
\delta m_{H^{\pm}}^{2}=\delta M_{A}^{2}+\delta M_{W}^{2}\end{equation}
and,
\noindent \begin{equation}
\delta Z_{H^{-}H^{+}}=\sin^{2}\beta \, \,\dZ{\cHe}  
+\cos^{2}\beta \,\,\dZ{\cHz}. \end{equation}

The renormalization conditions are fixed by an appropriate
renormalization scheme. For the mass counter terms on-shell conditions
are used, resulting in:
\begin{align}
\label{rMSSM:mass_osdefinition}
  \dMZsq = \re \se{ZZ}(\MZ^2), \quad \dMWsq = \re \se{WW}(\MW^2),
  \quad \de\MA^2 = \re \se{AA}(\MA^2). 
\end{align}
For the gauge bosons in Eq. \ref{rMSSM:mass_osdefinition} $\Si$ denotes the transverse part of the corresponding self-energy. 
Since the renormalized tadpole configurations are chosen to vanish in all orders,
their counter terms follow from $T_{\{h,H\}} + \de T_{\{h,H\}} = 0$: 
\begin{align}
  \dtadh = -{\tadh}, \quad \dtadH = -{\tadH}~. 
\end{align}
For the remaining renormalization constants, i.e. $\de\tb$, $\dZ{\cHe}$
and $\dZ{\cHz}$, the most convenient
choice is a $\drbar$ renormalization where:
\begin{subequations}
\label{rMSSM:deltaZHiggsTB}
\begin{align}
  \dZ{\cHe} &= \dZ{\cHe}^{\drbarm}
       \; = \; - \KKL \re \Sip_{HH \; |\al = 0} \KKR^{\rm div}, \\[.5em]
  \dZ{\cHz} &= \dZ{\cHz}^{\drbarm} 
       \; = \; - \KKL \re \Sip_{hh \; |\al = 0} \KKR^{\rm div}, \\[.5em]
  \dtanb &= -\edz (\dZ{\cHz} - \dZ{\cHe}) = \dtanb^{\drbarm}~.
\end{align}
\end{subequations}
The corresponding renormalization scale, $\mudim$, is set to 
$\mudim = \mt$ in all numerical evaluations.

\bigskip
With the previous equations we can calculate the corrections to the Higgs bosons masses. If we vary all the relevant MSSM parameters we can get a corrected light Higgs mass up to 135GeV, what is perfectly compatible with the observed Higgs mass around 125GeV.

The above procedure will be used in Chapter \ref{higgsmasssquark} to calculate the corrections to the Higgs bosons masses coming from the supersymmetric particles with sfermion mixing.

For phenomenological purposes, it is convenient to estimate in advance the approximate value for $M_h$ without going through the whole renormalization procedure. In \cite{Heinemeyer:2004ms} a formula can be found for the leading $m_t^4$ corrections up to two-loops which works reasonably well ${\cal O}$ (5GeV) for moderate and low $\tan \beta$:

\beq
M_h^2&=&m_h^{2,{\rm tree}}+m_h^{2,\alpha}+m_h^{2,\alpha \alpha_s}\label{mhcorrecsapprox}\\
m_h^{2,\alpha}&=&\frac{3}{2}\frac{G_F\sqrt{2}}{\pi^2}M_t^4\left\{-{\rm ln}\left(\frac{M_t^2}{M_S^2}\right)+\frac{X_t^2}{M_S^2}\left(1-\frac{1}{12}\frac{X_t^2}{M_S^2}\right)\right\}\\
m_h^{2,\alpha \alpha_s}&=&-3\frac{G_F\sqrt{2}}{\pi^2}\frac{\alpha_s}{\pi}M_t^4\Bigg\{-{\rm ln^2}\left(\frac{M_t^2}{M_S^2}\right)\\
&&-\left(2+\frac{X_t^2}{M_S^2}\right){\rm ln}\left(\frac{M_t^2}{M_S^2}\right)-\frac{X_t}{M_S}\left(2-\frac{1}{4}\frac{X_t^3}{M_S^3}\right)\Bigg\}
\eeq
where $X_t=A_t-\mu \cot \beta$, $A_t=A^u_{33}={\cal A}^u_{33}/y_{t}$, where $y_{t}$ is the top Yukawa, and $M_S$ is the relevant average mass, often defined as the geometric mean between $m_{\tilde U_{L33}}$ and $m_{\tilde U_{R33}}$. All of the input parameters are on-shell quantities. 

\begin{figure}[ht!]
\hspace{1.0cm}
\includegraphics[width=150mm]{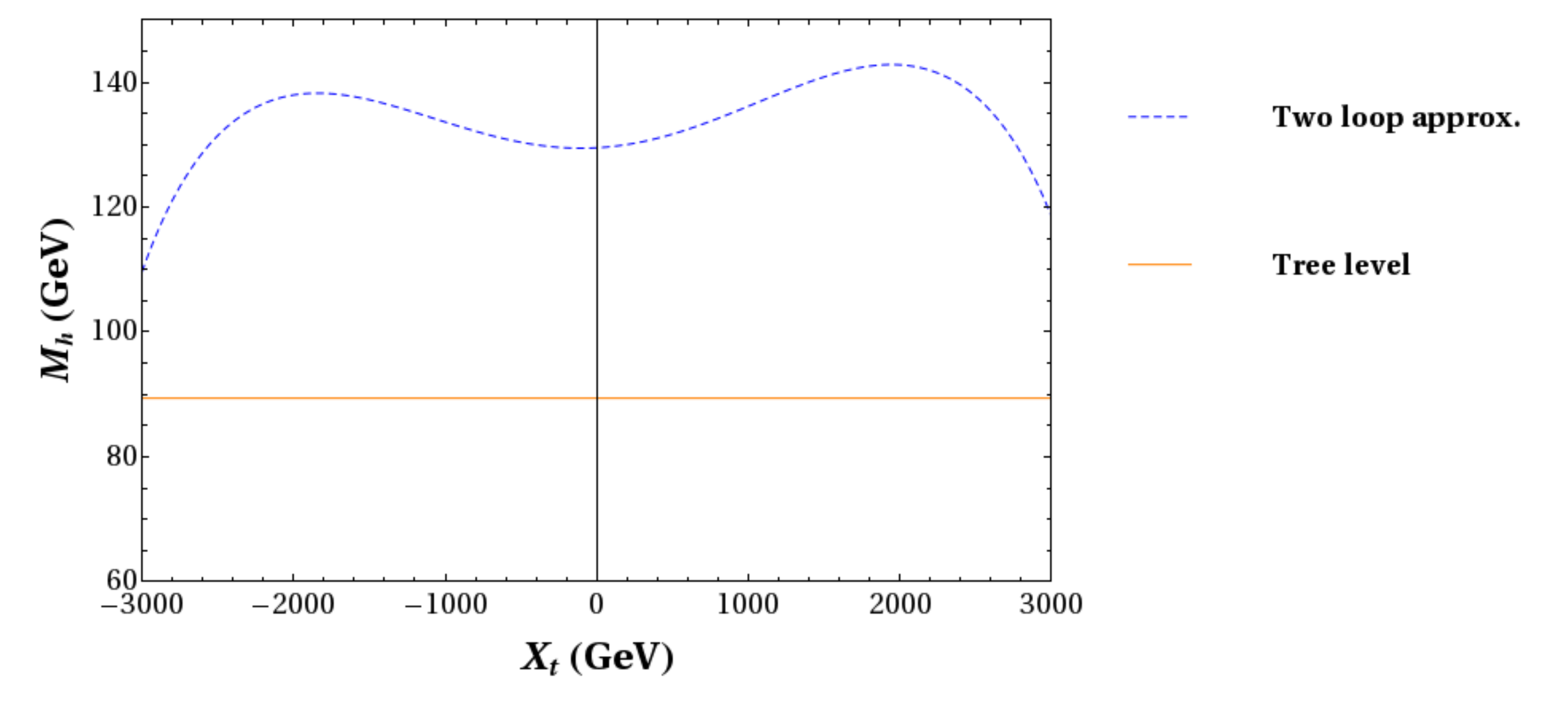}
\caption{Light Higgs boson mass calculated at tree level (orange line) and using the two loop approximation of Equation \ref{mhcorrecsapprox} (blue dashed curve) with respect to $X_t$, with the following values: $M_t=173.07$ GeV, $M_S=M_A=1000$ GeV, $\tan \beta=10$. One-loop contributions from other sectors lower the light Higgs boson mass by about 10 GeV}
\label{plot-mhcorrecs}
\end{figure}

In Figure \ref{plot-mhcorrecs} we show the values of the light Higgs boson mass calculated at tree level and with Equation \ref{mhcorrecsapprox} with respect to $X_t$. It can be seen the relevance of the loop contributions.

\subsubsection{The chargino/neutralino sector}
\label{particles-mssm-charginos-neutralinos}

Once the Higgs sector is studied 
we describe the supersymmetric partners of the Higgs bosons: the Higgsinos. Since the Higgsinos have the same quantum numbers as the electroweak gauginos, and there are mixing terms in the Lagrangian between them, then they mix giving rise to the physical particles in the final mass basis. Therefore we will study them together.
The electrical charge is a well-defined quantum number so the charged particles and the neutral ones will not mix between them. The neutral Higgsinos ($\stilde {\cal H}_1^0$ and $\stilde {\cal H}_2^0$) mix with the neutral gauginos ($\stilde B$ and $\stilde W^0$) forming the so called neutralinos (${\tilde \chi}^0_i$, $i=1,2,3,4$). The charged Higgsinos ($\stilde {\cal H}_2^+$ and $\stilde {\cal H}_1^-$) and the charged winos ($\stilde W^+$ and $\stilde W^-$) will originate the charginos (${\tilde \chi}^\pm_i$, $i=1,2$).

According to the MSSM Lagrangian, the charginos have the following mass matrix in the interaction basis $\psi^\pm = (\stilde W^+,\, \stilde {\cal H}_2^+,\, \stilde W^- ,\, \stilde {\cal H}_1^- )$:

\beq
{\bf M}_{{\tilde \chi}^\pm}
&=& \begin{pmatrix}{\bf 0}&{\bf X}^T\cr
             {\bf X} &{\bf 0}\end{pmatrix} ,
\phantom{xx}
\eeq

\beq
{\bf X} &=& \begin{pmatrix}M_2 & g v_u\cr
                     g v_d & \mu \cr \end{pmatrix}
\>=\> \begin{pmatrix}M_2 & \sqrt{2} \sbeta\, M_W\cr
               \sqrt{2} \cbeta\, M_W & \mu \cr \end{pmatrix}.
\label{charginomassmatrix}
\eeq

This matrix is diagonalized using two $2\times 2$ unitary matrices ${\bf U}$ and ${\bf V}$ that rotate the basis with these relations:

\beq
\begin{pmatrix} \stilde {\chi}^+_1\cr
         \stilde { \chi}^+_2\end{pmatrix} = {\bf V}
\begin{pmatrix}\stilde W^+\cr
         \stilde {\cal H}_2^+\end{pmatrix},\quad\qquad\>\>\>\>\>\>
\begin{pmatrix} \stilde {\chi}^-_1\cr
         \stilde {\chi}^-_2\end{pmatrix} = {\bf U}
\begin{pmatrix}\stilde W^-\cr
         \stilde {\cal H}_1^-\end{pmatrix},\phantom{xxx}
\eeq
obtaining a diagonal matrix for each sub-matrix:

\beq
{\bf U}^* {\bf X} {\bf V}^{-1} =
\begin{pmatrix}m_{{\tilde \chi}^\pm_1} & 0\cr
              0   & m_{{\tilde \chi}^\pm_2}\end{pmatrix},
\eeq
being these the eigenvalues for the full mass squared matrix:

\beq
{\bf V} {\bf X}^\dagger {\bf X} {\bf V}^{-1} =
{\bf U}^* {\bf X} {\bf X}^\dagger {\bf U}^{T} =
\begin{pmatrix}m^2_{{\tilde \chi}^\pm_1} & 0 \cr 0 & m^2_{{\tilde \chi}^\pm_2}\end{pmatrix}.
\eeq

The neutralinos have this mass matrix:
\beq
{\bf M}_{{\tilde \chi}^0} \,=\, \begin{pmatrix}
  M_1 & 0 & -g' v_d/\sqrt{2} & g' v_u/\sqrt{2} \cr
  0 & M_2 & g v_d/\sqrt{2} & -g v_u/\sqrt{2} \cr
  -g' v_d/\sqrt{2} & g v_d/\sqrt{2} & 0 & -\mu \cr
  g' v_u/\sqrt{2} & -g v_u/\sqrt{2}& -\mu & 0 \cr \end{pmatrix}.
\label{preneutralinomassmatrix}
\eeq
in the electroweak basis $\psi^0 = (\stilde B, \stilde W^0, \stilde 
{\cal H}_1^0, \stilde {\cal H}_2^0)$.
The diagonalization is performed through a rotation ${\stilde \chi}^0_i = {\bf N}_{ij} \psi^0_j$
, that fulfils 
\beq
{\bf N}^* {\bf M}_{{\tilde \chi}^0} {\bf N}^{-1}
\, =\, {\rm diag}(m_{{\tilde \chi}^0_1}, m_{{\tilde \chi}^0_2}, m_{{\tilde \chi}^0_3}, m_{{\tilde \chi}^0_4}).
\label{diagmN}
\eeq
The lightest neutralino ${\tilde \chi}^0_1$ turns out to be the LSP, and is one of the favourite candidates to dark matter.

\subsubsection{The gluino sector}
\label{particles-mssm-gluinos}
The gluinos $\stilde g$ are the supersymmetric partners of the gluons. As we can see in Table \ref{tab:gauge} they have spin 1/2, and they keep the same gauge numbers as their standard partners as it happens with the other particles. They are octet fermions, and since there are no other SUSY particles in this representation, they do not mix with other particles.

If SUSY were an exact symmetry they would be massless as the gluons, but since it is broken, it can be found a term in the soft SUSY breaking Lagrangian that give mass to them, the term $M_3$. Since they have not been seen in any collider yet, it can be set a lower bound to its mass (see Section \ref{exp-searches-mssm-particles}); although this value depends on the specific values of the other supersymmetric parameters, at present one can say that the gluino mass should be above $\sim{\cal O}$ (1 TeV).

\subsubsection{The squark sector}
\label{particles-mssm-squarks}

As shown in table \ref{tab:chiral}, the supersymmetric partners of the quarks are the squarks $({\stilde u}_L\>\>\>{\stilde d}_L )$, ${\stilde u}^*_R$, ${\stilde d}^*_R$ (and the same for the other two generations). We recall again that the squarks are $0$ spin bosons. The sub-indices $L$ and $R$ of the squarks are just a reminder of the chirality of the corresponding fermionic partners. The squarks are scalars, and thus have no chirality.

In order to proceed with the squarks, let us look first to the partner quarks. Starting with our Lagrangian, expressed in the interaction basis, we know that the different families of quarks will mix with each other. Therefore the final mass states will be different from these starting ones. To diagonalize the mass matrix, the basis should be rotated from the $SU(2)$ (interaction) eigenstate basis, $q^{\rm int}_{L,R}$, to the (physical) mass  eigenstate basis, 
$q_{L,R}^{\rm phys}$, by unitary transformations, $V^{u,d}_{L,R}$:

\begin{equation}
\VL u^{\rm phys}_{L,R} \\ c^{\rm phys}_{L,R} \\ t^{\rm phys}_{L,R} \VR =
V^u_{L,R} \VL u^{\rm int}_{L,R} \\ c^{\rm int}_{L,R} \\ t^{\rm int}_{L,R} \VR~,~~~~
\VL d^{\rm phys}_{L,R} \\ s^{\rm phys}_{L,R} \\ b^{\rm phys}_{L,R} \VR =
V^d_{L,R} \VL d^{\rm int}_{L,R} \\ s^{\rm int}_{L,R} \\ b^{\rm int}_{L,R} \VR~,
\end{equation}
such that the quark Yukawa coupling and mass matrices in the physical basis are:
\begin{eqnarray}
V^u_L Y^{u*}V^{u\dagger}_R&=&{\rm diag}(y_u,y_c,y_t)=
{\rm diag}\left(\frac{m_u}{v_2},\frac{m_c}{v_2},\frac{m_t}{v_2}\right),
\\
V^d_L Y^{d*}V^{d\dagger}_R&=&{\rm diag}(y_d,y_s,y_b)=
{\rm diag}\left(\frac{m_d}{v_1},\frac{m_s}{v_1},\frac{m_b}{v_1}\right).
\end{eqnarray} 
The quark flavour mixing that these rotations imply, and that is measured in the experiments, will be encoded in the Cabibbo-Kobayashi-Maskawa (CKM) matrix, defined as usual as, 
\begin{equation}
\VCKM= V^u_L V^{d\dagger}_L.
\end{equation}

\bigskip
Since the quarks and squarks belong to the same superfield, the same terms that we had to diagonalize for the quarks will be present for the part of the Lagrangian of the squarks, and thus the same rotation will be needed for them. The squarks are then rotated, parallel to the quarks, from the interaction basis, ${\tilde q}^{\rm int}_{L,R}$ to what is called the Super-CKM (SCKM) basis, ${\tilde q}_{L,R}$, by
\begin{equation}
\label{squarksrotsckm}
\VL \tilde u_{L,R} \\ \tilde c_{L,R} \\ \tilde t_{L,R} \VR =
V^u_{L,R} \VL \tilde u^{\rm int}_{L,R} \\ \tilde c^{\rm int}_{L,R} \\ \tilde t^{\rm int}_{L,R} \VR~,~~~
\VL \tilde d_{L,R} \\ \tilde s_{L,R} \\ \tilde b_{L,R} \VR =
V^d_{L,R} \VL \tilde d^{\rm int}_{L,R} \\ \tilde s^{\rm int}_{L,R} \\ \tilde b^{\rm int}_{L,R} \VR~.
\end{equation}

\bigskip
The minimal version of our supersymmetric model with respect to the flavour, the so-called minimal flavour violation hypothesis \cite{D'Ambrosio:2002ex}, defines the $\VCKM$ matrix as the only source of flavour violation for the squark sector, as will be explained with more detail in Section \ref{mfvhip}. This MFV case will imply that the squarks in the Super-CKM basis are already the physical mass eigenstates. However in this study we will go beyond the MFV hypothesis and explore the limits of flavour in a more general context of Non Minimal Flavour Violation. This implies that in NMFV one has to perform a second rotation to reach the squark physical basis. The general parametrization of flavour mixing for the squarks in this NMFV case which is our chosen scenario will be defined in Section \ref{paramnmfv} and the mass matrices for that case will also be defined there.

\subsubsection{The slepton sector}
\label{particles-mssm-sleptons}

The sleptons $({\stilde \nu}\>\>{\stilde e}_L )$ and ${\stilde e}^*_R$ (and the same for the other two generations), are the $0$ spin bosons supersymmetric partners of the leptons, as shown in table \ref{tab:chiral}.

When dealing with the squarks we saw how to arrive to the mass eigenstate basis by first performing a rotation from the eigenstate basis to an intermediate basis called SCKM, parallel to the rotation needed to diagonalize the quarks. When considering the leptons, the neutrinos in the MSSM as in the SM have no mass, thus this first rotation is not needed. Therefore when dealing with our chosen NMFV scenarios that incorporate general slepton flavour mixing, the sleptons will be diagonalized by one-step rotation matrices that will take them from the interaction eigenstate basis to the final mass basis. This will be treated carefully in Section \ref{paramnmfv}. A brief introduction about neutrino masses and mixing in the lepton sector, will be presented in Section \ref{smflavourlepton}.

\subsection{Experimental searches of MSSM particles and constraints}
\label{exp-searches-mssm-particles}

\begin{figure}[ht!]
\begin{center}
\includegraphics[width=210mm,angle=270]{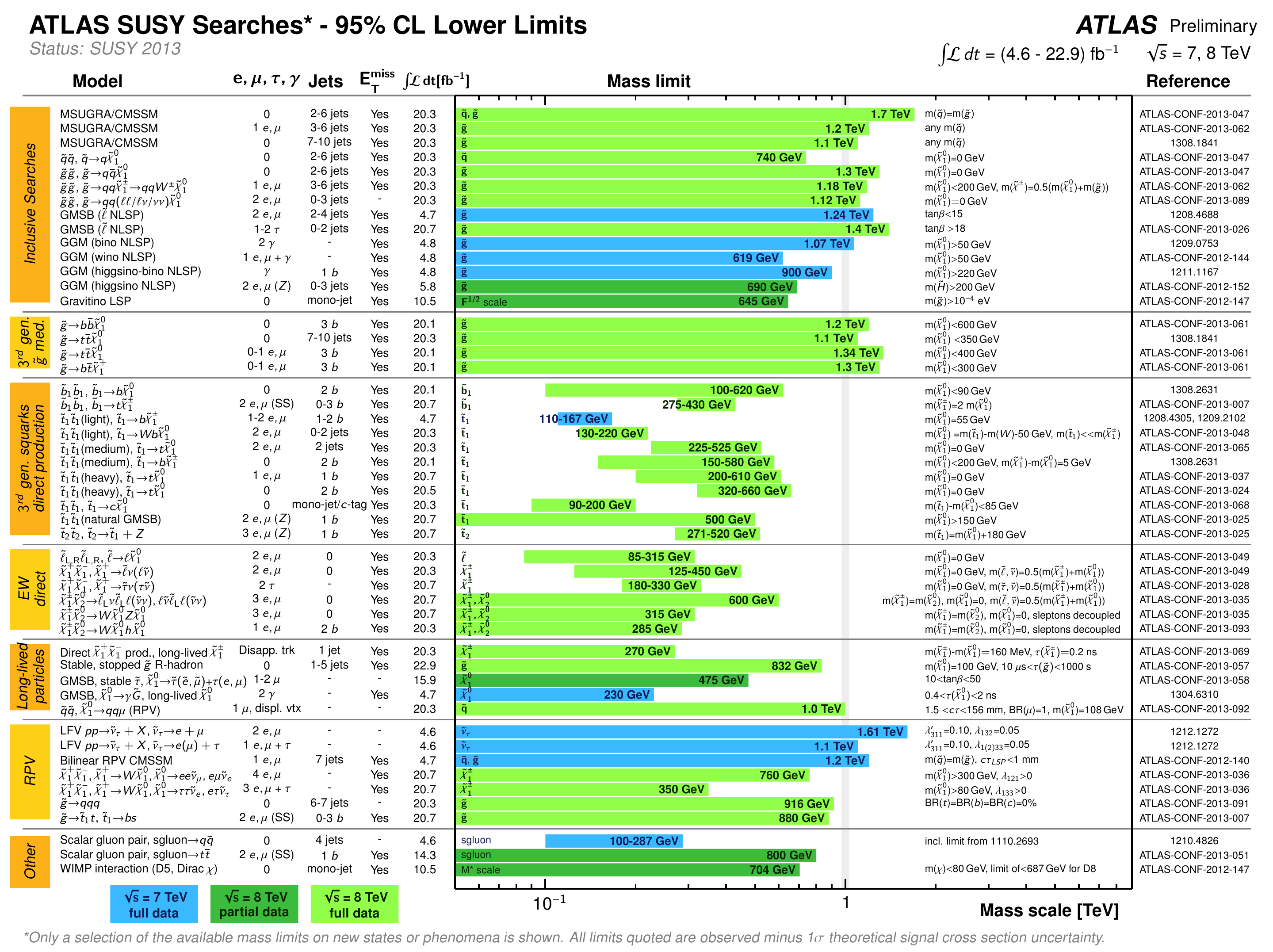}
\caption{Overall summary of SUSY searches in ATLAS. Plot taken from \cite{atlassusysearches}}
\label{atlassearchessusy1}
\end{center}
\end{figure}

In Figure \ref{atlassearchessusy1} and the top row of \ref{atlascmssearchessusy} there can be seen some of the present bounds for the SUSY masses from the sparticles searches by ATLAS \cite{atlassusysearches}. In the plot of Fig. \ref{atlassearchessusy1} it is collected a summary of the mass reach of a selection of searches for the 7 and 8 TeV data with a exclusion at the 95\% CL. In the top left plot of Fig. \ref{atlascmssearchessusy} is presented the 95\% CL exclusion limits for 8 TeV analyses for a simplified model where a pair of gluinos decays promptly via off-shell stop to four top quarks and two lightest neutralinos (LSP), in the gluino-neutralino mass plane. In the top right plot of Fig. \ref{atlascmssearchessusy} is shown the exclusion limits at 95\% CL from 8 TeV data from electroweak production of charginos and neutralinos and the decay via four different modes, in the chargino-neutralino mass plane (the details are in \cite{atlassusysearcheschargneutr}).

From these graphs, and taking in account than these limits are obtained under specific assumptions for each analysis, we see how the lower bound for the mass of the gluinos is around 1 TeV or even larger, and the one for the third generation squarks is located around 500 GeV (in some specific models not shown here is possible to have lighter squarks as for example stops around 200 GeV). In the plot of top left of Fig. \ref{atlascmssearchessusy} we see that depending on the mass of the LSP the lower bound for the gluino mass varies from 0.9 to 1.4 TeV. In the top right of Fig. \ref{atlascmssearchessusy} also a dependence on the LSP mass is shown for the chargino mass lower bound, varying it from 300 to 620 GeV, considering than the lightest chargino is degenerate with the second lightest neutralino, and a electroweak production of the LSP.

\begin{figure}[ht!]
\begin{center}
\includegraphics[width=210mm,angle=270]{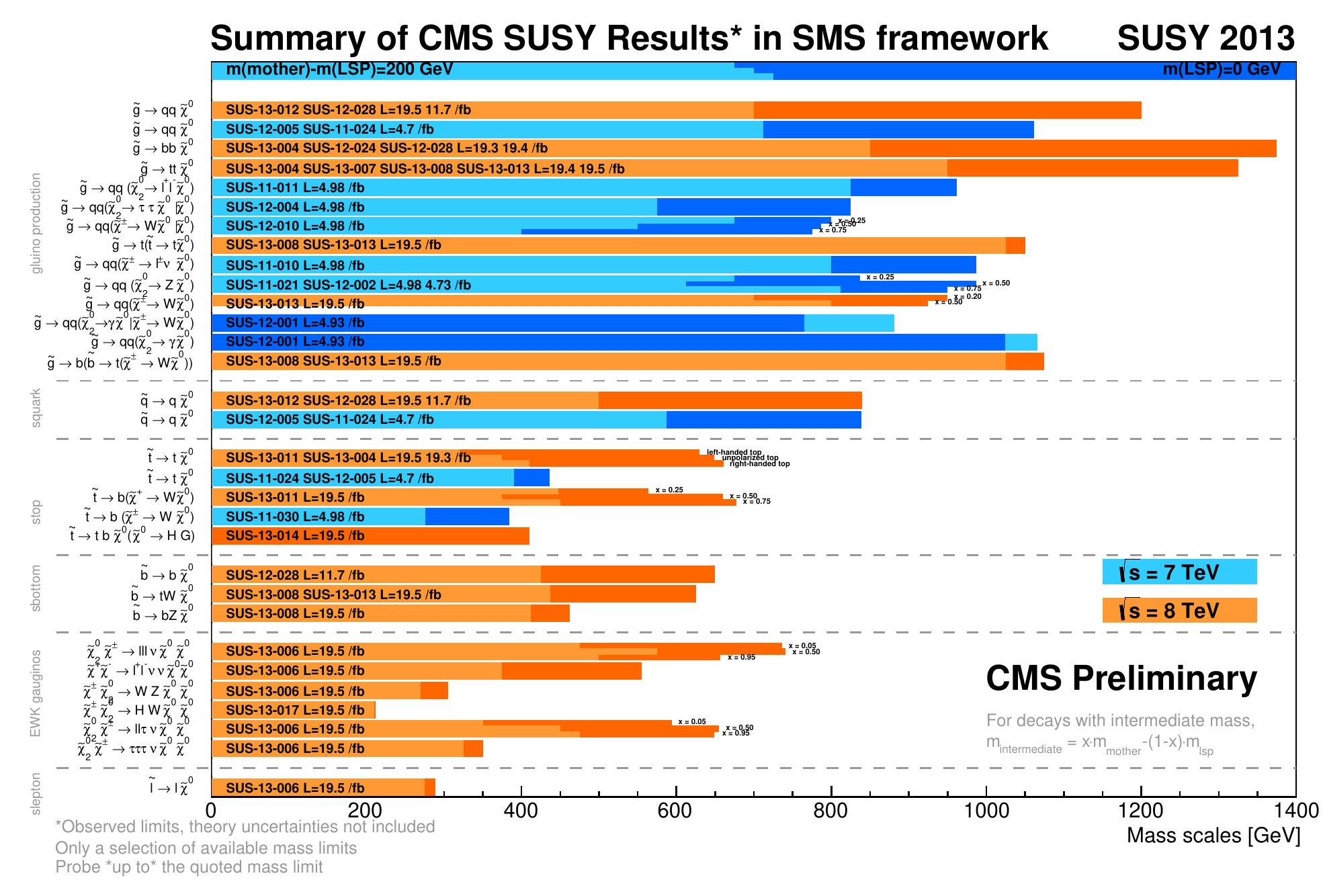}
\caption{Summary of SUSY searches for R-parity conserving scenarios in CMS. Plot taken from \cite{cmssusysearches}}
\label{cmssearchessusy1}
\end{center}
\end{figure}

\begin{figure}[ht!]
\begin{center}
\begin{tabular}{cc}
\includegraphics[width=80mm]{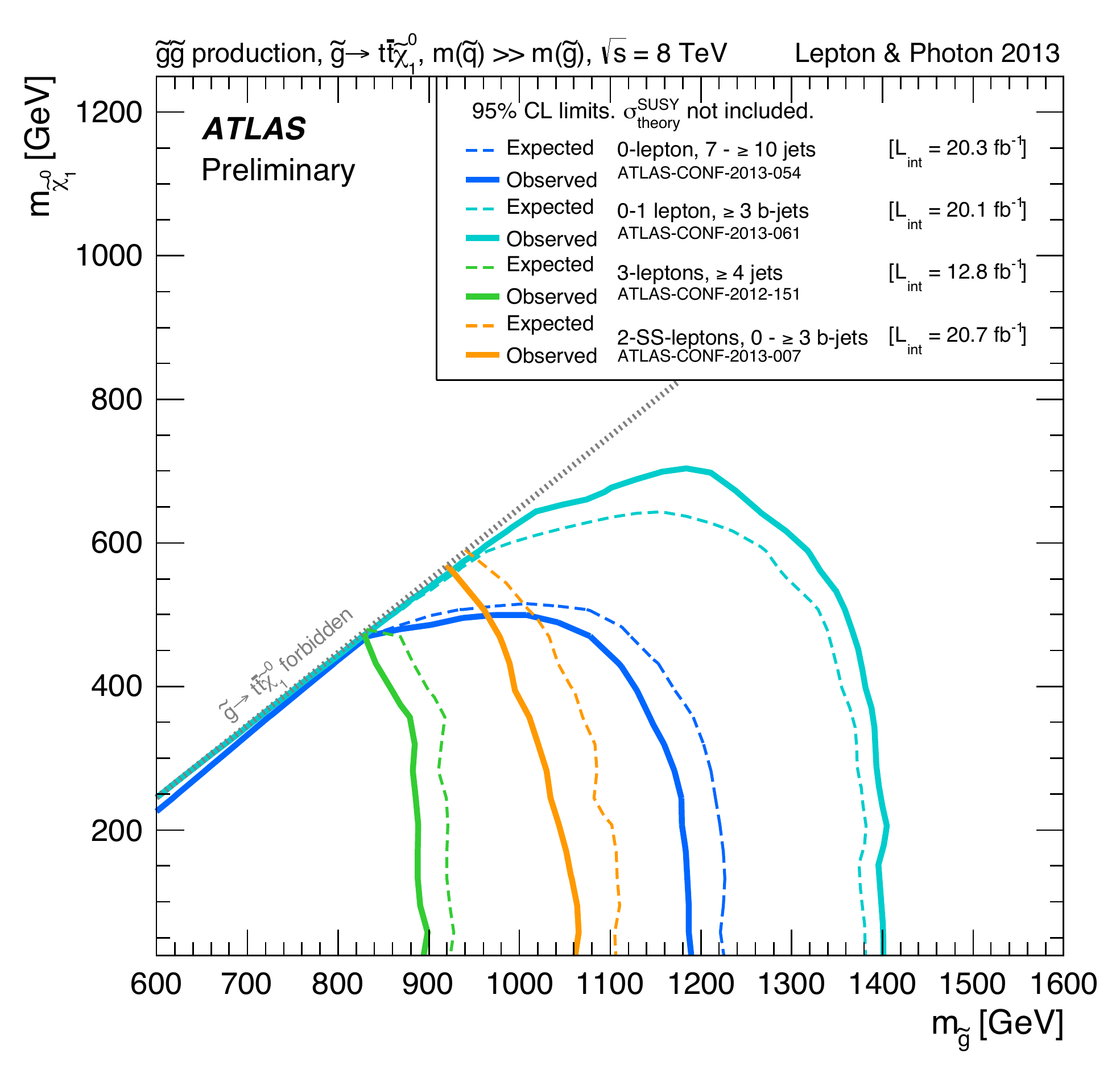} &
\includegraphics[width=80mm,height=78mm]{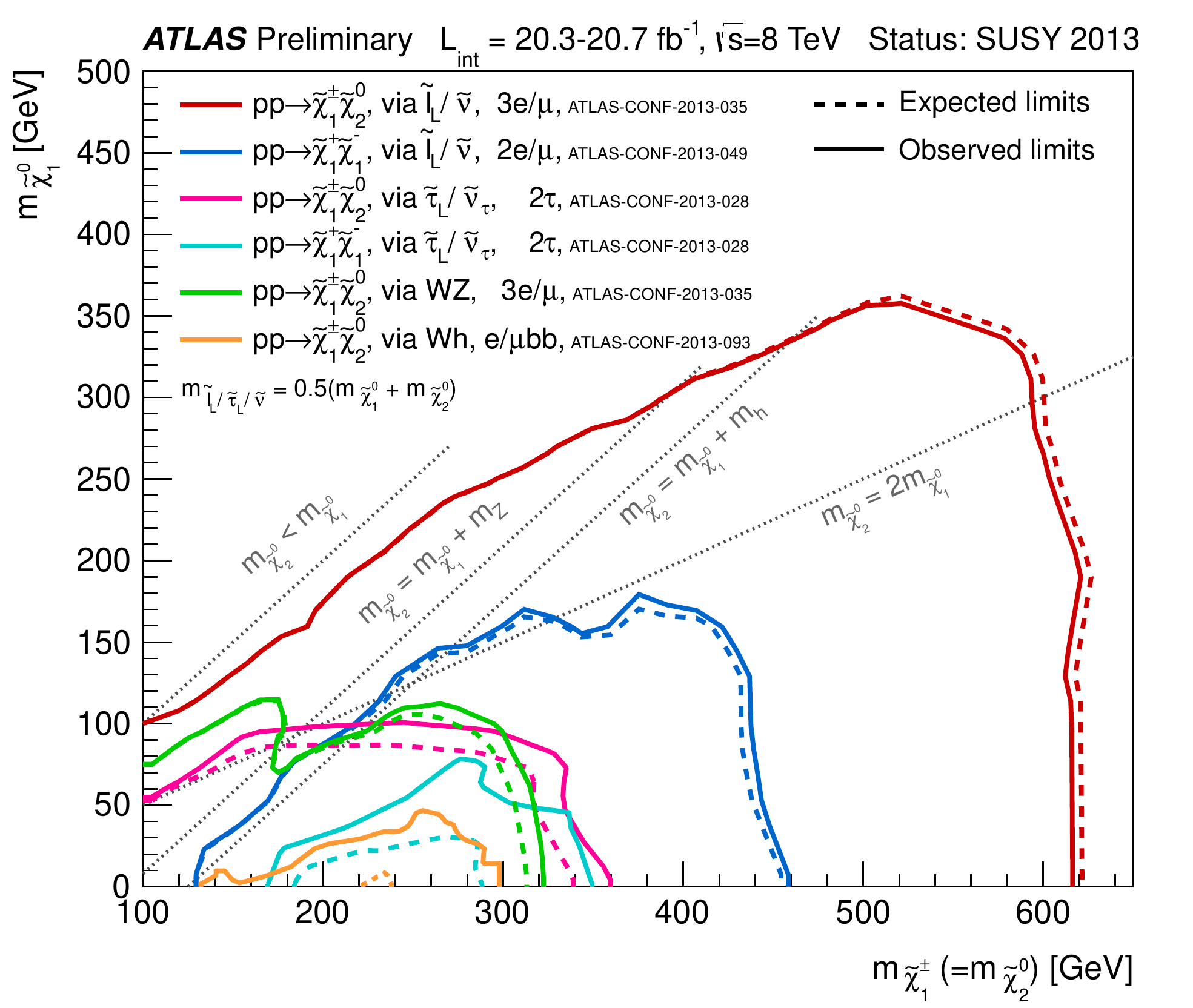} \\
\includegraphics[width=80mm]{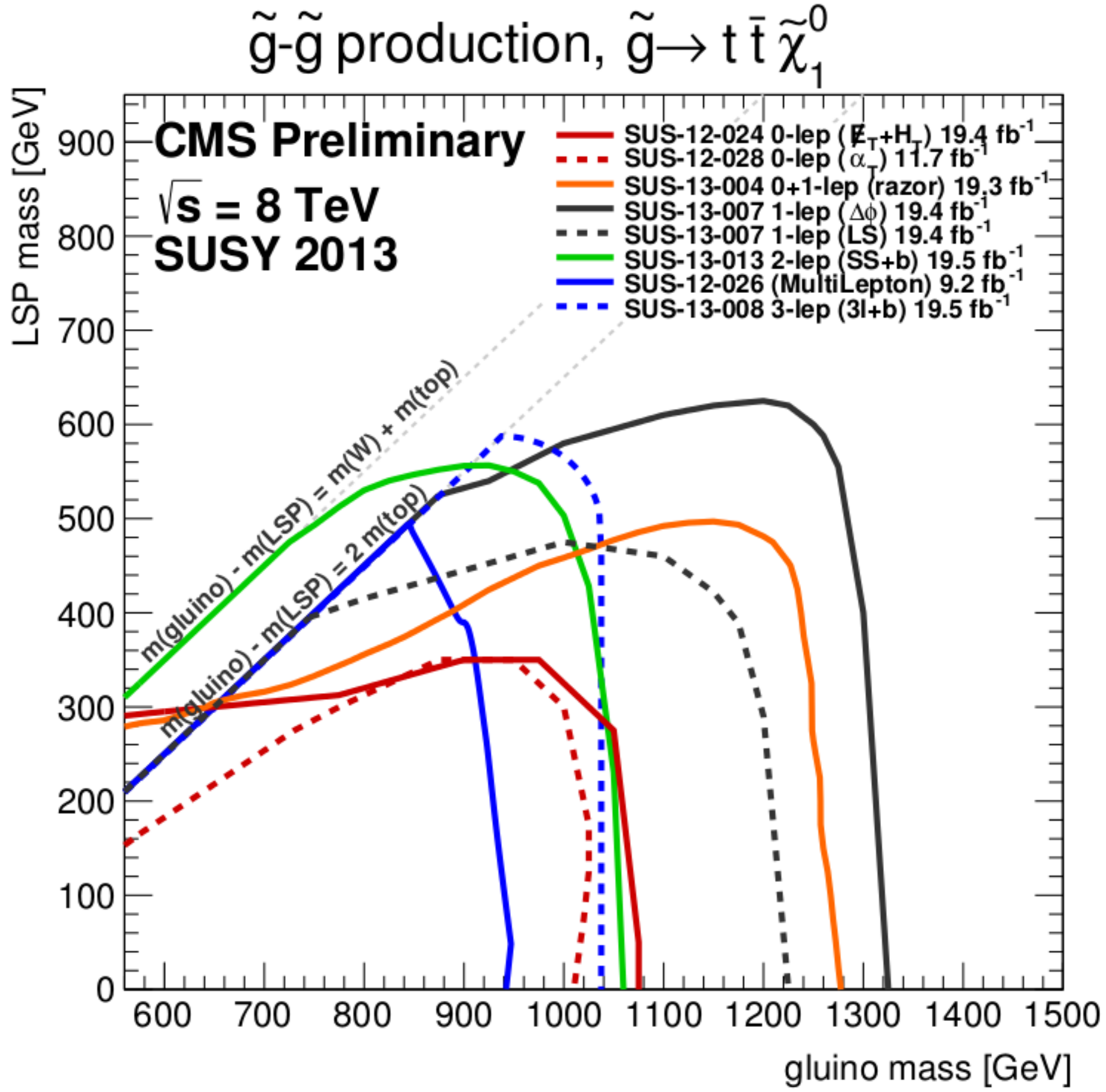} &
\includegraphics[width=80mm,height=80mm]{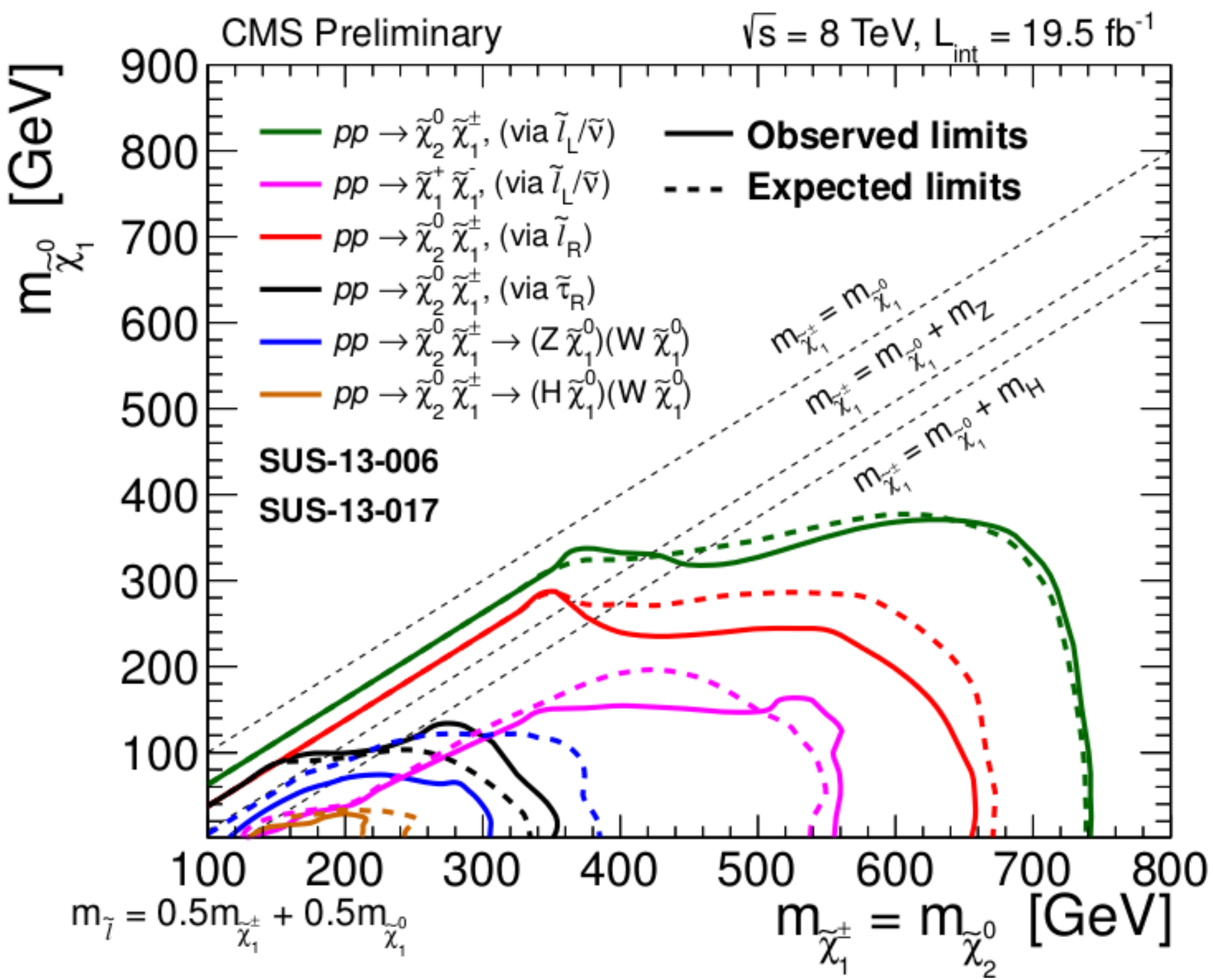} \\
\end{tabular}
\caption{SUSY searches in ATLAS (top row) and CMS (bottom row): Exclusion limits for gluino-mediated stop production decaying to a top and a neutralino (top left). Electroweak chargino-neutralino production (top right). Exclusion limits for gluino-mediated stop production decaying to a top and a neutralino (bottom left). Electroweak chargino-neutralino production (bottom right). More details in the text. Plots taken from \cite{atlassusysearches} and \cite{cmssusysearches}}
\label{atlascmssearchessusy}
\end{center}
\end{figure}

In Figure \ref{cmssearchessusy1} and the bottom row of Figure \ref{atlascmssearchessusy} there are presented the mass exclusion limits obtained by CMS in the search of sparticles \cite{cmssusysearches}. The plot in Figure \ref{cmssearchessusy1} shows a selection of mass exclusion limits for different sparticles in R-parity conserving scenarios for the 7 and 8 TeV data, with the conditions for the mother particle and the LSP: m(LSP) = 0 GeV (right bands) or m(mother) - m(LSP) = 200 GeV (left bands). The bottom left graph  of Figure \ref{atlascmssearchessusy} shows the exclusion limits for the 8 TeV analyses for a simplified model where a pair of gluinos decays promptly via off-shell stop to four top quarks and two lightest neutralinos (LSP), in the gluino-neutralino mass plane. In the bottom right plot of Figure \ref{atlascmssearchessusy} there can be seen the exclusion limits from 8 TeV data from electroweak production of charginos and neutralinos and the decay via six different modes, in the chargino-neutralino mass plane.

The results from CMS are very similar to the ones from ATLAS. The higher limits are in ATLAS as compared with CMS a bit heavier for some specific gluino cases, and a bit lower in some electroweak productions of charginos and neutralinos.

With respect to the Higgs bosons, since we are assuming that the discovered Higgs boson is the neutral lightest one of the MSSM, the other 4 Higgs bosons should still be found in the experiments. The charged Higgs have been searched for example through its decay to $\tau \nu$ \cite{Aad:2012tj,Aad:2012rjx,Chatrchyan:2012vca}, and through the decay to $c \bar s$ \cite{Aad:2013hla}, excluding masses at the 95\% CL from 90 to 160 GeV approximately. The heavy neutral Higgs bosons have been also searched, for example through the decays into a pair $\mu^+\mu^-$ or $\tau^+\tau^-$, excluding a significant part of the $M_A-\tan \beta$ parameter space (see results in \cite{Aad:2012cfr,Aad:2011rv,Chatrchyan:2012vp,Chatrchyan:2011nx,Aaij:2013nba}).

\chapter{The flavour of particles: a window to new physics}
\label{flavour}
Before entering into technical details, the word flavour already captures the core of the issue. Flavour suggests a quality of the objects that make them different one from each other, a different way to feel and interact with each of them, and behind it, a capricious arbitrariness in its existence and its diversity. Let us see how particles taste.

In the spectrum of particle physics, we find copies of the elementary fermions with the same SM gauge quantum numbers but with different masses. This leads to the concept of flavour, a new feature of particles. Within the SM we find six different flavours of quarks, up, charm and top flavours having electric charge $+2/3e$ and down, strange and bottom flavours having $-1/3e$. For leptons, there are three flavours, $e$, $\mu$ and $\tau$ with $-e$ electric charge, and three flavours $\nu_e$, $\nu_\mu$ and $\nu_\tau$ with zero charge. This opens a set of questions as: why the number of flavours is precisely this? what is the origin of flavour? why these mass and mixing patterns of leptons and quarks? is the mass the only difference between generations? why the rates of flavour changing neutral current processes are so much suppressed in Nature? These questions and others remain unanswered but have been studied exhaustively during many years, allowing us to know the structure of flavour of the SM  at present with precision and also its phenomenology. As other unknown answers, they open a path for new physics, and of course for the search of supersymmetry. In the following sections it will be shortly reviewed some flavour issues in the Standard Model and we will also comment on flavour beyond the SM.

\section{Flavour in the Standard Model}
\label{smflavour}

The fermions in the SM appear as 3 families of particles, with the same gauge properties, but with different masses. The model to describe it needs at least 13 parameters if we treat the neutrinos as massless: 9 fermion masses, 3 mixing angles and 1 complex phase; and one then adds 3 masses, 3 mixing angles and 1 or 3 phases if we treat the neutrinos as massive Dirac or Majorana particles respectively. 

We will start with a short review of the basic features of flavour in the quark sector, and after it we will focus on the lepton sector.

\subsection{Flavour in the quark sector}
\label{smflavourquark}

The mixing between the different flavour generations comes from the non-diagonality of the mass matrices in the interactions basis. This is due to the Yukawa interactions between the Higgs
bosons once they get a VEV and the quarks with non-diagonal Yukawa matrices. The relevant Lagrangian is:


\noindent \begin{equation}
\mathcal{L}_{YW}=-\bar{\cal Q}^{\rm int}_{L} Y^{u}\tilde \Phi {\cal U}^{\rm int}_{R}-\bar{\cal Q}^{\rm int}_{L} Y^{d} \Phi {\cal D}^{\rm int}_{R} +h.c.\end{equation}
where $Y^u$ and $Y^d$ are $3\times 3$ Yukawa coupling matrices, $\Phi=\left(\begin{array}{c} \phi^+ \\ \phi^0 \end{array}\right)$ is the SM Higgs doublet,  $\tilde \Phi=i\sigma_2\Phi^*=\left(\begin{array}{c} \phi^{0*} \\ -\phi^- \end{array}\right)$ and the interaction eigenstates are:

\noindent \begin{equation}
  {\cal Q}^{\rm int}_{L}=\left(\begin{array}{c} {\cal U}^{\rm int}_{L} \\ {\cal D}^{\rm int}_{L} \end{array}\right) \quad;\quad   {\cal U}^{\rm int}_{R}=\left(\begin{array}{c} u^{\rm int}_{R} \\ c^{\rm int}_{R} \\ t^{\rm int}_{R} \end{array}\right) \quad;\quad  {\cal D}^{\rm int}_{R}=\left(\begin{array}{c} d^{\rm int}_{R} \\ s^{\rm int}_{R} \\ b^{\rm int}_{R} \end{array}\right)
\end{equation}
where
\noindent \begin{equation}
 {\cal U}^{\rm int}_{L}=\left(\begin{array}{c} u^{\rm int}_{L} \\ c^{\rm int}_{L} \\ t^{\rm int}_{L} \end{array}\right) \quad;\quad  {\cal D}^{\rm int}_{L}=\left(\begin{array}{c} d^{\rm int}_{L} \\ s^{\rm int}_{L} \\ b^{\rm int}_{L} \end{array}\right).
\end{equation}
To obtain the mass eigenstates one rotates the quarks to the physical mass
basis with the following rotations:

\begin{equation}
\VL u^{\rm phys}_{L,R} \\ c^{\rm phys}_{L,R} \\ t^{\rm phys}_{L,R} \VR =
V^u_{L,R} \VL u^{\rm int}_{L,R} \\ c^{\rm int}_{L,R} \\ t^{\rm int}_{L,R} \VR~,~~~~
\VL d^{\rm phys}_{L,R} \\ s^{\rm phys}_{L,R} \\ b^{\rm phys}_{L,R} \VR =
V^d_{L,R} \VL d^{\rm int}_{L,R} \\ s^{\rm int}_{L,R} \\ b^{\rm int}_{L,R} \VR~,
\end{equation}
and then one obtains the diagonal quark masses and Yukawa couplings:

\begin{eqnarray}
\frac{v}{\sqrt 2}{\rm diag}\left(m_u,m_c,m_t\right)&=&{\rm diag}(y_u,y_c,y_t)=
V^u_L Y^{u*}V^{u\dagger}_R,
\\
\frac{v}{\sqrt 2}{\rm diag}\left(m_d,m_sm_b\right)&=&{\rm diag}(y_d,y_s,y_b)=
V^d_L Y^{d*}V^{d\dagger}_R.
\end{eqnarray} 
where $v=\langle \phi^0 \rangle$ is the VEV of the SM Higgs boson.

With respect to the electroweak forces, the photon and the $Z$ boson mediated currents conserve flavour, but the $W^\pm$ ones connect the different generations. This can be seen in the Lagrangian after the change of basis as follows. The relevant Lagrangian of the interactions between the gauge bosons and the
fermions in the SM is given by:

\noindent \begin{eqnarray}
\mathcal{L}_{\psi} & = & \sum_{\psi}i\bar{\psi}\gamma^{\mu}D_{\mu}\psi=\sum_{\psi}i\bar{\psi}\gamma^{\mu}\left(\partial_{\mu}-igT\cdot W_{\mu}-ig^{\prime}Y B_{\mu}\right)\psi\nonumber \\
 & = & i\bar{\psi}\gamma^{\mu}\left\{ \left(\begin{array}{cc}
\partial_{\mu} & 0\\
0 & \partial_{\mu}\end{array}\right)+\left(\begin{array}{cc}
0 & \frac{-ig}{2}\left(W_{1\mu}-iW_{2\mu}\right)\\
\frac{-ig}{2}\left(W_{1\mu}+iW_{2\mu}\right) & 0\end{array}\right)\right.\nonumber \\
 &  & \left.+\left(\begin{array}{cc}
\frac{-ig}{2}W_{3\mu} & 0\\
0 & \frac{ig}{2}W_{3\mu}\end{array}\right)+\left(\begin{array}{cc}
-ig^{\prime}Y B_{\mu} & 0\\
0 & -ig^{\prime}Y B_{\mu}\end{array}\right)\right\} \psi\end{eqnarray}

After breaking the electroweak symmetry the physical states are

\noindent \begin{equation}
W_{\mu}^{\pm}\equiv\frac{1}{\sqrt{2}}\left(W_{\mu}^{1}\mp iW_{\mu}^{^{2}}\right)\quad;\quad Z_{\mu}\equiv\cos\theta_{W}W_{\mu}^{^{3}}-\sin\theta_{W}B_{\mu}\quad;\quad A_{\mu}\equiv\sin\theta_{W}W_{\mu}^{^{3}}+\cos\theta_{W}B_{\mu}\end{equation}
where the gauge couplings are $g=\frac{e}{\sin\theta_{W}}$, $g^{\prime}=\frac{e}{\cos\theta_{W}}$. In the electroweak basis for the quark states the weak interaction Lagrangian is then:

\noindent \begin{eqnarray}
\mathcal{L}^{\rm int}_{W}&=&\frac{g}{\sqrt 2}\left\{ W^{+}_\mu\bar{\cal{U}}^{\rm int}_{L}\gamma_{\mu}{\cal{D}}^{\rm int}_{L} +  W^{-}_\mu\bar{\cal{D}}^{\rm int}_{L}\gamma_{\mu}{\cal{U}}^{\rm int}_{L} \right\} +\nonumber\\
& & \frac{g}{c_W}Z_{\mu}\left\{ g^u_L \bar{\cal{U}}^{\rm int}_{L}\gamma_{\mu}{\cal{U}}^{\rm int}_{L} +  g^u_R \bar{\cal{U}}^{\rm int}_{R}\gamma_{\mu}{\cal{U}}^{\rm int}_{R} + g^d_L \bar{\cal{D}}^{\rm int}_{L}\gamma_{\mu}{\cal{D}}^{\rm int}_{L} +  g^d_R \bar{\cal{D}}^{\rm int}_{R}\gamma_{\mu}{\cal{D}}^{\rm int}_{R}  \right\}
\end{eqnarray}
where  $g^u_L=(\frac{1}{2}-\frac{2}{3}s_w^2)$,  $g^u_R=-\frac{2}{3} s_w^2 $, $g^d_L=(-\frac{1}{2}+\frac{1}{3} s_w^2) $ and $g^d_R=\frac{1}{3} s_w^2 $. Finally, after the rotation to the mass basis the neutral currents remain flavour diagonal, but charged currents change, giving rise to intergenerational
interactions due to the non-diagonal matrix $V_{CKM}$: 

\noindent \begin{eqnarray}
\mathcal{L}^{\rm int}_{W}&=&\frac{g}{\sqrt 2}\left\{ W^{+}_\mu\bar{\cal{U}}^{\rm phys}_{L}\gamma_{\mu}V_{CKM}{\cal{D}}^{\rm phys}_{L} +  W^{-}_\mu\bar{\cal{D}}^{\rm phys}_{L}\gamma_{\mu}V^+_{CKM}{\cal{U}}^{\rm phys}_{L} \right\} +\nonumber\\
& & \frac{g}{c_W}Z_{\mu}\left\{ g^u_L \bar{\cal{U}}^{\rm phys}_{L}\gamma_{\mu}{\cal{U}}^{\rm phys}_{L} +  g^u_R \bar{\cal{U}}^{\rm phys}_{R}\gamma_{\mu}{\cal{U}}^{\rm phys}_{R} + \right. \nonumber\\
& & \left. \phantom{\frac{g}{c_W}Z_{\mu}\{} g^d_L \bar{\cal{D}}^{\rm phys}_{L}\gamma_{\mu}{\cal{D}}^{\rm phys}_{L} +  g^d_R \bar{\cal{D}}^{\rm phys}_{R}\gamma_{\mu}{\cal{D}}^{\rm phys}_{R}  \right\},
\end{eqnarray}
where,
\noindent \begin{equation}
\VCKM= V^u_L V^{d\dagger}_L.
\end{equation}
This change of flavour among generations in the SM is therefore encoded in the Cabibbo-Kobayashi-Maskawa (CKM) matrix \cite{Cabibbo:1963yz,Kobayashi:1973fv} $V_{CKM}$.

Although this flavour changing appears in the SM at tree level only in the charged
currents, at one loop level it is also present in the neutral currents.
However, it is a very tiny effect due to the GIM-mechanism \cite{Glashow:1970gm},
which leads to suppression factors of flavour changing neutral currents
(FCNC) given by the small ratio of the involved squared mass quark differences divided by the
squared $W^{\pm}$ mass. For the third generation this ratio is not so small,
but is compensated by the corresponding entries in the CKM matrix involving the mixings between the third and
the other generations that are small.\\

\subsubsection{The CKM matrix}
\label{ckmatrix}

The CKM matrix is a unitary $3\times 3$ matrix, and it is  parametrized in a standard way as:

\begin{equation}
V_{\rm CKM}=
\left(\begin{array}{ccc}
V_{ud}&V_{us}&V_{ub}\\
V_{cd}&V_{cs}&V_{cb}\\
V_{td}&V_{ts}&V_{tb}
\end{array}\right)=
\left(\begin{array}{ccc}
c_{12}c_{13}&s_{12}c_{13}&s_{13}e^{-i\delta}\\ -s_{12}c_{23}
-c_{12}s_{23}s_{13}e^{i\delta}&c_{12}c_{23}-s_{12}s_{23}s_{13}e^{i\delta}&
s_{23}c_{13}\\ s_{12}s_{23}-c_{12}c_{23}s_{13}e^{i\delta}&-s_{23}c_{12}
-s_{12}c_{23}s_{13}e^{i\delta}&c_{23}c_{13}
\end{array}\right)\,,
\end{equation}
where $c_{ij}=\cos\theta_{ij}$, $s_{ij}=\sin\theta_{ij}$
($i,j=1,2,3$), $\theta_{12}$, $\theta_{13}$ and $\theta_{23}$ are the 3 real mixing angles, and $\delta$ is the $\cp$ violation phase. $c_{ij}$
and $s_{ij}$ can all be chosen to be positive and $\delta$ may vary in
the range $0\le\delta\le 2\pi$. 

Given that $\theta_{12}$, $\theta_{13}$ and $\theta_{23}$ are known to be small from experiment, there is a hierarchy among the CKM elements, so it is common to use another parametrization that reflects this. This is the Wolfenstein version~\cite{Wolfenstein:1983yz} defined by
\begin{equation}
V_{\rm CKM}=
\left(\begin{array}{ccc}
1-{\lambda^2\over 2}&\lambda&A\lambda^3(\varrho-i\eta)\\ -\lambda&
1-{\lambda^2\over 2}&A\lambda^2\\ A\lambda^3(1-\varrho-i\eta)&-A\lambda^2&
1\end{array}\right)
+{\cal{O}}(\lambda^4)\,,
\end{equation}
being it an expansion on $\lambda\sim |V_{us}|\approx 0.22$. This is a more phenomenological parametrization, using in its structure information about
the relative sizes of the different terms of the matrix.

These last set of parameters can be chosen in different ways, but the standard definition is the following~\cite{Buras:1994ec}:
\begin{equation}
\lambda\equiv s_{12}\,,
\qquad
A \lambda^2\equiv s_{23}\,,
\qquad
A \lambda^3 (\varrho-i \eta)\equiv s_{13} e^{-i\delta}
\end{equation}
and therefore
\begin{equation}
\varrho=\frac{s_{13}}{s_{12}s_{23}}\cos\delta,
\qquad
\eta=\frac{s_{13}}{s_{12}s_{23}}\sin\delta.
\end{equation}

The elements of the CKM matrix are obtained through different measurements, for a review and references to each measurement see for instance \cite{Beringer:1900zz:CKM}. In Table \ref{tab:ckmelements} can be found a summary of all the elements and their related processes:

$|V_{ud}|$ is obtained from superallowed (between two members of an isospin multiplet) $0^+\rightarrow 0^+$ nuclear decays which is theoretically clean. They depend only on the vector part of the weak interaction, and therefore taking into account the Conserved Vector Current (CVC) hypothesis, the value should not depend on the specific nuclei involved and its nuclear structure values and uncertainties. This has been measured for example in the PIBETA experiment \cite{Pocanic:2003pf}.
The CKM element can be also obtained from the neutron lifetime.

$|V_{us}|$ is extracted from different measurements: without relying in other CKM elements from various $K\to\pi\ell\nu$ decays (to ensure that lifetimes and form factors are not affecting the value) or from  $\tau \to K \nu_\tau$; from the ratio $|V_{us}|/|V_{ud}|$ from the decays  $K\to \mu\nu/\pi\to\mu\nu$ and  $\tau\to K\nu/\tau \to \pi\nu$. Measurements can be found for example from KLOE \cite{Ambrosino:2005fw} or BaBar \cite{Aubert:2009qj}. 

$|V_{cd}|$ can be obtained from $D \to \pi l \nu$ (as in CLEO \cite{Besson:2009uv} and Belle \cite{Widhalm:2006wz})and from the most precise neutrino interactions as $\nu N \to c \mu^+ \mu^- X$ (in CHORUS \cite{KayisTopaksu:2005je}).

$|V_{cs}|$ is measured from $D_S$ decays in leptons as $D_s\to \tau\nu$ and $D_s\to \mu\nu$, as the average made from the measurements of CLEO, BaBar and Belle \cite{hfag:charm10}. To calculate the value of the CKM element, other quantities are needed, as for example the decay constant $f_{D_s}$, calculated using lattice QCD, and whose uncertainty dominates the error. Another way to obtain the element is through the semileptonic decay $D \to K l \nu$, as measured by BaBar \cite{Aubert:2007wg}, Belle \cite{Widhalm:2006wz} and CLEO \cite{Besson:2009uv}.

$|V_{cb}|$ can be determined from exclusive and inclusive semileptonic decays of $B$ mesons to charm, being the inclusive ones the most precise. A review of the measurements and an average can be read in \cite{Kowalewski:vcbub}.

$|V_{ub}|$ can be obtained as the previous element, from semileptonic $B$ decays (see the same reference \cite{Kowalewski:vcbub}), but also from the $B\to \tau\nu$ decay.

$|V_{td}|$ and $|V_{ts}|$ can not be measured from the equivalent decays of the previous cases to the top case. In general these are not likely to be precisely measured in tree level decays, but can be obtained from $B-\bar B$ oscillations or rare decays of $B$ or $K$ mesons, with tops in the internal loops. Ratios of processes involving the two elements are measured to eliminate hadronic uncertainties present in the measurement of each CKM element by itself. From these ratios it can be extracted the ratio of the CKM elements. $B\rightarrow X_s\gamma$ and $B_s\rightarrow\mu^+\mu^-$ are sensitive to the product of $|V_{ts}|$ and another element. Other decays could be used, but with much less precision.

$|V_{tb}|$ can be measured directly from the single top quark production cross section as measured by D0 \cite{Abazov:2011pt}, CDF \cite{cdf:vtb} and CMS \cite{Chatrchyan:2011vp}. Bounds to its value can be obtained from the top decay into W.

\begin{table}[h!]
\begin{center}
\vspace{1 cm}
\begin{tabular}{|c|c|}
\hline
CKM  & Process \\
\hline
$|V_{ud}|$ & $0^+\to 0^+$ transitions  \\
                    \hline
$|V_{us}|$ & $K\to\pi\ell\nu$ \\
                 &  $\tau \to K \nu_\tau$\\
                 \hline
$|V_{us}|/|V_{ud}|$                 &  $K\to \mu\nu/\pi\to\mu\nu$\\
                 &  $\tau\to K\nu/\tau \to \pi\nu$\\
                 \hline
$|V_{cd}|$ &  $D \to \pi l \nu$\\
                 &  $\nu N \to c \mu^+ \mu^- X$\\
                   \hline
$|V_{cs}|$   & $D_s\to \tau\nu$\\
                   & $D_s\to \mu\nu$\\
                   & $D \to K l \nu$\\
                   \hline
$|V_{ub}|$ & semileptonic $B$ decays \\
                  & $B\to \tau\nu$\\
                  \hline
$|V_{cb}|$ & semileptonic $B$ decays\\
\hline
$|V_{td}|$ and $|V_{ts}|$  & $\Delta M_{B_d}$ \\ 
                 & $\Delta M_{B_s}$\\
                 & rare $B$ and $K$ decays\\
                 \hline
$|V_{tb}|$ 
      & top production\\
\hline
\end{tabular}
\caption{CKM elements and main processes where their values are measured \label{tab:ckmelements}}
\end{center}
\end{table}

\subsubsection{The unitarity triangles}
\label{unittriangl}

\begin{figure}[ht!]
\begin{center}
\vspace{1 cm}
\includegraphics[width=145mm]{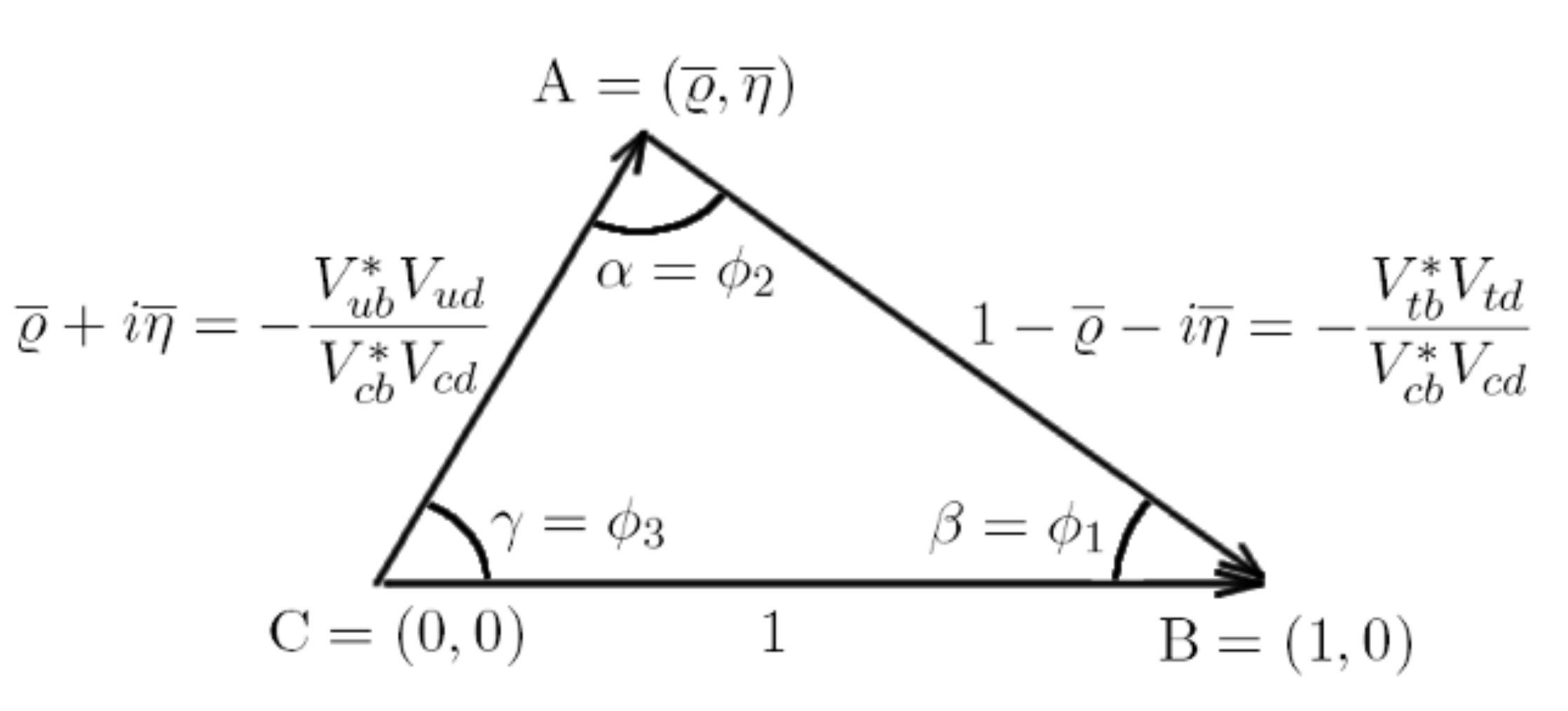}
\caption{Sketch of the unitarity triangle}
\label{uttriangle}
\end{center}
\end{figure}

Although we can not predict the values of the CKM elements, we know that if there is no new physics beyond the SM the matrix should be unitary, and therefore
checking its unitarity is a way to look for physics beyond the SM. The unitarity conditions are usually represented in what is called unitarity triangles. These conditions imply:

\begin{eqnarray}\label{utrels}
\sum_{\alpha=1}^3 V_{\alpha i}V_{\alpha j}^*  = 0~,~~i\ne j, \\
\sum_{i=1}^3 V_{\alpha i}V_{\beta i}^*  = 0~,~~\alpha\ne \beta.
\end{eqnarray}
where Greek subscripts run over the up-type quarks $u,c$ and
$t$, while Latin ones run over the down-type quarks $d,s$ and $b$. From these equations it follows for example:

\begin{equation}\label{utrelation}
V_{ud}^{}V_{ub}^* + V_{cd}^{}V_{cb}^* + V_{td}^{}V_{tb}^* =0.
\end{equation}
This condition (and the others six of \ref{utrels}) can be represented in the complex plane by a triangle as the one in figure \ref{uttriangle}
where 
\begin{eqnarray}
\overline\varrho+i\overline\eta\equiv\overrightarrow{\mathrm{CA}}=-\frac{V^*_{ub}V_{ud}}{V^*_{cb}V_{cd}},\\
\overrightarrow{\mathrm{AB}}=-\frac{V^*_{tb}V_{td}}{V^*_{cb}V_{cd}}=1-\overline\varrho-i\overline\eta\,,\nonumber\\
\overrightarrow{\mathrm{CB}}=1\,.
\end{eqnarray}
and thus the sides and angles of the triangles are determined by the CKM elements, and obtained from experiments. If we find that any of these triangles do not close that would be a signal of new physics. The areas of all the six triangles are the same, and it is a measurement of $\cp$ violation 

\begin{figure}[ht!]
\begin{center}
\begin{tabular}{c}
\includegraphics[width=106mm]{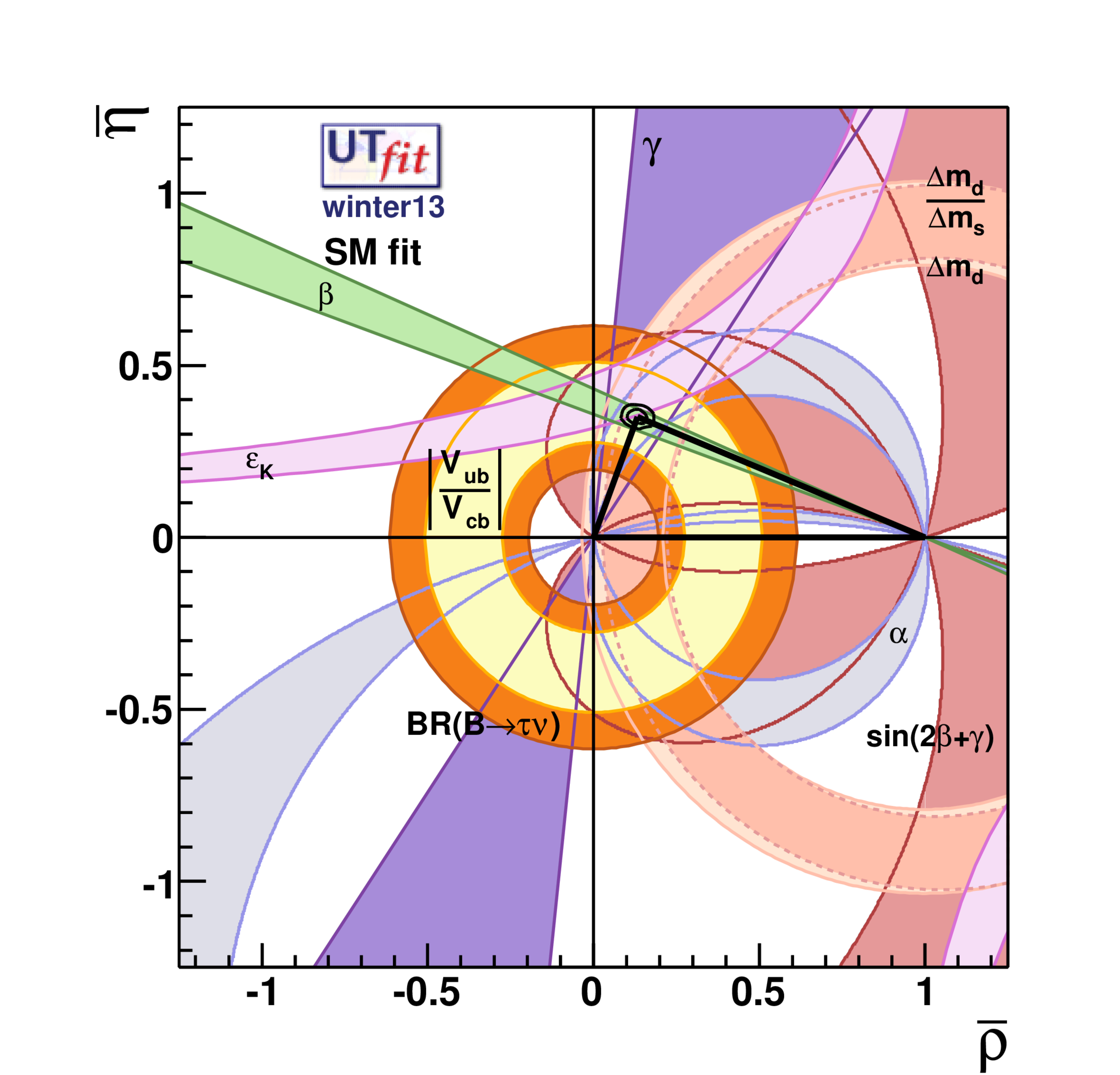}   \\
\includegraphics[width=98mm]{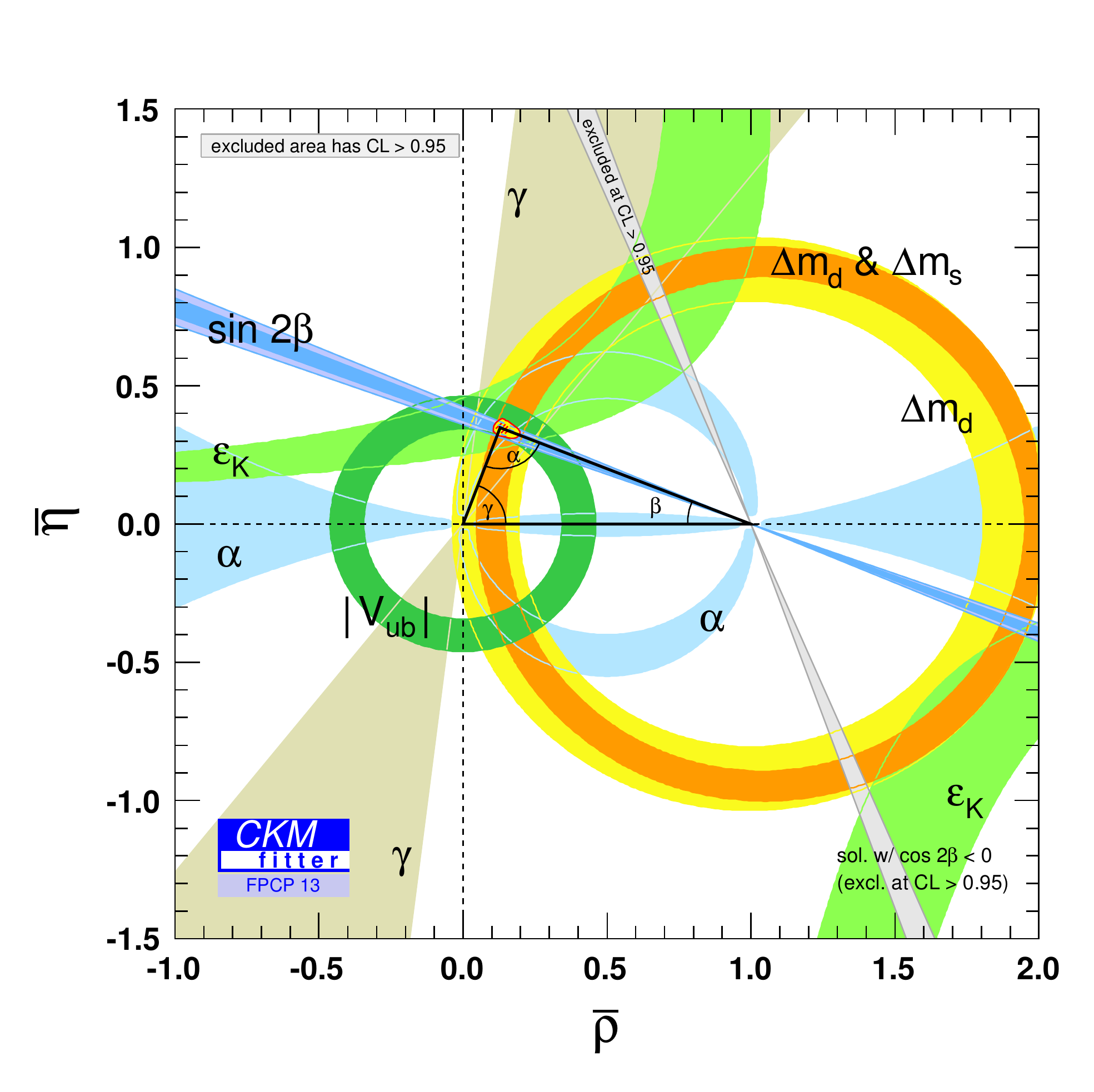}  \\
\end{tabular}
\caption{Unitarity triangles in the $\overline\varrho-\overline\eta$ plane from the {\bf UT}{\it fit} \cite{utfitres2013} (top) and CKMfitter \cite{ckmfitterres2013} (bottom) collaborations.}
\label{unittrianglescollab}
\end{center}
\end{figure}

The triangles can be determined by many different experimental measurements. Combining all these data with their different sources, methods, sizes, and errors is a complex work. The two main groups that do it are the {\bf UT}{\it fit} \cite{utfitweb} and the CKMfitter \cite{ckmfitterweb}. The first one uses a Bayesian approach to combine the data (explained for example in \cite{Ciuchini:2000de}) while the second one uses a frequentist approach with a scheme to treat the theoretical systematics called Rfit (explained in \cite{Charles:2004jd}). The last results for the two collaborations are collected in Figure \ref{unittrianglescollab} (\cite{utfitres2013} Winter 2013/Pre-Moriond 2013 for {\bf UT}{\it fit}; and \cite{ckmfitterres2013} FPCP13 for CKMfitter). The results for {\bf UT}{\it fit} can be found in \cite{utfitres2013}. The last results and inputs used for CKMfitter can be found in \cite{ckmfitterres2013sum}.

The plots from figure \ref{unittrianglescollab} illustrate the constraints in the $\overline\varrho-\overline\eta$ plane from various measurements and the global fit result. The shaded 95\% CL regions are shown and one can see that they all overlap consistently around the global fit region, leaving not many space for new physics.

For completeness, we review in the following how the angles of the triangle are obtained (see for instance \cite{Derkach:2013da} about this topic).

The angle $\alpha$ is obtained from charmless hadronic $B$ decays. This angle can not be measured directly, it should be extracted from measurements where other penguin diagrams not related with $\alpha$ appear. This is done for example in the $B\rightarrow\pi\pi$ system, as explained in \cite{Charles:2006vd}, by measuring six different quantities related with the three processes $B^0\rightarrow\pi^+\pi^-$,  $B^0\rightarrow\pi^0\pi^0$, and $B^+\rightarrow\pi^+\pi^0$ and then, by using the approximate isospin symmetry, combining them to extract the angle. 
From the decays $B\rightarrow\rho\rho$ and $B^0\rightarrow\rho^0\pi^0$ it can be also extracted, using basically the same method.

The angle $\gamma$ is one of the parameters of the fit known with less precision. It can be obtained from tree dominated decays $B\rightarrow D K$. These measurements have the advantage of having almost zero theoretical uncertainty when the hadronic parameters needed are also obtained from the data. The main techniques for measuring $\gamma$ are the GWS method \cite{Gronau:1990ra,Gronau:1991dp} using $D^0$ decays to $\cp$ eigenstates, the ADS method \cite{Atwood:1996ci,Atwood:2000ck} similar but where the $\cp$ asymmetry between the amplitudes involved is larger than in the last method, and the Dalitz plots analyzes used for example by Belle in \cite{Abe:2008wya} being the precision on the angle dominated by these last ones.

The angle $\beta$ can be obtained directly by $b\rightarrow c \bar c s$ decays (as $B^0\rightarrow {J\mskip -3mu/\mskip -2mu\psi\mskip 2mu} K^0_S$) or by $b \rightarrow q \bar q s$ penguin-dominated decay modes. The first ones are theoretically and experimentally very clean channels. These have been measured for example by LHCb \cite{Aaij:2012ke} (the most precise measurement of $\sin 2\beta$ at a hadron collider) or Belle \cite{Adachi:2012et}. The seconds have been measured for example by BaBar \cite{Lees:2011nf}. 
The measurement of $\sin 2\beta$ has been dominated by the $B$ factories, but it is currently being matched by the LHCb.

In summary, the current values of the CKM elements, obtained from a global fit using all the available measurements and imposing the SM constraints, are collected in the following matrix \cite{Beringer:1900zz:CKM}:
\begin{equation}
V_{\rm CKM}=
\left(\begin{array}{ccc}
0.97427 \pm 0.00015 & 0.22534 \pm 0.00065 & 0.00351^{+0.00015}_{-0.00014} \\
0.22520 \pm 0.00065 & 0.97344 \pm 0.00016 & 0.0412^{+0.0011}_{-0.0005}  \\
0.00867^{+0.00029}_{-0.00031} & 0.0404^{+0.0011}_{-0.0005} & 0.999146^{+0.000021}_{-0.000046}
\end{array}\right).
\end{equation}

\subsection{Flavour in the lepton sector}
\label{smflavourlepton}

The SM considers the neutrinos as massless particles, and all SM interactions preserve Lepton Flavour number. Therefore the SM predicts zero rates for all the Lepton Flavour Violating (LFV) processes to all orders in perturbation theory. However we know from experiments that the Lepton Flavour number is violated in the neutrino sector since the neutrinos oscillate, and thus models of new physics beyond the SM should be proposed to generate neutrino masses and oscillations in flavour to be in agreement with data.

The small mass of the neutrinos could be generated by introducing right-handed neutrinos (sterile neutrinos, without gauge interactions) and by adding small Dirac (via tiny neutrino Yukawa couplings) or/and Majorana mass terms in the Lagrangian, but then a new problem of fine-tuning appears, since these terms should produce unnaturally small values for the mass of the neutrinos. There are mechanisms that solve this issue as for instance the seesaw mechanism \cite{Minkowski:1977sc,GellMann:1980vs,VanNieuwenhuizen:1979hm,Yanagida:1979as,Sawada:1979gf,Glashow:seesaw,Mohapatra:1979ia,Barbieri:1979ag,Marshak:1980yc,Cheng:1980qt,Magg:1980ut,Lazarides:1980nt,Schechter:1980gr,Mohapatra:1980yp,Ma:1998dx}. The seesaw mechanism introduces two different natural mass scales: a very heavy Majorana mass $m_M$ from the high energy scale theory (usually set at $m_M \sim {\cal O} (10^{14}-10^{15} GeV)$) and a Dirac mass of the order of the electroweak scale ($m_D\sim Y_\nu\langle \phi \rangle$), that produce then very heavy (at the $m_M$ scale) and very light (at or below the eV scale) neutrinos in the final mass basis. However, for the purpose of the present work, we will not assume any particular model nor mechanism for neutrino mass generation.

About the LFV processes, they have been observed only in neutrino oscillations, but not in charged leptons. When
extending the SM to include neutrino masses and neutrino mixings in
agreement with the observed experimental values~\cite{pdg}, LFV
processes with external charged leptons of different generations can
then occur via one-loop diagrams with neutrinos in the internal
propagators, but the predicted rates are extremely tiny, far from being
in the foreseeable future reachable experimentally, due to the small masses of the
neutrinos. Therefore, a potential future
measurement of any of these 
(charged) LFV processes will be a clear signal of new physics and will
provide interesting information on the involved flavour mixing, as well
as on the underlying origin for this mixing (for a review see, for instance, \cite{Kuno:1999jp}). 
Thus, LFV processes provide one of the most 
challenging probes to physics beyond the SM of particle
physics, and in particular to new physics involving non-vanishing flavour
mixing between the three generations. In the rest of this work, we will use the term LFV to refer specifically to the charged lepton flavour violating processes, as it is common in the literature.

Without focusing in any particular model to produce the neutrino masses and mixings, that is beyond the scope of this work, we know from the experiments where $\nu$ oscillations have been assumed that the neutrinos get rotated from the interaction basis to the mass basis with a matrix equivalent to the CKM matrix of the quark sector, called here the Pontecorvo-Maki-Nakagawa-Sakata (PMNS) matrix \cite{Pontecorvo:1957cp,Pontecorvo:1967fh,Maki:1962mu}, whose values will be summarized below.

\bigskip

The mass states of the neutrinos get rotated to the flavour states by the PMNS matrix  \cite{Pontecorvo:1957cp,Pontecorvo:1967fh,Maki:1962mu} as follows:

\begin{equation}
\nu^{\rm int}_L = V_{\rm PMNS} \nu^{\rm phys}_L
\end{equation}
This matrix is usually parametrized in terms of 3 mixing angles $\theta_{12}$, $\theta_{13}$ and $\theta_{23}$, and 1 or 3 phases depending if the neutrinos are Dirac or Majorana fermions respectively:

\begin{eqnarray}
V_{\rm PMNS}&=&
\left(\begin{array}{ccc}
V_{e1} & V_{e2} & V_{e3} \\
V_{\mu 1} & V_{\mu 2} & V_{\mu 3} \\
V_{\tau 1} & V_{\tau 2} & V_{\tau 3}
\end{array}\right)\\
&=&
\left(\begin{array}{ccc}
c_{12} c_{13} & s_{12} c_{13} & s_{13} e^{-i \delta}  \\
- c_{12} s_{23} s_{13} e^{i \delta} - s_{12} c_{23} & -s_{12} s_{23} s_{13} e^{i \delta} + c_{12} c_{23} & s_{23} c_{13} \\
- c_{12} c_{23} s_{13} e^{i \delta} + s_{12} s_{23} & -s_{12} c_{23} s_{13} e^{i \delta} - c_{12} s_{23} & c_{23} c_{13} 
\end{array}\right)
\left(\begin{array}{ccc}
1 & 0 & 0 \\
0 & e^{i \frac{\alpha_{21}}{2}} & 0 \\
0 & 0 & e^{i \frac{\alpha_{31}}{2}} \\
\end{array}\right).
\nonumber\end{eqnarray}
where $s_{ij}=\sin \theta_{ij}$, $c_{ij}=\cos \theta_{ij}$, the angles $\theta_{ij}$ go from 0 to $\pi/2$ and the phases from 0 to $2\pi$. $\delta$ is the Dirac $\cp$ violation phase, and $\alpha_{ij}$ the Majorana $\cp$ violation phases. We are using here the same notation for the mixing angles and delta as in the quark case, but they should obviously not been confused. Notice also that, in contrast to the quark case where flavour states refer to the mass states, in the case of neutrinos flavour states refer to the interaction states. The other important parameters are the mass differences, as we will see later. Only two of the mass squared differences are independent parameters, that will be taken as $\Delta m_{21}^2$ and $\Delta m_{31}^2$. We will not enter here in the description of the topics related with the absolute value of the masses.

The mixing between neutrino flavours that imply the PMNS matrix has been observed experimentally through the neutrino oscillations in solar, atmospheric, accelerator and reactor neutrinos \cite{Cleveland:1998nv,Fukuda:1996sz,Abdurashitov:2009tn,Anselmann:1992um,Hampel:1998xg,Altmann:2005ix,Fukuda:2002pe,Ahmad:2001an,Ahmad:2002jz,Fukuda:1998mi,Ashie:2004mr,Eguchi:2002dm,Araki:2004mb}. Neutrinos produced by one of these sources in a specific flavour state are found in another state at a certain distance. As the distance is varied, the proportion of neutrinos in each flavour oscillates. The oscillation is produced by the fact that, as we have already said, the flavour states are not mass states (eigenstates of the Hamiltonian), and therefore evolve differently in time. Depending on the source the initial flavour state is different: solar ($\nu_e$), atmospheric ($\nu_\mu$, $\bar \nu_\mu$), accelerator ($\nu_\mu$) and reactor ($\bar \nu_e$) neutrinos. The probability of transition of the initial neutrinos after some distance depends mainly on the elements of the PMNS matrix, the mass differences between the mass states, the energy of the neutrinos produced and the distance at which the measurement is done. And the transitions happen only if at least two neutrinos are not degenerate and the mixing angles are not zero. Since the value of $V_{e3}$ is small, the mass and mixing angles related with the transitions ``12'' and ``23'' are responsible respectively of the transitions of the solar and atmospheric neutrinos, and is common to label them with these words (e.g. $\theta_{23}=\theta_{atm}$). It is also common to study the oscillations just considering the two main flavours involved on it, to simplify the equations.

The transition probability between two different flavours $l$ and $l^\prime$ is given by \cite{Beringer:1900zz:PMNS}:

\begin{equation}
P(\nu_l\rightarrow\nu_{l^\prime})=\sum_{j} |V_{l^\prime j}|^2|V_{lj}|^2+2 \sum_{j>k}|V_{l^\prime j}V_{lj}^*V_{lk}V_{l^\prime k}^*|\cos(\frac{\Delta m_{jk}^2}{2p}L-\phi_{l^\prime l;jk})
\end{equation}
where $\phi_{l^\prime l;jk}={\rm arg}(V_{l^\prime j}V_{lj}^*V_{lk}V_{l^\prime k}^*)$, and the result is the same for $P(\bar \nu_l\rightarrow\bar\nu_{l^\prime})$ just changing the sign of $\phi_{l^\prime l;jk}$.

When considering just two flavours in the transition: $|\nu_l\rangle=\cos\theta|\nu_1\rangle+\sin\theta|\nu_2\rangle$ and $|\nu_l^\prime\rangle=-\sin\theta|\nu_1\rangle+\cos\theta|\nu_2\rangle$, the previous equation gets simplified to:

\begin{equation}
P^{2\nu}(\nu_l\rightarrow\nu_{l^\prime})=\frac{1}{2}\sin^22\theta(1-\cos\frac{\Delta m^2 L}{2p})
\label{probtrans2neu}\end{equation}
where $\Delta m^2=m^2_2-m^2_1>0$, and it is often defined the oscillation length $L_0=4\pi p/\Delta m^2$.

\begin{figure}[ht!]
\begin{tabular}{cc}
\hspace{-0.5cm}
\includegraphics[width=80mm]{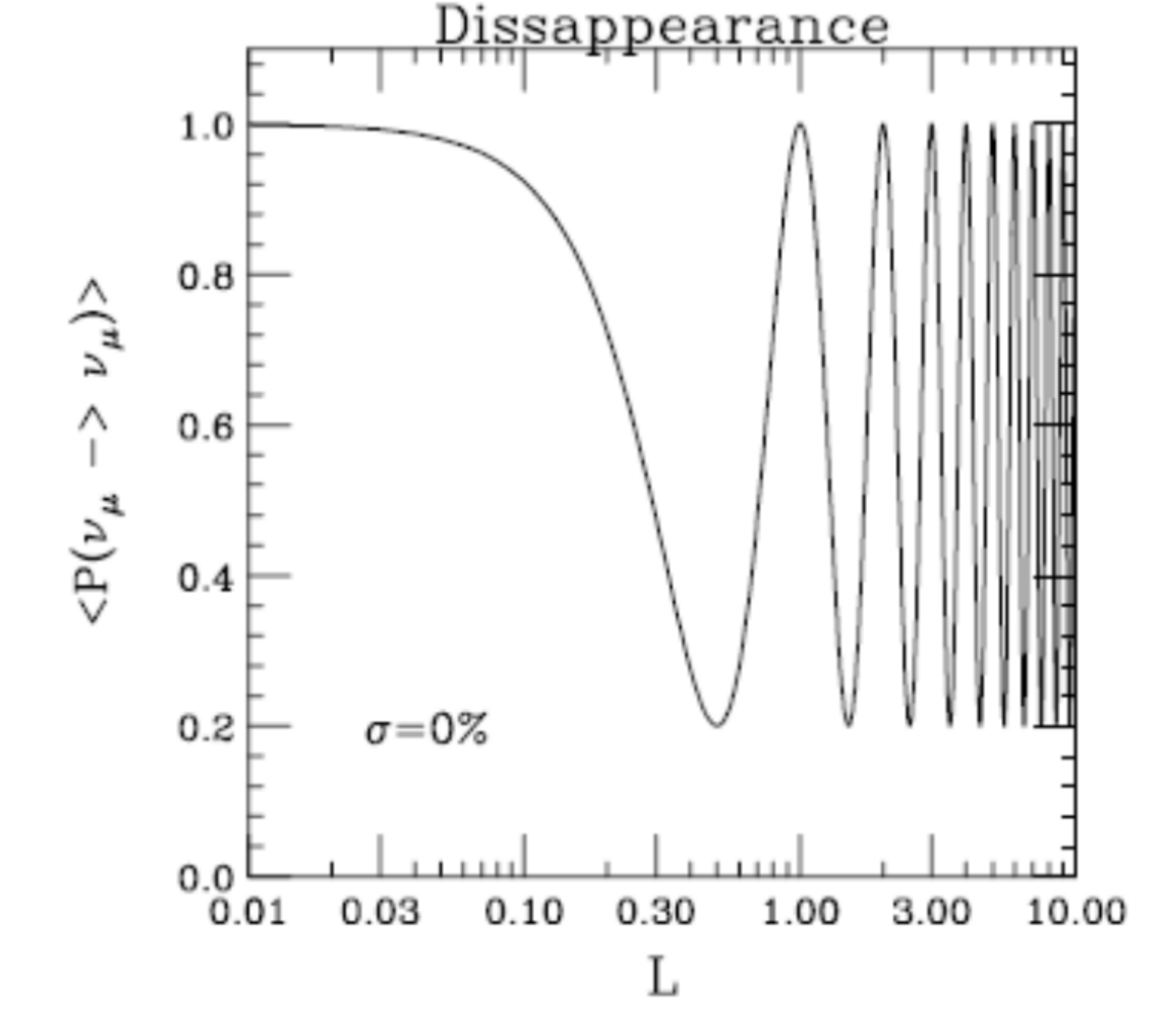} &
\includegraphics[width=80mm]{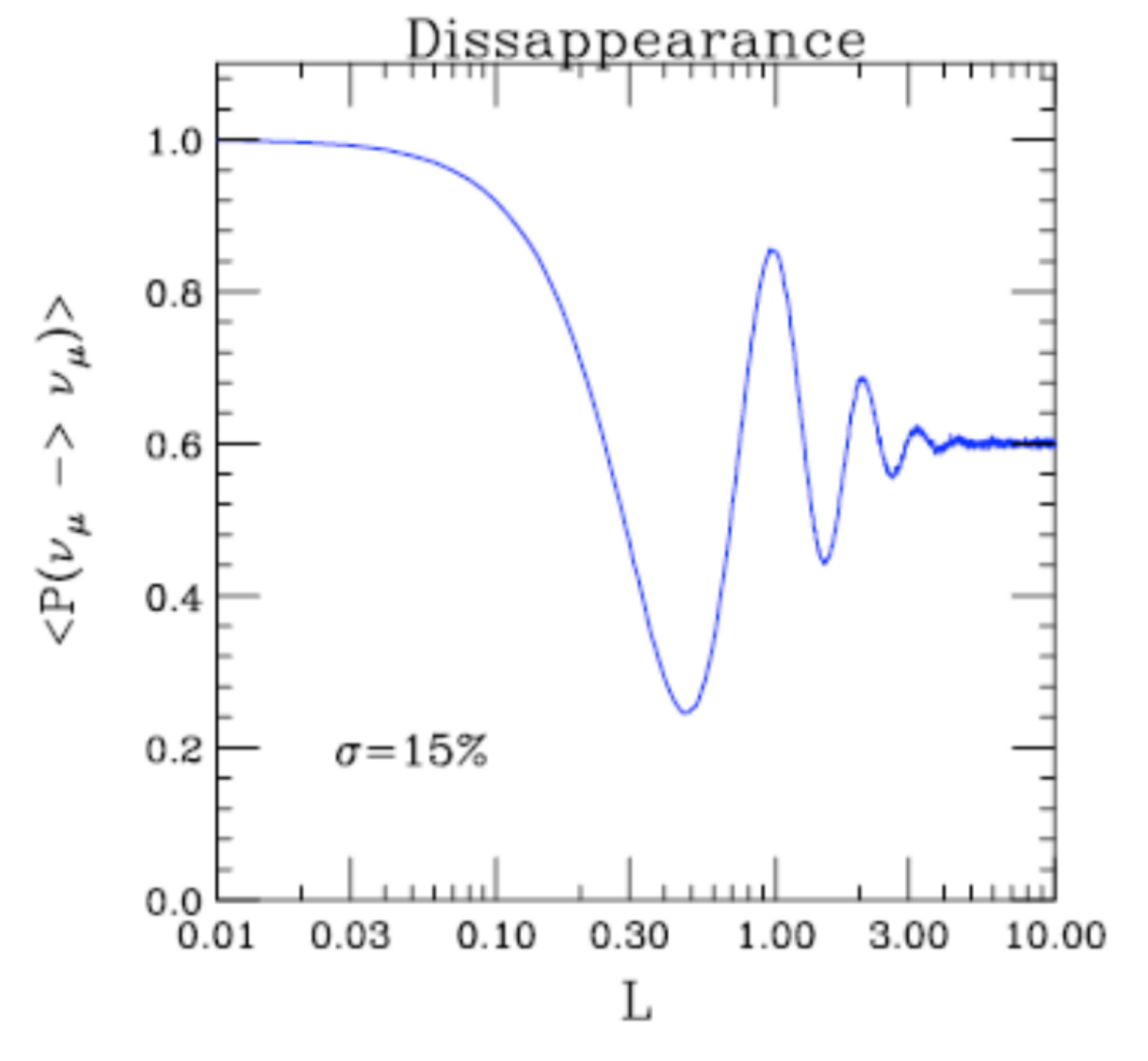} \\
\end{tabular}
\caption{Probability of finding muon neutrinos at distance $L$ after they are produced for two flavours with a mixing angle around 0.55, for neutrinos with a given energy that fulfils $L_0=1$ (left) or with energies spread around the previous value (right). Plots from \cite{Barenboim:neutrinos}}
\label{neuosc}
\end{figure}

Then setting experiments at the appropriate distances (related with the energy of the neutrinos produced) the transitions can be observed, and the angles obtained. As an example see Figure \ref{neuosc} where in the left plot we see the probability of finding a muon neutrino of a given energy at distance $L$ after it is produced: $P^{2\nu}(\nu_l\rightarrow\nu_l)=1-P^{2\nu}(\nu_l\rightarrow\nu_{l^\prime})$. In the right plot we find the same probability for detecting neutrinos with energy spread around the previous value.

The previous equations are valid for neutrino oscillations in vacuum. When neutrinos are travelling through matter, it has an effect over them and modifies the transition probabilities. The incoherent elastic scattering (the state after the scattering is not the same as the initial one) and quasielastic scattering (the energy transferred to the medium through the scattering is small) with the matter are processes not relevant to the neutrino transitions observed. The scattering for the $\nu_\mu\leftrightarrow\nu_\tau$ transition produces also negligible effects through the Sun and the Earth \cite{Botella:1986wy}. We are just left with the coherent elastic scattering for the $\nu_e\leftrightarrow\nu_{\mu,\tau}$ transition. This process produce differences in the refraction indices of the neutrinos involved, and the effect modifies the Equation \ref{probtrans2neu} changing $\theta$ and $L_0$ for $\theta_m$ and $L_m$ (see equations in \cite{Beringer:1900zz:PMNS}) that are related to the previous values and the density of electrons in the matter. These effects are important and should be taken into account for the experiments.

\bigskip
$\boldsymbol{\theta_{12}}$ and $\boldsymbol{\Delta m_{12}}$. Solar neutrinos have been observed in different experiments by the interaction of the neutrinos with the medium of the experiment: Homestake, SAGE, GALLEX, GNO, Kamiokande and Super-Kamiokande, Borexino and SNO. In Homestake the neutrinos transmuted chlorine into radioactive argon whose decay was then observed; SAGE, GALLEX and GNO observed the transformation of gallium into germanium; Kamiokande and Super-Kamiokande observed the Cherenkov light produced by the scattering between the neutrinos and the electrons of the water contained in the detector (with the advantage of being able of measuring the direction, and hence the origin of the Sun); Borexino detected that last scattering in a mixture of ultra pure liquid scintillator; SNO observed the interaction of the neutrinos with deuterium, producing electrons or neutrons.

The calculation of the number of neutrinos produced by the Sun is done through the standard solar model (see for instance \cite{Bahcall:2005va,Bahcall:2004pz,PenaGaray:2008qe} and references in \cite{Bahcallweb}). What at the beginning of the observations was called the solar neutrino problem, as the Sun produced less neutrinos than what was expected, was resolved by the neutrino oscillations. The SNO experiment, different to the other commented, was sensitive not only to electron neutrinos but to all flavours. This feature, combined with the precision measurement by the other experiments, especially Super-Kamiokande, confirmed the flavour oscillation \cite{Ahmad:2001an,Ahmad:2002jz,Aharmim:2005gt,Aharmim:2008kc}]. These observations were combined with the data from the KamLAND experiment, to obtain the best values for the mixing angle and the mass difference \cite{Gando:2010aa}. The KamLAND experiment is a liquid scintillator detector situated underground, in the same mine as Kamiokande, to reduce the effects from cosmic rays. KamLAND, instead of solar neutrinos, measured the well known flux from antineutrinos coming from nuclear reactors situated at a distance around 100 Km. KamLAND not only contributed to measure the flavour transition, but was able to observe the oscillation as the ratio distance/energy varied \cite{Araki:2004mb,Abe:2008aa,Gando:2010aa}.

$\boldsymbol{\theta_{23}}$ and $\boldsymbol{\Delta m_{23}}$. The oscillation of atmospheric neutrinos was observed first by Super-Kamiokande \cite{Fukuda:1998mi}. The experiment detected muon and electron neutrinos from the zenith angle, and observed how part of the muon neutrinos travelling trough the Earth disappeared (presumably oscillating into tau neutrinos), while that was not happening for the muon neutrinos from above. This effect was not present for the electron neutrinos. The hypothesis was confirmed by long-baseline experiments observing neutrinos from accelerators like K2K (KEK-to-Kamioka) \cite{Ahn:2006zza} and MINOS \cite{Adamson:2011ig}. The first produced neutrinos in the KEK-PS accelerator situated 250 Km away from the detector, with a mean energy of 1.3 GeV; and the second in the NuMI facility at Fermilab, 753 Km away, with a variable energy around 17 GeV.

$\boldsymbol{\theta_{13}}$. After many years having only bounds, finally in the last years some experiments have been able to measure $\theta_{13}$. These are Double Chooz \cite{Abe:2011fz}, Daya Bay \cite{An:2012eh} and RENO \cite{Ahn:2012nd}. The three studied $\bar \nu_e$ disappearance from reactors by using liquid scintillators.

The current experimental values for the angles and the mass differences, as obtained from \cite{Fogli:2012ua}, can be found in table \ref{tab:pmnselements}.

\begin{table}[h!]
\begin{center}
\vspace{1 cm}
\begin{tabular}{|c|c|c|}
\hline
Parameter & Best fit value & $3\sigma$ range \\
\hline
$\Delta m^2_{12}$ & $7.54\times 10^{-5} $ eV$^2$ & $ (6.99,8.18)\times 10^{-5}$ eV$^2$ \\
$\Delta m^2 (NH)$ & $2.43\times 10^{-3}$ eV$^2$ & $ (2.19,2.62)\times 10^{-3}$ eV$^2$ \\
$\Delta m^2 (NH)$ & $2.42\times 10^{-3}$ eV$^2$ & $ (2.17,2.61)\times 10^{-3}$ eV$^2$ \\
$\sin^2 \theta_{12} $ & $3.07\times 10^{-1}$ & $ (2.59,3.59)\times 10^{-1}$ \\
$\sin^2 \theta_{23} (NH)$ & $3.86\times 10^{-1}$ & $ (3.31,6.37)\times 10^{-1}$ \\
$\sin^2 \theta_{23} (IH)$ & $3.92\times 10^{-1}$ & $ (3.35,6.63)\times 10^{-1}$ \\
$\sin^2 \theta_{13} (NH)$ & $2.41\times 10^{-2}$ & $ (1.69,3.13)\times 10^{-2}$\\
$\sin^2 \theta_{13} (IH)$ & $2.44\times 10^{-2}$ & $ (1.71,3.15)\times 10^{-2}$ \\
\hline
\end{tabular}
\caption{Current experimental values for the angles and the mass differences of the PMNS matrix as obtained from \cite{Fogli:2012ua}. $\Delta m^2=|m_3^2-\frac{m_1^2+m_2^2}{2}|$ according to\cite{Fogli:2005cq}. Values valid for normal hierarchy (NH) or inverted hierarchy (IH) when stated. \label{tab:pmnselements}}
\end{center}
\end{table}

\section{Experimental searches of rare processes in flavour physics}
\label{bfactories}

The most important experiments at the moment with respect to the flavour changing processes in the quark sector are the $B$ factories and the LHCb. There are also important searches related to the first and second generation, but these mixings are very restricted by the past century experiments leaving not much space for new physics, that is why we will pay more attention here to the third generation. These experiments focus mainly in $B$ physics observables which turn out to be very good laboratories for the search of new physics. With respect to the study of LFV processes, besides the previously commented, there are MEG, SINDRUM II, and some other proposed experiments as COMET, Mu2e or PRISM.

{\bf Belle} \cite{webbelle} is an experiment situated in the KEKB $e^+e^-$ collider in Tsukuba (Japan). The first version run from 1999 to 2010 and collected over 1000 fb$^{-1}$ of data. It was designed for the observation of time-dependent $\cp$ violation in $B$ mesons, but it was able of conducting important researches in other topics, specially, but not only, in $B$ physics. Other topics covered were hadron spectroscopy, two-photon physics and $\tau$ physics. A review of Belle achievements can be found in \cite{Brodzicka:2012jm}. Most of the data was obtained with a luminosity around the $\Upsilon(4S)$ resonance, that is the best energy for the production of $B \bar B$ pairs (the $\Upsilon$ meson is formed by a b quark and its antiquark).

Measuring the decay rates and asymmetries between the decays of the two $B$ mesons for the different modes, it was possible to measure the different angles of the unitarity triangle. The decays used are the ones explained in the end of the section above. The equations for the rates and asymmetries can be found in \cite{Brodzicka:2012jm}. The first angle to be measured was $\beta$ and then $\alpha$ was obtained. This second angle was more difficult to be measured due to theoretical uncertainties from the contributions of penguin diagrams. The last studies of $\alpha$ using the full data set are: \cite{Adachi:2013mae} from $B\rightarrow\pi\pi$ and from $B\rightarrow\rho^0\rho^0$ is \cite{Adachi:2012cz}.  $\gamma$ was theoretically clean using only tree-level processes, but needed much more data than the other two. The last value for $\gamma$ from Belle can be found in \cite{Trabelsi:2013uj}.

$b \to s$ decays mediated by penguins were studied in Belle, searching for new physics inside the loops. For example, there were checked  $b \to s \bar{q} q$ induced decays as $B^0 \to \eta' K^0_S$ and $B^0 \to \phi K^0_S$. Other processes studied were  $B^- \to \ell^- \bar{\nu}_\ell$. These can be measured with high precision, and also calculated with high precision in the SM, so they are a good place to look for new physics. The calculation involves the parameter $f_B$ that has to be calculated by lattice QCD, so it is also a good test of it. Other interesting decay studied looking for new physics was $B \to D^{(*)} \tau \nu$, a channel that gave interesting results in BaBar.

Rare $B$ decays have also been studied to probe new physics beyond the SM. $B$ decays into hadrons not containing charm quarks are a good example. These are very suppressed in the SM, by the CKM elements (in the decays to the u quark) or by penguins (in the rest of the decays), and thus they are very interesting. Also radiative penguin decays as \bsg, or electroweak penguin decays as $b\to s(d) l^+l^-$, not allowed at tree level in the SM, have been observed. \bsg\ has been studied not only inclusively, but exclusively through channels like $B\to K\rho\gamma$, $B\to K\phi\gamma$ and $B\to K\pi\gamma$. These are promising modes for the next phase of the experiment. The inclusive measurement of $b\to s l^+ l^-$  is not easy experimentally, but the theoretical prediction is clean.
In general these decays suffer from large theoretical uncertainties, so it is common to measure ratios or asymmetries to eliminate them. 
These decays will be studied in the next sections in more detail, where we will see their potential to find new physics.

Tau physics has been another topic explored in Belle. The main issues studied were lepton flavour violating decays, $\cp$ violation in the charged lepton sector, the electric dipole moment of the $\tau$, and SM precision measurements. Tau physics is interesting since it is the only lepton whose mass allow it to decay into hadrons, and then is possible to study QCD with a cleaner system. The large mass of the tau also allows it to decay in many different LFV decays. Belle has studied 46 LFV tau decay modes using the whole set of data and found no evidence of it. LFV physics will be studied in detail in the following chapters. Two of the decays considered in our study $\tau \rightarrow \mu \gamma$ and $\tau \rightarrow e \gamma$ were studied in Belle \cite{Hayasaka:2007vc} looking for opposite $\tau^+\tau^-$ pairs, that decay to the searched decay in one side (the signal-side) with an invariant mass of the $\tau$ particle, and to a SM decay in the other side (the tag-side). It was also studied $\tau \rightarrow lll$ \cite{Hayasaka:2010np}. Other LFV searches are for example in \cite{Seon:2011ni}.

The measurements of Belle are consistent with the SM, leaving around a 10\% of space beyond the SM amplitudes allowing for new physics.

The next phase of Belle is called Belle-II \cite{webbelle2,Abe:2010sj}, and will be running in the SuperKEKB collider, starting with the physics in 2016. The luminosity at this Super-$B$ factory is $8\times 10^{35}$cm$^{-2}$s$^{-1}$, 50 times the peak luminosity achieved in KEKB. This corresponds to an annual integrated luminosity of 8 ab$^{-1}$ assuming 100 days of operation. A review about the physics in this Super-$B$ factory can be read in \cite{Aushev:2010bq}.

\bigskip
The {\bf BaBar} experiment \cite{webbabar} operated in the asymmetric $e^+e^-$ collider PEP-II at the SLAC National Accelerator Laboratory in California (USA). The PEP-II collider worked from 1999 to 2008. Designed to study the $\cp$ violation, BaBar worked at a centre of mass energy equal to the mass of the $\Upsilon(4S)$ resonance, as Belle. Apart from $\cp$ violation and the test of the CKM model, it produced interesting research on other fields as $\tau$ physics, heavy quarks physics, including bottomonium (bound states formed by a b quark and an anti-b) and charmonium physics, $D$ mesons decays and new physics searches. 

The chosen $\Upsilon(4S)$ resonance is a good choice in terms of signal to background ratio. The only problem is that the $B \bar B$ pairs created from its decay are created with very low momenta in the frame of the resonance, and hence time-dependent measurements are difficult. The asymmetric energy of the $e^+$ and $e^-$ beams produces a large boost of the resonance with respect to the laboratory frame, solving this problem.

The peak luminosity of PEP-II was $1.2\times 10^{34}$cm$^{-2}$s$^{-1}$. BaBar collected 430fb$^{-1}$ of data around the $\Upsilon(4S)$ resonance.

The angles of the unitarity triangle were measured in BaBar, by the decays explained at the end of the subsection \ref{unittriangl}. The study of $\alpha$ using the full data set from the $B\rightarrow\pi\pi$ decay can be found in \cite{Lees:2012kx}. The last value of $\gamma$ also with the whole collection of data was published in \cite{Lees:2013zd}. The angles show good consistency with the SM.

$\cp$ violation was confirmed in BaBar, and following the CPT theorem, this implies ${\cal T}$ violation. However this last is more difficult to be seen directly. As an example, $\cp$ violation was observed in the $B \to K \pi$ decay, but the   ${\cal T}$ reverse decay  $K \pi \to B$ that should be observed to measure the asymmetry was impossible to observe. It was not possible to prepare the initial state $K \pi$, and hadronic interactions would make difficult in any case to observe the asymmetry. However, a method was proposed by Bernabeu et al. \cite{Banuls:1999aj,Bernabeu:2012ab}, using the entanglement of the two $B$ mesons produced from the $\Upsilon(4S)$ resonance. This  ${\cal T}$ violation was finally observed in BaBar \cite{Lees:2012uka} with a significance equivalent to $14\sigma$.

Two interesting channels studied in BaBar were $B \to D^{(*)} \tau \nu$ and $B \to \tau \nu$. These are tree level decays mediated by a $W^\pm$ in the SM, so it could be mediated by a charged Higgs in the case of the MSSM. Thus the branching fractions, and the kinematic distributions (in the first decay) are sensitive to an enlarged Higgs sector as the one in the MSSM.
The decay  $B \to \tau \nu$ was measured \cite{Lees:2012ju}, and it was found to reach a $2.4\sigma$ discrepancy with the SM. The decay $B \to D^{(*)} \tau \nu$ was also measured \cite{Lees:2012xj} and the results were $3.4\sigma$ higher than in the SM. The combination of the last data of Belle with the data of BaBar, shown in \cite{bdtaunu}, enlarge the discrepancy with the SM to $4.8\sigma$. New results and analysis of data are awaited for confirmation.
Other interesting rare decays as \bsg\ were also measured \cite{Lees:2012wg}.

With respect to LFV, $B$ factories are $\tau$ factories too: $\sigma(e^+e^-\rightarrow \tau^+\tau^-)\approx \sigma(e^+e^-\rightarrow b \bar b)$, so it is interesting to study the tauonic channel (also for the large $\tau$ mass compared with other leptons, that increase some branching ratios). In BaBar, as in Belle, LFV decays were studied as for example the ones that will be studied in this work $\tau \rightarrow \mu \gamma$ and $\tau \rightarrow e \gamma$ \cite{Aubert:2009ag}, and $\tau$ in three leptons \cite{Lees:2010ez}; and also other decays as the ones in \cite{BABAR:2012aa}

\bigskip
The {\bf LHCb} experiment \cite{weblhcb} is a forward spectrometer located at the LHC collider, aimed to study heavy flavour physics. Its main object is to study $\cp$ violating observables and rare decays. It is designed to use the decays of b and c hadrons to obtain its measurements. In the LHC there are produced pairs of b and anti-b quarks, basically via gluon fusion, gluon splitting and flavour excitation. These pairs are produced mainly in the forward or backward direction, being this the reason for the chosen design of the experiment. The acceptance of the detector is for particles with a pseudorapidity between 2 and 5.3. The experiment is designed to operate at a luminosity of $4\times 10^{32}$cm$^{-2}$s$^{-1}$, lower than the nominal luminosity of the LHC, and the one used for the two main general detectors. This decrease is produced by reducing the overlap between the beams. Until the moment, the detector collected 3 fb$^{-1}$ of data. 

Between the observations of the LHCb one can find the $\cp$ violation in the decays of $B^0_s$ mesons \cite{Aaij:2013iua}, never observed before, the study of $B^{0} \to K^{*0} \mu^+ \mu^-$ \cite{Aaij:2011aa}, and the most precise measurement to date of $B^{0}_{s}$-$\bar{B}^{0}_{s}$ mixing \cite{Aaij:2013mpa}, an important observable for our present work that will be studied in next chapters with more detail.

One of the last successes of the LHCb (together with CMS) was the measurement of $B^0_s \to \mu^+ \mu^-$ \cite{Aaij:2013aka}. The value obtained was compatible with the SM. This observable is very relevant. It is highly suppressed in the SM, similarly to other FCNC process and also being helicity suppressed (because of the structure of the interaction, to conserve the helicity some final helicity states are suppressed), so it is a good probe for possible new physics, and it has a very precise measurement in the SM. It is sensitive to scalar and pseudoscalar contributions of new theories. It has direct implications in the MSSM space of parameters, and therefore will be used through this work as one of our reference observables.
The experiment also gave bounds on $B^0_{(s)} \to \mu^+ \mu^- \mu^+ \mu^-$ \cite{Aaij:2013lla} and also in the LFV observable $B^0_{(s)} \rightarrow e^{\pm}\mu^{\mp}$ \cite{Aaij:2013cby}.
With respect to charm physics, LHCb presented bounds to $D^0 \to \mu^+ \mu^-$ \cite{Aaij:2013cza}, also a FCNC helicity suppressed observable, and to $D^+ \to \pi^\pm \mu^+ \mu^\mp$  \cite{Aaij:2013sua}, giving the world best upper limits to all of them.

The study of the $\tau$ decays has some disadvantage in the LHCb as compared to the $B$ factories, not being able to do $\tau$ tagging, however the  inclusive $\tau$ cross section is much larger, $79.5\pm 8.3\mu b$ in LHCb as compared with 0.919 nb in the $B$ factories, compensating the former fact. Rare $\tau$ decays have been explored as $\tau \rightarrow \mu \mu \mu$ \cite{Chrzaszcz:2013uz} or the ones of \cite{Aaij:2013fia} and \cite{Aaij:2013cby}.

The next upgrade of the LHCb in 2018 \cite{lhcbupgrade} will allow to carry on precision studies on flavour. For example, it will be possible to determine  $B^0_s \to \mu^+ \mu^-$ down to a 10\% of the SM value, to measure the CKM angle $\gamma$ with an error smaller than 1 degree, or search $\cp$ violation in charm below the $10^{-4}$ level. Some expected values for the precision of different observables can be found in  \cite{lhcbupgrade}.

$\phantom{x}$

{\bf MEG} \cite{webmeg} is an experiment located at the Paul Scherrer Institute (PSI) in Switzerland, optimized to detect the decay $\mu^+\rightarrow e^+ \gamma$. It consists in an intense positive muon beam of $3\times 10^7 \mu^+/s$, that is stopped at a thin target inside of COBRA, a superconducting magnet, aiming to produce positrons with an helicoidal trajectory and a signal in the drift chamber, and photons that would be detected in a liquid Xenon scintillator detector outside of the magnet. 

The experiment has been running since 2008, with for example $22\times 10^{22}$ stopped muons during 2012, and has produced the best bound in the world for the searched decay \cite{Adam:2013mnn} $\br(\mu \to e \gamma) < 5.7 \times 10^{-13}$. An upgrade is proposed \cite{Baldini:2013ke} with a goal of reaching a sensitivity of $6 \times 10^{-14}$. In Chapter \ref{phenoflavourslep} will be studied this decay and the predictions of SUSY for it.

The research of rare $\mu$ decays by this experiment in PSI was the continuation of the {\bf SINDRUM II} experiment. This detector was a magnetic solenoidal spectrometer designed to search for $\mu-e$ conversion in nuclei, holding the current best bound for the conversion in Gold $\CR(\mu-e, {\rm Au}) < 7.0 \times 10^{-13}$ \cite{Bertl:2006up}. The muons were produced by a 590 MeV beam of protons hitting a carbon target. Inside the spectrometer was located the gold target that was hit by the muons. More details about the experiment can be found on \cite{Bertl:2006up} and \cite{bertl:talk}.

The next future experiments for studying rare $\mu$ decays are {\bf COMET} and {\bf Mu2e}. COMET \cite{lettercomet} will be located in Tokai in Japan and will use the J\_PARC proton beam. It will study the  $\mu-e$ conversion in aluminium atoms with a sensitivity of $10^{-17}$ at 90\% CL in the second phase. The first physics run of phase-I is due to 2017 and the second phase physics run in 2021. Mu2e \cite{webmu2e,Abrams:2012er}  will be located in Fermilab and will study the same conversion with the same sensitivity as COMET. The initial preliminary results are expected in 2021. After them, the next experiment designed is {\bf PRISM} \cite{Barlow:2011zza}. PRISM will use a fixed-field alternating gradient accelerator \cite{Witte:2012zza} for rendering monochromatic the muon beam obtained as in COMET. Having all the muons with the same energy will exploit the sharpness of the peak in the energy distribution of the electron decay, improving the signal/background ratio. It could improve the sensitivity of COMET until $10^{-18}$. Another proposed experiment is  {\bf Mu3e} \cite{webmu3e,Blondel:2013ia}, also in PSI, whose aim is to measure the decay $\mu \rightarrow e e e$  with an expect sensitivity of $10^{-15}$ and $10^{-16}$ in the first and second phases respectively.

Finally, for our selected particular LFV processes of relevance in the present work, we have collected a summary of their present bounds in the forthcoming Section \ref{sec:expbounds}.

\section{The flavour of new physics}
\label{npflavour}

In the previous section, we have seen how the flavour physics is a good window for the search of new physics. In the next sections we will shortly comment on some techniques and hypothesis related with the searches of new physics beyond the Standard Model in flavour observables, and in particular for the searches of SUSY in mesonic observables.

\subsection{The effective Hamiltonian approach and the scale of new physics}
\label{effhamiltnp}

Using the full supersymmetric theory for doing predictions is a difficult task for $B$ physics observables so we will use an easier way to attack the problem: the method of the Operator Product Expansion (OPE). The OPE is a way to Taylor expand the Hamiltonian of the theory in what are called effective Hamiltonians, and perform the predictions using the most relevant terms until some order. The processes we are interested in involve different energy scales, the idea behind the expansion is working with a Hamiltonian with the degrees of freedom of the energy of the process, where the high energy degrees of freedom have been integrated out and their effects are collected in some coefficients. These coefficients are calculated at high energy, and they are evolved to the low energy using the proper Renormalization Group Equations. We comment briefly on this expansion in the following:

Our goal is to calculate an approximation to the $S$ matrix element between some general initial and final states $\langle F \vert S \vert I \rangle$ using the full theory. Instead of the full Hamiltonian we will use an effective Hamiltonian:

\begin{equation}
{\cal H}_W=4\frac{G_F}{\sqrt{2}}  \sum_i C_i (\mu)  O_i(\mu)
\end{equation}
where $G_F$ is the Fermi constant $G_F/\sqrt{2}=g^2/8 M_W^2$, the $C_i$ are called Wilson coefficients, and $O_i$ are the low energy local operators, both renormalized at a scale $\mu$. To satisfy our goal it should be fulfilled the following relation:
\begin{equation}
\langle F  \vert i S \vert I \rangle =4\frac{G_F}{\sqrt{2}}
\sum_i C_i (\mu) \langle F \vert O_i(\mu) \vert I\rangle + \dots\,, \label{due}
\end{equation}

The terms of the expansion not shown are suppressed by inverse powers of the high energy scale, the scale where the full theory operates. 

The operators $O_i$ are a complete basis of the OPE, containing all the operators producing the transition between $I$ and $F$, with the proper dimensions and the same quantum numbers as the original element of the full theory that connects the two states, and respecting all the symmetries of the full theory. One must also take into account that when running the RGE to low energy, new operators can be generated to add to the first set.

Once the basis of operators is obtained, one has to obtain the Wilson coefficients. This is done by first ``matching'' the value of the amplitudes at high energy using the full theory and the effective Hamiltonian (up to terms suppressed by inverse powers of the high energy scale) as in Eq. \ref{due}, and second running the RGE from the matching scale down to the renormalization scale $\mu$.
The Wilson coefficients then encode (and depend only on) the information of the high energy (low distances) physics, and act as effective couplings in our new Hamiltonian.

Once the coefficients and operators are obtained, one can calculate the transition between the two chosen states. For that one has to calculate matrix elements of the new operators between the states. For transitions between hadronic states as the ones we are interested in this work, this has to be done by non-perturbative methods as lattice QCD, or they have to be measured in some process and then used the obtained values for other processes.

\bigskip
When understanding the SM as an effective theory, one can study the effects of physics beyond the SM by using higher dimensional operators not contained in the SM. These operators are suppressed by inverse powers of the scale of the new physics. As the SM predictions match better the experiments, the space for these new operators get reduced, meaning in general terms, moving away the new physics scale to higher energies.

For example, one can check the scale for new physics from the amplitudes of meson-antimeson mixings as $K^0$--$\overline{K^0}$, $B_d$--$\overline{B}_d$, and $B_s$--$\overline{B}_s$. The amplitudes of these $\Delta F=2$ transitions, where the quark flavour quantum numbers change by two units, go in the SM approximately as

\begin{equation}
 \mathcal{A}_{\rm SM}^{\Delta F=2} \approx
   \frac{ G_F^2 m_t^2 }{16 \pi^2} \left(V_{ti}^* V_{tj} \right)^2
\times
   \langle \overline{M} |  (\overline{Q}_{Li} \gamma^\mu Q_{Lj} )^2  | M \rangle
\times    F\left(\frac{M_W^2}{m_t^2}\right),
\end{equation}
where $F$ is a loop function of $\mathcal{O}(1)$, $V_{ij}$ are the CKM elements and $Q_{Li}$ are the corresponding quarks involved.

One can compare this with the amplitude produced by the new physics (NP) effective operators
\begin{equation}\label{eq:qlql}
\Delta \mathcal{L}^{\Delta F=2} = \sum_{i\not=j} \frac{c_{ij}}{\Lambda^2}
(\overline{Q}_{Li} \gamma^\mu Q_{Lj} )^2~,
\end{equation}
where the $c_{ij}$ are dimensionless couplings and $\Lambda$ is the new physics scale. 

As shown in \cite{Isidori:2010kg}, the condition $|\mathcal{A}^{\Delta F=2}_{\rm NP}| <  |\mathcal{A}^{\Delta F=2}_{\rm SM} |$ can be translated into lower bounds on the new physics scale
\begin{eqnarray}
\Lambda > \frac{ 4.4~{\rm TeV} }{| V_{ti}^* V_{tj}|/|c_{ij}|^{1/2}  }
\sim \left\{ \begin{array}{l}
1.3\times 10^4~{\rm TeV} \times |c_{sd}|^{1/2} \!\!\!\!\!\!\! \\
5.1\times 10^2~{\rm TeV} \times |c_{bd}|^{1/2} \!\!\!\!\!\!\! \\
1.1\times 10^2~{\rm TeV} \times |c_{bs}|^{1/2} \!\!\!\!\!\!\!
\end{array}
\right.
\label{eq:bound}
\end{eqnarray}

Therefore, in order to have the new physics at $\Lambda$ reachable by the present experiments the couplings $c_{ij}$ of the new operators connecting flavours of different generations should be very small, as seen in the last equation. In other words, these imply strong constraints on the flavour mixing coefficients.

The issue that new physics with general flavour structures usually tend to overcome easily the flavour constraints of the experiments is a well known problem. In the literature this problem is usually evaded by assuming the minimal flavour violation hypothesis on the new physics. 

\subsection{The minimal flavour violation hypothesis}
\label{mfvhip}
The minimal flavour violation hypothesis assumes that the Yukawa couplings (as in the SM) are the only source of flavour violation also in the new physics. 

This hypothesis can be studied using the effective field theory approach as before. To do that one first defines the flavour symmetry that is only broken by the Yukawa couplings. This global symmetry, the largest one commuting with the gauge group, would be:

\begin{equation}
G_{\rm Flavour} \equiv {\rm SU}(3)^3_q \otimes  {\rm SU}(3)^2_\ell
\otimes  {\rm U}(1)_B \otimes {\rm U}(1)_L \otimes {\rm U}(1)_Y \otimes {\rm U}(1)_{\rm PQ} \otimes {\rm U}(1)_{E_R}~,
\end{equation}
where
\begin{eqnarray}
{\rm SU}(3)^3_q     &=& {\rm SU}(3)_{Q_L}\otimes {\rm SU}(3)_{U_R} \otimes {\rm SU}(3)_{D_R}~,  \\
{\rm SU}(3)^2_\ell  &=&  {\rm SU}(3)_{L_L} \otimes {\rm SU}(3)_{E_R}~.
\end{eqnarray}

The $B$, $L$ and $Y$ charges can be identified with baryon number, lepton number and hypercharge respectively. These three charges are not broken by the Yukawa interactions. The last two ${\rm U}(1)$ charges are a Peccei-Quinn symmetry and a rotation of the right-handed singlets. More details about these groups can be found for instance in \cite{D'Ambrosio:2002ex}.

The usual method to implement the MFV hypothesis is to promote the Yukawa couplings as spurions. Instead of parameters, the spurions are defined as (non-physical) fields that transform in the proper way to conserve the symmetry that as parameters they break. They can then be used to construct all the possible symmetric terms of the Lagrangian, and after it they are promoted back to the original parameters, giving us all the possible symmetry breaking terms. In the case of the $G_{\rm Flavour}$ symmetry above, that means Yukawa couplings as fictitious fields transforming as:
\begin{equation}
{Y}^{u} \sim (3, \bar 3,1)_{{\rm SU}(3)^3_q}~,\qquad
{Y}^{d} \sim (3, 1, \bar 3)_{{\rm SU}(3)^3_q}~,\qquad
{Y}^{e} \sim (3, \bar 3)_{{\rm SU}(3)^2_\ell}~.
\end{equation}

With these dynamical fields the $G_{\rm Flavour}$ symmetry is then satisfied for every higher dimensional operator one can build with these fields in this spurion language, or what is the same, for every operator whose flavour structure is only determined by the Yukawa couplings.

In \cite{D'Ambrosio:2002ex} the possible new physics effective operators using this MFV approach were studied in the one and two Higgs doublets cases, and listed some of them. It is found how the scale of new physics is largely reduced comparing with the general hypothesis we saw before, and stays at $\mathcal{O}(10)$TeV. 

The MFV hypothesis can also be applied to supersymmetry, and more specifically to define the structure of the soft SUSY breaking terms that are the terms of the Lagrangian with an unknown flavour structure. This would translate into the following terms \cite{D'Ambrosio:2002ex}:
\begin{eqnarray}
m_{\tilde Q}^2  &=& {\tilde m}^2 \left( a_1 1 \hspace{-.085cm}{\rm l} 
+b_1 Y^{u} Y^{u\dagger} +b_2 Y^{d} Y^{d\dagger} 
+b_3 Y^{d} Y^{d\dagger} Y^{u} Y^{u\dagger}
+b_4 Y^{u} Y^{u\dagger} Y^{d} Y^{d\dagger} 
 \right)~,
\\
m_{\tilde U}^2 &=& {\tilde m}^2 \left( a_2 1 \hspace{-.085cm}{\rm l}
+b_5 Y^{u\dagger} Y^{u}  \right)~,\\
m_{\tilde D}^2 &=& {\tilde m}^2 \left( a_3 1 \hspace{-.085cm}{\rm l}
+b_6 Y^{d\dagger} Y^{d}  \right)~,\\
{\bar {\cal  A}}^{u} &=& A\left( a_4 1 \hspace{-.085cm}{\rm l}
+b_7 Y^{d} Y^{d\dagger}  \right) Y^{u}~,\\
{\bar {\cal  A}}^{d} &=& A\left( a_5 1  \hspace{-.085cm}{\rm l}
+b_8 Y^{u} Y^{u\dagger}  \right) Y^{d}~.
\end{eqnarray}
where $\tilde m$ and $A$ are generic SUSY soft mass scales, $a_i$ and $b_i$ are dimensionless numerical coefficients, and $1 \hspace{-.085cm}{\rm l}$ is the $3\times 3$
identity matrix. 

Within the MSSM most of the calculations for flavour observables are done by implementing the MFV hypothesis, therefore by requiring that, as in the SM, the only origin for flavour mixing be the CKM matrix in the quark sector (and the PMNS matrix in the neutrino sector if massive neutrinos are included), and flavour violation is exclusively mediated by the Yukawa interactions. 

In the SM, we saw how there were no flavour changing neutral currents at tree level, and the charged currents were mediated by the $W^{\pm}$ bosons (that contributed to the FCNC at loop level). In the MSSM the charged currents are also mediated by the charged Higgs $H^{\pm}$ and the charginos ${\tilde \chi}^\pm$, that contribute also to the loop-level neutral currents. As we saw before, in the standard sector the FCNC were very suppressed by the GIM mechanism. In the supersymmetric sector, the GIM mechanism will be replaced by a super-GIM mechanism, being the rates for FCNC processes proportional to the ratio between the differences of squared squarks masses and chargino/charged Higgs masses. 

In the rest of the work however, we will not use this MFV hypothesis for SUSY but we will assume instead a more general hypothesis, named Non Minimal Flavour Violation (NMFV), where the origin of the flavour violation is not needed to be specified and is assumed to be something else in addition to the Yukawa couplings. Then, to deal with this NMFV hypothesis without inquiring the specific origin for flavour violation one simply assumes generic soft SUSY breaking terms for the MSSM, with off-diagonal in flavour entries and study the sfermion mixing that these terms produce as well as their phenomenological consequences. This will be introduced in the next chapter and we will study later their present constraints and future prospects.

In the NMFV scenario, and in addition to the previously commented FCNC originated by charged currents, gluinos and neutralinos will also contribute to the FCNC. Usually the dominant contribution in the NMFV case comes from the gluinos as they are strongly interacting, therefore one expects sizeable contributions to FCNC processes in the NMFV scenarios. 

\chapter{General sfermion flavour mixing and selected scenarios in the MSSM}
\label{paramnmfvscen}

In this chapter we will depart from the basic supersymmetric model described in the previous chapters and introduce the NMFV framework to study the sfermion flavour mixing to its whole extent. This will be done by using a general parametrization of the sfermion mixing, that will be described in Section \ref{paramnmfv}, and a proper set of SUSY scenarios, described in Section \ref{scenarios}. 

This parametrization will allow us to carry out a general study of the phenomenological consequences of the sfermion flavour mixing treating all the mixings between different generations and the relevant observables on equal footing in the same framework.
The generality of the parametrization, and the absence of approximations such as the MIA, will allow us to study complex situations as combinations of multiple mixings, or effects not appearing in a perturbative first order approximation of these mixings parameters.

The different SUSY scenarios that we propose to evaluate the observables, are chosen in order to understand the effects of the different values of parameters of the MSSM and to explore the phenomenologically different regions of the space of parameters. The scenarios reflect the change in the experimental situation after the LHC started taking data, and all the calculations have been updated to be compatible with the last results of the experiments. The most challenging situation for the discovery of supersymmetry of having a heavy SUSY spectrum will also be studied in some scenarios, to find possible experimental windows that could overcome that fact and show supersymmetric effects on observables despite the heaviness of SUSY.

Both the parametrization described in this chapter and the proposed SUSY scenarios that are appropriate for our forthcoming studies of Flavour Violating processes have been represented in our papers \cite{AranaCatania:2011ak}, \cite{Arana-Catania:2013nha} and \cite{Arana-Catania:2013xma}.

\section{Parametrization of general sfermion mixing}
\label{paramnmfv}

Our goal with this work is to study the frontiers of flavour mixing in the MSSM from a general point of view, and describe the phenomenology it implies in some important observables. To this end, we have to take a big leap beyond the simplified hypothesis of the MFV, and describe a general framework for NMFV where all the possible flavour mixing parameters of the model can be brought into play, and study their consequences. This general sfermion flavour mixing situation will be parametrized in this section.

In previous sections it was introduced the fact that supersymmetry, at low energies must be a broken symmetry, and there were also introduced the soft SUSY breaking terms in the MSSM. These terms of equation \ref{MSSMsoft} introduce new matrices in flavour space, denoted as ${\bar {\cal  A}}_{ij}$ and $m_{ij}^2$, whose detailed structure is unknown for us. There could be new high-scale symmetries that could make these terms to be flavour diagonal at high energy, and then we would be in the MFV hypothesis we commented in the previous chapters, having the CKM and the PMNS matrices as the unique responsible of flavour violation. Nevertheless, as we have already said, in this work we will explore a more general hypothesis, the non-minimal flavour violation, where these terms have a general flavour structure, and explore how constrained is this structure by the experiments.

It is common to study supersymmetry at very high energy where simplification hypothesis are natural, and the number of parameters is greatly reduced, and then use the RGE equations to obtain a low-scale model. This will be commented later on the scenarios of Section \ref{scenarios}, but in this work in general we will focus on the low energy model (where  our SUSY scale should be understood as low energy as compared with the Planck energy), and study the general parameter space without considering any hypothesis about how these parameters are generated.

\subsection{The squark sector with general flavour mixing}
\label{sec:nmfv-squarks}

In Section \ref{particles-mssm-squarks} the squark sector was introduced within the MFV hypothesis. The mass matrices of the squarks were rotated from the electroweak basis to the SCKM basis, with a rotation parallel to the one made to the quarks, and the mass matrices of both kind of particles were diagonal in this last basis. Now we explore a more general hypothesis, were after rotating to the SCKM basis, the squark mass matrices are not yet diagonal because of the non-diagonal soft SUSY breaking terms. By the squarks rotation of Eqs. \ref{squarksrotsckm} one gets the relevant soft-SUSY-breaking Lagrangian \ref{MSSMsoft} transformed from:
\begin{align}
    \mathcal{L}_{\rm soft} 
 &= -  \tilde {\cal U}^{{\rm int}*}_i m_{\tilde U ij}^2 \, \tilde {\cal U}^{\rm int}_j
   - \tilde {\cal D}^{{\rm int}*}_i m_{\tilde D ij}^2 \tilde {\cal D}^{\rm int}_j
   - \tilde {\cal Q}^{{\rm int}\dagger}_i m_{\tilde Q ij}^2 \tilde {\cal
 Q}^{\rm int}_j  \non \\
  &-\left[ \tilde {\cal Q}^{\rm int}_i {\bar {\cal  A}}^{u}_{ij} \tilde {\cal U}^{{\rm int}*}_j {\cal H}_2 
         - \tilde {\cal Q}^{\rm int}_i {\bar   {\cal A}}^{d}_{ij} \tilde {\cal D}^{{\rm int}*}_j {\cal H}_1 \; + \; \text{h.c.} 
    \right]  
\label{eq:lsoft}
\end{align}
to:
\begin{align}
\label{eq:lsoft-superCKM}
   &\mathcal{L}_{\rm soft}
 = -  \tilde {\cal U}_{Ri}^* m_{\tilde U_R ij }^2 \tilde {\cal U}_{Rj} 
   - \tilde {\cal D}_{Ri}^* m_{\tilde D_R ij}^2 \tilde {\cal D}_{Rj}
-  \tilde {\cal U}_{Li}^* m_{\tilde U_L ij}^2 \tilde {\cal U}_{Lj} 
   - \tilde {\cal D}_{Li}^* m_{\tilde D_L ij}^2 \tilde {\cal D}_{Lj} \\
 &\quad
-  \left[ \tilde {\cal U}_{Li} {\cal A}^u_{ij} \tilde {\cal U}^*_{Rj} {\cal H}_2^0
- \tilde {\cal D}_{Li} (\VCKM)_{ki} {\cal A}^u_{kj} \tilde {\cal U}^*_{Rj} 
{\cal H}_2^+  
 -  \tilde {\cal U}_{Li} (\VCKM^*)_{ik} {\cal A}^d_{kj} \tilde {\cal D}^*_{Rj} 
{\cal H}_1^-
+ \tilde {\cal D}_{Li} {\cal A}^d_{ij} \tilde {\cal D}^*_{Rj} {\cal H}_1^0 +  \text{h.c.}
    \right], \nonumber 
\end{align}
where we have used calligraphic capital letters for the
squark fields with generation indices,
\begin{equation}
\tilde {\cal U}^{{\rm int}}_{1,2,3}=\tilde u^{{\rm int}}_R,\tilde c^{{\rm int}}_R,\tilde t^{{\rm int}}_R; 
\tilde {\cal D}^{{\rm int}}_{1,2,3}=\tilde d^{{\rm int}}_R,\tilde s^{{\rm int}}_R,\tilde b^{{\rm int}}_R;
\tilde {\cal Q}^{{\rm int}}_{1,2,3}=(\tilde u^{{\rm int}}_L \, \tilde d^{{\rm int}}_L)^T, (\tilde c^{{\rm int}}_L\, \tilde s^{{\rm int}}_L)^T, (\tilde t^{{\rm int}}_L \, \tilde b^{{\rm int}}_L)^T  ;
\end{equation}
\begin{equation}
\tilde {\cal U}_{L1,2,3}=\tilde u_L,\tilde c_L,\tilde t_L; 
\tilde {\cal D}_{L1,2,3}=\tilde d_L,\tilde s_L,\tilde b_L;
\tilde {\cal U}_{R1,2,3}=\tilde u_R,\tilde c_R,\tilde t_R; 
\tilde {\cal D}_{R1,2,3}=\tilde d_R,\tilde s_R,\tilde b_R;
\end{equation}
and ($q=u, d$)
\begin{equation}
{\cal A}^{q} = V^{q}_L {\bar {\cal A}}^q V^{q \dagger}_R, 
 m_{\tilde U_R}^2 = V_R^u  m_{\tilde U}^2 V_R^{u \dagger}, 
 m_{\tilde D_R}^2 = V_R^d  m_{\tilde D}^2 V_R^{d \dagger}, 
 m_{\tilde U_L}^2 = V_L^u  m_{\tilde Q}^2 V_L^{u \dagger}, 
 m_{\tilde D_L}^2 = V_L^d  m_{\tilde Q}^2 V_L^{d \dagger}. 
\label{eq:su2}
\end{equation}
The usual procedure to introduce general flavour mixing in the squark sector is to include the non-diagonality in flavour space at this stage, namely, in the Super-CKM basis. Thus, one usually writes the $6\times 6$ non-diagonal mass matrices, ${\mathcal M}_{\tilde u}^2$ and ${\mathcal M}_{\tilde d}^2$, referred to the Super-CKM basis, being ordered respectively as $(\SupL, \SchaL, \StopL, \SupR, \SchaR, \StopR)$ and  $(\SdownL, \SstrL, \SbotL, \SdownR, \SstrR, \SbotR)$, and write them in terms of left- and right-handed blocks $M^2_{\tilde q \, AB}$ ($q=u,d$, $A,B=L,R$), which are non-diagonal $3\times 3$ matrices,
\begin{equation}
{\mathcal M}_{\tilde q}^2 =\left( \begin{array}{cc}
M^2_{\tilde q \, LL} & M^2_{\tilde q \, LR} \\ 
M_{\tilde q \, LR}^{2 \, \dagger} & M^2_{\tilde q \,RR}
\end{array} \right), \qquad \tilde q= \tilde u, \tilde d.
\label{eq:blocks-matrix}
\end{equation} 
 where:
 \begin{alignat}{5}
 M_{\tilde u \, LL \, ij}^2 
  = &  m_{\tilde U_L \, ij}^2 + \left( m_{u_i}^2
     + (T_3^u-Q_u\sin^2 \theta_W ) M_Z^2 \cos 2\beta \right) \delta_{ij},  \notag\\
 M^2_{\tilde u \, RR \, ij}
  = &  m_{\tilde U_R \, ij}^2 + \left( m_{u_i}^2
     + Q_u\sin^2 \theta_W M_Z^2 \cos 2\beta \right) \delta_{ij} \notag, \\
  M^2_{\tilde u \, LR \, ij}
  = &  \left< {\cal H}_2^0 \right> {\cal A}_{ij}^u- m_{u_{i}} \mu \cot \beta \, \delta_{ij},
 \notag, \\
 M_{\tilde d \, LL \, ij}^2 
  = &  m_{\tilde D_L \, ij}^2 + \left( m_{d_i}^2
     + (T_3^d-Q_d \sin^2 \theta_W ) M_Z^2 \cos 2\beta \right) \delta_{ij},  \notag\\
 M^2_{\tilde d \, RR \, ij}
  = &  m_{\tilde D_R \, ij}^2 + \left( m_{d_i}^2
     + Q_d\sin^2 \theta_W M_Z^2 \cos 2\beta \right) \delta_{ij} \notag, \\
  M^2_{\tilde d \, LR \, ij}
  = &  \left< {\cal H}_1^0 \right> {\cal A}_{ij}^d- m_{d_{i}} \mu \tb \, \delta_{ij}.
\label{eq:SCKM-entries}
\end{alignat}
with, $i,j=1,2,3$, $Q_u=2/3$, $Q_d=-1/3$, $T_3^u=1/2$ and $T_3^d=-1/2$. $\theta_W$ is the weak angle, $M_Z$ is the $Z$ gauge boson mass, and $(m_{u_1},m_{u_2}, m_{u_3})=(m_u,m_c,m_t)$, $(m_{d_1},m_{d_2}, m_{d_3})=(m_d,m_s,m_b)$. It should be noted that the non-diagonality in flavour comes from the values of $m_{\tilde U_L \, ij}^2$, $m_{\tilde U_R \, ij}^2$, $m_{\tilde D_L \, ij}^2$, $m_{\tilde D_R \, ij}^2$, ${\cal A}_{ij}^u$ and ${\cal A}_{ij}^d$ for $i \neq j$.

The next step is to rotate the squark states from the Super-CKM basis, 
${\tilde q}_{L,R}$, to the
physical basis, ${\tilde q}^{\rm phys}$. If we set the order in the Super-CKM basis as above, and in the physical basis as
${\tilde u}_{1,..6}$ and ${\tilde d}_{1,..6}$, respectively, these last rotations are given by two $6 \times 6$ matrices, $R^{\tilde u}$ and $R^{\tilde d}$, 
\BE
\VL  \tiu_{1} \\ \tiu_{2}  \\ \tiu_{3} \\
                                    \tiu_{4}   \\ \tiu_{5}  \\\tiu_{6}   \VR
  \; = \; R^{\tiu}  \VL \SupL \\ \SchaL \\\StopL \\ 
  \SupR \\ \SchaR \\ \StopR \VR ~,~~~~
\VL  \tid_{1} \\ \tid_{2}  \\  \tid_{3} \\
                                   \tid_{4}     \\ \tid_{5} \\ \tid_{6}  \VR             \; = \; R^{\tid}  \VL \SdownL \\ \SstrL \\ \SbotL \\
                                      \SdownR \\ \SstrR \\ \SbotR \VR ~,
\label{newsquarks}
\end{equation} 
yielding the diagonal mass-squared matrices as follows,
\BEA
{\rm diag}\{m_{\tiu_1}^2, m_{\tiu_2}^2, 
          m_{\tiu_3}^2, m_{\tiu_4}^2, m_{\tiu_5}^2, m_{\tiu_6}^2 
           \}  & = &
R^{\tiu}  \;  {\cal M}_{\tiu}^2   \; 
 R^{\tiu \dagger}    ~,\\
{\rm diag}\{m_{\tid_1}^2, m_{\tid_2}^2, 
          m_{\tid_3}^2, m_{\tid_4}^2, m_{\tid_5}^2, m_{\tid_6}^2 
          \}  & = &
R^{\tid}  \;   {\cal M}_{\tid}^2   \; 
 R^{\tid \dagger}    ~.
\EEA 

\bigskip
As commented before, we will introduce the non-diagonal terms in flavour space in the Super-CKM basis and then rotate to the mass basis. These squark flavour mixings are usually described in terms of a set of
dimensionless parameters $\deXYij$ ($X,Y=L,R$; $i,j=t,c,u$ or $b,s,d$),
which for simplicity in the computations are frequently  considered
within the Mass Insertion Approximation~\cite{Hall:1985dx}. We
will not use here this approximation, but instead we will solve exactly
the diagonalization of the squark mass matrices.   

In the numerical part of the present study we will restrict ourselves
  to the case where there are flavour mixings exclusively between the second and third squark generation. These mixings are known to produce the largest flavour violation effects in $B$ meson physics since their size are usually governed by the third generation quark masses.


The terms that will produce generation mixing in the squark mass matrices \ref{eq:SCKM-entries} are then parametrized as follows,
\noindent \begin{equation}  
m^2_{\tilde U_L}= \left(\begin{array}{ccc}
 m^2_{\tilde U_{L11}} & 0 & 0\\
0 & m^2_{\tilde U_{L22}}  & \delta_{23}^{LL} m_{\tilde U_{L22}}m_{\tilde U_{L33}}\\
0 & \delta_{23}^{LL} m_{\tilde U_{L22}}m_{\tilde U_{L33}}& m^2_{\tilde U_{L33}} \end{array}\right),\label{matdeltas1}\end{equation}

\noindent \begin{equation}
v_2 {\cal A}^u  =\left(\begin{array}{ccc}
0 & 0 & 0\\
0 & 0 & \delta_{ct}^{LR} m_{\tilde U_{L22}}m_{\tilde U_{R33}}\\
0 & \delta_{ct}^{RL} m_{\tilde U_{R22}}m_{\tilde U_{L33}} & m_{t}A_{t}\end{array}\right),\end{equation}

\noindent \begin{equation}  
m^2_{\tilde U_R}= \left(\begin{array}{ccc}
 m^2_{\tilde U_{R11}} & 0 & 0\\
0 & m^2_{\tilde U_{R22}}  & \delta_{ct}^{RR} m_{\tilde U_{R22}}m_{\tilde U_{R33}}\\
0 & \delta_{ct}^{RR} m_{\tilde U_{R22}}m_{\tilde U_{R33}}& m^2_{\tilde U_{R33}} \end{array}\right),\end{equation}

\noindent \begin{equation}
m^2_{\tilde D_L} =\VCKM^{\dagger}m^2_{\tilde U_L}\VCKM,
\label{eq:relac-mu2ll-md2ll-su2}\end{equation}

\noindent \begin{equation}
v_1 {\cal A}^d   =\left(\begin{array}{ccc}
0 & 0 & 0\\
0 & 0 & \delta_{sb}^{LR} m_{\tilde D_{L22}}m_{\tilde D_{R33}} \\
0 & \delta_{sb}^{RL} m_{\tilde D_{R22}}m_{\tilde D_{L33}} & m_{b}A_{b}\end{array}\right),\end{equation}

\noindent \begin{equation}  
m^2_{\tilde D_R}= \left(\begin{array}{ccc}
 m^2_{\tilde D_{R11}} & 0 & 0\\
0 & m^2_{\tilde D_{R22}}  & \delta_{sb}^{RR} m_{\tilde D_{R22}}m_{\tilde D_{R33}}\\
0 & \delta_{sb}^{RR} m_{\tilde D_{R22}}m_{\tilde D_{R33}}& m^2_{\tilde D_{R33}} 
\end{array}\right).\label{matdeltas6}\end{equation}

It is worth mentioning that the relation between the two soft squark mass matrices in the `Left' sector (\ref{eq:relac-mu2ll-md2ll-su2}) is due to SU(2) gauge
invariance, and it can be derived from the two last relations of Eq. \ref{eq:su2}. This relation between the non-diagonal terms of these two squark mass matrices is the reason why it is introduced just one $\delta_{23}^{LL}$ instead of two independent deltas $\delta_{ct}^{LL}$ and $\delta_{sb}^{LL}$ in the matrices $m^2_{\tilde U_L}$ and $m^2_{\tilde D_L}$.  
In all this work, for simplicity, we are assuming that all $\deABij$
parameters are real, therefore, hermiticity of the squared mass matrices implies $\delta_{ij}^{AB}= \delta_{ji}^{BA}$.

It should be noted that in the `Left-Right' sector we have 
normalized the trilinear couplings at low energies as ${\cal A}^q_{ij}= y_{q_i} A^q_{ij}$  (with $A^u_{33}=A_t$ and $A^d_{33}=A_b$) and we have neglected the $A_i$ couplings of the first and second generations. Finally, it should also be noted that the dimensionless parameters $\deXYij$ defining the non-diagonal entries in flavour space 
$(i \neq j)$ are normalized respect the  geometric mean of the corresponding diagonal squared soft masses. For instance, 
\begin{eqnarray}
&&\delta^{LL}_{23}= m^2_{\tilde U_{L23}}/(m_{\tilde U_{L22}}m_{\tilde U_{L33}}), 
\delta^{RR}_{ct}= m^2_{\tilde U_{R23}}/(m_{\tilde U_{R22}}m_{\tilde U_{R33}}),
~~~\nonumber\\
&&\delta^{LR}_{ct}= (v_2 {\cal A}^u)_{23}/(m_{\tilde U_{L22}}m_{\tilde U_{R33}}), 
\delta^{RL}_{ct}= (v_2 {\cal A}^u)_{32}/(m_{\tilde U_{R22}}m_{\tilde U_{L33}}),
~~~{\rm etc}.
\label{deltasdefs}
\end{eqnarray}     

\subsection{The slepton sector with general flavour mixing}
\label{sec:slepsector}

The slepton sector will be explored with the same general approach we used in the squark sector in the previous section. We will consider general flavour mixing in the slepton sector at low energy, without assuming any high-energy hypothesis for the generation of the relevant terms.
As in the case of squarks, this NMFV framework goes beyond the minimal flavour violation that is always induced by the Yukawa couplings. As it happened with the quarks, the rotation from the interaction basis to the mass basis of the SM particles induces some minimum amount of flavour violation in the sparticles that can not be avoided. In the quarks, this violation was encoded in the CKM matrix. In the leptons/sleptons it will be induced by the PMNS matrix
of the neutrino sector and transmitted by the tiny neutrino Yukawa
couplings. However, we will ignore these latter here because of their smallness. 

The most general hypothesis for flavour
mixing in the slepton sector assumes, as in the squark case, a mass matrix that is not diagonal
in flavour space, both for charged sleptons and sneutrinos. 
In the charged slepton sector we have a $6 \times 6$ mass matrix,
since there are six electroweak interaction eigenstates, 
${\tilde l}_{L,R}$ with $l=e, \mu, \tau$. For the sneutrinos 
we have a $3 \times 3$ mass matrix, since within the MSSM we have
only three electroweak interaction eigenstates, ${\tilde \nu}_{L}$ with
$\nu=\nu_e, \nu_\mu, \nu_\tau$. 

The non-diagonal entries in this $6 \times 6$ general matrix for charged
sleptons can be described in a model-independent way in terms of a set
of 
dimensionless parameters $\deABij$ ($A,B=L,R$; $i,j=1,2,3$, 
$i \neq j$), where $L,R$ refer to the 
``left-'' and ``right-handed'' SUSY partners of the corresponding
leptonic degrees of freedom, and $i,j$
indices run over the three generations. Due to the small size of the slepton Yukawa couplings, in the present case of slepton mixing, we
assume that the $\deABij$'s  provide the unique origin
of LFV processes with potentially measurable rates. Notice that we are using, for shortness, the same generic notation for the deltas  $\deABij$ in both sectors, squarks and sleptons, but they are obviously different parameters and should not be confused along this work.   

One usually starts  with the non-diagonal $6 \times 6$ slepton squared
mass matrix referred to the electroweak interaction basis,  
 that we order here as    $(\SelL, \SmuL, \StauL, \SelR, \SmuR,
 \StauR)$, and write this matrix in terms of left- and right-handed
 blocks $M^2_{\tilde l \, AB}$  
 ($A,B=L,R$), which are non-diagonal $3\times 3$ matrices,
\begin{equation}
{\mathcal M}_{\tilde l}^2 =\left( \begin{array}{cc}
M^2_{\tilde l \, LL} & M^2_{\tilde l \, LR} \\ 
M_{\tilde l \, LR}^{2 \, \dagger} & M^2_{\tilde l \,RR}
\end{array} \right),
\label{eq:slep-6x6}
\end{equation} 
 where:
 \begin{alignat}{5}
M_{\tilde l \, LL \, ij}^2 
  = &  m_{\tilde L \, ij}^2 + \left( m_{l_i}^2
     + (-\frac{1}{2}+ \sin^2 \theta_W ) M_Z^2 \cos 2\beta \right) \delta_{ij},  \notag\\
 M^2_{\tilde l \, RR \, ij}
  = &  m_{\tilde E \, ij}^2 + \left( m_{l_i}^2
     -\sin^2 \theta_W M_Z^2 \cos 2\beta \right) \delta_{ij} \notag, \\
  M^2_{\tilde l \, LR \, ij}
  = &  v_1 {\cal A}_{ij}^l- m_{l_{i}} \mu \tb \, \delta_{ij},
\label{eq:slep-matrix}
\end{alignat}
with flavour indices $i,j=1,2,3$ corresponding to the first, second and
third generation respectively; and $(m_{l_1},m_{l_2},
m_{l_3})=(m_e,m_\mu,m_\tau)$ are the lepton masses.
It should be noted that the non-diagonality in flavour comes
exclusively from the soft SUSY-breaking parameters, that could be
non-vanishing for $i \neq j$, namely: the masses $m_{\tilde L \, ij}$
for the slepton $SU(2)$ doublets, $(\tilde \nu_{Li}\,\,\, \tilde
l_{Li})$, the masses $m_{\tilde E \, ij}$ for the slepton $SU(2)$
singlets, $(\tilde l_{Ri})$, and the trilinear couplings, ${\cal
  A}_{ij}^l$.   

In the sneutrino sector there is, correspondingly, a one-block $3\times
3$ mass matrix, that is referred to the $(\tinu_{eL}, \tinu_{\mu L},
\tinu_{\tau L})$ electroweak interaction basis: 
\begin{equation}
{\mathcal M}_{\tilde \nu}^2 =\left( \begin{array}{c}
M^2_{\tilde \nu \, LL}   
\end{array} \right),
\label{eq:sneu-3x3}
\end{equation} 
 where:
\begin{equation} 
  M_{\tilde \nu \, LL \, ij}^2 
  =   m_{\tilde L \, ij}^2 + \left( 
      \frac{1}{2} M_Z^2 \cos 2\beta \right) \delta_{ij},   
\label{eq:sneu-matrix}
\end{equation} 
 
It should also be noted that, again, due to $SU(2)_L$ gauge invariance
the same soft masses $m_{\tilde L \, ij}$ enter in both the slepton and
sneutrino $LL$ mass matrices. 
If neutrino masses and neutrino flavour mixings (oscillations) were taken into account, the soft SUSY-breaking parameters for the
sneutrinos would differ from the corresponding ones for charged
sleptons by a rotation with the PMNS matrix. However, taking the neutrino masses and oscillations 
into account in the SM leads to LFV effects that are extremely small. For instance, in $\mu \to e \gamma$  they are of \order{10^{-47}} in the SM case
with Dirac neutrino masses around 1~eV and maximal mixing~\cite{Kuno:1999jp,DiracNu}, and of \order{10^{-40}} in case of
Majorana neutrinos~\cite{Kuno:1999jp,MajoranaNu}. Consequently we do not expect
large effects from the inclusion of neutrino masses here and it is safe to ignore them.

The general slepton flavour mixing is introduced via the
non-diagonal terms in the soft breaking slepton mass matrices and
trilinear coupling matrices, which are defined here as:

\noindent \begin{equation}  
m^2_{\tilde L}= \left(\begin{array}{ccc}
 m^2_{\tilde L_{1}} & \delta_{12}^{LL} m_{\tilde L_{1}}m_{\tilde L_{2}} & \delta_{13}^{LL} m_{\tilde L_{1}}m_{\tilde L_{3}} \\
 \delta_{21}^{LL} m_{\tilde L_{2}}m_{\tilde L_{1}} & m^2_{\tilde L_{2}}  & \delta_{23}^{LL} m_{\tilde L_{2}}m_{\tilde L_{3}}\\
\delta_{31}^{LL} m_{\tilde L_{3}}m_{\tilde L_{1}} & \delta_{32}^{LL} m_{\tilde L_{3}}m_{\tilde L_{2}}& m^2_{\tilde L_{3}} \end{array}\right),\label{mLL}\end{equation}

\noindent \begin{equation}
v_1 {\cal A}^l  =\left(\begin{array}{ccc}
m_e A_e & \delta_{12}^{LR} m_{\tilde L_{1}}m_{\tilde E_{2}} & \delta_{13}^{LR} m_{\tilde L_{1}}m_{\tilde E_{3}}\\
\delta_{21}^{LR}  m_{\tilde L_{2}}m_{\tilde E_{1}} & m_\mu A_\mu & \delta_{23}^{LR} m_{\tilde L_{2}}m_{\tilde E_{3}}\\
\delta_{31}^{LR}  m_{\tilde L_{3}}m_{\tilde E_{1}} & \delta_{32}^{LR}  m_{\tilde L_{3}} m_{\tilde E_{2}}& m_{\tau}A_{\tau}\end{array}\right),\label{v1Al}\end{equation}

\noindent \begin{equation}  
m^2_{\tilde E}= \left(\begin{array}{ccc}
 m^2_{\tilde E_{1}} & \delta_{12}^{RR} m_{\tilde E_{1}}m_{\tilde E_{2}} & \delta_{13}^{RR} m_{\tilde E_{1}}m_{\tilde E_{3}}\\
 \delta_{21}^{RR} m_{\tilde E_{2}}m_{\tilde E_{1}} & m^2_{\tilde E_{2}}  & \delta_{23}^{RR} m_{\tilde E_{2}}m_{\tilde E_{3}}\\
\delta_{31}^{RR}  m_{\tilde E_{3}} m_{\tilde E_{1}}& \delta_{32}^{RR} m_{\tilde E_{3}}m_{\tilde E_{2}}& m^2_{\tilde E_{3}} \end{array}\right).\label{mRR}\end{equation}

In all this work, for simplicity, we are assuming that all $\deABij$
parameters are real, therefore, hermiticity of 
${\mathcal M}_{\tilde l}^2$ and 
${\mathcal M}_{\tilde \nu}^2$ implies $\delta_{ij}^{AB}= \delta_{ji}^{BA}$.
Besides, in order to avoid extremely large off-diagonal matrix entries
we restrict ourselves to $|\deABij| \leq 1$. It is worth to recall again that our parametrization of the off-diagonal in flavour space entries in the above mass matrices is purely phenomenological and does not rely on any specific assumption on the origin of the MSSM soft mass parameters. In particular, it should be noted that our parametrization for the $LR$ and $RL$ squared mass entries connecting different generations (i.e. for $i \neq j$) assumes a similar generic form as for the $LL$ and $RR$ entries. For instance, $M^2_{\tilde l \, LR \, 23}= \delta_{23}^{LR} m_{\tilde L_{2}}m_{\tilde E_{3}}$. This implies that our hypothesis for the trilinear off-diagonal couplings ${\cal A}^l_{ij}$ with $i \neq j$ (as derived from Eq.(\ref{v1Al})) is one among other possible definitions considered in the literature. In particular, it is related to the usual assumption $M^2_{\tilde l \, LR \, ij} \sim v_1 M_{\rm SUSY}$ by setting ${\cal A}^l_{ij} \sim {\cal O}(M_{\rm SUSY})$, where $v^2=v_1^2+v_2^2=2\frac{M_W^2}{g^2}$ and $M_{\rm SUSY}$ is a typical SUSY mass scale, as it is done for instance in Ref.~\cite{Crivellin:2011jt}. It should be also noted that the diagonal entries in Eq.(\ref{v1Al}) have been normalized as it is usual in the literature, namely, by factorizing out the corresponding lepton Yukawa coupling: ${\cal A}^l_{ii}= y_{l_i} A^l_{ii}$, with $A^l_{11}=A_e$, $A^l_{22}=A_\mu$ and $A^l_{33}=A_\tau$. Finally, it should be mentioned that our choice in Eqs.(\ref{mLL}), (\ref{v1Al}) and (\ref{mRR}) is to normalize the non-diagonal in flavour entries with respect to the geometric mean of the corresponding diagonal squared soft masses. For instance, 
\begin{eqnarray}
&&\delta^{LL}_{23}= (m^2_{\tilde L})_{23}/(m_{\tilde L_{2}}m_{\tilde L_{3}}), \,\,\,\,\, 
\delta^{RR}_{23}= (m^2_{\tilde E})_{23}/(m_{\tilde E_{2}}m_{\tilde E_{3}}),
~~~\nonumber\\
&&\delta^{LR}_{23}= (v_1 {\cal A}^l)_{23}/(m_{\tilde L_{2}}m_{\tilde E_{3}}), \,\,\,\,\,
\delta^{RL}_{23}=\delta^{LR}_{32}= (v_1 {\cal A}^l)_{32}/(m_{\tilde L_{3}}m_{\tilde E_{2}}).
\label{deltas23defs}
\end{eqnarray}

The next step is to rotate the sleptons and sneutrinos from the electroweak interaction basis to the physical mass eigenstate basis, 
\BE
\VL  \til_{1} \\ \til_{2}  \\ \til_{3} \\
                                    \til_{4}   \\ \til_{5}  \\\til_{6}   \VR
  \; = \; R^{\til}  \VL \SelL \\ \SmuL \\\StauL \\ 
  \SelR \\ \SmuR \\ \StauR \VR ~,~~~~
\VL  \tinu_{1} \\ \tinu_{2}  \\  \tinu_{3}  \VR             \; = \; R^{\tinu}  \VL \tinu_{eL} \\ \tinu_{\mu L}  \\  \tinu_{\tau L}   \VR ~,
\label{newsleptons}
\end{equation} 
with $R^{\til}$ and $R^{\tinu}$ being the respective $6\times 6$ and
$3\times 3$ unitary rotating matrices that yield the diagonal
mass-squared matrices as follows, 
\begin{eqnarray}
{\rm diag}\{m_{\tilde l_1}^2, m_{\tilde l_2}^2, 
          m_{\tilde l_3}^2, m_{\tilde l_4}^2, m_{\tilde l_5}^2, m_{\tilde l_6}^2 
           \}  & = &
R^{\tilde l}   {\cal M}_{\tilde l}^2    
 R^{\tilde l \dagger}    ~, \label{sleptons}\\
{\rm diag}\{m_{\tilde \nu_1}^2, m_{\tilde \nu_2}^2, 
          m_{\tilde \nu_3}^2  
          \}  & = &
R^{\tilde \nu}     {\cal M}_{\tilde \nu}^2    
 R^{\tilde \nu \dagger}    ~.
\label{sneutrinos} 
\end{eqnarray}
The physics must not depend on the ordering of the masses. However,
  in our numerical analysis we work with mass ordered states, 
$m_{\til_i} \le m_{\til_j}$ for $i < j$ and $m_{\tinu_k} \le  m_{\tinu_l}$ 
for $k < l$.


\section{Selected SUSY scenarios}
\label{scenarios}

The choice of the scenarios to study is a crucial point in our research. The most general version of the MSSM has too many parameters to understand the implications of varying all of them in the physical observables we can measure. If we do not want to lose the predictability of the theory and be able to handle it, we must assume new relations that reduce the number of parameters to a reasonable amount. In the following, we will comment some usual ways to reduce the number of parameters. However, the parameters related with the flavour mixing, will be maintained in the most general case, since it is their consequences what we want to understand in depth.


\subsection{Constrained MSSM scenarios}
\label{cmssm}

For the studies in Chapters \ref{sec:Bphysics} and \ref{higgsmasssquark} we will choose as two of the possible SUSY scenarios, with a much more reduced number of parameters respect to the MSSM, the so-called Constrained MSSM (CMSSM) and Non Universal Higgs Mass (NUHM) scenarios, which are defined by the following input parameters (see~\cite{AbdusSalam:2011fc} and references therein):  
 \begin{eqnarray}
 {\rm CMSSM}:& ~~~~~m_0, m_{1/2}, A_0, {\rm sign}(\mu), \tb 
 \nonumber \\
 {\rm NUHM}:& ~~~~~m_0,  m_{1/2}, A_0, {\rm sign}(\mu), \tb, 
 m_{H_1}, m_{H_2},
 \end{eqnarray} 
where, $A_0$ is the universal trilinear coupling, $m_0$, $m_{1/2}$, $m_{H_1}$, $m_{H_2}$, are the universal scalar mass,
gaugino mass, and $m_{H_1}$ and $m_{H_2}$ Higgs masses, respectively, at the GUT scale,   ${\rm sign}(\mu)$ is the sign of the $\mu$ parameter and again $\tb =v_2/v_1$.  The soft Higgs masses in the NUHM are usually parametrized as 
$m^2_{H_{1,2}}=(1+\delta_{1,2})m_0^2$, such that by taking $\delta_{1,2}=0$ one 
moves from the NUHM to the CMSSM. 

It should be remarked that the condition of universal squark soft masses is fulfilled just at the GUT scale. When 
running with the RGE these soft mass matrices from the GUT scale down to the relevant low energy, they will generically turn non-diagonal in flavour. For instance, in CMSSM and other SUSY-GUT scenarios the
flavour changing deltas in the squark sector go (in the leading logarithmic approximation)
as
$\de^{LL}_{23} \simeq   -\frac{1}{8\pi^2}\frac{(3m_0^2+A_0^2)}{\tilde
  m^2}(Y^{q\dagger}Y^q)_{23}\log(\frac{M_{{\rm GUT}}}{M_{{\rm EW}}})$
($\tilde m^2$ is usually taken as the geometric mean of the involved
flavour diagonal squared squark mass matrix entries, see Eq.
\ref{deltasdefs}),
whereas the $LR$, $RL$ and $RR$ deltas are
suppressed instead by small mass ratios, $\sim \frac{(m_q A_0)}{\tilde
  m^2}$ and
$\sim \frac{(m_q^2)}{\tilde m^2}$, respectively. Furthermore, in these
scenarios the mixing involving the first generation squarks is
naturally suppressed by the smallness of the corresponding Yukawa
couplings. Overall, in these scenarios like in other MFV cases the non-diagonal terms are exclusively generated in this running via the Yukawa couplings, hence their size will be governed by off-diagonal terms in the CKM (or PMNS matrix in the slepton case), and therefore they can be safely neglected at low energy. Contrary, in NMFV scenarios, the universal hypothesis in the sfermions mass matrices is by definition not fulfilled at low energies. In these scenarios that will be the core of our work, we will depart from the low energy model with MFV, by imposing non vanishing values for the various $\deXYij$.
 
To perform the running of the soft parameters from the GUT scale down to low energy, that we set here around 1 TeV,  we solve numerically the one-loop RGEs with the code {\it SPHENO}~\cite{SPheno}. 
The diagonalization of all the mass matrices is
performed with the program
\fh~\cite{feynhiggs,mhiggslong,mhiggsAEC,mhcMSSMlong}.  

\begin{figure}[H]
\begin{center}
\includegraphics[width=145mm]{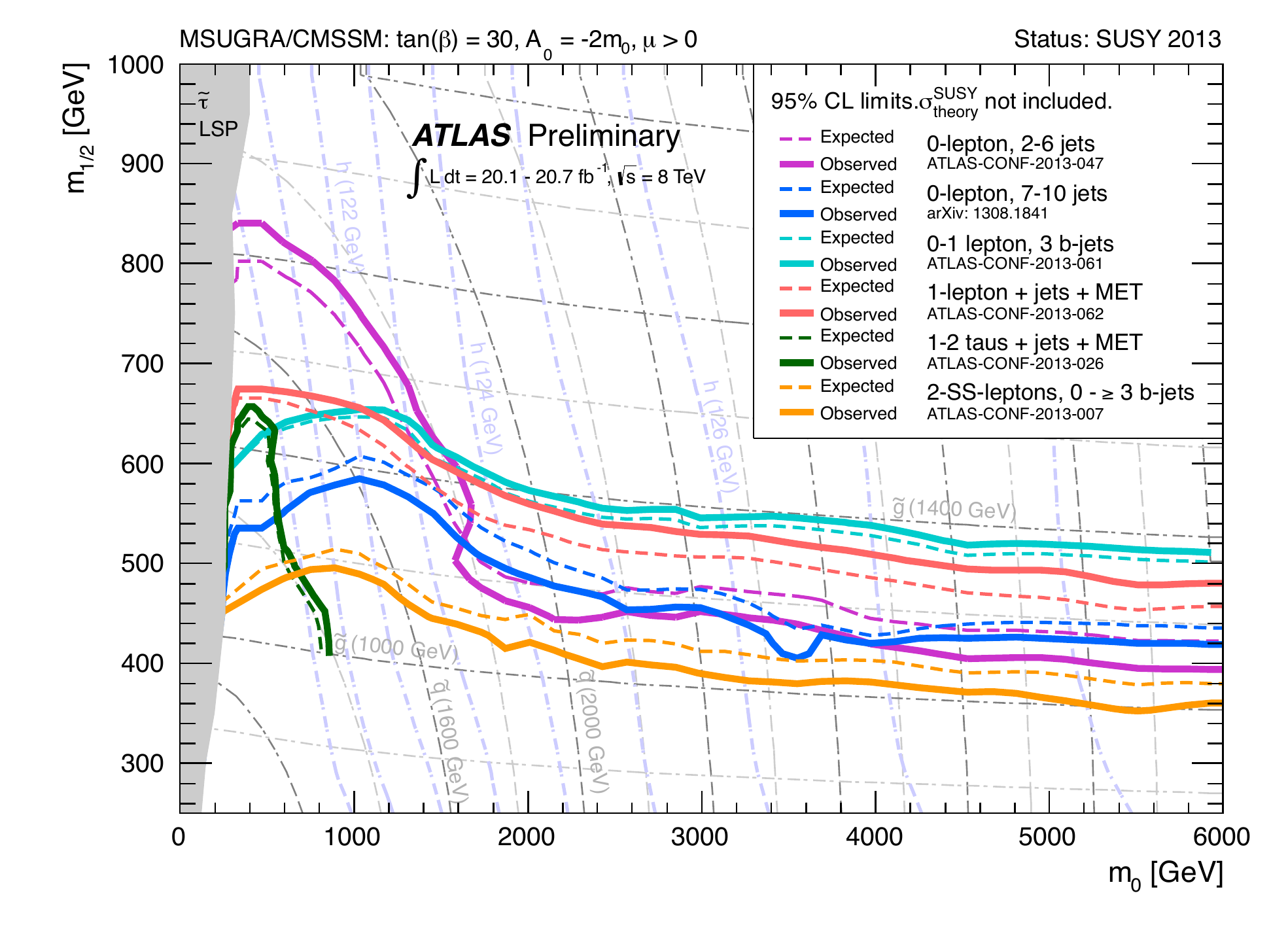}\\
\includegraphics[width=145mm]{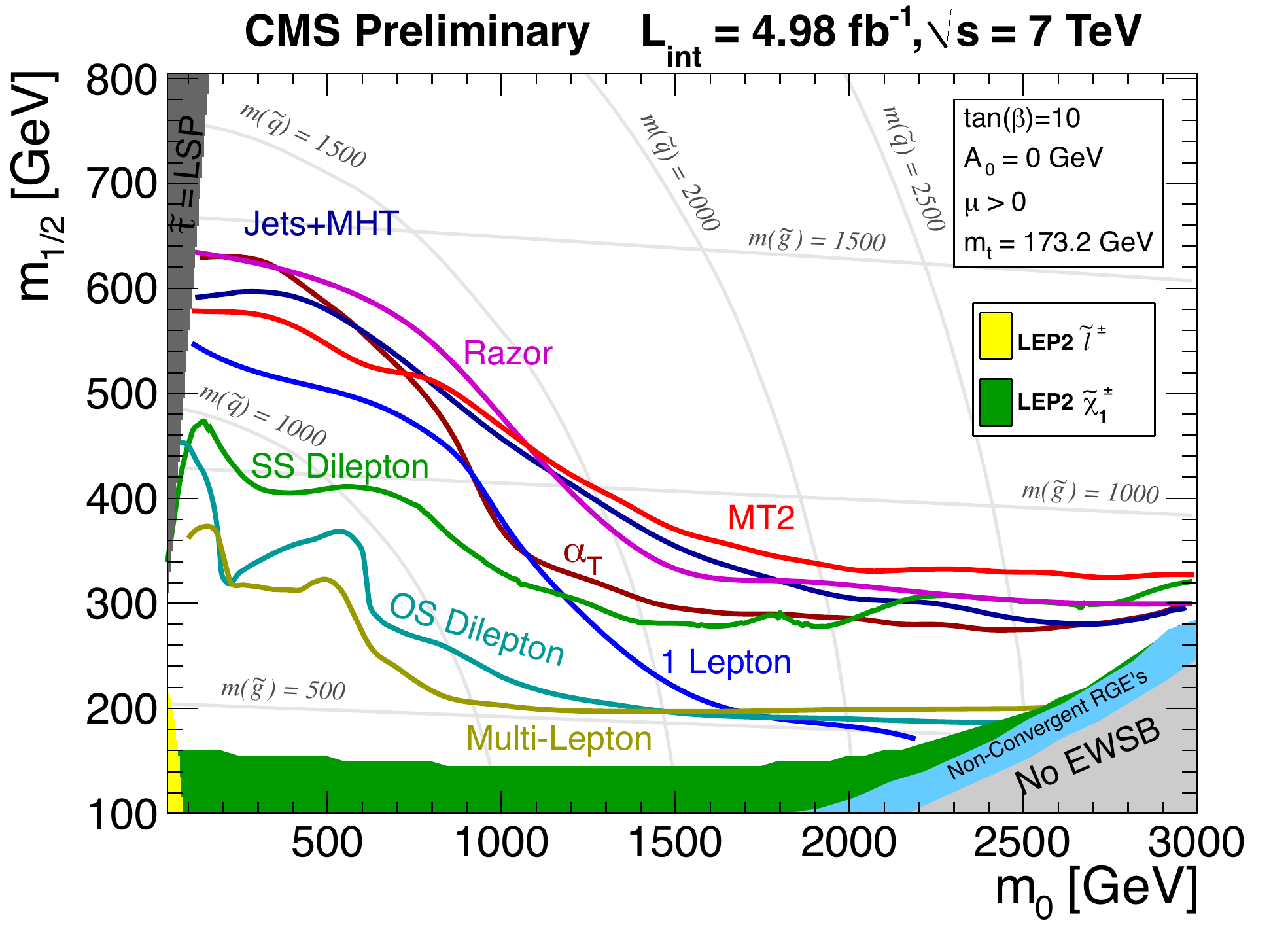}
\caption{Exclusion plots at the 95\% CL in the $m_0-m_{1/2}$ plane for CMSSM/MSUGRA for Atlas using the 8 TeV data, with $\tan \beta=30$, $A_0=-2m_0$ and $\mu>0$ (top) and for CMS using the 7 TeV data, with $\tan \beta=10$, $A_0=0$ and $\mu>0$ (bottom). Figures taken from  \cite{atlassusysearches,cmssusysearches}}
\label{cmmsexclusion}
\end{center}
\end{figure}

In Figure \ref{cmmsexclusion} there can be seen the present exclusion limits in the $m_0-m_{1/2}$ plane for the CMSSM/MSUGRA scenario for ATLAS and CMS using respectively the 8 and 7 TeV data (the other parameters are given in the figure) \cite{atlassusysearches,cmssusysearches}.

\subsubsection{Specific CMSSM points}
\label{frameworka}

The selected scenarios in specific constrained SUSY scenarios, CMSSM and NUHM, with input parameters $m_0$, $m_{1/2}$, $A_0$, $\tb$, sign($\mu$), $\delta_{1,2}$, are summarized in table \ref{points} \footnote{We adopt here the definition in terms of the GUT-scale input parameters, while the original definition in  \cite{Allanach:2002nj}
was based on the weak-scale parameters.}, and supplemented with $\deXYij$ as
 described above. Regarding CMSSM, we have chosen six SPS benchmark points~\cite{Allanach:2002nj}, SPS1a, SPS1b, SPS2, SPS3, SPS4, and SPS5 and one  more point with 
very heavy spectra (VHeavyS). Regarding NUHM, we have chosen a point with heavy SUSY spectra and light Higgs sector (HeavySLightH) and a point (BFP) that was proven in the moment of the study \cite{Buchmueller:2011ki} to give the best fit to the set of low energy observables. For later reference, needed in our posterior analysis of the Higgs mass corrections, we also include in the table the corresponding MSSM Higgs masses,  computed with \fh~\cite{feynhiggs,mhiggslong,mhiggsAEC,mhcMSSMlong} and with all flavour changing deltas switched off, i.e., $\deXYij=0$.     
It should be noted that our studies for these points were performed before the last data of the LHC, so some of these points (the SPS ones) are already excluded by the LHC. The study will be presented here in any case to illustrate with well known examples the phenomenology of the flavour physics and the effects on the Higgs mass corrections. The SPSX points are especially relevant since they have been studied at length in the literature. However, we will also present an updated study with currently allowed points and scenarios, that will be described in the next section. Both studies will be compared to understand also the effect of the last experiments in our constraints of the $\deXYij$ parameters and in the MSSM.

\begin{center}
\begin{table}[h!]
\centering
\begin{tabular}{|c|c|c|c|c|c|c|c|c|c|c|}
\hline 
\multicolumn{11}{|c|}{Scenarios for Constrained MSSM} \\
\hline
points & $m_{1/2}$ & $m_0$ & $A_0$ & $\tb$ &  $\delta_1$ & $\delta_2$ &$m_{h}$ &$m_{H}$ &$M_{A}$ &$m_{H^{\pm}}$ \\\hline 
SPS1\,a & 250 & 100 & -100 & 10 &  $0$ & $0$ &112 &394 &394&402 \\
 SPS1\,b & 400 & 200 & 0 & 30 &  $0$ & $0$  &116 &526 &526 &532\\
 SPS2 &  300 & 1450 & 0 & 10 &  $0$ & $0$  &115 &1443 &1443 &1445 \\
 SPS3 &  400 & 90 & 0 & 10 &   $0$ & $0$   &115 &573 &572 &578\\
 SPS4 &  300 & 400 & 0 & 50 &  $0$ & $0$   &114 &404 &404 &414\\
 SPS5 &  300 & 150 & -1000 & 5 &  $0$ & $0$   &111 &694 &694 &698\\
VHeavyS &  800 & 800 & -800 & 5 &  $0$ & $0$   &120 &1524 &1524 &1526\\
HeavySLightH &600 &600 &0 & 5 &   $-1.86$ & $+1.86 $ &114 &223&219 &233\\
BFP &530 &110 &-370 & 27 &   $-84.7$ & $-84.7$ &120 &507 &507 &514\\
\hline
\end{tabular}
\caption{Values of $m_{1/2}$, $m_0$, $A_0$, $\tb$, $\delta_{1}$, $\delta_{2}$ and Higgs boson masses $m_{h}$, $m_{H}$, $M_{A}$ and $m_{H^{\pm}}$ for the pre-LHC CMSSM points considered in the analysis. All parameters with mass dimension are in GeV, and ${\rm sign}(\mu)>0$ for all points. For the computations of the Higgs boson masses in this table all the  $\deXYij$ are set to zero.\label{points}}
\end{table}
\end{center}

\subsection{Phenomenological MSSM}

We will also consider in this work (in particular for Chapters \ref{phenoflavourslep} and \ref{lfvhiggsdecaysslepton}) scenarios directly at the lower effective SUSY breaking scale, where a much larger amount of parameters are available, and then set some relations between them to reduce their number. In that case we will not consider theoretical arguments for setting the relations, but our interest will be to obtain a phenomenologically useful model. The studied models are called ``phenomenological MSSM scenarios (pMSSM)'' in the literature (see, for instance, \cite{Arbey:2012dq,Bechtle:2012jw,AbdusSalam:2011fc}). The assumptions for these models are the following: the breaking parameters are real, the masses and trilinear couplings appearing in the breaking terms are the same for the first and second generation at the electroweak symmetry breaking scale, and all these masses and trilinear couplings are diagonal in flavour. Being this the standard definition, we will not respect this last condition, and we will introduce non-zero elements for the non-diagonal terms. But before introducing them, the pMSSM will reduce the parameters of our model to just 22:

\begin{itemize} 
\vspace*{-3mm}
\item[--] $\tan \beta$: the ratio of the vevs;\vspace*{-3mm} 

\item[--] $M_A$: the pseudoscalar Higgs mass;\vspace*{-3mm}

\item[--] $\mu$: the Higgs--Higgsino (supersymmetric) mass parameter (with both
signs);\vspace*{-3mm} 

\item[--] $M_1, M_2, M_3$: the bino, wino and gluino mass parameters;\vspace*{-3mm}

\item[--] $m_{\tilde{Q}_1}, m_{\tilde{U}_1}, m_{\tilde{D}_1},  m_{\tilde{L}_1},
 m_{\tilde{E}_1}$: the first (=second) generation sfermion mass
 parameters;\vspace*{-3mm}

\item[--]  $A_u, A_d, A_e$: the first (=second) generation diagonal trilinear
couplings;\vspace*{-3mm}

\item[--] $m_{\tilde{Q}_3}, m_{\tilde{U}_3}, m_{\tilde{D}_3},  m_{\tilde{L}_3},
 m_{\tilde{E}_3}$: the third generation sfermion mass parameters;\vspace*{-3mm}

\item[--] $A_t, A_b, A_\tau$: the third generation diagonal trilinear
couplings.\vspace*{-3mm}
\end{itemize} 

In this pMSSM framework, we consider several possibilities for the
MSSM parameters, leading to simple patterns of SUSY masses with specific 
relations among them and where the number of input parameters is strongly
reduced.  The selected scenarios lead to predictions of $\Mh$ that are compatible with present data over a large part of
  the parameter space. We have also required that all our points lead to a prediction
of the anomalous magnetic moment of the muon in the MSSM that can fill
the present discrepancy between the Standard Model prediction and the
experimental value. Specifically, we use \citeres{Bennett:2006fi} and 
\cite{Davier:2010nc} to extract the size of this discrepancy,
see also \citere{gm2-Jegerlehner}:
\begin{equation}
(g-2)_\mu^{\rm exp}-(g-2)_\mu^{\rm SM}= (30.2 \pm 9.0) \times 10^{-10}.
\label{gminus2}  
\end{equation}
We then require that the SUSY contributions from charginos and neutralinos in the MSSM to one-loop level, 
$(g-2)_\mu^{\rm SUSY}$  be within the interval defined by $3 \sigma$ around the central value in \refeq{gminus2}, namely:    
\BE
 (g-2)_\mu^{\rm SUSY} \in  (3.2 \times 10^{-10},57.2 \times 10^{-10}) 
\label{gminus2interval}  
\EE 

We will consider two kind of scenarios in the pMSSM in growing order of number of parameters: simplified pMSSM scenarios with common parameters (covered in Sections \ref{sec:f2} and \ref{pmssmvheavysusy}) and specific points (covered in Section \ref{frameworkc}).
 
\subsubsection{pMSSM-4 scenarios}
\label{sec:f2}

To simplify the analysis of
the upper bounds of the deltas, we will focus in scenarios where the
mass scales that are relevant for our studied processes are all set to common mass scales. 

Regarding the LFV processes, the mass scale will be generically 
called here $m_{\rm SUSY-EW}$. 
This implies setting the slepton soft masses, the gaugino soft
masses, $M_2$ and $M_1$  and the $\mu$ parameter in terms of  this 
$m_{\rm SUSY-EW}$. It should also be noted that these same mass
parameters are the relevant ones for $(g-2)_\mu$. The remaining
relevant parameter in both LFV and $(g-2)_\mu$ is $\tb$, and the
LFV analysis of Chapter \ref{phenoflavourslep} will be performed in the ($m_{\rm SUSY-EW}$, $\tb$) plane.

Regarding the Higgs boson, we are  interested in choices of the MSSM parameters that
lead to a prediction of $\Mh$ that is compatible with LHC data, so we have 
to set the corresponding relevant mass parameters for this
observable. These are mainly the squark soft masses and trilinear soft
couplings, with particular relevance of those parameters of the third
generation squarks. All these squark mass scales will be set, in this
framework, relative to one single mass scale, $m_{\rm SUSY-QCD}$.  

Since we wish to explore a wide range in $\tb$,
from 5 to 60, $\MA$ is fixed to $1000 \gev$ to ensure the agreement
with the present bounds in the $(\tb, \MA)$ plane from LHC
searches~\cite{CMS-PAS-HIG-12-050}. 
 Finally, to reduce even further the number of
input parameters we will assume an approximate GUT relation among
the gaugino soft masses, $M_2=2 M_1\, =\,M_3/4$ and the $\mu$ parameter
will be set equal to $M_2$. Regarding the trilinear couplings, they will
all be set to zero except those of the stop and sbottom sectors, being
relevant for $\Mh$, and that will be simplified to $A_t=A_b$. In
summary, our scenarios in this pMSSM framework are set in terms of four input
parameters:  $m_{\rm SUSY-EW}$, $m_{\rm SUSY-QCD}$, $M_2$ and
$\tb$.
We refer to our scenarios
here as ``pMSSM-4'', 
indicating the number of free parameters. These kind of scenarios have
the advantage of reducing considerably the number of input parameters
respect to the MSSM and, consequently, making easier the analysis of
their phenomenological implications.

For the forthcoming numerical analysis we consider the following
specific pMSSM-4 mass patterns:   
\begin{itemize}
\item[{\bf (a)}]
\BEA
 m_{\tilde L}&=&m_{\tilde E}=m_{\rm SUSY-EW}\nonumber \\ 
 M_2&=& m_{\rm SUSY-EW} \nonumber \\ 
 m_{\tilde Q}&=&m_{\tilde U}=m_{\tilde D}=m_{\rm SUSY-QCD}\nonumber \\ 
 A_t&=&1.3 \, m_{\rm SUSY-QCD} \nonumber \\
 m_{\rm SUSY-QCD}&=& 2 \, m_{\rm SUSY-EW}
\label{Sa}
\EEA
\item[{\bf (b)}] 
\BEA
 m_{\tilde L}&=&m_{\tilde E}=m_{\rm SUSY-EW}\nonumber \\ 
 M_2&=& m_{\rm SUSY-EW}/5 \nonumber \\ 
 m_{\tilde Q}&=&m_{\tilde U}=m_{\tilde D}=m_{\rm SUSY-QCD}\nonumber \\ 
 A_t&=&m_{\rm SUSY-QCD} \nonumber \\
 m_{\rm SUSY-QCD}&=& 2 \, m_{\rm SUSY-EW}
\label{Sb}
\EEA
\item[{\bf (c)}] 
\BEA
 m_{\tilde L}&=&m_{\tilde E}=m_{\rm SUSY-EW}\nonumber \\ 
 M_2&=& 300 \,\,{\rm GeV} \nonumber \\ 
 m_{\tilde Q}&=&m_{\tilde U}=m_{\tilde D}=m_{\rm SUSY-QCD}\nonumber \\ 
 A_t&=&m_{\rm SUSY-QCD} \nonumber \\
 m_{\rm SUSY-QCD}&= & m_{\rm SUSY-EW}
\label{Sc}
\EEA 
\item[{\bf (d)}] 
\BEA
  m_{\tilde L}&=&m_{\tilde E}=m_{\rm SUSY-EW}\nonumber \\ 
 M_2&=& m_{\rm SUSY-EW}/3 \nonumber \\ 
 m_{\tilde Q}&=&m_{\tilde U}=m_{\tilde D}=m_{\rm SUSY-QCD}\nonumber \\ 
 A_t&=&m_{\rm SUSY-QCD} \nonumber \\
 m_{\rm SUSY-QCD}&=& m_{\rm SUSY-EW}
\label{Sd}
\EEA 
\end{itemize}
Where we have simplified the notation for the soft sfermion masses, by
using $m_{\tilde L}$ for 
$m_{\tilde L}=m_{\tilde L_{1}}=m_{\tilde L_{2}}=m_{\tilde L_{3}}$, etc.
In the forthcoming numerical analysis of the maximum allowed values of
the deltas within these scenarios that will be done in Chapter \ref{phenoflavourslep}, the most relevant parameters $m_{\rm
  SUSY-EW} \equiv m_{\rm SUSY}$ and $\tb$ will be varied within the
intervals: 
\BEA
500 \,\,{\rm GeV} \leq &m_{\rm SUSY}&\leq 2500 \,\,{\rm GeV} \nonumber \\
5 \leq &\tb& \leq 60
\EEA 
Due to the particular mass patterns chosen above, scenario (a) will deal
with approximately equally heavy sleptons and charginos/neutralinos and
with doubly heavy squarks; same for scenario (b) but with 1/5 lighter
charginos/neutralinos; scenario (c) with equally heavy sleptons and
squarks and charginos/neutralinos close to 300 GeV and scenario (d) with  
 1/3 lighter charginos/neutralinos.
The values of $\At$ have been selected to ensure that 
$\Mh \sim 125 - 126 \gev$ over large parts of the ($\msusy$, $\tb$) plane.


\subsubsection{pMSSM-4 scenarios with very heavy SUSY}
\label{pmssmvheavysusy}

The results in this section have been published in \cite{Arana-Catania:2013xma}. 

A specific variation of the previous pMSSM-4 scenarios will be defined for the Chapter \ref{lfvhiggsdecaysslepton} to study the LFV Higgs decays. In this part of the work, only the slepton sector will be relevant, so only one common SUSY mass scale will be defined, called $m_\text{SUSY}$. This $m_\text{SUSY}$ will be considered very heavy, in the interval (0.5 TeV, 10 TeV). In particular for these scenarios we choose the following setting for the relevant mass parameters:
\begin{eqnarray}
m_{\tilde L} &=& m_{\tilde E} = m_\text{SUSY} \,, \\ \label{msleptons}
\mu &=& M_2 = a \,m_\text{SUSY} \,, \label{M2mu}
\end{eqnarray}
where $a$ is a constant coefficient that we will fix in the next sections to different values, namely, $a = \frac{1}{5}$, $\frac{1}{3}$, 1.
We also set an approximate GUT inspired relation for the gaugino masses:
\begin{equation}
M_2 = 2 M_1 = M_3/4 \,.
\end{equation}
And the same notation is used for the common soft masses as in the previous section. The trilinear couplings in the slepton sector have been fixed here to the generic SUSY mass scale, $A_\mu=A_e=A_\tau = m_\text{SUSY}$. We have checked that other choices with non-vanishing values for any of the couplings of the first and second generation do not alter the conclusions of the study.

Regarding the non-diagonal trilinear couplings in the slepton sector we have also assumed a rather simple but realistic setting by relating them with the single soft SUSY-breaking mass scale, $m_\text{SUSY}$. Specifically, we assume the following linear relation:
\begin{eqnarray}
 {\cal A}^l_{23} &=& {\tilde \delta}^{LR}_{23} m_\text{SUSY} \,, \,\,\,\,\,  
 {\cal A}^l_{32} = {\tilde \delta}^{LR}_{32} m_\text{SUSY} \,,
 \label{trilinear23}
\end{eqnarray} 
where the new  dimensionless parameters ${\tilde \delta}^{LR}_{23}$ and ${\tilde \delta}^{LR}_{32}$ are trivially related to the previously introduced ones ${\delta}^{LR}_{23}$ and ${\delta}^{LR}_{32}$ of Eq.(\ref{deltas23defs}) by:
\begin{equation}
{\delta}^{LR}_{23}= \left( \frac{v_1}{m_\text{SUSY}} \right) {\tilde \delta}^{LR}_{23}\,, \,\,\,\,\,
{\delta}^{LR}_{32}= \left( \frac{v_1}{m_\text{SUSY}} \right) {\tilde \delta}^{LR}_{32}.
\label{LRversusLRtilde}
\end{equation}
It is clear from Eq.(\ref{LRversusLRtilde}) that ${\delta}^{LR}_{23}$ and ${\delta}^{LR}_{32}$  scale with $m_\text{SUSY}$ as $\sim \frac{1}{m_\text{SUSY}}$. Therefore, in the forthcoming analysis of the LFV observables, whenever  the decoupling behaviour of these observables with large $m_\text{SUSY}$ be explored we will use instead the more suited parameters ${\tilde \delta}^{LR}_{23}$ and ${\tilde \delta}^{LR}_{32}$, which can be kept fixed to a constant value while taking large $m_\text{SUSY}$ values. 
 
Concerning the size of the flavour violating trilinear couplings that are of relevance here, ${\cal A}^l_{23}$ and ${\cal A}^l_{32}$, there are well-known theoretical upper bounds that arise from  vacuum stability. If any of these trilinear couplings is too large, the MSSM scalar potential develops a charge and/or colour breaking (CCB) minimum deeper than the Standard-Model-like (SML) local minimum or an unbounded from below (UFB) direction in the field space \cite{Casas:1996de}. Then the requirement of the absence of 
these dangerous CCB minima or UFB directions implies specific upper bounds on the size of the non-diagonal trilinear couplings, and consequently on the size of the flavour changing deltas. For the case of interest here the upper bound from stability can be written simply as \cite{Casas:1996de}:
\begin{equation}
|{\cal A}^l_{23}| \leq y_{\tau} \sqrt{m_{\tilde L_{2}}^2+m_{\tilde E_{3}}^2+m_1^2},
\end{equation}  
and similarly for ${\cal A}^l_{32}$. 
Here, 
\begin{equation} 
y_{\tau}=\frac{gm_\tau}{\sqrt{2}M_W \cos \beta}
\end{equation} 
is the Yukawa coupling of the tau lepton, and the squared soft mass $m_1^2$ can be written as:
\begin{equation} 
m_1^2= (m^2_A+M_W^2+M_Z^2 \sin^2\theta_W) \sin^2\beta- \frac{1}{2}M_Z^2.
\end{equation}  
In our simplified scenarios here for the slepton, gaugino and Higgs sectors, with just three MSSM input parameters,  $m_\text{SUSY}$, $\tan\beta$, and $m_A$, and by considering Eqs.(\ref{trilinear23}) and (\ref{LRversusLRtilde}), the previous bound implies in turn the following bound on ${\delta}^{LR}_{23}$, and correspondingly on 
${\tilde \delta}^{LR}_{23}$ (and similar bounds for $2 \leftrightarrow 3$):
\begin{equation}  
| {\delta}^{LR}_{23}|\leq \frac{m_\tau}{m_\text{SUSY}} \sqrt{2+\frac{m_1^2}{m^2_\text{SUSY}}}\,\,,\,\,\,
|{\tilde \delta}^{LR}_{23}| \leq y_{\tau}\sqrt{2+\frac{m_1^2}{m^2_\text{SUSY}}}.
\label{delta23bounds}
\end{equation} 
For example, if we take $m_\text{SUSY}=m_A=1$ TeV we get upper bounds for $|{\tilde \delta}^{LR}_{23}|$ of ${\cal O}(0.1)$ in the low $\tan\beta$ region close to 5, and of ${\cal O}(1)$ in the large $\tan\beta$ region close to 50. These correspond to an upper bound on  $|{\delta}^{LR}_{23}|$ of $\sim 0.0035$ that is nearly independent on $\tan\beta$  and it gets weaker for larger $m_\text{SUSY}$ values due to the scaling factor 
($\frac{m_\tau}{m_\text{SUSY}}$) in Eq.(\ref{delta23bounds}).

However, the reliability of these bounds have been questioned in the literature because of the fact that the existence of deeper minima than the SML local minimum does not necessarily imply a problem whenever the lifetime of this false SML vacuum is sufficiently long. In this later case, other theoretical upper bounds based on metastability then apply. Indeed, by demanding that the lifetime of the whole observable universe staying at the SML vacuum be longer than the age of the universe the constraints on the flavour changing deltas get much more relaxed~\cite{Park:2010wf}. According to  \cite{Park:2010wf} the upper bounds on ${\delta}^{LR}_{23}$ from metastability, in contrast to the limits from stability, turn out to be independent on the Yukawa coupling, they do not decouple for asymptotically large ${m_\text{SUSY}}$ and they are dependent on $\tan\beta$. For instance, for $\tan\beta= 3$ and ${m_\text{SUSY}}=5$ TeV the metastability limit on ${\delta}^{LR}_{23}$ gets weaker than its stability bound by a factor of 40, whereas for $\tan\beta= 30$ and ${m_\text{SUSY}}=5$ TeV it is weaker by a smaller factor of 4, leading to approximate upper bounds of $|{\delta}^{LR}_{23}|\leq 0.02$ and $|{\delta}^{LR}_{23}|\leq 0.002$ respectively. This translates into an upper bound of  
about $|{\tilde \delta}^{LR}_{23}| \leq 2-3$ for $3 \leq \tan\beta \leq 30$ and 
${m_\text{SUSY} }\leq 10$ TeV.
The numerical estimates of these metastability bounds are done considering each delta separately (i.e setting the other deltas to zero value) and for a specific assumption of the relevant Euclidean action providing the decay probability via tunnelling of the metastable vacuum into the global minimum. Changing the input value for the Euclidean action can increase notably the maximum allowed value up to almost doubling it, leading to roughly $|{\tilde \delta}^{LR}_{23}| \leq 4-6$. The effects on these bounds from switching on more than one delta at the same time have not been considered yet in the literature but they could relevantly modify these bounds. For the present work, and given the uncertainty in all these estimates of the upper bounds from stability and metastability arguments, we will consider a rather generous interval when performing the numerical estimates of branching ratios and event rates. Concretely we will choose several examples for ${\tilde \delta}^{LR}_{23}$ of very different size that will be taken within the wide interval $|{\tilde \delta}^{LR}_{23}| \leq 10$. This corresponds to $|{\delta}^{LR}_{23}|\leq 0.009$ for the particular values of $m_\text{SUSY} =$ 5 TeV  and $\tan\beta =$ 40.

On the other hand, the values of the soft masses of the squark sector are irrelevant for LFV processes, except in the present case of LFV MSSM Higgs bosons decays that will be studied in Chapter \ref{lfvhiggsdecaysslepton} where these parameters enter in the prediction of the radiatively corrected Higgs boson masses. Since we want to identify the discovered boson with the lightest MSSM Higgs boson, we will set these parameters to values which give a prediction of $m_h$ compatible with the LHC data. Specifically, in Chapter \ref{lfvhiggsdecaysslepton}, we fix them to the particular values $m_{\tilde Q} =$ $m_{\tilde U} =$ $m_{\tilde D} =$ $A_t =$ $A_b =$ 5 TeV, which we have checked provide a value for $m_h$ that lies within the LHC-favoured range [121 GeV, 127 GeV] for all the MSSM parameter space considered here. 

In summary, the input parameters of our pMSSM-4 scenarios with very heavy SUSY are: the mass of the pseudoscalar Higgs boson, $m_A$, the ratio of the two Higgs vacuum expectation values, $\tan\beta$, the generic SUSY mass scale, $m_\text{SUSY}$, the value of $a$, and the four relevant delta parameters for the particular studied LFV Higgs decays into $\tau \mu$, $\delta_{23}^{LL}$, $\delta_{23}^{RR}$, $\delta_{23}^{LR}$ and $\delta_{23}^{RL}$ (or ${\tilde \delta}^{LR}_{23}$ and ${\tilde \delta}^{RL}_{23}$ alternatively to the two latter), which we vary within the following intervals:
\begin{itemize}
\item 200 GeV $\leq m_A \leq$ 1000 GeV,
\item 1 $\leq \tan\beta \leq$ 60,
\item 0.5 TeV $\leq m_\text{SUSY} \leq$ 10 TeV,
\item $-1 \leq$ $\delta_{23}^{LL}$, $\delta_{23}^{RR}$ $\leq 1$,
\item $-10 \leq$ ${\tilde \delta}_{23}^{LR}$, ${\tilde \delta}_{23}^{RL}$ $\leq 10$,\\
(or, equivalently, $|{\delta}^{LR}_{23}|, |{\delta}^{RL}_{23}|\leq 0.009$ for $m_\text{SUSY} =$ 5 TeV  and $\tan\beta =$ 40). 

\end{itemize}


\subsubsection{Specific pMSSM points}
\label{frameworkc}

In this framework, we have selected six specific points in the MSSM
parameter space, S1,...,S6, as examples of points that are allowed by
present data, including recent LHC searches and the measurements of the
muon anomalous magnetic moment. In \refta{tab:spectra} the
values of the various MSSM parameters as well as the values of the
predicted MSSM mass spectra are summarized. They were evaluated with
the program \fh~\cite{feynhiggs,mhiggslong,mhiggsAEC,mhcMSSMlong}. For simplicity, and to reduce the number of
independent MSSM input parameters in these specific pMSSM points we have assumed equal soft masses for
the sleptons of the first and second generations (similarly for the
squarks),  equal soft masses for the left and right slepton sectors
(similarly for the squarks, where $\tilde Q$ denotes the the
``left-handed'' squark sector, whereas $\tilde U$ and $\tilde D$ denote
the up- and down-type parts of the ``right-handed'' squark sector)
and also equal trilinear couplings for
the stop, $A_t$,  and sbottom squarks, $A_b$. In the slepton sector we
just consider the stau trilinear coupling, $A_\tau$. The other trilinear
sfermion couplings are set to zero. Regarding the soft
SUSY-breaking parameters for the gaugino
masses, $M_i$ ($i=1,2,3$),  we assume again an approximate GUT inspired relation. The
pseudoscalar Higgs mass $\MA$, and the $\mu$ parameter are also taken as
independent input parameters. In summary, the six points S1,...,S6 are
defined in terms of the following subset of ten input MSSM parameters: 

\BEA
m_{\tilde L_1} &=& m_{\tilde L_2} \; ; \; m_{\tilde L_3} \; 
(\mbox{with~} m_{\tilde L_{i}} = m_{\tilde E_{i}}\,\,,\,\,i=1,2,3) \non \\
m_{\tilde Q_1} &=& m_{\tilde Q_2} \; ; \; m_{\tilde Q_3} \; 
(\mbox{with~} m_{\tilde Q_i} = m_{\tilde U_i} = m_{\tilde D_i}\,\,,\,\,i=1,2,3) 
                                                               \non \\
A_t&=&A_b\,\,;\,\,A_\tau \nonumber \\
M_2&=&2 M_1\, =\,M_3/4 \,\,;\,\,\mu \nonumber \\
\MA&\,\,;\,\, &\tb
\EEA

\begin{table}[h!]
\centering
\begin{tabular}{|c|c|c|c|c|c|c|}
\hline 
\multicolumn{7}{|c|}{Scenarios for Phenomenological MSSM} \\
\hline
 & S1 & S2 & S3 & S4 & S5 & S6 \\\hline
$m_{\tilde L_{1,2}}$& 500 & 750 & 1000 & 800 & 500 &  1500 \\
$m_{\tilde L_{3}}$ & 500 & 750 & 1000 & 500 & 500 &  1500 \\
$M_2$ & 500 & 500 & 500 & 500 & 750 &  300 \\
$A_\tau$ & 500 & 750 & 1000 & 500 & 0 & 1500  \\
$\mu$ & 400 & 400 & 400 & 400 & 800 &  300 \\
$\tb$ & 20 & 30 & 50 & 40 & 10 & 40  \\
$\MA$ & 500 & 1000 & 1000 & 1000 & 1000 & 1500  \\
$m_{\tilde Q_{1,2}}$ & 2000 & 2000 & 2000 & 2000 & 2500 & 1500  \\
$m_{\tilde Q_{3}}$  & 2000 & 2000 & 2000 & 500 & 2500 & 1500  \\
$A_t$ & 2300 & 2300 & 2300 & 1000 & 2500 &  1500 \\\hline
$m_{\tilde l_{1}}-m_{\tilde l_{6}}$ & 489-515 & 738-765 & 984-1018 & 474-802  & 488-516 & 1494-1507  \\
$m_{\tilde \nu_{1}}-m_{\tilde \nu_{3}}$& 496 & 747 & 998 & 496-797 & 496 &  1499 \\
$m_{{\tilde \chi}_1^\pm}-m_{{\tilde \chi}_2^\pm}$  & 375-531 & 376-530 & 377-530 & 377-530  & 710-844 & 247-363  \\
$m_{{\tilde \chi}_1^0}-m_{{\tilde \chi}_4^0}$& 244-531 & 245-531 & 245-530 & 245-530  & 373-844 & 145-363  \\
$M_{h}$ & 126.6 & 127.0 & 127.3 & 123.1 & 123.8 & 125.1  \\
$M_{H}$  & 500 & 1000 & 999 & 1001 & 1000 & 1499  \\
$M_{A}$ & 500 & 1000 & 1000 & 1000 & 1000 & 1500  \\
$M_{H^\pm}$ & 507 & 1003 & 1003 & 1005 & 1003 & 1502  \\
 $m_{\tilde u_{1}}-m_{\tilde u_{6}}$& 1909-2100 & 1909-2100 & 1908-2100 & 336-2000 & 2423-2585 & 1423-1589  \\
$m_{\tilde d_{1}}-m_{\tilde d_{6}}$ & 1997-2004 & 1994-2007 & 1990-2011 & 474-2001 & 2498-2503 &  1492-1509 \\
$m_{\tilde g}$ &  2000 & 2000 & 2000 & 2000 & 3000 &  1200 \\
\hline
\end{tabular}
\caption{Selected points in the pMSSM parameter space (upper part)
and their corresponding spectra for $\deXYij=0$ (lower part). 
All mass parameters and trilinear couplings are given in GeV.} 
\label{tab:spectra}
\end{table}

The specific values of these ten MSSM parameters in \refta{tab:spectra},
to be used in the forthcoming analysis of flavour violation, are chosen to provide
different  
patterns in the various sparticle masses, but all leading to rather
heavy spectra, thus they are naturally in agreement with the
absence of SUSY signals at LHC. In particular  
all points lead to rather heavy squarks and gluinos above $1200\gev$ and
heavy sleptons above $500\gev$ (where the LHC limits would also
  permit substantially lighter scalar leptons). 
The values of $\MA$ within the interval
$(500,1500)\gev$, $\tb$ within the interval $(10,50)$ and a large
$A_t$ within $(1000,2500)\gev$ are fixed such that a light Higgs boson
$h^0$ within the LHC-favoured range $(123,127)\gev$ is obtained%
\footnote{
The uncertainty takes into account experimental uncertainties as well as
theoretical uncertainties, where the latter would permit an even larger
interval. However, restricting to the chosen 
$\pm 2 \gev$ gives a good impression of the allowed parameter space.
}%
. It
should also be noted that the large chosen values of $\MA \ge 500$
GeV place the Higgs sector of these specific points in the so called decoupling
regime\cite{Haber:1989xc},  
where the couplings of $h^0$ to gauge bosons and fermions are close to
the SM Higgs couplings, and the heavy $H^0$ couples like the
pseudoscalar $A^0$, and all heavy Higgs bosons are close in mass.
Increasing $\MA$  the heavy
Higgs bosons tend to decouple from low energy physics and the light
$h^0$ behaves like $H_{\rm SM}$. This type of MSSM Higgs sector seems
to be in good agreement with recent LHC data\cite{LHCHiggs,Chatrchyan:2013lba,CMS-PAS-HIG-12-050}. 
We have checked with the code {\it HiggsBounds}~\cite{higgsbounds}
  that the Higgs sector is in agreement with the LHC searches (where S3
  is right ``at the border'').
Particularly, the so far absence of gluinos at
LHC, forbids too low $M_3$ and, therefore, given the  assumed GUT
relation, forbids also a too low $M_2$. Consequently, the
values of $M_2$ and $\mu$ are fixed as to get gaugino masses compatible
with present LHC bounds. 

As in the previous framework, the scenarios selected here lead to 
predictions of $(g-2)_\mu$ and
$\Mh$ that are compatible with present data over a large part of
  the parameter space.

Our estimate of $(g-2)_\mu^{\rm SUSY}$ for the six S1,...,S6 points with
the code {\it SPHENO}~\cite{SPheno} is
(where \fh\ gives similar results), respectively,   
\BE
 (15.5 \, (\mbox{S1}), \, 13.8 \, (\mbox{S2}), \, 15.1 \, (\mbox{S3}),
  16.7 \, (\mbox{S4}), \, 6.1 \, (\mbox{S5}), \, 7.9 \, (\mbox{S6}))\, 
\times 10^{-10}
\EE 
which are clearly within the previous allowed interval.
The relatively low values of $(g-2)_\mu^{\rm SUSY}$ are due to the relatively heavy slepton
spectrum that was chosen. However, they are well within the preferred
interval. These specific points will be considered in Chapters \ref{sec:Bphysics}, \ref{higgsmasssquark} and \ref{phenoflavourslep}.


\subsection{Effects of flavour mixing in the sfermion spectrum}
\label{effectsmixingspect}

All the mass spectra predictions presented before are for the MFV framework where all the non-diagonal entries of the SUSY breaking terms are set to zero, namely they are valid for $\deXYij=0$. In the rest of the work we will study the effect of non-vanishing flavour mixing deltas on different observables, but before jumping into it, we will give a glimpse of the effect of flavour mixing on the sfermion masses within the proposed scenarios themselves. 

\begin{figure}[h!] 
\centering
\hspace*{-8mm} 
{\resizebox{17.3cm}{!} 
{\begin{tabular}{cc} 
\includegraphics[width=13.3cm,height=17.2cm,angle=270]{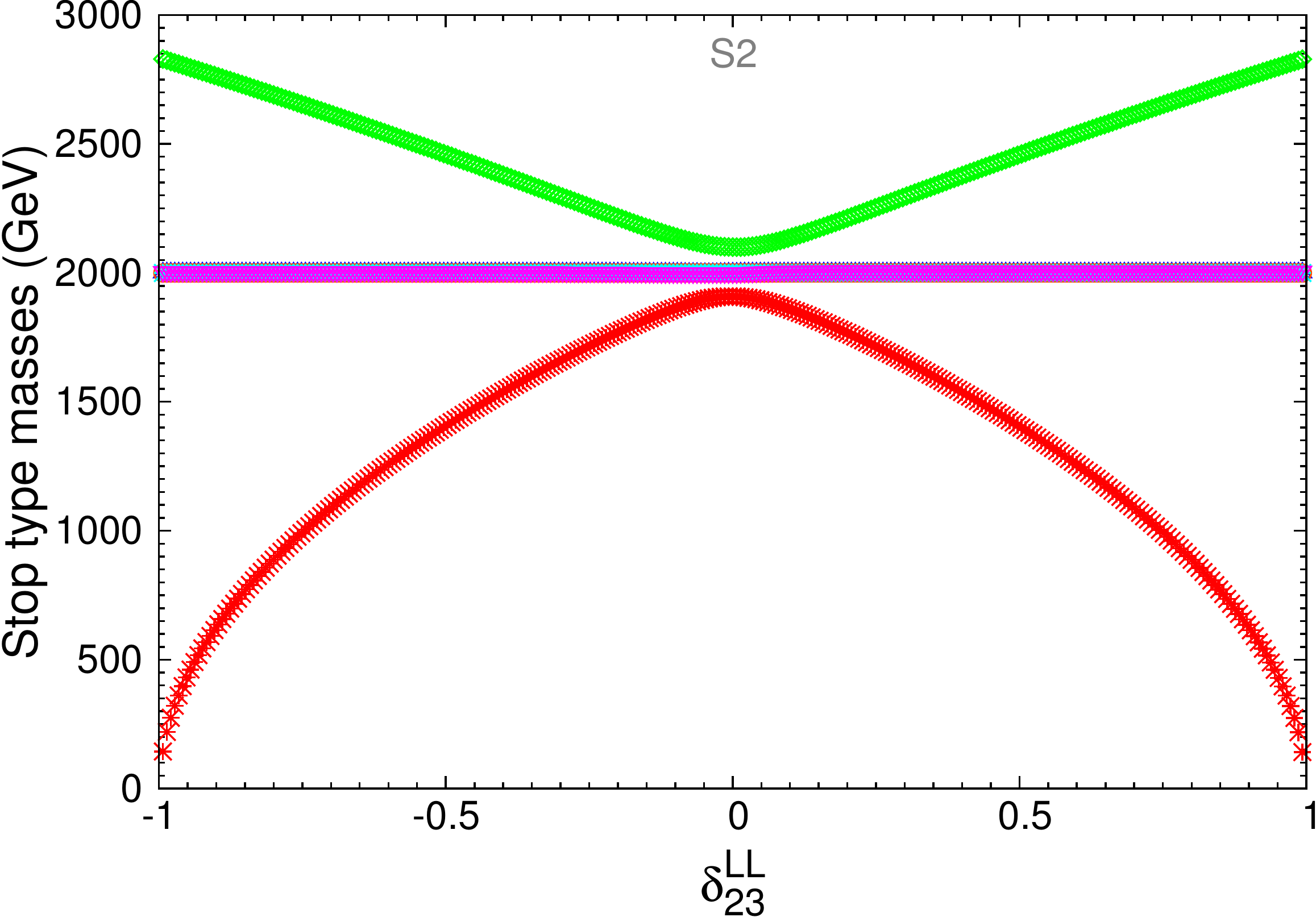}& 
\includegraphics[width=13.3cm,height=17.2cm,angle=270]{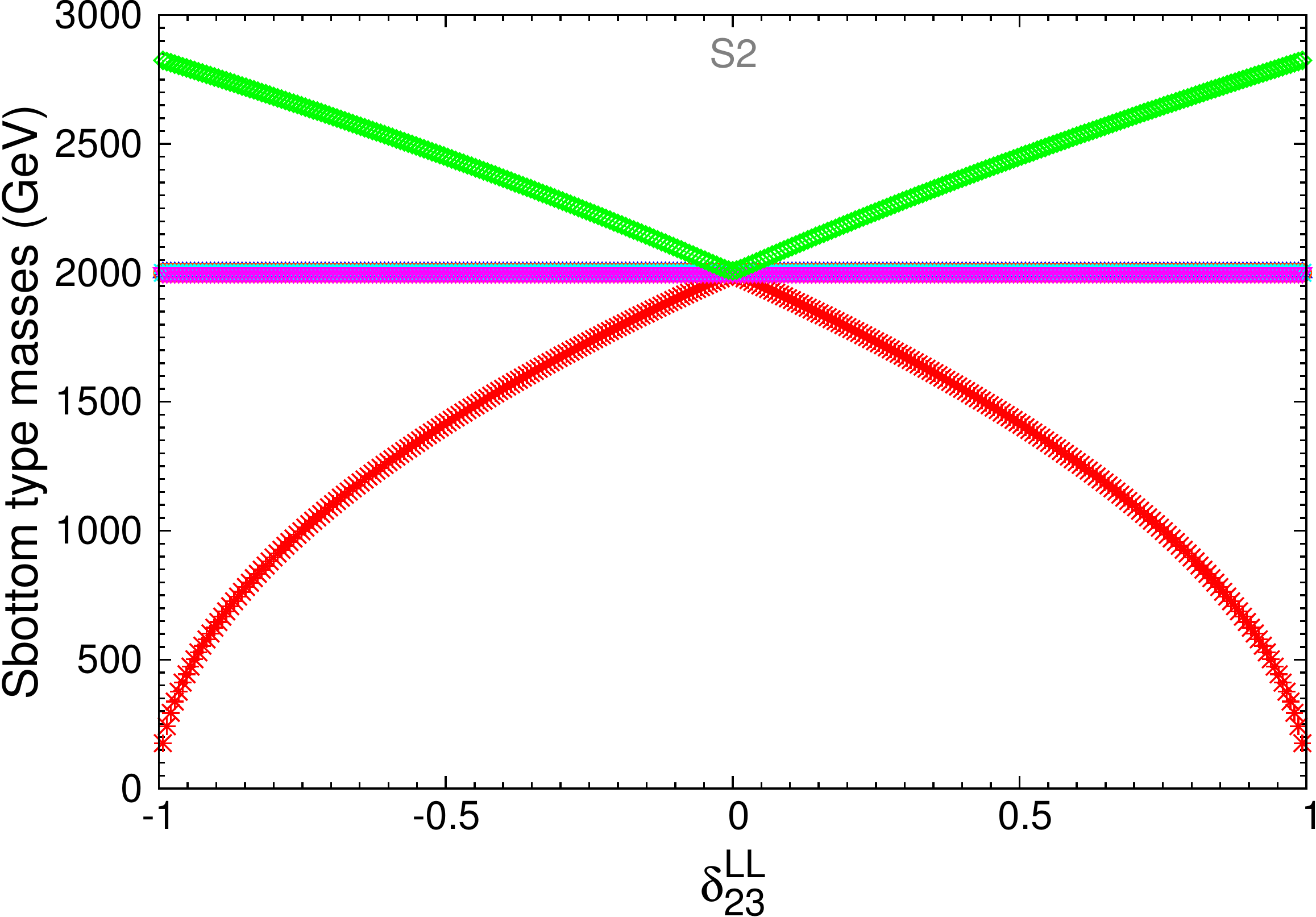}\\ 
\includegraphics[width=13.3cm,height=17.2cm,angle=270]{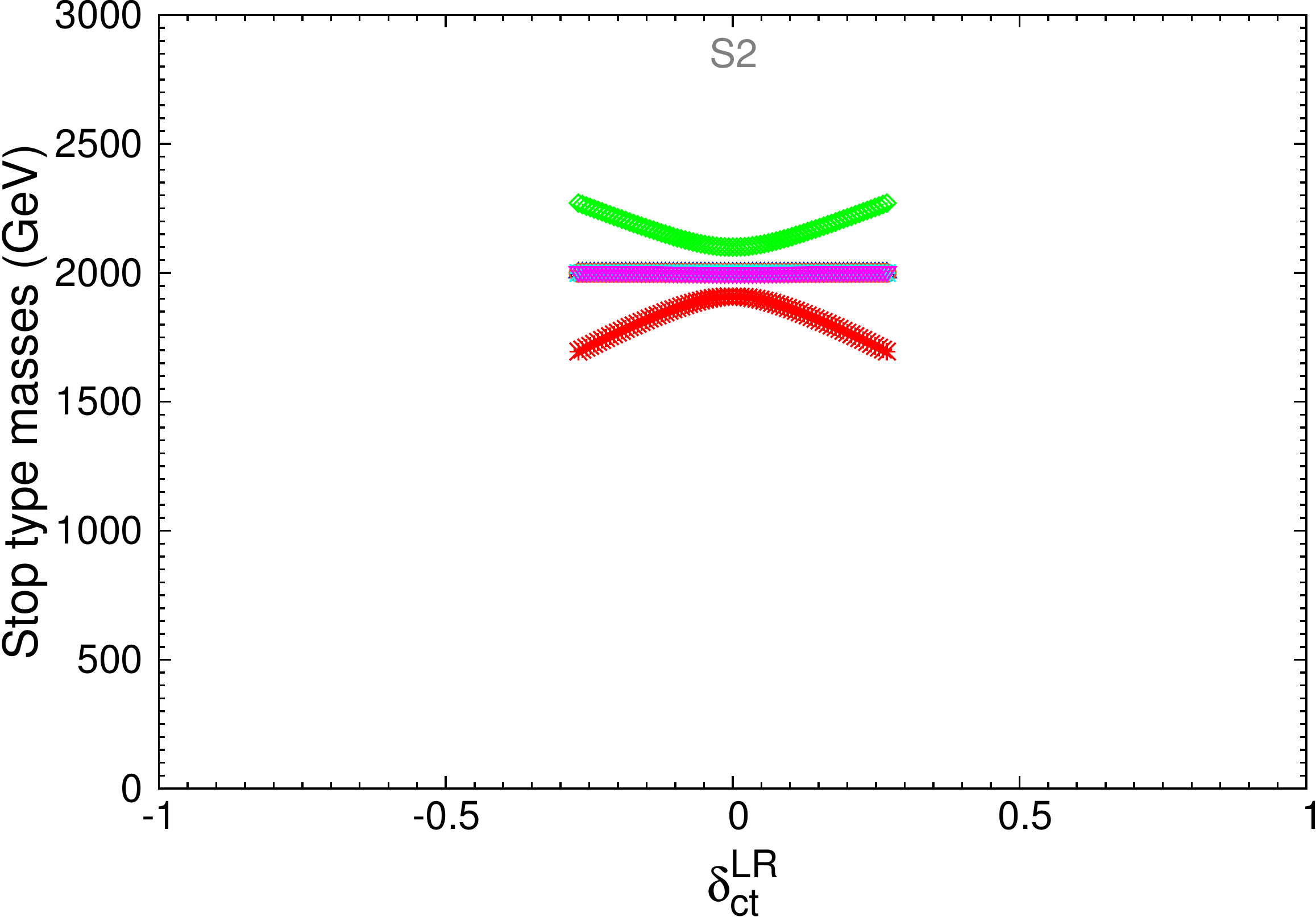}&
\includegraphics[width=13.3cm,height=17.2cm,angle=270]{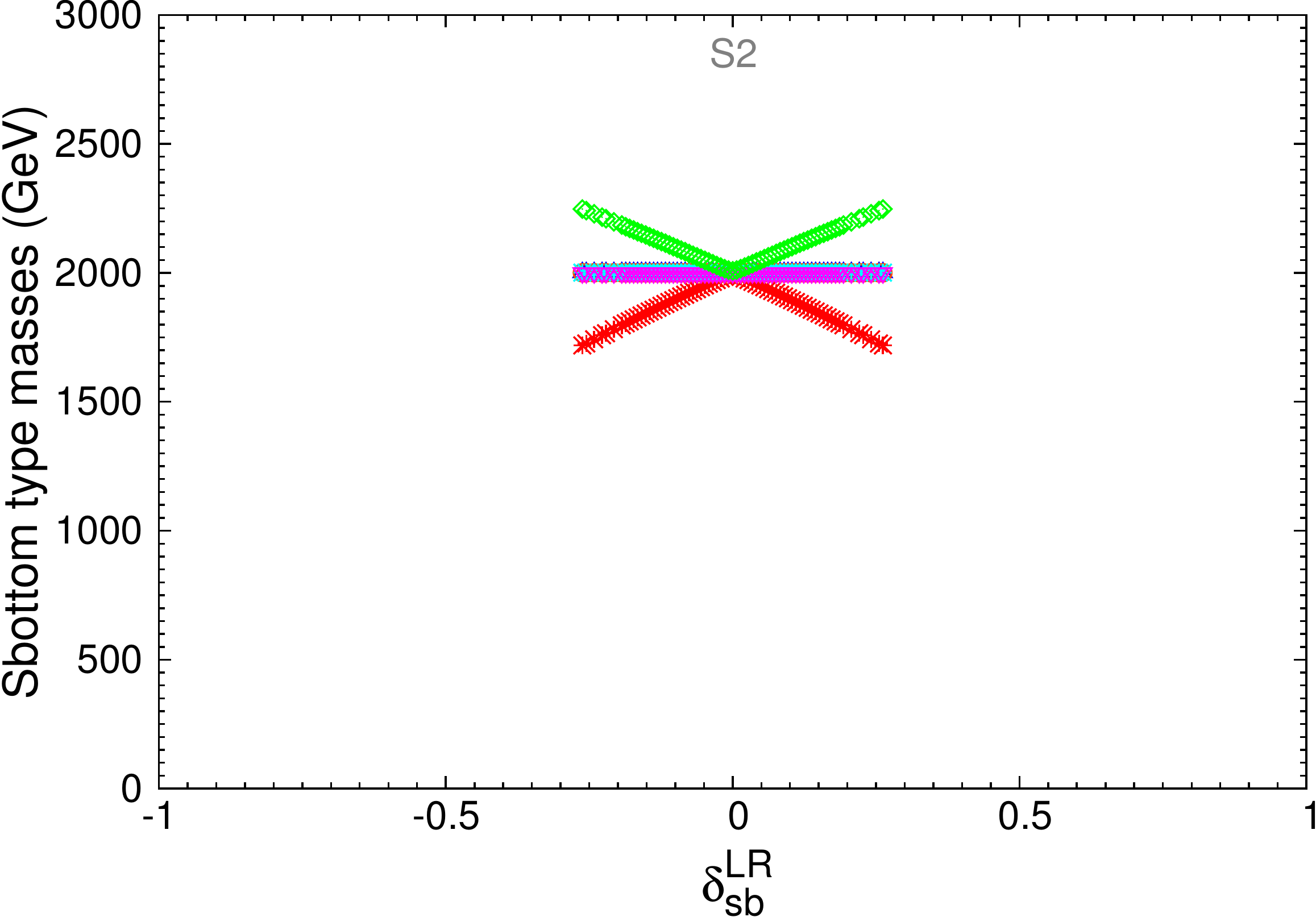}\\ 
\includegraphics[width=13.3cm,height=17.2cm,angle=270]{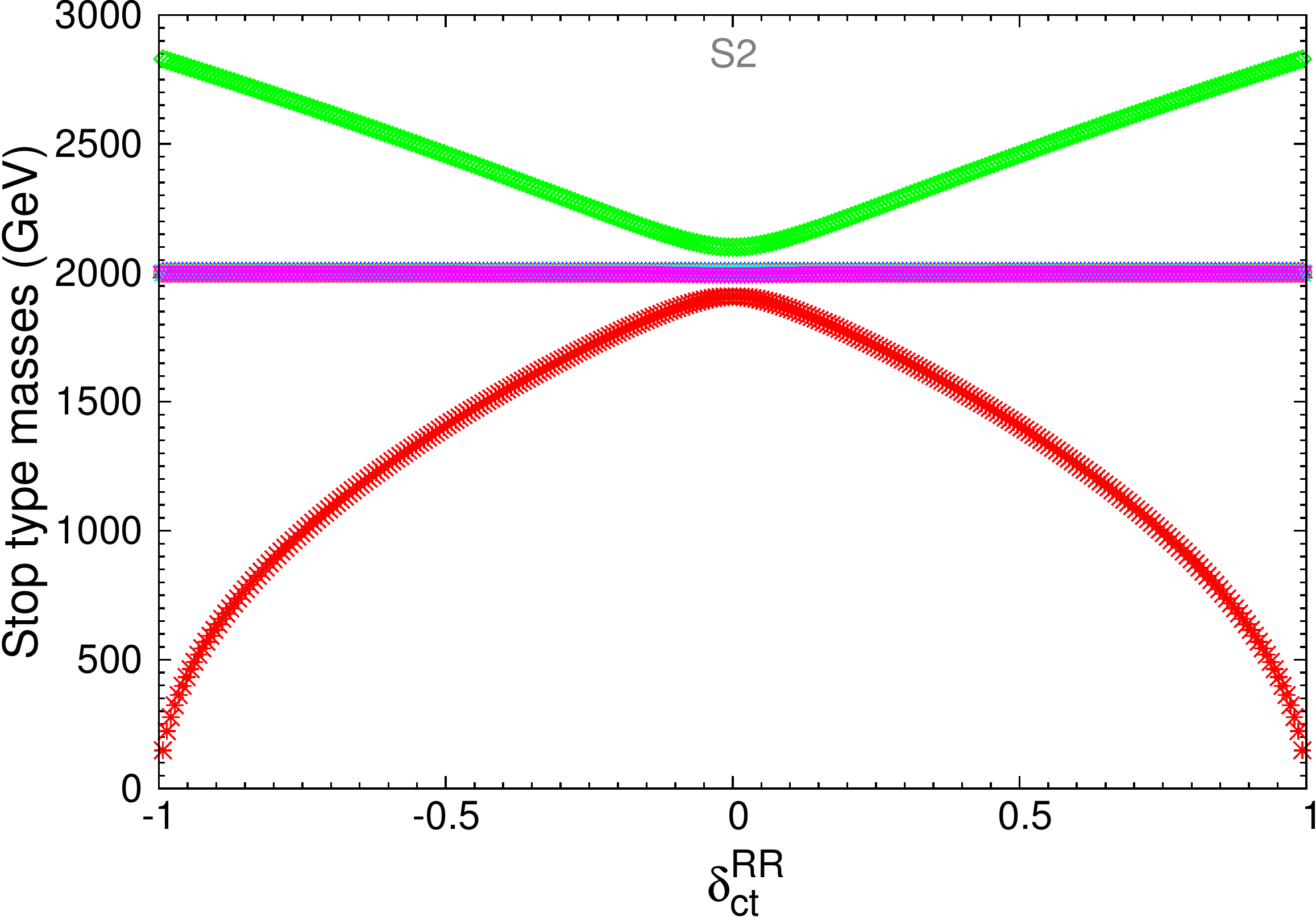}& 
\includegraphics[width=13.3cm,height=17.2cm,angle=270]{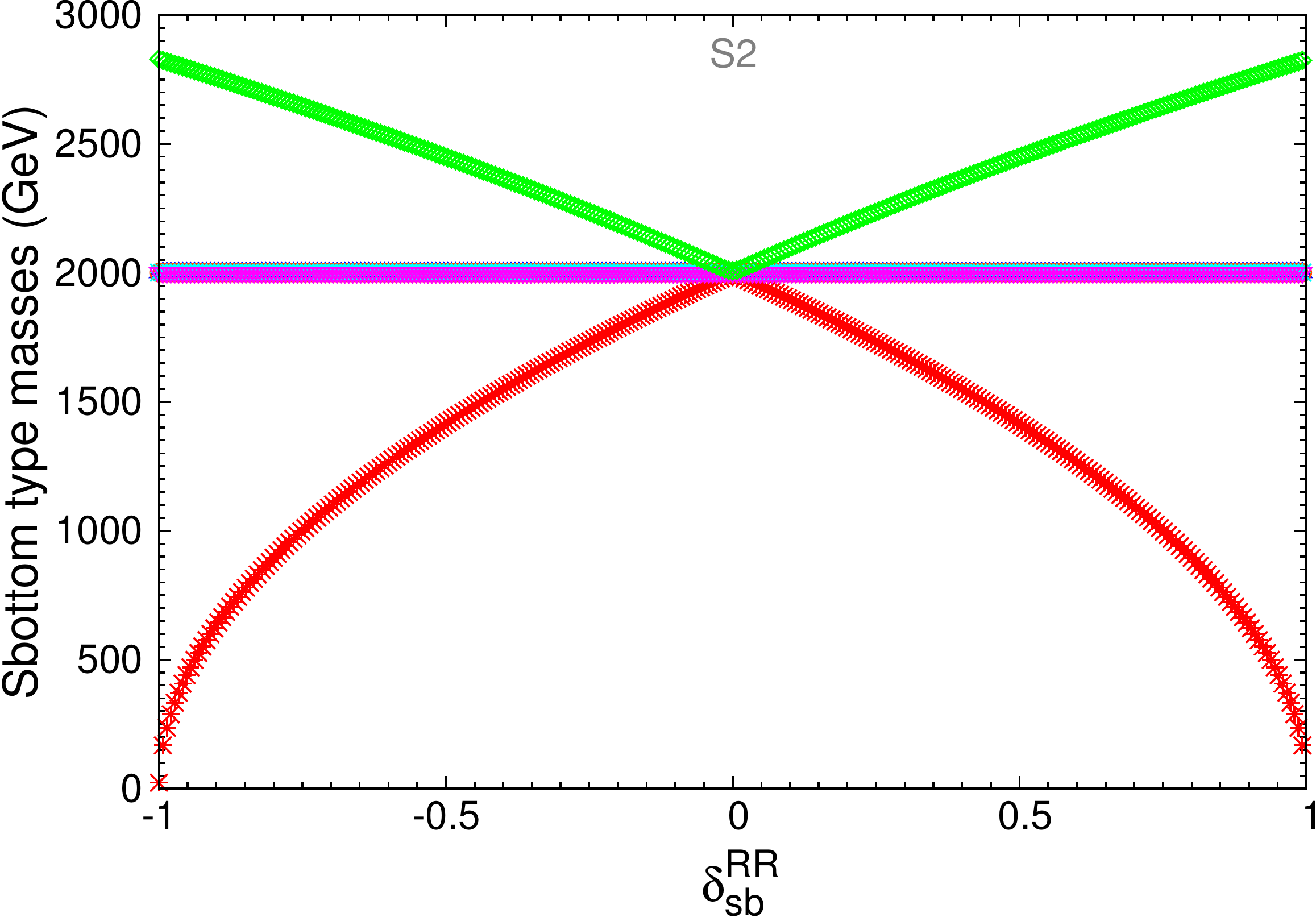}\\ 
\end{tabular}}}
\caption{Values of the six squark masses of the top type (left column) or bottom type (right column) for the relevant NMFV deltas for each type, for the S2 point defined in \ref{tab:spectra}.}   
\label{figsquarkspect}
\end{figure}

As we saw in Chapter \ref{paramnmfv}, the deltas enter in the sfermion mass and rotation matrices, and therefore they will have an effect on the sfermion spectrum and on the spectrum of all particles that get loop corrected by the sfermions. About these loop corrections, the effects in the spectrum of the Higgs sector will be studied with detail in Chapter \ref{higgsmasssquark}. Now we can show briefly the effects on the sfermion spectrum. For this illustration we have selected here in particular the squark sector (for sleptons a similar illustration could be done). We have chosen one of the previous points, the S2 point of table \ref{tab:spectra}, and set one by one the different $\deXYij$ parameters of the squark mass matrices (see Eqs. \ref{matdeltas1} to \ref{matdeltas6}) to values between -1 and 1. The value of the six physical squark masses as we vary the deltas are shown in Figure \ref{figsquarkspect}. In the left plots are shown the stop squark masses, and in the right plots the sbottom squark masses. Only the relevant deltas for each type of masses are shown, and the results for the deltas of the changed chirality (i.e. $\delta^{RL}_{ij}$ instead of $\delta^{LR}_{ij}$) are not shown since the results are equal.

It can be seen that most of the squarks remain almost degenerated (they are overimposed in the horizontal lines) except two, that get important positive and negative corrections from the deltas. In fact, the negative corrections can drive the value of the squark masses to values below the experimental bound around 500 GeV for the third generation (see Section \ref{exp-searches-mssm-particles}), and thus since the sparticles have not been observed, these can set bounds for the flavour mixing parameters. However, as we will see in the next chapters, the bounds coming from $B$ physics, LFV and the Higgs masses are much more restrictive than these ones.
Besides, in the middle plots of figure \ref{figsquarkspect} we can also see that not all the values of the deltas are represented. Large values of certain deltas can generate large corrections to other particles masses, like the Higgs boson masses, and turn them negative. Obviously these leading to negative mass squared points are not shown in the plots. In particular for this S2 point, the light Higgs boson mass is the particle that will get large corrections from the deltas $\delta^{LR/RL}_{ct}$ and $\delta^{LR/RL}_{sb}$ and may turn non-physical. The size of these corrections will be studied in full detail in Section \ref{sec:numanal}. 

\chapter{Squark flavour mixing effects on $B$ physics} 
\label{sec:Bphysics}

In this section we will analyze how far we can go in the flavour space in our Non-Minimal Flavour Violating SUSY scenarios. The limits on squark mixing in this chapter of it will be set by $B$-Physics observables. 
Since we are mainly interested
in the phenomenological consequences of the flavour mixing between the
third and second generations, as it was explained when it was introduced the general flavour parametrization we are using in Section \ref{sec:nmfv-squarks}
, we will focus%
\footnote{
We have checked that electroweak precision observables, where NMFV
  effects enter, for instance, via $\De\rho$~\cite{mhNMFVearly}, do not
  lead to relevant additional constraints on the allowed parameter space. Our results on this constraint are in agreement with \citere{Cao1}.}%
~on the following three B
meson observables: 1) Branching ratio of the radiative $B$ decay \bsg, 
2) Branching ratio of the $B_s$ muonic decay \bmm, 
and 3) $B_s-{\bar B_s}$ mass difference \dmbs.  Another
$B$ observable of interest in the present context is \bsll.
However, we have not included this in our study, because the
predicted rates in NMFV-SUSY scenarios for this observable are closely
correlated with those from \bsg\ due to the dipole
operators dominance in the photon-penguin diagrams mediating \bsll\ 
decays.  It implies that the restrictions on the flavour
mixing $\deXYij$ parameters from \bsll\ are also expected to be correlated with those
from the radiative decays.    

The evaluation of the observables is based on the BPHYSICS subroutine included in the SuFla code~\cite{Isidori:2006pk,sufla}. 
For the evaluation of \dmbs\
we have incorporated into this subroutine additional contributions from the
one-loop gluino boxes~\cite{Baek:2001kh} which are known to be very relevant
in the context of NMFV 
scenarios~\cite{Foster:2005wb,Gabbiani:1996hi,Becirevic:2001jj}. 

The scenarios and points that will be worked on here were presented in Section \ref{scenarios}. The summary of the relevant features for our analysis of these three B
meson observables is given in the following. 

The results in this chapter have been published in \cite{AranaCatania:2011ak} and \cite{Arana-Catania:updatedsquark}. 

\section{Implications for \boldmath{\bsg}}

The relevant effective
Hamiltonian for this decay is given in terms of the Wilson coefficients $C_i$ and operators $O_i$ by:

\noindent \begin{equation}
\mathcal{H}_{\rm eff}=-\frac{4G_{F}}{\sqrt{2}}\VCKM^{ts*}\VCKM^{tb}\sum_{i=1}^{8}(C_{i}O_{i}+C'_{i}O'_{i}). 
\end{equation}
Where the primed operators can be obtained from the unprimed ones by replacing 
$L \leftrightarrow R$.
The complete list of operators can be found, for instance, in \cite{Gambino:2001ew}.  
In the context of SUSY scenarios with the MSSM particle content and MFV, only two of these operators get relevant contributions, the so-called photonic dipole operator $O_{7}$ and gluonic dipole operator
$O_{8}$ given, respectively, by:

\noindent \begin{equation}
O_{7}=\frac{e}{16\pi^{2}}m_{b}\left(\bar{s}_{L}\sigma^{\mu\nu}b_{R}\right)F_{\mu\nu},\end{equation}

\noindent \begin{equation}
O_{8}=\frac{g_{3}}{16\pi^{2}}m_{b}\left(\bar{s}_{L}\sigma^{\mu\nu}T^{a}b_{R}\right)G_{\mu\nu}^{a}.\end{equation}
We have omitted the colour indices here for brevity.
Within NMFV also $O'_{7,8}$ have to be taken into account:

\noindent \begin{equation}
\label{opo7p}
O'_{7}=\frac{e}{16\pi^{2}}m_{b}\left(\bar{s}_{R}\sigma^{\mu\nu}b_{L}\right)F_{\mu\nu},\end{equation}

\noindent \begin{equation}
\label{opo8p}
O'_{8}=\frac{g_{3}}{16\pi^{2}}m_{b}\left(\bar{s}_{R}\sigma^{\mu\nu}T^{a}b_{L}\right)G_{\mu\nu}^{a}.\end{equation}

The Wilson coefficients at the SUSY scale are obtained as usual by the 
matching procedure of the proper matrix element being computed from the 
previous 
effective Hamiltonian to the corresponding matrix elements being computed 
from the specific SUSY model, the NMFV-MSSM in our case. 
These Wilson coefficients encode, therefore, the contributions to 
${\rm BR}(B \to X_s \ga$) from the loops of the SUSY and Higgs sectors of the 
MSSM with NMFV. The effects from squark flavour mixings that are parametrized by the 
$\deXYij$, are included in this observable via the squark physical 
masses and rotation matrices, given in Section \ref{sec:nmfv-squarks}, that appear in 
the computation of the matrix element at the one loop level and, therefore,
are also encoded in the Wilson coefficients. The explicit expressions
for these  coefficients in the MSSM, in terms of the particles and sparticles physical basis, can
be found, for instance, in
refs.~\cite{Bertolini:1990if,Cho:1996we,Degrassi:2000qf}. 
We have included in our analysis the most relevant loop contributions to
the Wilson coefficients, concretely: 1) loops with Higgs bosons, 2)
loops with charginos and 3) loops with gluinos. It should be noted that, at one loop order, the gluino
loops do not contribute in MFV scenarios, but they are very relevant
(dominant in many cases) in the present NMFV scenarios.   
 
Once the Wilson coefficients are known at the effective low energy SUSY scale, they
are evolved with the proper Renormalization Group Equations down
to the typical low-energy scale for this process. As a consequence of this running the
previous operators mix and the corresponding Wilson coefficients
get involved in the computation of the $B \to X_s \ga$
decay rate. The RGE-running is done in two steps: The first one is from the SUSY scale down to the electroweak scale, and the second one is from this electroweak scale down to the B-physics scale.  For the first step, we use the LO-RGEs for the relevant Wilson coefficients as in
\cite{Degrassi:2000qf} and fix six active quark flavours in this
running. For the second running we use the NLO-RGEs as in \cite{Hurth:2003dk} and fix, correspondingly, five active quark flavours. For the charged Higgs sector we use the NLO formulas for the
Wilson coefficients as in \cite{Ciuchini:1997xe}.  

 The resummation of scalar induced large $\tb$ effects is
 performed, as usual,  by the effective Lagrangian approach that
 parametrizes the one-loop generated effective couplings between the
  ${\cal H}_2$ Higgs doublet and the down type quarks in softly broken SUSY
 models~\cite{Hall:1993gn}. A necessary condition to take into account
 all $\tb$-enhanced terms in flavour changing amplitudes is the
 diagonalization of the down-type quark mass matrix in the presence of
 these effective couplings \cite{Carena:2000uj,Carena:1999py,Isidori:2001fv}. A summary of this effective Lagrangian formalism
 for the resummation of large $\tb$ effects in the three $B$
 observables of our interest, within the context of MFV scenarios, can be found in
 \cite{Buras:2002vd}. We follow here the treatment of
 \cite{Isidori:2002qe} where the resummation of large $\tb$
 effects via effective Lagrangians is generalised to the case where the
 effective ${\bar d}^i_R d^j_L {\cal H}^0_2$ coupling contains also non-minimal
 sources of flavour mixing. It should be noted that the most relevant scalar induced
 large $\tb$ effects for the present work are those generated by
 one-loop diagrams with gluino-sbottom and chargino-stop inside the
 loops. The large $\tb$ resummation effects and the relevance of other chirally enhanced corrections
to FCNC processes within the NMFV context have recently been
studied exhaustively also in \cite{Crivellin1,Crivellin2}
(previous studies can be found, for instance, in
\citeres{Okumura:2002wa,Okumura:2003hy,Foster:2004vp}). 

The total branching ratio for this decay is finally estimated by adding
the new contributions from the SUSY and Higgs sectors to the  
SM rate. More specifically, we use Eq.42 of \cite{Hurth:2003dk} for
the estimate of  \bsg\ in terms of the ratios of 
the Wilson coefficients $C_{7,8}$ and  $C'_{7,8}$ (including all the mentioned new contributions)
divided by the corresponding $C_{7,8}^{\rm SM}$ in the SM. 

For the numerical estimates of \bsg\ we use the FORTRAN
subroutine BPHYSICS (modified as to include the contributions from  $C'_{7,8}$ which were not included in its original version) included in the SuFla code, that
incorporates all the above mentioned ingredients\cite{Isidori:2006pk,sufla}. Subleading NLO 
MSSM corrections were evaluated in~\cite{Degrassi:2007kj,Degrassi:2006eh}. However, their effect on our evaluations would be minor.


\section{Implications for \boldmath{\bmm}}

The relevant effective Hamiltonian for this process is \cite{Chankowski:2000ng,Bobeth:2002ch}: 

\noindent \begin{equation}
\mathcal{H}_{\rm eff}=-\frac{G_{F}\alpha}{\sqrt{2} \pi}\VCKM^{ts*}\VCKM^{tb}\sum_{i} (C_{i}O_{i}+C'_{i} O'_{i}), 
\end{equation}
where the operators $O_i$ are given by:
\begin{align}
{O}_{10}&=\left(\bar{s}\ga^{\nu}P_Lb\right)\left(\bar{\mu}\ga_{\nu}\ga_5\mu\right),
& {O}_{10}^{\prime}&=\left(\bar{s}\ga^{\nu}P_Rb\right)\left(\bar{\mu}\ga_{\nu}\ga_5\mu\right),\nonumber\\
 {O}_{S}&=m_b\left(\bar{s}P_Rb\right)\left(\bar{\mu}\mu\right),
& {O}_{S}^{\prime}&=m_s\left(\bar{s} P_Lb \right)\left(\bar{\mu}\mu\right),\nonumber\\
 {O}_{P}&=m_b\left(\bar{s} P_Rb \right)\left(\bar{\mu}\ga_5\mu\right),
& {O}_{P}^{\prime}&=m_s\left(\bar{s} P_Lb \right)\left(\bar{\mu}\ga_5\mu\right).\label{bsm:Ops}
\end{align}
We have again omitted the colour indices here for brevity.

In this case, the RG running is straightforward since the anomalous dimensions of 
the above involved operators are zero, 
and the 
prediction for the decay rate is simply expressed by: 
\begin{align}
\bmm &= \frac{G_F^2\alpha^2 m_{B_s}^2 f_{B_s}^2\tau_{B_s}}{64 \pi^3}\lvert \VCKM^{ts*}\VCKM^{tb}\rvert^2\sqrt{1-4\hat{m}_{\mu}^2}
\nonumber\\
&\times\left[\left(1-4\hat{m}_{\mu}^2\right)\lvert F_S\rvert^2+\lvert F_P+2\hat{m}_{\mu}^2 F_{10}\rvert^2\right],
\label{bsm:br}
\end{align}
where $\hat{m}_{\mu}=m_{\mu}/m_{B_s}$ and the 
$F_i$ are given by
\begin{align}
F_{S,P}&=m_{B_s}\left[\frac{C_{S,P}m_b-C_{S,P}^{\prime}m_s}{m_b+m_s}\right],
&F_{10}=C_{10}-C_{10}^{\prime}.
\nonumber
\end{align}

Within the SM the most relevant operator is $O_{10}$ as the Higgs mediated 
contributions to $O_{S,P}$ can be safely neglected. It should be noted that the 
contribution from $O_{10}$ to the decay rate is helicity suppressed by a 
factor of  $\hat{m}_{\mu}^2$ since the $B_s$ meson has spin zero. 
In contrast, in SUSY scenarios the scalar and pseudo-scalar operators, 
$O_{S,P}$, can be very important. This happens particularly at large $\tb
\gsim 30$ where the contributions to $C_S$ and $C_P$ from neutral Higgs
penguin diagrams can become large and dominate the branching ratio,
because in this case the branching ratio grows with $\tb$ as
${\tan}^6\beta$.  The studies in the literature of these MSSM
Higgs-penguin contributions to ${\rm BR}(B_s\to \mu ^ + \mu^-)$ have focused
on both MFV \cite{Babu:1999hn,Bobeth:2001sq,Isidori:2001fv} and NMFV  
scenarios
\cite{Chankowski:2000ng,Isidori:2002qe,Foster:2004vp,Foster:2005wb}.  
In both cases the rates for ${\rm BR}(B_{s}\to\mu^{+}\mu^{-})$ at large $\tan
\beta$ can be enhanced by a few orders of magnitude compared with the
prediction in the SM, therefore providing an optimal window for SUSY
signals.

In the present context of SUSY-NMFV, with no preference for large $\tb$ 
values, there are in general three types of one-loop diagrams that contribute 
to the previous $C_i$ Wilson coefficients for this $B_s \to \mu^+ \mu^-$
decay: 1) Box diagrams, 2) $Z$-penguin diagrams and 3) neutral Higgs
boson $H$-penguin diagrams, where $H$ denotes the three neutral MSSM Higgs bosons. In our numerical estimates
we have included what are known to be the dominant contributions to
these three types of diagrams \cite{Chankowski:2000ng}: chargino
contributions to box and Z-penguin diagrams and chargino and gluino
contributions to $H$-penguin diagrams.   

Regarding the resummation of large $\tb$ effects we have followed
the same effective Lagrangian formalism as explained in the previous
case of $B \to X_s \ga$. In the case of contributions from
$H$-penguin diagrams to  
$B_s \to \mu^+ \mu^-$ this resummation is very relevant and it has been
generalised to NMFV-SUSY scenarios in \cite{Isidori:2002qe}. 

For the numerical estimates we use again the BPHYSICS subroutine included in the SuFla code~\cite{Isidori:2006pk,sufla} which
incorporates all the ingredients that we have pointed out above.   


\section{Implications for \boldmath{\dmbs}} 

The relevant effective Hamiltonian for $B_s-{\bar B_s}$ mixing and, hence, for 
the $B_s/{\bar B_s}$ mass difference \dmbs\ is:

\begin{equation}
\mathcal{H}_{\rm eff}=
\frac{G_F^2}{16\pi^2}M_W^2 
\left(\VCKM^{tb*}{}\VCKM^{ts}\right)^2
\sum_{i}C_i O_i.
\label{Ham}
\end{equation}
In the SM the most relevant operator is:
\begin{align}
O^{VLL}&=
(\bar{b}^{\alpha}\ga_{\mu}P_L s^{\alpha})(\bar{b}^{\beta}\ga^{\mu}P_L s^{\beta}).
\label{SMOps}
\end{align}
Where we have now written explicitly the colour indices.
 In scenarios beyond the SM, as the present NMFV MSSM, other operators are also
 relevant:
\begin{align}
O^{LR}_{1}&=(\bar{b}^{\alpha}\ga_{\mu}P_L s^{\alpha})(\bar{b}^{\beta}\ga^{\mu}P_R s^{\beta}),
&O^{LR}_{2}&=(\bar{b}^{\alpha}P_L s^{\alpha})(\bar{b}^{\beta}P_R s^{\beta}),\label{Ops1}\\
O^{SLL}_{1}&=(\bar{b}^{\alpha}P_L s^{\alpha})(\bar{b}^{\beta}P_L s^{\beta}),
&O^{SLL}_{2}&=(\bar{b}^{\alpha}\sigma_{\mu\nu}P_L s^{\alpha})(\bar{b}^{\beta}\sigma^{\mu\nu}P_L s^{\beta}),\label{Ops2}
\end{align}
and the corresponding operators $O^{VRR}$ and
$O^{SRR}_{i}$ that can be obtained by replacing $P_L \leftrightarrow P_R$
  in~\eqref{SMOps} and~\eqref{Ops2}.
The mass difference $\Delta M_{B_s}$ is then evaluated by taking the matrix
element
\begin{align}
\Delta M_{B_s}=2\lvert\langle\bar{B}_s\lvert\mathcal{H}_{\rm eff}\rvert B_s\rangle\rvert,
\label{delmb}
\end{align}
where $\langle\bar{B}_s\lvert\mathcal{H}_{\rm eff}\rvert B_s\rangle$ is given
by
\begin{align}
\langle\bar{B}_s\lvert\mathcal{H}_{\rm eff}\rvert B_s\rangle=&
\frac{G_F^2}{48\pi^2}M_W^2 m_{B_s} f^2_{B_s}
\left(\VCKM^{tb*} \VCKM^{ts}\right)^2
\sum_{i}P_i C_i\left(\mu_W\right).
\label{matel}
\end{align}
Here $m_{B_s}$ is the $B_s$ meson mass, and
$f_{B_s}$ is the $B_s$ decay constant.
The coefficients $P_i$ contain the effects due to RG running between
the electroweak scale $\mu_W$ and $m_b$ as well as the relevant hadronic matrix element.
 We use the coefficients $P_i$ from the lattice calculation \cite{Becirevic:2001xt}:   
\begin{align}
P^{VLL}_1=&0.73,
&P^{LR}_1=&-1.97,
&P^{LR}_2=&2.50,
&P^{SLL}_1=&-1.02,
&P^{SLL}_2=&-1.97.
\label{pcoef}
\end{align}
The coefficients
$P^{VRR}_1$, etc.,~may be obtained from those above by simply exchanging 
$L \leftrightarrow R$.

In the present context of SUSY-NMFV, again with no preference for large $\tb$ 
values, and besides the SM loop contributions, there are in general three types of one-loop diagrams that contribute 
to the previous $C_i$ Wilson coefficients for $B_s-{\bar B_s}$ mixing: 
1) Box diagrams, 2) $Z$-penguin diagrams and 3) double Higgs-penguin diagrams.
In our numerical estimates we have included what are known to be the dominant 
contributions to these three types of diagrams in scenarios with non-minimal flavour violation (for a review see, for instance, \cite{Foster:2005wb}): gluino contributions to box
diagrams, chargino contributions to box and Z-penguin diagrams, and 
chargino and gluino contributions to double $H$-penguin diagrams. As in the previous observables, 
the total prediction for  $\Delta M_{B_s}$ includes, of course, the SM contribution.
 
Regarding the resummation of large $\tb$ effects we have followed again the effective Lagrangian formalism, generalised to NMFV-SUSY scenarios \cite{Isidori:2002qe}, as in the previous cases of $B \to X_s \ga$ and $B_s \to \mu^+ \mu^-$. It should be noted that, in the case of $\Delta M_{B_s}$, the resummation of large $\tb$ effects is very relevant for the double $H$-penguin contributions, which grow very fast with $\tb$.

\begin{figure}[tp]
\begin{center}
\psfig{file=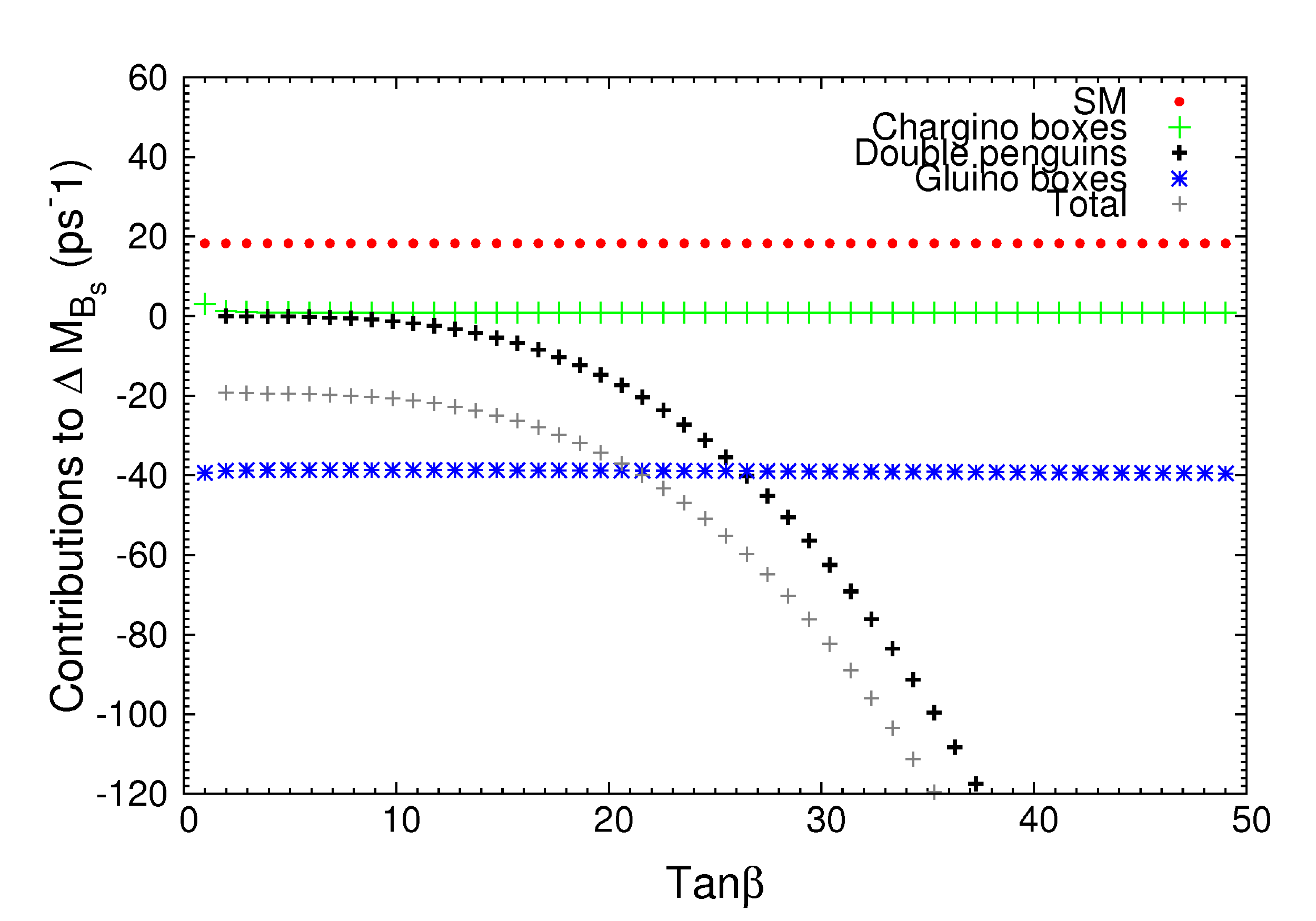,scale=0.45,angle=270,clip=}
\caption{Relevant contributions to $\Delta M_{B_s}$ in NMFV-SUSY scenarios as a function of $\tb$. They include: SM, Double Higgs penguins, gluino boxes and chargino boxes. The total prediction for $\Delta M_{B_s}$ should be understood here as $\Delta M_{B_s}=|{\rm Total}|$. The parameters are set to $\delta^{LL}_{23}=\delta^{RR}_{sb}=0.1$,$m_{\tilde q}=500 \gev$, $A_t= -m_{\tilde q}$, 
$m_{\tilde g}=\sqrt{2}m_{\tilde q}$, $\mu=m_{\tilde q}$, and $M_A= 300 \gev$. The other flavour changing deltas are set to zero.}
\label{fig:deltabs-allcontributions}
\end{center}
\end{figure} 
For the numerical estimates we have modified the BPHYSICS subroutine included in the SuFla code~\cite{Isidori:2006pk,sufla} which incorporates all the ingredients that we have pointed out above, except the contributions from gluino boxes. These contributions are known to be very important for $B_s-{\bar B_s}$ mixing in SUSY scenarios with non-minimal flavour violation \cite{Gabbiani:1996hi,Becirevic:2001jj,Foster:2005wb} and therefore they must be included into our analysis of $\Delta M_{B_s}$.  Concretely, we have incorporated them into the BPHYSICS subroutine by adding the full one-loop formulas for the gluino boxes of \cite{Baek:2001kh} to the other above quoted contributions that were already included in the code. In order to illustrate the relevance of these gluino contributions in our analysis of $\Delta M_{B_s}$, we show in Fig.\ref{fig:deltabs-allcontributions} each separate contribution as a function of $\tb$ in a particular example with $\delta^{LL}_{23}=\delta^{RR}_{sb}=0.1$, that we have chosen for comparison with \cite{Foster:2005wb}. The other flavour changing deltas are set to zero, and the other relevant MSSM parameters are set to $m_{\tilde q}=500 \gev$, $A_t= -m_{\tilde q}$, 
$m_{\tilde g}=\sqrt{2}m_{\tilde q}$, $\mu=m_{\tilde q}$, and $M_A= 300 \gev$, as in Fig.24 of \cite{Foster:2005wb}.   We clearly see in Fig.\ref{fig:deltabs-allcontributions} that it is just in the very large $\tb$ region where double Higgs- penguins dominate. For moderate and low $\tb$ values, $\tb \leq 20$ (which is a relevant region for our numerical analysis,
see below) the gluino boxes fully dominates the SUSY corrections to
$\Delta M_{B_s}$ and compete with the SM contributions. Our numerical
estimate in this plot is in complete agreement with the results in
\cite{Foster:2005wb} (see, in particular, Fig.24 of this reference) which
analyzed and compared in full detail these corrections.    
 

\section{Numerical results on constraints from \boldmath{$B$} observables}

In the following sections we present our numerical results for
the three $B$ observables in the NMFV-SUSY scenarios  
and a discussion on the allowed values for the flavour changing deltas,
$\deXYij$. 

The discussion will be structured in two different sections. The first one will be devoted to the CMSSM scenarios, defined in Section \ref{frameworka}, that contains the classical SPS benchmark points that have been studied exhaustively during years before the starting of the LHC and that we choose here as illustrative examples. These points will show us how the observables behave with respect to the different parameters through a general set of scenarios. However, when the first buzz of the collisions came out of the LHC all these scenarios grew old immediately, and new scenarios were required to accommodate the new experimental situation. In the second section a set of presently allowed scenarios with heavier SUSY particles defined in \ref{frameworkc} will be covered, taking into account the fact that SUSY has not appeared yet in the LHC.

In the next sections the predictions for \bsg, \bmm\ and \dmbs\ 
versus the various $\deXYij$ will be shown.
For this analysis, we have assumed that just one at a time
of these deltas is not vanishing. Results for two non-vanishing deltas
will be shown later.  The following seven flavour changing deltas are
considered: $\delta^{LL}_{23}$, $\delta^{LR}_{ct}$, $\delta^{LR}_{sb}$,
$\delta^{RL}_{ct}$, 
$\delta^{RL}_{sb}$, $\delta^{RR}_{ct}$ and $\delta^{RR}_{sb}$; and we
have explored delta values within the interval $-1<\deXYij<1$. In all
plots, the predictions for $\deXYij=0$ represent our estimate of the  
corresponding observable in the MFV case. This will allow us to compare
easily the results in the two kind of scenarios, NMFV and MFV.

\subsection{Numerical results for pre-LHC allowed scenarios and pre-LHC $B$ data}
\label{numresframeworka}

\begin{figure}[h!] 
\centering
\hspace*{-8mm} 
{\resizebox{17.3cm}{!} 
{\begin{tabular}{cc} 
\includegraphics[width=13.2cm,height=17.2cm,angle=270]{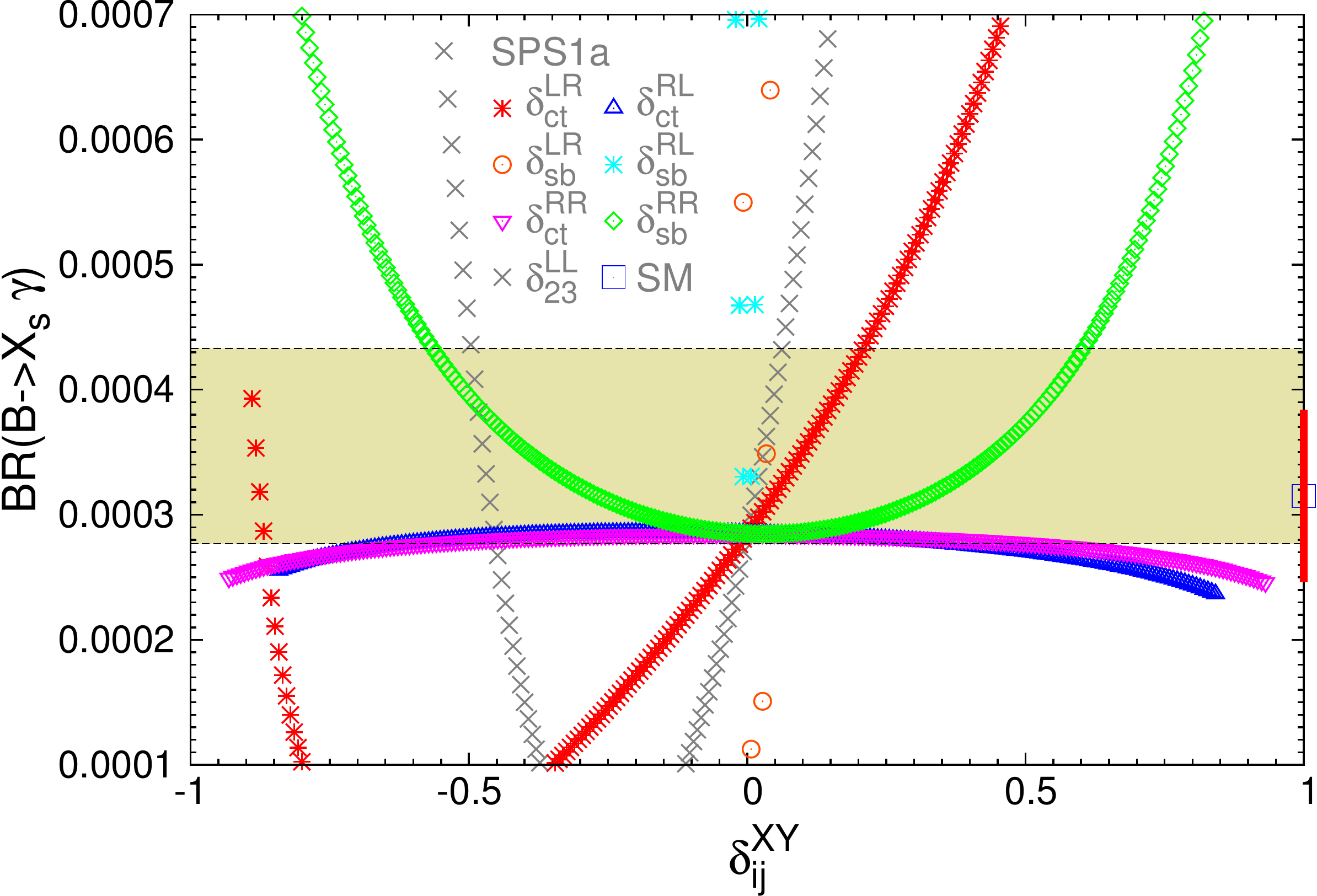}& 
\includegraphics[width=13.2cm,height=17.2cm,angle=270]{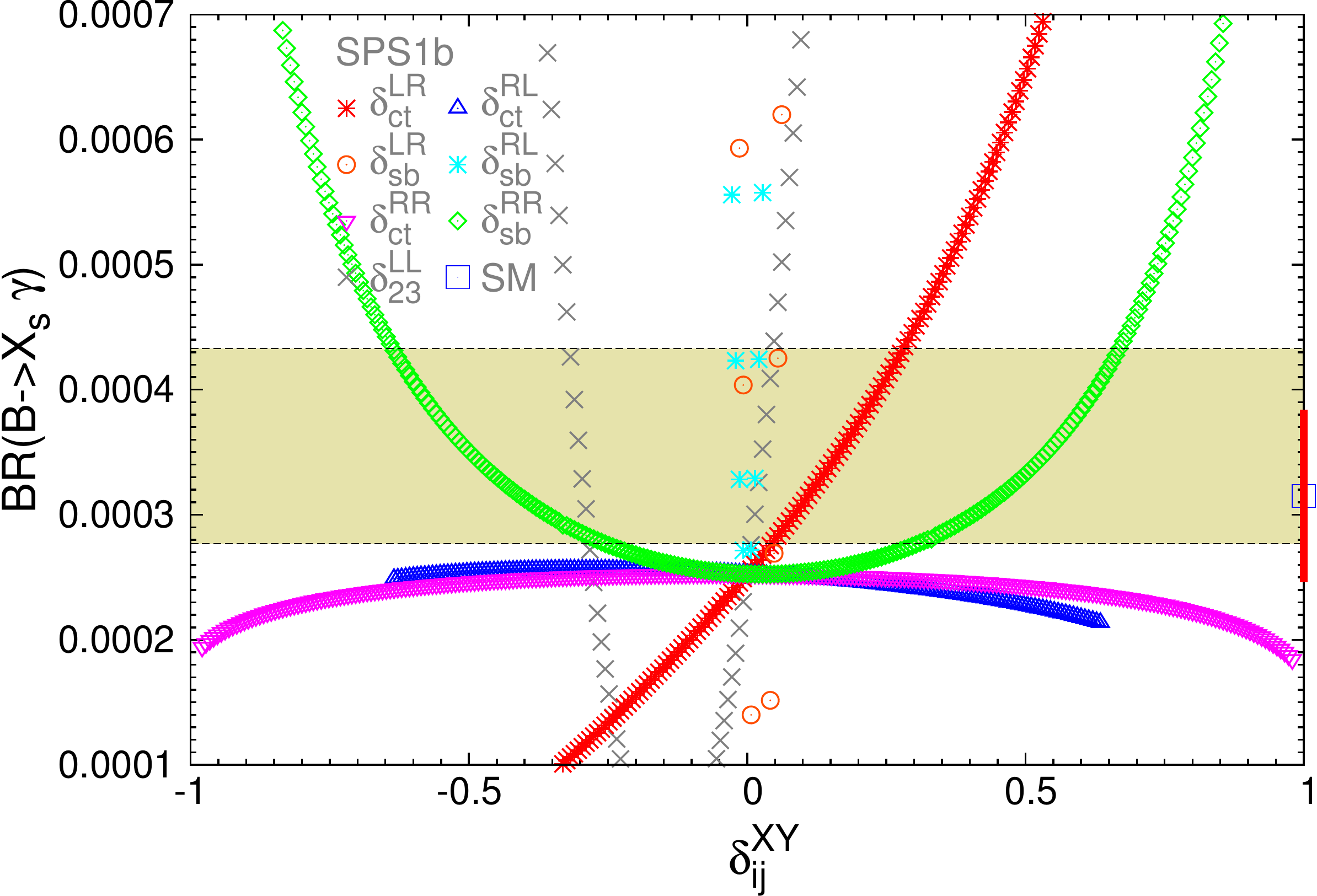}\\ 
\includegraphics[width=13.2cm,height=17.2cm,angle=270]{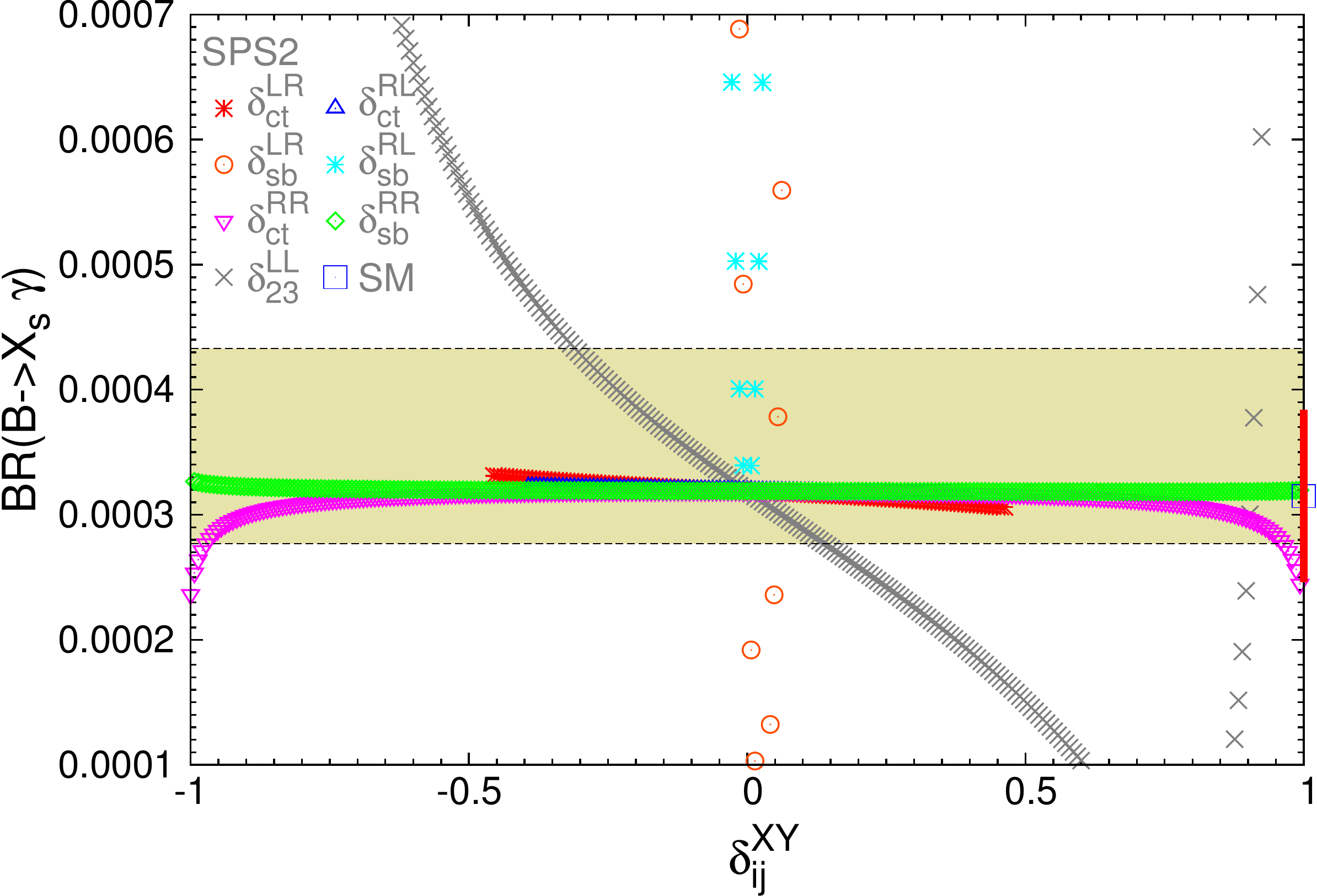}&
\includegraphics[width=13.2cm,height=17.2cm,angle=270]{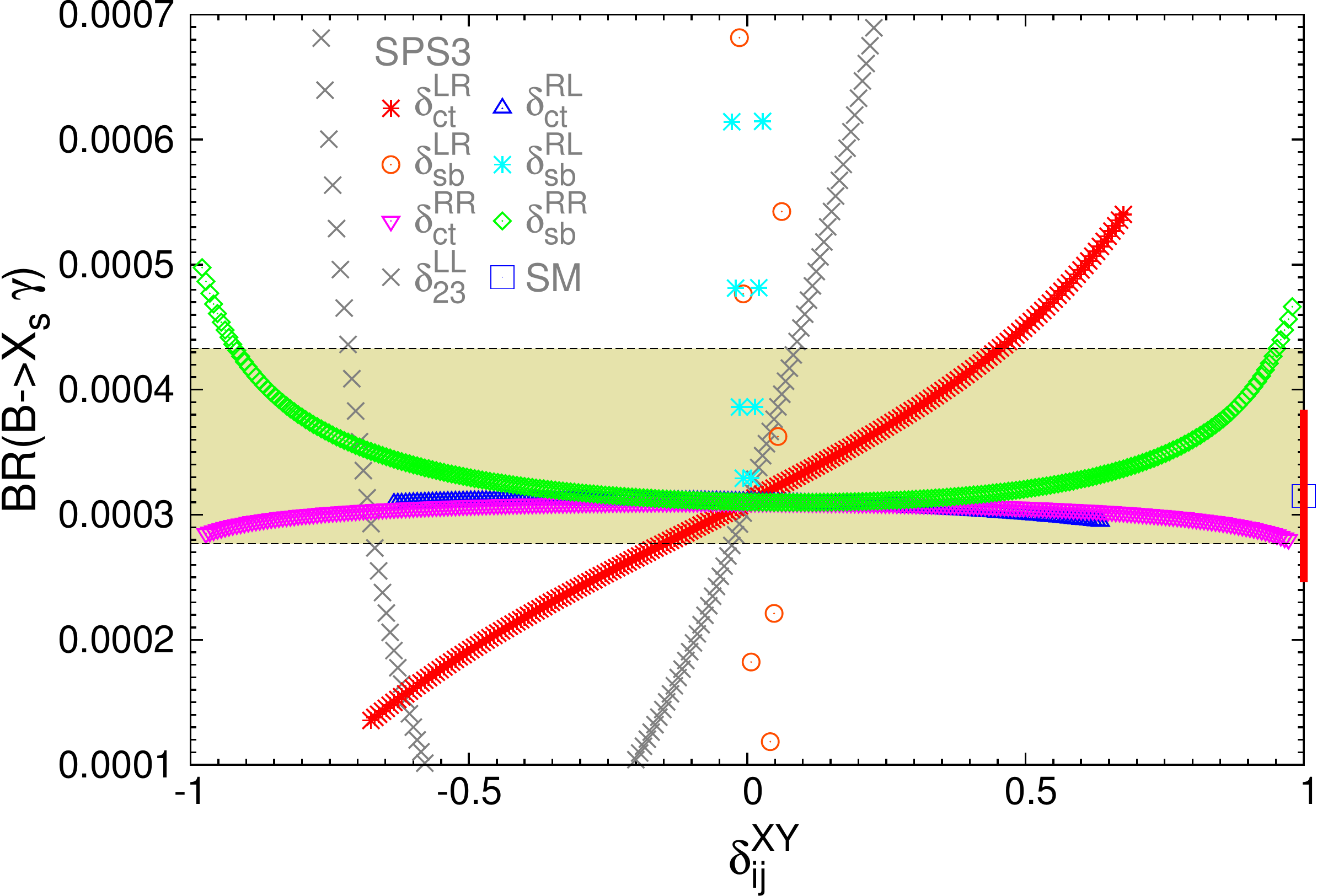}\\ 
\includegraphics[width=13.2cm,height=17.2cm,angle=270]{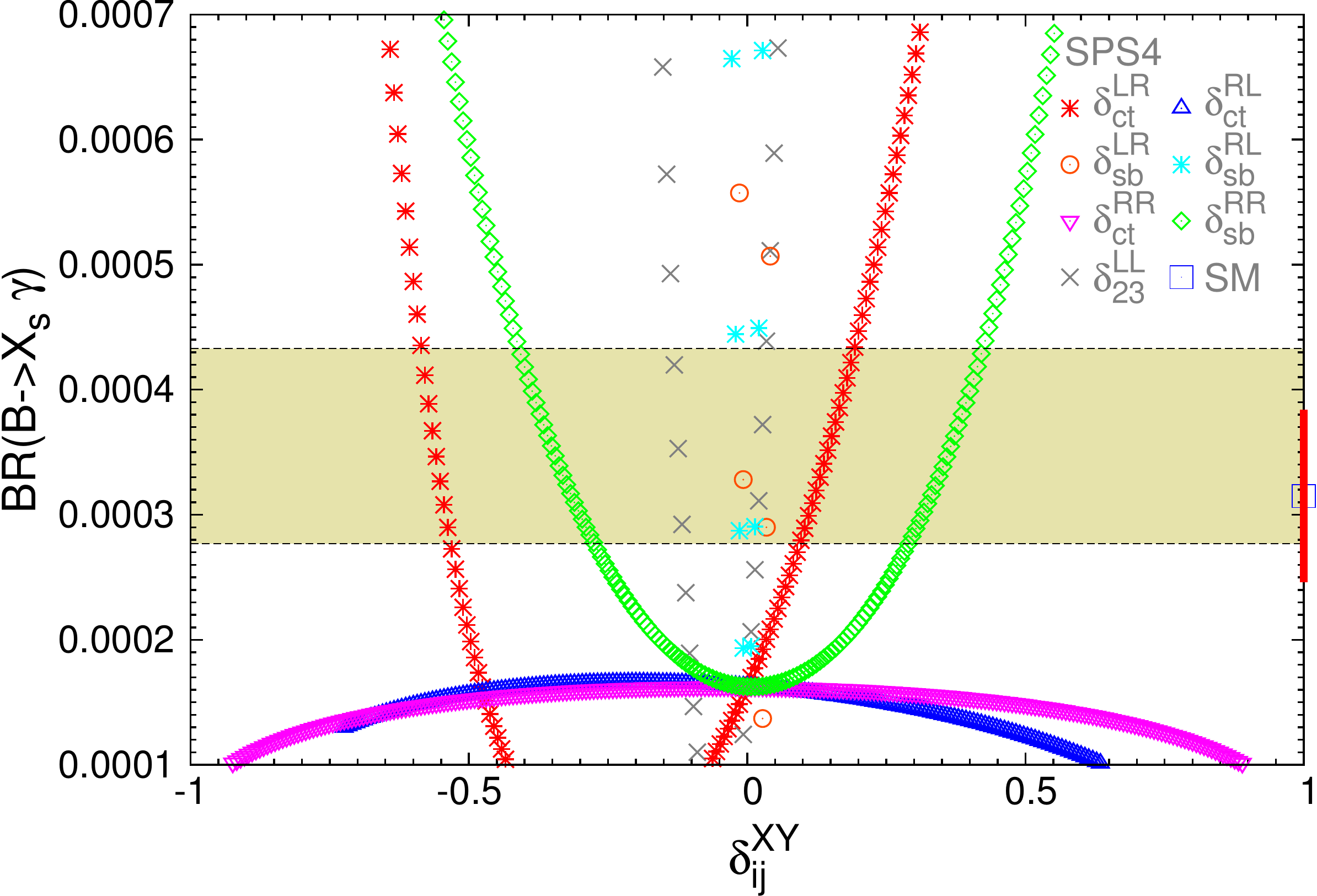}& 
\includegraphics[width=13.2cm,height=17.2cm,angle=270]{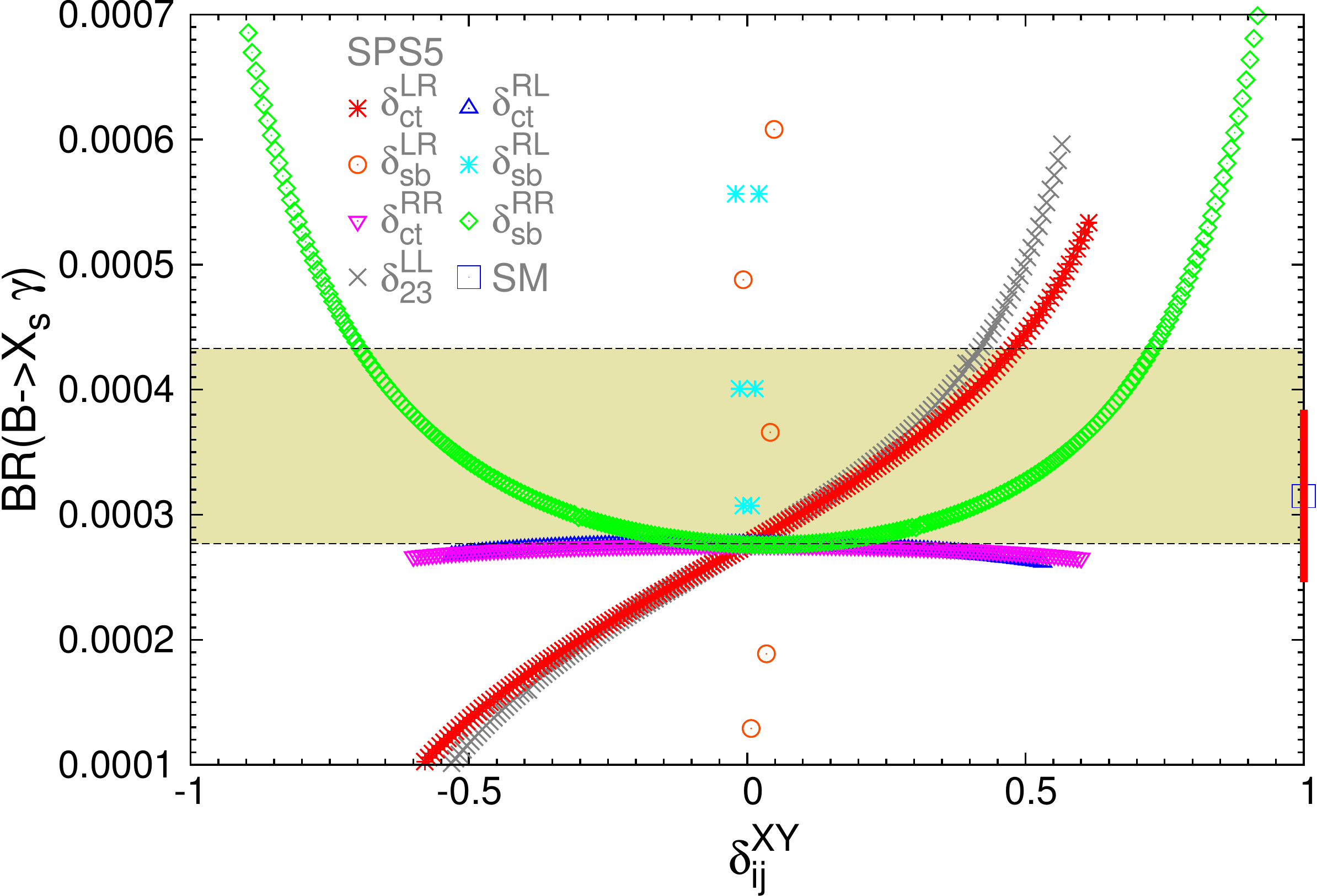}\\ 
\end{tabular}}}
\caption{Sensitivity to the NMFV deltas in \bsg\ for the SPSX points of table
  \ref{points}. The experimental pre-LHC allowed $3\sigma$ area is the horizontal
  coloured band. The pre-LHC SM prediction and the theory uncertainty $\Dtheo(\bsg)$
  (red bar) are displayed on the right axis.}  
\label{figbsgamma}
\end{figure}

\begin{figure}[h!] 
\centering
\hspace*{-8mm} 
{\resizebox{17.3cm}{!} 
{\begin{tabular}{cc} 
\includegraphics[width=13.2cm,height=17.2cm,angle=270]{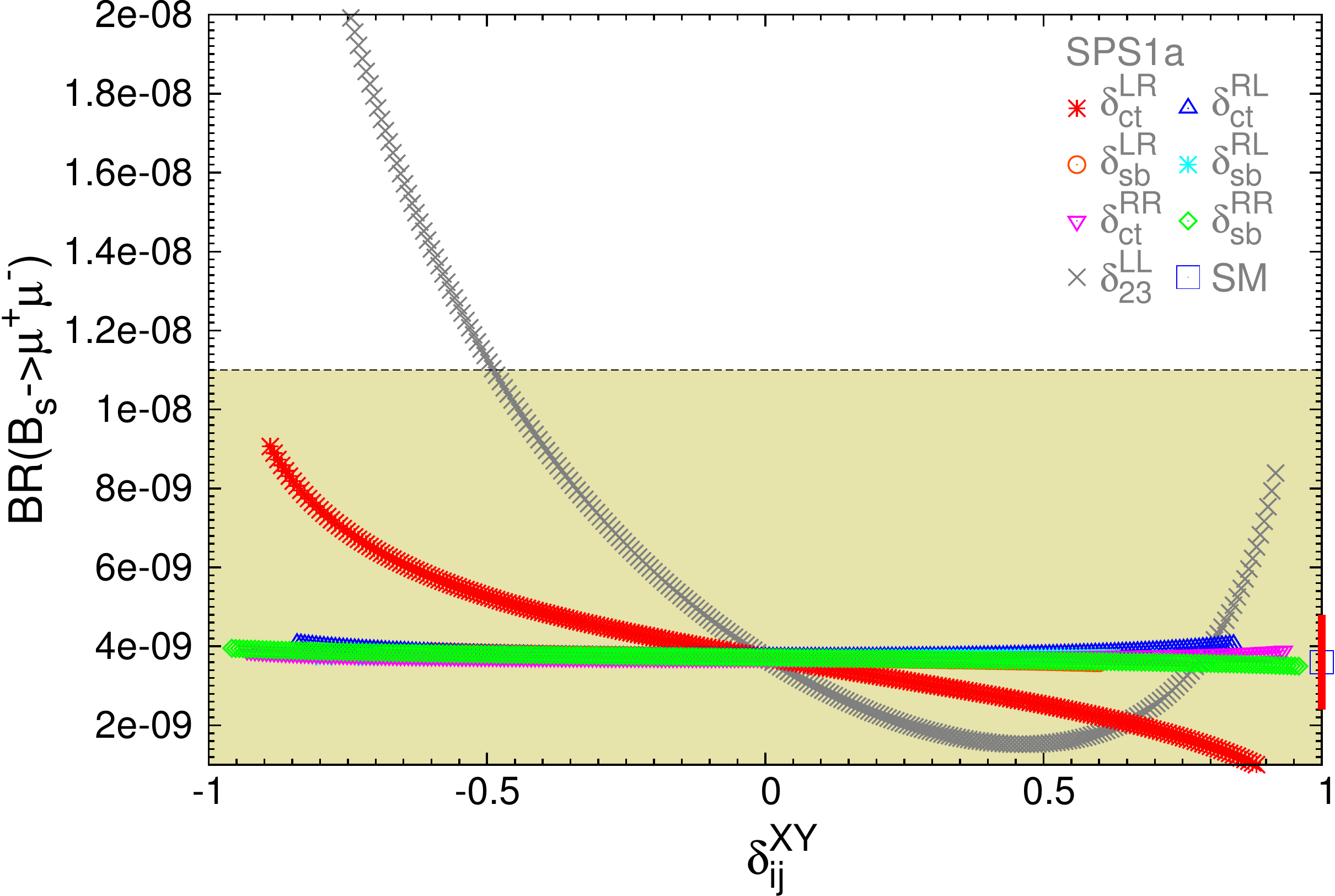}& 
\includegraphics[width=13.2cm,height=17.2cm,angle=270]{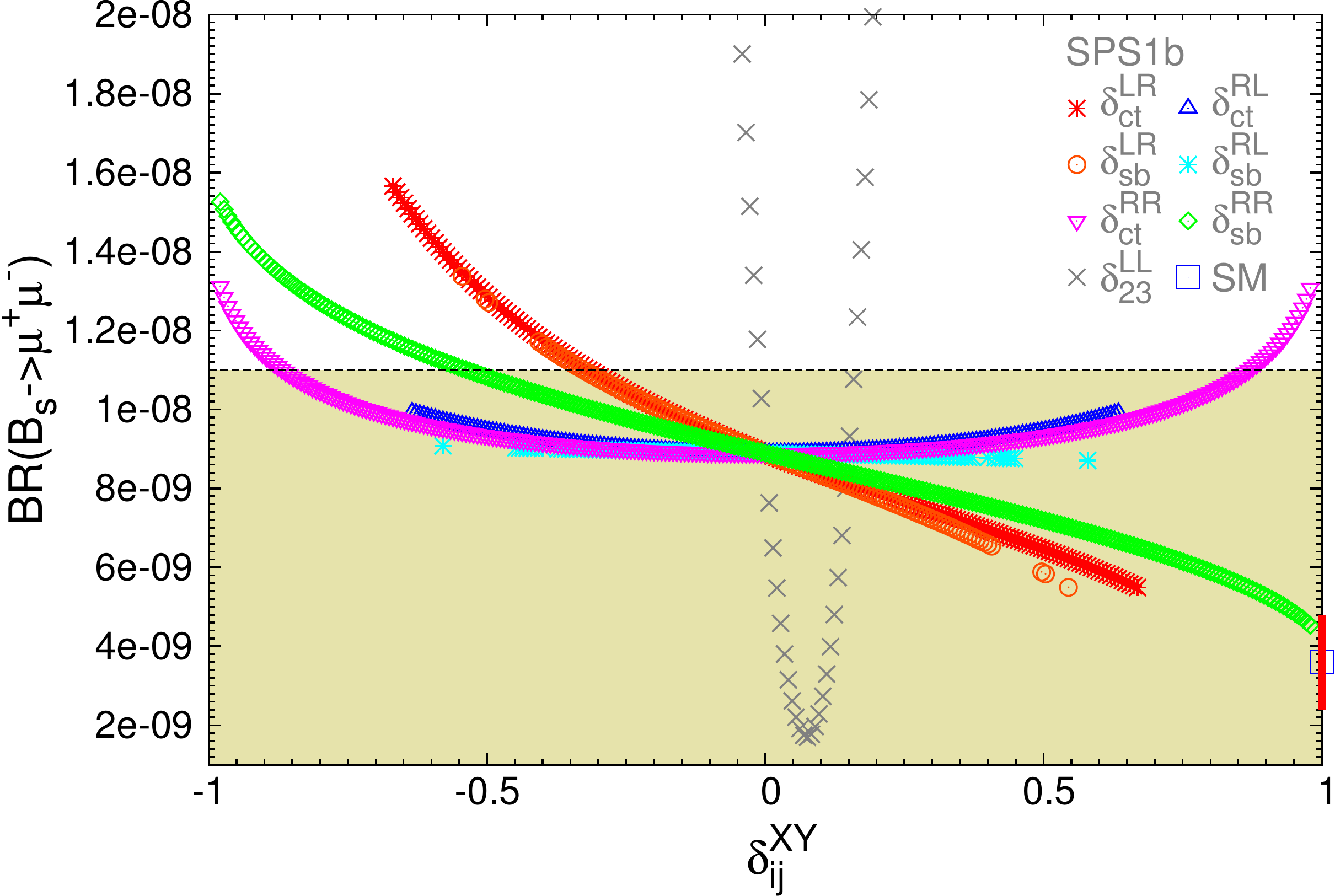}\\ 
\includegraphics[width=13.2cm,height=17.2cm,angle=270]{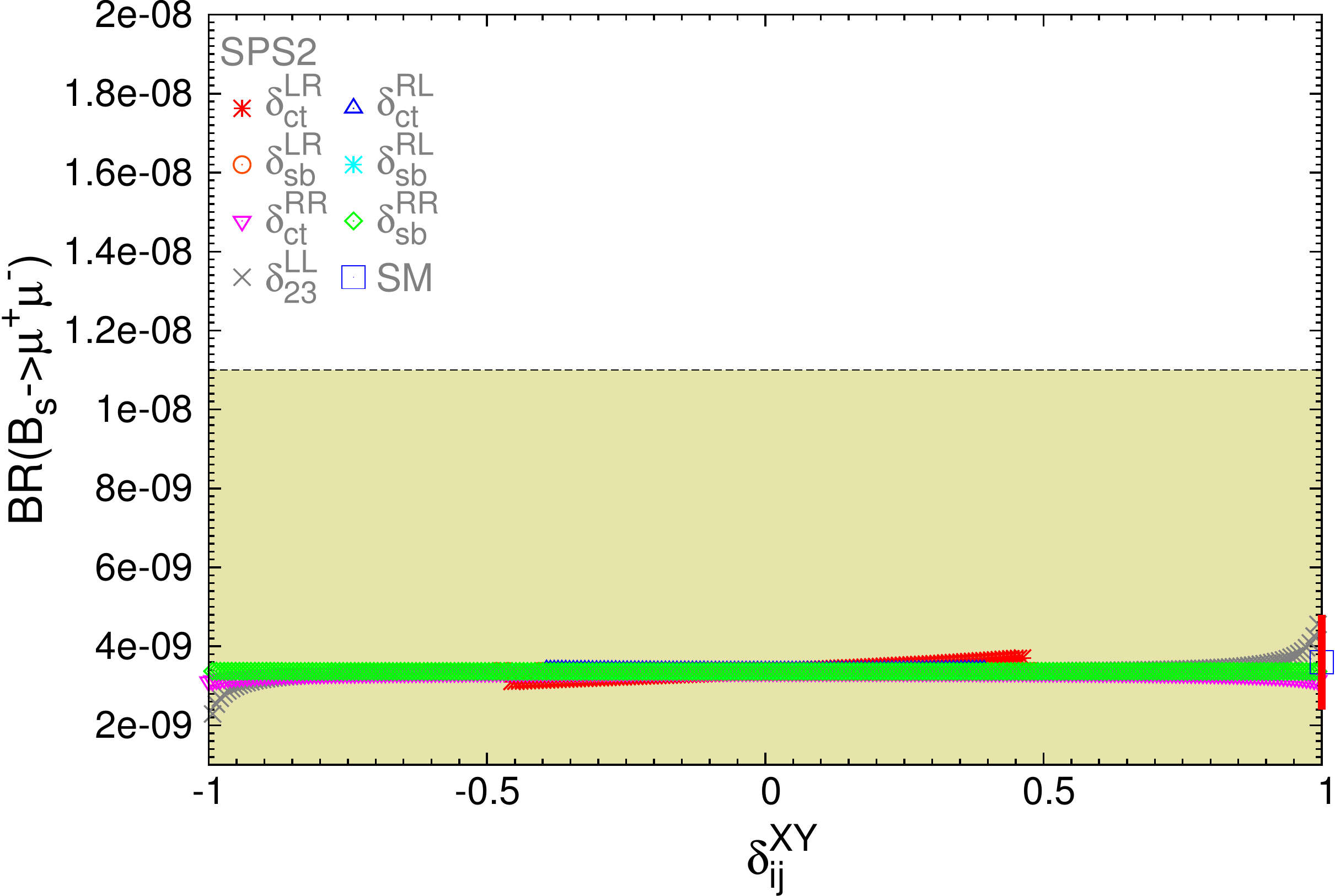}&
\includegraphics[width=13.2cm,height=17.2cm,angle=270]{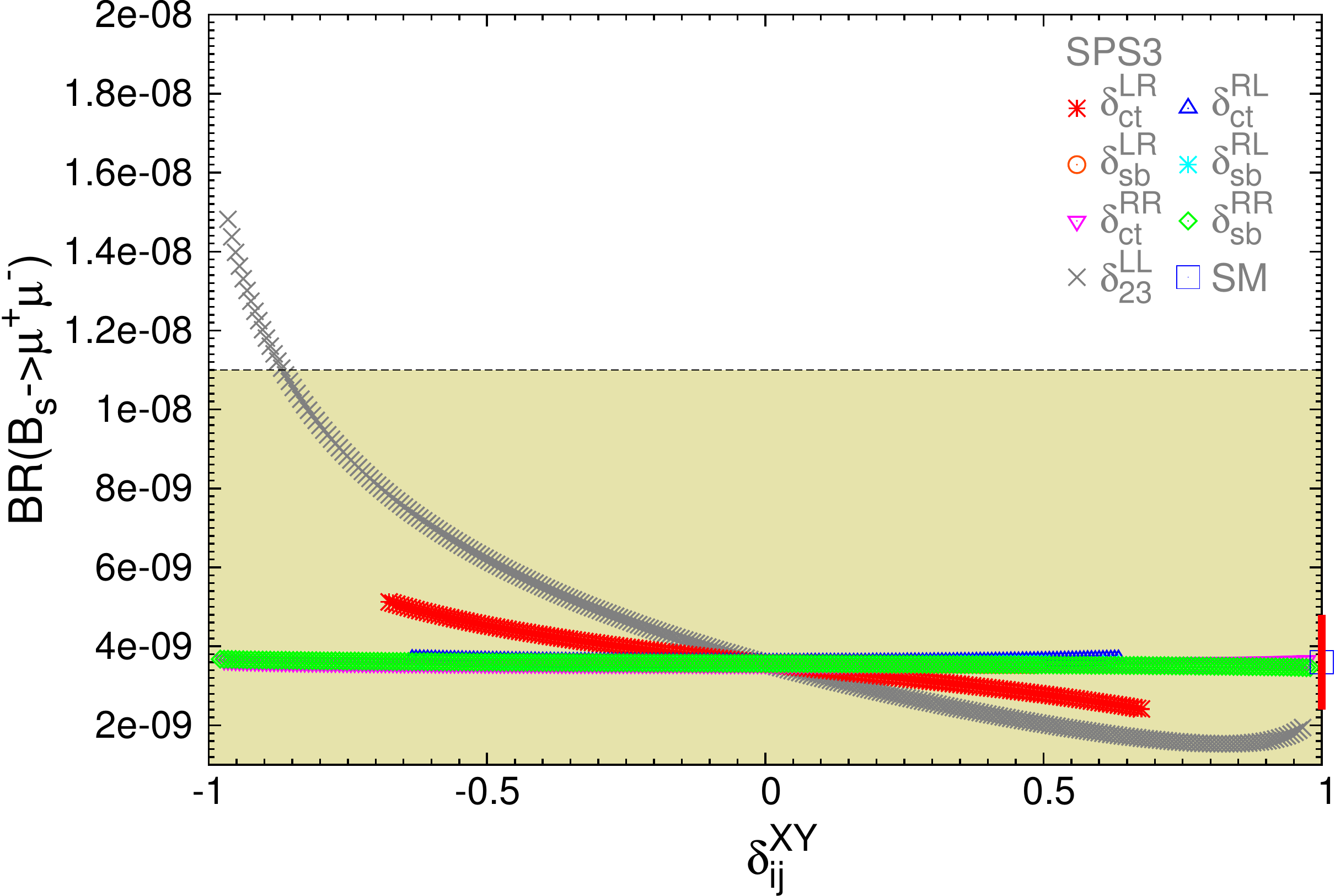}\\ 
\includegraphics[width=13.2cm,height=16.9cm,angle=270]{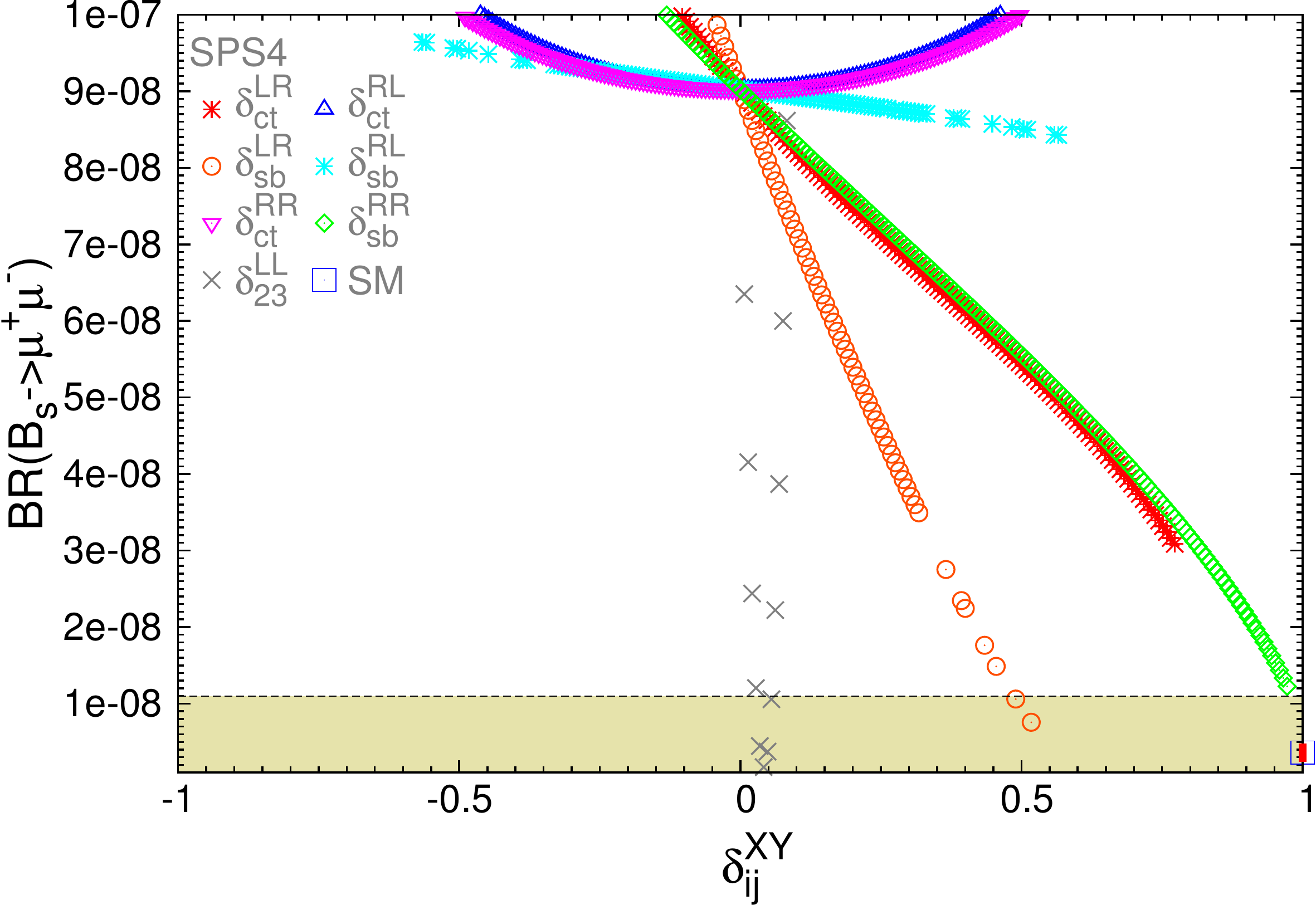}& 
\includegraphics[width=13.2cm,height=17.2cm,angle=270]{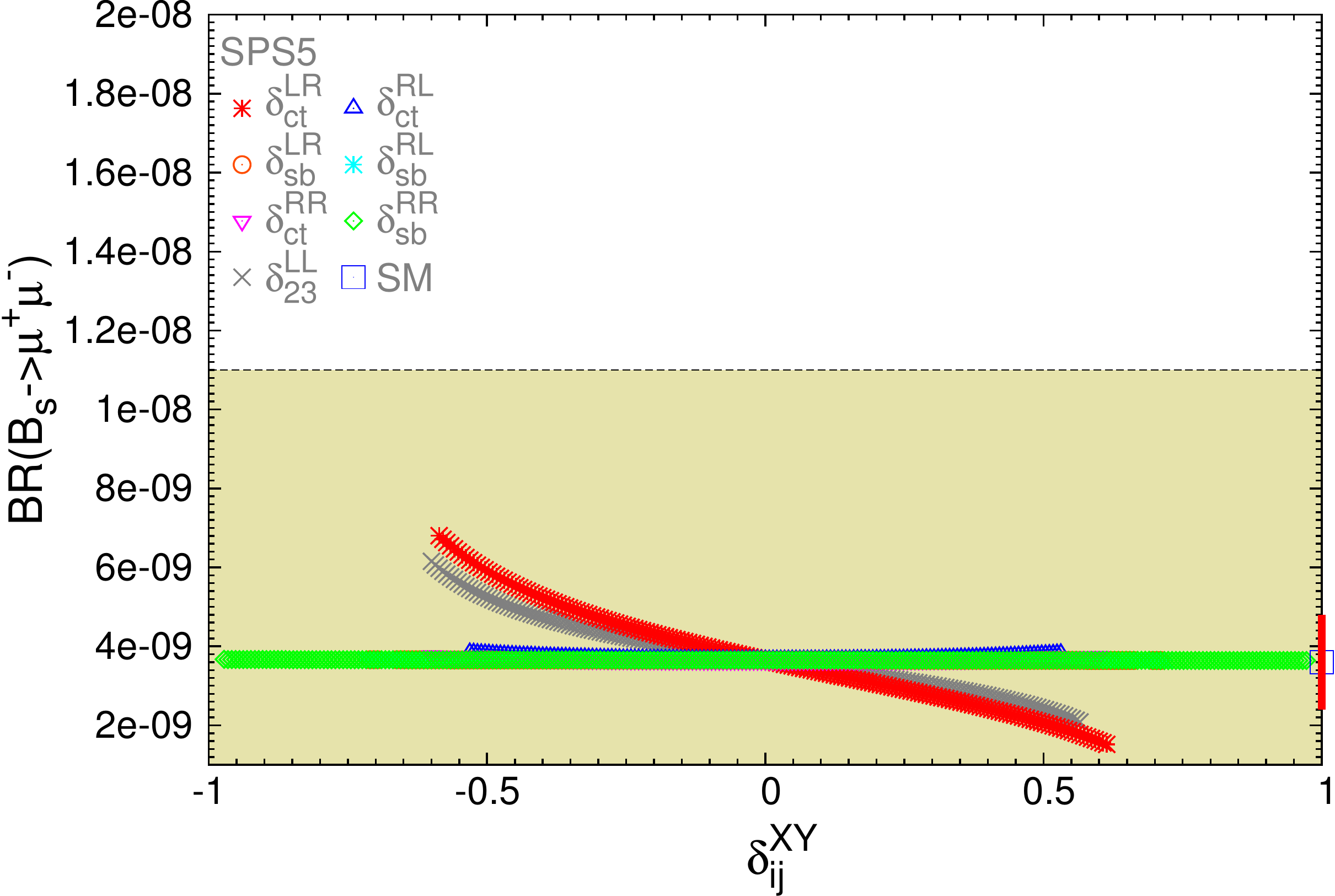}\\ 
\end{tabular}}}
\caption{Sensitivity to the NMFV deltas in ${\rm BR}(B_s \to \mu^+ \mu^-)$ for
  the SPSX points of table \ref{points}.  The experimental allowed region by
  the $95\% ~{\rm CL}$ pre-LHC bound is the horizontal coloured area. The pre-LHC SM prediction
  and the theory uncertainty \Dtheo(\bmm) (red bar) are displayed on the right
  axis.}   
\label{figbmumu}
\end{figure}

\begin{figure}[h!] 
\centering
\hspace*{-8mm} 
{\resizebox{17.3cm}{!} 
{\begin{tabular}{cc} 
\includegraphics[width=13.2cm,height=17.2cm,angle=270]{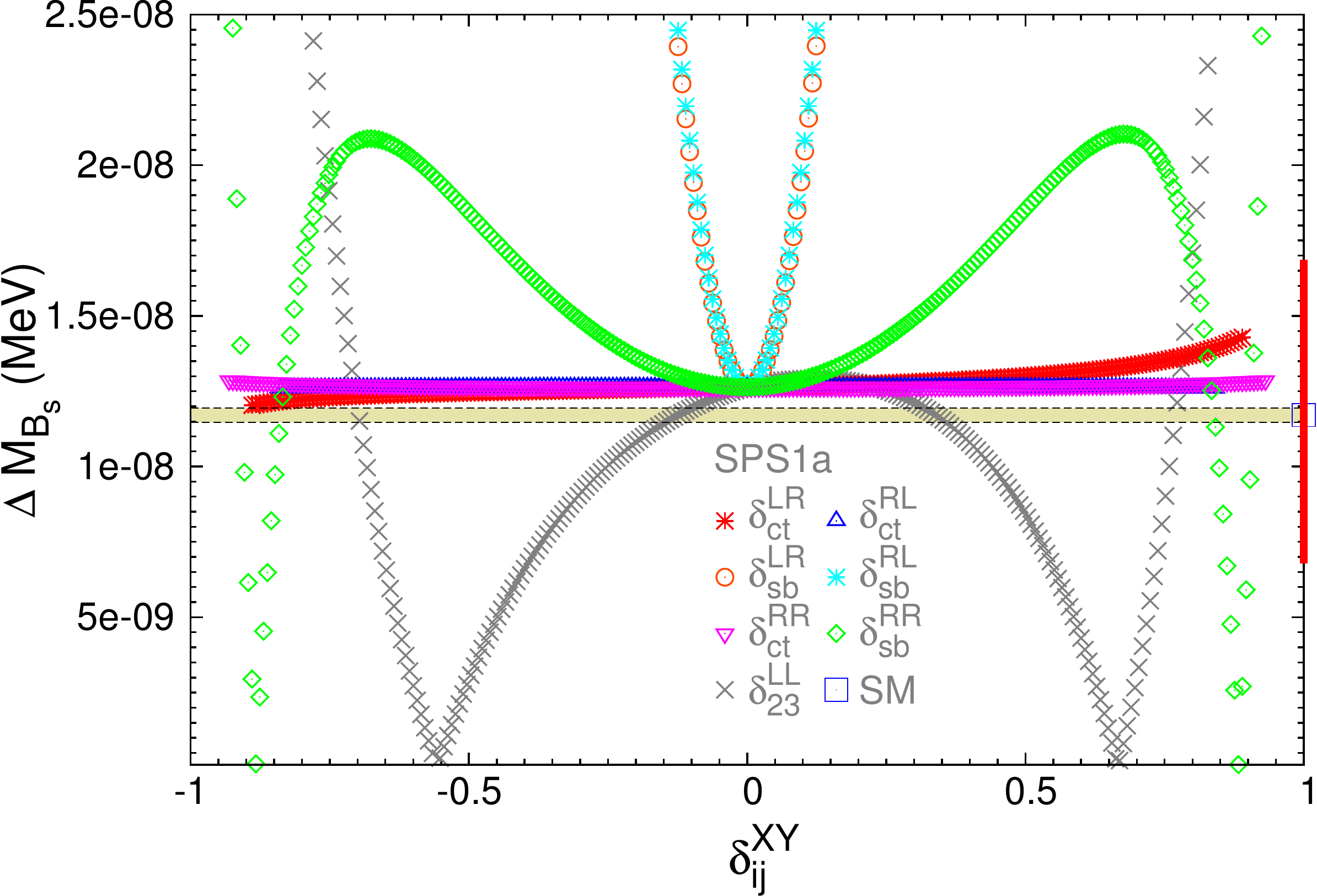}& 
\includegraphics[width=13.2cm,height=17.2cm,angle=270]{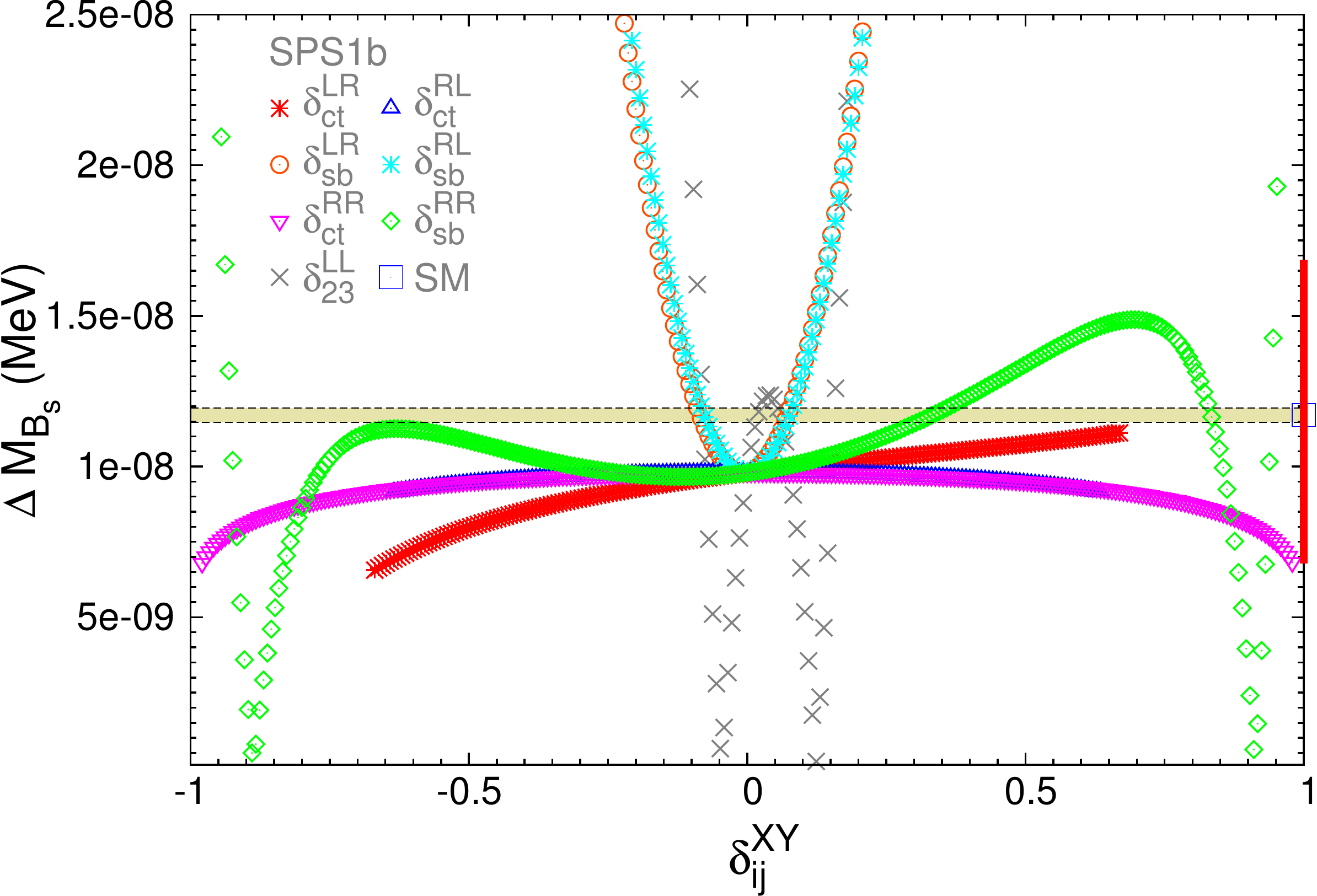}\\ 
\includegraphics[width=13.2cm,height=17.2cm,angle=270]{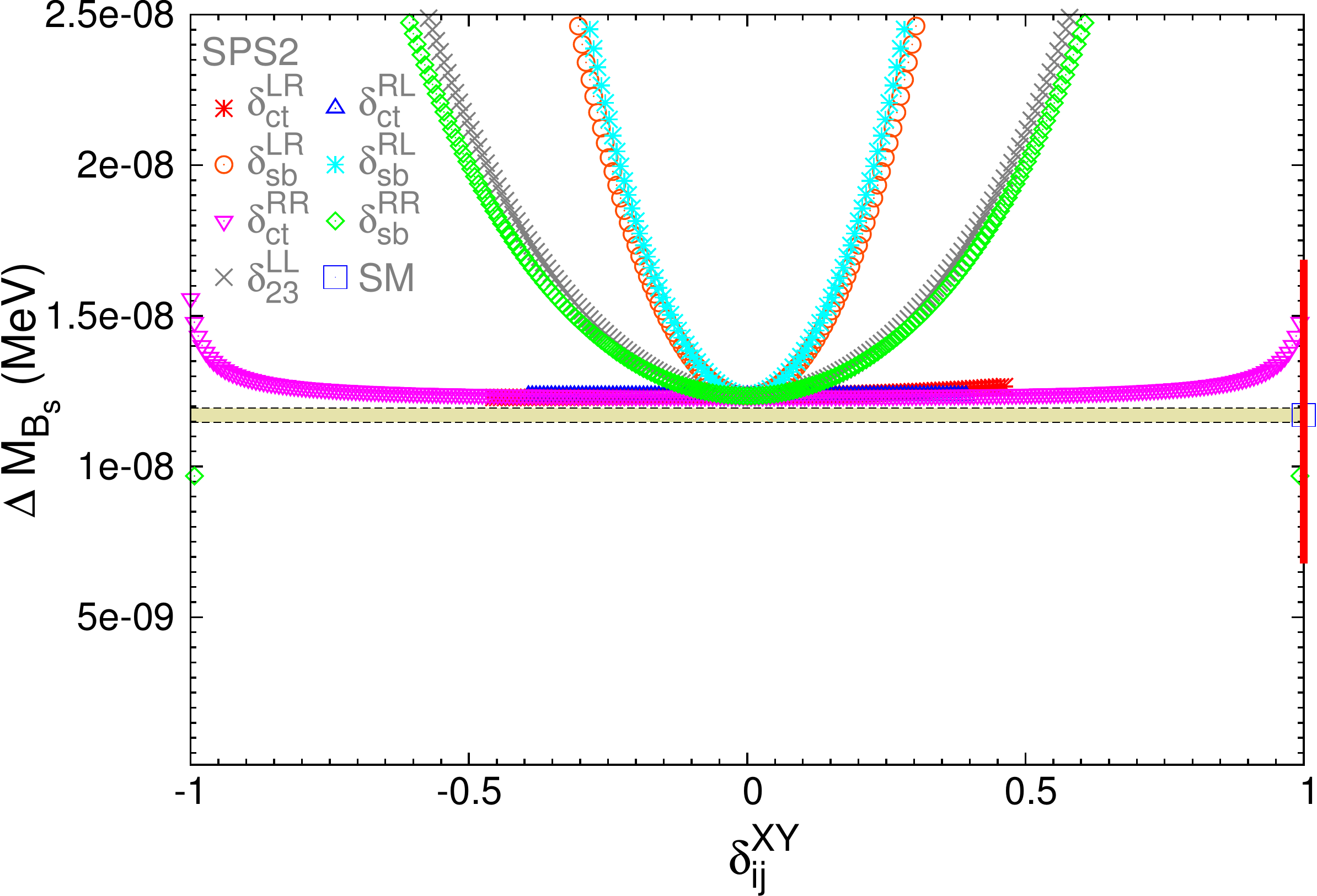}&
\includegraphics[width=13.2cm,height=17.2cm,angle=270]{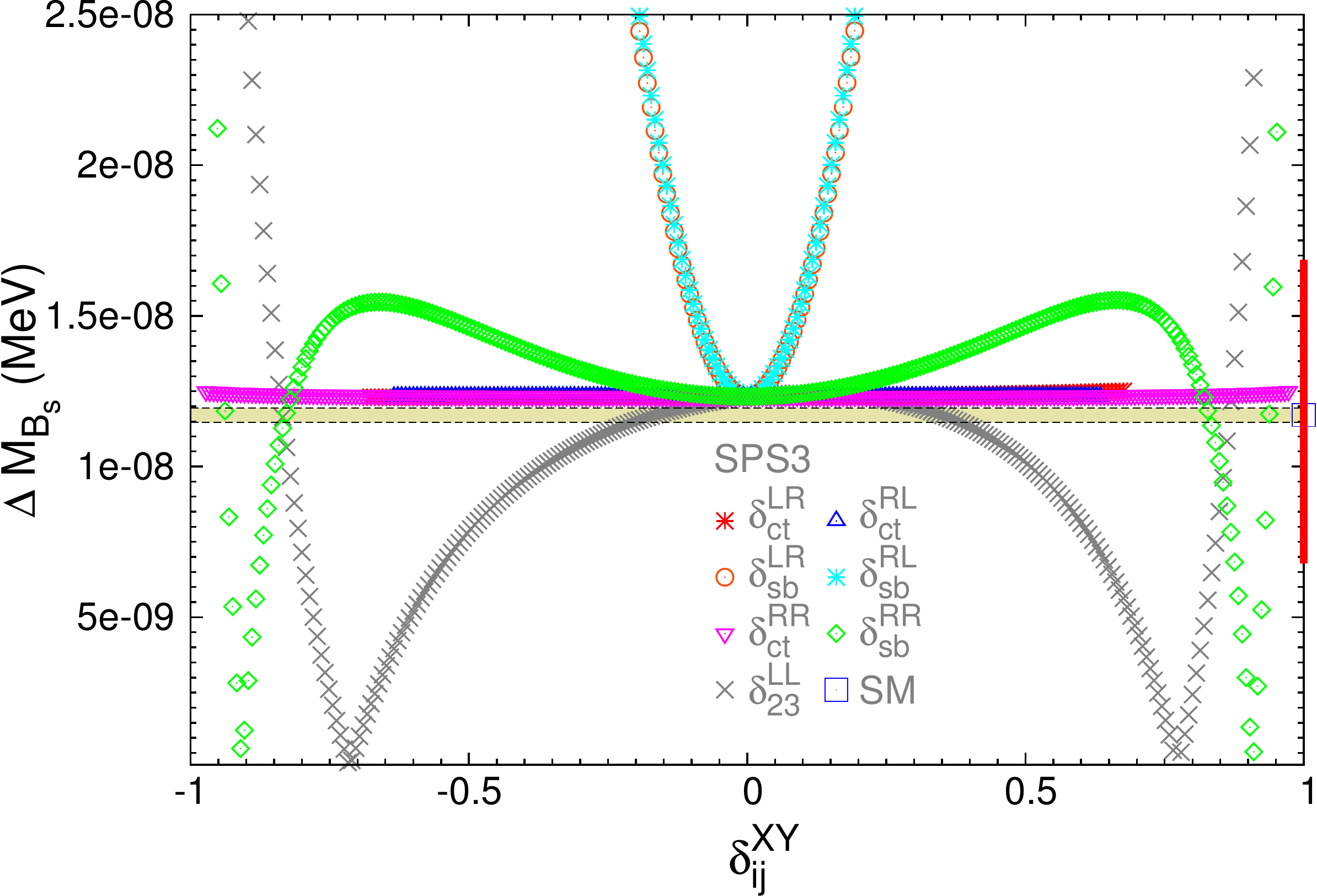}\\ 
\includegraphics[width=13.2cm,height=17.2cm,angle=270]{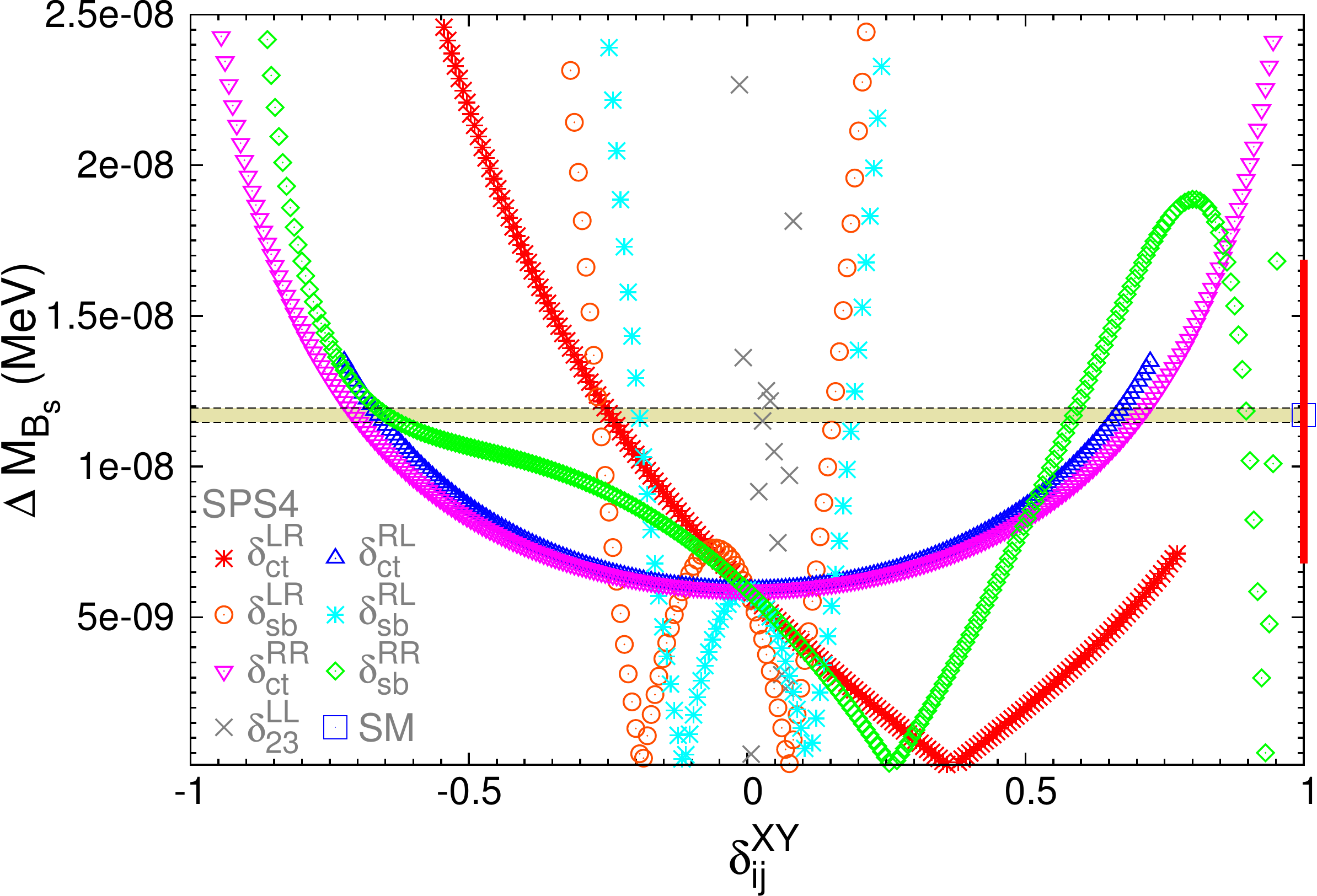}& 
\includegraphics[width=13.2cm,height=17.2cm,angle=270]{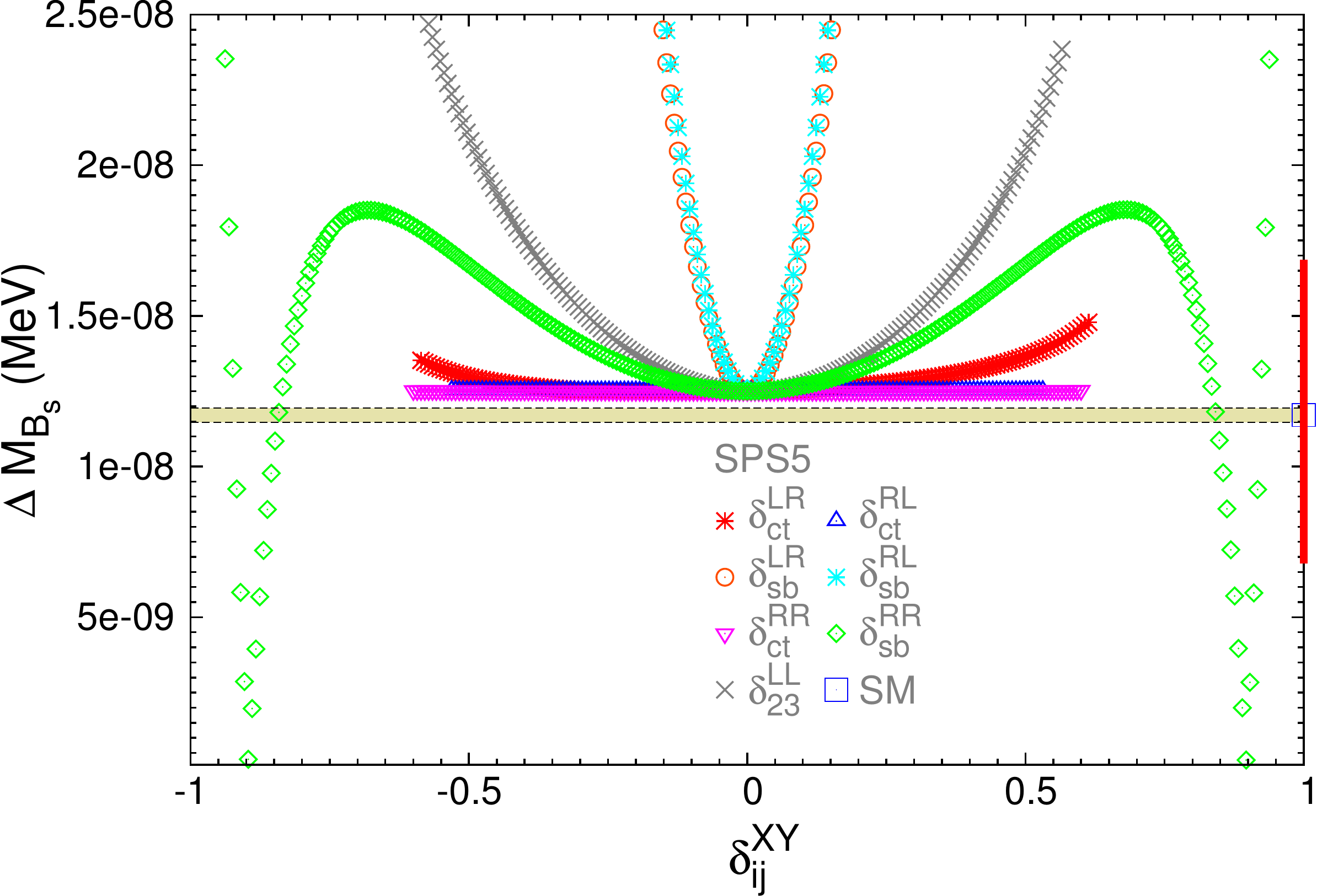}\\ 
\end{tabular}}}
\caption{Sensitivity to the NMFV deltas in $\Delta M_{B_s}$ for the SPSX
  points of table \ref{points}.  The experimental pre-LHC allowed $3\sigma_{\rm exp}$ area is
  the horizontal coloured band. The pre-LHC SM prediction and the theory uncertainty
  \Dtheo(\dmbs) (red bar) are displayed on the right axis.}  
\label{figdeltams}
\end{figure}

For the first study we analyze the results of the six SPS points of the CMSSM scenarios, described in Section \ref{frameworka}. The predictions for \bsg, \bmm\ and \dmbs\ 
versus the various $\deXYij$, for the six selected SPS points, are displayed respectively in  \reffis{figbsgamma}, \ref{figbmumu} and \ref{figdeltams}.  It should be noted
that some  of the predicted lines in these plots do not expand
along the full interval $-1<\deXYij<1$, and they are restricted to a smaller 
interval. This is due to the fact, as already said in the previous chapter, that for some regions of
the parameter space a too large delta value can generate very large corrections to the mass of the light Higgs boson, and even lead to a negative mass squared value. 
These problematic
points are consequently not shown in our plots.    

\medskip
The experimental values and the SM predictions for the three observables at the moment we first studied them, before the LHC, were the following:

\medskip
For \bsg\ the experimental measurement of this observable \cite{Nakamura:2010zzi,Asner:2010qj} (where we added the various contributions to the experimental error in quadrature), and its prediction within the SM \cite{Misiak:2009nr} were: 

\noindent \begin{equation}
\bsg_{\rm exp}=(3.55 \pm 0.26)\times10^{-4}
\label{bsgamma-exp}
\end{equation}

\noindent \begin{equation}
\bsg_{\rm SM}= (3.15 \pm 0.23)\times10^{-4}
\label{bsgamma-SM}
\end{equation}

 In the case of \bmm\ the experimental upper bound for this observable \cite{CMSLHCb}, and the prediction within the SM \cite{Buras:2009if} were: 
\noindent \begin{equation}
\bmm_{\rm exp} < 1.1 \times 10^{-8}\,\,\,\, (95\% ~{\rm CL}) 
\footnote{In the present moment the experimental situation has changed drastically and now there is a measurement instead of a bound. This will be discussed in the next section \ref{numresafterlhc}}
\label{bsmumu-exp}
\end{equation}

\noindent \begin{equation}
\bmm_{\rm SM}= (3.6\pm 0.4)\times 10^{-9}
\label{bsmumu-SM}
\end{equation}

Finally, the experimental measurement \cite{Nakamura:2010zzi} of \dmbs\ (we added again the various contributions to the experimental error in quadrature), and its prediction within the SM (using the NLO expression of \cite{Buras:1990fn} and the error estimate of \cite{Golowich:2011cx}) were:
\noindent \begin{align}
\label{deltams-exp}
{\dmbs}_{\rm exp} &= (117.0 \pm 0.8) \times 10^{-10} \mev~, \\
{\dmbs}_{\rm SM} &= (117.1^{+17.2}_{-16.4}) \times 10^{-10} \mev~.
\label{deltams-SM}
\end{align} 

We have included in the right vertical axis of the figures, for
comparison, the respective SM prediction for each observable. The error bars displayed are
the corresponding SM uncertainties as explained above, expanded with 3${\sigma}_{\rm SM}$ errors. 
The shadowed horizontal bands in the case of \bsg\ and
\dmbs\ are their corresponding experimental measurements 
 in (\ref{bsgamma-exp}), and (\ref{deltams-exp}), expanded with
3${\sigma}_{\rm exp}$ errors. In the case of \bmm\ the shadowed area 
corresponds to the allowed region by the upper bound in (\ref{bsmumu-exp}).


\bigskip
The main conclusions extracted from these figures for the three $B$
observables are summarized as follows: 

\begin{itemize}
\item \bsg:
\begin{itemize}
\item[-] Sensitivity to the various deltas:

We find strong sensitivity to $\delta^{LR}_{sb}$, $\delta^{RL}_{sb}$, $\delta^{LL}_{23}$, $\delta^{RR}_{sb}$ and $\delta^{LR}_{ct}$, in all the studied points, for both high and low $\tb$ values.
The order found from the highest to the lowest sensitivity is, generically:  1) $\delta^{LR}_{sb}$ and $\delta^{RL}_{sb}$ the largest, 2)   $\delta^{LL}_{23}$ the next, 3) $\delta^{LR}_{ct}$ and $\delta^{RR}_{sb}$ the next to next, and 4) slight sensitivity to $\delta^{RR}_{ct}$ and $\delta^{RL}_{ct}$. 

\item[-] Comparing the predictions with the pre-LHC experimental data:

If we focus on the five most relevant deltas, $\delta^{LR}_{sb}$,
$\delta^{RL}_{sb}$, $\delta^{LL}_{23}$, $\delta^{RR}_{sb}$ and
$\delta^{LR}_{ct}$, we see clearly that tiny deviations from zero (i.e.,
deviations from MFV) in these deltas, and specially in the first three,
produce sizeable shifts  
in  \bsg, and in many cases take it out of the pre-LHC experimental
allowed band.    
Therefore, it is obvious from these plots that \bsg\ set stringent bounds on the deltas (when varying just one delta), which were particularly tight on  $\delta^{LR}_{sb}$, $\delta^{RL}_{sb}$, $\delta^{LL}_{23}$, $\delta^{RR}_{sb}$, and  $\delta^{LR}_{ct}$, indeed for all the studied SPS points. There are just two exceptions, where the predicted central values of \bsg\ are already outside the experimental band 
in the MFV case (all deltas set to zero), and assuming one of these three most relevant deltas to be non-vanishing, the prediction 
moves inside the experimental band. This happens, for instance, in  the points SPS4 and SPS1b that have the largest $\tb$ values of 50 and 30 respectively.  Interestingly, it means that some points of the CMSSM, particularly those with large $\tb$ values, that would have been excluded in the context of MFV, could be recovered as compatible with data within NMFV-SUSY scenarios. 
\item[-] Intervals of $\deXYij$  allowed by pre-LHC data:

In order to conclude on the pre-LHC allowed delta intervals we have assumed that our
predictions of \bsg\ within SUSY scenarios have a somewhat
larger theoretical error $\Dtheo(\bsg)$ than the SM prediction 
$\Dtheo_{\rm SM}(\bsg)$
given in (\ref{bsgamma-SM}). As a very conservative value we use
$\Dtheo(\bsg) = 0.69 \times 10^{-4}$. 
A given $\deXYij$ value is then considered to be allowed
by data if the corresponding interval, defined by $\bsg \pm \Dtheo(\bsg)$,
intersects with the experimental band. It corresponds to 
adding linearly the experimental uncertainty and the MSSM theoretical
uncertainty. In total a predicted ratio in the
interval  
\noindent \begin{align}
\label{bsglinearerr}
2.08\times 10^{-4} < \bsg\ < 5.02\times 10^{-4},
\end{align} 
is regarded as allowed. 
Our results for these allowed intervals in this pre-LHC situation are summarized in the third column of table
\ref{tableintervals}. In this table we see again that the less constrained
parameters by \bsg\ were $\delta^{RL}_{ct}$,
and $\delta^{RR}_{ct}$. Therefore,
except for the excluded SPS4 case, these two deltas could be sizeable,
$|\deXYij|$ larger than ${\cal O}(0.1)$, and compatible with \bsg\ data.    
\end{itemize}
\end{itemize}

\begin{itemize}
\item
\bmm:

\begin{itemize}
\item[-] Sensitivity to the various deltas:

We find significant sensitivity to the NMFV deltas in scenarios with very large $\tb$, as it is the case of SPS4 and SPS1b. 
This sensitivity is clearly due to the Higgs-mediated contribution that, grows as $\tan^6\beta$. The largest sensitivity is to $\delta^{LL}_{23}$. In the case of SPS4, there is also significant sensitivity to $\delta^{LR}_{sb}$, $\delta^{RR}_{sb}$ and $\delta^{LR}_{ct}$. In the SPS1b scenario there is also found some sensitivity to $\delta^{LR}_{sb}$,  $\delta^{RR}_{ct}$, $\delta^{RR}_{sb}$ and $\delta^{LR}_{ct}$.
  
\item[-] Comparing the predictions with the pre-LHC experimental data:

\reffi{figbmumu} clearly shows that most of the $|\deXYij| \leq 1 $ explored values were allowed by pre-LHC experimental data on ${\rm BR}(B_s \to \mu^+ \mu^-)$. It is in the points with very large $\tb$, i.e SPS4 and SPS1b,  where there are some relevant differences between the MFV and the NMFV predictions. First, all predictions for MFV scenarios except for SPS4, are inside the pre-LHC experimental allowed area. In the case of SPS1b, the comparison of the NMFV predictions with data constrained specially $\delta^{LL}_{23}$, but also $\delta^{LR}_{sb}$,  $\delta^{RR}_{ct}$, $\delta^{RR}_{sb}$ and $\delta^{LR}_{ct}$. In the case of SPS4 some input non-vanishing values of  $\delta^{LL}_{23}$, $\delta^{LR}_{sb}$ or $\delta^{RR}_{sb}$ can place the prediction inside the pre-LHC experimental allowed area. In the case of the SPS1a and SPS3 scenarios some constraints for $\delta^{LL}_{23}$ could be found.
 
\item[-] Intervals of $\deXYij$  allowed by pre-LHC data:

As in the previous observable, we assume here that our predictions for 
\bmm\ have a slightly larger error as the SM prediction in
(\ref{bsmumu-SM}), where, however, the theory uncertainty is very small in
comparison with the pre-LHC experimental bound. We choose 
$\Dtheo(\bmm) = 0.12 \times 10^{-8}$. 
Then, a given $\deXYij$ value is allowed by data if the predicted interval,
defined by $\bmm + \Dtheo(\bmm)$, intersects
the experimental area. The upper line of the experimental area in this case is
set by the $95\% ~{\rm CL}$ upper bound given in (\ref{bsmumu-exp}).  It
implies that for a predicted ratio to be allowed it must fulfil: 
\noindent \begin{equation}
\label{bsmmlinearerr}
\bmm\ < 1.22 \times 10^{-8}
\footnote{As commented before, in the present moment the experimental situation changed drastically and now there is a measurement instead of a bound. This will be discussed in the next section \ref{numresafterlhc}}.
\end{equation}

The results for the pre-LHC allowed $\deXYij$ intervals are collected in the fourth column of table \ref{tableintervals}. We conclude from this table that, except for scenarios with large $\tanb \geq 30$, like SPS4 and SPS1b, the size of the deltas could be sizeable, $|\deXYij|$ larger than ${\cal O}(0.1)$, and compatible with \bmm\ data. 
\end{itemize}
\end{itemize}

\begin{itemize}
\item \dmbs:

\begin{itemize}
\item[-] Sensitivity to the various deltas:

Generically, we find strong sensitivity to most of the NMFV deltas in all the studied points, including those with large and low $\tb$ values. The pattern of the $\Delta M_{B_s}$ predictions as a function of the various
$\deXYij$ differs substantially for each SPS point. This is mainly because in this observable there are two large competing contributions, the double Higgs penguins and the gluino boxes, with very different behaviour with  $\tb$, as we have seen in \reffi{fig:deltabs-allcontributions}.  
In the case of SPS4 with extremely large $\tb= 50$, the high sensitivity to all deltas is evident in this figure. In the case of SPS5 with low $\tb= 5$, there is important sensitivity to all deltas, except  $\delta^{RR}_{ct}$, $\delta^{LR}_{ct}$ and $\delta^{RL}_{ct}$. 
Generically, for all the studied points, we find the highest sensitivity to 1) $\delta^{LR}_{sb}$, $\delta^{RL}_{sb}$ and $\delta^{LL}_{23}$; 2)   $\delta^{RR}_{sb}$ the next, 3) $\delta^{LR}_{ct}$ the next to next; and 4)
the lowest sensitivity is to  $\delta^{RL}_{ct}$ and $\delta^{RR}_{ct}$. Consequently, these last two were the less constrained ones by the pre-LHC data.
 
\item[-] Comparing the predictions with the pre-LHC experimental data:

In this case, the pre-LHC experimental allowed $3 \sigma_{\rm exp}$ band is already very narrow, and all the central predictions at $\deXYij=0$, i.e. for MFV scenarios,  lay indeed outside this band. However, if we assume again that our predictions suffer of a similar large uncertainty as the SM prediction, given in (\ref{deltams-SM}), then the MFV predictions are all compatible with pre-LHC data except for SPS4. It is also obvious from this figure that the predictions within NMFV, as compared to the MFV ones, lie quite generically outside the experimental allowed band, except for the above commented deltas with low sensitivity.  

\item[-] Intervals of $\deXYij$  allowed by pre-LHC data:

We consider again, that a given $\deXYij$ value is allowed by pre-LHC  \dmbs\
data if the predicted interval $\dmbs \pm \Dtheo(\dmbs)$, intersects the  
experimental band. It corresponds to adding linearly the experimental uncertainty 
and the theoretical uncertainty. Given the controversy on the realistic 
size of the theoretical error in the estimates of $\Dtheo(\dmbs)$ in the MSSM (see, for instance,~\cite{refLunghi}), 
we choose a very conservative value for the theoretical error in our estimates, 
considerably larger than the SM value in (\ref{deltams-SM}), of
$\Dtheo(\dmbs)=51 \times 10^{-10}$ MeV.  This  
implies that a predicted mass difference in the interval
\begin{align}
\label{deltabslinearerr}
63\times 10^{-10} < \dmbs\ {\rm (MeV)} < 168.6\times 10^{-10},
\end{align} 
is regarded as allowed.

The pre-LHC allowed intervals for the deltas that are obtained with this requirement are collected in the fifth column of table \ref{tableintervals}. As we have already commented, the restrictions on the b-sector parameters from $\Delta M_{B_s}$ were very strong, and in consequence, there were narrow intervals allowed for, 
$\delta^{LR}_{sb}$, $\delta^{RL}_{sb}$, and $\delta^{LL}_{23}$. In the case of 
 $\delta^{RR}_{sb}$ there were indeed sequences of very narrow allowed intervals, which in some cases reduce to just a single allowed value. The parameters that showed a largest allowed interval, with sizeable $|\deXYij|$, larger than ${\cal O}(0.1)$, were $\delta^{RR}_{ct}$, $\delta^{RL}_{ct}$ and $\delta^{LR}_{ct}$. 
\end{itemize}

\end{itemize}

\clearpage
\newpage

\subsubsection*{Total allowed pre-LHC $\deXYij$ intervals}

\vspace{0.5cm}
We finally summarise in table \ref{tabdeltasummary} the total allowed
intervals for all the NMFV deltas, $\deXYij$, in these pre-LHC scenarios, where we have required
compatibility with the pre-LHC data of the three considered $B$
observables, \bsg, ${\rm BR}(B_s \to \mu^+ \mu^-)$ and $\Delta
M_{B_s}$. It is obvious, from the previous discussion, that the most
restrictive observables were \bsg\ and $\Delta M_{B_s}$, leading to a
pattern of allowed delta intervals which was clearly the intersect of
their two corresponding intervals. The main conclusion from this table
is that, except for SPS4 (the point SPS4 was practically excluded), the
NMFV deltas in the top-sector could be sizeable $|\de_{ct}^{XY}|$  larger
than ${\cal O}(0.1)$ and still compatible with $B$ data. In particular,
$\delta^{RL}_{ct}$, and $\delta^{RR}_{ct}$ were the less constrained
parameters, and to a lesser extent also $\delta^{LR}_{ct}$. The
parameters on the bottom-sector were, in contrast, quite constrained. The
most tightly constrained were clearly  $\delta^{LR}_{sb}$ and
$\delta^{RL}_{sb}$, specially the first one with just some singular
allowed values: either positive and of the order of  $3-5 \times
10^{-2}$, or negative and with a small size of the order of $-7 \times
10^{-3}$; for the second the limits were around  $2 \times 10^{-2}$ for
both positive and negative values. $\delta^{RR}_{sb}$ was the less
constrained parameter in the bottom sector, with larger allowed
intervals of  $|\delta^{RR}_{sb}| \lsim 0.4-0.9$ depending on the
scenario. 

All SPS points are defined with a positive~$\mu$ value. We have
checked the effect of switching the sign of~$\mu$. While the numerical
results are changing somewhat, no qualitative change can be
observed. Consequently, confining ourselves to positive~$\mu$ does
not constitute a general restriction of our analysis. Similar
observations are made in the Higgs-sector analysis later.

The intervals allowed by $B$ data that we have presented above  will be
of interest for a following study in this work in the next chapter, where we will 
explore the size of the radiative corrections to the MSSM Higgs masses
within these NMFV-MSSM scenarios and we will require compatibility with
$B$ data. In the final analysis of these corrections, we will use the
constraints from $B$ data as extracted from two non-vanishing deltas. As
expected, these constraints vary significantly respect to the ones with
just one non-vanishing delta.   

\newpage
\renewcommand{\arraystretch}{1.55}
\begin{table}[H]
\begin{table}[H]
\begin{center}
\resizebox{16cm}{!} {
\begin{tabular}{|c|c|c|c|c|} \hline
 & & ${\rm BR}(B\rightarrow X_s\gamma)$ & ${\rm BR}(B_s\rightarrow\mu^+\mu^-)$ & $\dmbs$  \\ \hline
$\delta^{LL}_{23}$ & \begin{tabular}{c} SPS1a \\ SPS1b \\ SPS2 \\ SPS3 \\ SPS4 \\ SPS5 \end{tabular} &  
\begin{tabular}{c}
(-0.51:-0.43) (-0.034:0.083) \\ (-0.33:-0.27)  (-0.014:0.062) \\ (-0.43:0.34)   (0.90:0.92) \\ (-0.73:-0.65) (-0.083:0.12) \\ (-0.14:-0.11) (0.0069:0.034) \\ (-0.26:0.50) \end{tabular}&  \begin{tabular}{c} (-0.53:0.92) \\ (-0.014:0.16) \\ (-0.99:0.99) \\ (-0.90:0.97) \\ (0.028:0.055) \\ (-0.60:0.57) \end{tabular}&  \begin{tabular}{c} (-0.73:-0.65) (-0.41:0.55) (0.73:0.79) \\ (-0.090:-0.069) (-0.021:0.097) (0.14:0.17) \\ (-0.37:0.37) \\ (-0.86:-0.79) (-0.56:0.66) (0.83:0.89) \\ (-0.0069)(0.021:0.055)(0.076) \\ (-0.37:0.39) 
 \end{tabular}  \\ \hline
$\delta^{LR}_{ct}$  & \begin{tabular}{c} SPS1a \\ SPS1b \\ SPS2 \\ SPS3 \\ SPS4 \\ SPS5 \end{tabular}    
& \begin{tabular}{c} 
(-0.89:-0.86) (-0.12:-0.097) (-0.062:0.28) \\ (-0.083:0.36) \\ (-0.46:0.46) \\ (-0.43:0.61) \\ (-0.61:-0.51) (0.041:0.23) \\ (-0.27:0.58) \end{tabular} & \begin{tabular}{c} (-0.89:0.89) \\ (-0.44:0.67) \\ (-0.46:0.46) \\ (-0.68:0.68) \\ excluded \\ (-0.59:0.61) \end{tabular} & \begin{tabular}{c} (-0.89:0.89) \\ (-0.67:0.67) \\ (-0.46:0.46) \\ (-0.68:0.68) \\ (-0.39:-0.021) (0.74:0.77) \\ (-0.59:0.61)
\end{tabular}  \\ \hline
$\delta^{LR}_{sb}$  & \begin{tabular}{c} SPS1a \\ SPS1b \\ SPS2 \\ SPS3 \\ SPS4 \\ SPS5 \end{tabular}    & 
\begin{tabular}{c} 
(0)(0.034) \\ (-0.0069:0) (0.048:0.055) \\ (-0.0069:0) (0.048:0.055) \\ (-0.0069:0) (0.048:0.055) \\ (-0.0069)(0.034) \\ (-0.0069:0) (0.041) \end{tabular} & \begin{tabular}{c} (-0.60:0.60) \\ (-0.43:0.54) \\ (-0.48:0.48) \\ (-0.61:0.61) \\ (0.49) \\ (-0.71:0.71) \end{tabular} & \begin{tabular}{c} (-0.076:0.076) \\ (-0.15:0.14) \\ (-0.19:0.19) \\ (-0.12:0.12) \\ (-0.29:-0.24) (-0.10:-0.014) (0.12:0.18) \\ (-0.090:0.090) 
\end{tabular}  \\ \hline
$\delta^{RL}_{ct}$  & \begin{tabular}{c} SPS1a \\ SPS1b \\ SPS2 \\ SPS3 \\ SPS4 \\ SPS5 \end{tabular}    & 
\begin{tabular}{c}
(-0.84:0.84) \\ (-0.63:0.63) \\ (-0.39:0.39) \\ (-0.63:0.63) \\ excluded \\ (-0.53:0.53) \end{tabular} 
& 
\begin{tabular}{c} 
(-0.84:0.84) \\ (-0.63:0.63) \\ (-0.39:0.39) \\ (-0.63:0.63) \\ excluded \\ (-0.53:0.53) \end{tabular} 
& 
\begin{tabular}{c}
(-0.84:0.84) \\ (-0.63:0.63) \\ (-0.39:0.39) \\ (-0.63:0.63) \\ (-0.72:-0.21) (0.21:0.72) \\ (-0.53:0.53)\end{tabular} \\ \hline
$\delta^{RL}_{sb}$  & \begin{tabular}{c} SPS1a \\ SPS1b \\ SPS2 \\ SPS3 \\ SPS4 \\ SPS5 \end{tabular}    &
\begin{tabular}{c} 
(-0.014:0.014) \\ (-0.021:0.021) \\ (-0.014:0.014) \\ (-0.021:0.021) \\ (-0.021:-0.014)(0.014:0.021) \\ (-0.014:0.014) \end{tabular} 
& 
\begin{tabular}{c}  (-0.71:0.71) \\ (-0.58:0.58) \\ (-0.55:0.55) \\ (-0.63:0.63) \\ excluded \\ (-0.72:0.72) \end{tabular} 
& 
\begin{tabular}{c}  (-0.069:0.069) \\ (-0.14:0.14) \\ (-0.17:0.17) \\ (-0.11:0.11) \\ (-0.21:-0.17) (0.16:0.21) \\ (-0.083:0.083)\end{tabular}  \\ \hline
$\delta^{RR}_{ct}$ & \begin{tabular}{c} SPS1a \\ SPS1b \\ SPS2 \\ SPS3 \\ SPS4 \\ SPS5 \end{tabular}   & \begin{tabular}{c} 
(-0.93:-0.67) (-0.64:0.93) \\ (-0.93:-0.61) (-0.56:0.90) \\ (-1.0:0.99) \\ (-0.97:0.97) \\ excluded \\ (-0.60:0.60) \end{tabular}     & \begin{tabular}{c} (-0.93:0.93) \\ (-0.95:0.94) \\ (-1.0:0.99) \\ (-0.97:0.97) \\ excluded \\ (-0.60:0.60) \end{tabular}     & \begin{tabular}{c} (-0.93:0.93) \\ (-0.98:0.98) \\ (-1.0:0.99) \\ (-0.98:0.97) \\ (-0.85:-0.22) (0.22:0.85) \\ (-0.60:0.60)
\end{tabular}   \\ \hline
$\delta^{RR}_{sb}$  & \begin{tabular}{c} SPS1a \\ SPS1b \\ SPS2 \\ SPS3 \\ SPS4 \\ SPS5 \end{tabular}    &
\begin{tabular}{c} 
(-0.65:0.68) \\ (-0.71:0.74) \\ (-0.99:0.99) \\ (-0.98:0.98) \\ (-0.45:-0.18) (0.19:0.46) \\ (-0.77:0.80) \end{tabular}  & \begin{tabular}{c} 
(-0.96:0.96) \\ (-0.73:0.98) \\ (-0.99:0.99) \\ (-0.98:0.98) \\ excluded \\ (-0.97:0.97) \end{tabular}  & \begin{tabular}{c} (-0.91:-0.90) (-0.86:-0.80) (-0.41:0.41) (0.81:0.86) (0.90:0.91) \\ (-0.94:-0.92) (-0.83:0.88)  (0.93:0.94) \\ (-0.99) (-0.39:0.39) (0.99) \\ (-0.94:-0.93) (-0.88:0.88) (0.93:0.94) \\ (-0.80:-0.028) (0.461:0.71) (0.86:0.91)  (0.94:0.95) \\ (-0.92) (-0.87:-0.78) (-0.51:0.51) (0.78:0.87)  (0.92)
\end{tabular}    \\ \hline
\end{tabular}}
\end{center}
\end{table}
\caption{Allowed delta intervals by ${\rm BR}(B\rightarrow X_s\gamma)$, ${\rm BR}(B_s\rightarrow\mu^+\mu^-)$ and $\dmbs$ for pre-LHC allowed scenarios and pre-LHC $B$ data. \label{tableintervals}}
\end{table}
\renewcommand{\arraystretch}{1.55}

\renewcommand{\arraystretch}{1.1}
\begin{table}[H]
\begin{center}
\resizebox{10.3cm}{!} {
\begin{tabular}{|c|c|c|} \hline
 & & Total allowed intervals \\ \hline
$\delta^{LL}_{23}$ & \begin{tabular}{c} SPS1a \\ SPS1b \\ SPS2 \\ SPS3 \\ SPS4 \\ SPS5 \end{tabular} &  
\begin{tabular}{c} 
(-0.034:0.083) \\ (-0.014:0.062) \\ (-0.37:0.34) \\ (-0.083:0.12) \\ (0.028:0.034) \\ (-0.26:0.39) \end{tabular} \\ \hline
$\delta^{LR}_{ct}$  & \begin{tabular}{c} SPS1a \\ SPS1b \\ SPS2 \\ SPS3 \\ SPS4 \\ SPS5 \end{tabular}    
& \begin{tabular}{c} 
(-0.89:-0.86) (-0.12:-0.097) (-0.062:0.28) \\ (-0.083:0.36) \\ (-0.46:0.46) \\ (-0.43:0.61) \\ excluded \\ (-0.27:0.58) \end{tabular}   \\ \hline
$\delta^{LR}_{sb}$  & \begin{tabular}{c} SPS1a \\ SPS1b \\ SPS2 \\ SPS3 \\ SPS4 \\ SPS5 \end{tabular}    & 
\begin{tabular}{c} 
(0)(0.034) \\ (-0.0069:0) (0.048:0.055) \\ (-0.0069:0) (0.048:0.055) \\ (-0.0069:0) (0.048:0.055) \\ excluded \\ (-0.0069:0) (0.041) \end{tabular}  \\ \hline
$\delta^{RL}_{ct}$  & \begin{tabular}{c} SPS1a \\ SPS1b \\ SPS2 \\ SPS3 \\ SPS4 \\ SPS5 \end{tabular}    & 
\begin{tabular}{c}
(-0.84:0.84) \\ (-0.63:0.63) \\ (-0.39:0.39) \\ (-0.63:0.63) \\ excluded \\ (-0.53:0.53) \end{tabular} 
  \\ \hline
$\delta^{RL}_{sb}$  & \begin{tabular}{c} SPS1a \\ SPS1b \\ SPS2 \\ SPS3 \\ SPS4 \\ SPS5 \end{tabular}    &
\begin{tabular}{c}  (-0.014:0.014) \\ (-0.021:0.021) \\ (-0.014:0.014) \\ (-0.021:0.021) \\ excluded \\ (-0.014:0.014) \end{tabular} 
  \\ \hline
$\delta^{RR}_{ct}$ & \begin{tabular}{c} SPS1a \\ SPS1b \\ SPS2 \\ SPS3 \\ SPS4 \\ SPS5 \end{tabular}   & \begin{tabular}{c} 
(-0.93:-0.67) (-0.64:0.93) \\ (-0.93:-0.61) (-0.56:0.90) \\ (-1.0:0.99) \\ (-0.97:0.97) \\ excluded \\ (-0.60:0.60) \end{tabular}    \\ \hline
$\delta^{RR}_{sb}$  & \begin{tabular}{c} SPS1a \\ SPS1b \\ SPS2 \\ SPS3 \\ SPS4 \\ SPS5 \end{tabular}    &
\begin{tabular}{c}  (-0.41:0.41)  \\ (-0.71:0.74) \\ (-0.99) (-0.39:0.39) (0.99) \\ (-0.94:-0.93) (-0.88:0.88) (0.93:0.94) \\ excluded \\ (-0.51:0.51) (0.78:0.80)
\end{tabular}    \\ \hline
\end{tabular}}  
\end{center}
\caption{Total allowed delta intervals by ${\rm BR}(B\rightarrow X_s\gamma)$, ${\rm BR}(B_s\rightarrow\mu^+\mu^-)$ and $\dmbs$ for pre-LHC allowed scenarios and pre-LHC $B$ data. \label{tabdeltasummary}}
\end{table}
\renewcommand{\arraystretch}{1.55}


\subsection{Numerical results for post-LHC allowed scenarios and post-LHC $B$ data}
\label{numresafterlhc}

Once we have a feeling about the behaviour of the observables with respect to the different $\deXYij$, we analyze the current experimental situation. This will be done using the present bounds from $B$ physics and through a study of the set of scenarios described in Section \ref{frameworkc}, defined to take into account the fact that SUSY did not appear yet in the LHC, and therefore the SUSY spectra should be heavier than what was thought in its first formulations. As will be shown in the following, this heaviness of SUSY will have some impact in the considered $B$ observables when comparing with the results of the previous section.

\begin{figure}[h!] 
\centering
\hspace*{-8mm} 
{\resizebox{17.3cm}{!} 
{\begin{tabular}{cc} 
\includegraphics[width=13.2cm,height=17.2cm,angle=270]{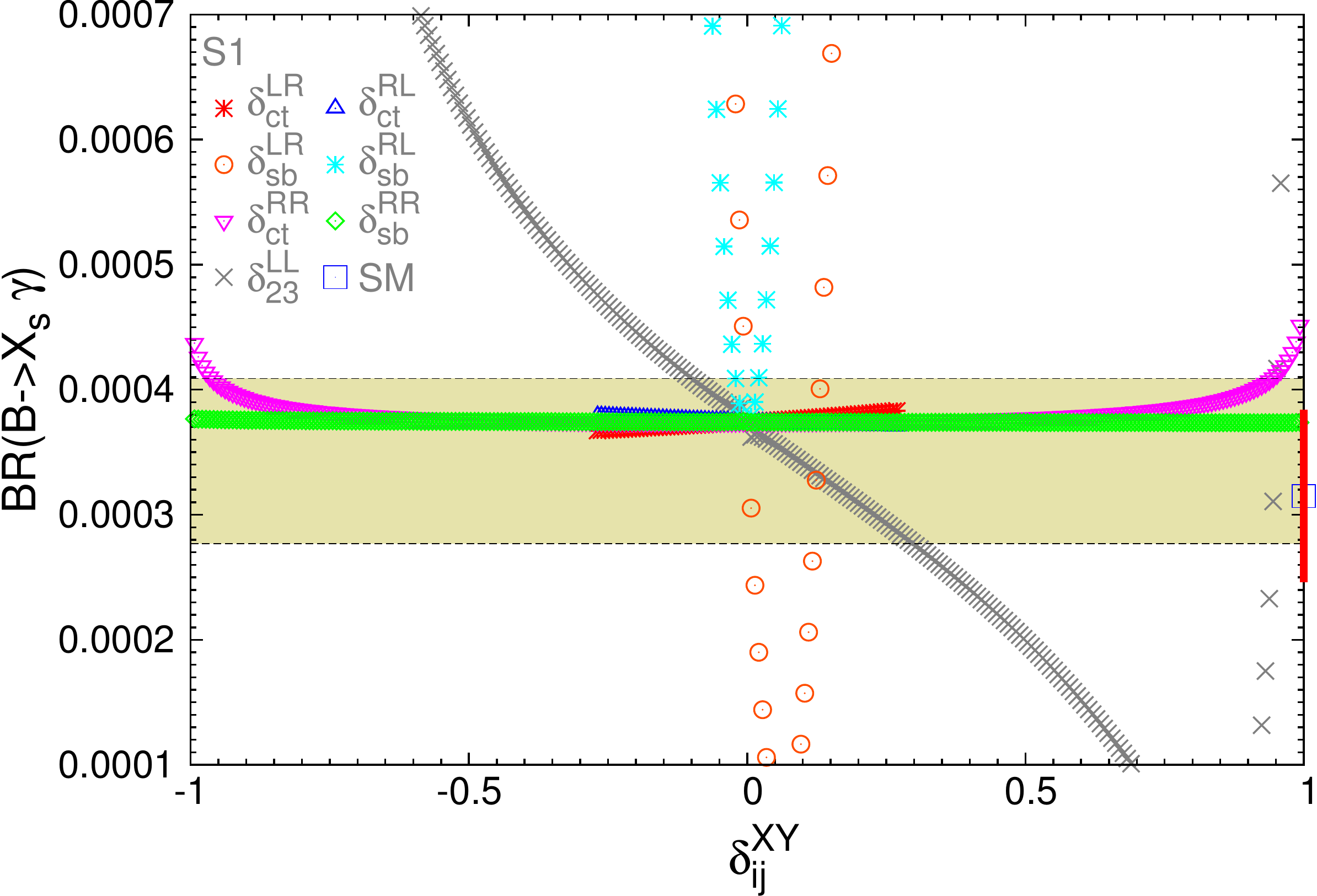}& 
\includegraphics[width=13.2cm,height=17.2cm,angle=270]{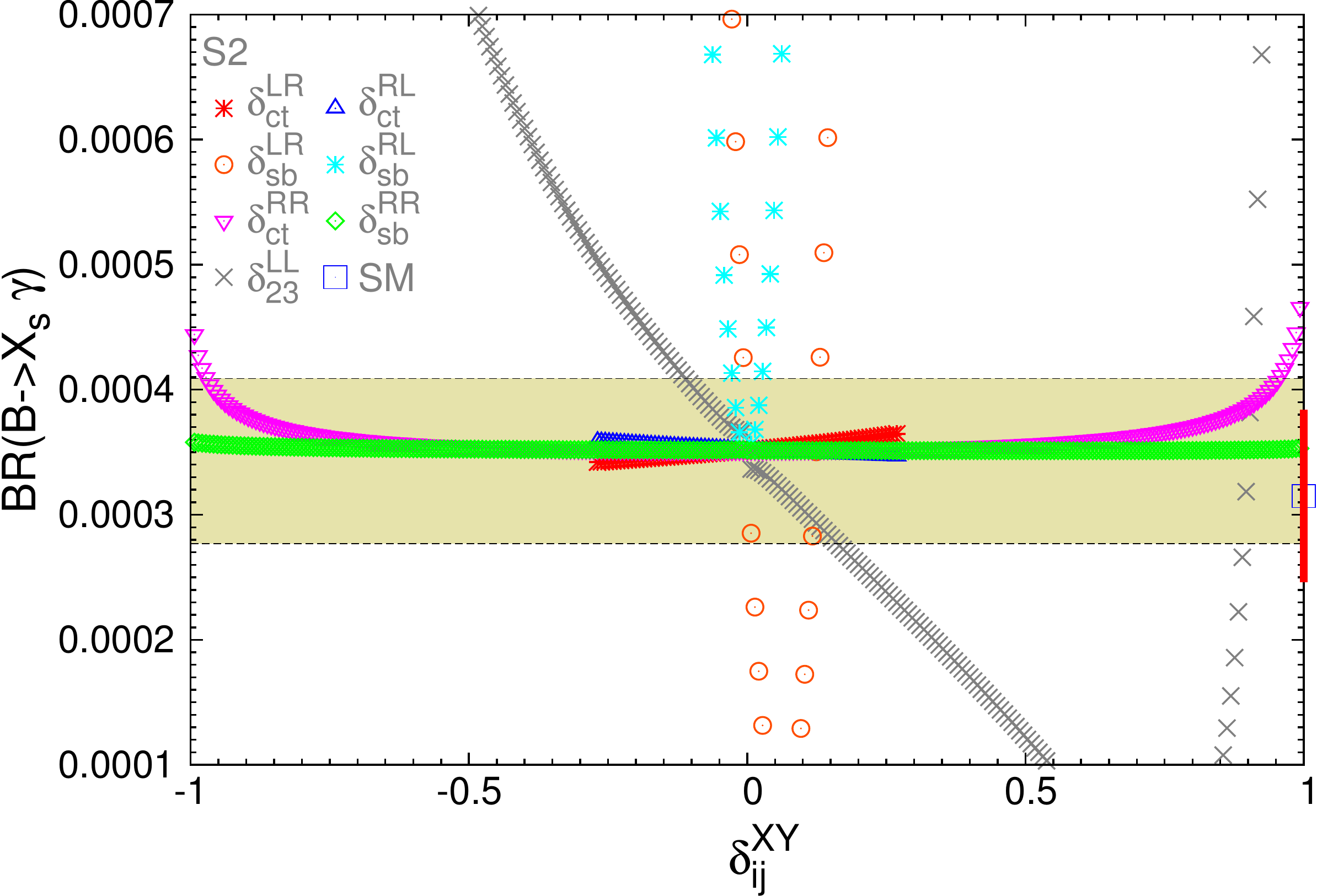}\\ 
\includegraphics[width=13.2cm,height=17.2cm,angle=270]{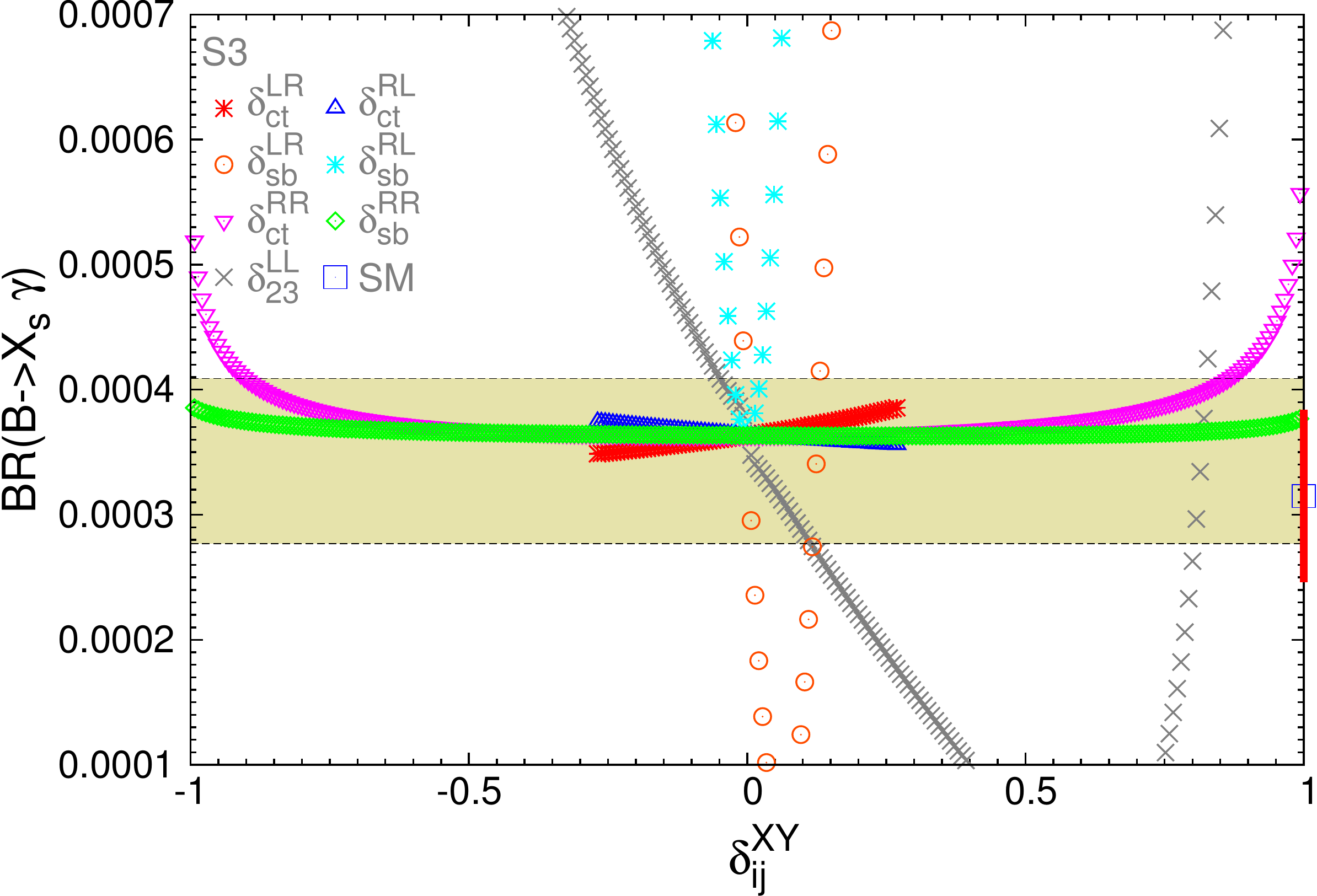}&
\includegraphics[width=13.2cm,height=17.2cm,angle=270]{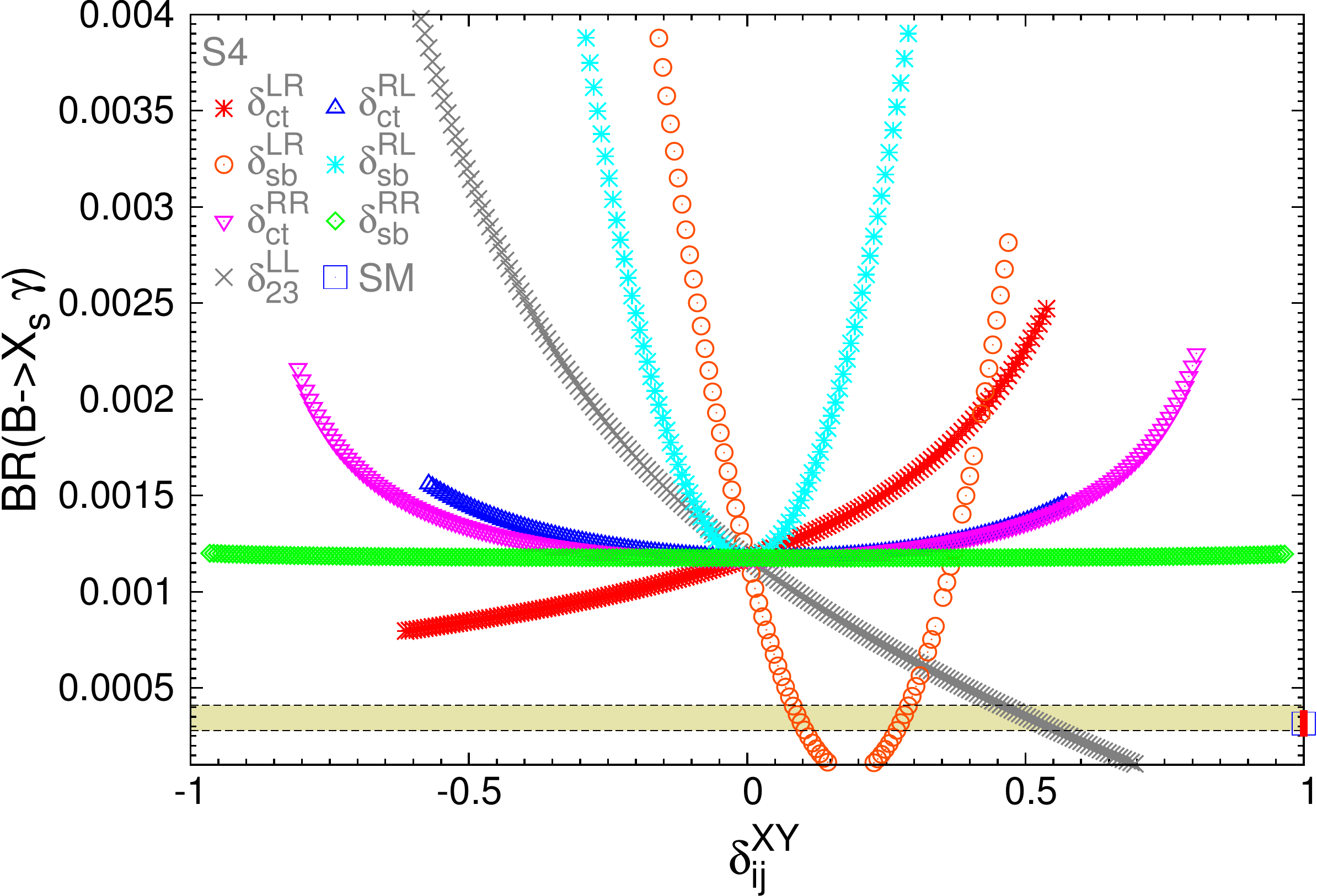}\\ 
\includegraphics[width=13.2cm,height=17.2cm,angle=270]{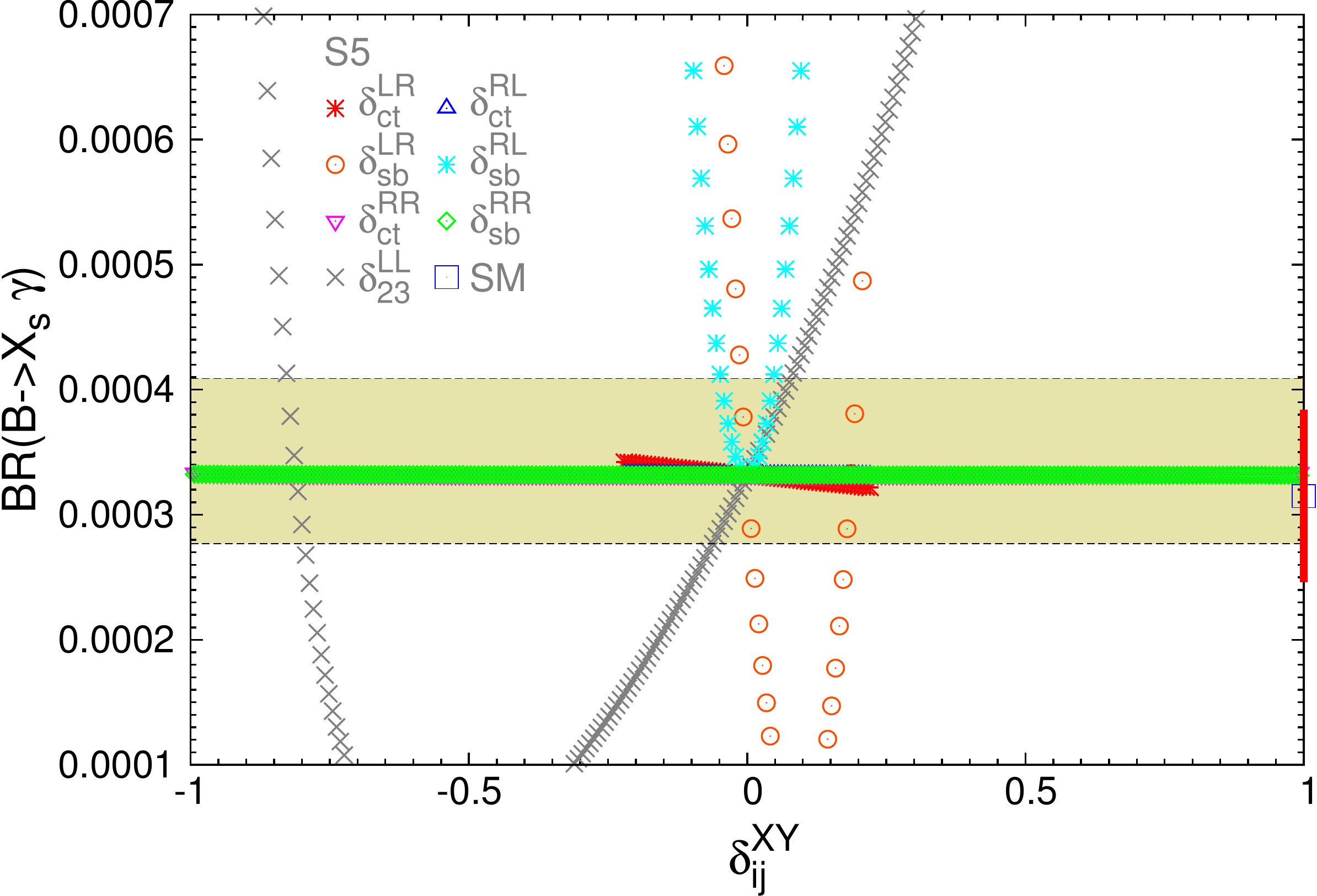}& 
\includegraphics[width=13.2cm,height=17.2cm,angle=270]{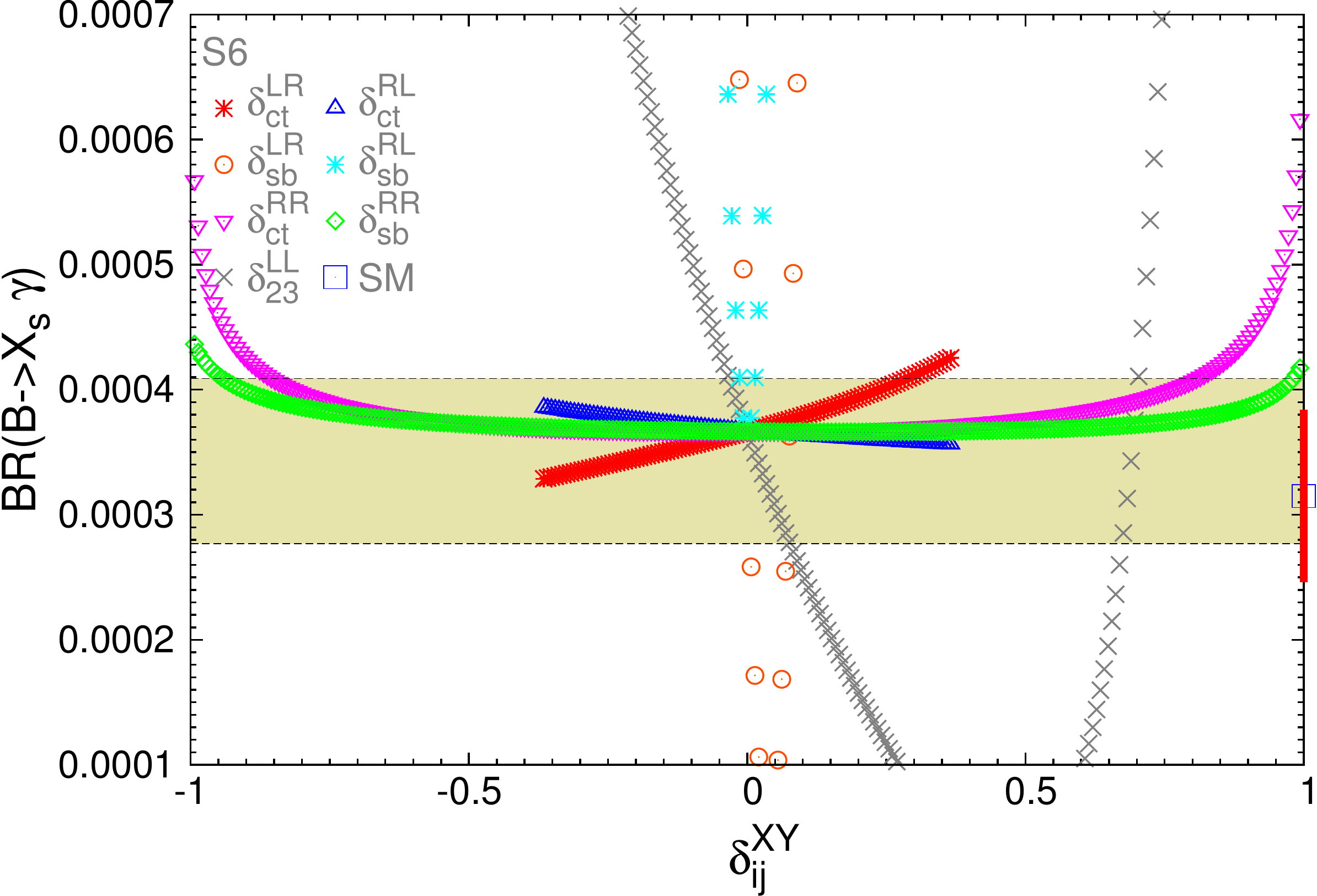}\\ 
\end{tabular}}}
\caption{Sensitivity to the NMFV deltas in \bsg\ for the six post-LHC scenarios of table \ref{tab:spectra}. The experimental allowed $3\sigma$ area is the horizontal
  coloured band. The SM prediction and the theory uncertainty $\Dtheo(\bsg)$
  (red bar) are displayed on the right axis.}  
\label{sfigbsgamma}
\end{figure}

\begin{figure}[h!] 
\centering
\hspace*{-8mm} 
{\resizebox{17.3cm}{!} 
{\begin{tabular}{cc} 
\includegraphics[width=13.2cm,height=17.2cm,angle=270]{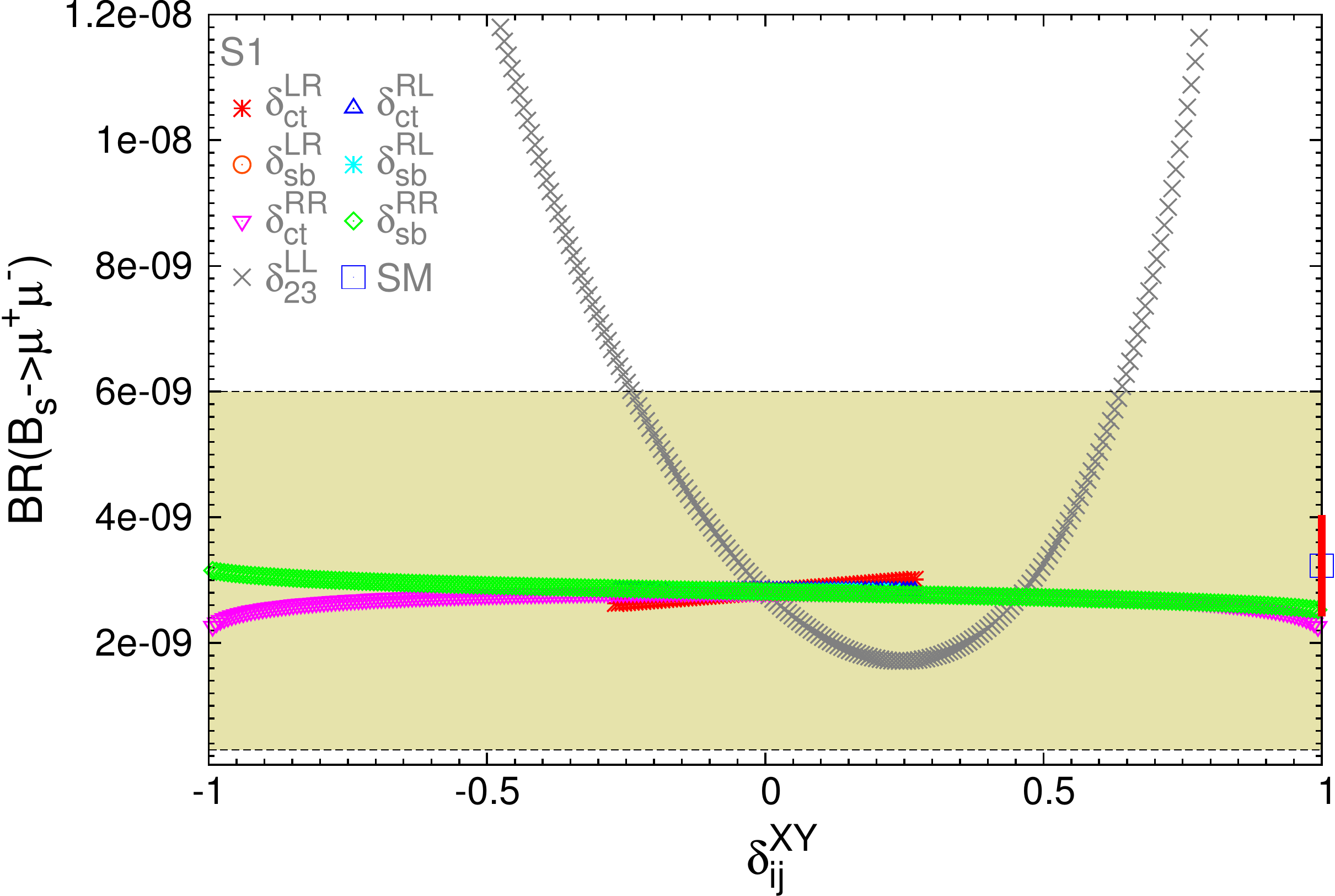}& 
\includegraphics[width=13.2cm,height=17.2cm,angle=270]{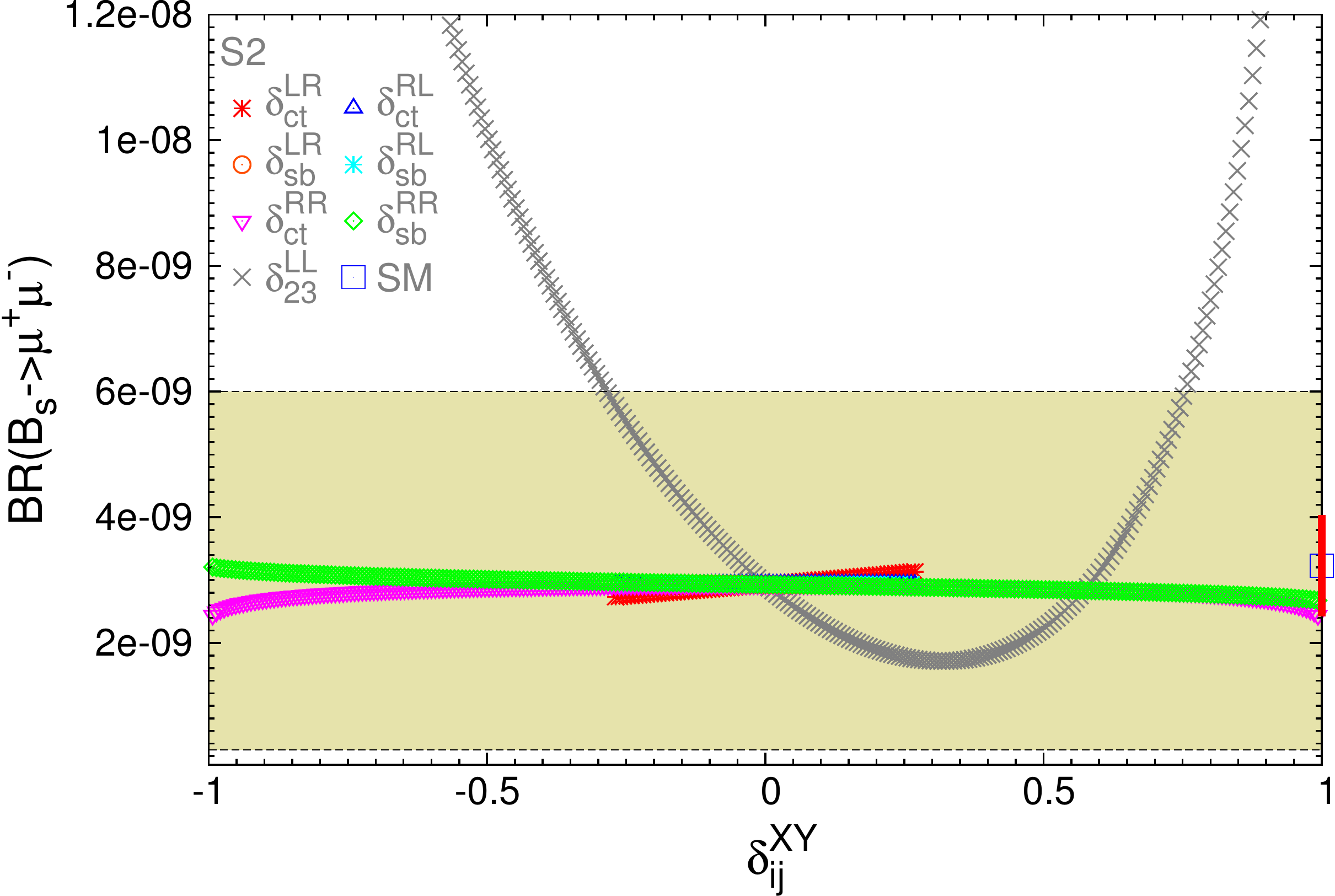}\\ 
\includegraphics[width=13.2cm,height=17.2cm,angle=270]{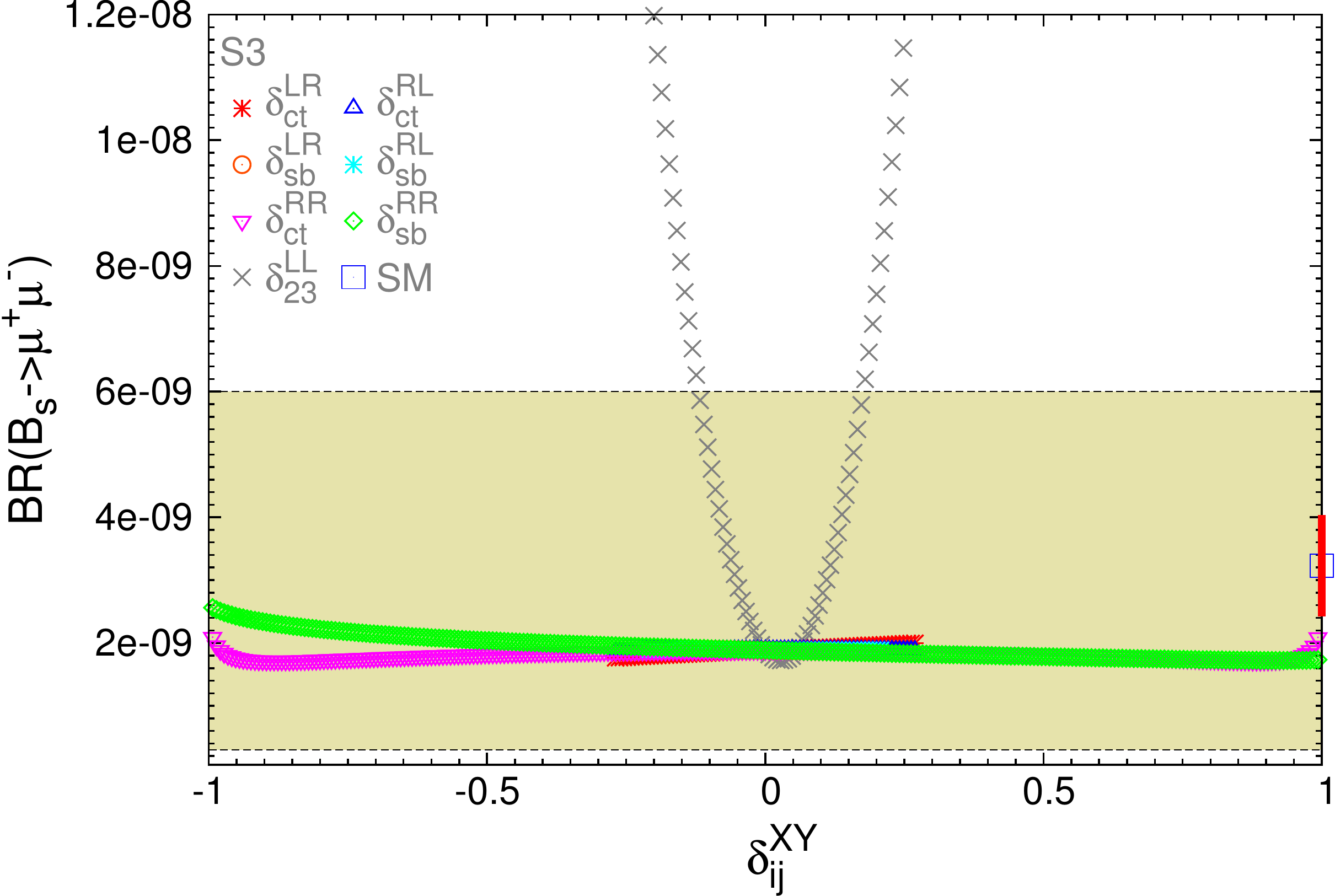}&
\includegraphics[width=13.2cm,height=17.2cm,angle=270]{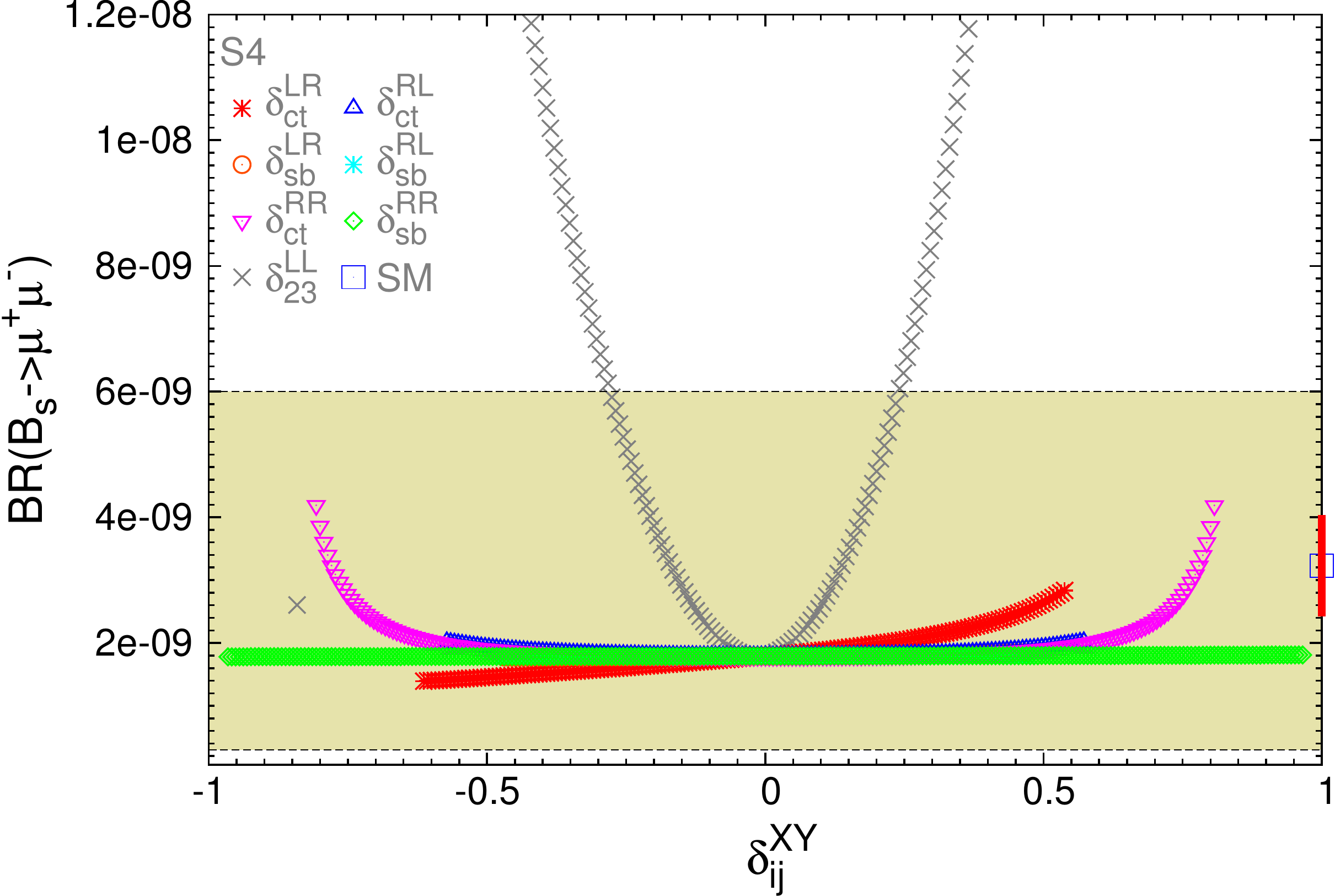}\\ 
\includegraphics[width=13.2cm,height=17.2cm,angle=270]{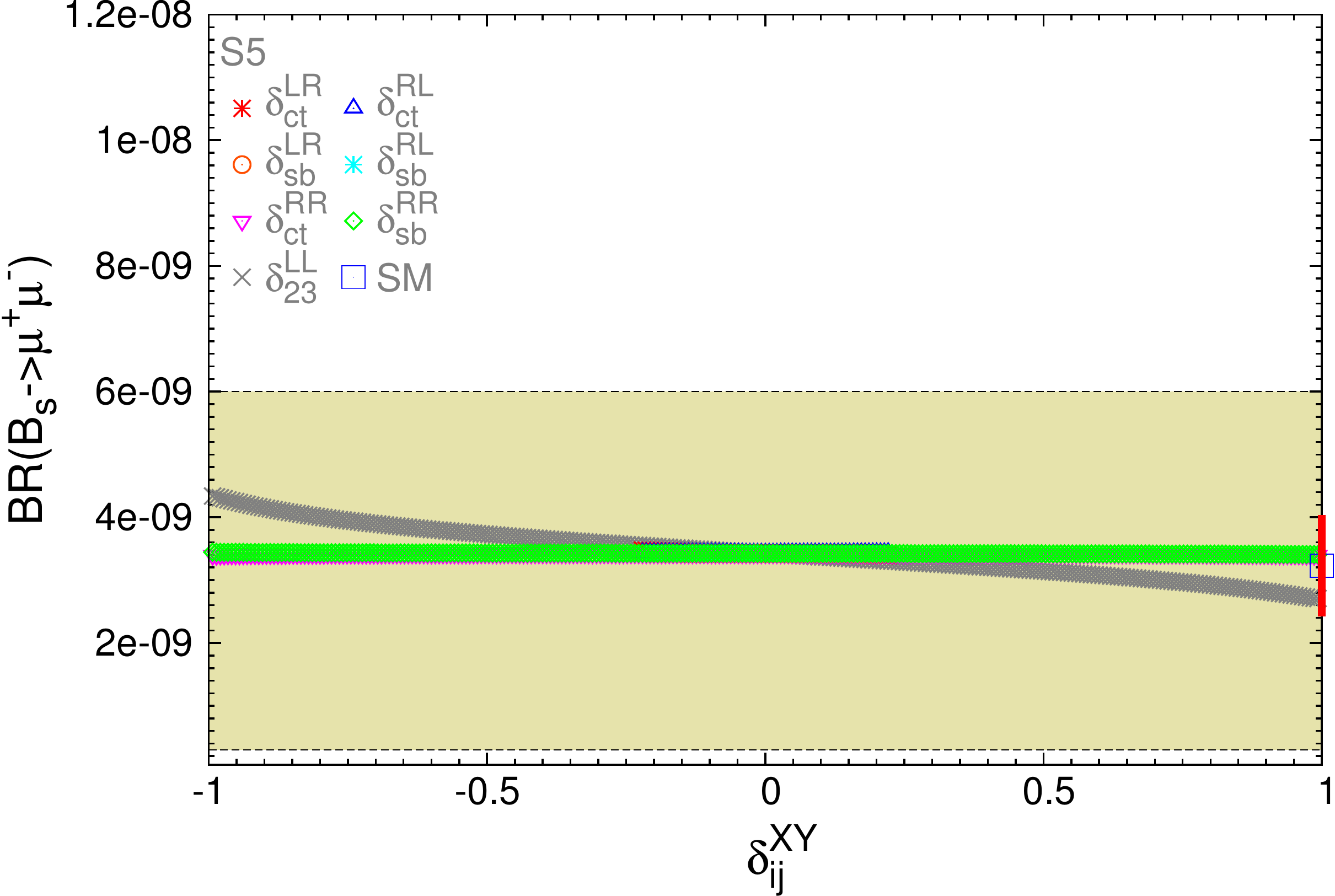}& 
\includegraphics[width=13.2cm,height=17.2cm,angle=270]{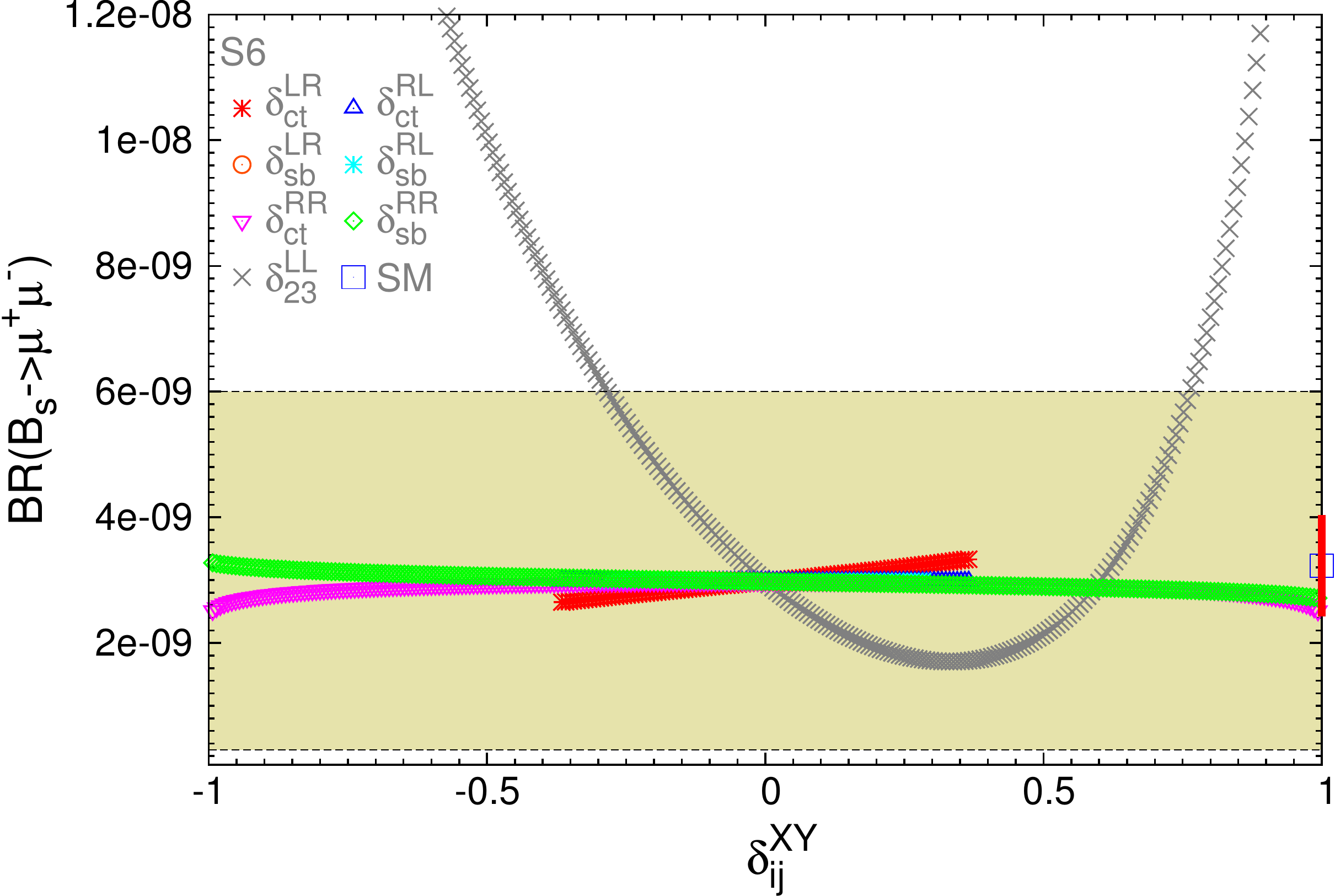}\\ 
\end{tabular}}}
\caption{Sensitivity to the NMFV deltas in ${\rm BR}(B_s \to \mu^+ \mu^-)$ for the six post-LHC scenarios of table \ref{tab:spectra}. The experimental allowed $3\sigma$ area is the horizontal  coloured band. The SM prediction and the theory uncertainty $\Dtheo(\bmm)$
  (red bar) are displayed on the right axis.}   
\label{sfigbmumu}
\end{figure}

\begin{figure}[h!] 
\centering
\hspace*{-8mm} 
{\resizebox{17.3cm}{!} 
{\begin{tabular}{cc} 
\includegraphics[width=13.2cm,height=17.2cm,angle=270]{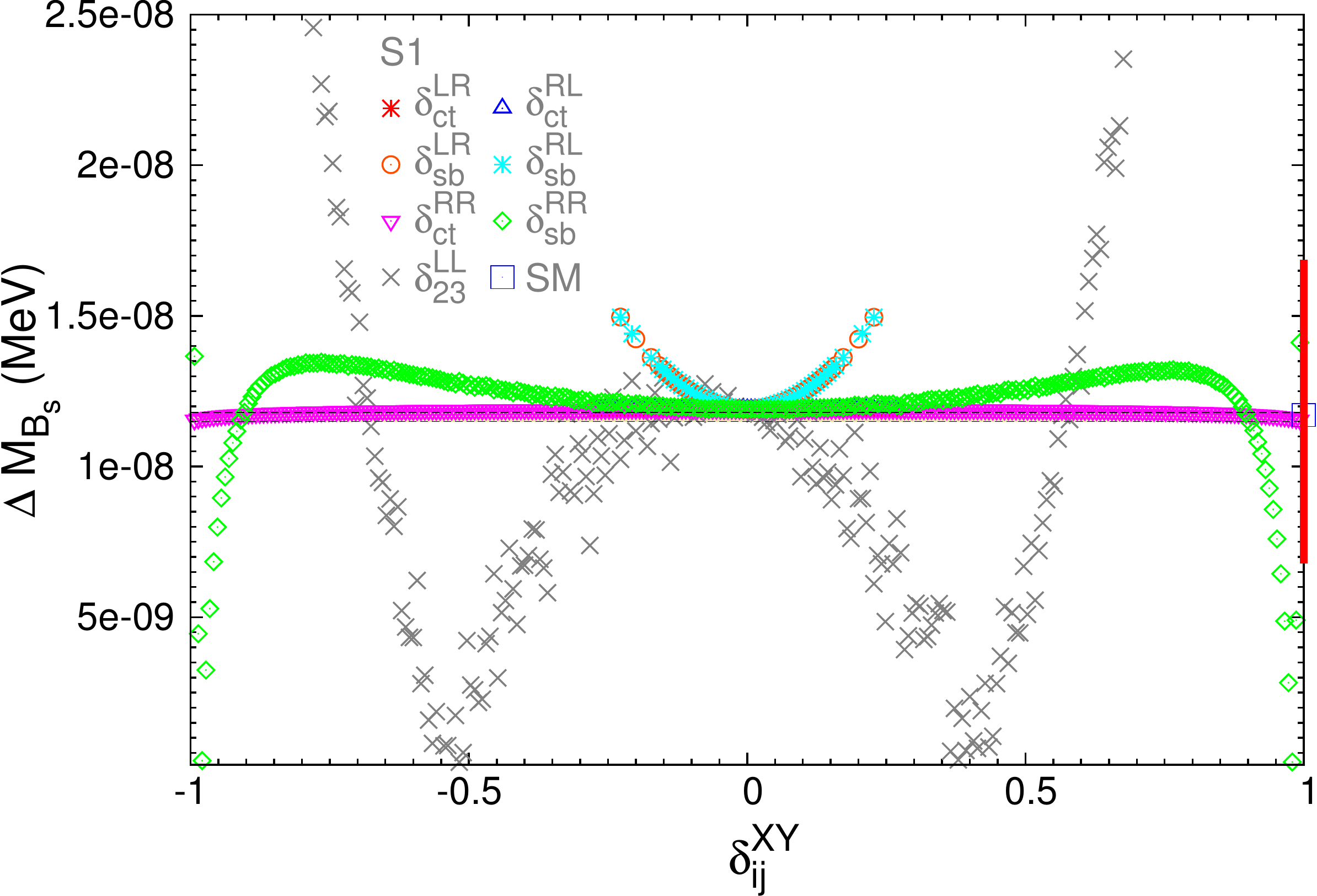}& 
\includegraphics[width=13.2cm,height=17.2cm,angle=270]{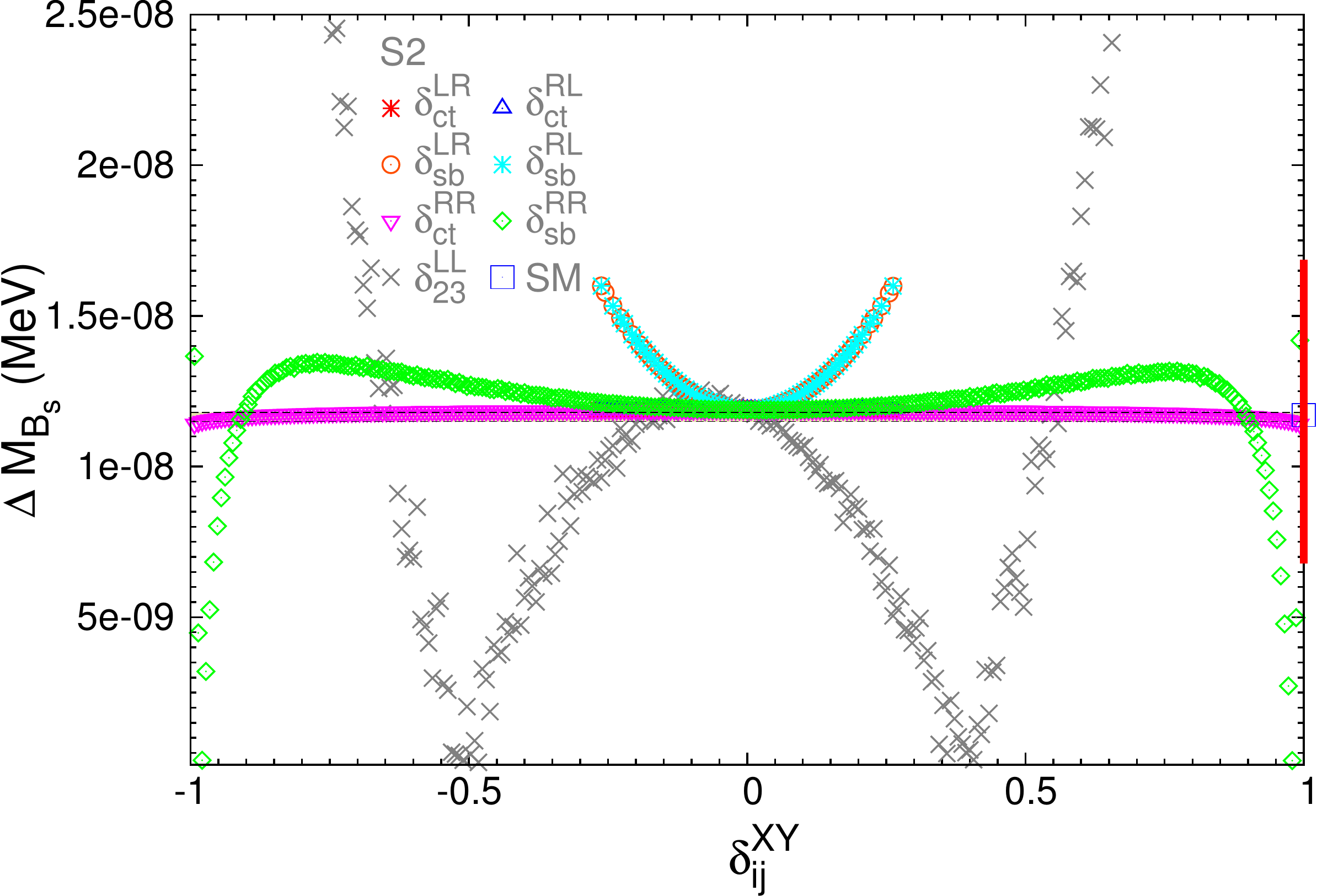}\\ 
\includegraphics[width=13.2cm,height=17.2cm,angle=270]{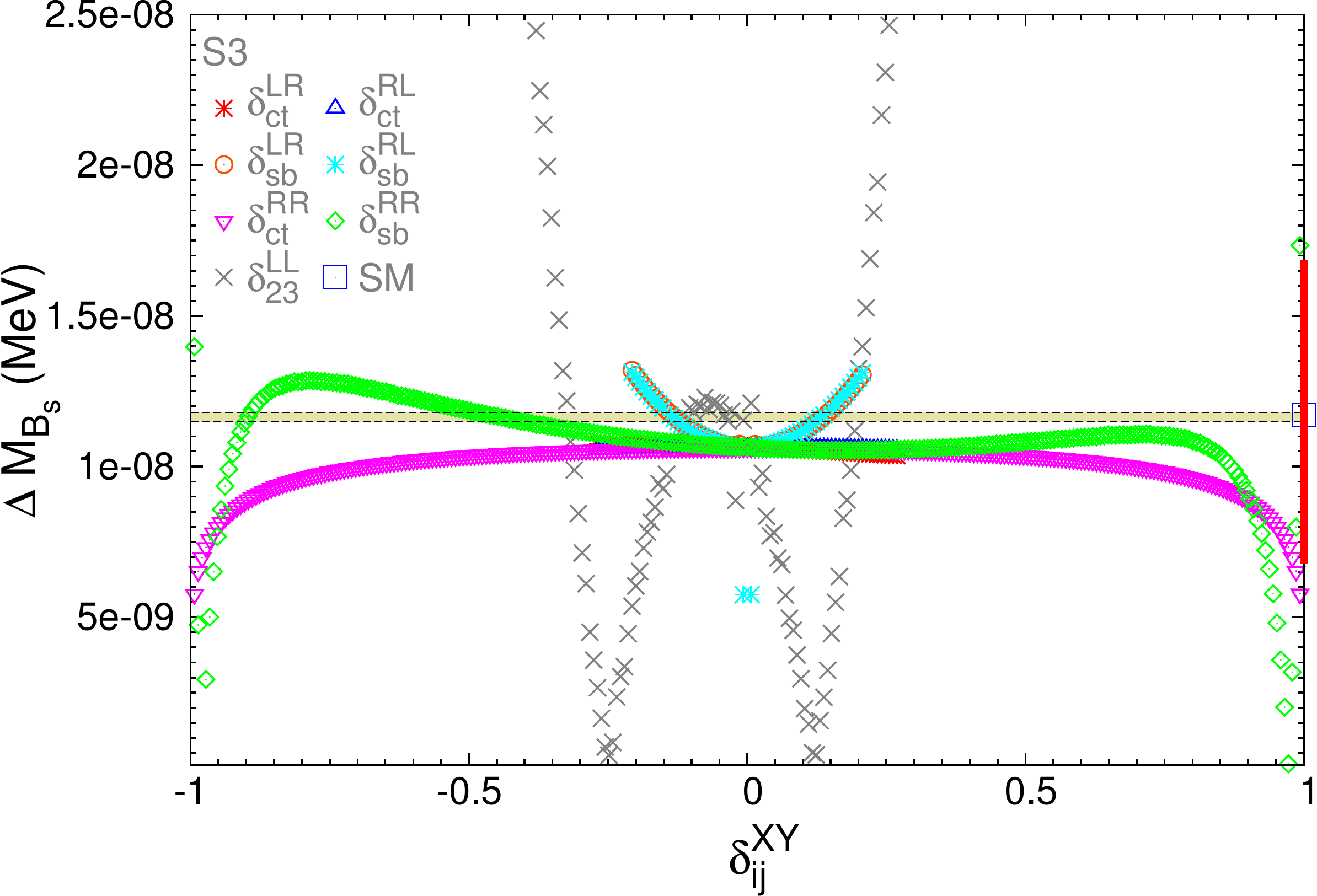}&
\includegraphics[width=13.2cm,height=17.2cm,angle=270]{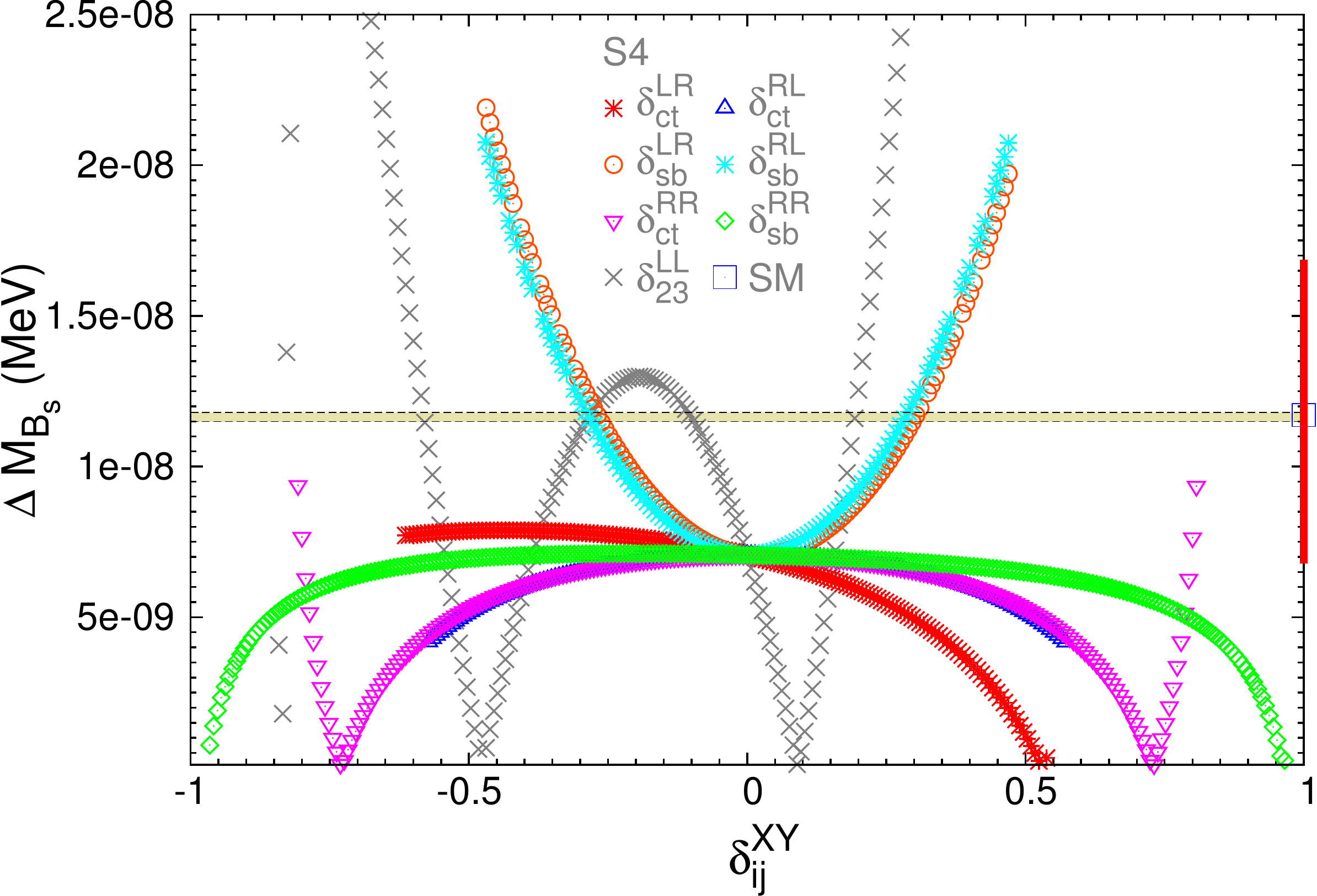}\\ 
\includegraphics[width=13.2cm,height=17.2cm,angle=270]{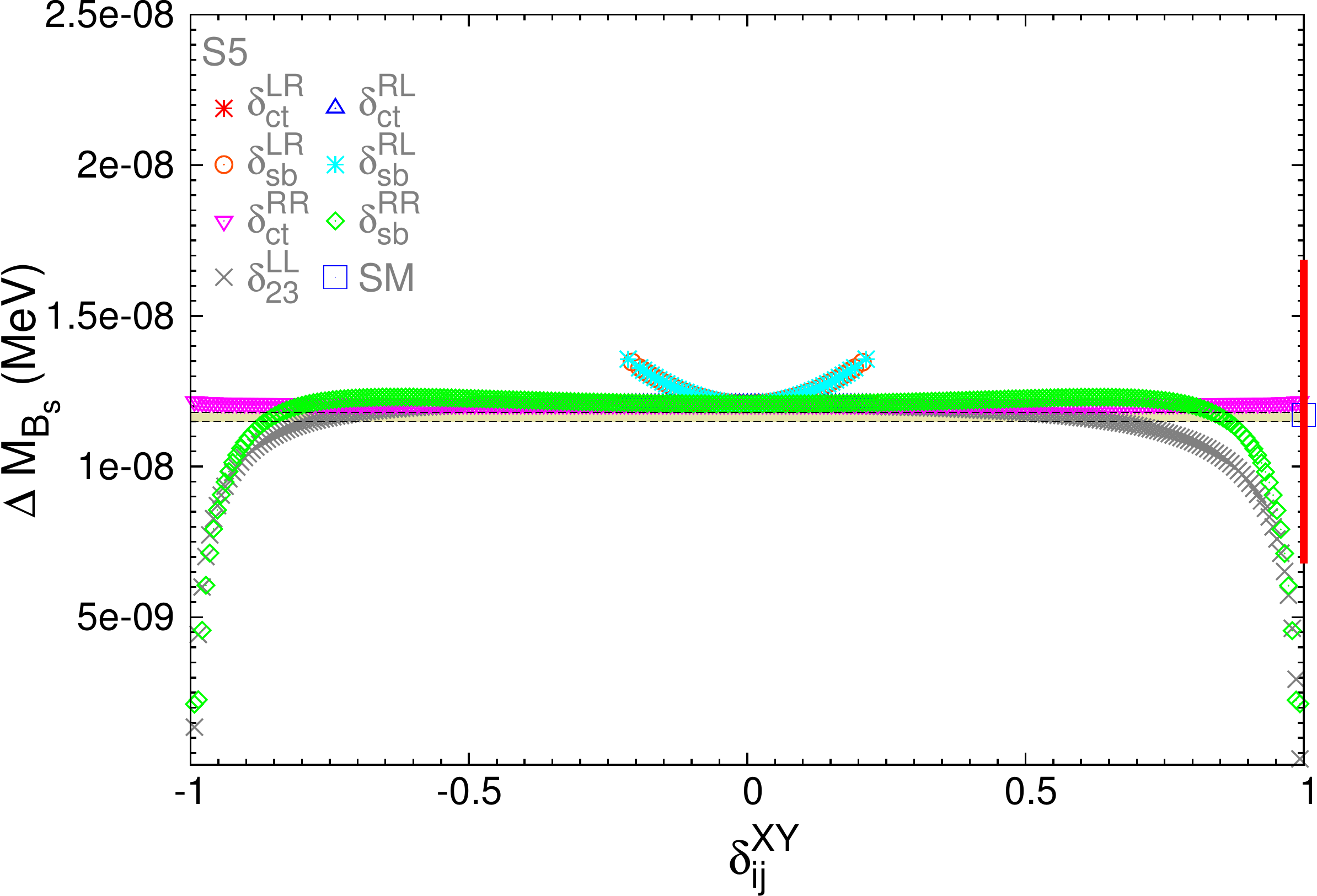}& 
\includegraphics[width=13.2cm,height=17.2cm,angle=270]{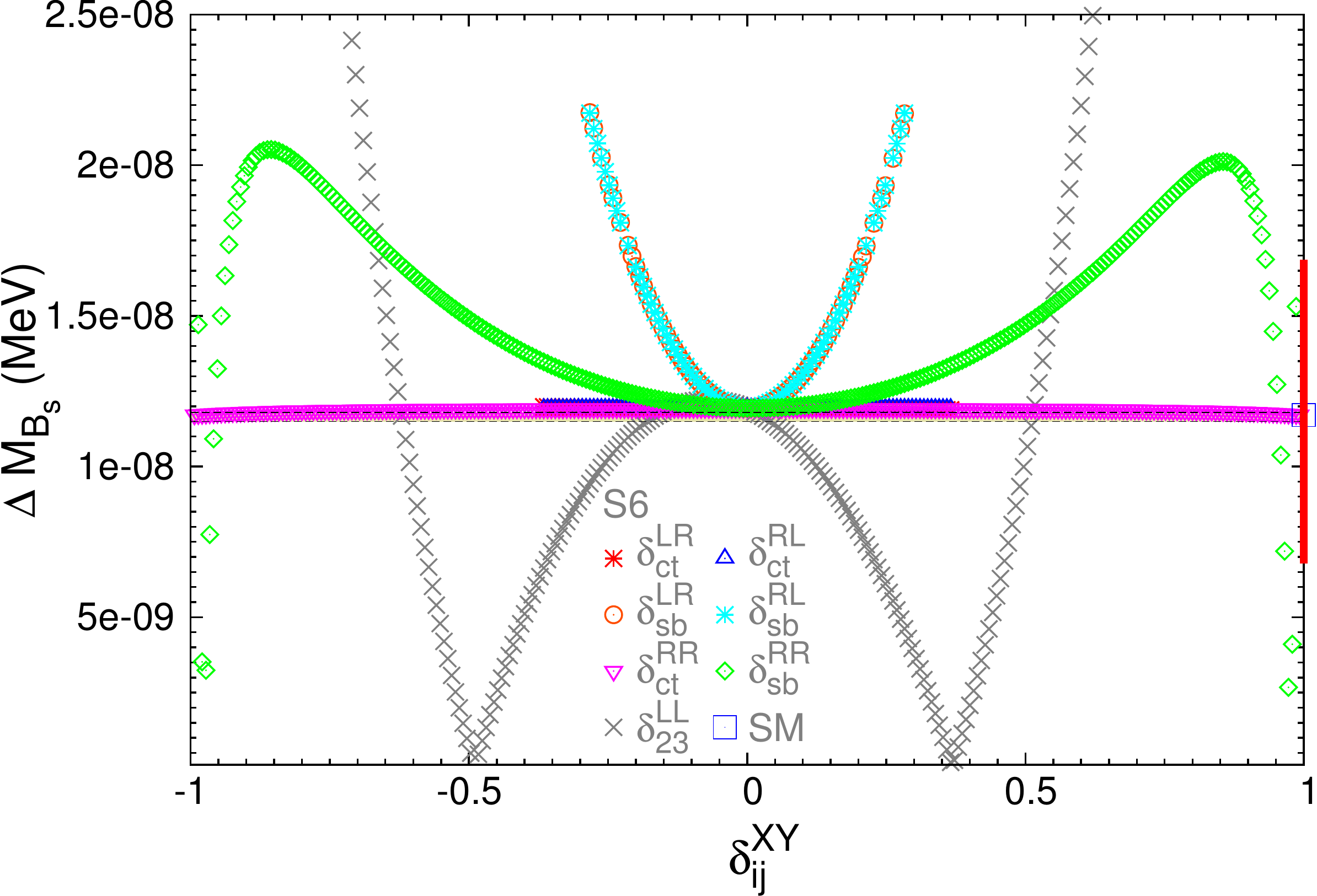}\\ 
\end{tabular}}}
\caption{Sensitivity to the NMFV deltas in $\Delta M_{B_s}$ for the six post-LHC scenarios of table \ref{tab:spectra}. The experimental allowed $3\sigma$ area is the horizontal coloured band. The SM prediction and the theory uncertainty $\Dtheo(\dmbs)$
  (red bar) are displayed on the right axis.}  
\label{sfigdeltams}
\end{figure}

In figures  \reffis{sfigbsgamma}, \ref{sfigbmumu} and \ref{sfigdeltams} the predictions for \bsg, \bmm\ and \dmbs\ are shown versus the different $\deXYij$, for the six post-LHC scenarios of Section \ref{frameworkc}.  It should be noted that again some  of the predicted lines in these plots do not expand
along the full interval $-1<\deXYij<1$, and they are restricted to a smaller interval. As previously, for some regions of the parameter space a too large delta value can generate very large corrections to the mass of the light Higgs boson, and its mass squared turns negative. These problematic points are consequently not included in our plots.    

\medskip
The present experimental values and the SM prediction for the three observables in the post-LHC period are the following:

\medskip
For \bsg\ the experimental measurement of this observable \cite{hfag:rad} (where we have added the various contributions to the experimental error in quadrature), and its prediction within the SM \cite{Misiak:2009nr} (that we take the same as in Eq. \ref{bsgamma-SM}) are: 

\noindent \begin{equation}
\bsg_{\rm exp}=(3.43 \pm 0.22)\times10^{-4}
\label{bsgamma-exp2}
\end{equation}

\noindent \begin{equation}
\bsg_{\rm SM}= (3.15 \pm 0.23)\times10^{-4}
\label{bsgamma-SM2}
\end{equation}

 In the case of \bmm\ there is indeed at present an experimental measurement \cite{Chatrchyan:2013bka}\cite{Aaij:2013aka} instead of the previous upper bound. This measurement and the present prediction within the SM \cite{Buras:2012ru} are: 
\noindent \begin{equation}
\bmm_{\rm exp} = (3.0^{+1.0}_{-0.9})\times 10^{-9}
\label{bsmumu-exp2}
\end{equation}

\noindent \begin{equation}
\bmm_{\rm SM}= (3.23\pm 0.27)\times 10^{-9}
\label{bsmumu-SM2}
\end{equation}

Finally, the experimental measurement \cite{hfag:pdg} of \dmbs\ (we have again added the various contributions to the experimental error in quadrature), and its prediction within the SM (that we take the same as in Eq. \ref{deltams-SM}) are:
\noindent \begin{align}
\label{deltams-exp2}
{\dmbs}_{\rm exp} &= (116.4 \pm 0.5) \times 10^{-10} \mev~, \\
{\dmbs}_{\rm SM} &= (117.1^{+17.2}_{-16.4}) \times 10^{-10} \mev~.
\label{deltams-SM2}
\end{align} 

We have included in the right vertical axis of the figures, for
comparison, the respective SM prediction for each observable. The error bars displayed are
the corresponding SM uncertainties as explained above, expanded with 3${\sigma}_{\rm SM}$ errors. 
The shadowed horizontal bands are the experimental measurements, expanded with
3${\sigma}_{\rm exp}$ errors.


\bigskip
The main conclusions extracted from these figures for the three $B$
observables are summarized as follows: 

\begin{itemize}
\item \bsg:
\begin{itemize}
\item[-] Sensitivity to the various deltas:

We find again strong sensitivity to $\delta^{LR}_{sb}$, $\delta^{RL}_{sb}$, $\delta^{LL}_{23}$, in all the studied points, for all the studied $\tb$ values. Some sensitivity to $\delta^{LR}_{ct}$ and $\delta^{RL}_{ct}$ can also be noticed. In the scenario S4 it can be found a significant sensitivity to all the deltas, except to $\delta^{RR}_{sb}$. This scenario has the lowest mass value for the squarks of the third generation and the trilinear of the top compared with the other scenarios, and large $\tb$; so it is reasonable to expect a stronger general effect since it decouples less than the others and has a large $\tb$ that is very relevant for these b-type observables. 
The order found from the highest to the lowest sensitivity is, generically:  1) $\delta^{RL}_{sb}$ and $\delta^{LR}_{sb}$ the largest, 2)   $\delta^{LL}_{23}$ the next, 3) the others $\deXYij$.

In contrast with the pre-LHC SUSY points studied in \ref{numresframeworka}, now we do not find here large sensitivity with respect to $\delta^{RR}_{sb}$ and $\delta^{LR}_{ct}$. It can be seen also that all the curves are less steep than these previous ones. This is also not a surprising effect since having a heavier SUSY spectrum in these post-LHC scenarios tends to soften the effects of SUSY particles and the predicted rates are closer to the SM ones.

\item[-] Comparing the predictions with the experimental $B$ data:
The deltas with the largest impact on the observable $\delta^{LR}_{sb}$ and $\delta^{RL}_{sb}$ take us out of the experimental allowed band already with tiny values for the deltas of ${\cal O}$ (0.01). This happens also for $\delta^{LL}_{23}$, but with values that can reach the half of the area studied depending on the point.
The U shape of the predicted branching ratio with some deltas make it possible for the observable to be in agreement with the experimental values for two disconnected regions. For the case of the $\delta^{LL}_{23}$ it can be seen that one of these regions lays around the maximum value considered for this $|\deXYij|$.
The rest of the deltas do not get relevant constraints, except $\delta^{RR}_{ct}$ for extreme large values close to $\pm 1$.

In the case of the point S4 whose MFV case (i.e. all deltas equal to zero) is excluded by the observable, it can be seen how non-zero values for  $\delta^{LR}_{sb}$ and $\delta^{LL}_{23}$ can drive us again into the allowed area. 

The fact that some excluded MFV points of the SUSY parameter space can be allowed in the NMFV case is an important lesson for understanding the new limits of our theory.

Comparing with the pre-LHC situation it can be seen that although the experimental band for $B$ decays became a bit narrower, the observable turned less restrictive, specially for some of the deltas whose constraints disappeared. The reason again is because in this post-LHC situation the SUSY spectrum is heavier.

\item[-] Intervals of $\deXYij$  allowed by data:

To conclude on the allowed delta intervals we assume that our
predictions of \bsg\ have a somewhat
larger theoretical error $\Dtheo(\bsg)$ than the SM prediction 
$\Dtheo_{\rm SM}(\bsg)$
given in (\ref{bsgamma-SM2}). As a very conservative value we use
$\Dtheo(\bsg) = 0.69 \times 10^{-4}$. 
A given $\deXYij$ value is then considered to be allowed
by data if the corresponding interval, defined by $\bsg \pm \Dtheo(\bsg)$,
intersects with the experimental band. It corresponds to 
adding linearly the experimental uncertainty and the MSSM theoretical
uncertainty. In total a predicted ratio in the
interval  
\noindent \begin{align}
\label{bsglinearerr2}
2.08\times 10^{-4} < \bsg\ < 4.78\times 10^{-4},
\end{align} 
is now regarded as allowed. 
Our results for these allowed intervals are summarized in the third column of table
\ref{tableintervals2}. The less constrained parameters are $\delta^{RR}_{sb}$ and $\delta^{RR}_{ct}$ as we saw when looking at the sensitivity. These deltas can reach $|\deXYij|$ of ${\cal O}(1)$. 
With respect to the pre-LHC situation the constraints on $\delta^{RR}_{sb}$ have almost disappeared.

\end{itemize}
\end{itemize}

\begin{itemize}
\item
\bmm:

\begin{itemize}
\item[-] Sensitivity to the various deltas:

We find only a relevant sensitivity to $\delta^{LL}_{23}$. Although large $\tb$ scenarios were studied, that we saw in the pre-LHC scenarios enlarged the sensitivity to the deltas, the fact that we are now setting also large $M_A$ (because of the bounds of the recent CMS searches ~\cite{CMS-PAS-HIG-12-050} for MSSM neutral Higgs bosons decaying to 
$\tau {\bar \tau}$ pairs in the so-called $m_h^\text{max}$ scenario \cite{Carena:2002qg}\cite{Carena:2013qia}) counteracts the enhancing effect of the large $\tb$. As we commented before, this sensitivity is due to the Higgs-mediated contribution growing as $\tan^6\beta$ and decreasing as $M_A$ grows. The sensitivity to the $\delta^{LL}_{23}$ disappears in the low $\tb$ scenario S5 as expected. Some sensitivity to $\delta^{RR}_{ct}$ is found in scenario S4 for large values of the delta. We recall that for some of the parameters this is the less heavy of the scenarios considered now. 
  
\item[-] Comparing the predictions with the experimental data:

The experimental data changed drastically in the last time since it went from having a bound to the observable to finally getting a measurement. However, since SUSY also got more constrained, specially the presently excluded low values of $M_A$ for the large $\tb$ case (that are the most relevant parameters here), the situation did not improve a lot respect to the pre-LHC situation. It got a bit more restrictive for some of the points in the case of $\delta^{LL}_{23}$, but now we have no constraints at all for the rest of the parameters coming from this observable. 

\item[-] Intervals of $\deXYij$  allowed by data:

We assume here that our predictions for 
\bmm\ within SUSY scenarios have a slightly larger error as the SM prediction in
(\ref{bsmumu-SM2}). Still the theory uncertainty is very small in
comparison with the experimental errors. We choose 
$\Dtheo(\bmm) = 0.81 \times 10^{-9}$. 
Then, a given $\deXYij$ value is allowed by data if the predicted interval,
defined by $\bmm \pm \Dtheo(\bmm)$, intersects
the experimental area. To be allowed, a point should lay in the following interval: 
\noindent \begin{align}
\label{bsmmlinearerr2}
\bmm\ < 6.81 \times 10^{-9}.
\end{align} 

The results for the allowed $\deXYij$ intervals are collected in the fourth column of table \ref{tableintervals2}. Values of $|\deXYij|$ of ${\cal O}(1)$ compatible with \bmm\ data can be achieved for all the deltas, except for $\delta^{LL}_{23}$ that get reduced to ${\cal O}(0.1)$.
\end{itemize}
\end{itemize}

\begin{itemize}
\item \dmbs:

\begin{itemize}
\item[-] Sensitivity to the various deltas:

The main sensitivity is found again for $\delta^{LR}_{sb}$, $\delta^{RL}_{sb}$ and $\delta^{LL}_{23}$. Some sensitivity is found also for the $\delta^{RR}_{sb}$ parameter, specially in the S6 scenarios where it is very relevant. For the scenario S4 it can be seen a relevant sensitivity to all the deltas. In contrast, in the scenario S5 the variation of the observable with respect to all the deltas almost disappear.
As we commented in the pre-LHC scenarios, different loop contributions are relevant for the observable, as shown in \reffi{fig:deltabs-allcontributions}, depending on the value of $\tb$. Comparing with these other scenarios, the general behaviour with the deltas is the same, although the general variation is a bit milder in this case. 

\item[-] Comparing the predictions with the experimental data:

Being now smaller the variation with the deltas, one profits however from the experimental allowed band that has been reduced even more in the last time, but unfortunately the theoretical errors for the predicted points are still huge (assuming than the SUSY errors are at least as large as the SM ones). Except for the case of the S4 scenario, that lies at the border of the allowed area, for the rest of the scenarios we get constraints from this observable only for $\delta^{LL}_{23}$, and large values of $\delta^{RR}_{sb}$. Except for  $\delta^{LL}_{23}$, we get softer bounds in this current allowed scenarios, as compared to the previous ones.

\item[-] Intervals of $\deXYij$  allowed by data:

As done previously, we consider that a given $\deXYij$ value is allowed by \dmbs\
data if the predicted interval $\dmbs \pm \Dtheo(\dmbs)$, intersects the  
experimental band. It corresponds to adding linearly the experimental uncertainty 
and the theoretical uncertainty. As we did before, and given the controversy on the realistic 
size of the theoretical error in the estimates of $\Dtheo(\dmbs)$ in the MSSM (see, for instance,~\cite{refLunghi}), 
we choose a very conservative value for the theoretical error in our estimates, 
considerably larger than the SM value in (\ref{deltams-SM2}), of
$\Dtheo(\dmbs)=51 \times 10^{-10}$ MeV.  This  
implies that a predicted mass difference in the interval
\begin{align}
\label{deltabslinearerr2}
63.9\times 10^{-10} < \dmbs\ {\rm (MeV)} < 168.9\times 10^{-10},
\end{align} 
is regarded as allowed.

The allowed intervals for the deltas are collected in the fifth column of table \ref{tableintervals2}. The main restrictions are set to the sb-type deltas as it was expected and to $\delta^{LL}_{23}$. We observe again for some scenarios the appearance of disconnected allowed regions due to the form of the variation with respect to the deltas.
Deltas of ${\cal O}(0.1)$ are allowed for all parameters, and reach up to ${\cal O}(1)$ for the $RR$ delta types.

\end{itemize}

\end{itemize}

\clearpage
\newpage

\subsubsection*{Total present allowed $\deXYij$ intervals and comparison with the pre-LHC situation}

\vspace{0.5cm}
To conclude this section, we collect in table \ref{tabdeltasummary2} the total allowed intervals for all the flavour parameters of the squark sector in the selected LHC points, being compatible with \bsg, ${\rm BR}(B_s \to \mu^+ \mu^-)$ and $\Delta M_{B_s}$.


The main restrictions to the flavour parameters come from \bsg\ and in some cases from $\Delta
M_{B_s}$ and not yet from the young measurement ${\rm BR}(B_s \to \mu^+ \mu^-)$.
The most restricted deltas are $\delta^{LR}_{sb}$ and $\delta^{RL}_{sb}$ that can reach $|\deXYij|$ of ${\cal O}(0.01)$, then $\delta^{LL}_{23}$, $\delta^{LR}_{ct}$ and $\delta^{RL}_{ct}$ can be one order of magnitude larger, with the first being more restricted than the last two, and finally the less restricted are $\delta^{RR}_{ct}$ and $\delta^{RR}_{sb}$ than in general could reach a size of ${\cal O}(1)$. 

We remark again the interesting fact that some of the presently excluded MFV points can be driven to the allowed area of the observables, by setting some of the deltas to a non-zero value.

It was to be expected to get larger constraints in the $sb$ sector since it is the one directly involved in the flavour changing observables we are considering. And it is also understandable that the $LR$ deltas become more restricted than the others, since they involve the trilinear couplings that are not just part of a mass term, but they enter straight in the relevant Feynman rules of the observables, and not only through the diagonalization of the mass matrices.

In general, the bounds on the squark mixing deltas are more relaxed in our scenarios S1-S6 compared to the set of benchmark scenarios that was analyzed before the LHC started operation. The scenarios investigated in the pre-LHC era contained relatively light SUSY particles, leading to relatively large corrections from NMFV effects. After the so far
unsuccessful search for BSM physics at the LHC, scalar quarks masses (in particular of the first and second generation) have substantially higher lower bounds. Benchmark scenarios that take this into account (as our S1-S6) naturally permit larger values for the NMFV deltas.

As we told in advance in the last section, once we know the allowed intervals of the flavour parameters these will be explored to learn the consequences they have in the radiative corrections to the just discovered Higgs boson.

\newpage
\renewcommand{\arraystretch}{1.2}
\begin{table}[H]
\begin{table}[H]
\begin{center}
\small
\resizebox{16cm}{!} {
\begin{tabular}{|c|c|c|c|c|} \hline
 & & ${\rm BR}(B\rightarrow X_s\gamma)$ & ${\rm BR}(B_s\rightarrow\mu^+\mu^-)$ & $\dmbs$  \\ \hline
$\delta^{LL}_{23}$ & \begin{tabular}{c} S1 \\ S2 \\ S3 \\ S4 \\ S5 \\ S6 \end{tabular} &  
\begin{tabular}{c}(-0.27:0.48) (0.94:0.95) \\ (-0.23:0.32) (0.88:0.91) \\ (-0.12:0.21) (0.79:0.83) \\ (0.41:0.61) \\ (-0.83:-0.78) (-0.14:0.14) \\ (-0.076:0.14) (0.66:0.71)
 \end{tabular}&  \begin{tabular}{c} (-0.28:0.66) \\ (-0.33:0.78) \\ (-0.13:0.19) \\ (-0.30:0.26) \\ (-0.99:0.99) \\ (-0.33:0.79) \end{tabular}&  \begin{tabular}{c} (-0.72:-0.63) (-0.41:0.28) (0.50:0.61) \\ (-0.69:-0.59) (-0.37:0.23) (0.50:0.59) \\ (-0.34:-0.30) (-0.19:0.06) (0.17:0.21) \\ (-0.61:-0.54) (-0.39:0.01) (0.15:0.23) \\ (-0.97:0.97) \\ (-0.66:-0.57) (-0.37:0.23) (0.46:0.56) \end{tabular}  \\ \hline
$\delta^{LR}_{ct}$  & \begin{tabular}{c} S1 \\ S2 \\ S3 \\ S4 \\ S5 \\ S6  \end{tabular}    
& \begin{tabular}{c} (-0.27:0.27) \\ (-0.27:0.27) \\ (-0.27:0.27) \\ excluded \\ (-0.22:0.22) \\ (-0.37:0.37) \end{tabular} & \begin{tabular}{c} (-0.27:0.27) \\ (-0.27:0.27) \\ (-0.27:0.27) \\ (-0.61:0.54) \\ (-0.22:0.22) \\ (-0.37:0.37)  \end{tabular} & \begin{tabular}{c} (-0.27:0.27) \\ (-0.27:0.27) \\ (-0.27:0.27) \\ (-0.61:0.13) \\ (-0.22:0.22) \\ (-0.37:0.37)
\end{tabular}  \\ \hline
$\delta^{LR}_{sb}$  & \begin{tabular}{c} S1 \\ S2 \\ S3 \\ S4 \\ S5 \\ S6 \end{tabular}    & 
\begin{tabular}{c} (-0.0069:0.014) (0.12:0.13) \\ (-0.0069:0.014) (0.11:0.13) \\ (-0.0069:0.014) (0.11:0.13) \\ (0.076:0.12) (0.26:0.30) \\ (-0.014:0.021) (0.17:0.19) \\ (0:0.0069) (0.069:0.076)
 \end{tabular} & \begin{tabular}{c} (-0.23:0.23) \\ (-0.26:0.26) \\ (-0.21:0.21) \\ (-0.47:0.47) \\ (-0.21:0.21) \\ (-0.28:0.28) \end{tabular} & \begin{tabular}{c} (-0.23:0.23) \\ (-0.26:0.26) \\ (-0.21:0.21) \\ (-0.39:0.42) \\ (-0.21:0.21) \\ (-0.20:0.20)
\end{tabular}  \\ \hline
$\delta^{RL}_{ct}$  & \begin{tabular}{c} S1 \\ S2 \\ S3 \\ S4 \\ S5 \\ S6 \end{tabular}    & 
\begin{tabular}{c}
(-0.27:0.27) \\ (-0.27:0.27) \\ (-0.27:0.27) \\ excluded \\ (-0.22:0.22) \\ (-0.37:0.37) \end{tabular} 
& 
\begin{tabular}{c} 
(-0.27:0.27) \\ (-0.27:0.27) \\ (-0.27:0.27) \\ (-0.57:0.57) \\ (-0.22:0.22) \\ (-0.37:0.37) \end{tabular} 
& 
\begin{tabular}{c}
(-0.27:0.27) \\ (-0.27:0.27) \\ (-0.27:0.27) \\ (-0.32:0.32) \\ (-0.22:0.22) \\ (-0.37:0.37) \end{tabular} \\ \hline
$\delta^{RL}_{sb}$  & \begin{tabular}{c} S1 \\ S2 \\ S3 \\ S4 \\ S5 \\ S6  \end{tabular}    &
\begin{tabular}{c} 
(-0.034:0.034) \\ (-0.034:0.034) \\ (-0.034:0.034) \\ excluded \\ (-0.062:0.062) \\ (-0.021:0.021) \end{tabular} 
& 
\begin{tabular}{c}(-0.23:0.23) \\ (-0.26:0.26) \\ (-0.21:0.21) \\ (-0.47:0.47) \\ (-0.21:0.21) \\ (-0.28:0.28)  \end{tabular} 
& 
\begin{tabular}{c}(-0.23:0.23) \\ (-0.26:0.26) \\ (-0.21:0.21) \\ (-0.40:0.40) \\ (-0.21:0.21) \\ (-0.20:0.20) \end{tabular}  \\ \hline
$\delta^{RR}_{ct}$ & \begin{tabular}{c} S1 \\ S2 \\ S3 \\ S4 \\ S5 \\ S6  \end{tabular}   & \begin{tabular}{c} 
(-0.99:0.99) \\ (-0.99:0.99) \\ (-0.98:0.97) \\ excluded \\ (-0.99:0.99) \\ (-0.96:0.94) \end{tabular}     & \begin{tabular}{c} (-0.99:0.99) \\ (-0.99:0.99) \\ (-0.99:0.99) \\ (-0.81:0.81) \\ (-0.99:0.99) \\ (-0.99:0.99) \end{tabular}     & \begin{tabular}{c} (-0.99:0.99) \\ (-0.99:0.99) \\ (-0.99:0.99) \\ (-0.81:-0.80) (-0.34:0.34) (0.80:0.81) \\ (-0.99:0.99) \\ (-0.99:0.99)
\end{tabular}   \\ \hline
$\delta^{RR}_{sb}$  & \begin{tabular}{c}  S1 \\ S2 \\ S3 \\ S4 \\ S5 \\ S6 \end{tabular}    &
\begin{tabular}{c} (-0.99:0.99) \\ (-0.99:0.99) \\ (-0.99:0.99) \\ excluded \\ (-0.99:0.99) \\ (-0.99:0.99)
 \end{tabular}  & \begin{tabular}{c} (-0.99:0.99) \\ (-0.99:0.99) \\ (-0.99:0.99) \\ (-0.97:0.97) \\ (-0.99:0.99) \\ (-0.99:0.99)
 \end{tabular}  & \begin{tabular}{c} (-0.96:0.96) \\ (-0.96:0.96) \\ (-0.96:0.94) \\ (-0.70:0.54) \\ (-0.97:0.97) \\ (-0.97:-0.94) (-0.63:0.64) (0.93:0.97)\end{tabular}    \\ \hline
\end{tabular}}
\end{center}
\end{table}
\caption{Allowed delta intervals by ${\rm BR}(B\rightarrow X_s\gamma)$, ${\rm BR}(B_s\rightarrow\mu^+\mu^-)$ and $\dmbs$ for post-LHC allowed scenarios and post-LHC $B$ data. \label{tableintervals2}}
\end{table}
\renewcommand{\arraystretch}{1.55}

\renewcommand{\arraystretch}{1.1}
\begin{table}[H]
\begin{center}
\resizebox{9.0cm}{!} {
\begin{tabular}{|c|c|c|} \hline
 & & Total allowed intervals \\ \hline
$\delta^{LL}_{23}$ & \begin{tabular}{c}  S1 \\ S2 \\ S3 \\ S4 \\ S5 \\ S6 \end{tabular} &  
\begin{tabular}{c} 
(-0.27:0.28) \\ (-0.23:0.23) \\ (-0.12:0.06) (0.17:0.19) \\ excluded \\ (-0.83:-0.78) (-0.14:0.14) \\ (-0.076:0.14) \end{tabular} \\ \hline
$\delta^{LR}_{ct}$  & \begin{tabular}{c}  S1 \\ S2 \\ S3 \\ S4 \\ S5 \\ S6 \end{tabular}    
& \begin{tabular}{c} 
(-0.27:0.27) \\ (-0.27:0.27) \\ (-0.27:0.27) \\ excluded \\ (-0.22:0.22) \\ (-0.37:0.37) \end{tabular}   \\ \hline
$\delta^{LR}_{sb}$  & \begin{tabular}{c}  S1 \\ S2 \\ S3 \\ S4 \\ S5 \\ S6 \end{tabular}    & 
\begin{tabular}{c} 
(-0.0069:0.014) (0.12:0.13) \\ (-0.0069:0.014) (0.11:0.13) \\ (-0.0069:0.014) (0.11:0.13) \\ (0.076:0.12) (0.26:0.30) \\ (-0.014:0.021) (0.17:0.19) \\ (0:0.0069) (0.069:0.076) \end{tabular}  \\ \hline
$\delta^{RL}_{ct}$  & \begin{tabular}{c}  S1 \\ S2 \\ S3 \\ S4 \\ S5 \\ S6 \end{tabular}    & 
\begin{tabular}{c}
(-0.27:0.27) \\ (-0.27:0.27) \\ (-0.27:0.27) \\ excluded \\ (-0.22:0.22) \\ (-0.37:0.37) \end{tabular} 
  \\ \hline
$\delta^{RL}_{sb}$  & \begin{tabular}{c}  S1 \\ S2 \\ S3 \\ S4 \\ S5 \\ S6 \end{tabular}    &
\begin{tabular}{c} (-0.034:0.034) \\ (-0.034:0.034) \\ (-0.034:0.034) \\ excluded \\ (-0.062:0.062) \\ (-0.021:0.021) \end{tabular} 
  \\ \hline
$\delta^{RR}_{ct}$ & \begin{tabular}{c}  S1 \\ S2 \\ S3 \\ S4 \\ S5 \\ S6 \end{tabular}   & \begin{tabular}{c} 
(-0.99:0.99) \\ (-0.99:0.99) \\ (-0.98:0.97) \\ excluded \\ (-0.99:0.99) \\ (-0.96:0.94)  \end{tabular}    \\ \hline
$\delta^{RR}_{sb}$  & \begin{tabular}{c}  S1 \\ S2 \\ S3 \\ S4 \\ S5 \\ S6 \end{tabular}    &
\begin{tabular}{c}  (-0.96:0.96) \\ (-0.96:0.96) \\ (-0.96:0.94) \\ excluded \\ (-0.97:0.97) \\ (-0.97:-0.94) (-0.63:0.64) (0.93:0.97)
\end{tabular}    \\ \hline
\end{tabular}}  
\end{center}
\caption{Total allowed delta intervals by ${\rm BR}(B\rightarrow X_s\gamma)$, ${\rm BR}(B_s\rightarrow\mu^+\mu^-)$ and $\dmbs$ for post-LHC allowed scenarios and post-LHC $B$ data. \label{tabdeltasummary2}}
\end{table}
\renewcommand{\arraystretch}{1.55}

\bigskip

\clearpage
\newpage

\chapter{Higgs boson mass corrections from flavour mixing in the squark sector}
\label{higgsmasssquark}

The discovery of the Higgs boson still leaves an important question opened: what kind of Higgs boson have we discovered? Is it the new boson a signal of a new supersymmetric world or just the last undiscovered piece of our old Standard Model? 
The currently ongoing precision measurements on the Higgs characteristics as the mass, the couplings and the decays will be crucial to answer these questions. In this chapter we will study the Higgs boson mass corrections generated from general squark flavour mixing. We will study the size of the Higgs boson radiative mass corrections depending on the different flavour mixing $\deXYij$ parameters of the squark sector and the MSSM parameters, and we will also set constraints on the possible size of the squark flavour mixing from the present bounds on the value of the Higgs mass itself.

The results in this chapter have been published in \cite{AranaCatania:2011ak} and \cite{Arana-Catania:updatedsquark}. 

\section{Status and framework for the Higgs mass corrections computation}
\label{sec:statusframeworkhiggscorrec}

As we commented when introducing the Higgs sector in Sections \ref{particles-mssm-higgs} and \ref{sec:renrMSSM}, higher-order contributions can give 
large corrections to the tree-level relations (see
e.g.\ \citere{reviews} for reviews) of this sector. In the MSSM%
\footnote{We concentrate in this work on the case with real parameters. For
  complex parameters see \citeres{mhcMSSMlong,mhcMSSM2L} and references
  therein.}
the status of higher-order corrections to the masses and mixing angles
in the neutral Higgs sector is quite advanced. 
The full one-loop and potentially all leading two-loop corrections are
known, see~\citere{mhiggsAEC} for a review. Most recently leading
three-loop corrections became available~\cite{mhiggsEP3l,mhiggsFD3l}. 

However, the impact of non-minimal flavour violation in the squark sector, on the
MSSM Higgs-boson masses and mixing angles, entering already at the 
one-loop level, has not been explored very deeply so far.
A one-loop calculation taking into account the $LL$-mixing between the
third and second generation of scalar up-type quarks has been performed
in \citere{mhNMFVearly}. A full one-loop calculation of the Higgs-boson
self-energies including all NMFV mixing terms had been implemented into
the Fortran code \fh~\cite{feynhiggs,mhiggslong,mhiggsAEC,mhcMSSMlong}, 
however no cross checks or numerical evaluations analyzing the
effects of the new mixing terms were performed.
Possible effects from NMFV on Higgs boson decays
were investigated in \cite{HdecNMFV,Curiel:2002pf,Curiel:2003uk,HdecNMFV3}. Within a similar context of NMFV
there have been also studied some effects of scharm-stop flavour mixing in top-quark
FCNC processes \cite{Cao1} and charged Higgs processes \cite{Dittmaier:2007uw} as well as the implications for LHC
\cite{Cao2,Dittmaier:2007uw}. 
Some previous studies on the induced radiative corrections on 
the Higgs mass from scharm-stop flavour mixing have also been performed in
\cite{Cao1,Cao2}, but any effects from mixing involving the first generation of scalar quarks have been
neglected. The numerical estimates in \cite{Cao1,Cao2} also neglect all the flavour mixings in the scalar down-type sector, except for those of $LL$-type that are induced from the scalar up-type sector by SU(2) invariance. In \cite{Cao2} they also consider one example with a particular numerical value of non-vanishing ${\tilde s}_L-{\tilde b}_R$ mixing.
 
We study in this chapter the consequences from NMFV in the squark sector
for the MSSM Higgs-boson spectrum, where our results are obtained
  in full generality, i.e.\ all generations in the scalar up- and
  down-type quark sector are included in our analytical results.
In the numerical analysis we focus particularly on the
flavour mixing between the second and third generations. Our estimates include all type of flavour mixings, $LL$, $LR$, $RL$, and $RR$. We devote special attention to the $LR/RL$
 sector. These kind of mixing effects are expected
to be sizeable, since they enter the off-diagonal $A$~parameters, which
appear directly in the coupling of the Higgs bosons to scalar quarks.

With respect to the scenarios, we will use the ones defined in Section \ref{scenarios}. Concerning the constraints from flavour observables we take into account the most up-to-date evaluations in the NMFV MSSM for \bsg, \bmm\ and \dmbs. 
These observables and the constraints coming from them were studied in detail in Chapter \ref{sec:Bphysics}. Here we analyze the one-loop contributions of NMFV to the
MSSM Higgs boson masses, focusing on the parameter space still allowed
by the experimental flavour constraints, as it was studied in this previous chapter. In this way the full possible
impact of NMFV in the MSSM on the Higgs sector can be explored. It will also be compared the situation previous to the LHC and after the LHC data.

\section{Analytical results for the MSSM corrected Higgs boson self energies}
\label{sec:analytical-results}

Following the detailed prescription for the computation of the one-loop corrected Higgs boson masses given in Sections \ref{particles-mssm-higgs} and \ref{sec:renrMSSM}, one finds the analytical expressions for these masses in terms of the renormalized self-energies which, in turn, are written in terms of the unrenormalized self-energies and tadpoles. To shorten the presentation of these analytical results, it is convenient to report here just on these unrenormalized self-energies and tadpoles. 

The relevant one-loop corrections have been evaluated 
with the help of 
\fa~\cite{feynarts} and  \fc~\cite{formcalc}.
For completeness the new Feynman rules included in the model file are listed
in the Appendix A.
All the results for the unrenormalized self-energies and tadpoles are collected in
Appendix B. We have shown explicitly just the relevant contributions
for the present study of the radiative corrections to the Higgs boson masses within NMFV scenarios in the squark sector, namely, the one-loop contributions from quarks
and squarks. The corresponding generic Feynman-diagrams for the
Higgs bosons self-energies, gauge boson
self-energy diagrams and tadpole diagrams are collected in the \reffi{figfdall} in Appendix B.
It should also be noticed that the contributions from the squarks are the only ones that differ from the usual ones of the MSSM with MFV.  
It should be noted also that the corrections from flavour mixing, which are the subject of our interest here,  are implicit in both the
$\VCKM$, and in the values of the rotation matrices, $R^{\tilde u}$, $R^{\tilde d}$, the squark masses, $m_{\tilde u_i}$, $m_{\tilde d_i}$ ($i=1,..,6$) and the non-diagonal in flavour trilinear couplings, that appear in these formulas of the unrenormalized self-energies and tadpoles and that have been introduced in Section \ref{sec:nmfv-squarks}. 

Finally, it is worth mentioning that we have checked the finiteness in our analytical results for the renormalized Higgs self-energies. It is obviously expected, but it is not a trivial check in the present scenarios with three generations of quarks and squarks and with flavour mixing. We have also checked that the analytical results of the self-energies in Appendix B agree with the formulas in \fh~\cite{feynhiggs,mhiggslong,mhiggsAEC,mhcMSSMlong}. Each one of the terms contained in the Appendix B was compared with the corresponding term in \fh. During this process and the check of the finiteness, discrepancies were found with the charged Higgs part of \fh, leading to an updated version of the code\footnote{
We especially thank T.~Hahn for his efforts put into this update.}.

\section{Numerical analysis of the Higgs mass corrections}
\label{sec:numanal}
In this section we present our numerical results for the radiative corrections to the Higgs bosons
 masses from squark flavour mixing within NMFV-SUSY scenarios. 

The results will be given in two different sections where we will compare the situation previously and following to the LHC data. In the first one, we will evaluate the CMSSM scenarios, defined in Section \ref{frameworka}, containing mainly the SPS points that have been largely studied and used as benchmark points in the evaluations before the LHC. Although these points are discarded by the experiments now, they illustrate the main features of the sensitivities on the Higgs bosons masses to the various delta parameters. Since in that pre-LHC situation the assumed scale of SUSY was lighter, therefore the effect of the SUSY corrections had more impact on the observables. In the second section, we will evaluate the corrections for the scenarios more appropriate for the present post-LHC situation as defined in Section \ref{frameworkc}, with heavier SUSY particles and taking into account the most up-to-date experimental measurements.

Since all one-loop corrections in the present NMFV scenario are common to the MSSM except for the corrections from squarks, which depend on the $\deXYij$ values,  we will focus just on the results of these corrections as a function of the flavour mixing parameters,  and present the differences with respect to the predictions within the MSSM. Correspondingly, we define: 
\begin{equation}
 \Dmphi (\deXYij) \equiv 
 \mphi^{\rm NMFV}(\deXYij) - \mphi^{\rm MSSM}, \quad \phi =h,\, H,\, H^{\pm}, 
\end{equation}
where $\mphi^{\rm NMFV}(\deXYij)$ and $\mphi^{\rm MSSM}$ have been calculated at the
one-loop level.
 It should be noted that  $\mphi^{\rm NMFV}(\deXYij=0) = \mphi^{\rm MSSM}$ and,
 therefore, by construction, $\Dmphi(\deXYij=0) = 0$, and $\Dmphi$ gives the
size of the one-loop NMFV contributions to $\mphi$. The numerical calculation of $\mphi^{\rm NMFV}(\deXYij)$ and $\mphi^{\rm MSSM}$ has been done 
with (the updated version of) \fh~\cite{feynhiggs,mhiggslong,mhiggsAEC,mhcMSSMlong}, which solves the eqs. \ref{eq:proppole} and \ref{eq:proppolech} for finding the positions of the poles of the propagator matrix. Previous results for $\De \mh (\de^{LL}_{23})$ can be found in~\cite{mhNMFVearly}. 


\subsection{Numerical results for pre-LHC scenarios}
\label{numresframeworkahiggs}

The numerical study of the Higgs bosons masses corrections will be performed first for the case when only one $\deXYij$ is different from zero, and later for the case with two non-vanishing $\deXYij$ where interferences can happen between both contributions.

\subsubsection{\boldmath{$\Dmphi$} versus one \boldmath{$\deXYij\neq 0$}}
\label{sec:numanalonedelta}

\begin{figure}[h!] 
\centering
\hspace*{-10mm} 
{\resizebox{17.3cm}{!} 
{\begin{tabular}{cc} 
\includegraphics[width=13.2cm,height=17.2cm,angle=270]{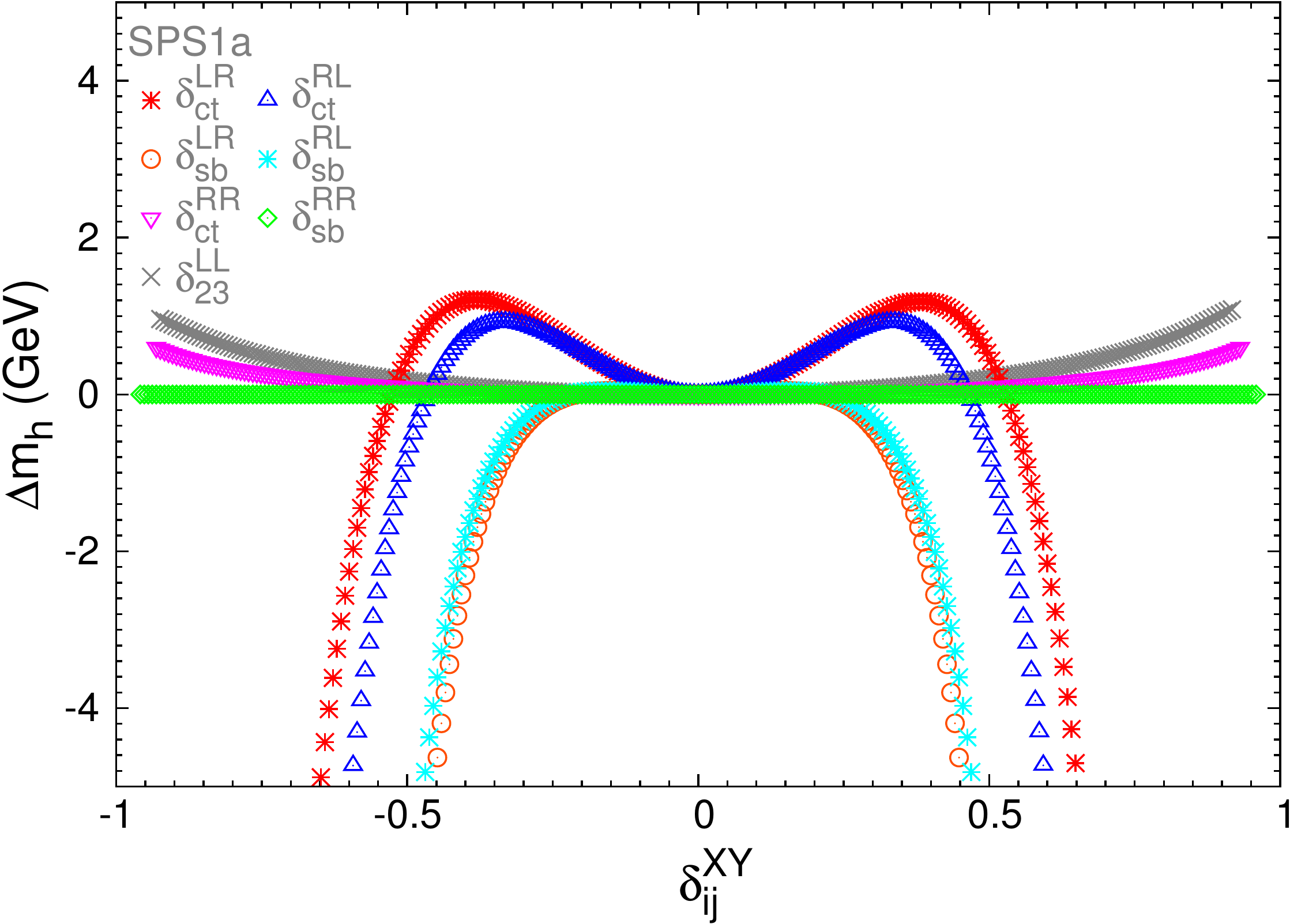}& 
\includegraphics[width=13.2cm,height=17.2cm,angle=270]{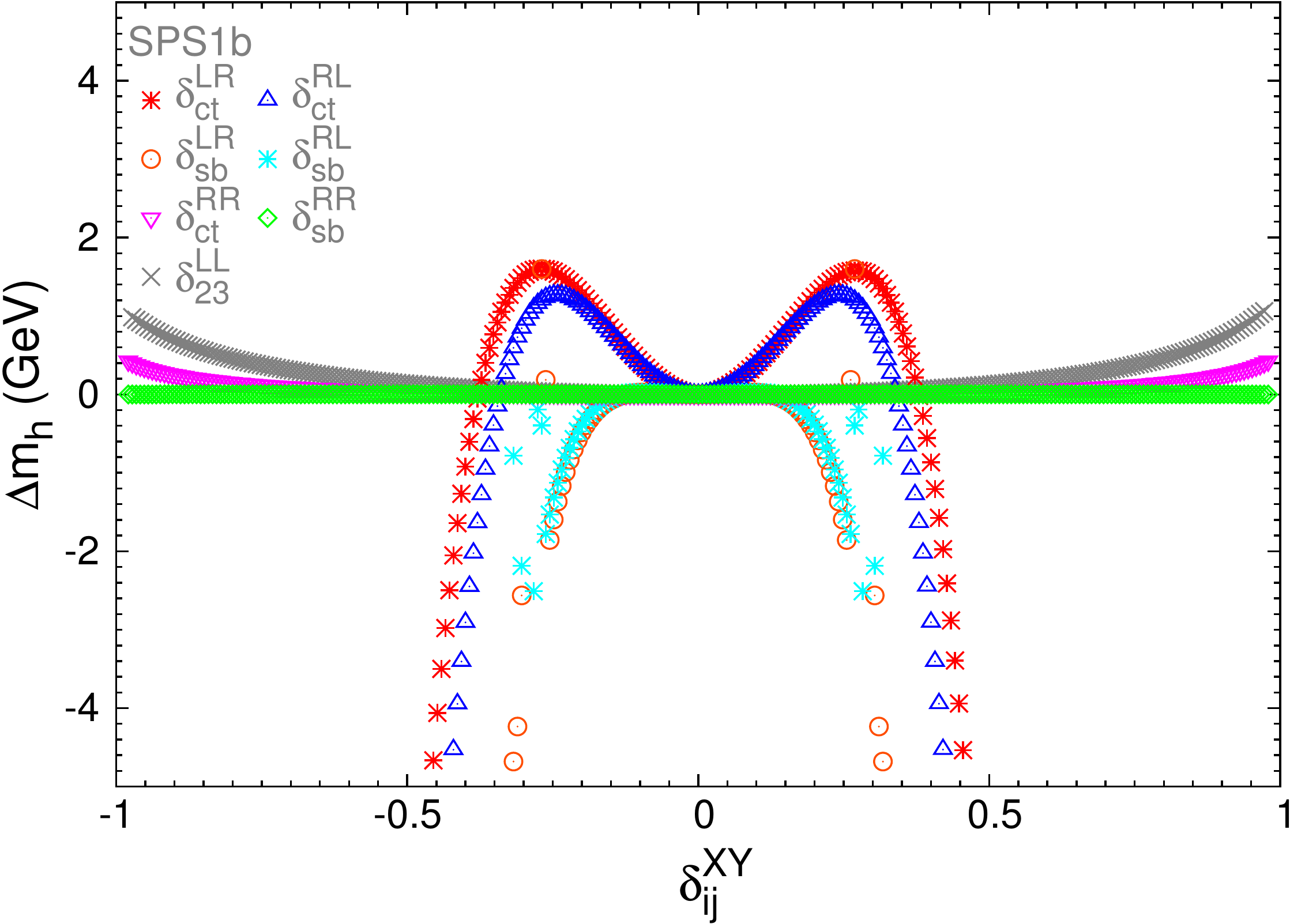}\\ 
\includegraphics[width=13.2cm,height=17.2cm,angle=270]{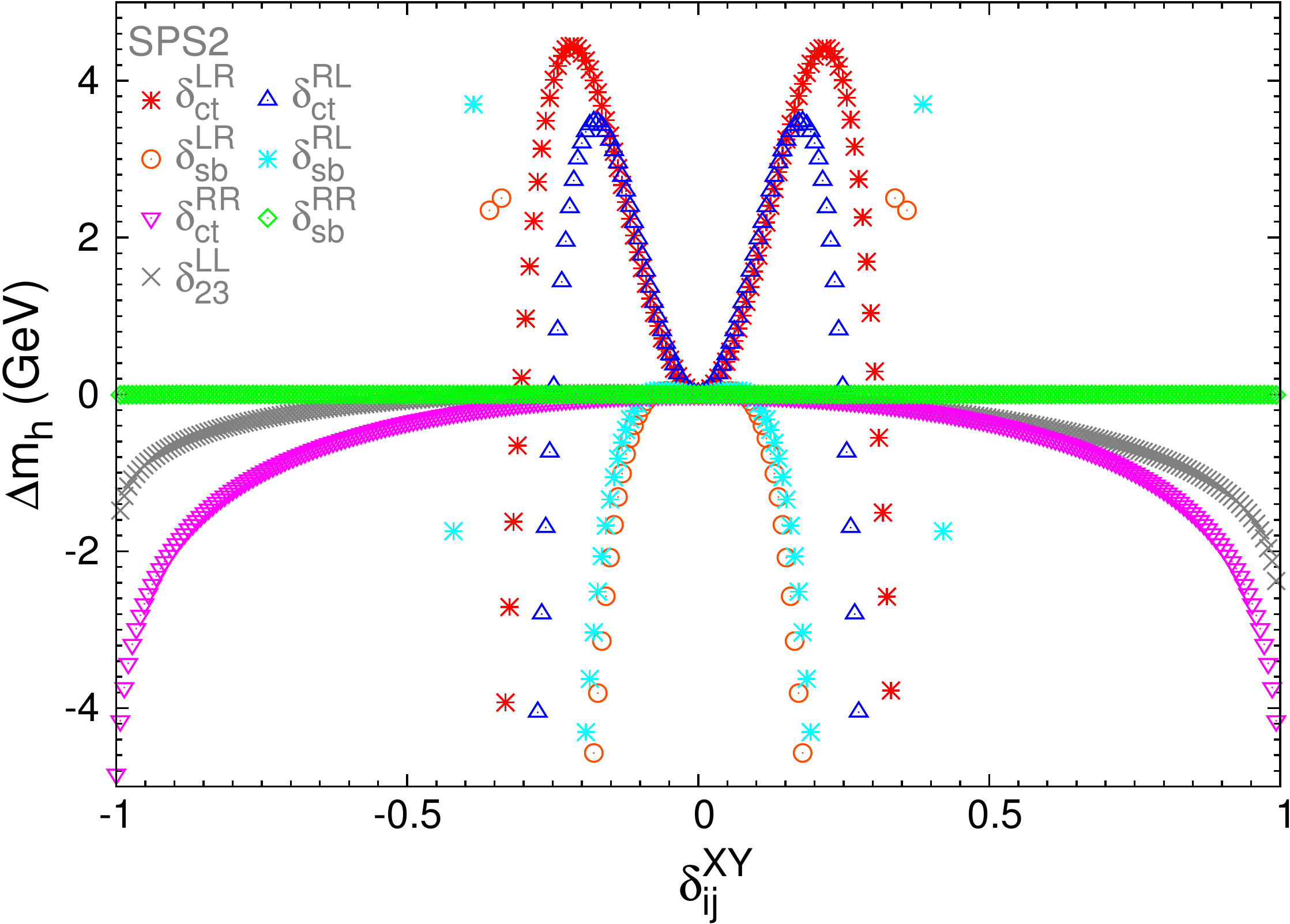}&
\includegraphics[width=13.2cm,height=17.2cm,angle=270]{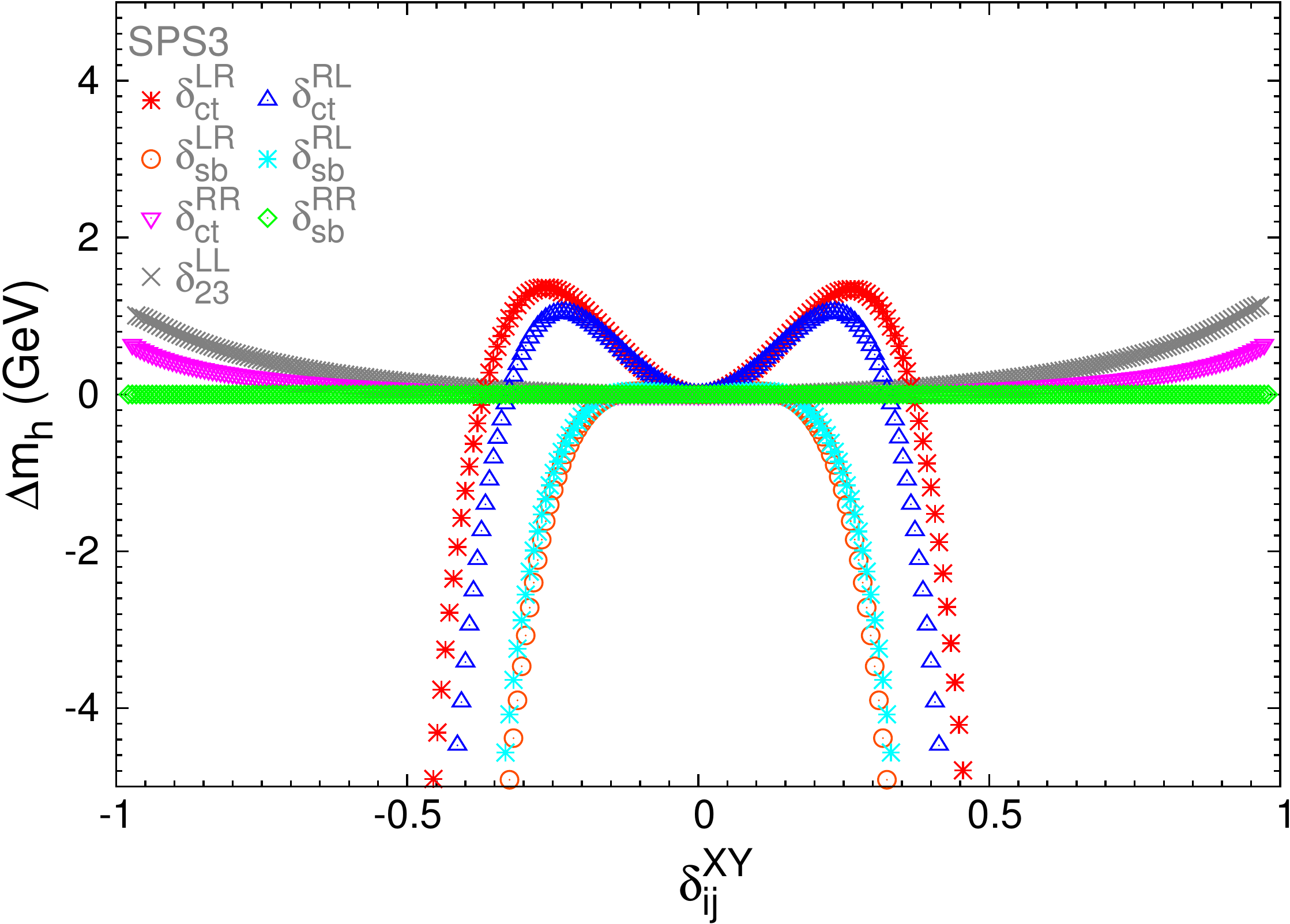}\\ 
\includegraphics[width=13.2cm,height=17.2cm,angle=270]{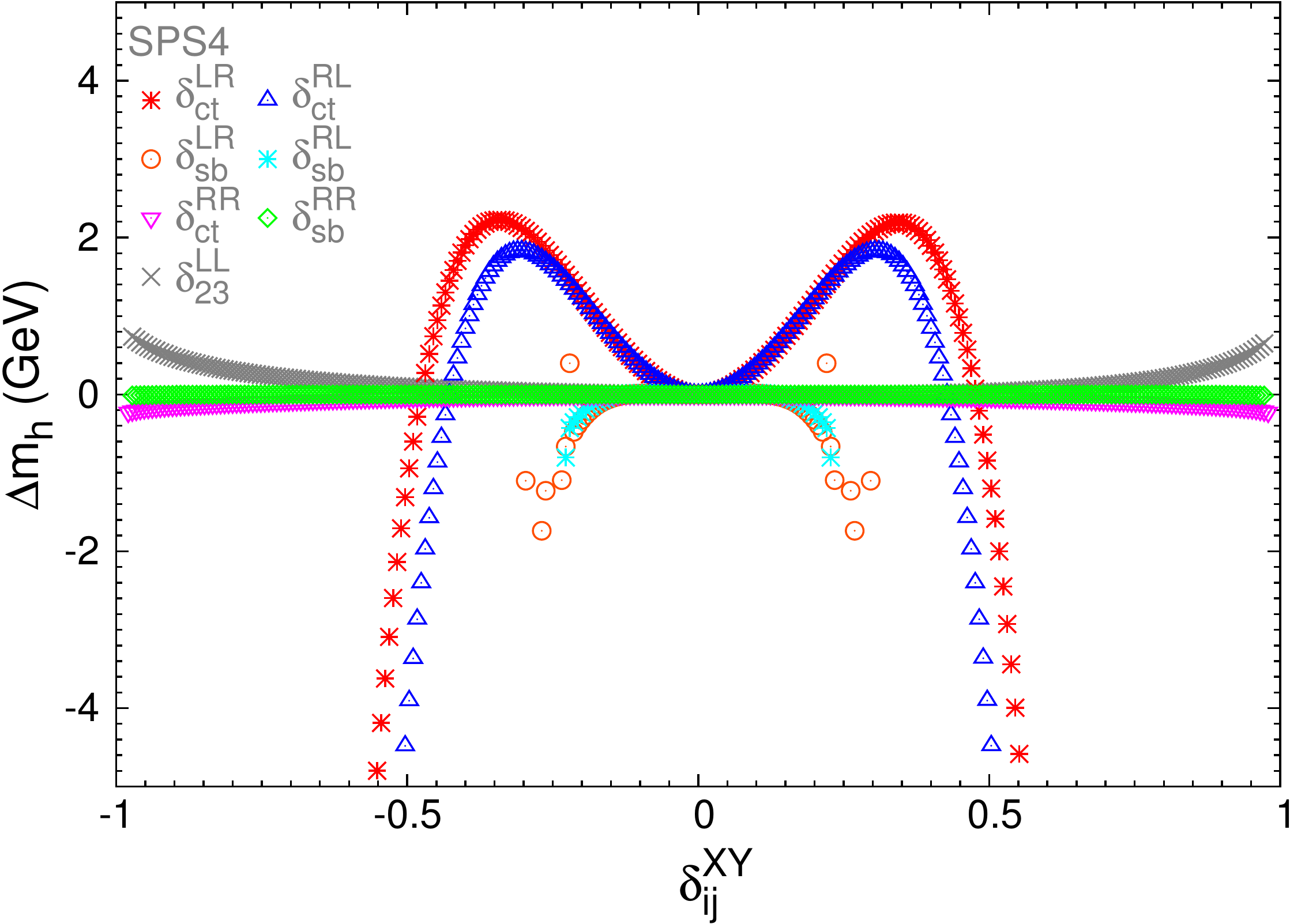}& 
\includegraphics[width=13.2cm,height=17.2cm,angle=270]{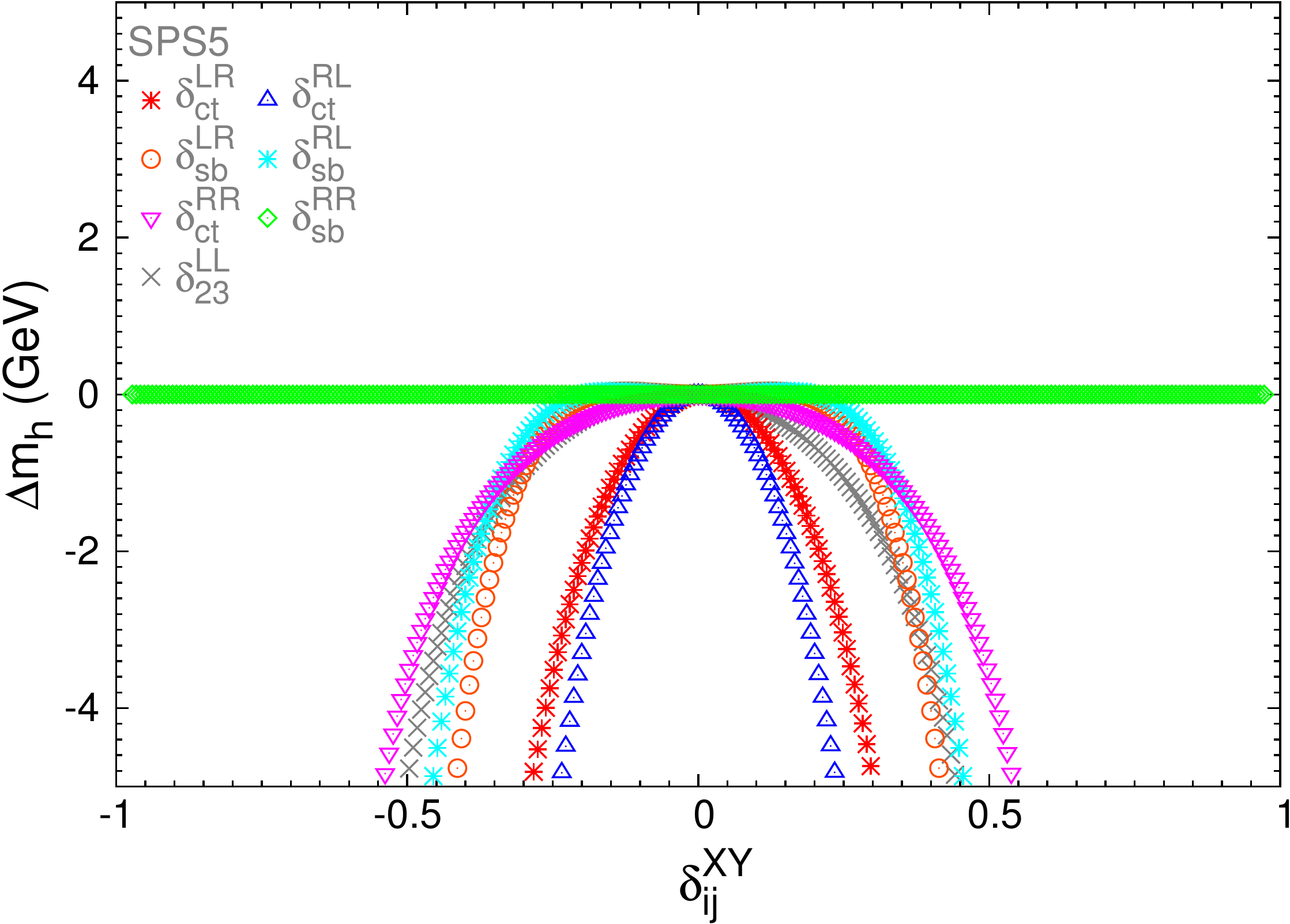}\\ 
\end{tabular}}}
\caption{Sensitivity to the NMFV deltas in $\Dmh$ for the SPSX points of table \ref{points}.}
 \label{figdeltamh0}
\end{figure}

\begin{figure}[h!] 
\centering
\hspace*{-10mm} 
{\resizebox{17.3cm}{!} 
{\begin{tabular}{cc} 
\includegraphics[width=13.2cm,height=17.2cm,angle=270]{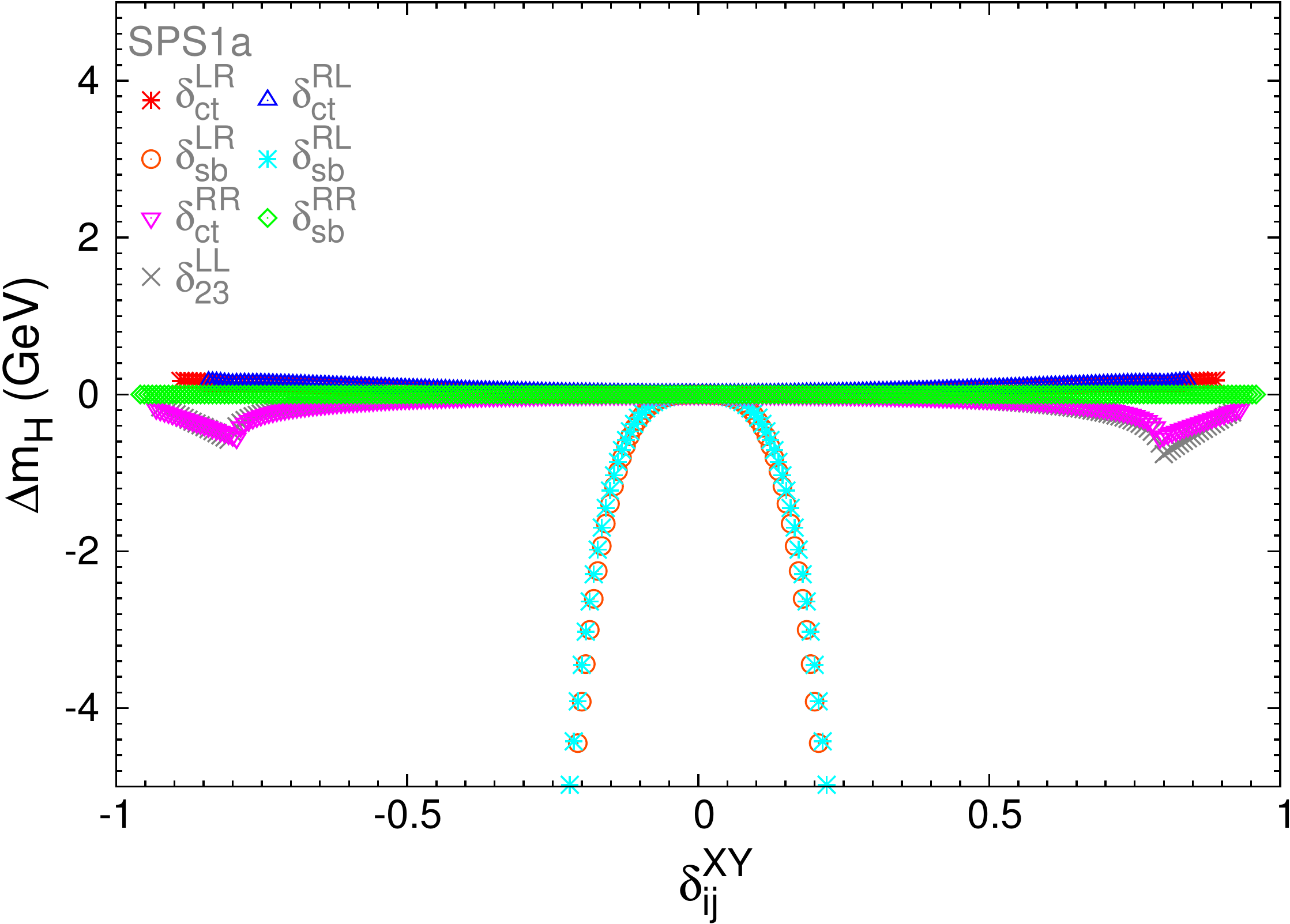}& 
\includegraphics[width=13.2cm,height=17.2cm,angle=270]{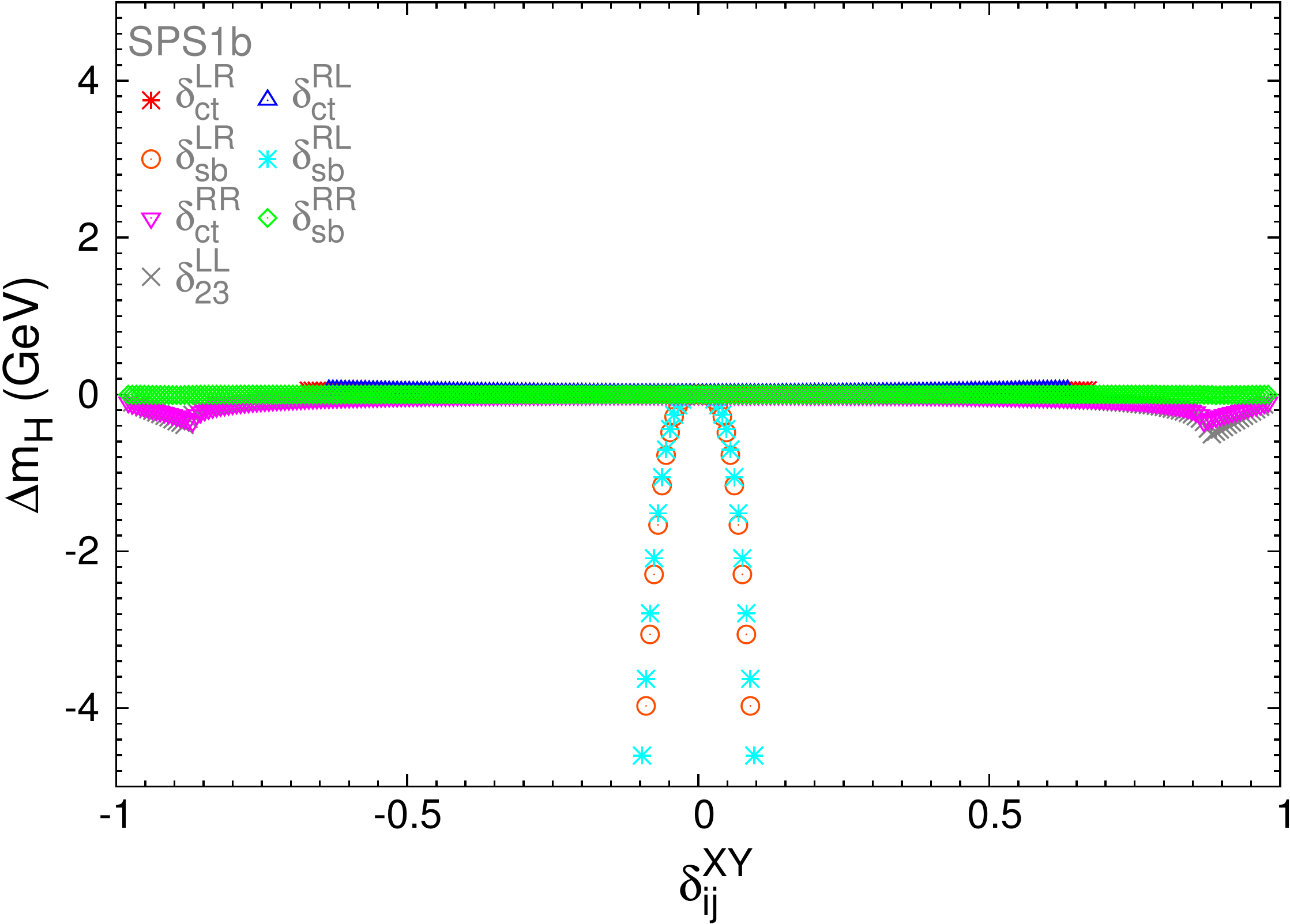}\\ 
\includegraphics[width=13.2cm,height=17.2cm,angle=270]{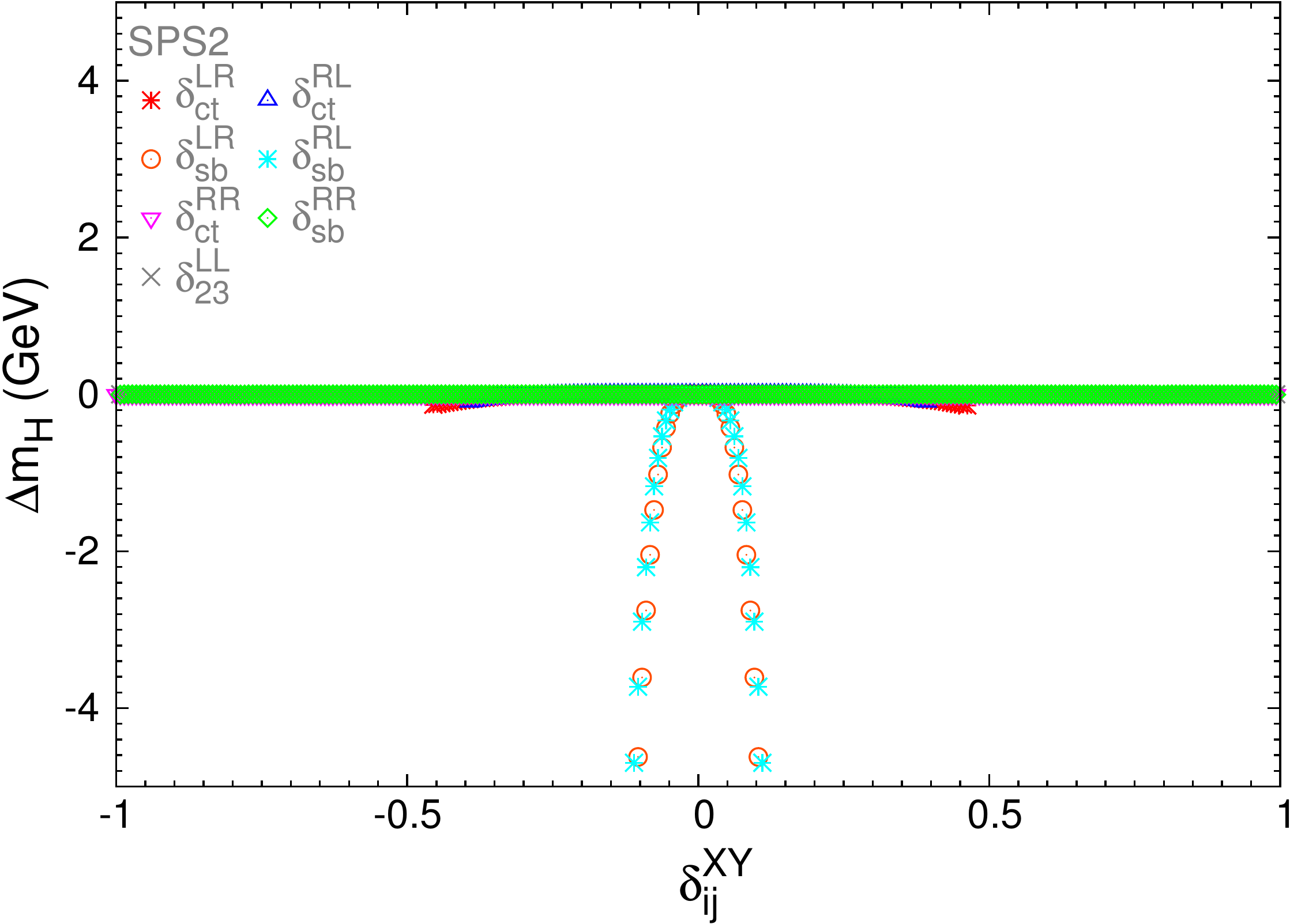}&
\includegraphics[width=13.2cm,height=17.2cm,angle=270]{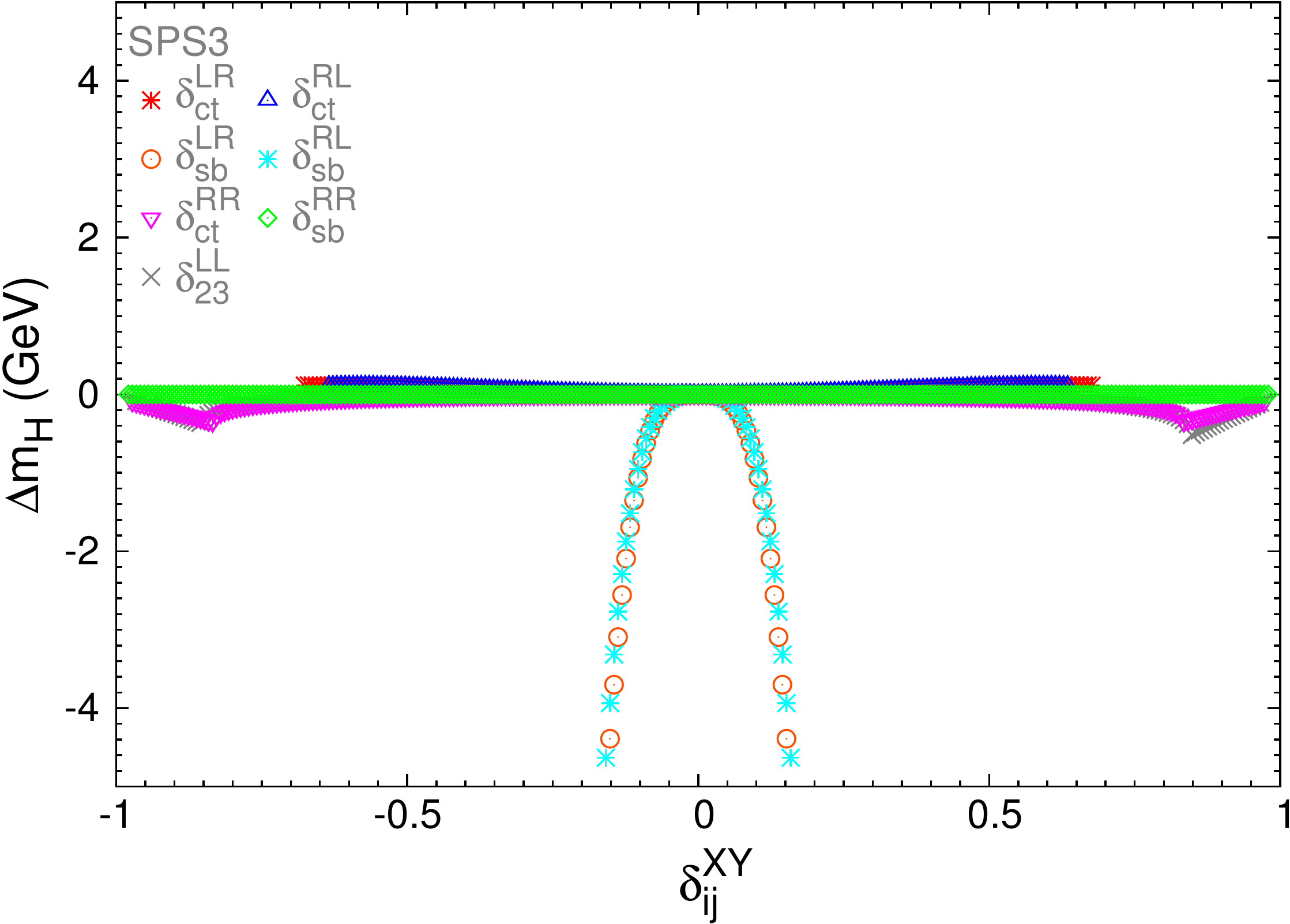}\\ 
\includegraphics[width=13.2cm,height=17.2cm,angle=270]{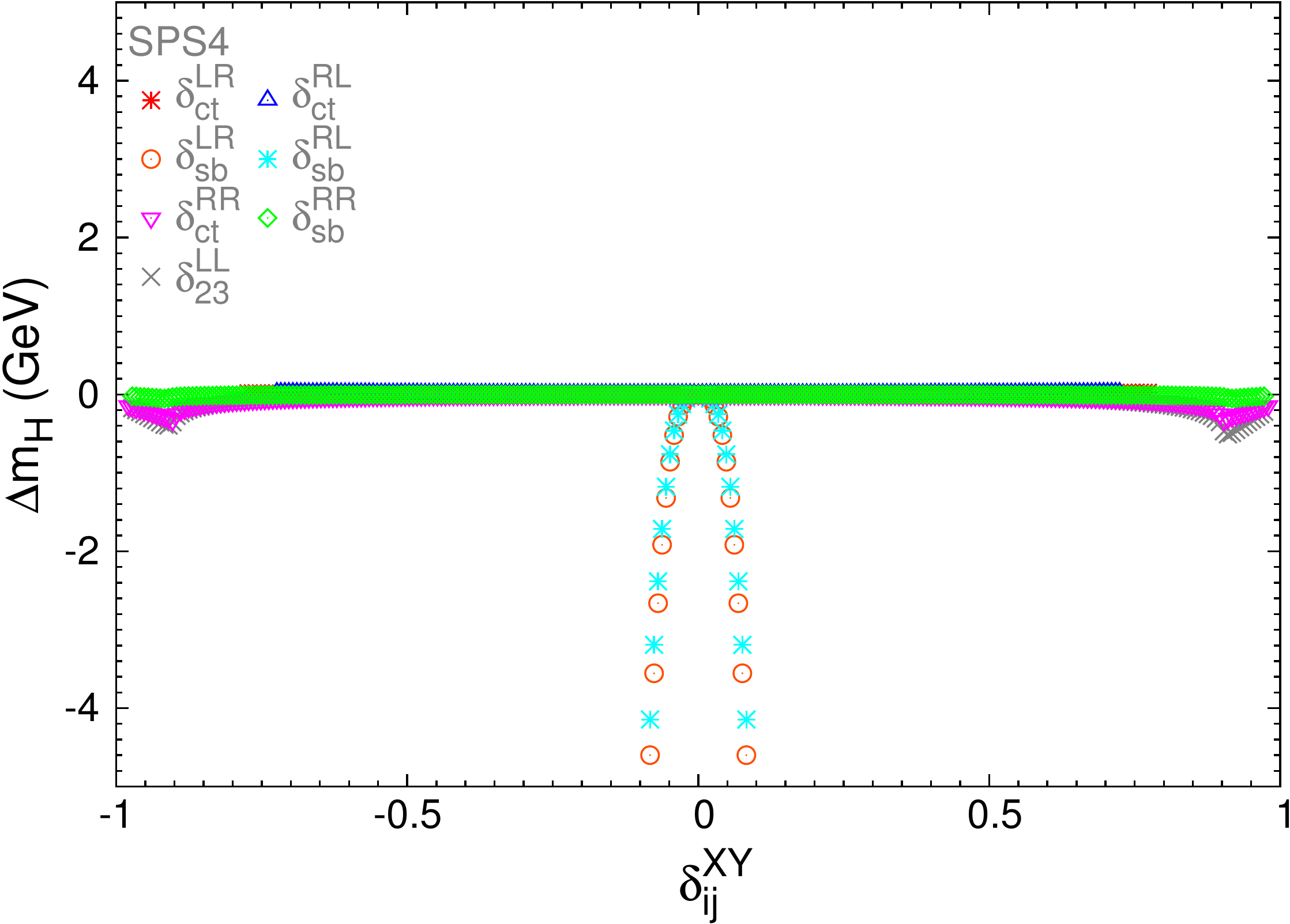}& 
\includegraphics[width=13.2cm,height=17.2cm,angle=270]{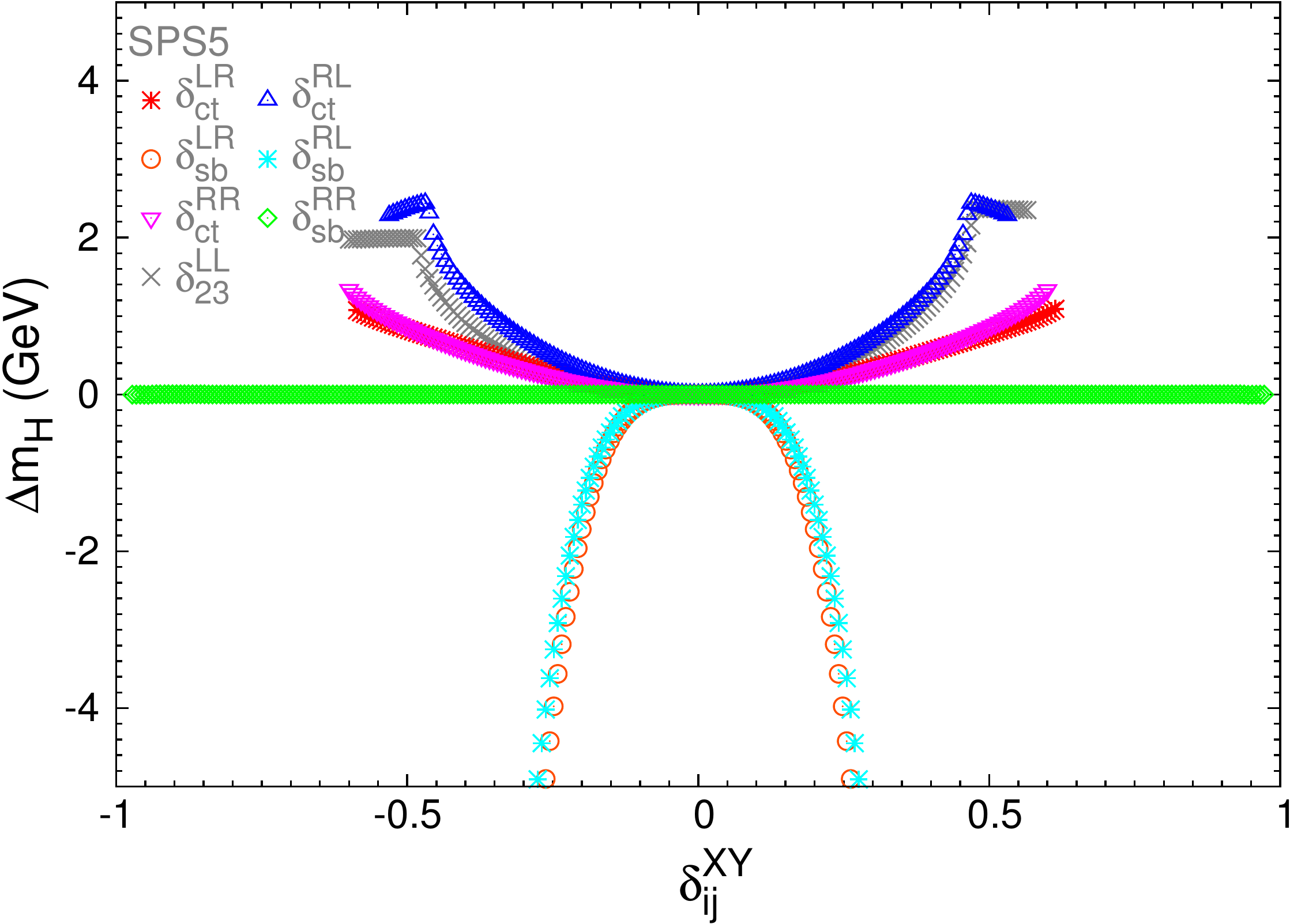}\\ 
\end{tabular}}}
\caption{Sensitivity to the NMFV deltas in $\DmH$ for the SPSX points of table \ref{points}.}
\label{figdeltamH0} 
\end{figure}

\begin{figure}[h!] 
\centering
\hspace*{-10mm} 
{\resizebox{17.3cm}{!} 
{\begin{tabular}{cc} 
\includegraphics[width=13.2cm,height=17.2cm,angle=270]{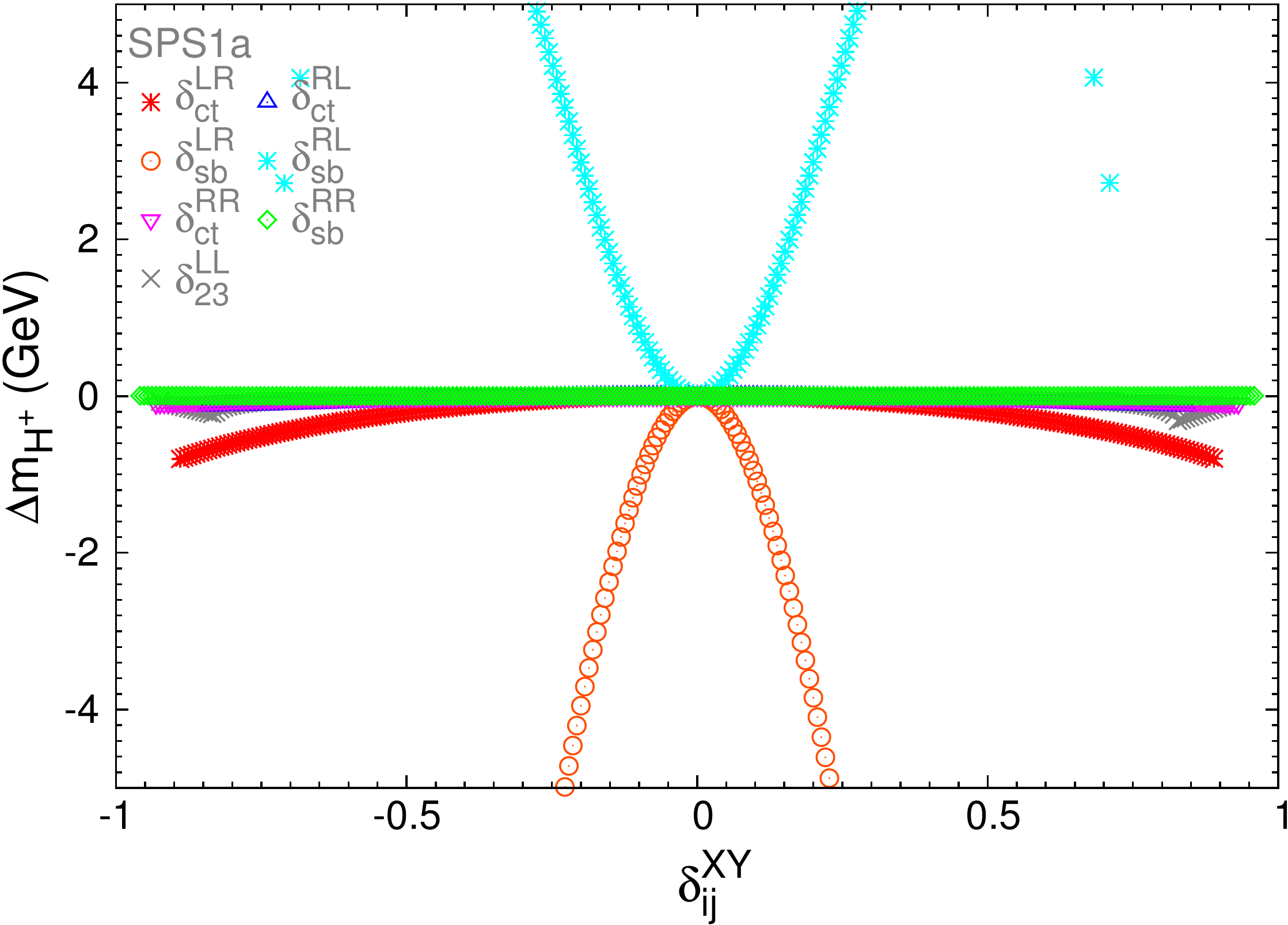}& 
\includegraphics[width=13.2cm,height=17.2cm,angle=270]{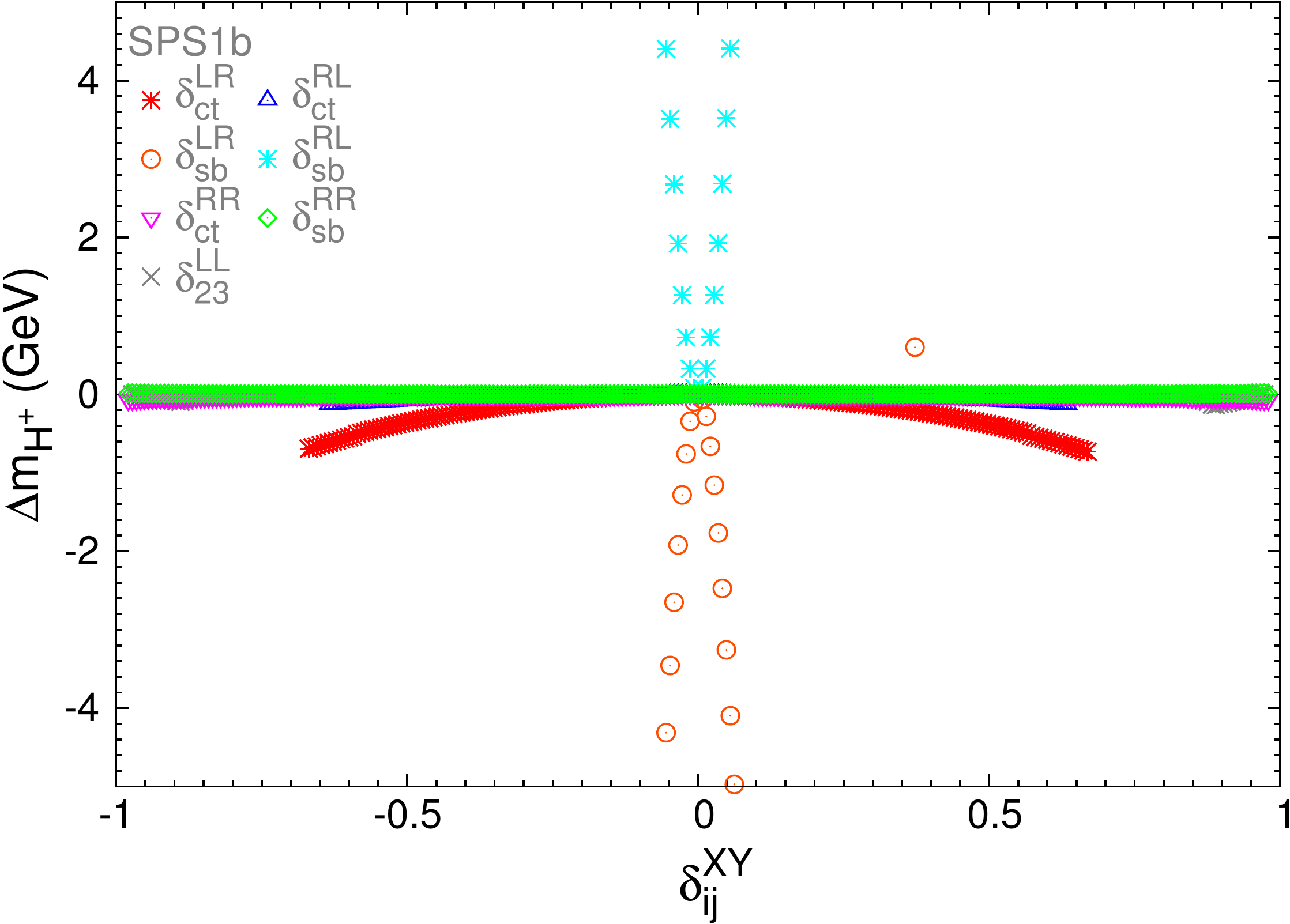}\\ 
\includegraphics[width=13.2cm,height=17.2cm,angle=270]{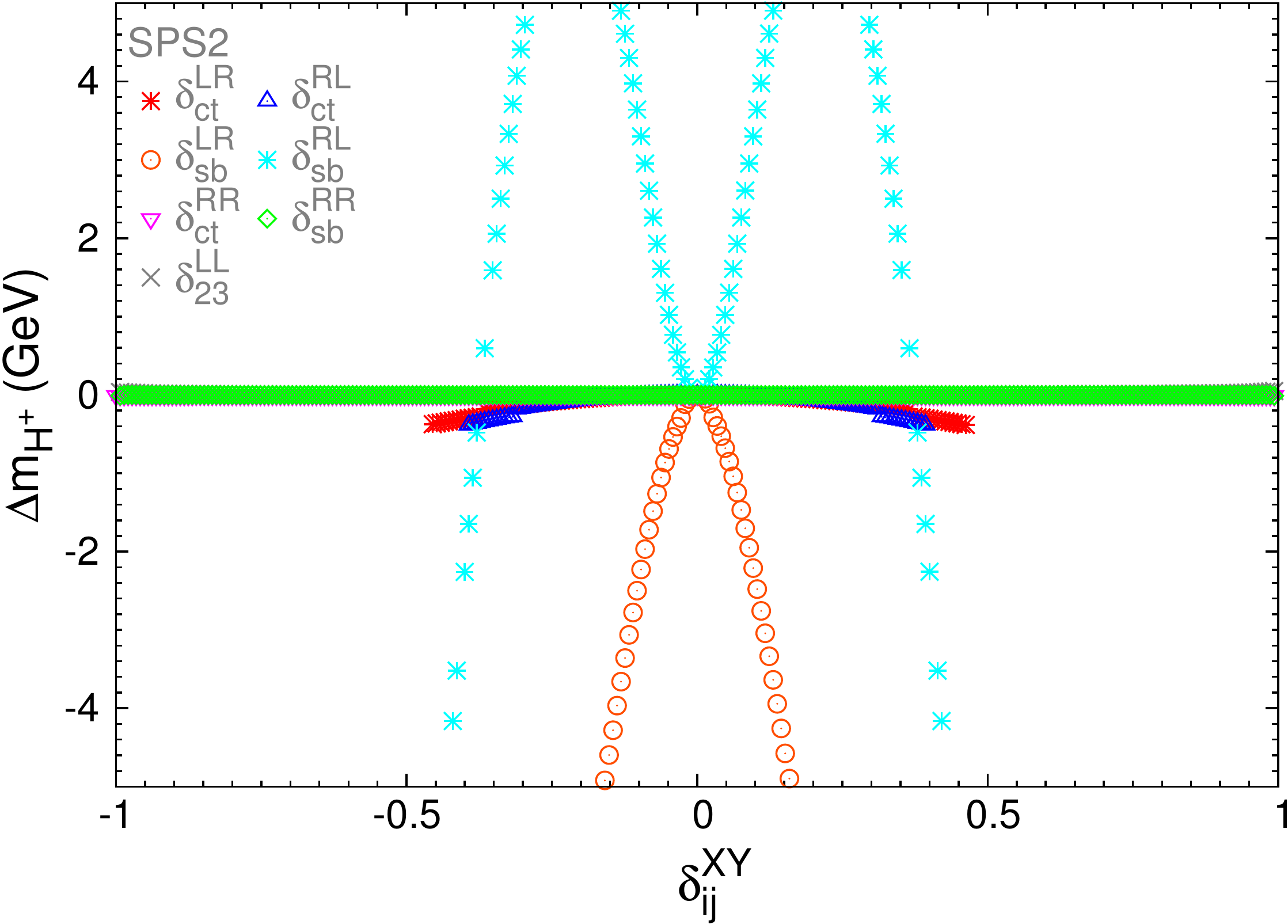}&
\includegraphics[width=13.2cm,height=17.2cm,angle=270]{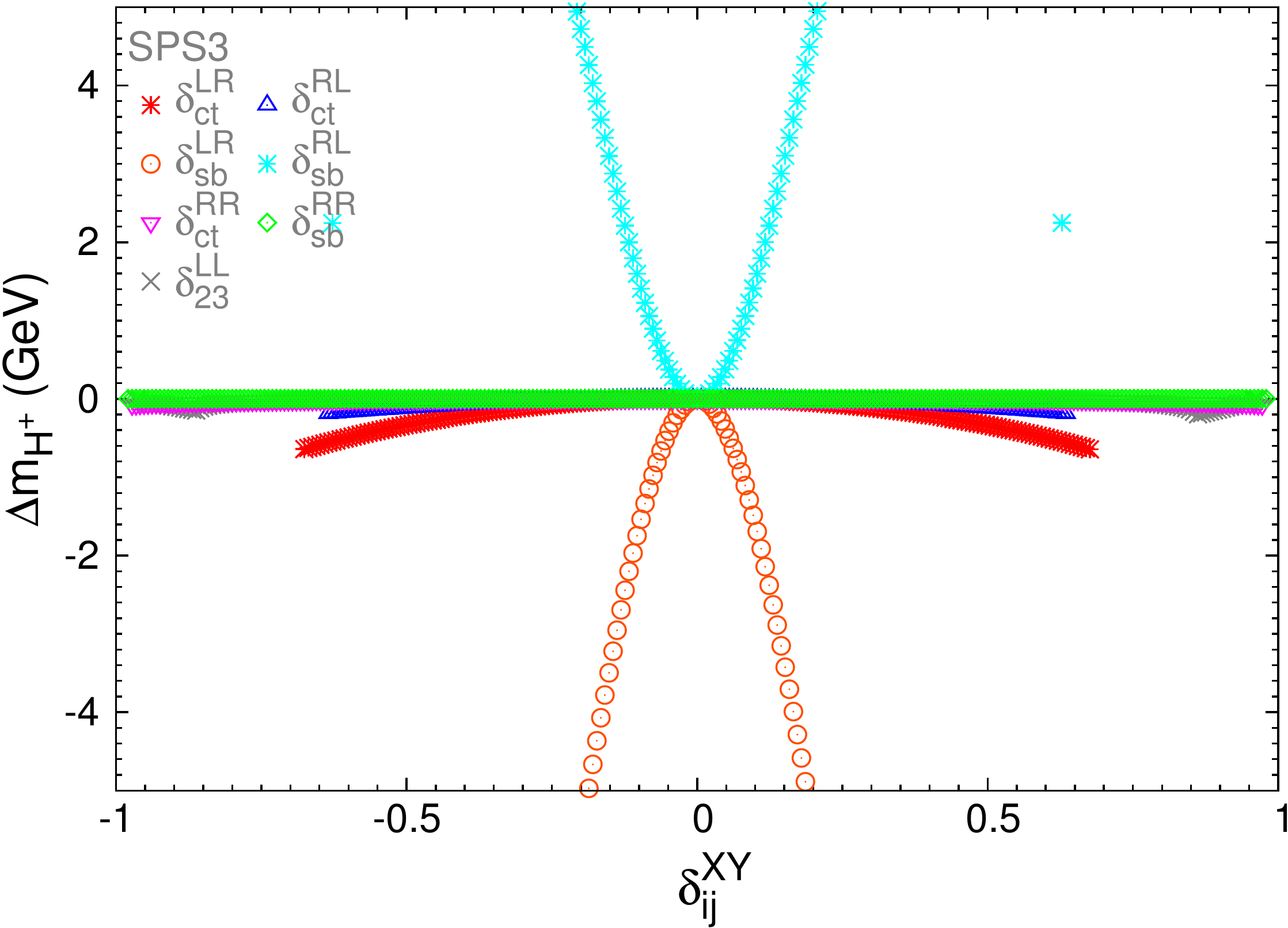}\\ 
\includegraphics[width=13.2cm,height=17.2cm,angle=270]{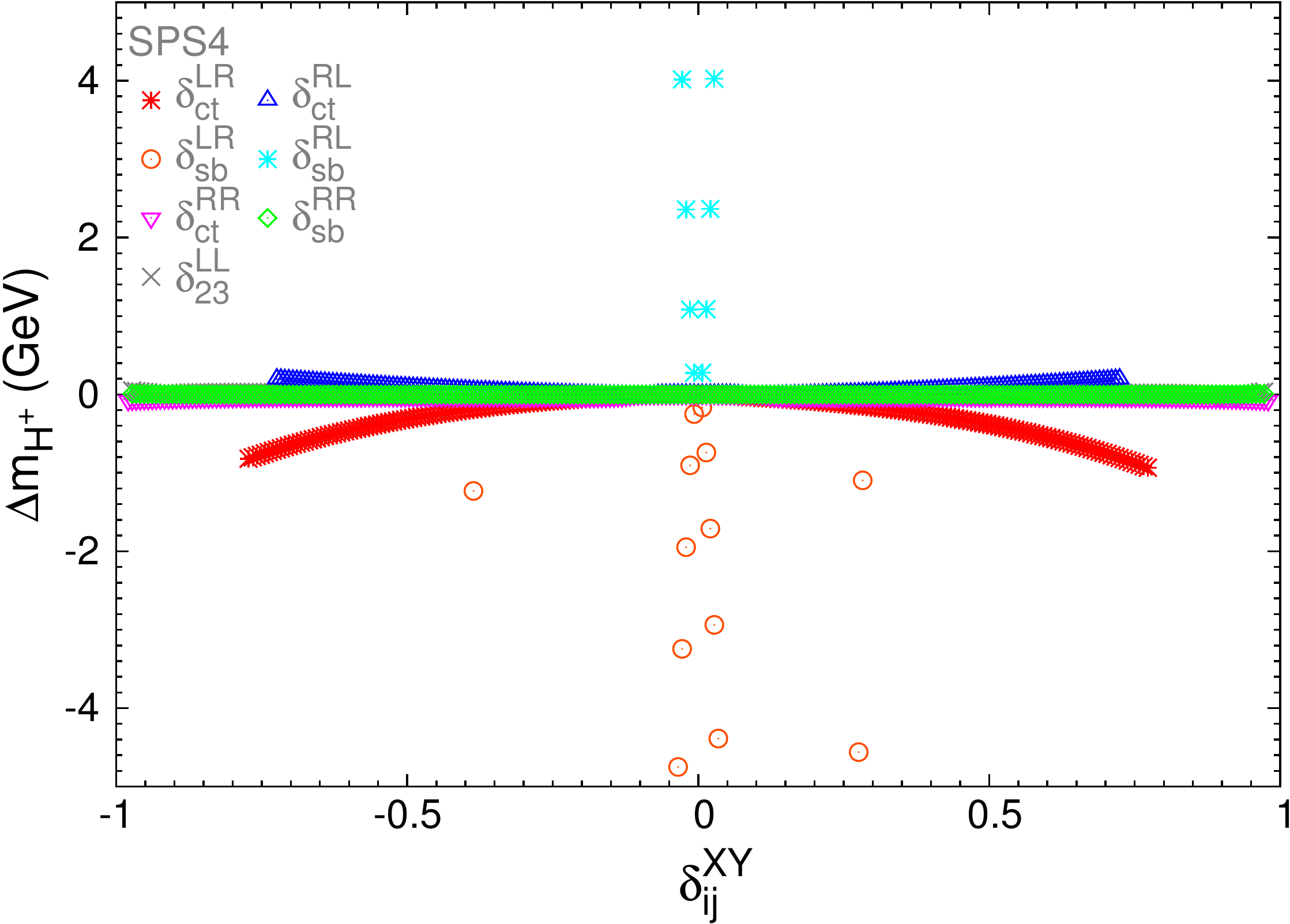}& 
\includegraphics[width=13.2cm,height=17.2cm,angle=270]{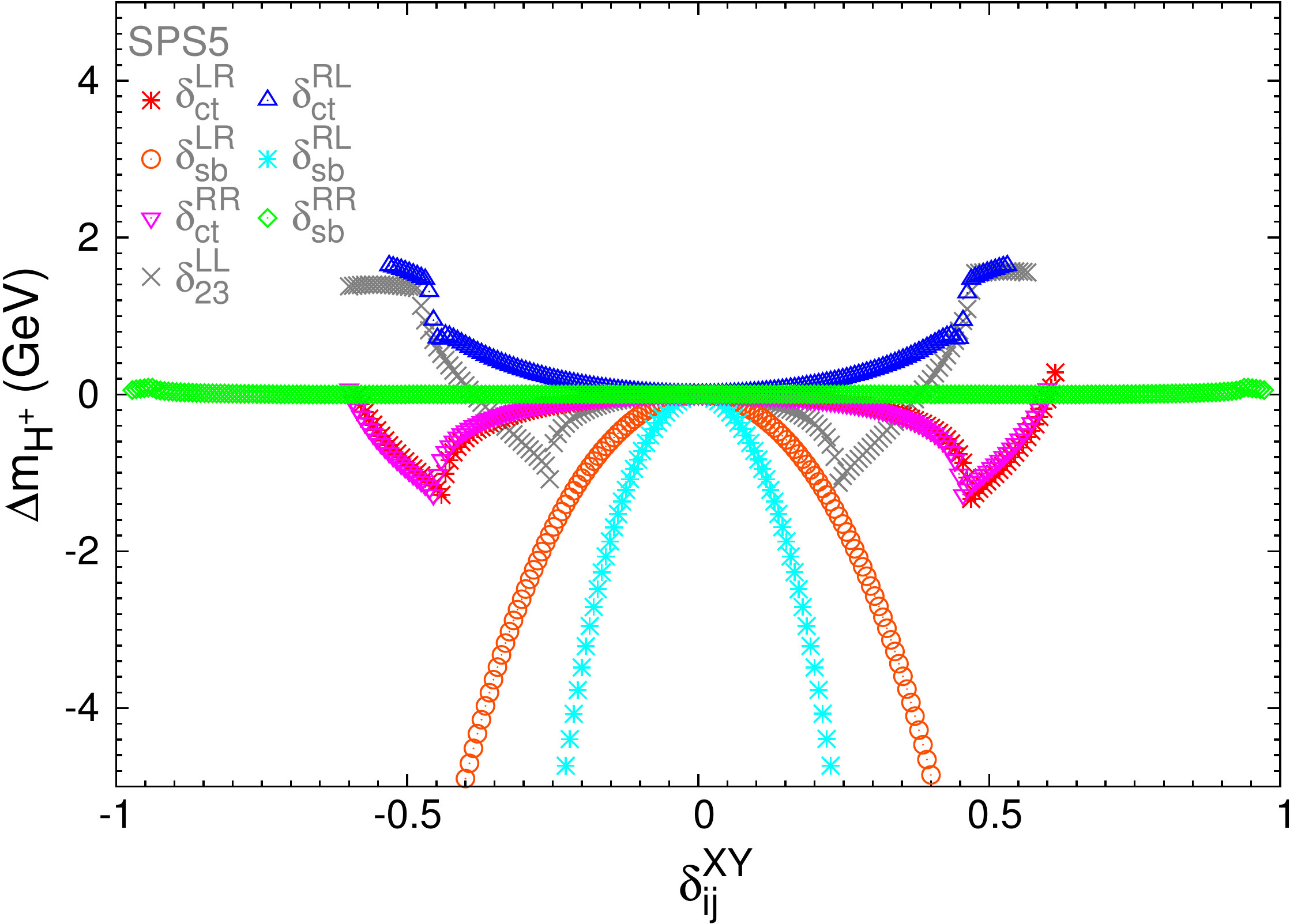}\\ 
\end{tabular}}}
\caption{Sensitivity to the NMFV deltas in $\DmHp$ for the SPSX points of table \ref{points}.}
 \label{figdeltamHp}
\end{figure}

We show in figs. \ref{figdeltamh0}, \ref{figdeltamH0} and
\ref{figdeltamHp} our numerical results for  
$\Dmh$, $\DmH$ and $\DmHp$, respectively, as functions of the seven
considered flavour changing deltas, $\de^{LL}_{23}$, $\de^{LR}_{ct}$,
$\de^{LR}_{sb}$, $\de^{RL}_{ct}$, $\de^{RL}_{sb}$, $\de^{RR}_{ct}$ and
$\de^{RR}_{sb}$,  
which we vary in the interval $-1 \leq \deXYij \leq 1$. In these plots
we have chosen the same six SPS points of table \ref{points}, as for the
previous study of  constraints from $B$ physics in the pre-LHC scenarios of \ref{numresframeworka}. We do
not consider yet the experimental bounds on the Higgs mass value, since we just want
to show here the general behaviour of the mass corrections with the deltas. These
experimental bounds will be taken into account in the next subsection.   
As in the study of the constraints, we have checked the impact of switching the sign of~$\mu$
  and found a small quantitative but no qualitative effect.

The main conclusions from these figures are the following:
\begin{itemize}
\item[-] General features: 

All mass corrections, $\Dmh$, $\DmH$ and $\DmHp$, are symmetric $\deXYij \to - \deXYij$, as expected. This feature is obviously different than in the previous plots of the $B$ observables. The sign of the mass corrections can be both positive and negative, depending on the particular delta value.  The size of the Higgs mass corrections, can be very large in some  
$\deXYij \neq 0$ regions, reaching values even larger than 10 GeV  
in some cases, at the central region with not very large delta values, 
$|\deXYij|< 0.5$. In fact, the restrictions from $B$ physics in this central region is crucial to get a reliable estimate of these effects.  

For low $\tb$, where the restrictions from $B$ physics to the deltas were less severe in the pre-LHC situation, the Higgs mass corrections are specially relevant. Particularly, $\Dmh$ turns out to be negative and large for $\tb =5$ (SPS5) for all deltas, except $\de^{RR}_{sb}$. For instance, at
$|\deXYij|\simeq 0.5$, the mass correction $\Dmh$ for SPS5 
is negative and $\gsim 5 \gev$ in all flavour changing deltas except $\de^{RR}_{sb}$ where the correction is negligible. In the case of $\DmH$ and $\DmHp$ the size of the correction at low $\tb$  is smaller, $\lsim 2 \gev$ in the central region, except for $\de^{LR}_{sb}$ and $\de^{RL}_{sb}$ that can also generate large  corrections $\gsim 5 \gev$. 

 In the cases with large 
$\tb$ (SPS4 and SPS1b), we also find large mass corrections but, as already said, they were much more limited by $B$ constraints. In particular,
for SPS4 all deltas were excluded, except for a very narrow window in $\de^{LL}_{23}$ (see table \ref{tabdeltasummary}).

In the cases with moderate $\tb=10$ (SPS1a, SPS2 and SPS3), we find large corrections $|\Dmh| \gsim 5 \gev$ in the central region 
of  $\de^{LR}_{sb}$, $\de^{RL}_{sb}$, $\de^{LR}_{ct}$ and $\de^{RL}_{ct}$. The other Higgs bosons get large corrections $|\DmH|,|\DmHp|\gsim 5 \gev$ in the deltas central region only for  
$\de^{LR}_{sb}$ and $\de^{RL}_{sb}$.
   
\item[-] Sensitivity to the various deltas:
 
We find very strong sensitivity in the three mass corrections $\Dmh$, $\DmH$ and $\DmHp$, to $\de^{LR}_{sb}$ and $\de^{RL}_{sb}$ for all the seven considered SPS points. 

In the case of $\Dmh$ there is also an important sensitivity to 
$\de^{LR}_{ct}$ and $\de^{RL}_{ct}$ in all the considered points. The
strong sensitivity to $LR$ and $RL$ parameters can be understood due to
the relevance of the $A$-terms in these Higgs mass corrections. It can
be noticed in the Feynman rules (i.e. see the coupling of two squarks
and one/two Higgs bosons in Appendix A) that the $A$-terms enter
directly into the couplings, and in some cases,
as in the couplings of down-type squarks to the $\cp$-odd Higgs boson,
enhanced by $\tb$. Therefore, considering the relationship between the
$A$-terms and these $LR$ and $RL$ parameters as is shown in
Eq. \ref{deltasdefs}, the strong sensitivity to these parameters can be
understood. A similar strong sensitivity to $\de^{LR}_{ct}$ in $\Dmh$
has been found in \cite{Cao1}. 

In SPS5 there is a noticeable sensitivity to all deltas except
$\de^{RR}_{sb}$. In other points, the effects of $\de^{LL}_{23}$,
$\de^{RR}_{ct}$  on $\Dmh$ are only appreciated at the large delta
region, close to $\pm 1$. For instance, in SPS2,  $\Dmh = -5 \gev$
for $\de^{RR}_{ct}= \pm 1$.

In the case of $\DmH$, apart from $\de^{LR}_{sb}$ and $\de^{RL}_{sb}$, there is only noticeable  sensitivity to other deltas in  SPS5. The same comment applies to $\DmHp$.
\end{itemize}

\clearpage
\newpage
\clearpage
\newpage
\clearpage
\newpage
\clearpage
\newpage

\subsubsection{\boldmath{$\Dmh$} versus two \boldmath{$\deXYij\neq 0$}}

 Our previous results on the Higgs mass corrections show that the corrections to the lightest Higgs mass $\Dmh$ were negative in many of the studied cases and could be very large for some regions of the flavour changing deltas which in this pre-LHC situation were still allowed by $B$ data. These negative and large
 mass corrections, could lead to a prediction for the corrected one-loop mass in these kind of NMFV-SUSY scenarios, $m_{h}^{\rm NMFV} \simeq m_{h}^{\rm MSSM} + \Dmh$, which were indeed too low and already excluded by the pre-LHC data~\cite{LEPHiggsSM,LEPHiggsMSSM}. Therefore, interestingly, the study of these mass corrections could be conclusive in the setting of additional restrictions on the size of some flavour changing deltas which otherwise were not bounded from $B$ data. As we will see in the post-LHC situation, both the corrections positive and negative can be relevant, and will lead to constraints in some of the deltas.

In order to explore further the size of these `dangerous' mass corrections, we have computed numerically the size of  $\Dmh$ as a function of two non-vanishing deltas and have looked for areas in these two dimensional plots that were allowed by $B$ data. We show in \reffis{figdoubledeltaBFP}, \ref{figdoubledeltaSPS3},  \ref{figdoubledeltaSPS2},  \ref{figdoubledeltaSPS5},  \ref{figdoubledeltaVHeavyS}, and \ref{figdoubledeltaHeavySLightH} the numerical results of the $\Dmh$ contour-lines in two dimensional plots, 
$(\de^{LL}_{23},\deXYij)$, for the respective pre-LHC points BFP, SPS2, SPS3, SPS5, VHeavyS and HeavySLightH of table \ref{points}.

\begin{figure}[H]  
\centering 
{\resizebox{13cm}{!}  
{\begin{tabular}{cccc}  
\includegraphics[width=13.2cm]{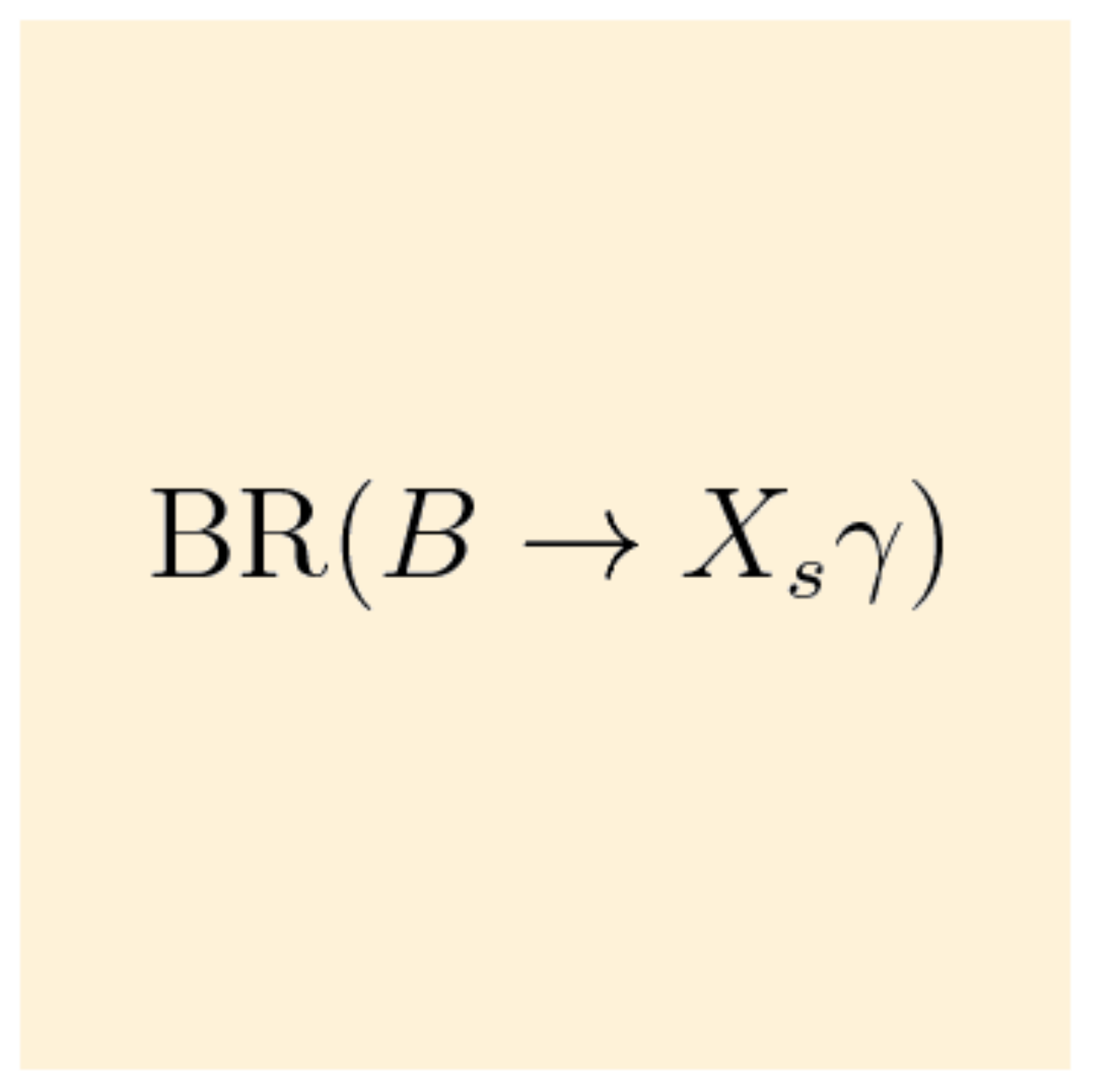}&  
\includegraphics[width=13.2cm]{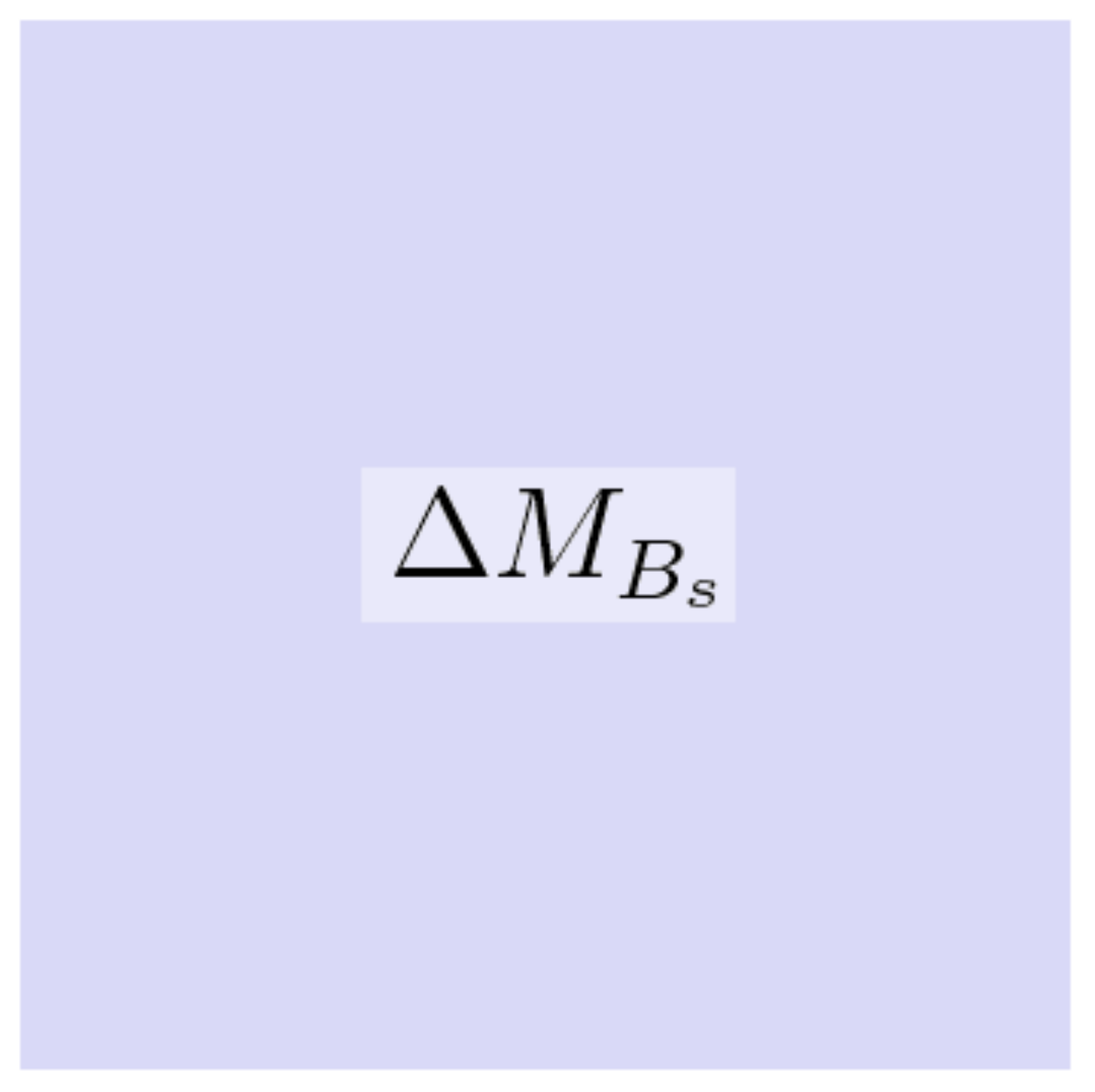}&  
\includegraphics[width=13.2cm]{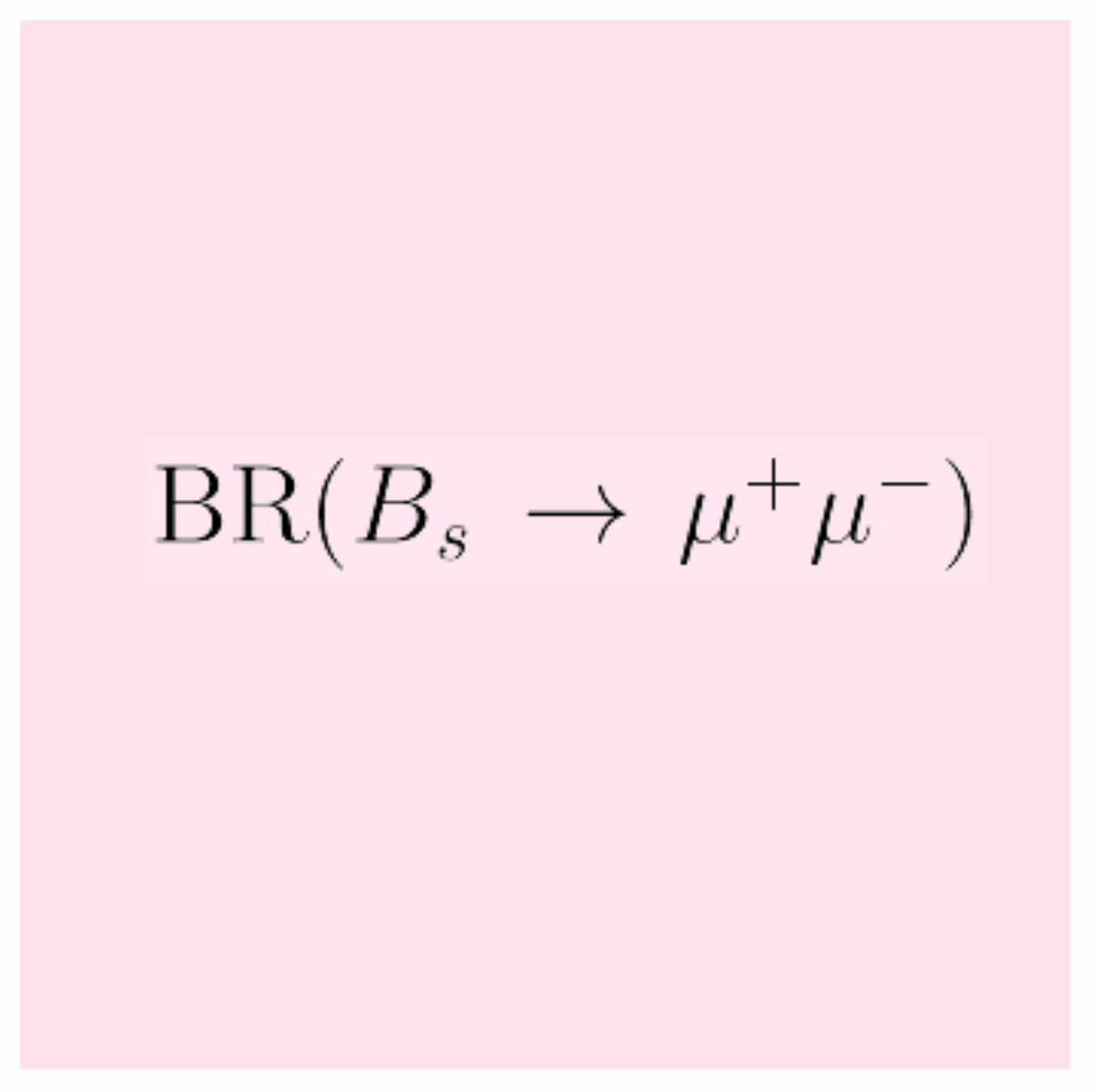}& 
\includegraphics[width=13.2cm]{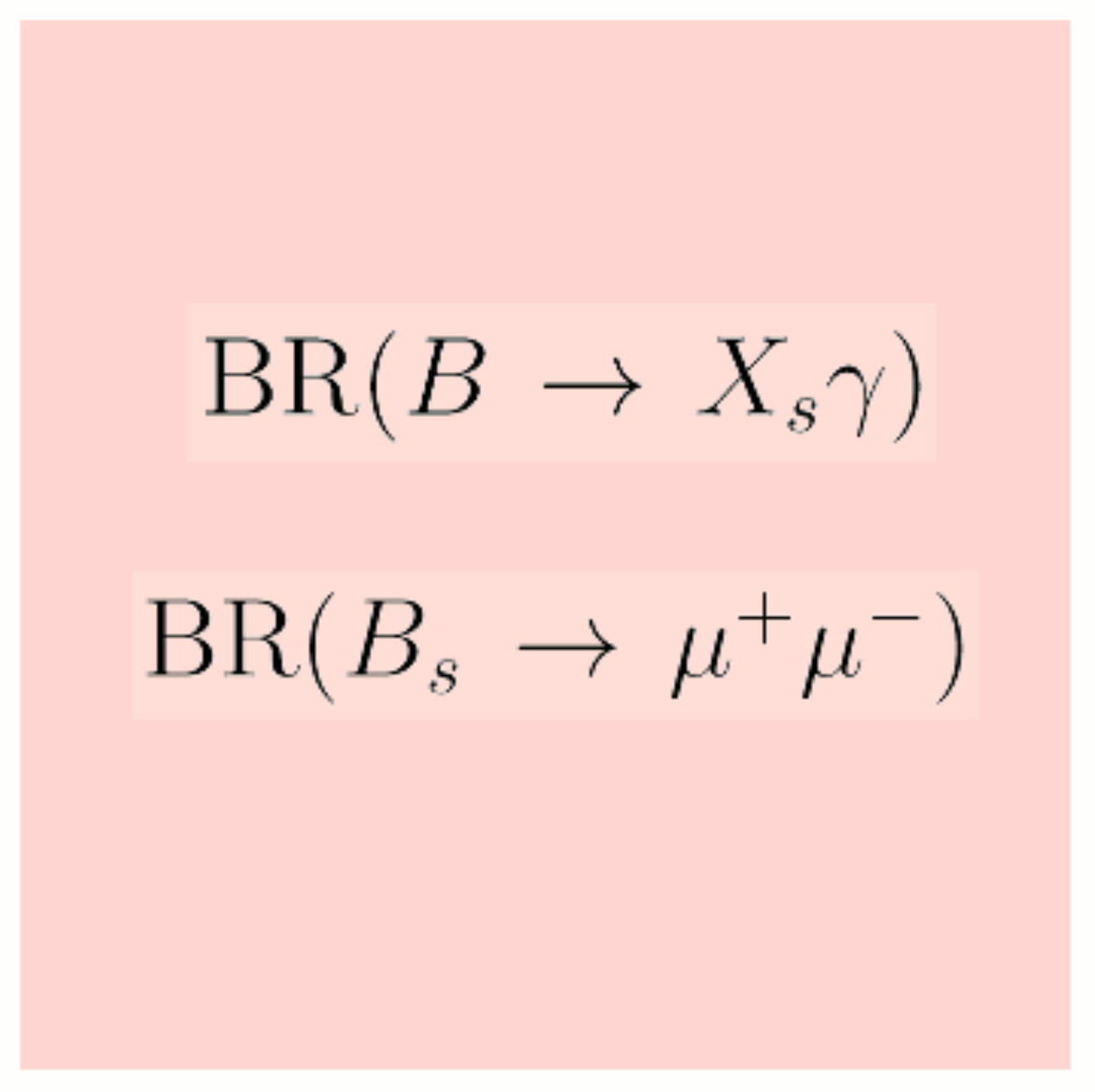}\\
\includegraphics[width=13.2cm]{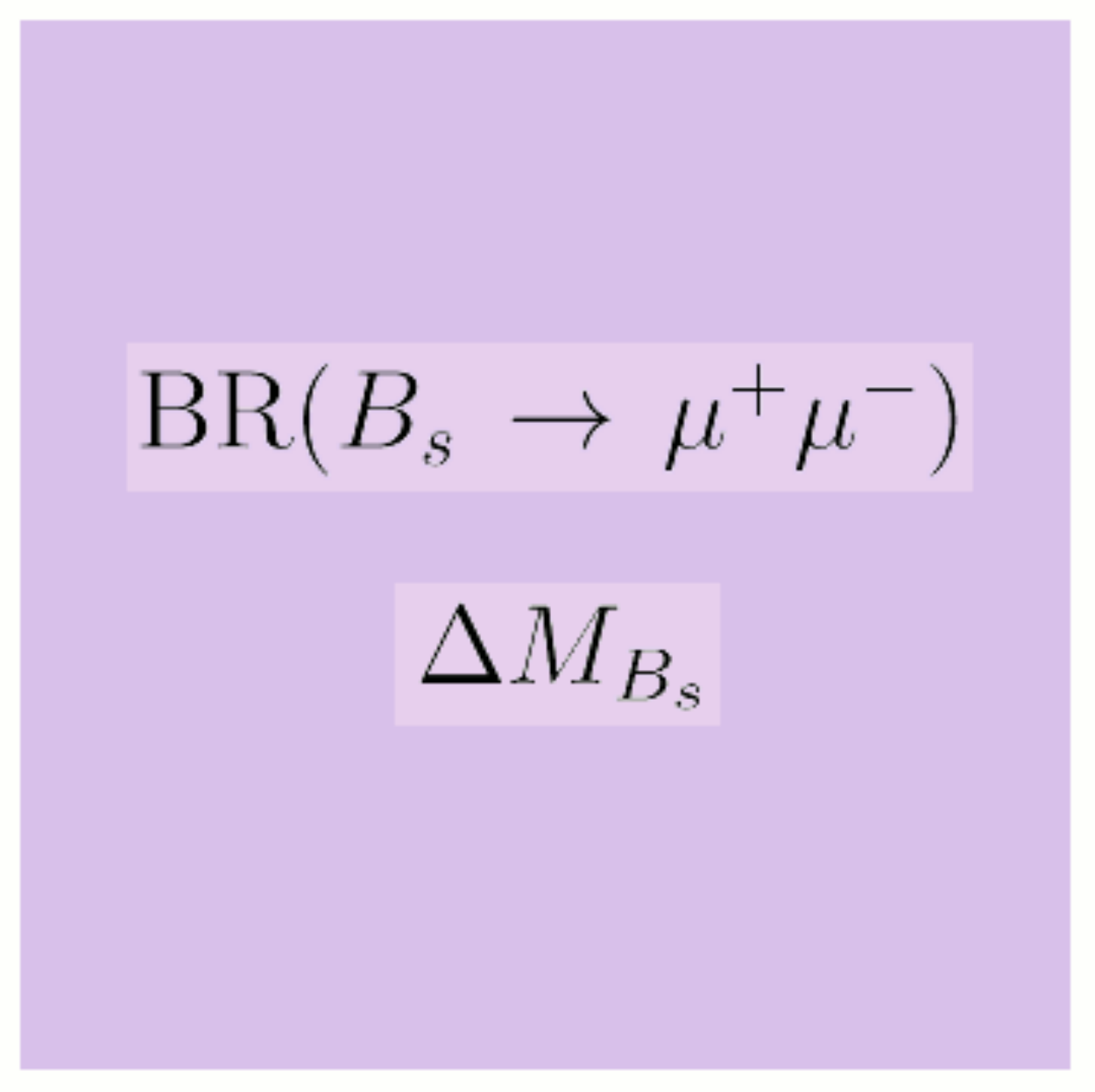}&  
\includegraphics[width=13.2cm]{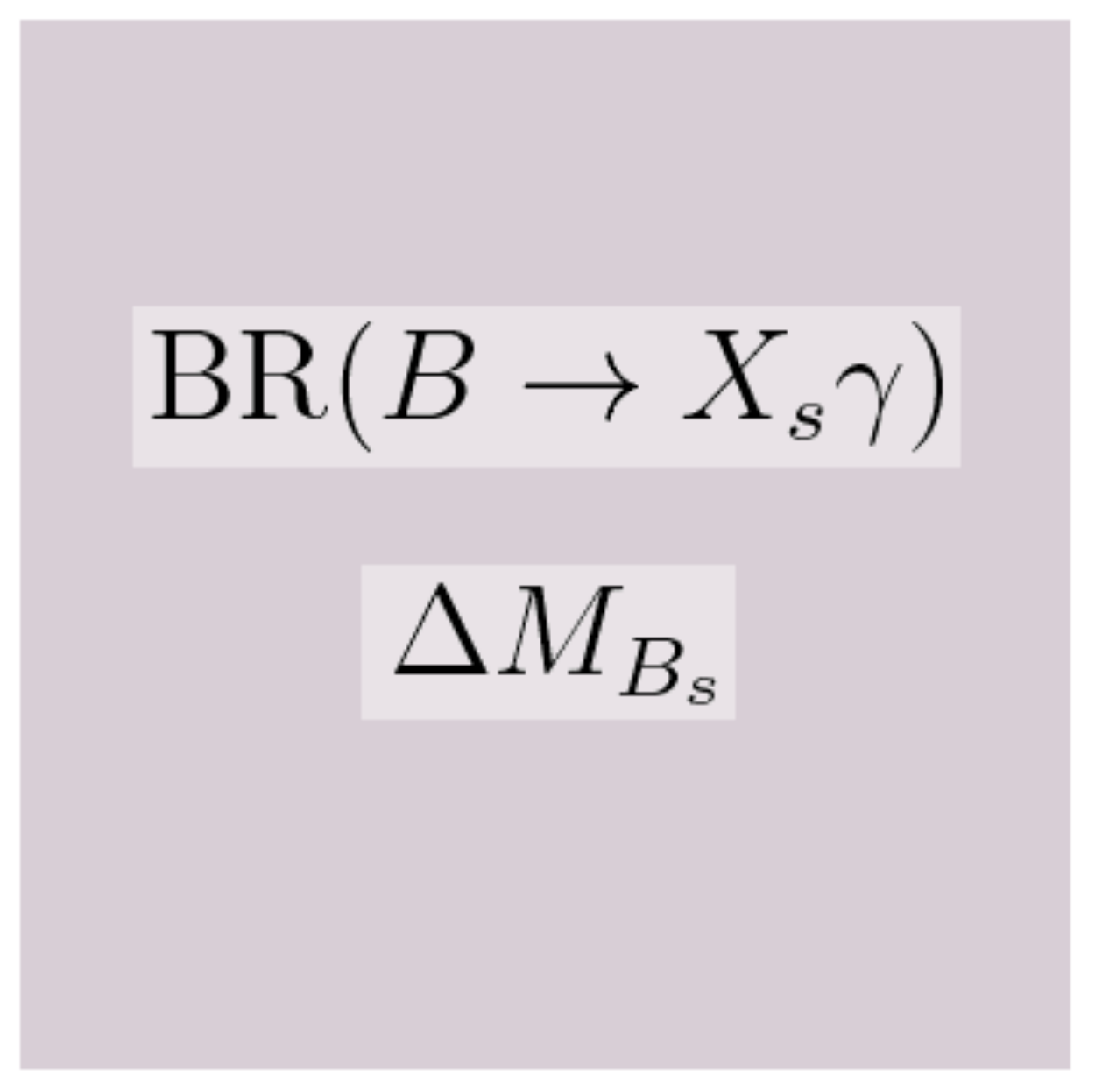}& 
\includegraphics[width=13.2cm]{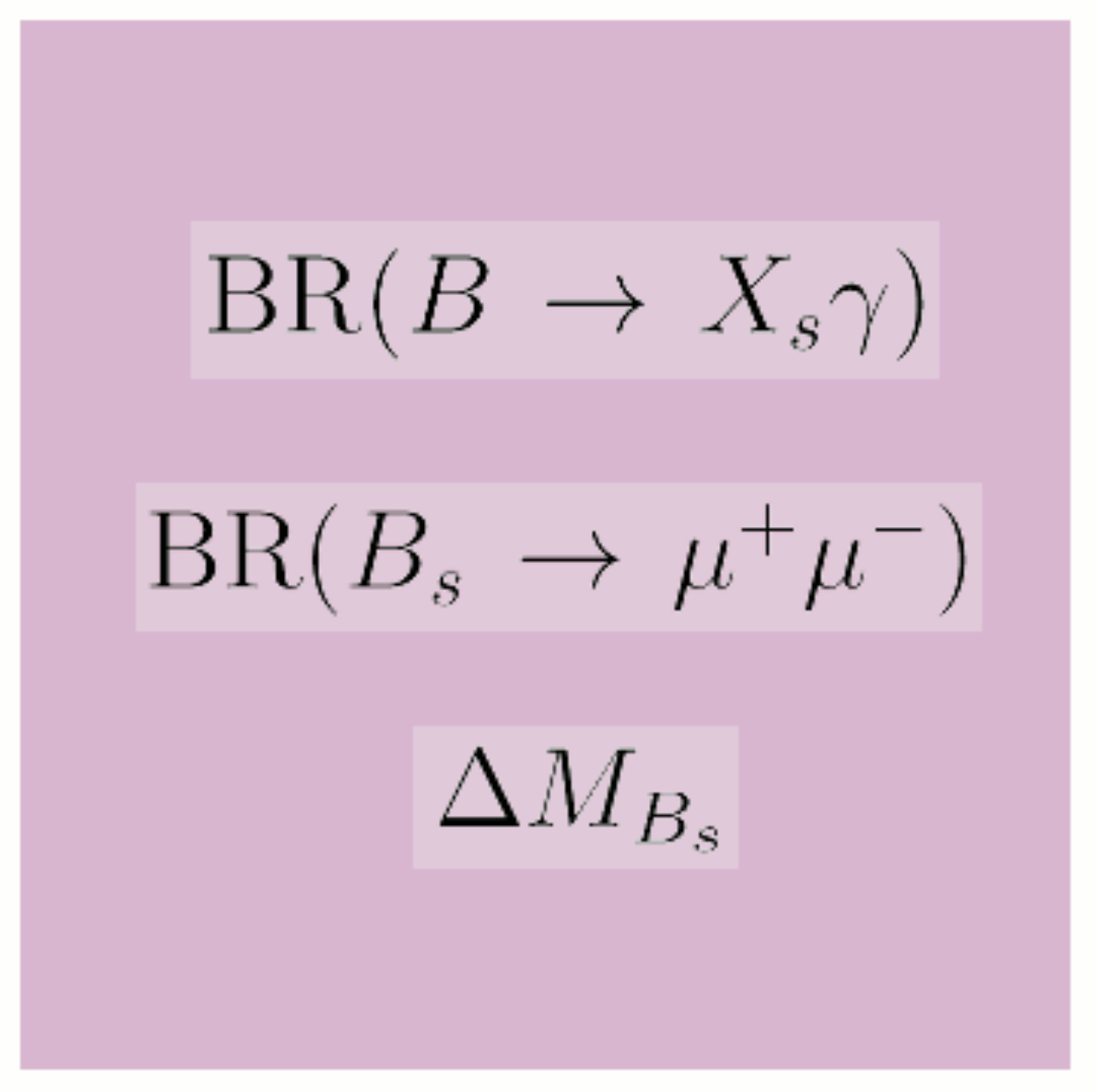}&
\includegraphics[width=13.2cm]{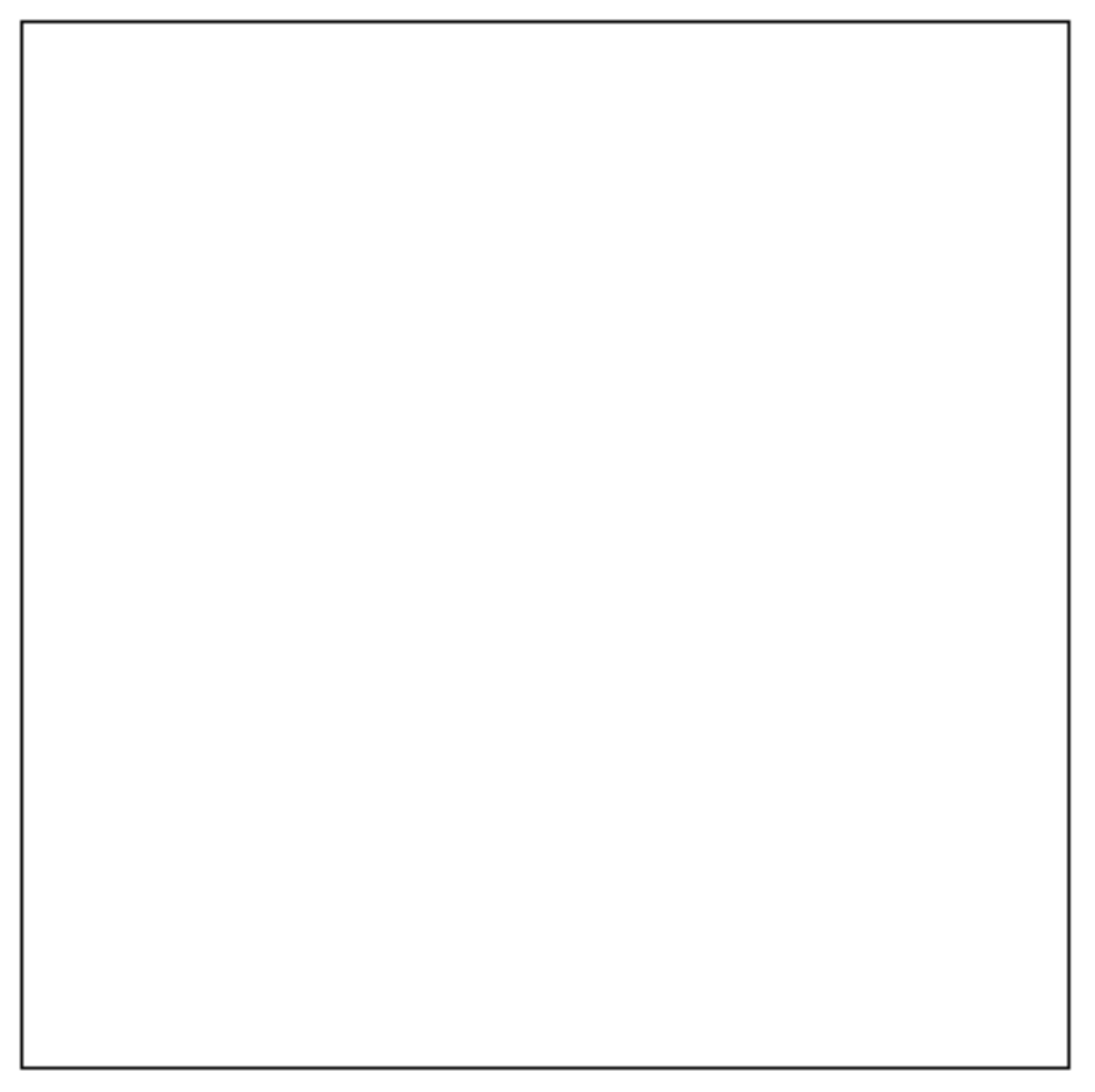}\\
\end{tabular}}} 
\caption{Legend for plots of Higgs mass corrections varying two deltas simultaneously displayed in figs. \ref{figdoubledeltaBFP}, \ref{figdoubledeltaSPS3}, \ref{figdoubledeltaSPS2}, \ref{figdoubledeltaSPS5}, \ref{figdoubledeltaVHeavyS} and \ref{figdoubledeltaHeavySLightH}. Each coloured area represents the disallowed region by the specified observable/s inside each box. A white area placed at the central regions of the mentioned figures represents a region allowed by the three $B$ observables. A white area placed outside the coloured areas represent regions of the parameter space that generate negative squared masses.  These problematic points are consequently not shown in our plots, as we did in the previous plots.
The discontinuous lines in those figures represent the contour lines for the $B$ observables corresponding to the maximum and minimum pre-LHC allowed values: dash-dot-dash for the upper bound of \bsg\ (Eq.  (\ref{bsglinearerr})), dot-dash-dot for the lower bound of \bsg\ (Eq. (\ref{bsglinearerr})), dashed line for the upper bound of $\dmbs$ (Eq. (\ref{deltabslinearerr})), a sequential three dotted line for  the lower bound of $\dmbs$ (Eq. (\ref{deltabslinearerr})), and a dotted line for the upper bound of \bmm\ (Eq. (\ref{bsmmlinearerr})).}  
\label{colleg} 
\end{figure} 

\clearpage
\newpage
\begin{figure}[h!] 
\centering
\hspace*{-10mm} 
{\resizebox{14.6cm}{!} 
{\begin{tabular}{cc} 
\includegraphics[width=13.2cm]{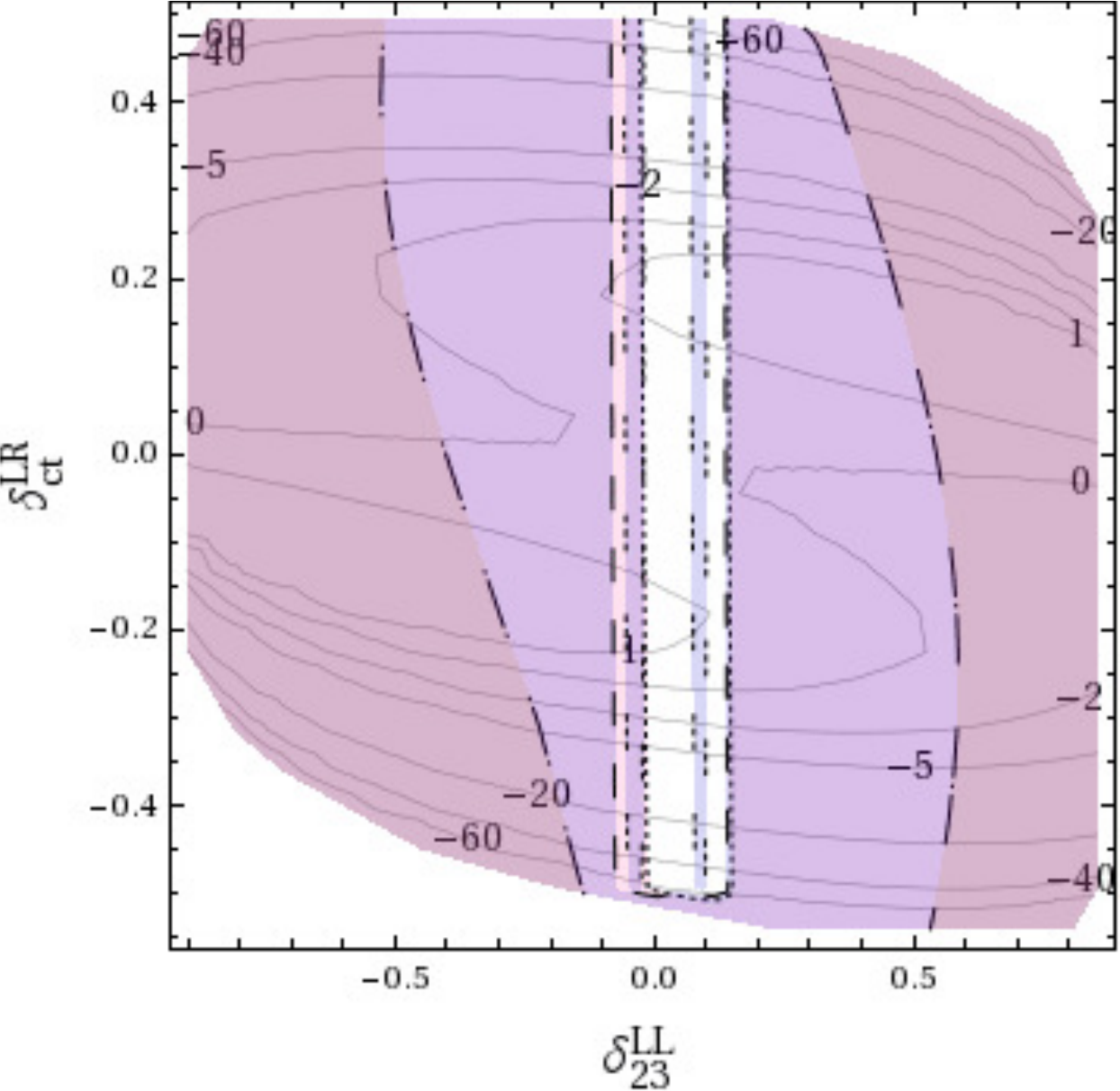}& 
\includegraphics[width=13.2cm]{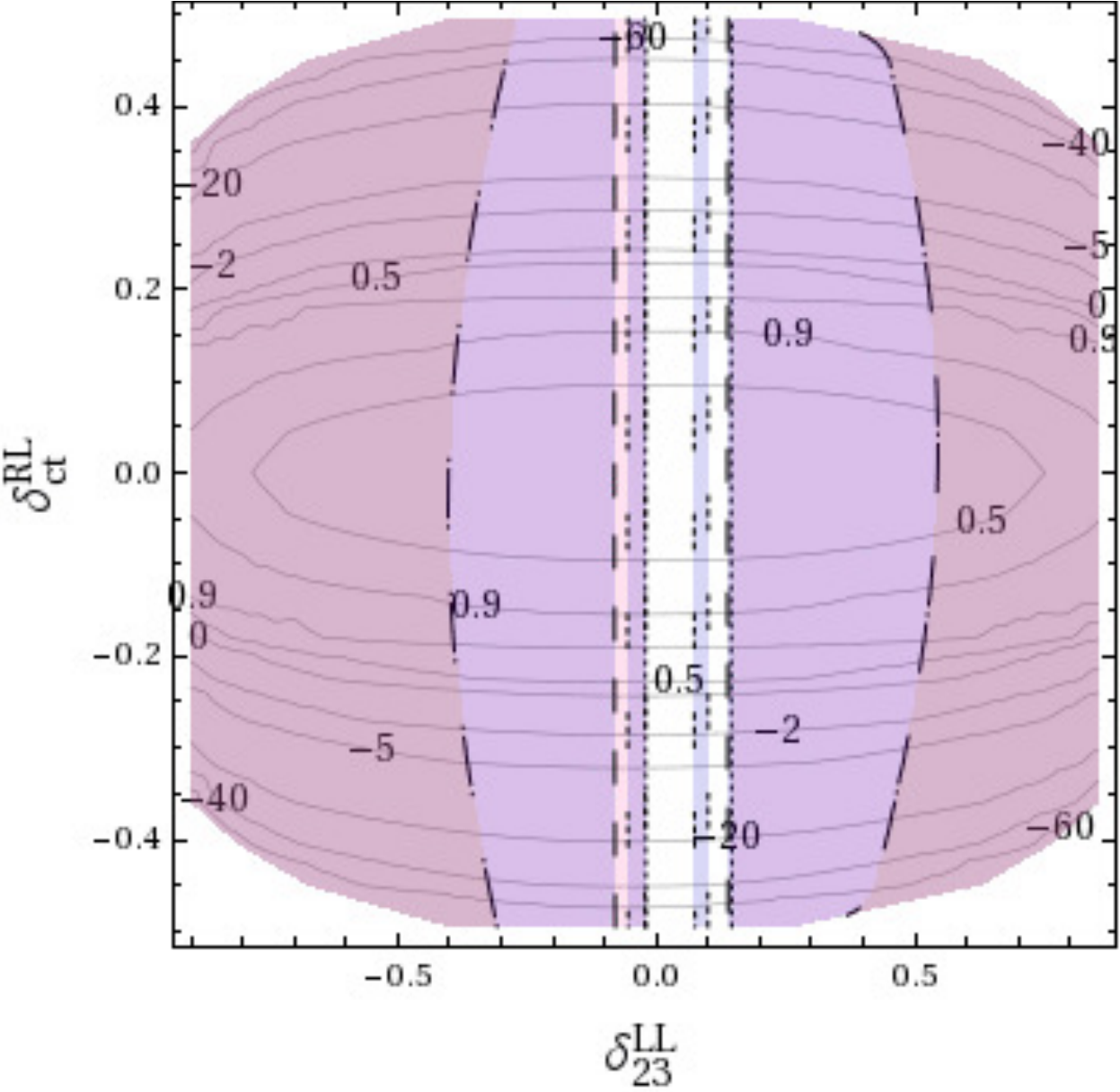}\\ 
\includegraphics[width=13.2cm]{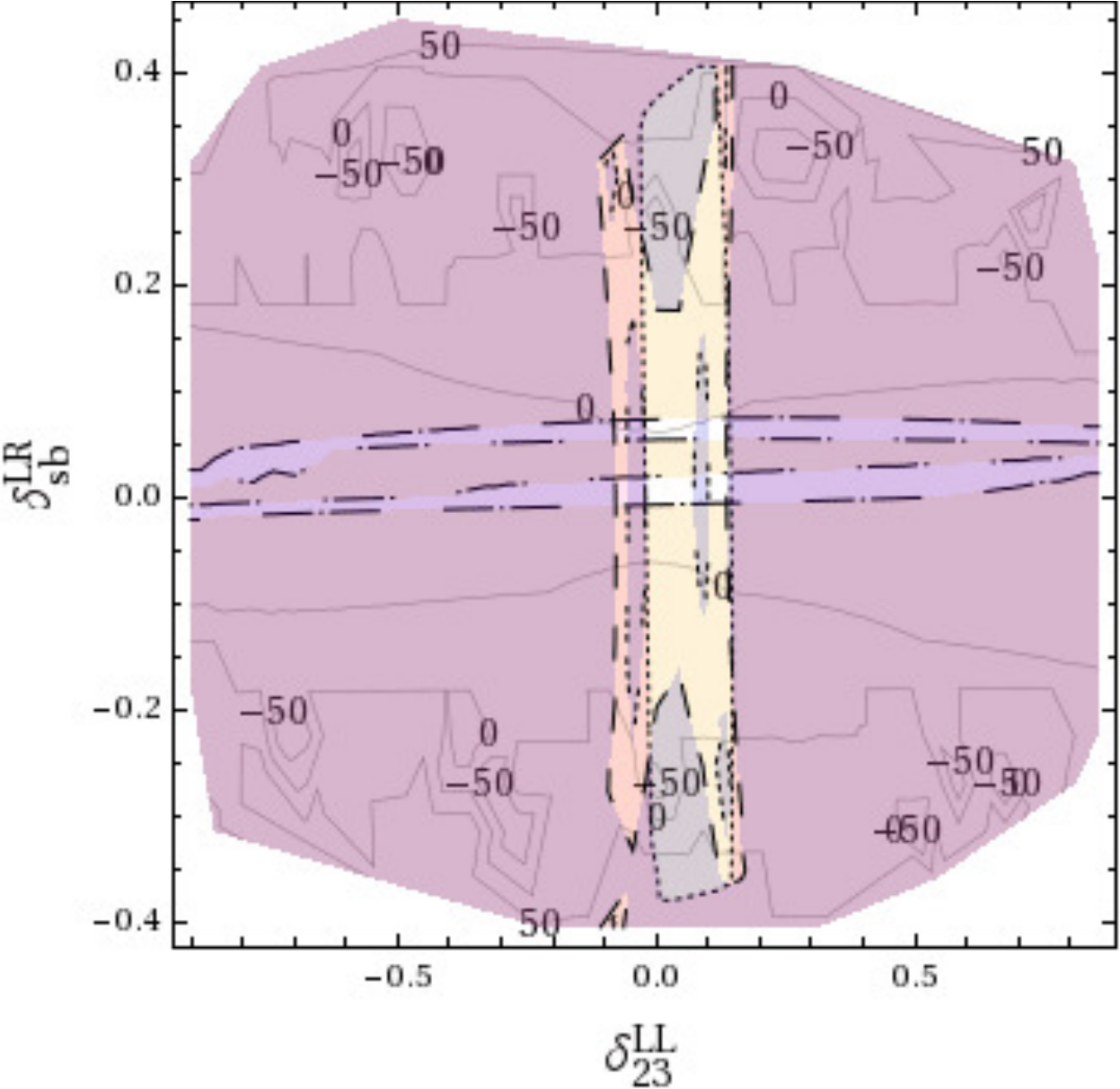}&
\includegraphics[width=13.2cm]{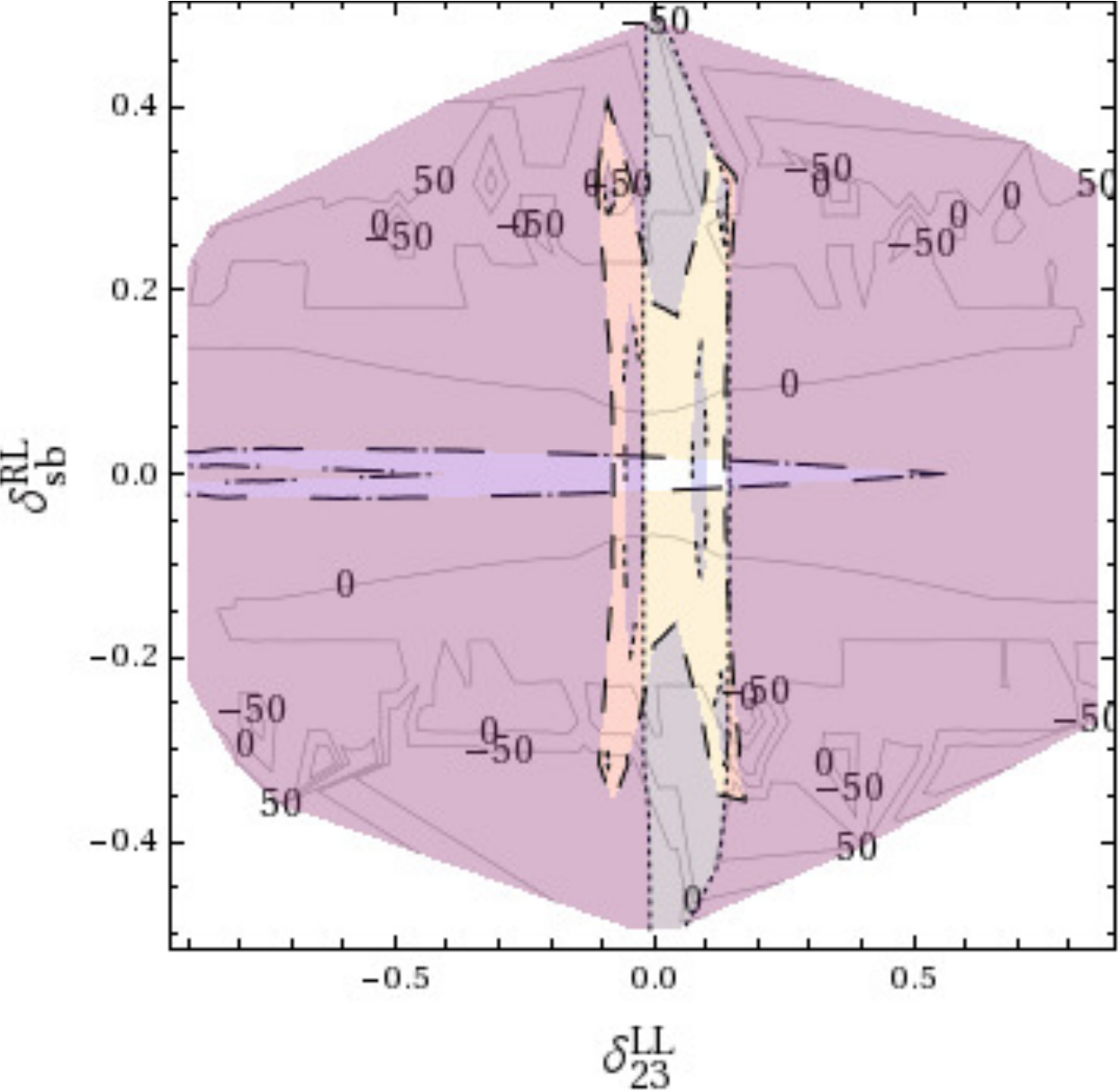}\\ 
\includegraphics[width=13.2cm]{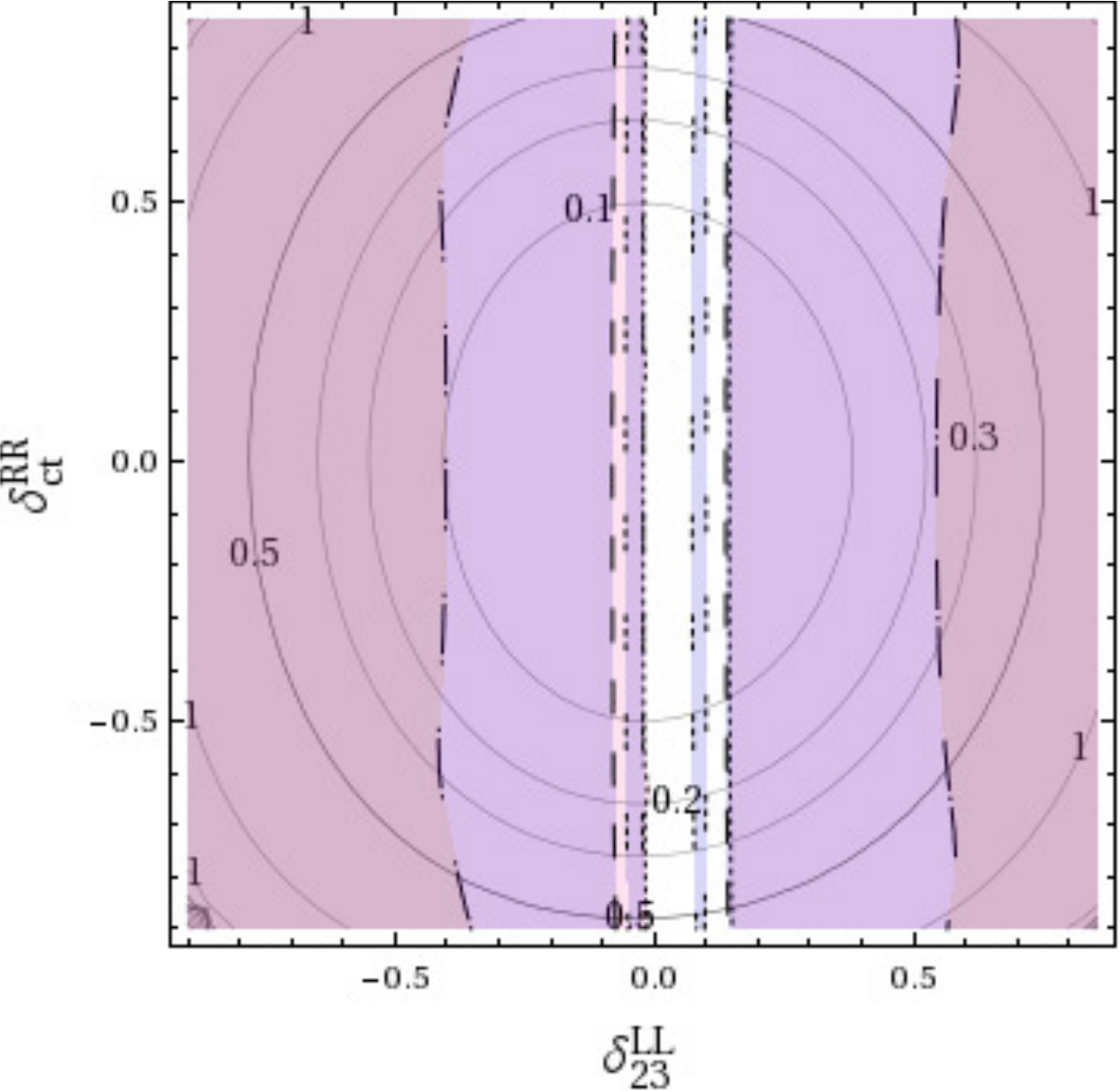}& 
\includegraphics[width=13.2cm]{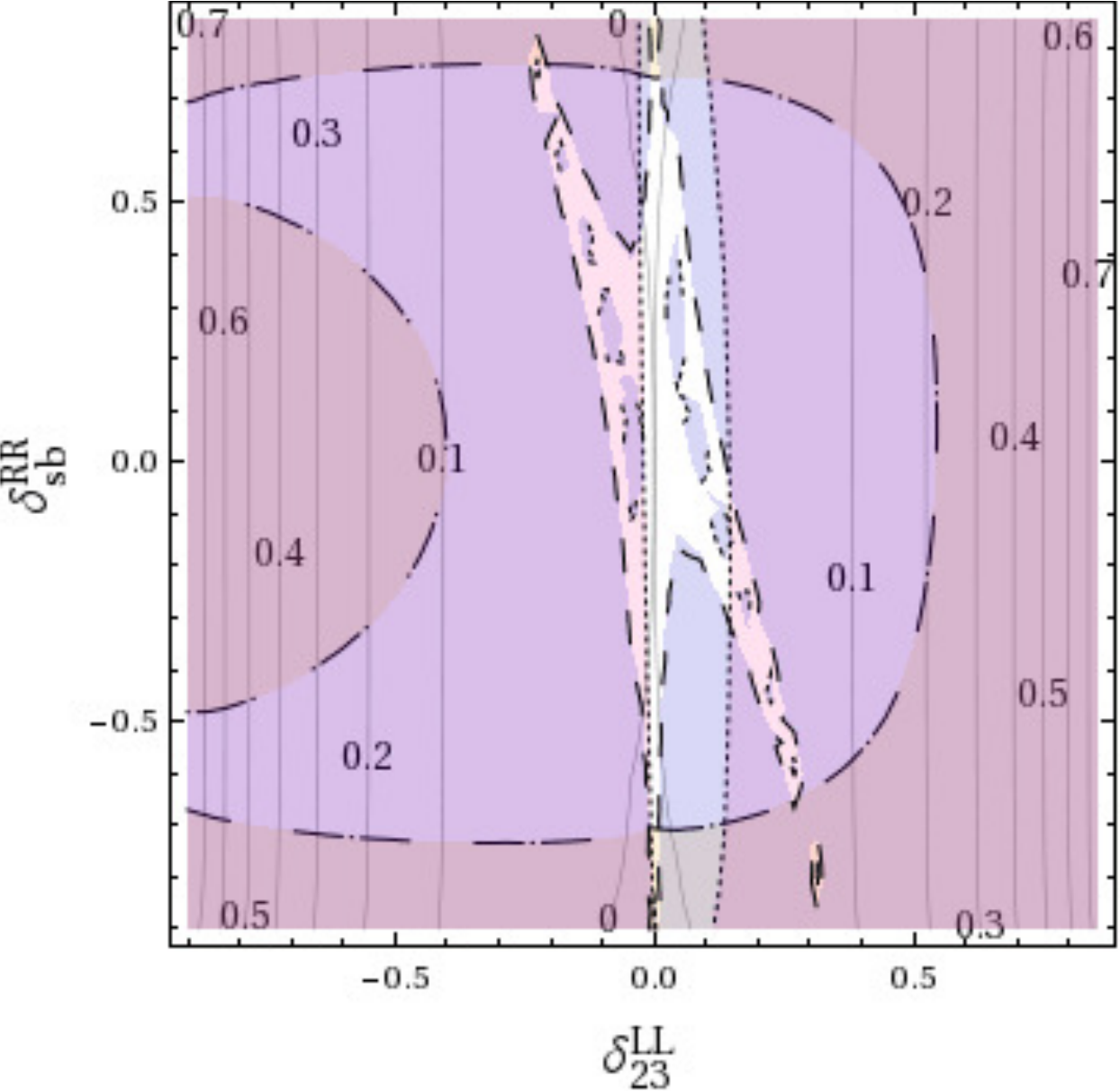}\\ 
\end{tabular}}}
\caption{$\Delta m_{h}$ (GeV) contour lines from our two deltas analysis for BFP. The colour code for the allowed/disallowed areas by pre-LHC $B$ data is given in Fig.\ref{colleg}.} 
 \label{figdoubledeltaBFP}
\end{figure}

\clearpage
\newpage
\begin{figure}[h!] 
\centering
\hspace*{-10mm} 
{\resizebox{14.6cm}{!} 
{\begin{tabular}{cc} 
\includegraphics[width=13.2cm]{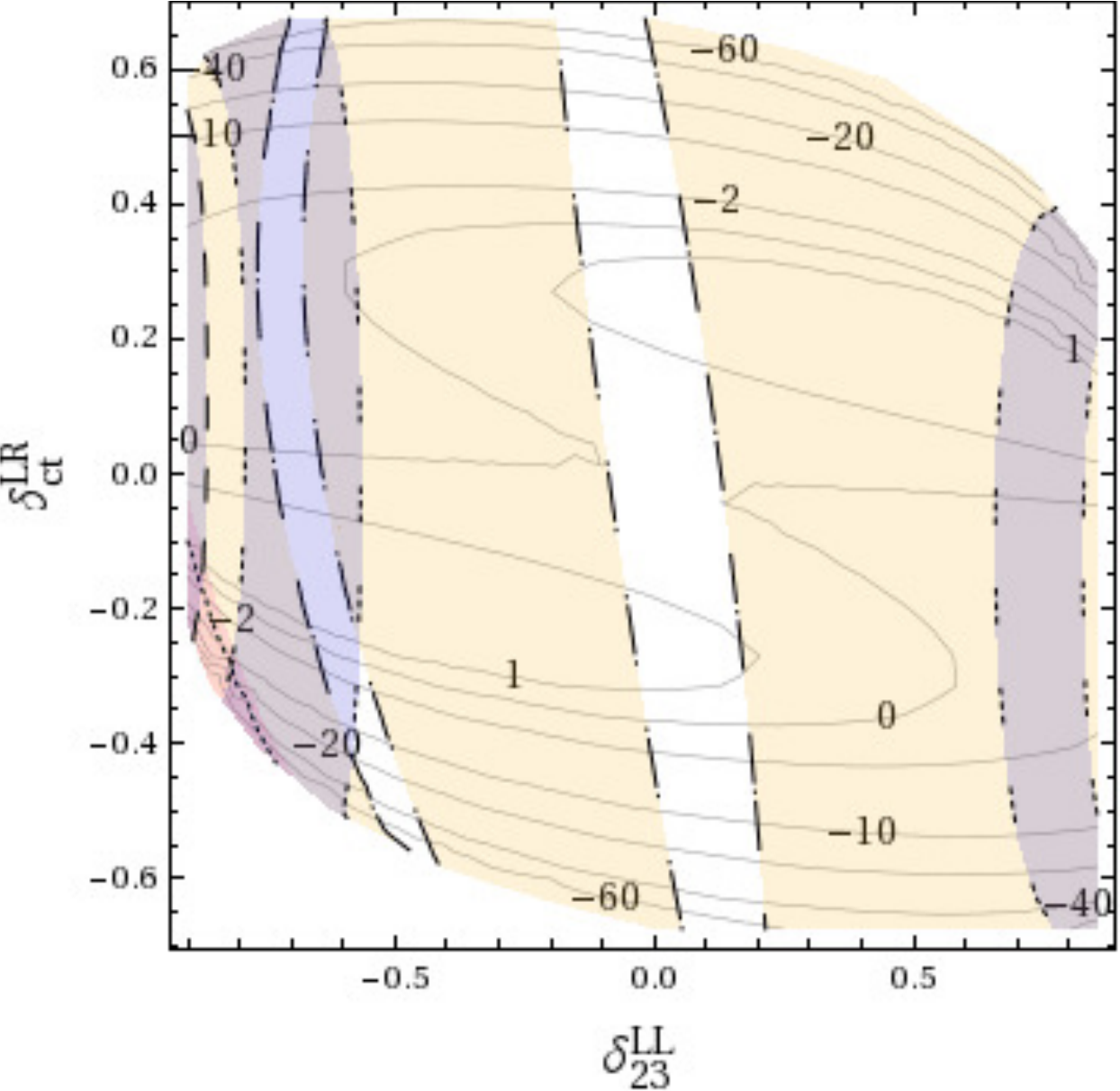}& 
\includegraphics[width=13.2cm]{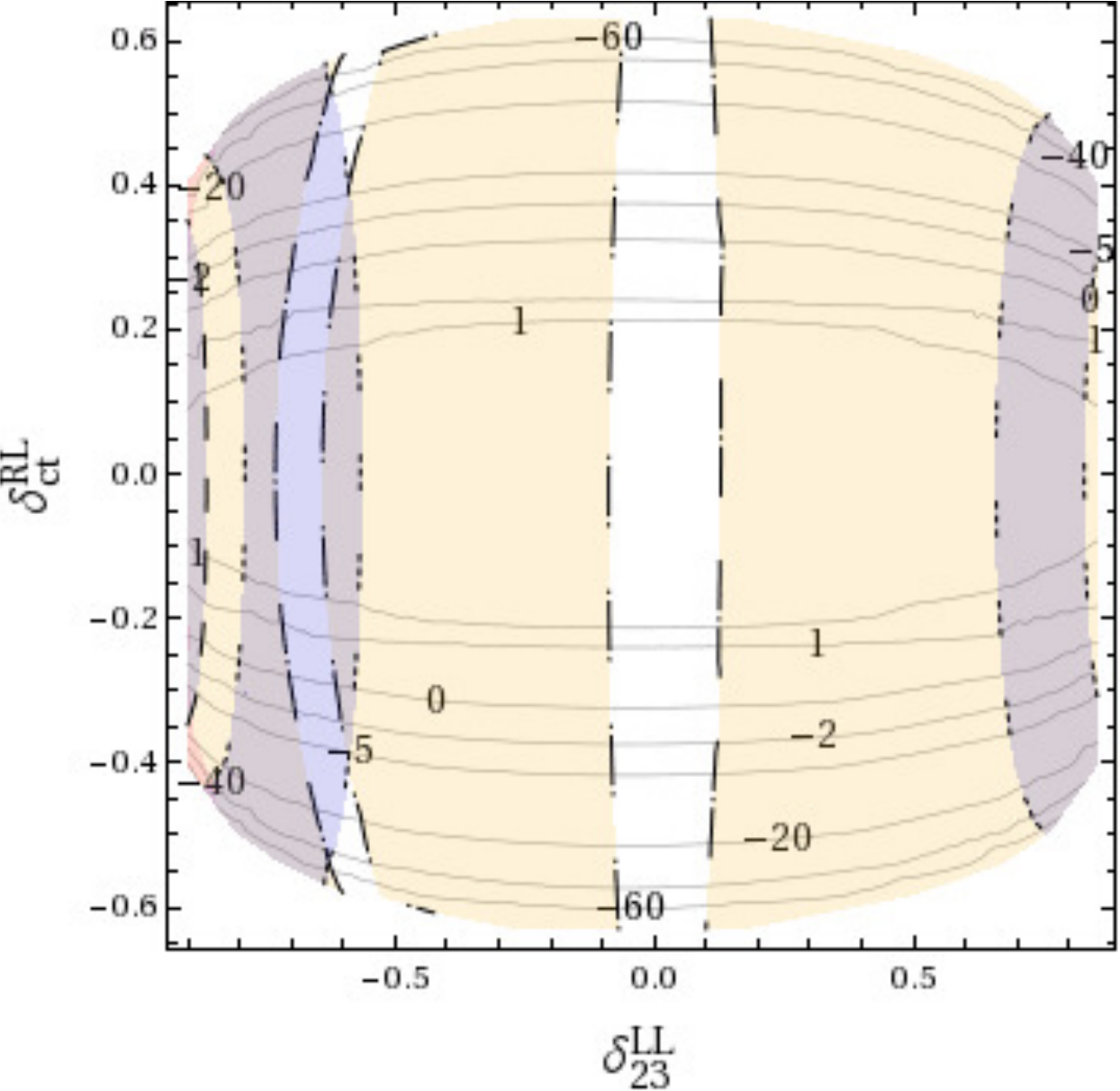}\\ 
\includegraphics[width=13.2cm]{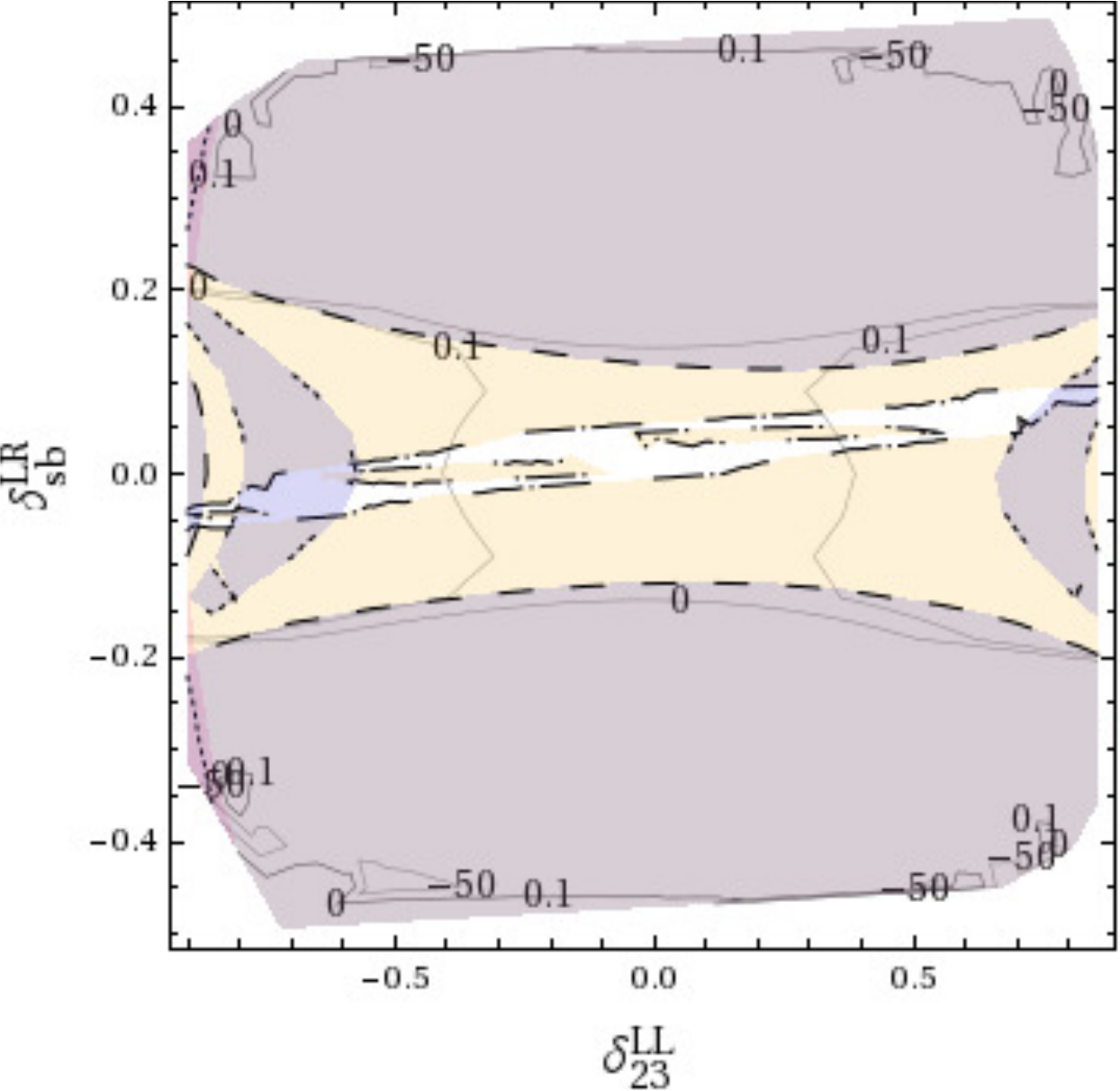}&
\includegraphics[width=13.2cm]{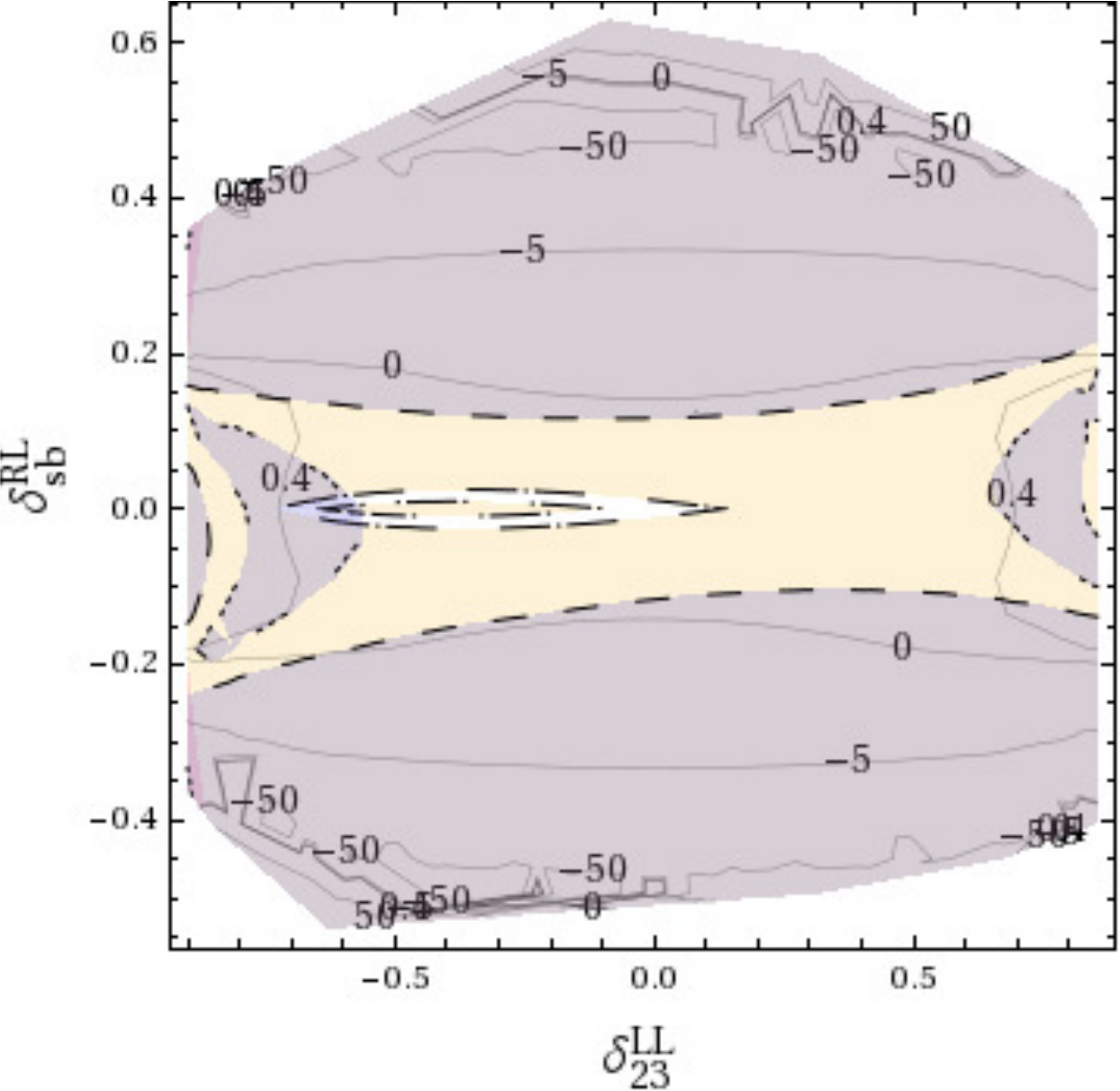}\\ 
\includegraphics[width=13.2cm]{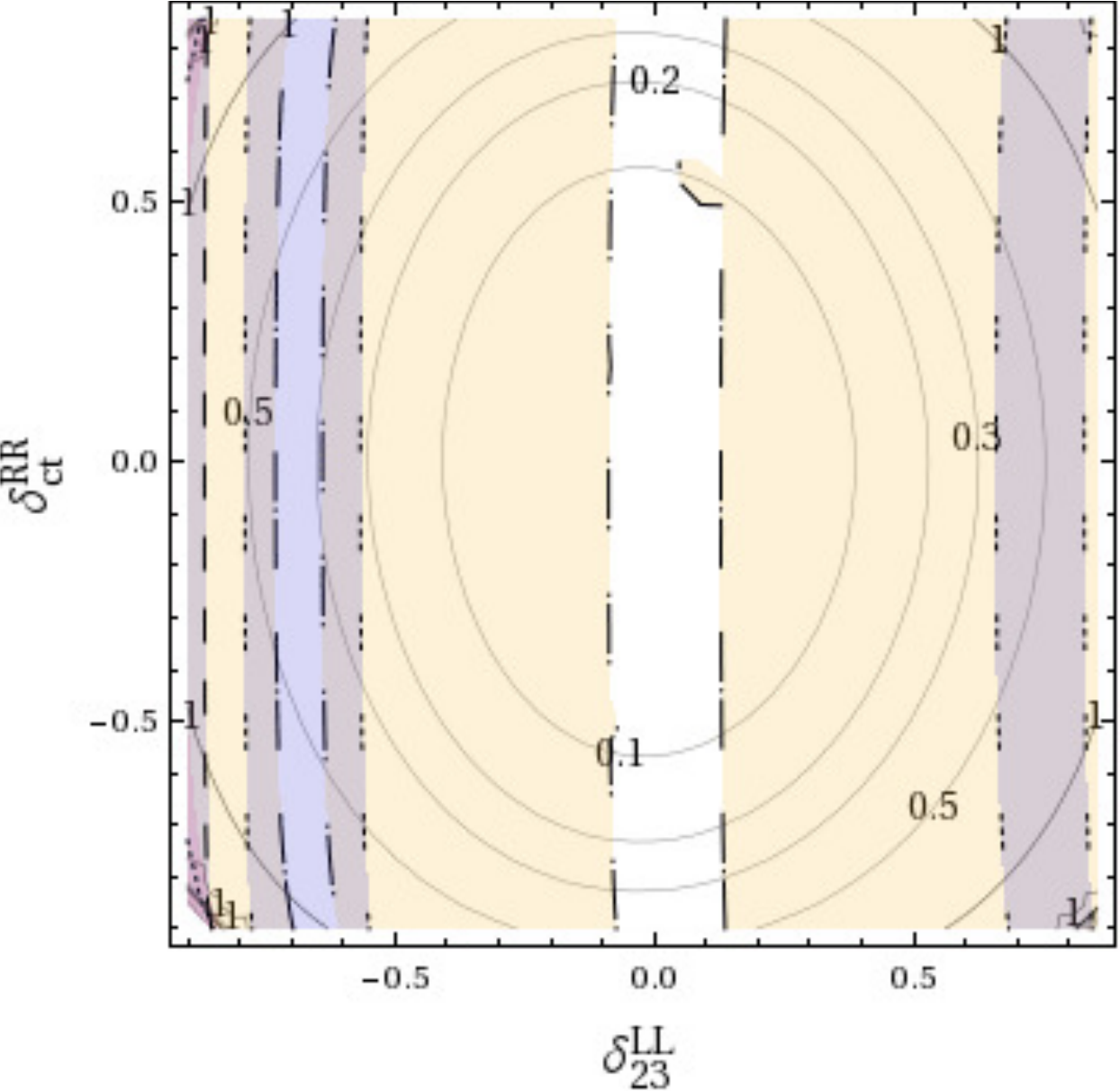}& 
\includegraphics[width=13.2cm]{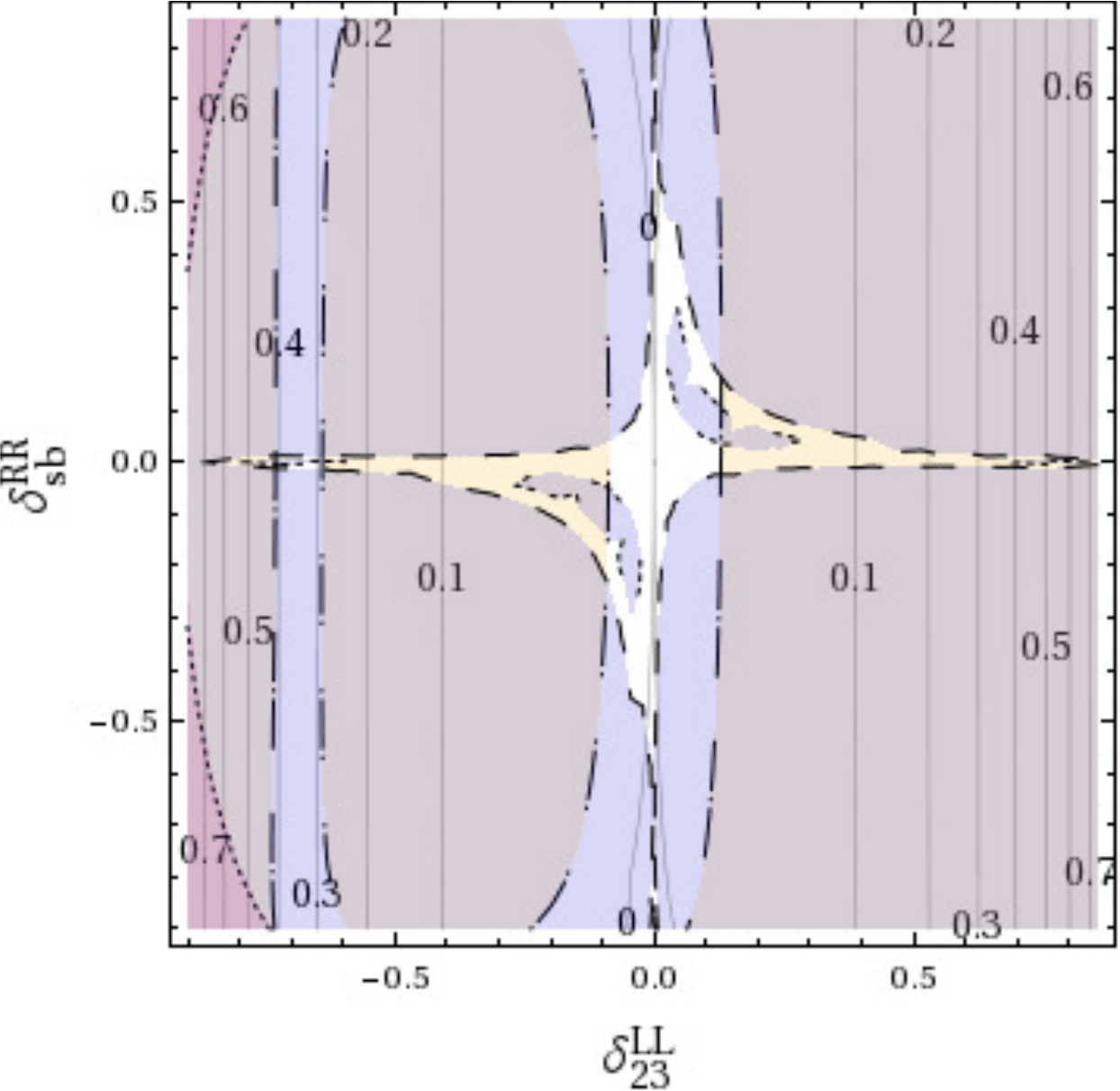}\\ 
\end{tabular}}}
\caption{$\Delta m_{h}$ (GeV) contour lines from our two deltas analysis for SPS3. The colour code for the allowed/disallowed areas by pre-LHC $B$ data is given in Fig.\ref{colleg}.} 
 \label{figdoubledeltaSPS3} 
\end{figure}
\clearpage
\newpage
\begin{figure}[h!] 
\centering
\hspace*{-10mm} 
{\resizebox{14.6cm}{!} 
{\begin{tabular}{cc} 
\includegraphics[width=13.2cm]{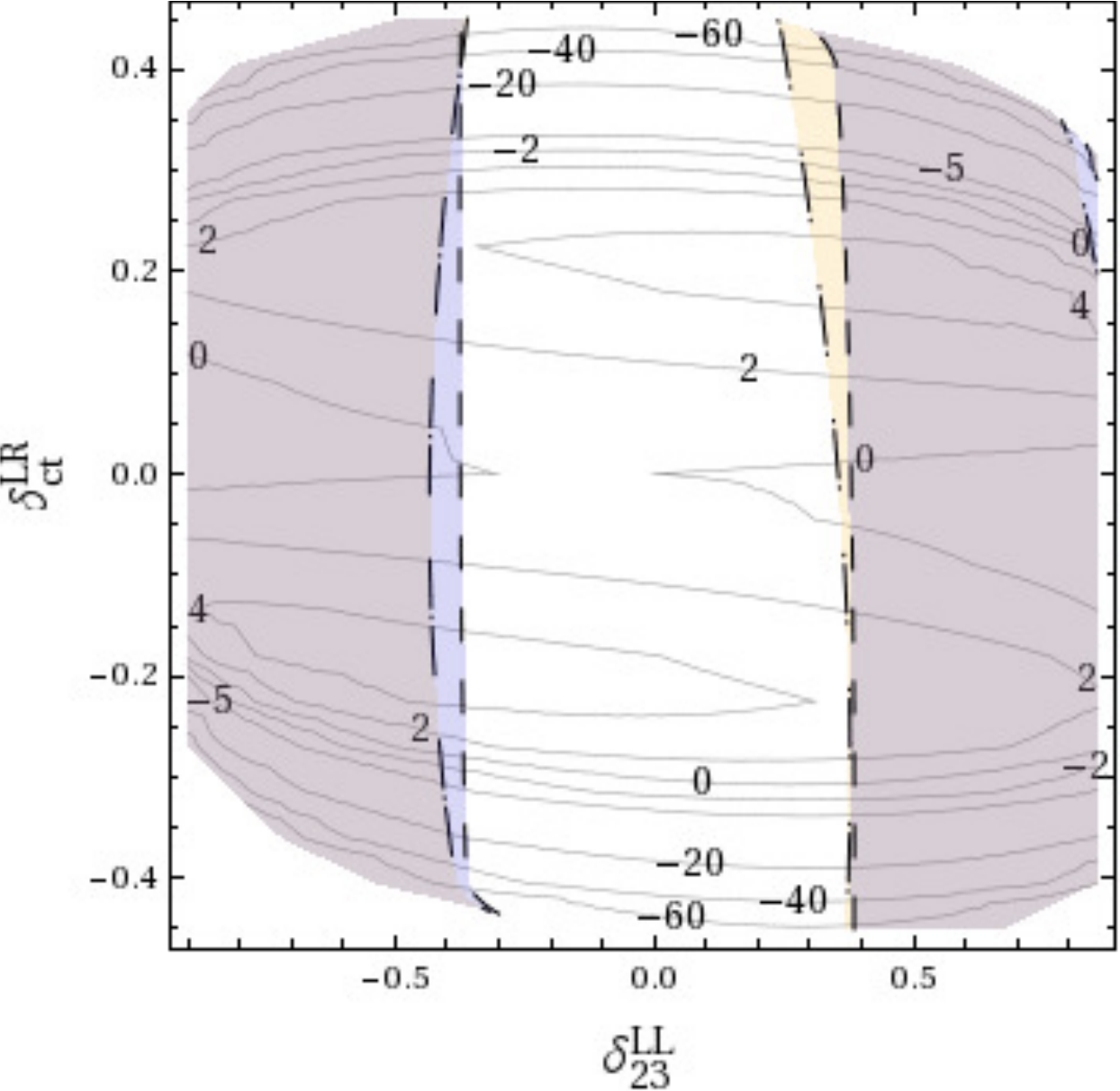}& 
\includegraphics[width=13.2cm]{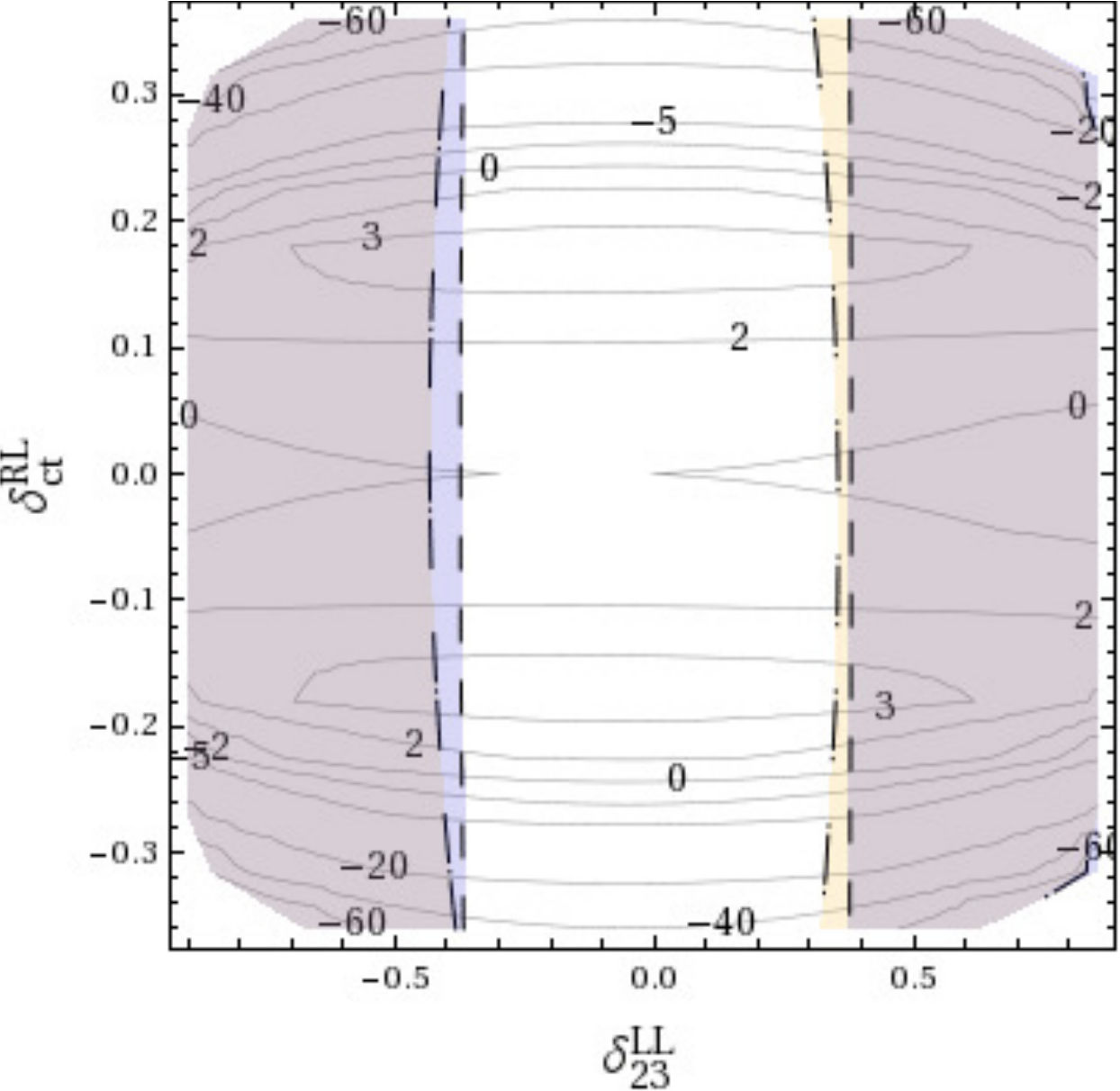}\\ 
\includegraphics[width=13.2cm]{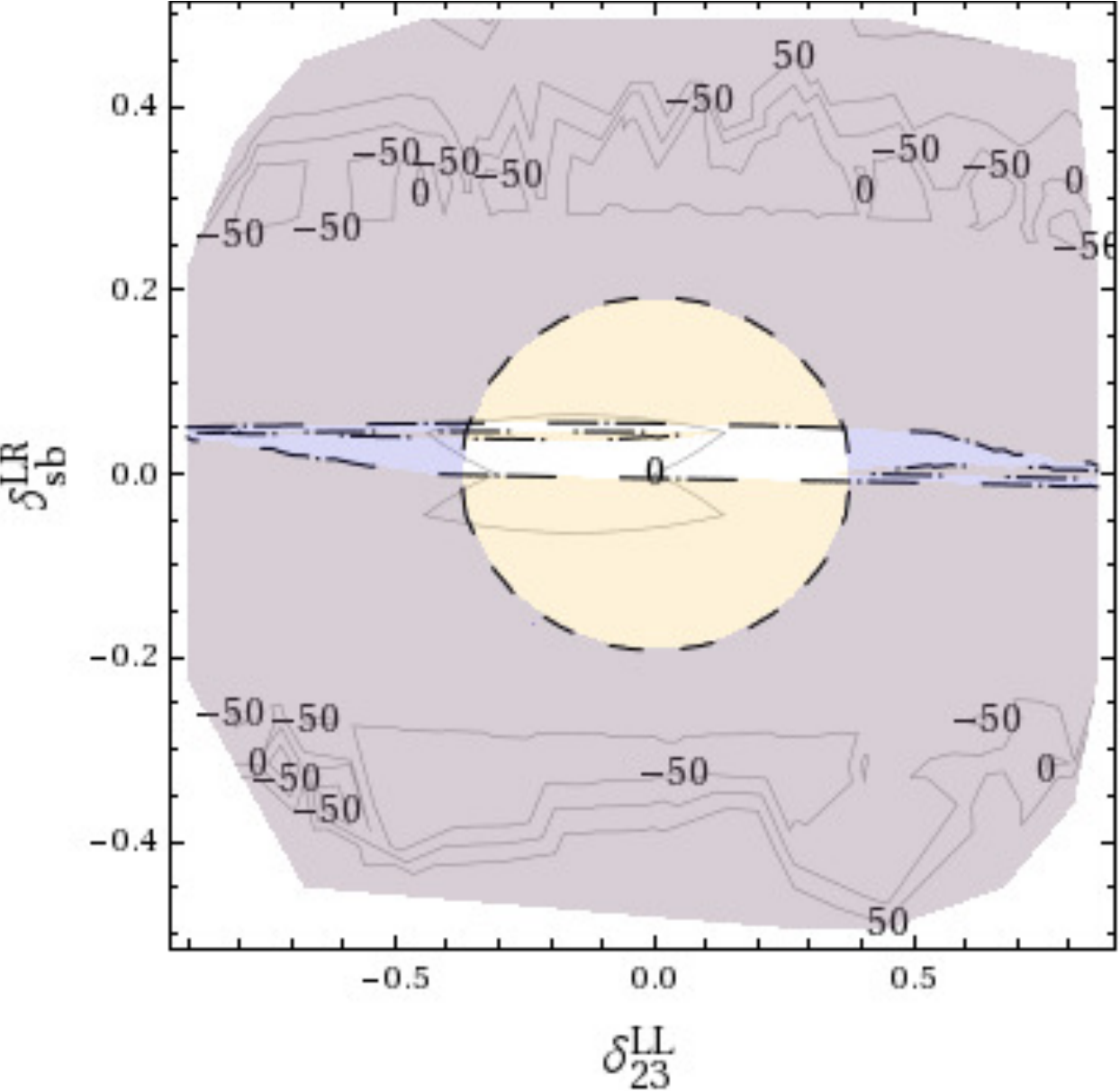}&
\includegraphics[width=13.2cm]{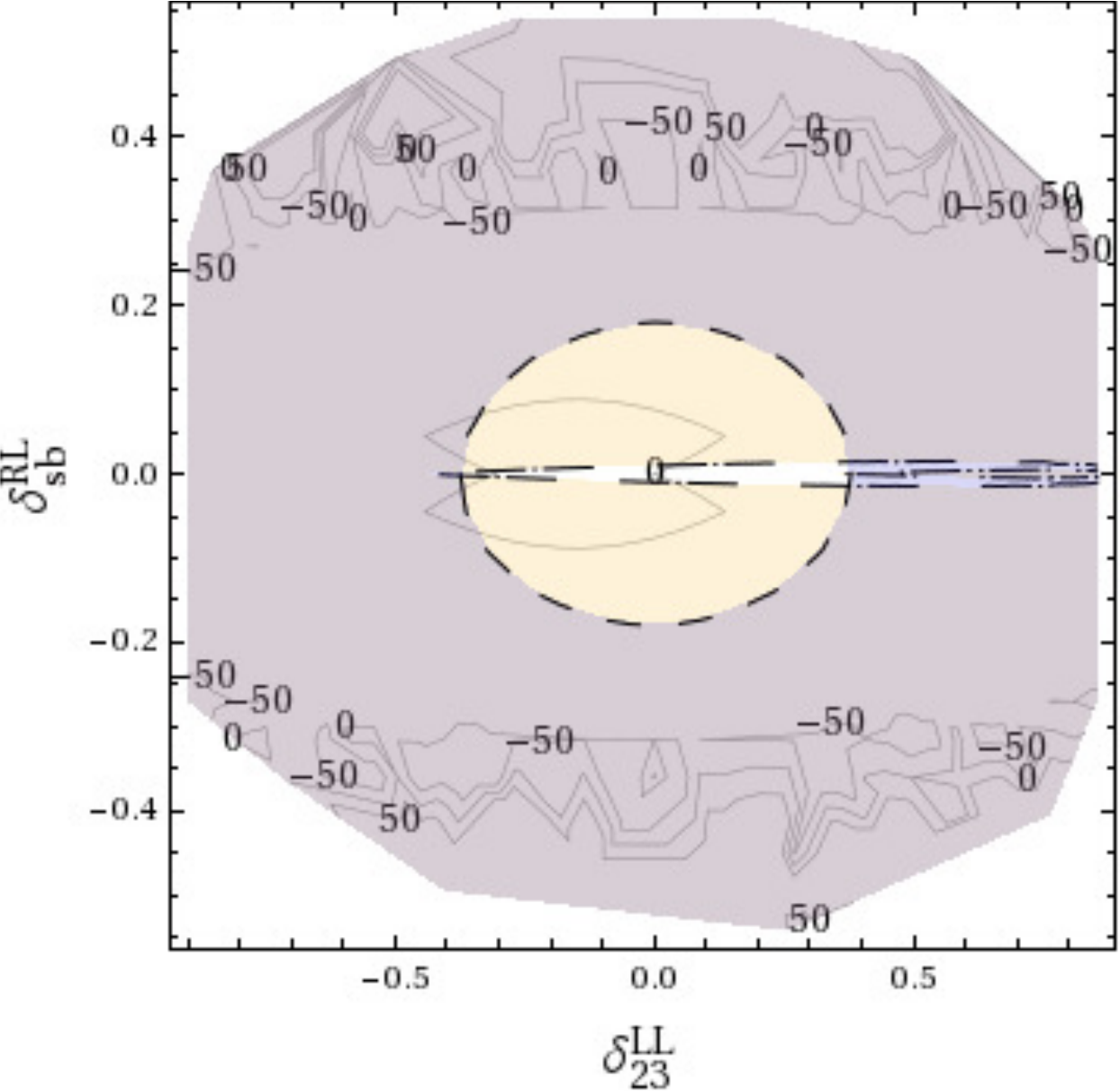}\\ 
\includegraphics[width=13.2cm]{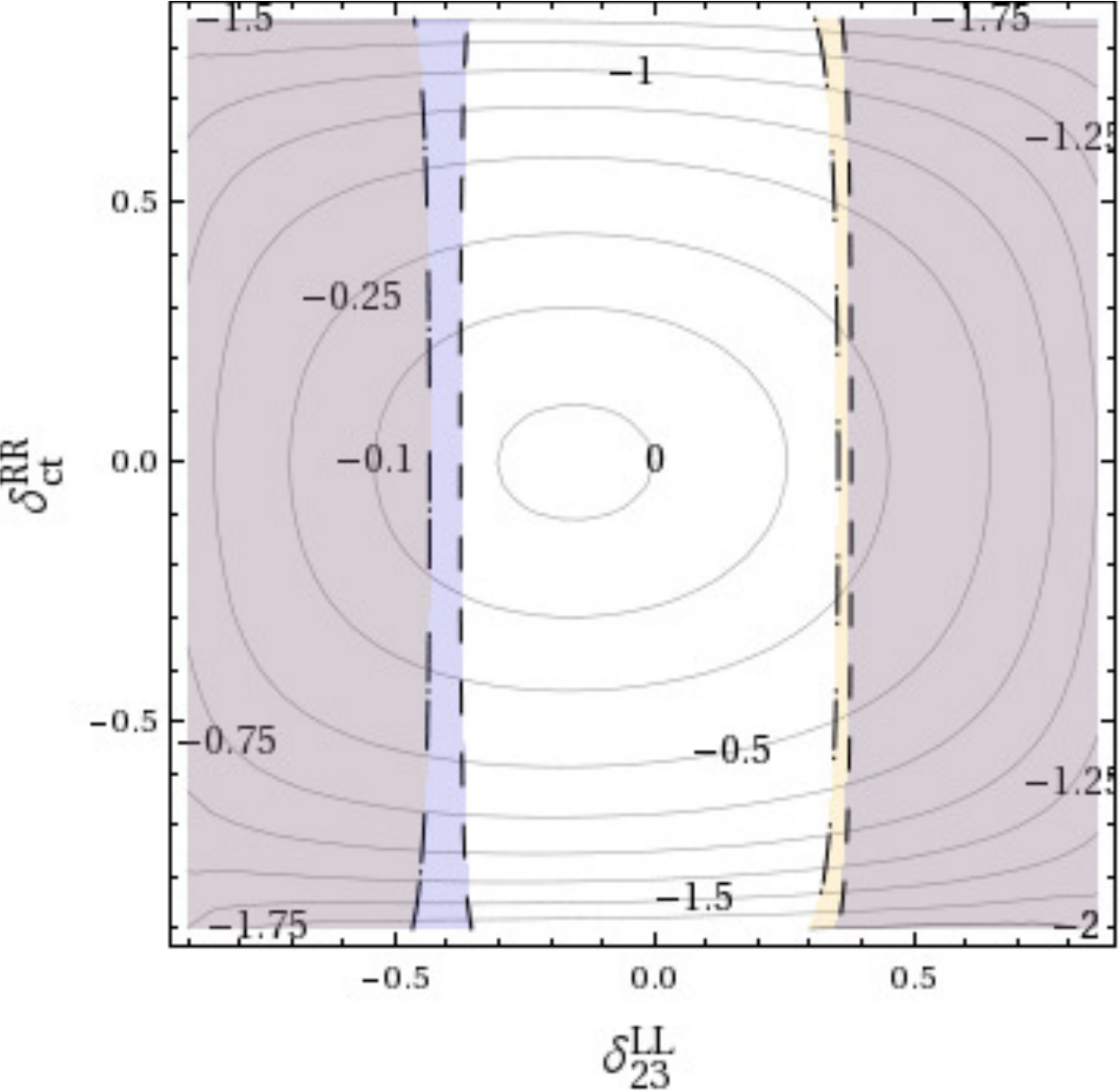}& 
\includegraphics[width=13.2cm]{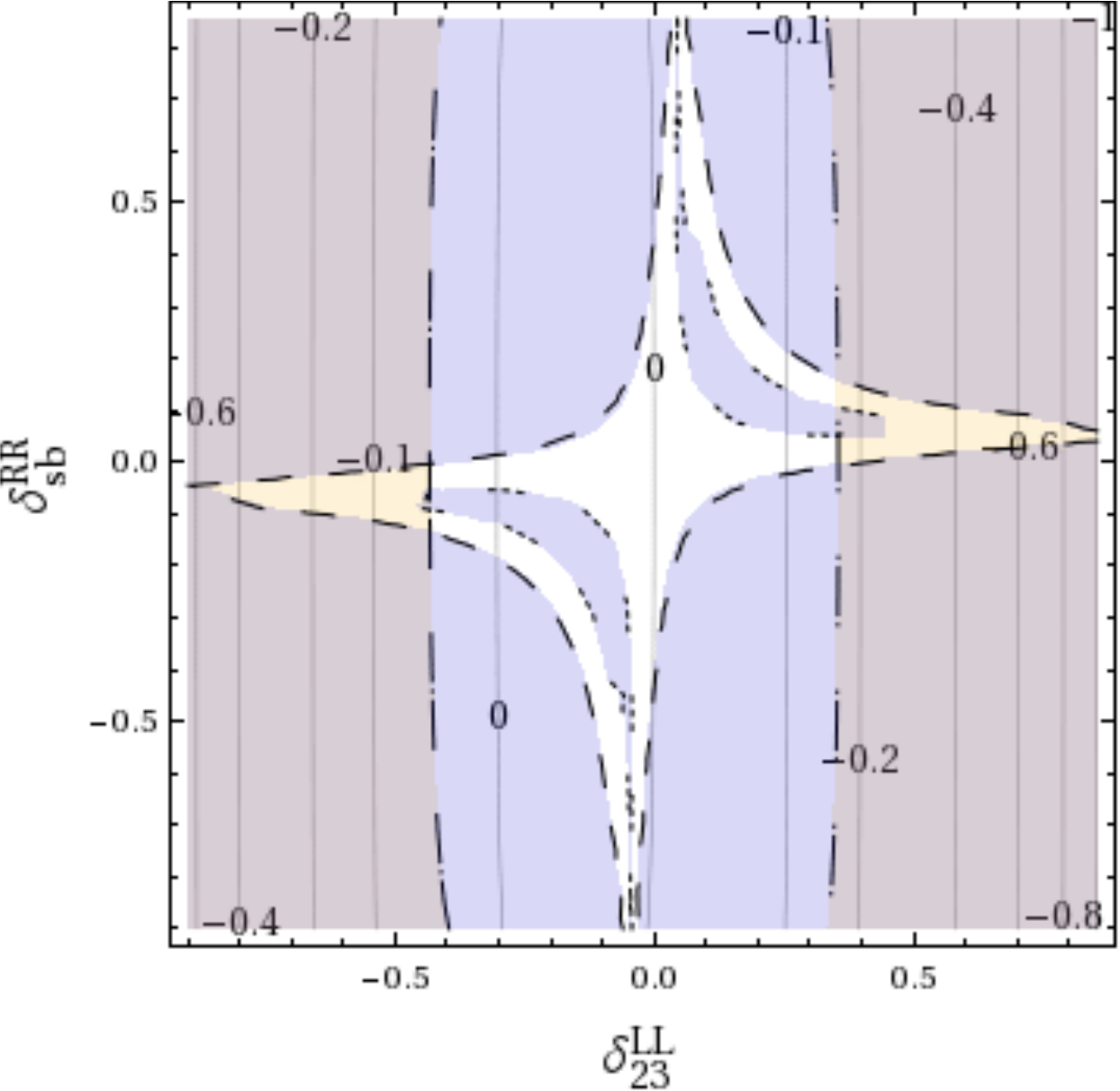}\\ 
\end{tabular}}}
\caption{$\Delta m_{h}$ (GeV) contour lines from our two deltas analysis for SPS2. The colour code for the allowed/disallowed areas by pre-LHC $B$ data is given in Fig.\ref{colleg}.} 
 \label{figdoubledeltaSPS2} 
\end{figure}
\clearpage
\newpage

\begin{figure}[h!] 
\centering
\hspace*{-10mm} 
{\resizebox{14.6cm}{!} 
{\begin{tabular}{cc} 
\includegraphics[width=13.2cm]{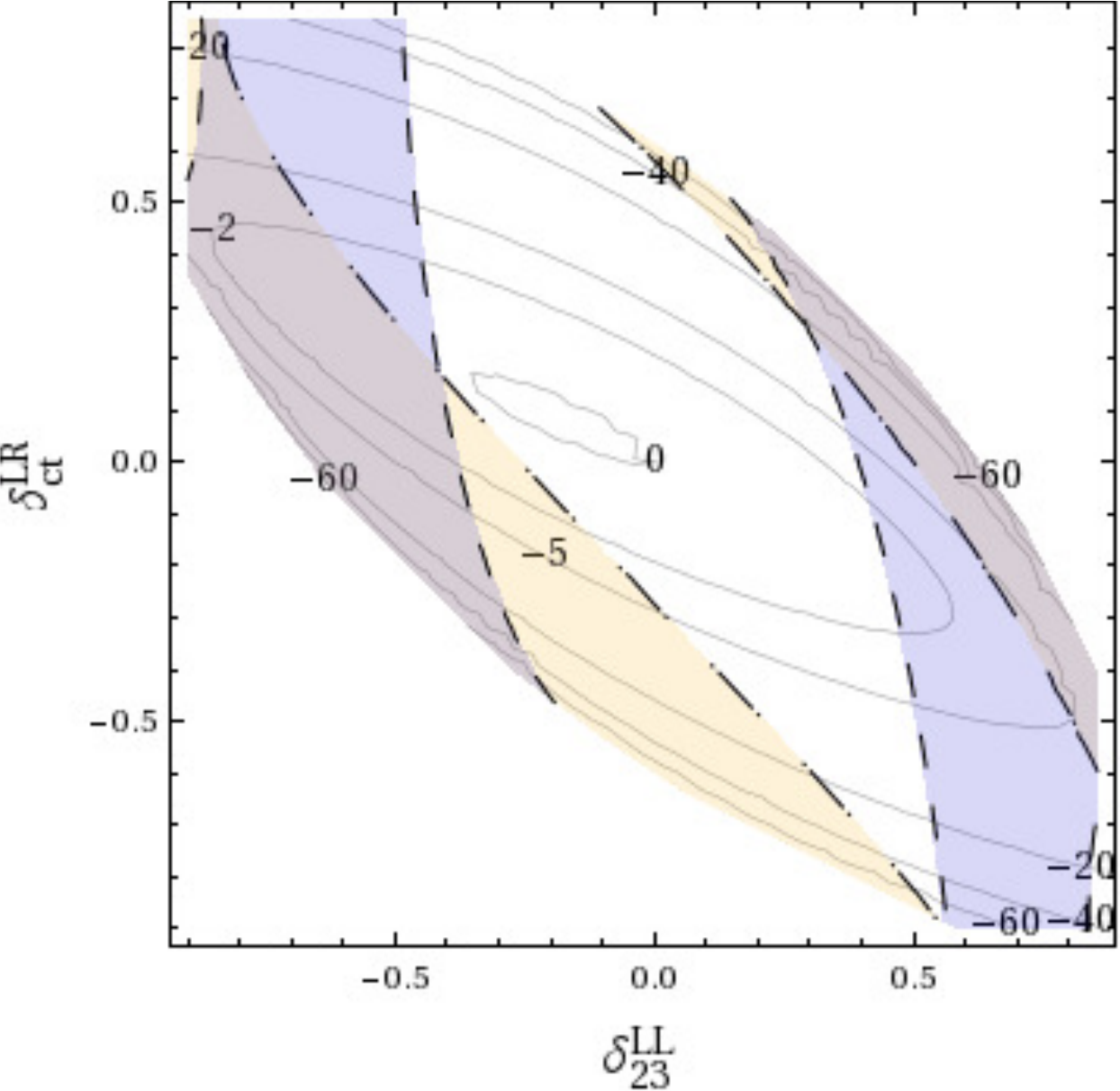}& 
\includegraphics[width=13.2cm]{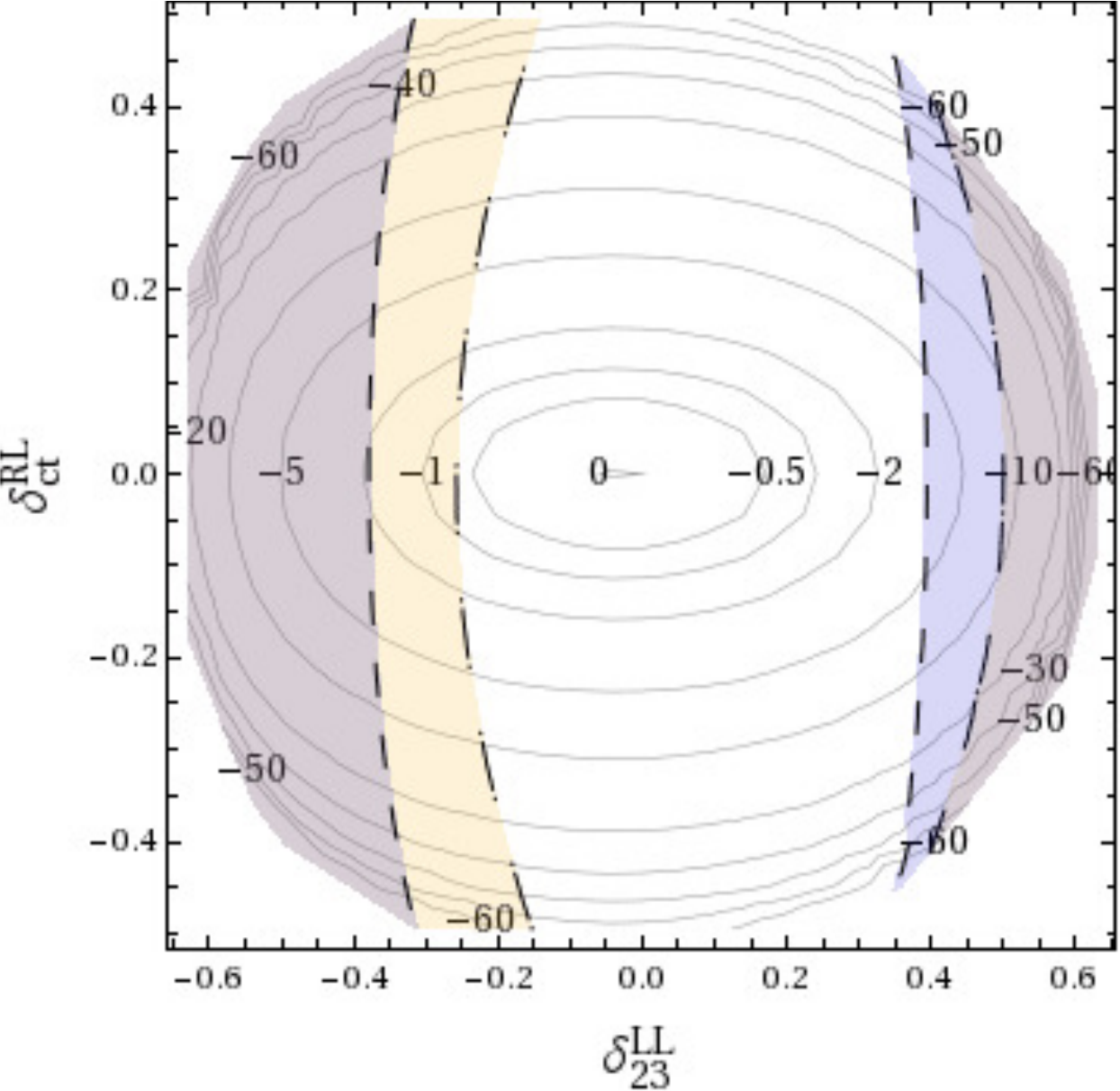}\\ 
\includegraphics[width=13.2cm]{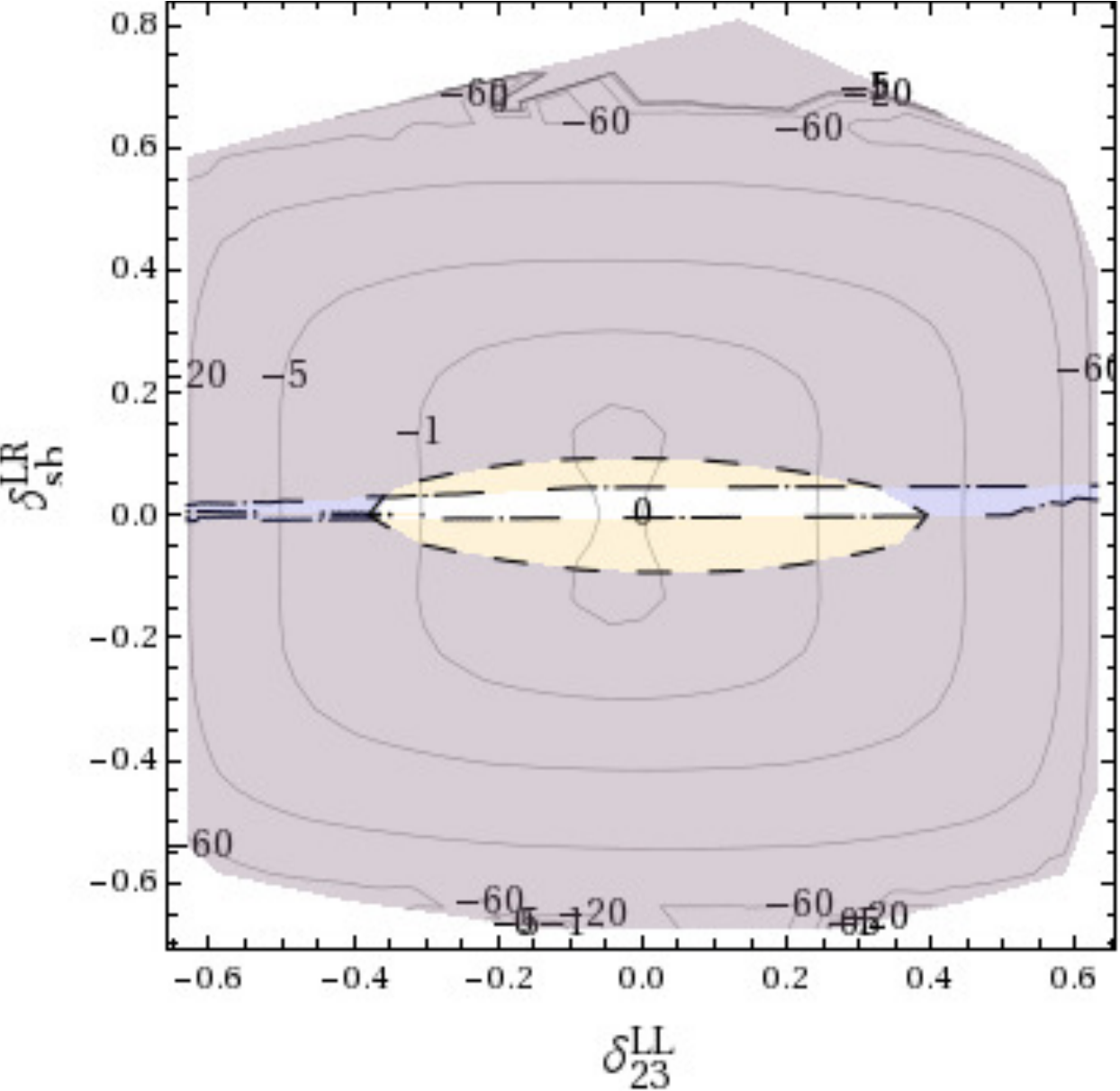}&
\includegraphics[width=13.2cm]{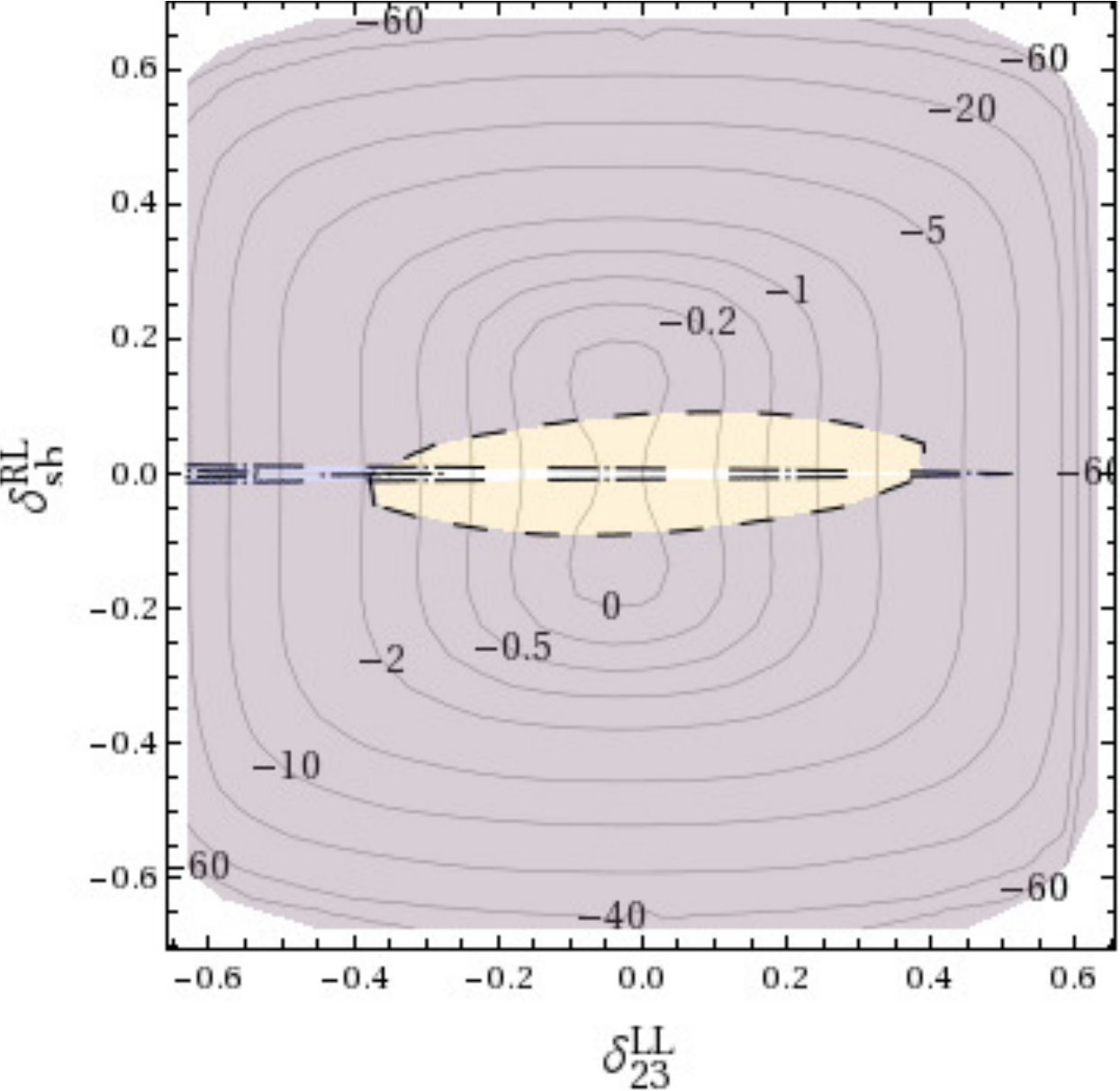}\\ 
\includegraphics[width=13.2cm]{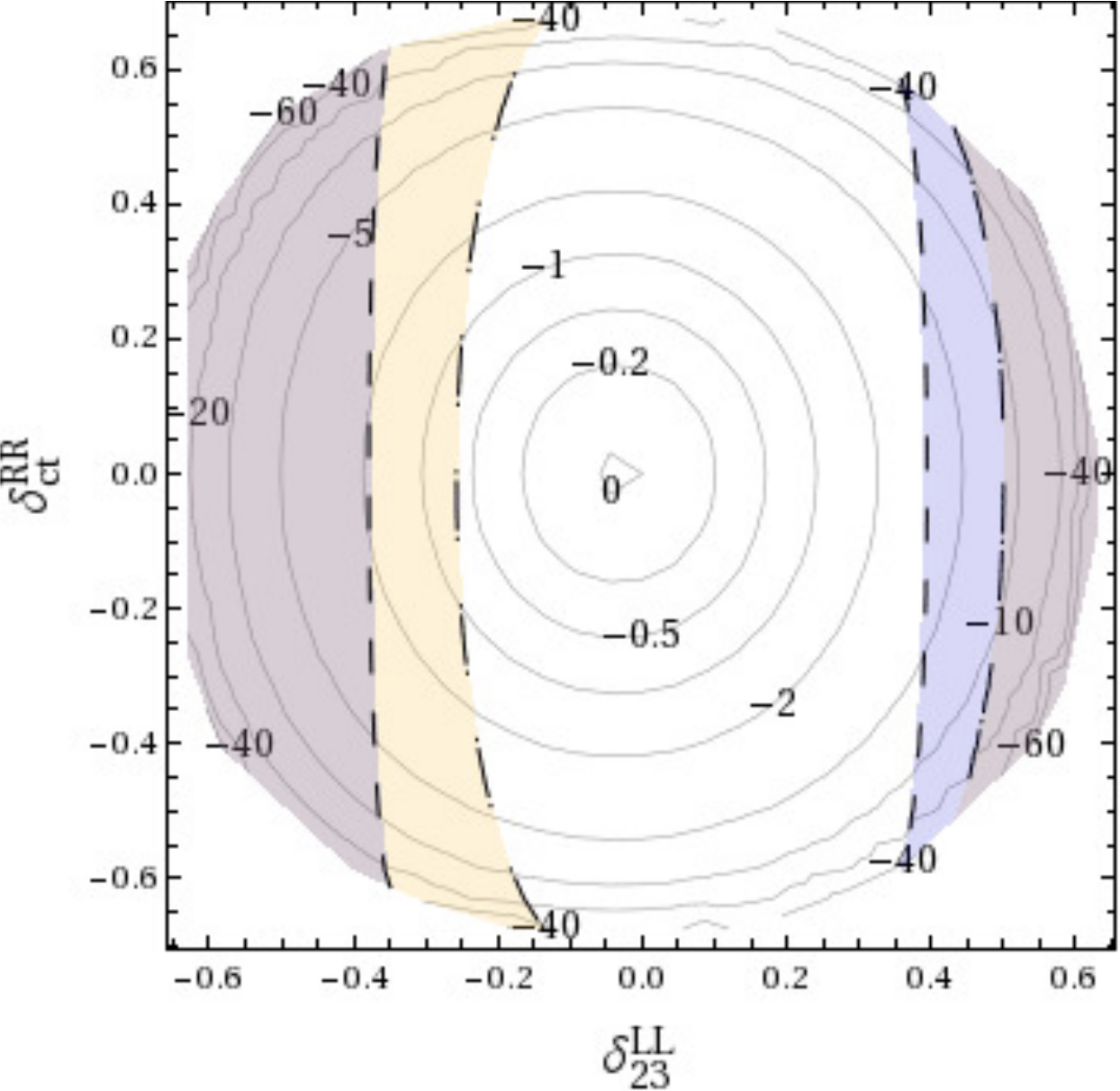}& 
\includegraphics[width=13.2cm]{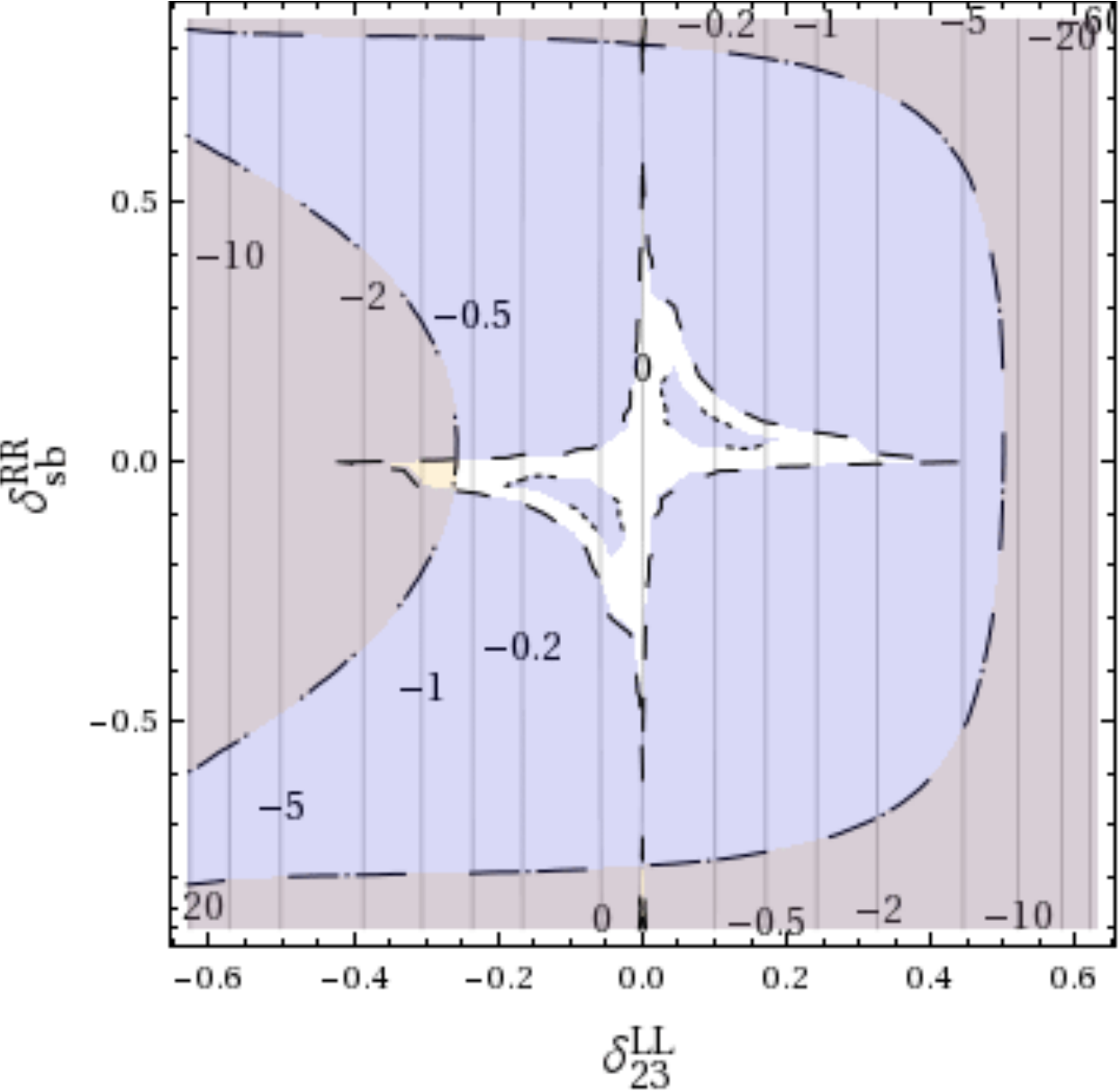}\\ 
\end{tabular}}}
\caption{$\Delta m_{h}$ (GeV) contour lines from our two deltas analysis for SPS5. The colour code for the allowed/disallowed areas by pre-LHC $B$ data is given in Fig.\ref{colleg}.} 
 \label{figdoubledeltaSPS5} 
\end{figure}
\clearpage
\newpage
\begin{figure}[h!] 
\centering
\hspace*{-10mm} 
{\resizebox{14.6cm}{!} 
{\begin{tabular}{cc} 
\includegraphics[width=13.2cm]{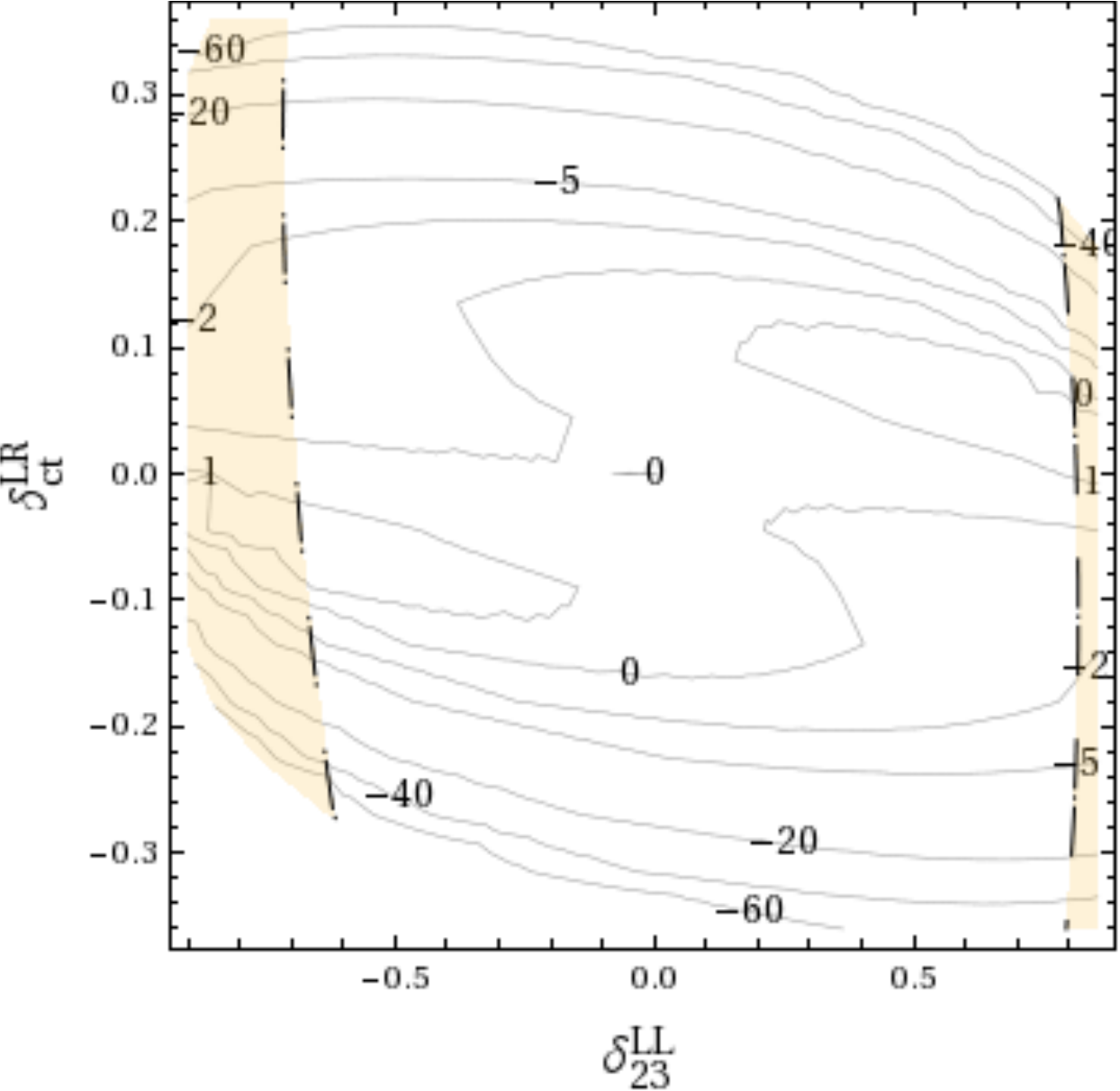}& 
\includegraphics[width=13.2cm]{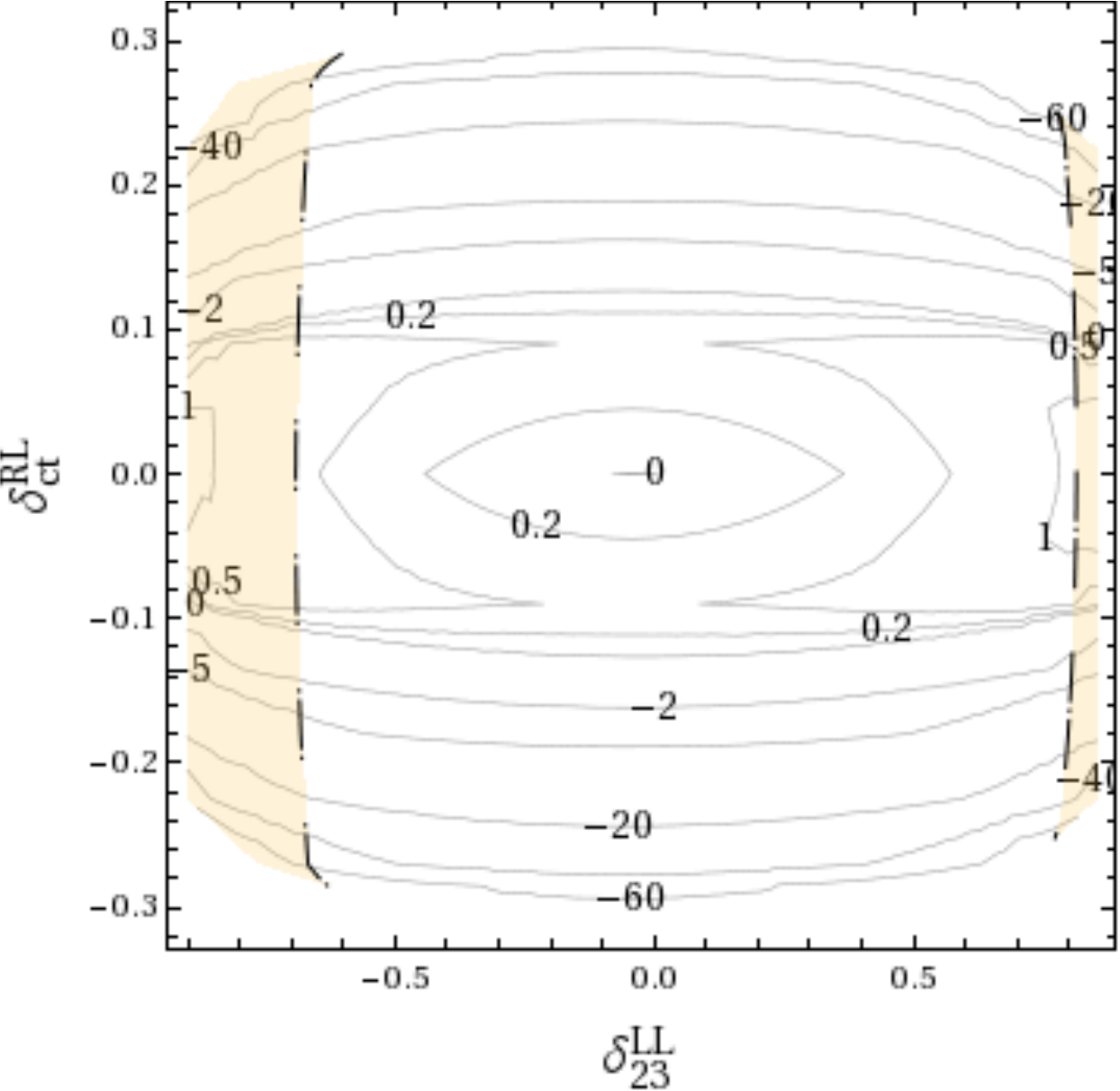}\\ 
\includegraphics[width=13.2cm]{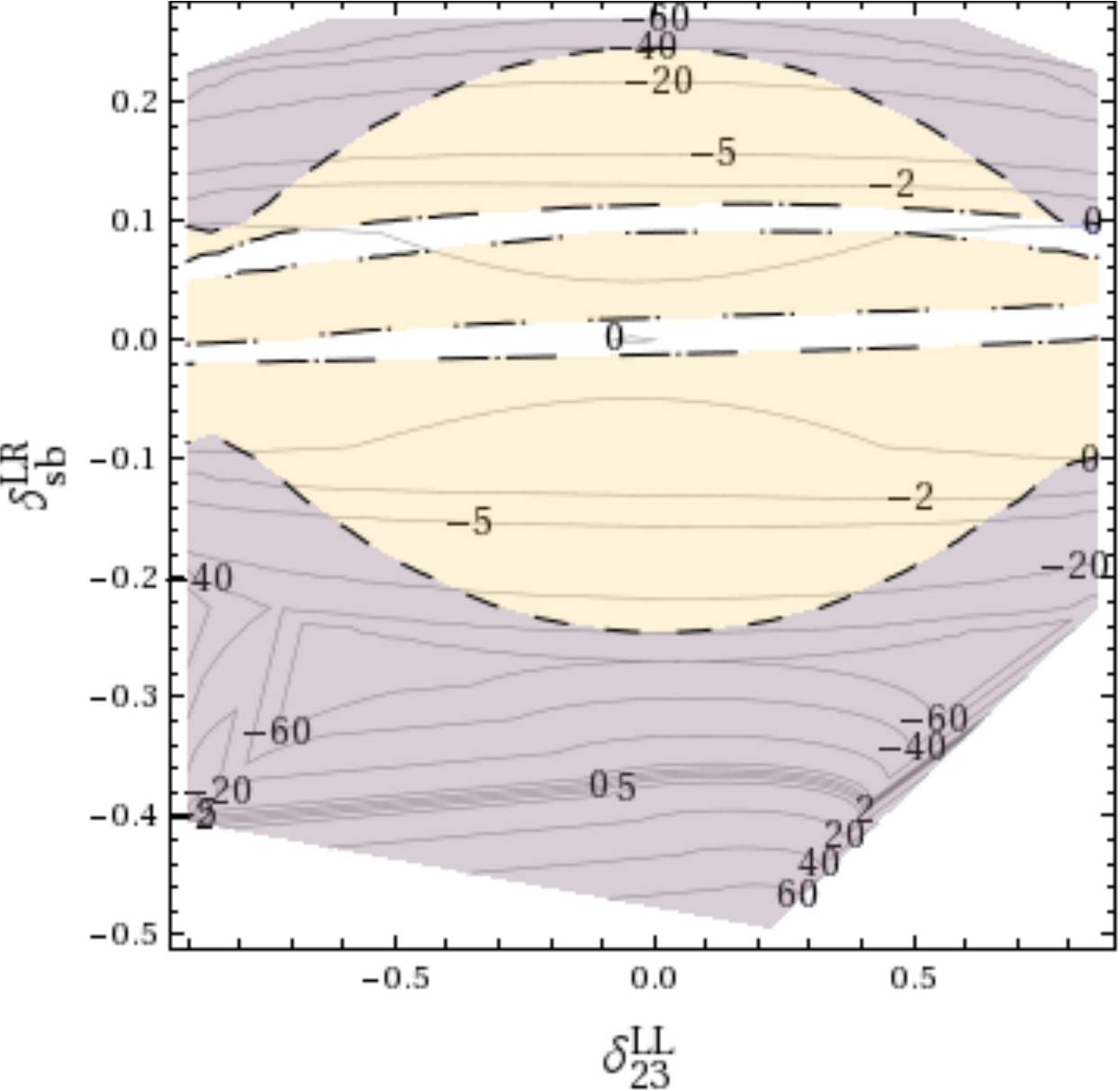}&
\includegraphics[width=13.2cm]{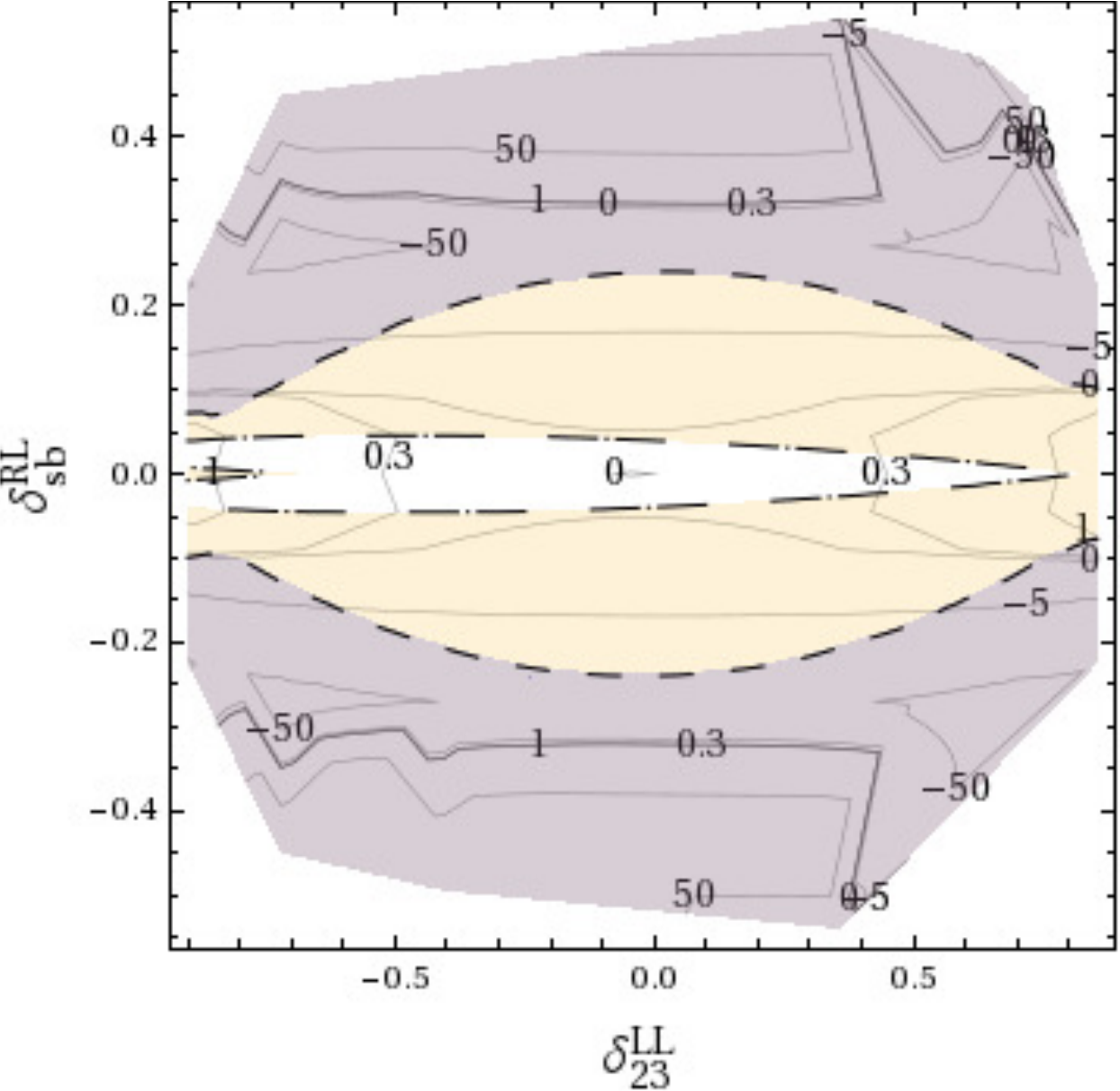}\\ 
\includegraphics[width=13.2cm]{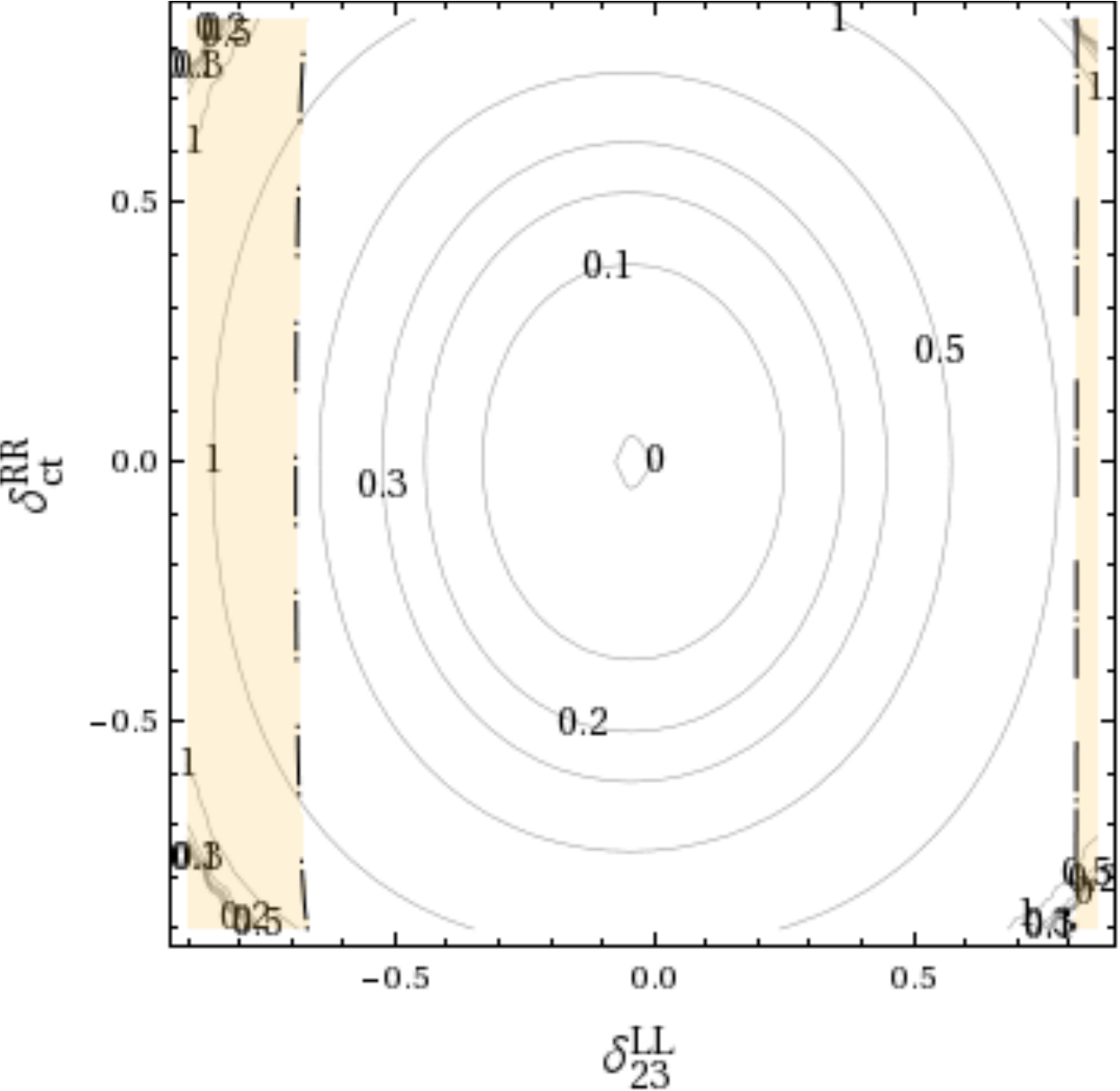}& 
\includegraphics[width=13.2cm]{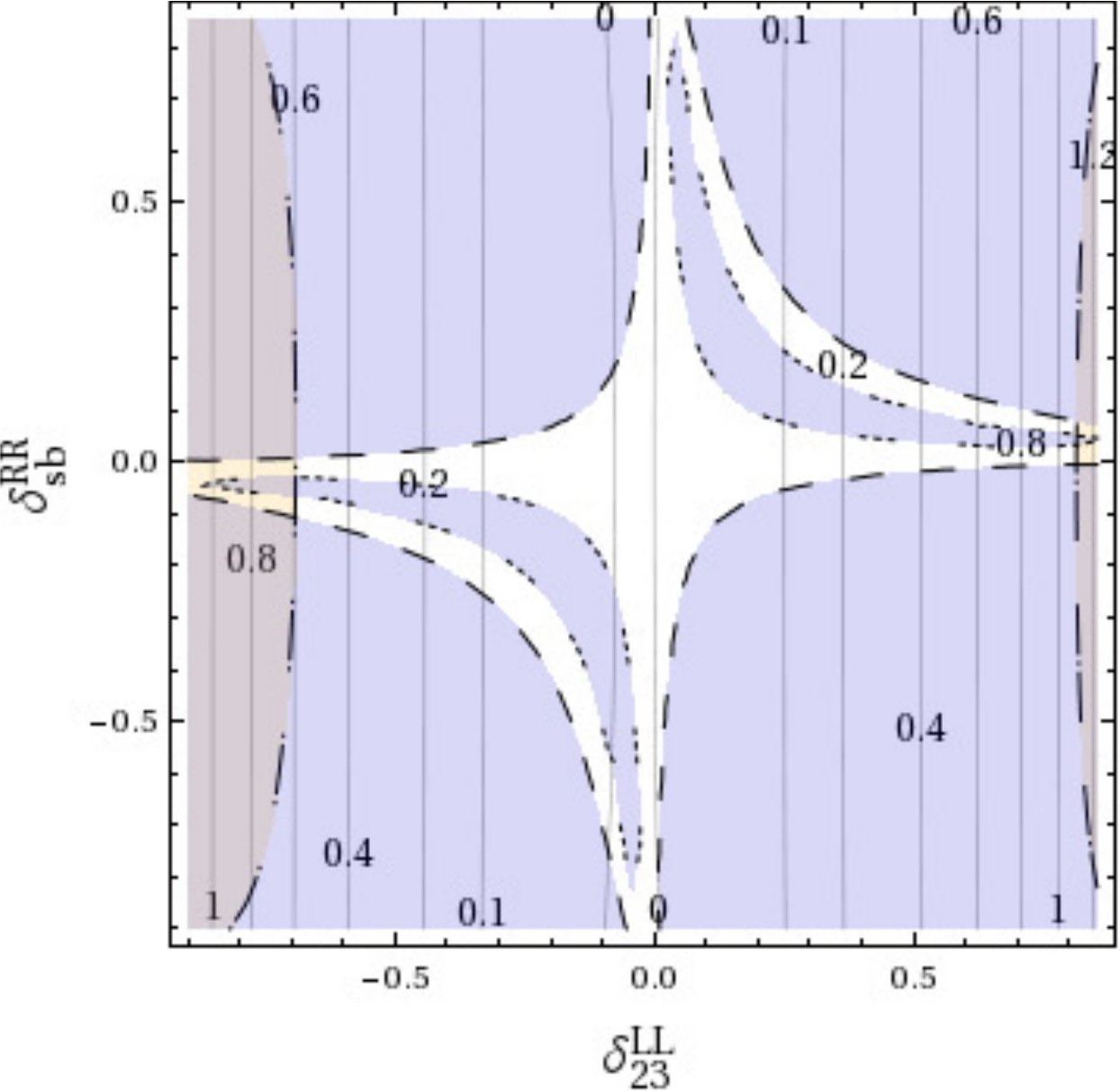}\\ 
\end{tabular}}}
\caption{$\Delta m_{h}$ (GeV) contour lines from our two deltas analysis for VHeavyS. The colour code for the allowed/disallowed areas by pre-LHC $B$ data is given in Fig.\ref{colleg}.} 
 \label{figdoubledeltaVHeavyS}
\end{figure}
\clearpage
\newpage
\begin{figure}[h!] 
\centering
\hspace*{-10mm} 
{\resizebox{14.4cm}{!} 
{\begin{tabular}{cc} 
\includegraphics[width=13.2cm]{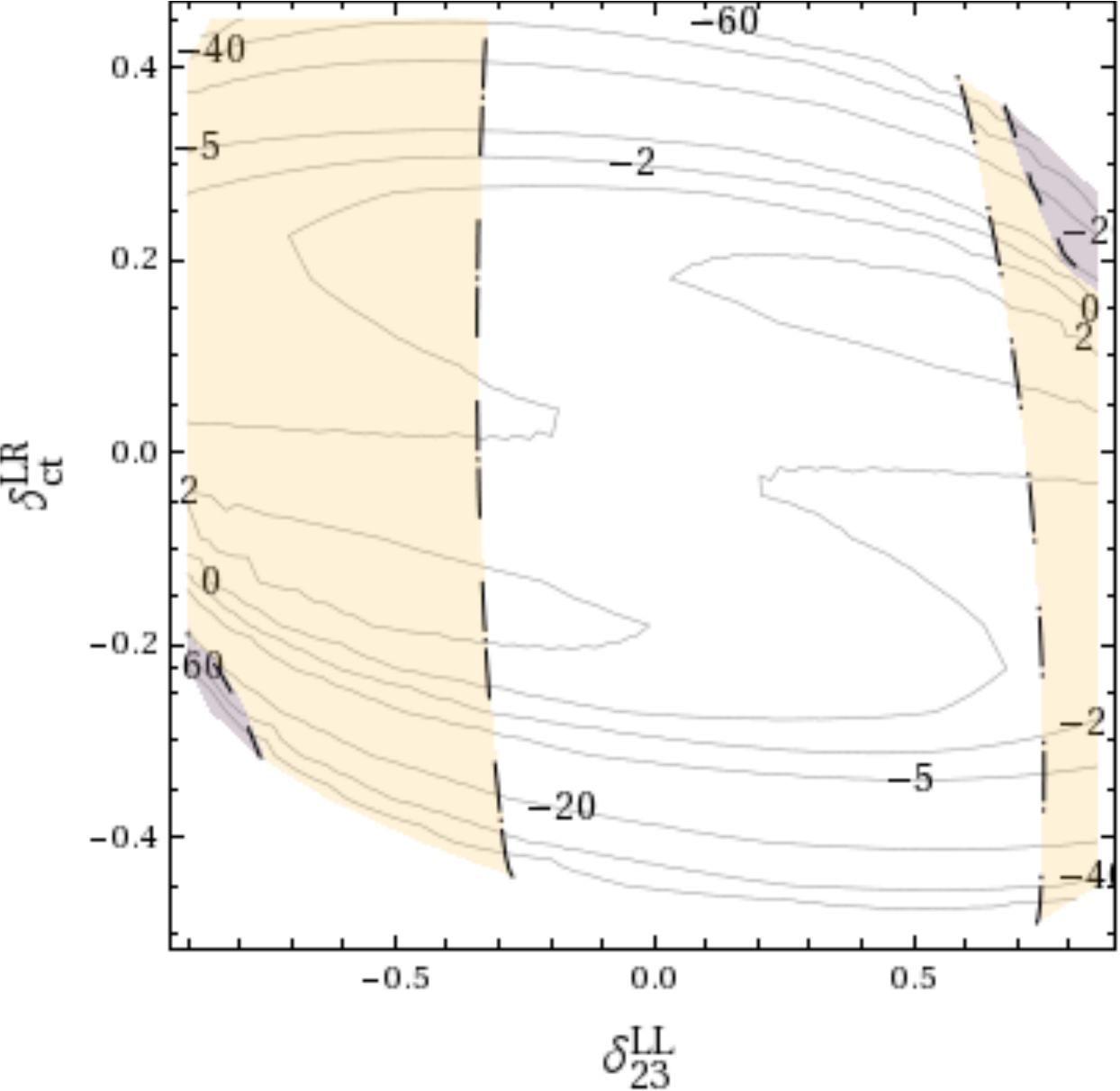}& 
\includegraphics[width=13.2cm]{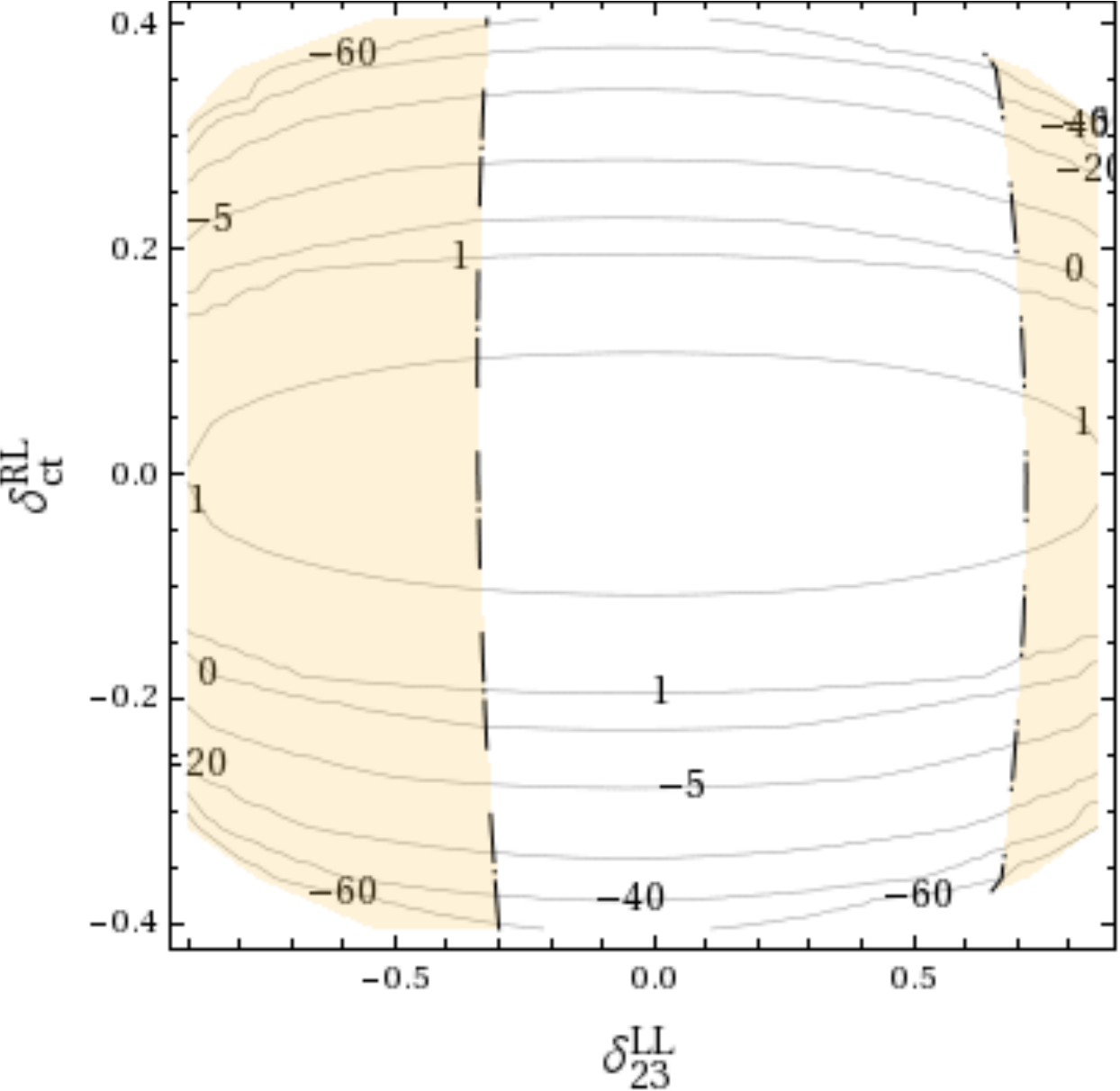}\\ 
\includegraphics[width=13.2cm]{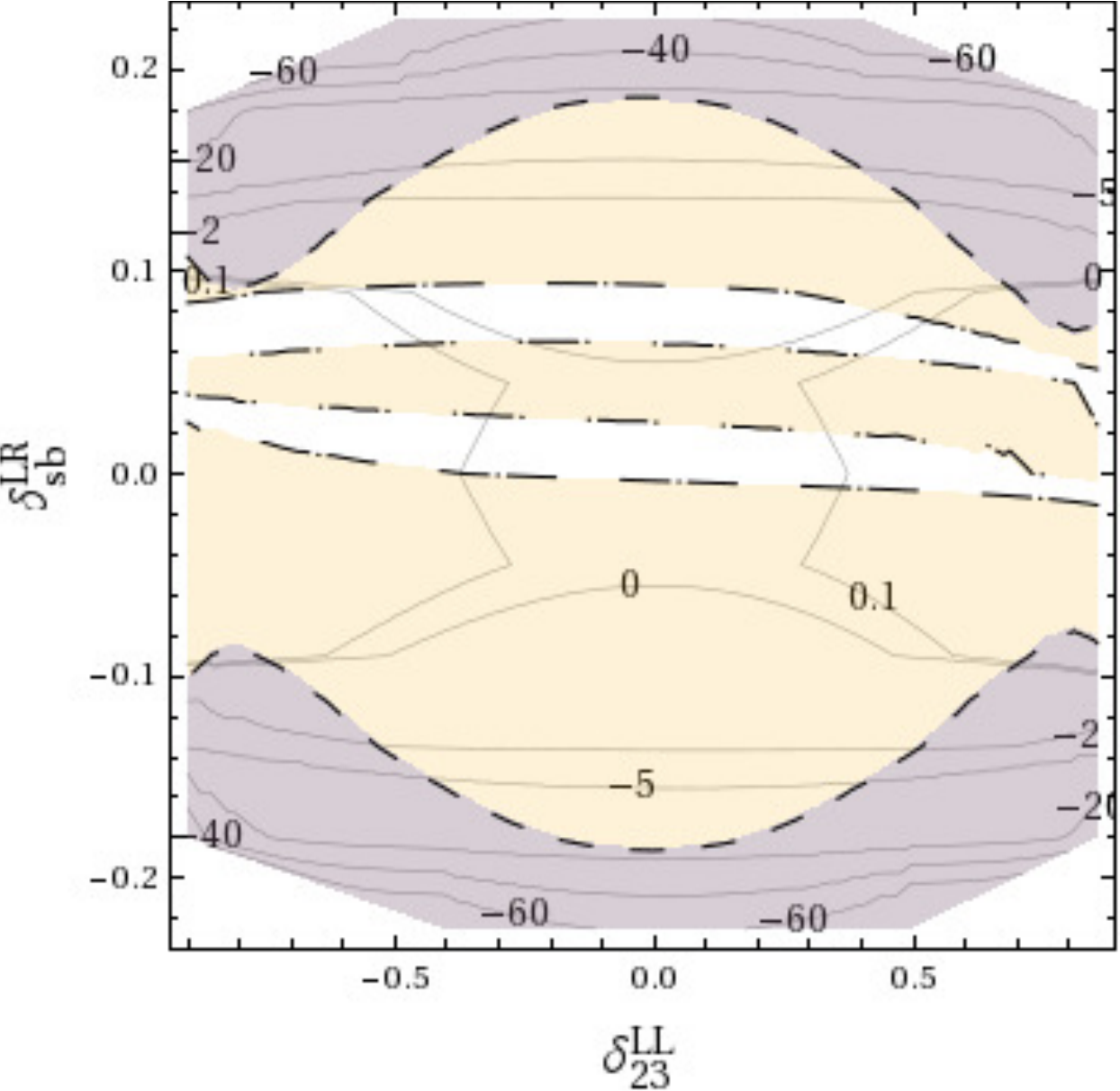}&
\includegraphics[width=13.2cm]{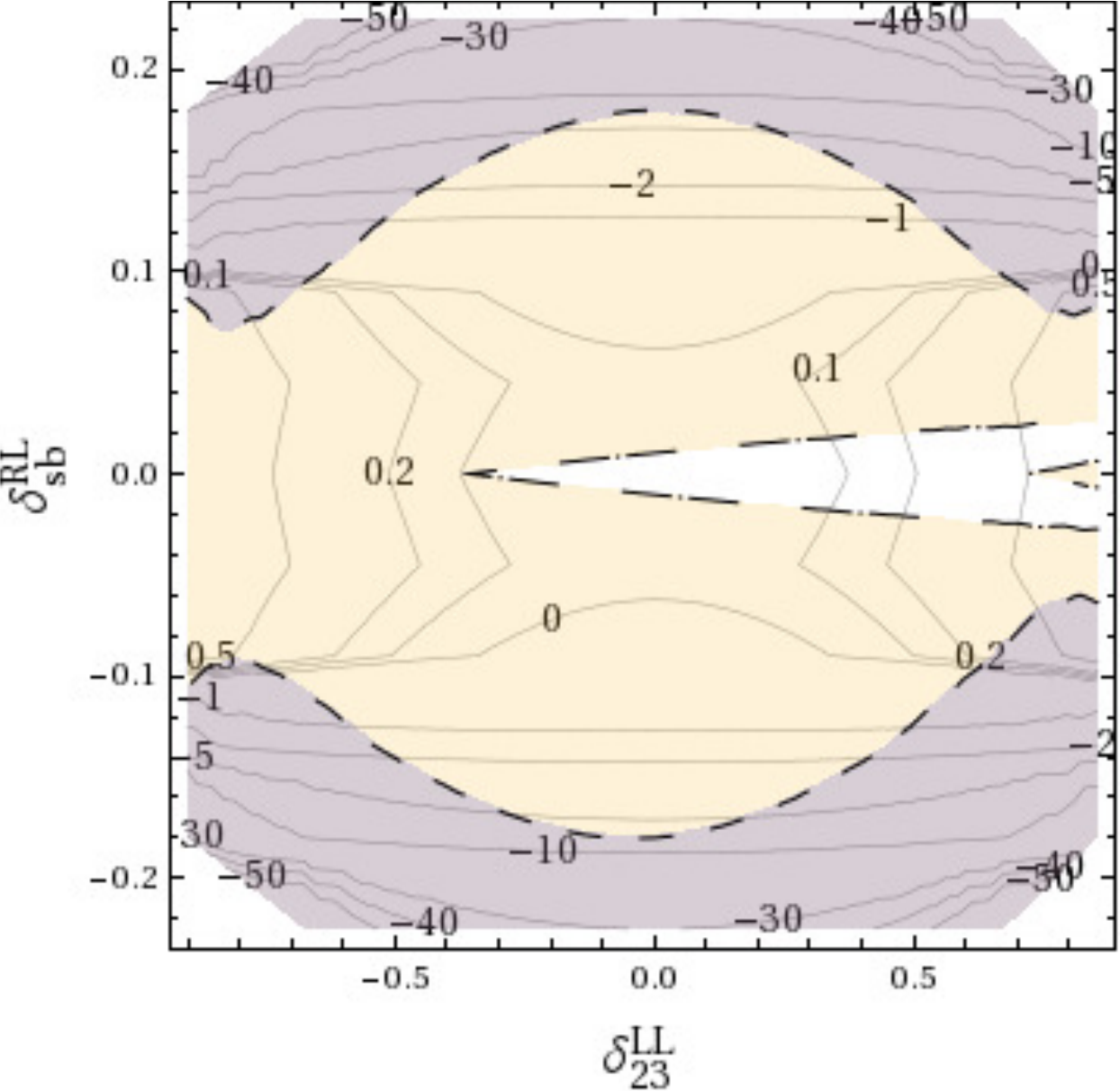}\\ 
\includegraphics[width=13.2cm]{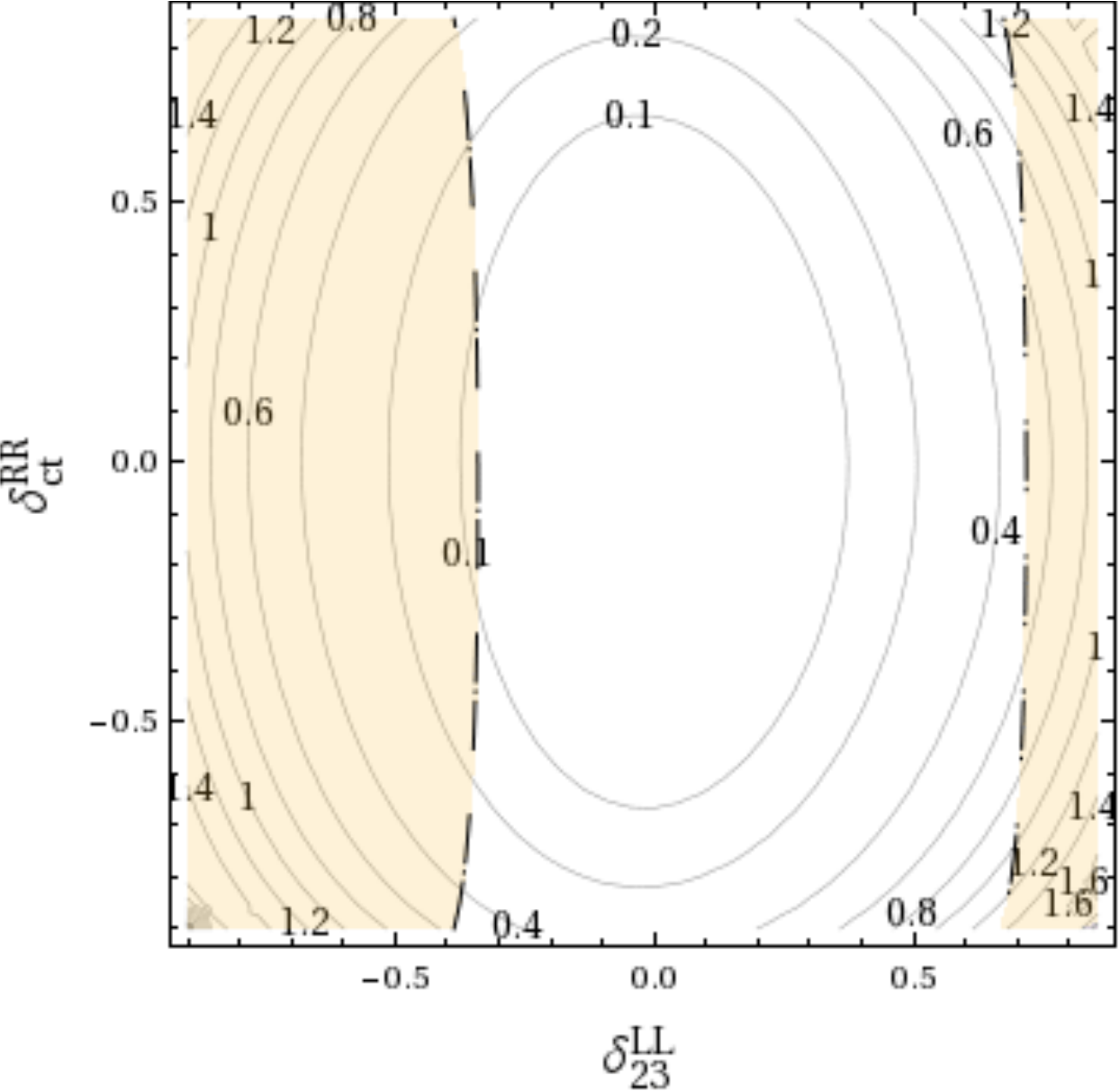}& 
\includegraphics[width=13.2cm]{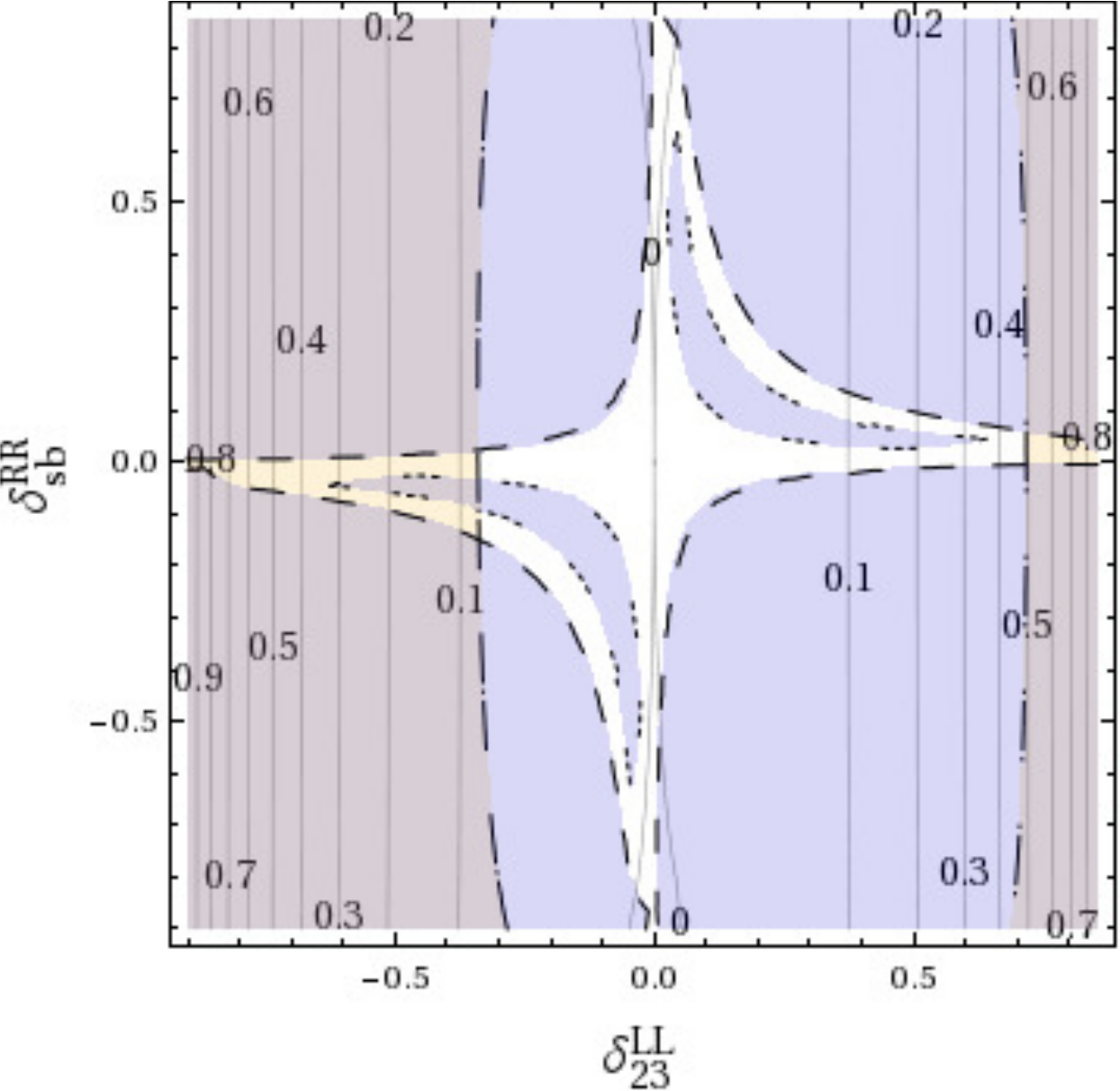}\\ 
\end{tabular}}}
\caption{$\Delta m_{h}$ (GeV) contour lines from our two deltas analysis for HeavySLightH. The colour code for the allowed/disallowed areas by pre-LHC $B$ data is given in Fig.\ref{colleg}.} 
\label{figdoubledeltaHeavySLightH}
\end{figure}
\clearpage
\newpage
We have chosen in all plots $\de^{LL}_{23}$ as one of these non-vanishing deltas mainly because of two reasons. First, because it is one of the most frequently studied flavour changing parameters in the literature and, therefore, a convenient reference parameter. 
Second, because there are several well motivated SUSY scenarios, where this parameter is allowed to take the largest value, as we explained in Section \ref{sec:nmfv-squarks}.

In these two-dimensional figures we have included the allowed/disallowed by
pre-LHC $B$ data areas that have been found by following the procedure explained in
Section \ref{sec:Bphysics}, and the allowed intervals are given in
eqs.(\ref{bsglinearerr}), (\ref{bsmmlinearerr}) and (\ref{deltabslinearerr}). The
colour code explaining the meaning of each coloured area and the codes for the
discontinuous lines are given in Fig.\ref{colleg}. Contour lines corresponding
to mass corrections above 60 GeV or below -60 GeV have not been represented. In several scenarios the plots involving $\de^{LR,RL}_{sb}$ 
show a seemingly abrupt behaviour for $|\de^{LR,RL}_{sb}| \gsim 0.3$,
corresponding to extremely large (one-loop) corrections to $\mh$.
 In general, in the case of very large one-loop corrections, in order to get a more stable result further higher order corrections
would be required, as it is known from the higher-order corrections to
$\mh$ in the MFV case (see, e.g., \citere{mhiggsAEC}).
However, we cannot explore this possibility here. On the other hand, in order to understand the behaviour of $\mh$ as a function of
$\de^{LR,RL}_{sb}$ a simple analytical formula would have to be
extracted from the general result. However, this is beyond the scope of
this thesis.

The main conclusions from these two dimensional figures are summarized in the following:

The points that have been chosen in these plots are quite representative of all the different patterns found. The plots for SPS1a (not shown here) manifest  similar patterns as those of SPS3. The plots for SPS1b (not shown here) manifest similar patterns as those of BFP. The plots for SPS4 are not included because they do not manifest any allowed areas by $B$ data.

The largest mass corrections $\Dmh$ found, being allowed by $B$ data occur in plots $(\de^{LL}_{23},\de^{LR}_{ct} )$ and  $(\de^{LL}_{23},\de^{RL}_{ct} )$. This applies to all the studied points.
They can be as large as $(-50,-100)$ GeV at $\de^{LR}_{ct}$ or $\de^{RL}_{ct}$ close to the upper and lower horizontal bands in these plots where $\de^{LR}_{ct}$ or $\de^{RL}_{ct}$ are close to $\pm 0.5$. Again these large corrections from the $LR$ and $RL$ parameters are due to the $A$-terms, as we explained
at the end of Section \ref{sec:numanalonedelta}.
Comparing the different plots, it can be seen that the size of the
allowed area by the pre-LHC $B$ data (the white area inside of the coloured regions)
can be easily understood basically in terms of $\tb$, and the heaviness
of the SUSY and Higgs spectra. Generically, the plots  with largest
allowed regions and with largest Higgs mass corrections correspond to
scenarios with low $\tb = 5$ 
and heavy spectra. Consequently, the cases of VHeavyS and HeavySLightH
 are the most interesting ones, exhibiting very large radiative corrections,
 resulting from the heavy SUSY spectra. In the case of HeavySLightH
 the large corrections are not only found for $\Dmh$, but also, though to a 
lesser extent, for the other Higgs bosons,
 $\DmH$ and $\DmHp$ (not shown here). Consequently, in this scenario
 the deltas were very restricted by the mass bounds, especially by
 $\mh$. 

There are also important corrections in the allowed areas of the two dimensional plots of   $(\de^{LL}_{23},\de^{RR}_{ct} )$ for some points, particularly for SPS5 (and to a lesser extent for SPS2). Here the corrections can be as large as -50 GeV in the upper and lower parts, i.e. for $\de^{RR}_{ct}$ close to $\pm 0.5$. In the case of SPS2 they can be up to -2 GeV for this same region.

As for the remaining two-dimensional plots they do not show relevant allowed areas where the mass corrections are interestingly large.

\subsection{Numerical results for post-LHC scenarios}
\label{numresafterlhchiggs}

After studying the phenomenology of the mass corrections with respect to the different $\deXYij$ varying the parameters of the model, we perform in this section a numerical evaluation taking into account the present experimental situation. The scenarios studied for this purpose are the ones described in Section \ref{frameworkc}. These scenarios are compatible with the LHC data and with all the current experimental bounds from $B$ physics and therefore have much heavier sparticles than the scenarios from the previous section.

\subsubsection{\boldmath{$\Dmphi$} versus one \boldmath{$\deXYij\neq 0$}}
\label{sec:numanalonedelta2}

\begin{figure}[h!] 
\centering
\hspace*{-10mm} 
{\resizebox{17.3cm}{!} 
{\begin{tabular}{cc} 
\includegraphics[width=13.2cm,height=17.2cm,angle=270]{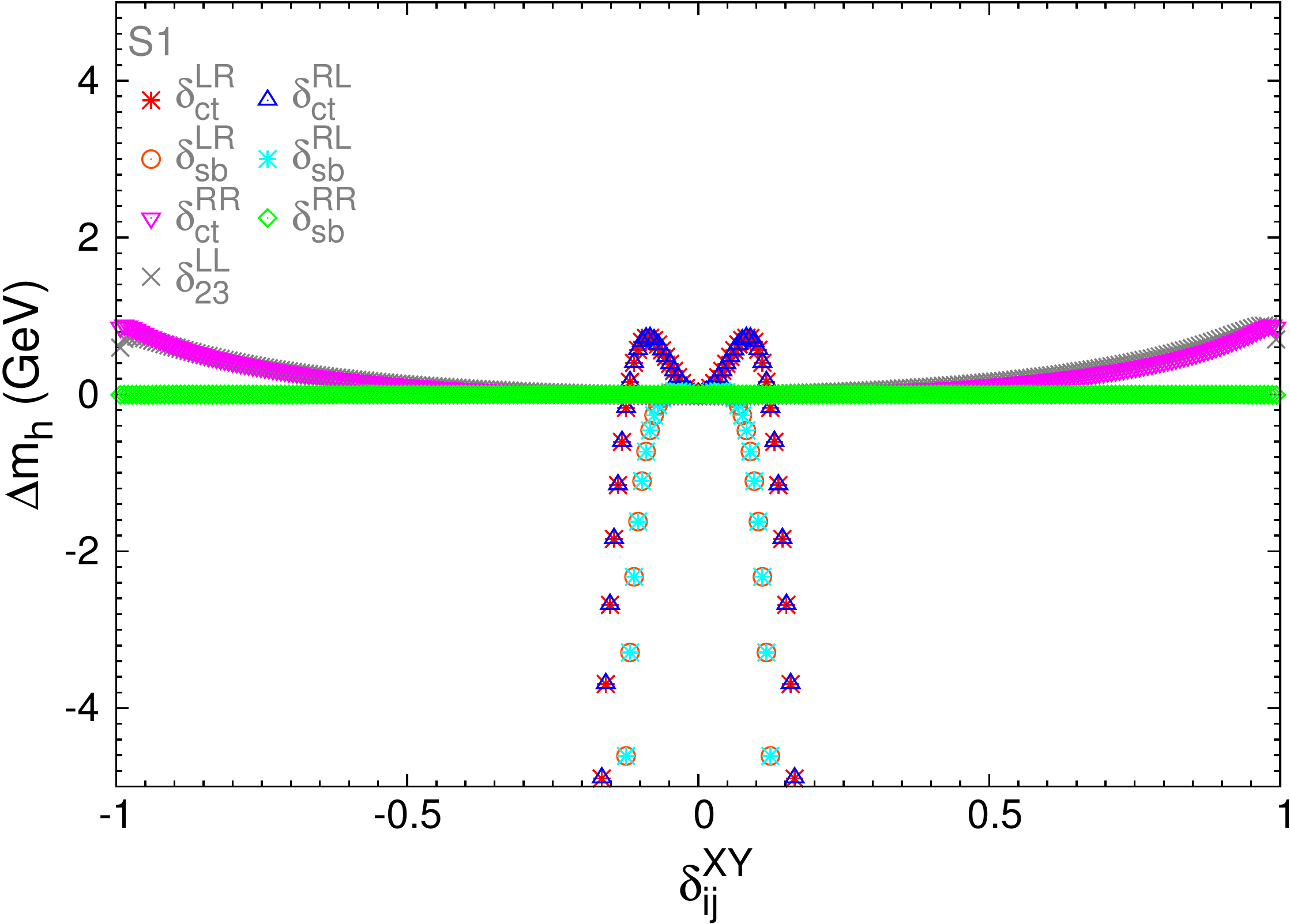}& 
\includegraphics[width=13.2cm,height=17.2cm,angle=270]{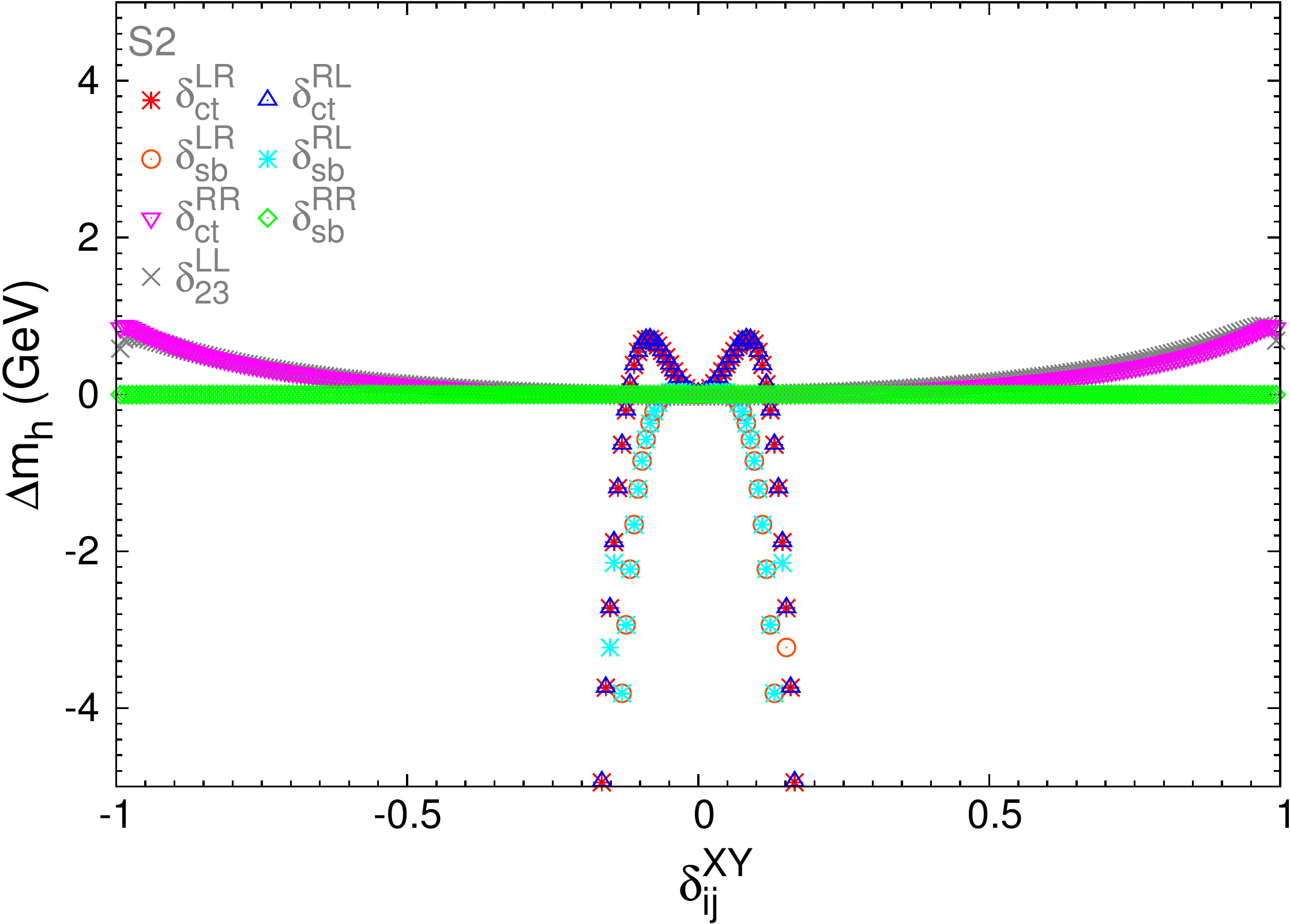}\\ 
\includegraphics[width=13.2cm,height=17.2cm,angle=270]{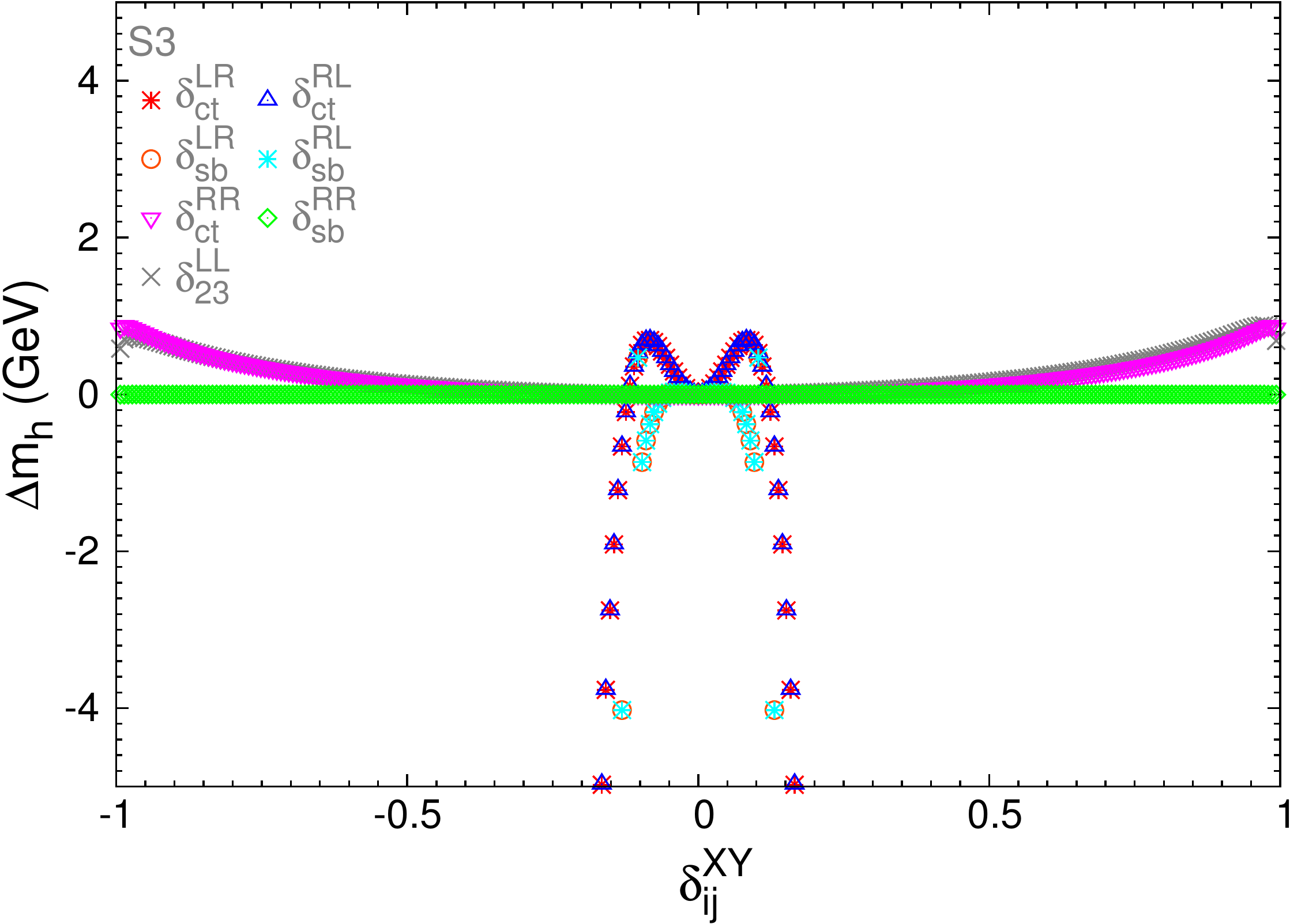}&
\includegraphics[width=13.2cm,height=17.2cm,angle=270]{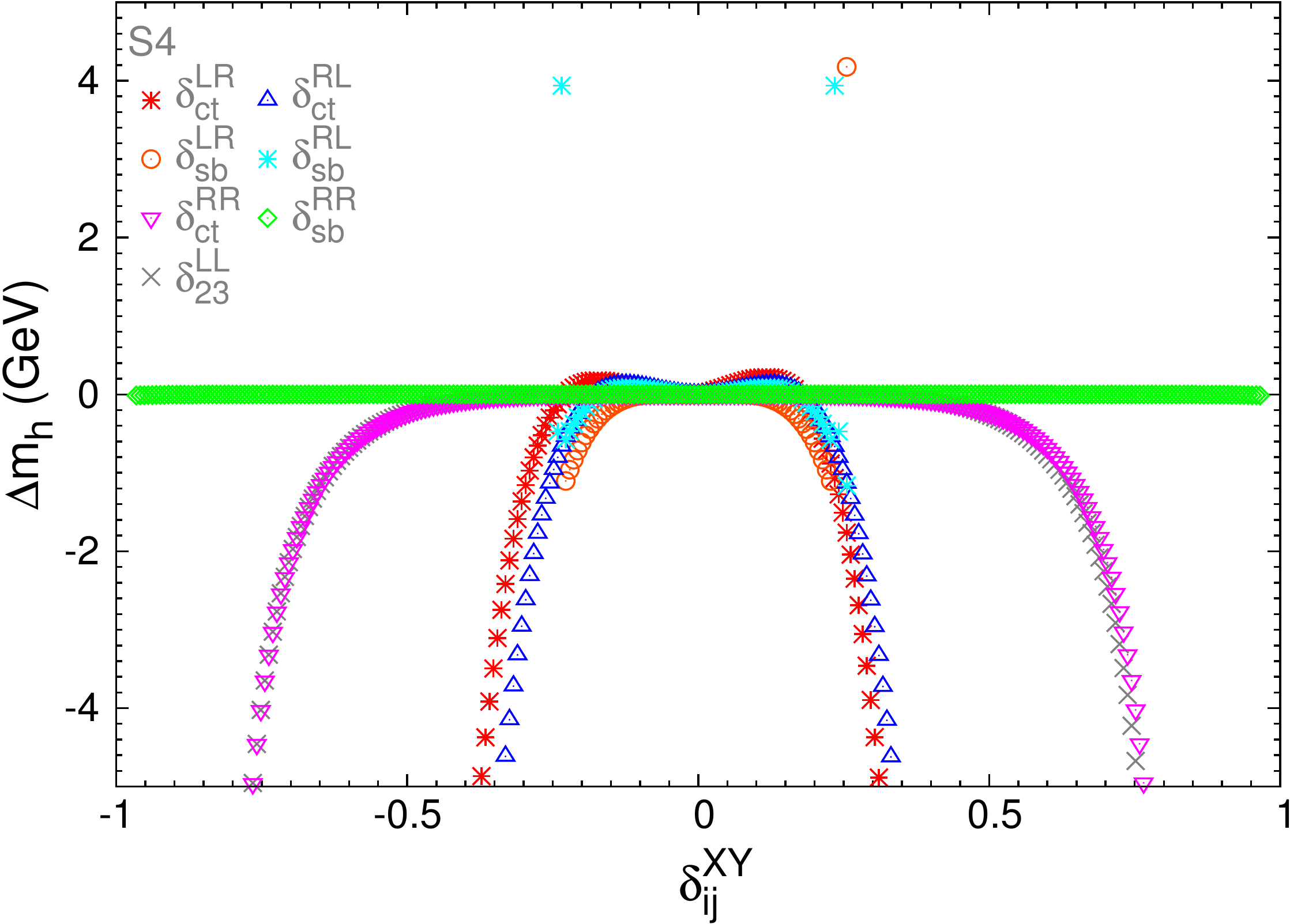}\\ 
\includegraphics[width=13.2cm,height=17.2cm,angle=270]{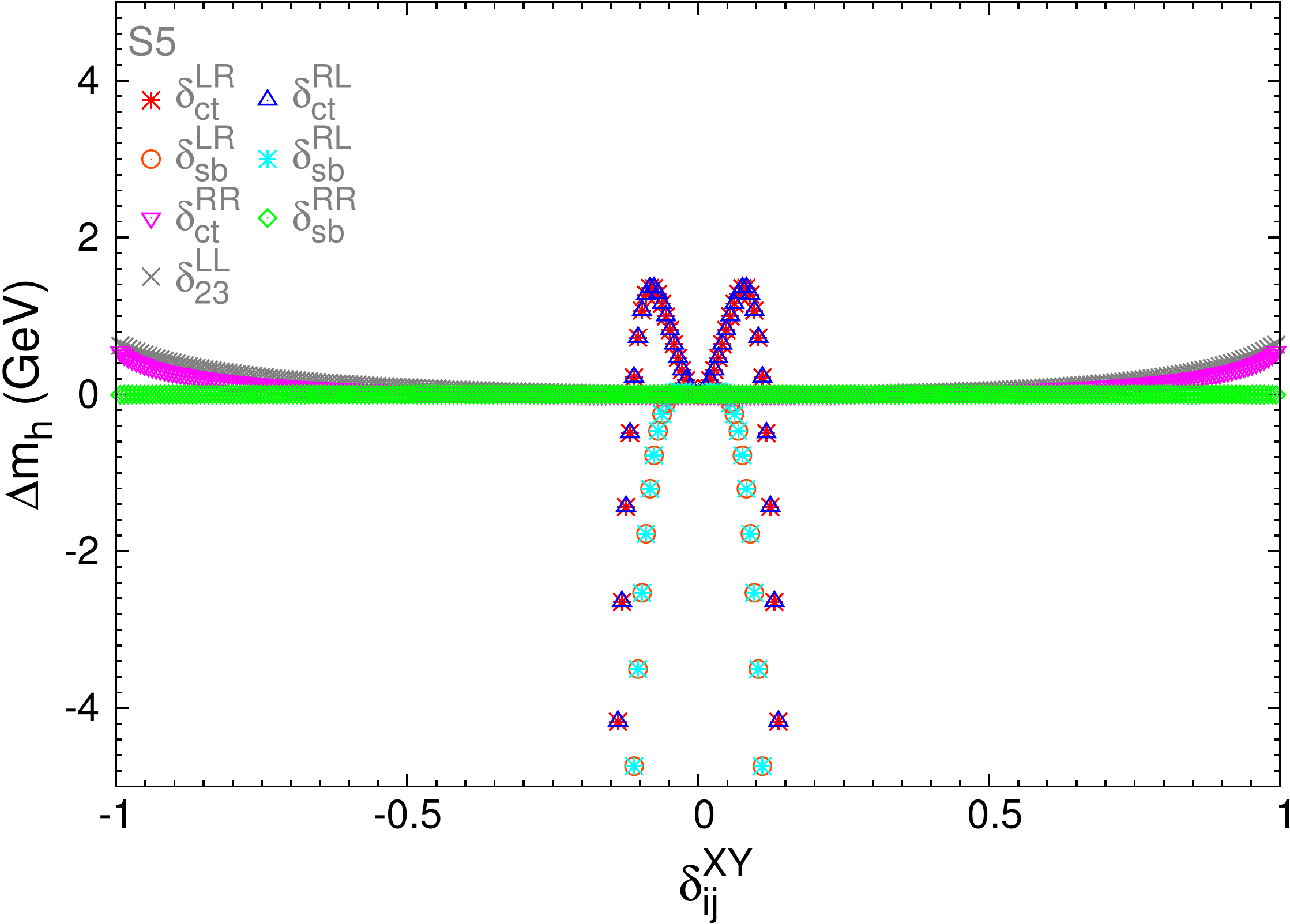}& 
\includegraphics[width=13.2cm,height=17.2cm,angle=270]{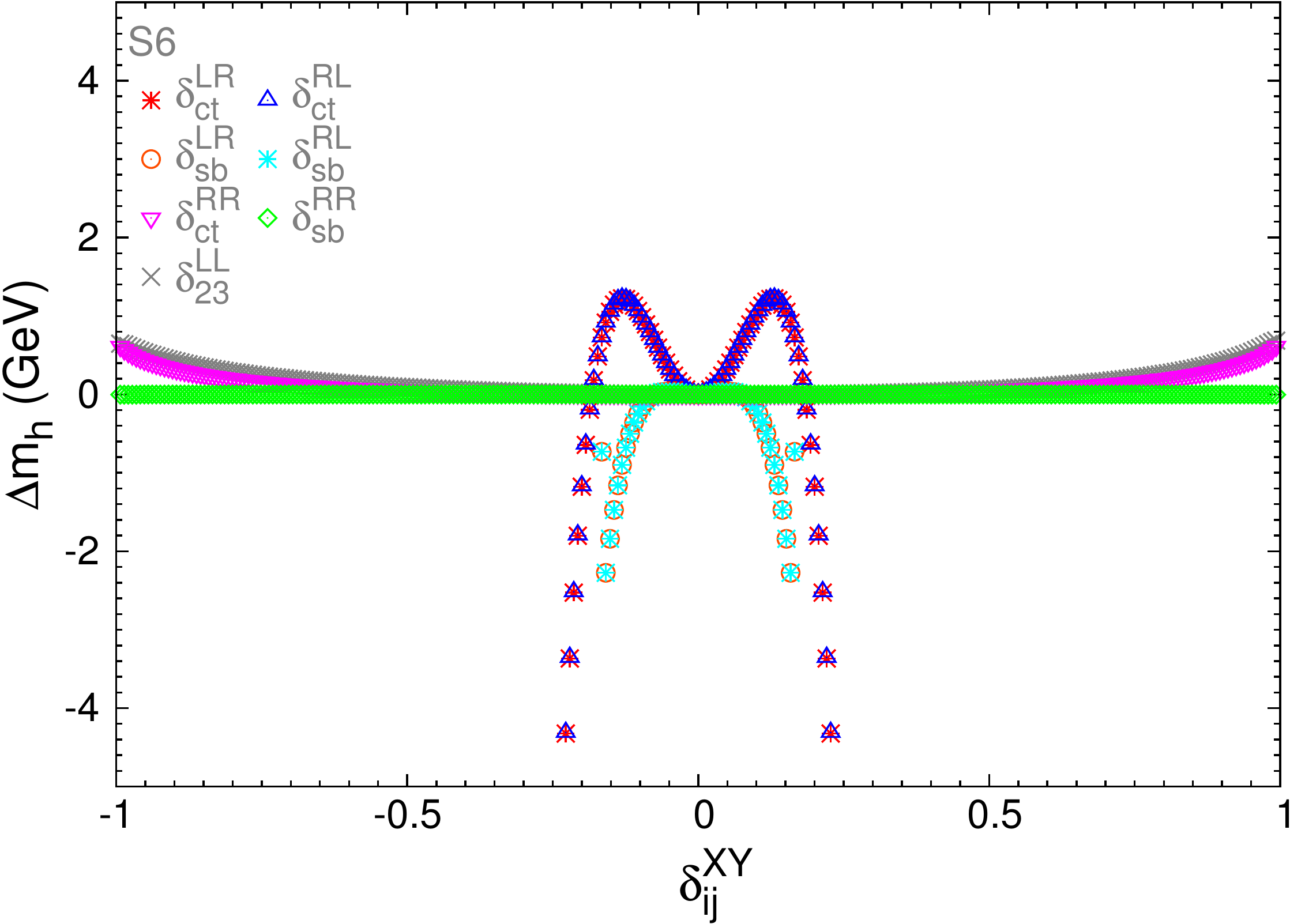}\\ 
\end{tabular}}}
\caption{Sensitivity to the NMFV deltas in $\Dmh$ for the six post-LHC scenarios of table \ref{tab:spectra}.}
 \label{figdeltamh02}
\end{figure}

\begin{figure}[h!] 
\centering
\hspace*{-10mm} 
{\resizebox{17.3cm}{!} 
{\begin{tabular}{cc} 
\includegraphics[width=13.2cm,height=17.2cm,angle=270]{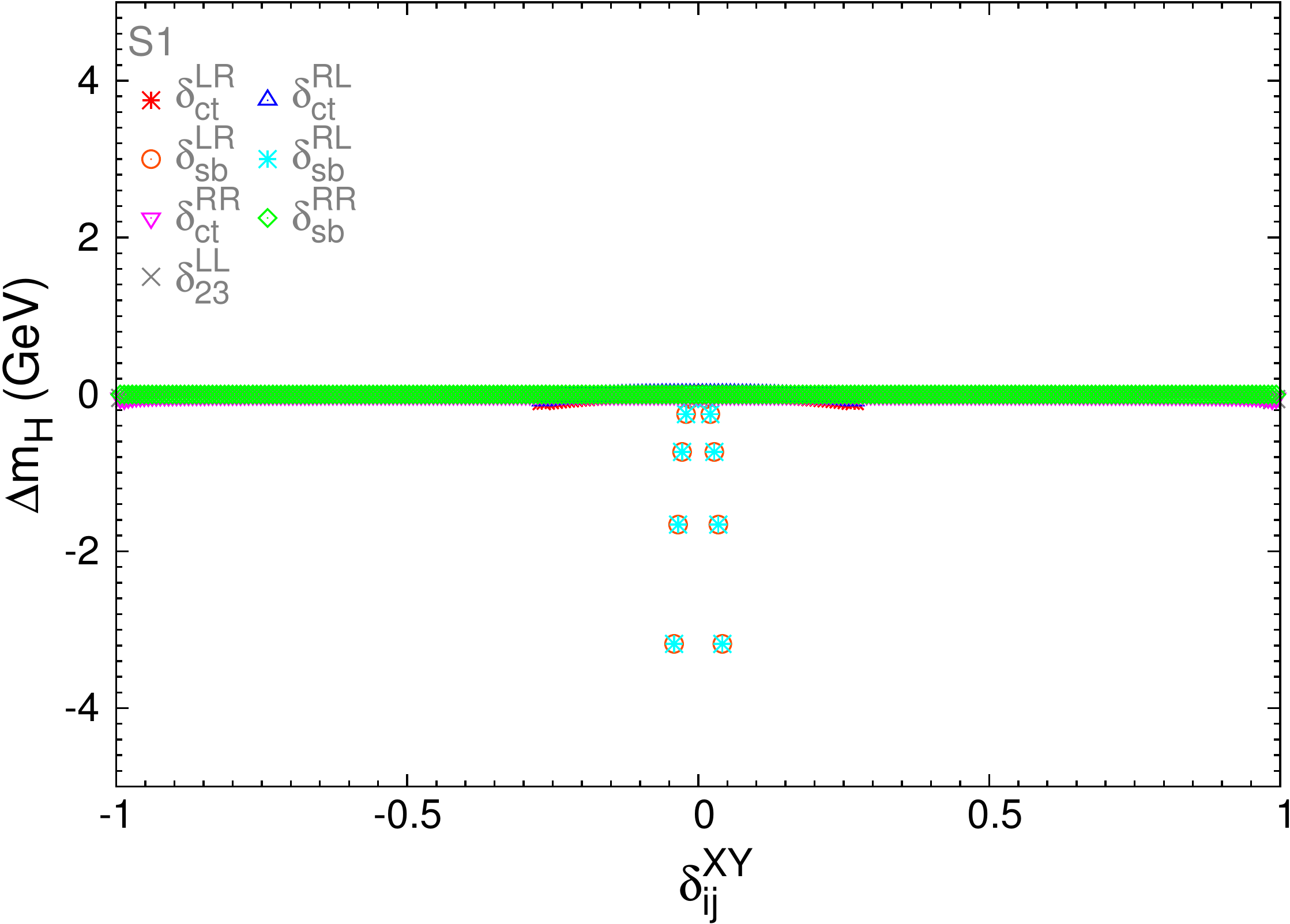}& 
\includegraphics[width=13.2cm,height=17.2cm,angle=270]{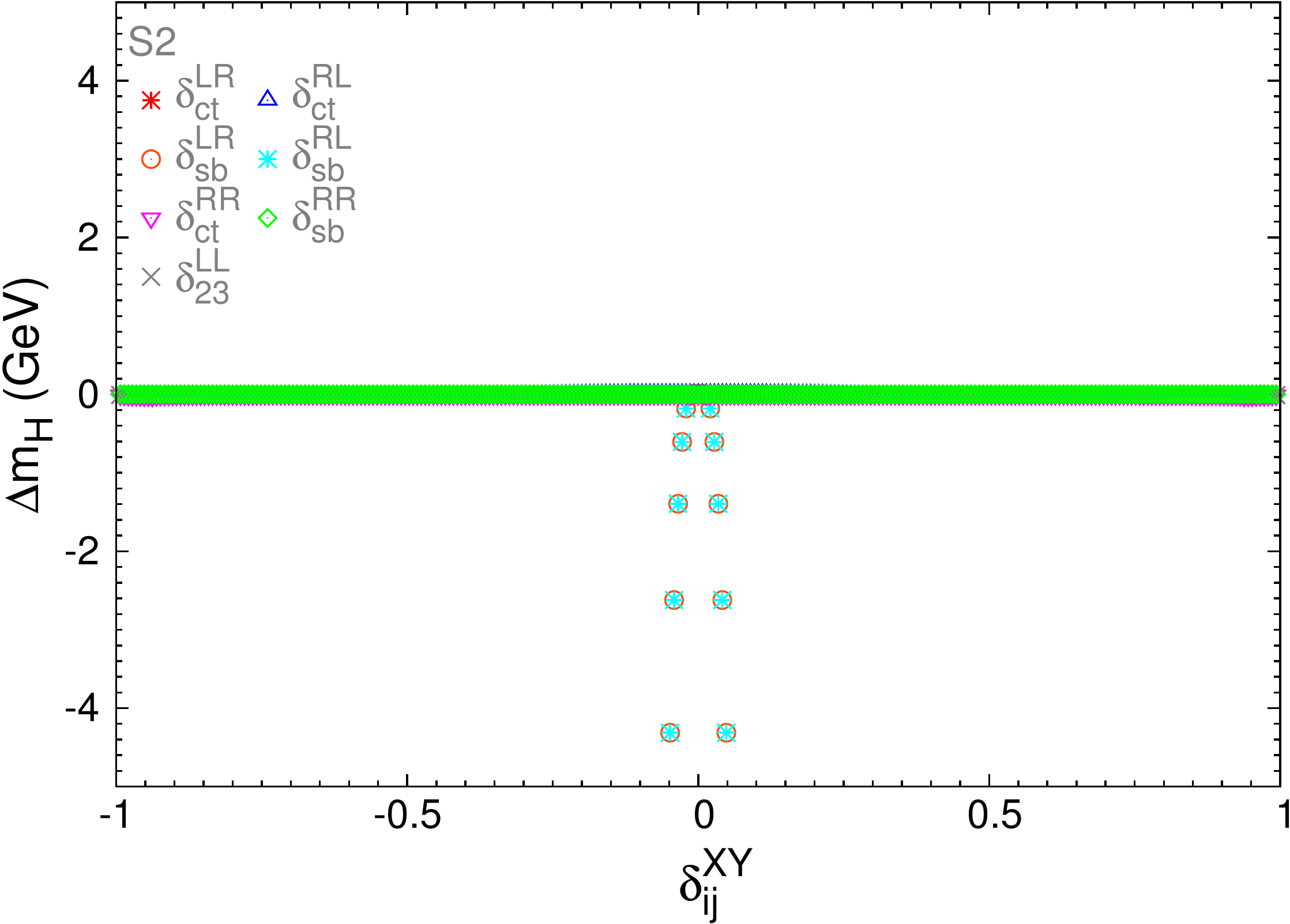}\\ 
\includegraphics[width=13.2cm,height=17.2cm,angle=270]{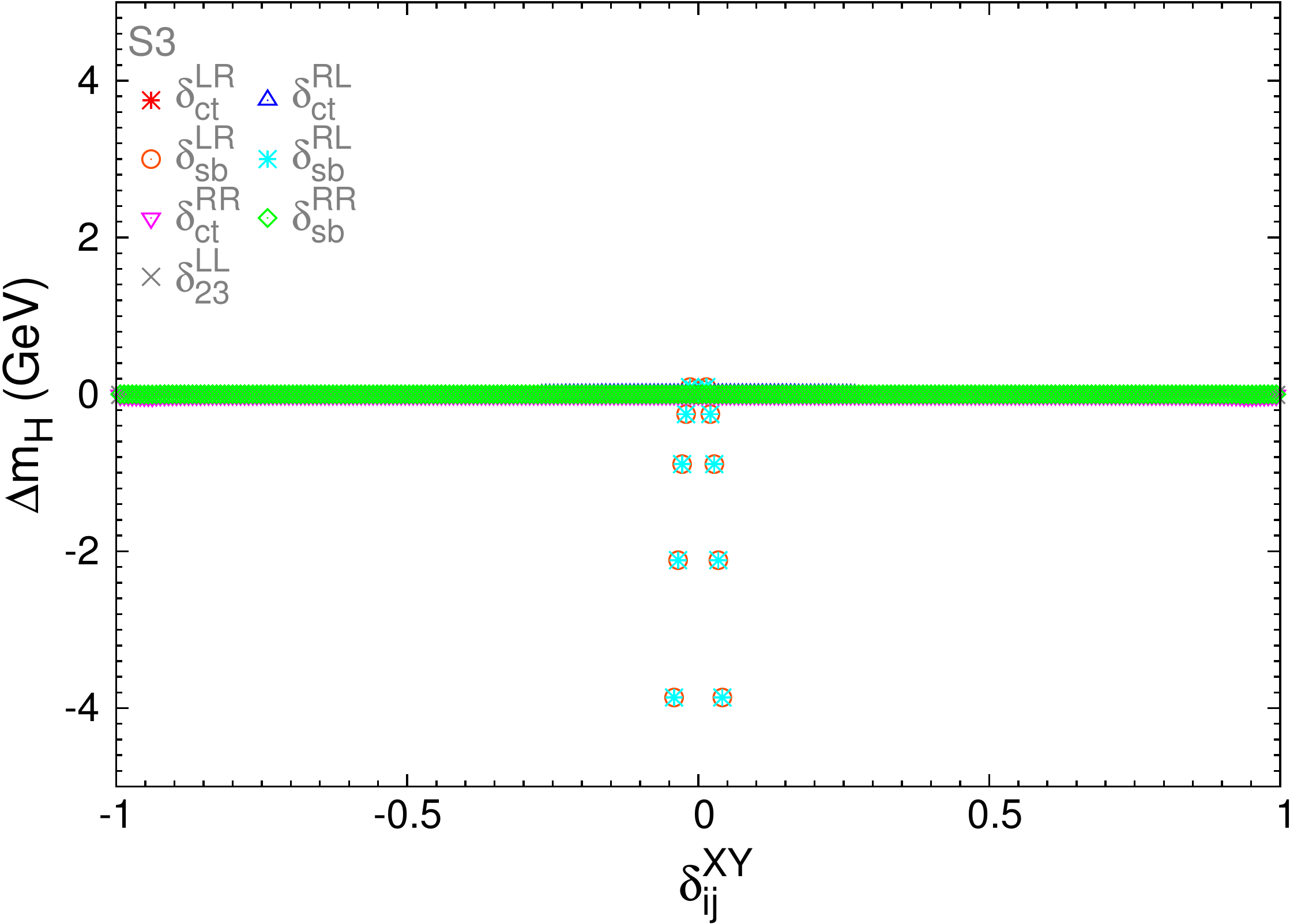}&
\includegraphics[width=13.2cm,height=17.2cm,angle=270]{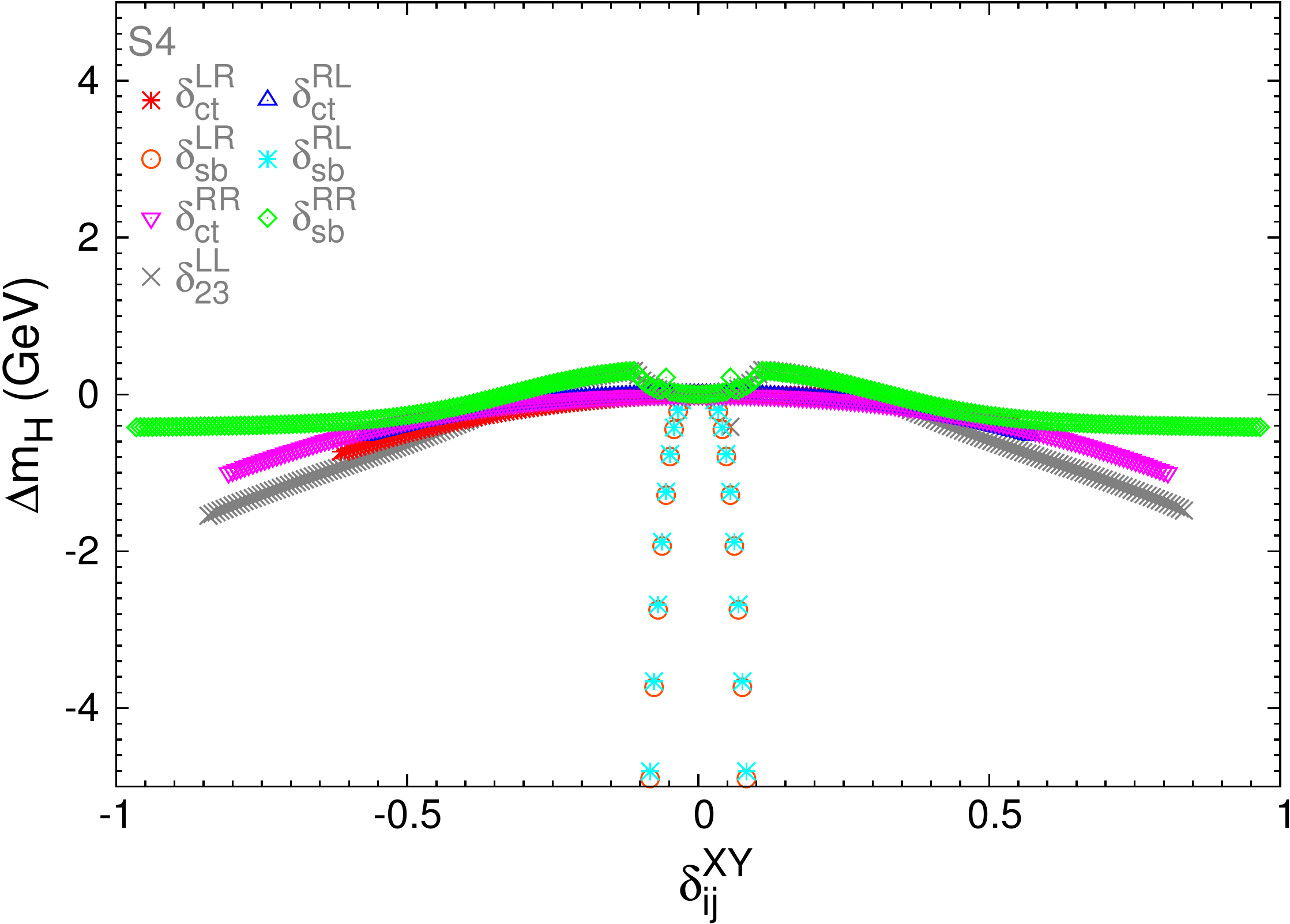}\\ 
\includegraphics[width=13.2cm,height=17.2cm,angle=270]{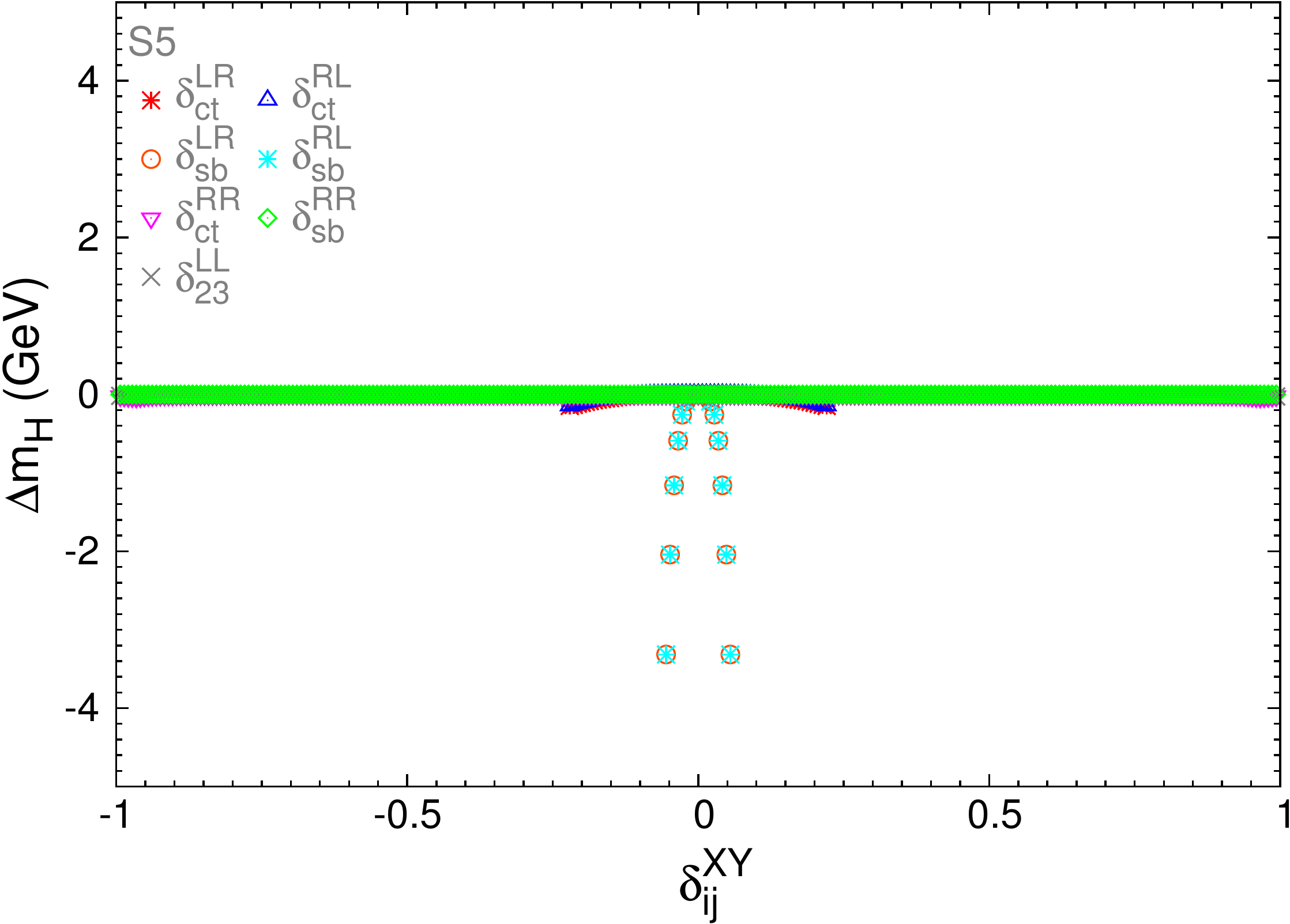}& 
\includegraphics[width=13.2cm,height=17.2cm,angle=270]{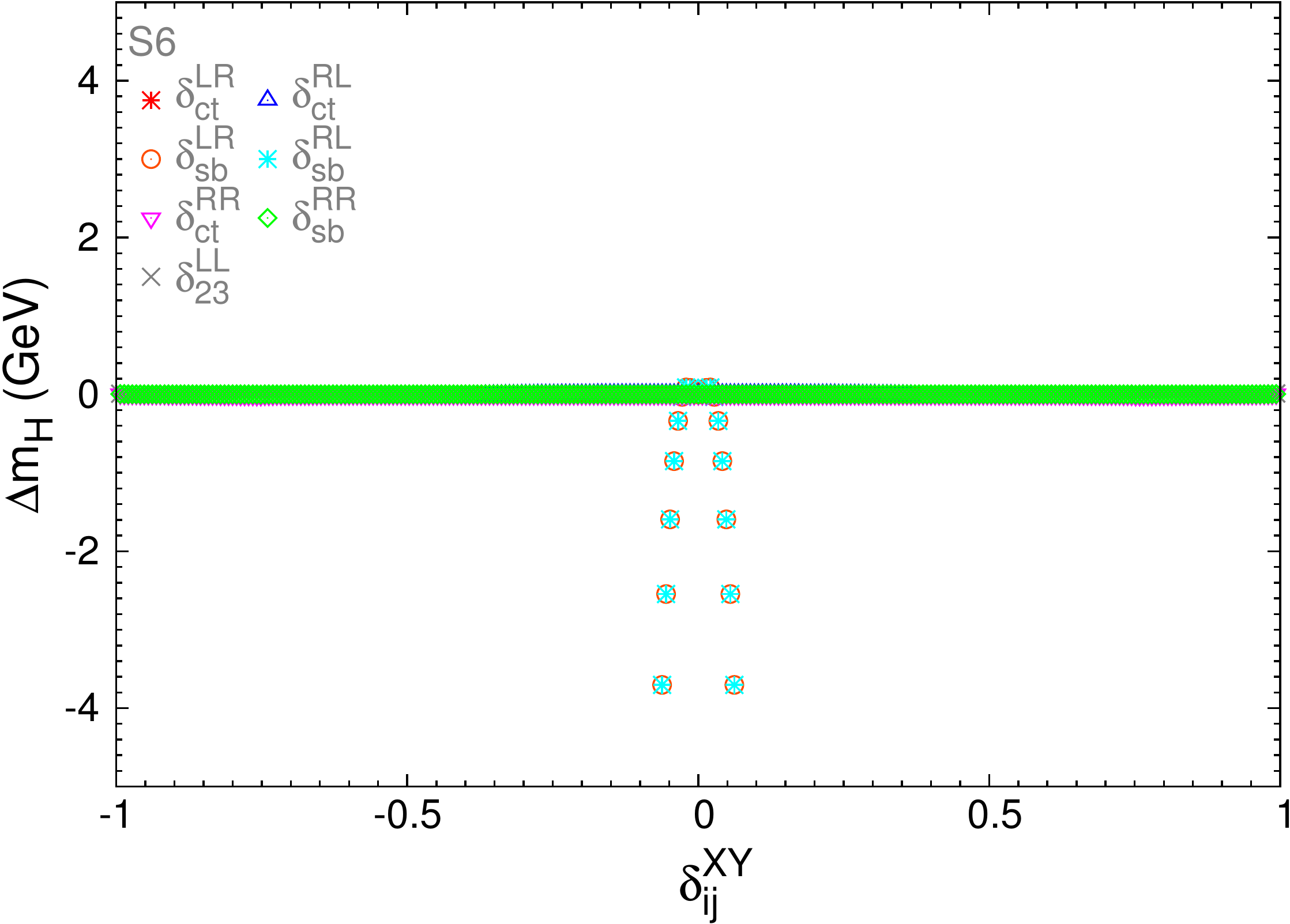}\\ 
\end{tabular}}}
\caption{Sensitivity to the NMFV deltas in $\DmH$ for the six post-LHC scenarios of table \ref{tab:spectra}.}
\label{figdeltamH02} 
\end{figure}

\begin{figure}[h!] 
\centering
\hspace*{-10mm} 
{\resizebox{17.3cm}{!} 
{\begin{tabular}{cc} 
\includegraphics[width=13.2cm,height=17.2cm,angle=270]{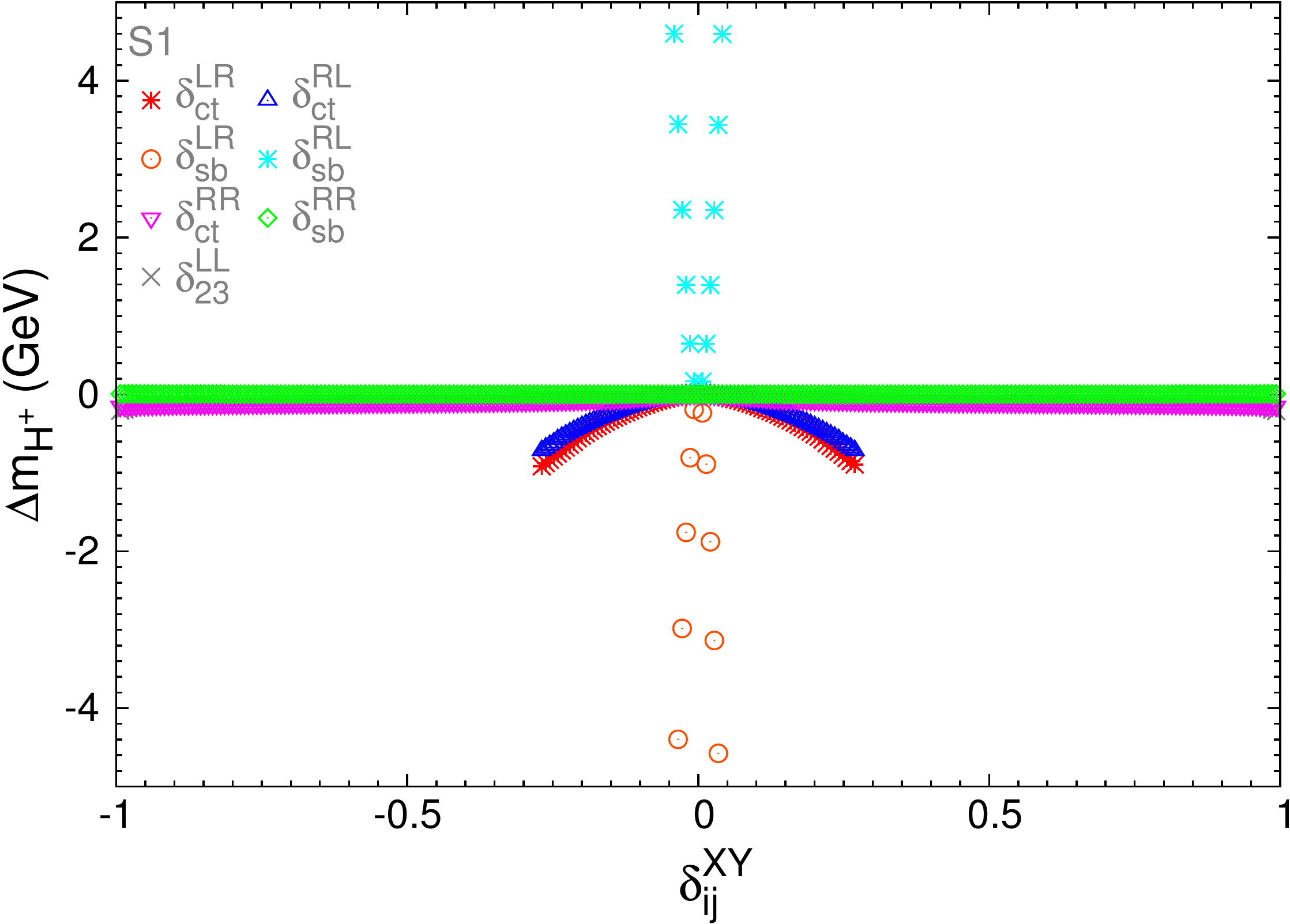}& 
\includegraphics[width=13.2cm,height=17.2cm,angle=270]{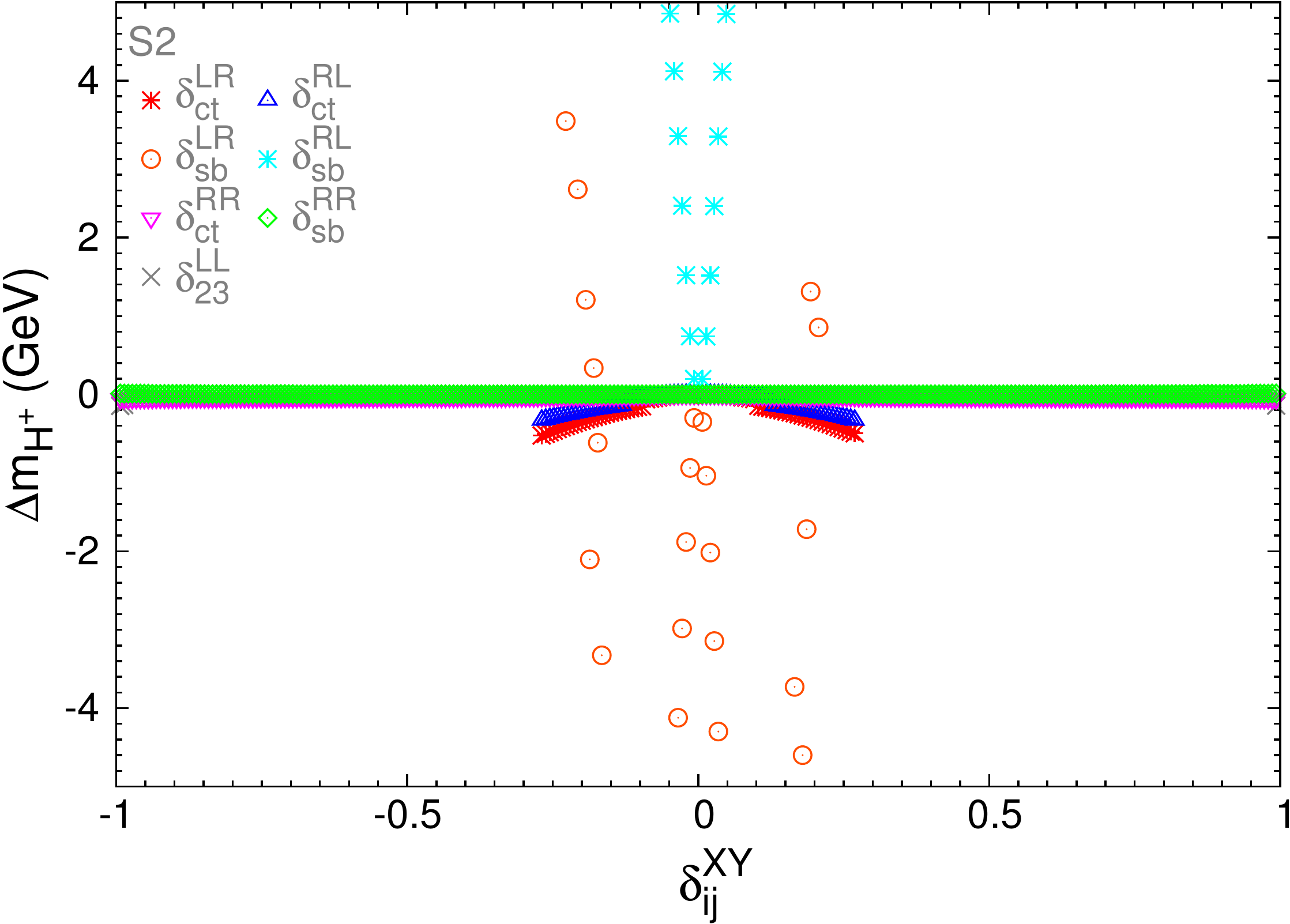}\\ 
\includegraphics[width=13.2cm,height=17.2cm,angle=270]{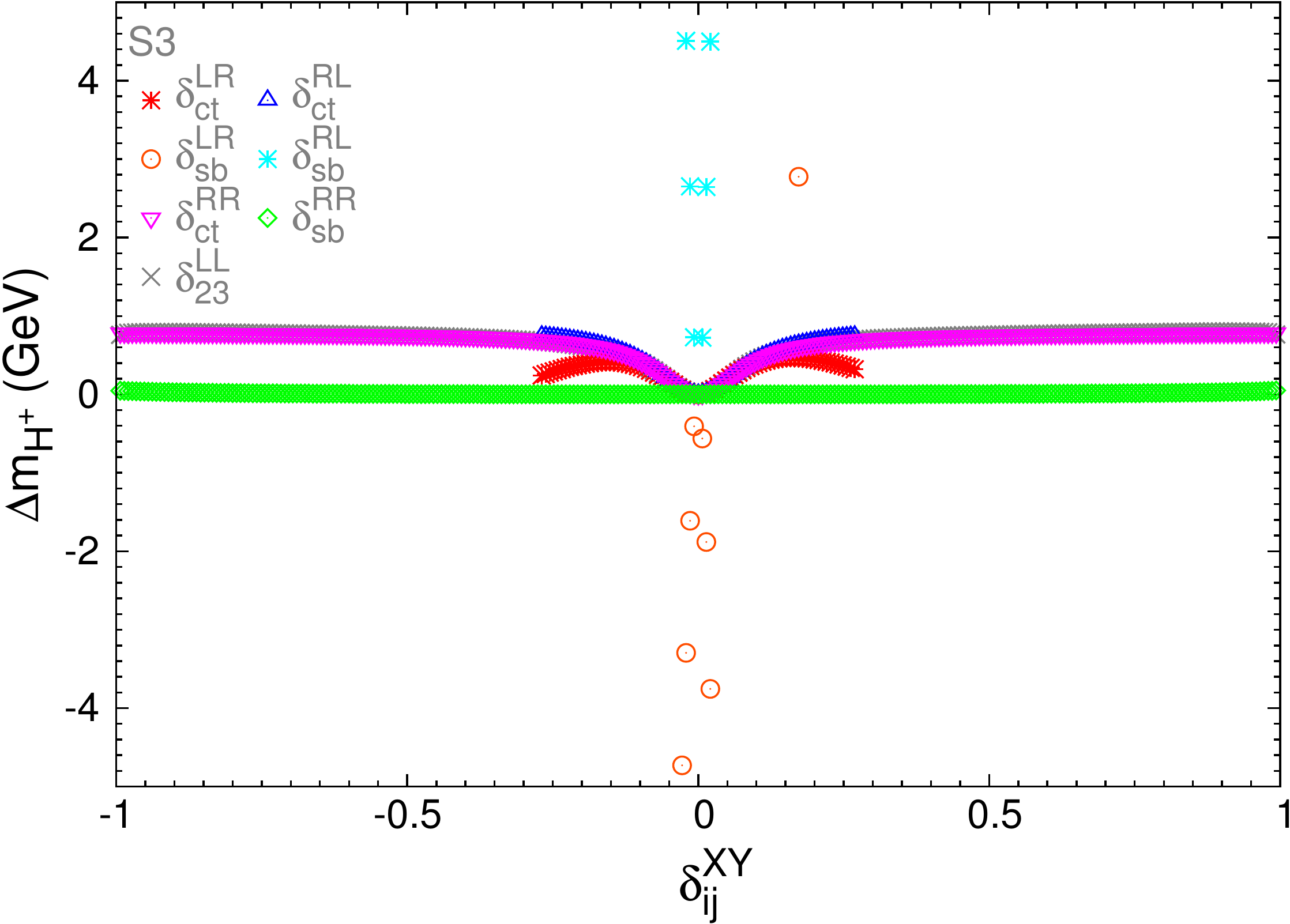}&
\includegraphics[width=13.2cm,height=17.2cm,angle=270]{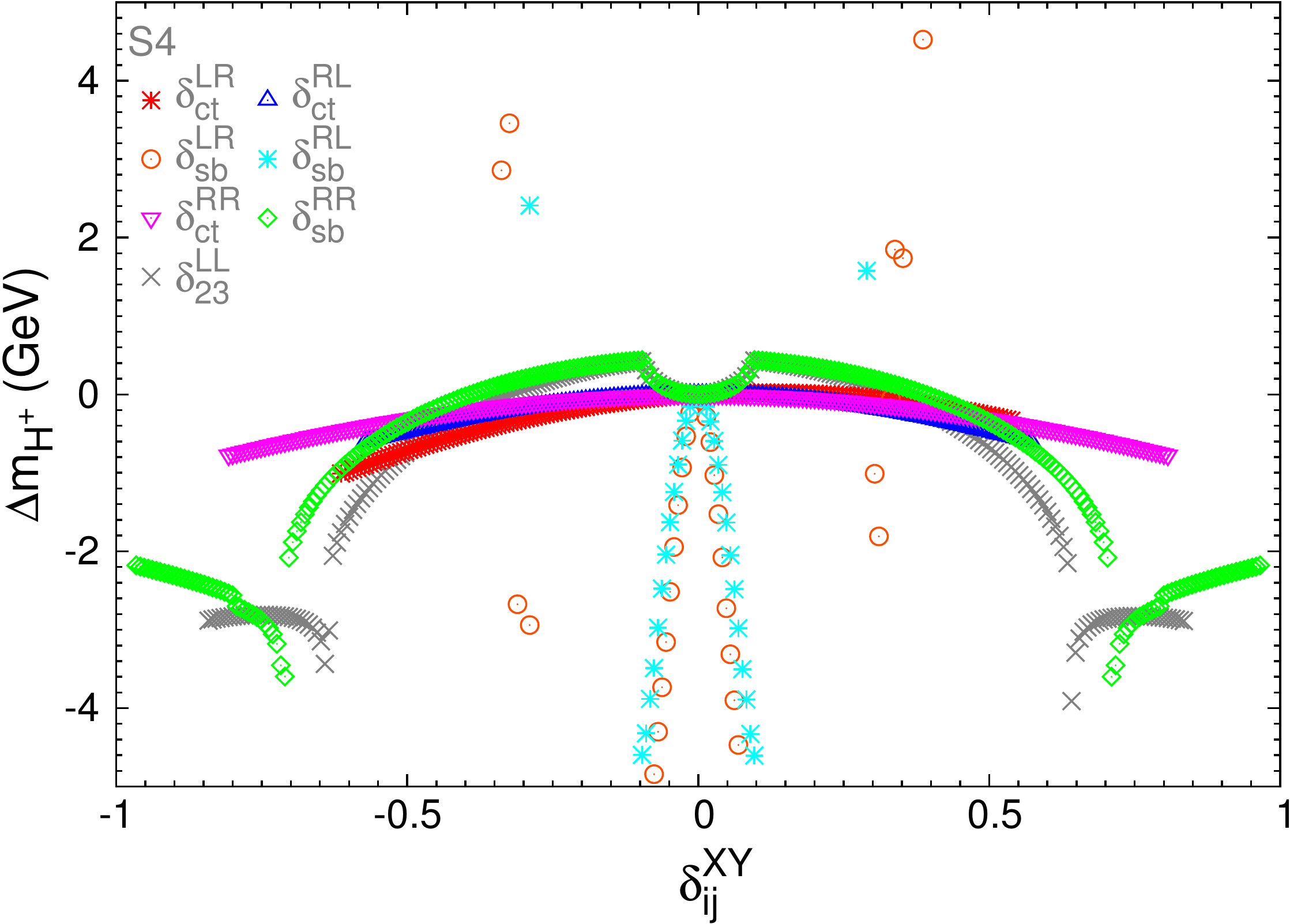}\\ 
\includegraphics[width=13.2cm,height=17.2cm,angle=270]{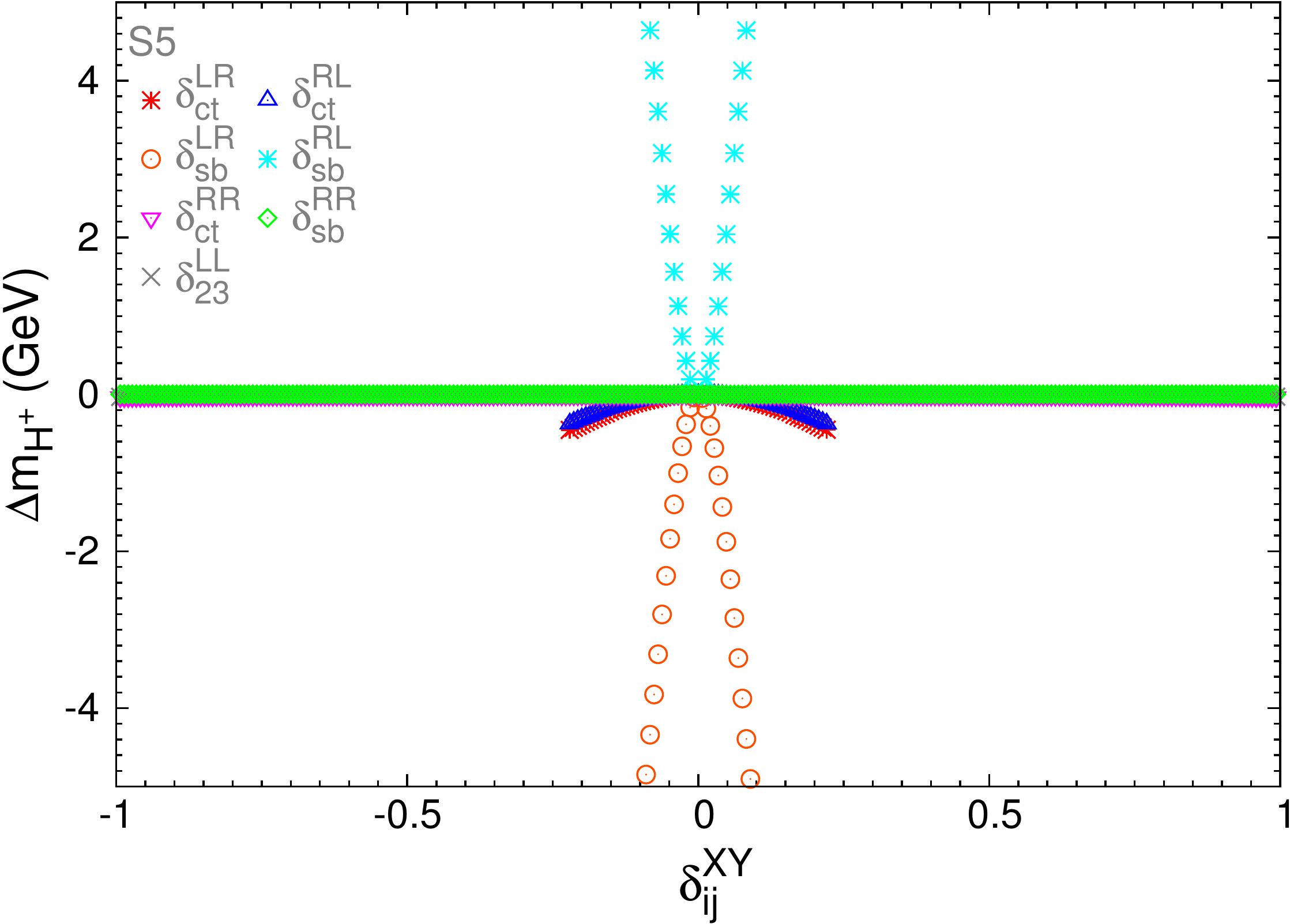}& 
\includegraphics[width=13.2cm,height=17.2cm,angle=270]{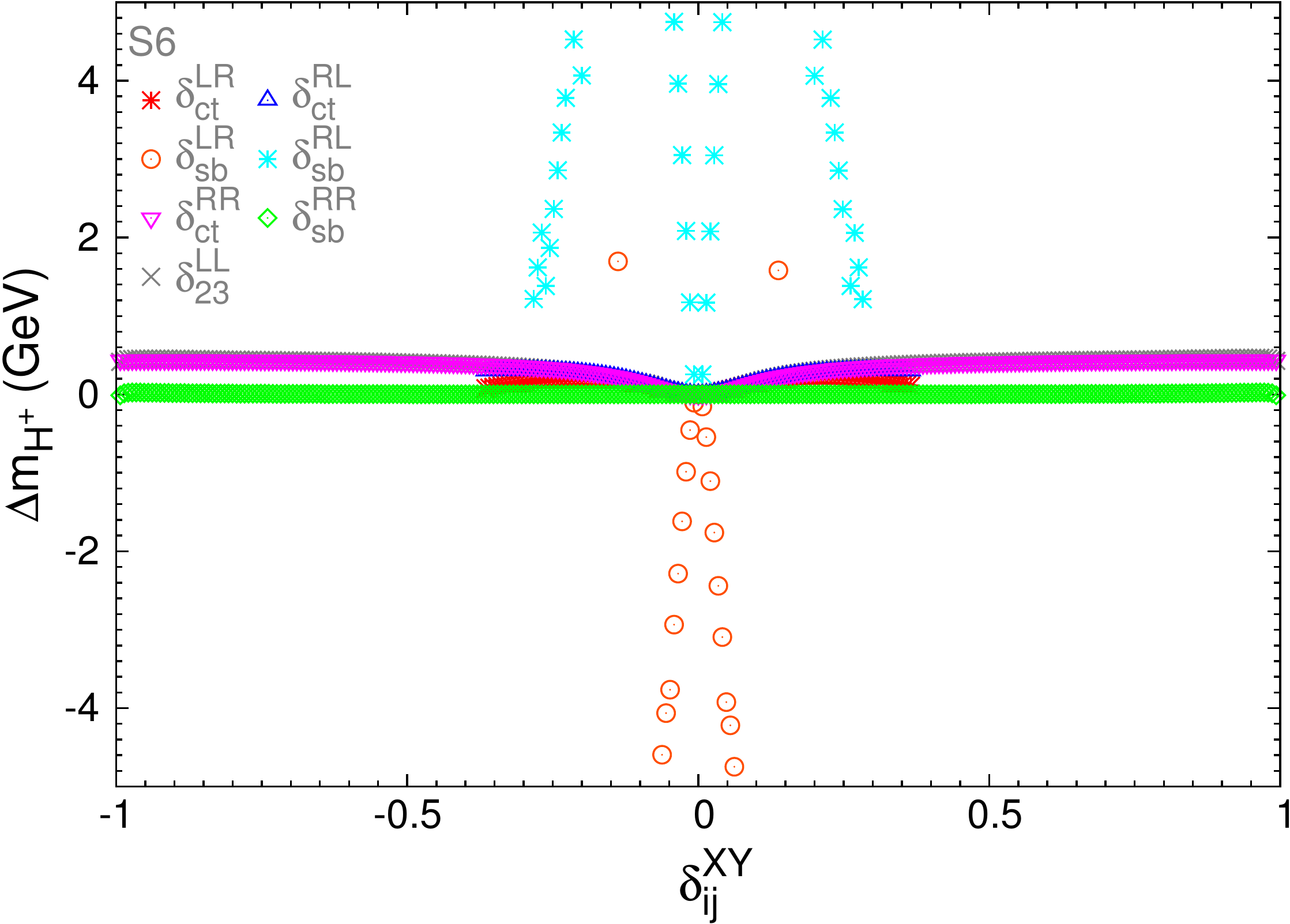}\\ 
\end{tabular}}}
\caption{Sensitivity to the NMFV deltas in $\DmHp$ for the six post-LHC scenarios of table \ref{tab:spectra}.}
 \label{figdeltamHp2}
\end{figure}

In figs. \ref{figdeltamh02}, \ref{figdeltamH02} and
\ref{figdeltamHp2} there can be seen our numerical results for  
$\Dmh$, $\DmH$ and $\DmHp$, respectively, as functions of the seven
considered flavour changing deltas, $\de^{LL}_{23}$, $\de^{LR}_{ct}$,
$\de^{LR}_{sb}$, $\de^{RL}_{ct}$, $\de^{RL}_{sb}$, $\de^{RR}_{ct}$ and
$\de^{RR}_{sb}$,  
which we vary in the interval $-1 \leq \deXYij \leq 1$. These plots have been done for the points of table \ref{tab:spectra} designed considering the most up-to-date experimental measurements. This same set of points was considered in Section \ref{numresafterlhc} for studying the current constraints from $B$ physics to the squark flavour mixing. Taking those into account here would lead to a reduced variation of the Higgs boson mass corrections.

The main conclusions from these new figures, taking into account what we learned from the pre-LHC points, are the following:
\begin{itemize}
\item[-] General features: 

The forms of all mass corrections remain the same as in the pre-LHC studied points, being symmetric and with positive and negative corrections depending on the value of the delta. The size of the corrections however has changed, for example the positive corrections to the light Higgs boson $\Dmh$ are staying in all scenarios below 2 GeV. In contrast, the negative corrections to this boson mass, and in general all the corrections to $\DmH$ and $\DmHp$ are much more steep, and reach large values with smaller values of delta than in the previous scenarios. For example, $\delta^{LR/RL}_{ct}$ can generate corrections to $\Dmh$ around 5 GeV with values of the deltas around 0.1, while in the pre-LHC scenarios that size of the corrections was reached for $\delta^{LR/RL}_{ct}$ with values around 0.5. As we commented before, the negative corrections $\Dmh$ coming from $\delta^{LR/RL}_{ct}$ can be very large, and even lead to a negative mass squared value. In these post-LHC scenarios this happens for values of the deltas around 0.3, while in the pre-LHC scenarios this happened for values between 0.5 and 0.6 depending on the point.

There are no big differences in the behaviour of the corrections between the different points, even having different values of $\tan \beta$, except for the point S4 that has very light squarks for the third generation. We find large corrections $|\Dmh| \gsim 5 \gev$ for $\de^{LR}_{sb}$, $\de^{RL}_{sb}$, $\de^{LR}_{ct}$ and $\de^{RL}_{ct}$ around 0.1 and 0.2 for all the points except S4. For $\DmH$ and $\DmHp$ these corrections are reached for deltas of 0.1 but only for $\de^{LR}_{sb}$ and $\de^{RL}_{sb}$. We find also $\Dmh$ corrections around 1 GeV for values of the deltas around 0.9 in the case of $\de^{RR}_{ct}$ and $\de^{LL}_{23}$, that could be relevant taking into account the future precision of the LHC and future experiments.
  
\item[-] Sensitivity to the various deltas:
 
We find again very strong sensitivity in the three mass corrections $\Dmh$, $\DmH$ and $\DmHp$, to $\de^{LR}_{sb}$ and $\de^{RL}_{sb}$ for all the considered points. 

In the case of $\Dmh$ there is also an important sensitivity to 
$\de^{LR}_{ct}$ and $\de^{RL}_{ct}$ in all the considered points, due to the same reasons we
explained in the pre-LHC case. Also some sensitivity to $\de^{RR}_{ct}$ and $\de^{LL}_{23}$ is found for large values of the deltas. For the S4 point, this sensitivity is much stronger reaching large corrections. 

For $\DmH$ in there can be seen some sensitivity to $\de^{RR}_{ct}$, and a general sensitivity to all deltas for S4.
 
\end{itemize}

\newpage
\clearpage
\newpage
\clearpage
\newpage


\chapter{Slepton flavour mixing effects on LFV observables}
\label{phenoflavourslep}


In this chapter we will perform a general study of all the flavour mixing parameters in the slepton sector, as they were introduced in Section \ref{sec:slepsector}. To this end, different LFV processes will be selected in the first section, relevant to each one of the flavour parameters. These processes are known to get important contributions in SUSY theories and will consist in radiative $l_j \to l_i \gamma$ \cite{Borzumati:1986qx,Casas:2001sr,Hisano:2009ae,Antusch:2006vw}, leptonic $l_j \to 3 l_i$ \cite{Arganda:2005ji,Antusch:2006vw} and semileptonic $\tau \to \mu \eta$ and $\tau \to e \eta$ LFV decays \cite{Arganda:2008jj,Herrero:2009tm}, and conversion rates of leptons in heavy nuclei \cite{Arganda:2007jw}. Although the radiative decays are usually the most constraining
LFV processes, the leptonic and semileptonic decays are also of interest
because they can be mediated by the MSSM Higgs bosons, therefore  giving
access to the Higgs sector parameters and, presumably, with different
sensitivities to the various flavour mixing parameters than those involved in the
radiative decays. Once the processes are selected, and with the scenarios that were defined in Section \ref{scenarios} we will study the sensitivity and the constraints coming from these observables to the flavour mixing parameters and the MSSM parameters, and analyze comparatively which processes are the most competitive ones.

Previous studies of general slepton 
mixing within the MSSM have already set upper bounds for the
values of some of these flavour mixing parameters from experimental LFV searches (for a review see, for instance,
\cite{Raidal:2008jk}). Most of these studies done within the Mass Insertion Approximation focused on the LFV radiative
decays~\cite{Masina:2002mv}. Others also took into account the leptonic LFV three body
decays \cite{Arganda:2005ji}, as well as the muon to electron
conversion in heavy nuclei\cite{Paradisi:2005fk}. There are also some 
studies that focus on the chirally-enhanced loop corrections that are 
induced in the MSSM in presence of general sources of lepton flavour 
violation~\cite{Crivellin:2010er,Crivellin:2011jt}. For a recent review on LFV see, for instance, \cite{Hirsch:2011zza}.

Our work will update these studies in the
light of recent data, both on the most relevant LFV processes
\cite{Adam:2013mnn,Aubert:2009ag,Bertl:2006up,Bellgardt:1987du,Hayasaka:2010np,Hayasaka:2010et,Miyazaki:2008mw}
and also in view of the collected data at LHC\cite{LHCHiggs,Chatrchyan:2013lba,LHCSusy,CMS-PAS-HIG-12-050}.
Besides, our full one-loop computations for all these LFV processes have the advantage over other studies in the literature
that they are done fully in the physical mass basis. Therefore we do not use the MIA and our results are valid in a
wider range of the sfermion flavour mixing parameters.
The selected scenarios are all compatible with
LHC data. In particular these scenarios have relatively
heavy SUSY spectra, which are naturally in
agreement with the present MSSM particle mass bounds (although
substantially lower masses, especially in the electroweak sector, are
allowed by LHC data). Furthermore the selected scenarios are chosen such
that the light $\cp$-even MSSM Higgs mass is around $125 - 126 \gev$ and
thus in agreement with the Higgs boson discovery~\cite{LHCHiggs}.
In addition we
require that our selected MSSM scenarios give a
prediction for the muon anomalous magnetic moment, $(g-2)_\mu$, in
agreement with current data~\cite{Bennett:2006fi}.

The results in this chapter have been published in \cite{Arana-Catania:2013nha}.


\section{Study of LFV processes} 
\label{sec:scenariosLFV}

The general slepton flavour mixing introduced in Section \ref{sec:slepsector} produce interactions
among mass eigenstates of different generations, therefore changing
flavour. 
In the physical basis for leptons $l_i$ ($i=1,2,3$), sleptons $\til_X$
($X=1,..,6$), sneutrinos $\tinu_X$ ($X=1,2,3$),  neutralinos ${\tilde
  \chi}_A^0$ ($A=1,2,3,4$), charginos ${\tilde \chi}_A^\pm$ ($A=1,2$)
and Higgs bosons,  
$H_p\,\,\, (p=1,2,3)= h^0, H^0, A^0$,  one gets generically
non-vanishing couplings for intergenerational interactions
like, for instance: 
${\tilde \chi}_A^0 l_i {\tilde l}_X$, 
${\tilde \chi}_A^\pm l_i {\tilde \nu}_X$, 
$Z \til_X \til_Y$, $H_p \til_X \til_Y$ and  $H_p \tinu_X \tinu_Y$. 
When these interactions appear in loop-induced processes they
can then mediate LFV processes
involving leptons of different flavours $l_i$ and $l_j$, with $i\neq j$,
in the external states. The dependence of the LFV rates for these
processes on the previously introduced  $\deABij$ parameters then
appears both in the values of the physical slepton and sneutrino masses,
and in the values of these intergenerational couplings via the rotation
matrices  $R^{\til}$ and $R^{\tinu}$. For the present work, we use the
set of Feynman rules for these and other relevant couplings among mass
eigenstates, as summarized in
\citeres{Arganda:2005ji,Arganda:2007jw}.


\subsection{Selected LFV processes}

Our selection of LFV processes is driven by the requirement that we wish
to determine the constraints on all the slepton flavour mixing parameters
by studying different kinds of one-loop LFV vertices involving $l_i$ and
$l_j$ with $i \neq j$ in the external lines. In particular we want to
study the sensitivity to the $\deABij$'s in the most relevant
(three-point) LFV one-loop vertices, which are: the vertex with a
photon, $(l_il_j\gamma)_{\rm 1-loop}$, the vertex with a $Z$ gauge
boson, $(l_il_jZ)_{\rm 1-loop}$ and the vertices with the Higgs
bosons, 
$(l_il_jh^0)_{\rm 1-loop}$, $(l_il_jH^0)_{\rm 1-loop}$ and
$(l_il_jA^0)_{\rm 1-loop}$. This leads us to single out some specific
LFV processes where these one-loop generated LFV vertices play a
relevant role. We have chosen the following subset of LFV processes, all
together involving these particular LFV one-loop vertices:  
\begin{itemize}
\item[1.-] Radiative LFV decays: $\mu \to e \gamma$, $\tau \to e \gamma$ and
$\tau \to \mu \gamma$. These are sensitive to the $\deABij$'s via the
  $(l_il_j\gamma)_{\rm 1-loop}$ vertices with a real photon. 
\item[2.-] Leptonic LFV decays: $\mu \to 3 e$, $\tau \to 3 e$ and $\tau
  \to 3 \mu$. These are sensitive to the $\deABij$'s via the
  $(l_il_j\gamma)_{\rm 1-loop}$ vertices with a virtual photon, via
  the $(l_il_jZ)_{\rm 1-loop}$ vertices with a virtual $Z$, and via the
  $(l_il_jh^0)_{\rm 1-loop}$, $(l_il_jH^0)_{\rm 1-loop}$ and
  $(l_il_jA^0)_{\rm 1-loop}$ vertices with  virtual Higgs bosons. 
\item[3.-] Semileptonic LFV tau decays: $\tau \to \mu \eta$ and $\tau
  \to e \eta$. These are sensitive to the $\deABij$'s via $(\tau \mu
  A^0)_{\rm 1-loop}$ and $(\tau e A^0)_{\rm 1-loop}$ vertices,
  respectively, with a virtual $A^0$, and via  
$(\tau \mu Z)_{\rm 1-loop}$ and $(\tau e Z)_{\rm 1-loop}$ vertices,
  respectively with a virtual $Z$. 
\item[4.-] Conversion of $\mu$ into $e$ in heavy nuclei: These are
  sensitive to the $\deABij$'s via the $(\mu e\gamma)_{\rm 1-loop}$
  vertex with a virtual photon, via the $(\mu e Z)_{\rm 1-loop}$
  vertex with a virtual $Z$, and via the $(\mu e h^0)_{\rm 1-loop}$ and
  $(\mu e H^0)_{\rm 1-loop}$ vertices with a virtual $h^0$ and $H^0$
  Higgs boson, respectively.  
\end{itemize}  
%

\begin{figure}[ht!]
\begin{center}
\psfig{file=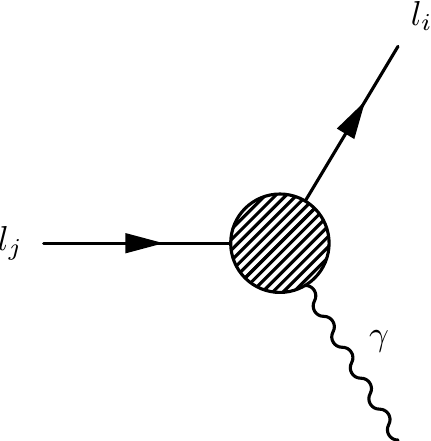 ,scale=0.83,clip=}\\
\vspace{1em}
\psfig{file=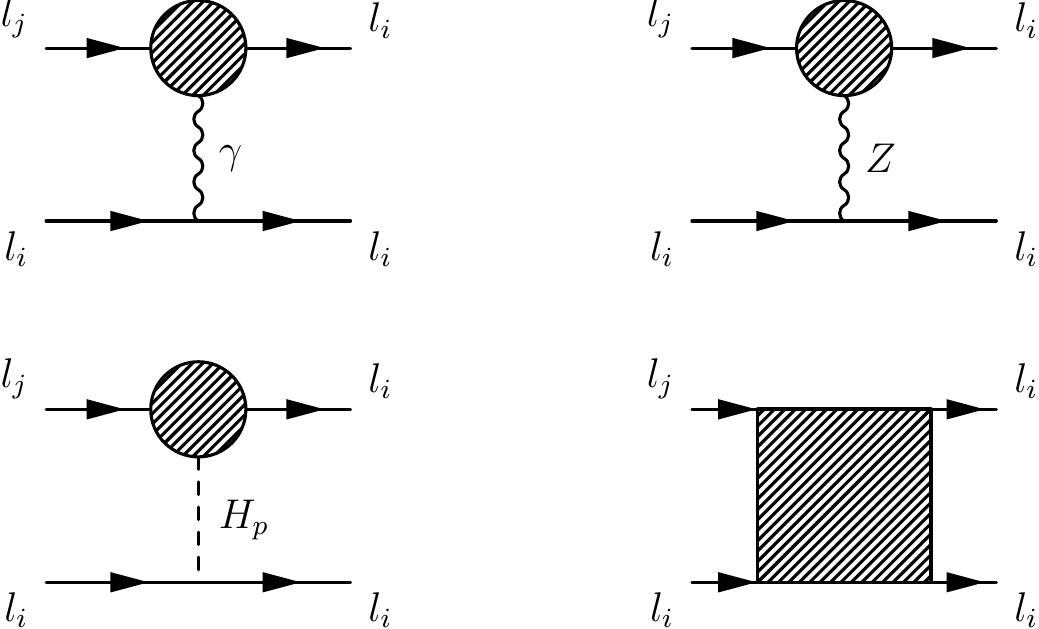,scale=0.83,clip=}\\
\vspace{1em}
\psfig{file=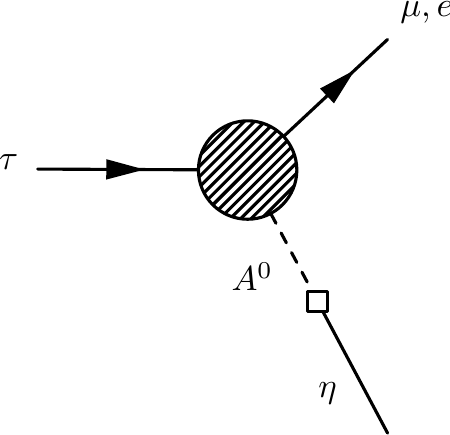,scale=0.83,clip=}
\psfig{file=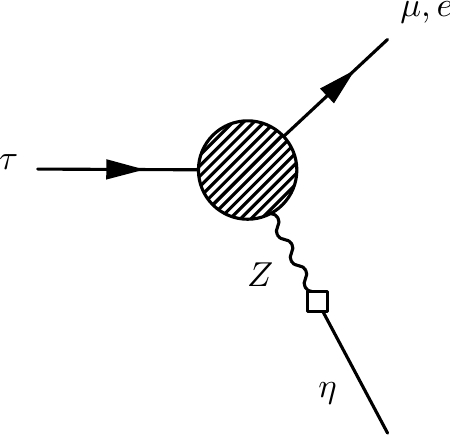,scale=0.83,clip=}\\
\vspace{1em}
\psfig{file=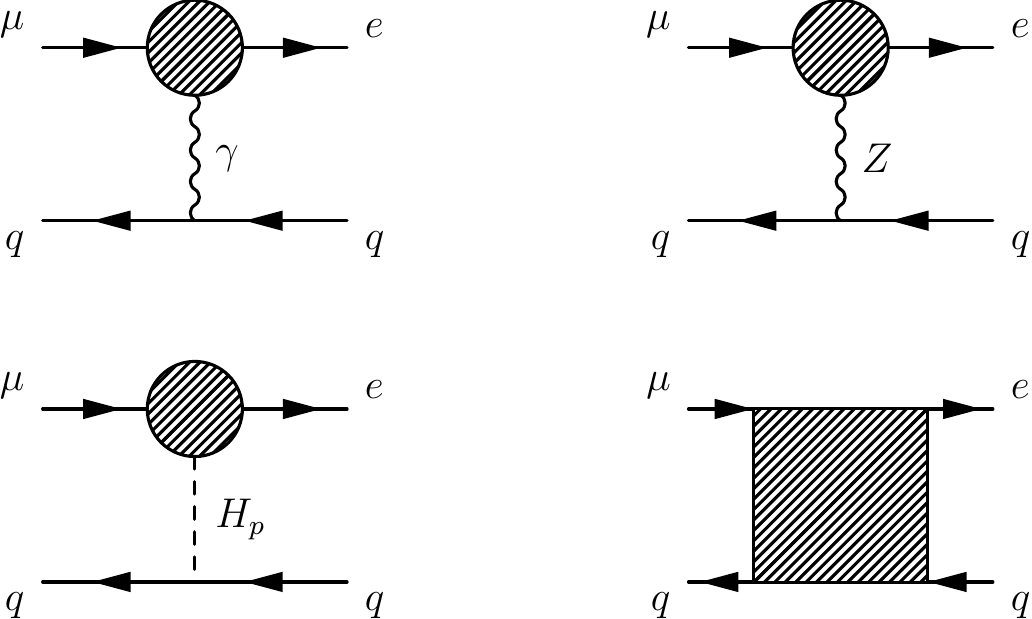,scale=0.83,clip=} 
\end{center}
\caption{Generic one-loop diagrams contributing to LFV processes: 
1) $l_j \to l_i \gamma$; 
2) $l_j \to 3 l_i$, 
  mediated by $\gamma$ and $Z$ gauge bosons, by $H_p=h^0,H^0,A^0$ Higgs
  bosons and by boxes;   
3) $\tau \to \mu \eta$ and $\tau \to e \eta$, mediated by $A^0$ Higgs boson and by $Z$ gauge
  boson;
4) $\mu-e$ conversion in nuclei,
  mediated by $\gamma$, and $Z$ gauge bosons, by $H_p=h^0,H^0$ Higgs
  bosons, and by boxes.}  
\label{diagramsLFV}
\end{figure} 

The generic one-loop diagrams contributing to all the LFV processes above, are 
summarized in Fig.\ref{diagramsLFV}. These include the $\gamma$-mediated
diagrams, the $Z$-mediated diagrams (the $Z$-penguin diagrams in $l_j \to 3 l_i$ were also studied in \cite{Hirsch:2012ax}), and the $h^0$, $H^0$ and
$A^0$-mediated diagrams. The generic one-loop box diagrams are also
shown in this figure. These also include the $\deABij$'s but their
sensitivities to these parameters are much lower than via the above
quoted three-point vertices. 
They are, however, included in our analytical results and in our
  numerical evaluation.

For our forthcoming numerical analysis of these LFV processes we have
implemented the full one-loop formulas into our private Fortran
  code. The analytical results are taken from various publications:
\citere{Arganda:2005ji} for $\br(l_j \to 3 l_i)$ and  
$\br(l_j \to l_i \gamma)$, \citere{Arganda:2008jj} for $\br(\tau \to \mu
\eta)$ and $\br(\tau \to e \eta)$, and \citere{Arganda:2007jw} for  
the $\mu - e$ conversion rate in heavy nuclei, relative to the muon
capture rate $\CR(\mu-e, {\rm Nuclei})$.  
Following the same procedure of \cite{Arganda:2008jj} we use Chiral
Perturbation Theory for the needed hadronization of quark bilinears
involved in  the quark-level $\tau \to \mu q q'$ and $\tau \to e q q'$
decays that lead the $\eta$ particle in the final state. Our treatment
of the heavy nuclei and the proper approximations to go from the LFV
amplitudes at the parton level to the LFV rates at the nuclear level are
described in \cite{Arganda:2007jw}. For brevity, we omit to
explicit here all 
these needed formulas for the computation of the LFV rates and refer
the reader to the above quoted references for the details.  

The list of specific one-loop diagrams contributing to the relevant
$(l_il_j\gamma)_{\rm 1-loop}$, $(l_il_jZ)_{\rm 1-loop}$
$(l_il_jh^0)_{\rm 1-loop}$, $(l_il_jH^0)_{\rm 1-loop}$ and
$(l_il_jA^0)_{\rm 1-loop}$ vertices can also be found in
\citeres{Arganda:2005ji,Arganda:2007jw}. 
The main contributions come from the loops with
  charginos/sneutrinos and with neutralinos/sleptons. This will be
  relevant for the analytical interpretation of our results below.


\subsection{The MIA basic reference formulas} 
\label{sec:MIA}

For completeness, and in order to get a better understanding of the
forthcoming full one-loop results leading to the maximal allowed deltas
and their behaviour with the relevant MSSM parameters, we include in this
section the main formulas for the LFV radiative decays within the Mass
Insertion Approximation that we take from
\citere{Paradisi:2005fk}. These are simple formulas and illustrate
clearly the qualitative behaviour of the LFV rates with all the deltas
and all the MSSM parameters.  The branching ratios of the radiative $l_j
\to l_i \gamma$ decays, with $ji=21$, $31$ and $32$, are:  
\BEA
{\rm BR}(l_j \to l_i \gamma) &=& \frac{\alpha}{4} 
(\frac{m_{l_j}^5}{\Gamma_{l_j}}) \left(|(A_{ij}^L)|^2+|(A_{ij}^R)|^2\right)
\label{BRs}
\EEA
where $\Gamma_{l_j}$ is the total ${l_j}$ width,  and the amplitudes, in the single delta insertion approximation, are given by~\cite{Paradisi:2005fk}:
\BEA
(A_{ij}^L)_{\rm MIA} &=& \frac{\alpha_2}{4 \pi} \Delta^{LL}_{ij}
\left[\frac{f_{1n}(a_{L2})+f_{1c}(a_{L2})}{m_{\tilde L}^4} + \frac{\mu M_2 \tb}{(M_2^2-\mu^2)} \frac{(f_{2n}(a_{L2},b_L)+f_{2c}(a_{L2},b_L))}{m_{\tilde L}^4}\right] \nonumber \\  
&+& \frac{\alpha_1}{4 \pi} \Delta^{LL}_{ij}
\left[\frac{f_{1n}(a_L)}{m_{\tilde L}^4}+\mu M_1 \tb \left(\frac{-f_{2n}(a_L,b_L)}
{m_{\tilde L}^4(M_1^2-\mu^2)} +\frac{2f_{2n}(a_L)}{m_{\tilde L}^4(m_{\tilde R}^2-m_{\tilde L}^2)}\right) \right] \nonumber \\ 
&+&\frac{\alpha_1}{4 \pi} \Delta^{LL}_{ij} \left[\frac{\mu M_1 \tb}{(m_{\tilde R}^2-m_{\tilde L}^2)^2}  
   \left(\frac{f_{3n}(a_R)}{m_{\tilde R}^2}-\frac{f_{3n}(a_L)}{m_{\tilde L}^2}\right) \right]
\nonumber \\ 
&+&\frac{\alpha_1}{4 \pi} \Delta^{LR}_{ij}\left[\frac{1}{(m_{\tilde L}^2-m_{\tilde R}^2)}
\left(\frac{M_1}{m_{l_j}}\right) \left(\frac{f_{3n}(a_R)}{m_{\tilde R}^2}-\frac{f_{3n}(a_L)}{m_{\tilde L}^2} \right)\right]
\EEA
and 
\BEA
(A_{ij}^R)_{\rm MIA} &=&
\frac{\alpha_1}{4 \pi} \Delta^{RR}_{ij}
\left[\frac{4f_{1n}(a_R)}{m_{\tilde R}^4}+\mu M_1 \tb \left(\frac{2f_{2n}(a_R,b_R)}
{m_{\tilde R}^4(M_1^2-\mu^2)} +\frac{2f_{2n}(a_R)}{m_{\tilde R}^4(m_{\tilde L}^2-m_{\tilde R}^2)}\right) \right] \nonumber \\ 
&+&\frac{\alpha_1}{4 \pi} \Delta^{RR}_{ij} \left[\frac{\mu M_1 \tb}{(m_{\tilde L}^2-m_{\tilde R}^2)^2}  
   \left(\frac{f_{3n}(a_L)}{m_{\tilde L}^2}-\frac{f_{3n}(a_R)}{m_{\tilde R}^2}\right) \right]
\nonumber \\ 
&+&\frac{\alpha_1}{4 \pi} \Delta^{RL}_{ij}\left[\frac{1}{(m_{\tilde R}^2-m_{\tilde L}^2)}
\left(\frac{M_1}{m_{l_j}}\right) \left(\frac{f_{3n}(a_L)}{m_{\tilde L}^2}-\frac{f_{3n}(a_R)}{m_{\tilde R}^2} \right)\right]~,
\EEA
where $\alpha_1=(5/3)(\alpha/\cos^2\theta_W)$,
$\alpha_2=(\alpha/\sin^2\theta_W)$, $a_{L2}=M_2^2/m_{\tilde L}^2$,
$a_L=M_1^2/m_{\tilde L}^2$, $a_R=M_1^2/m_{\tilde R}^2$,  
 $b_L=\mu^2/m_{\tilde L}^2$, $b_R=\mu^2/m_{\tilde R}^2$, 
 $\Delta^{AB}_{ij}=\delta^{AB}_{ij} m_{\tilde A}m_{\tilde B}$ and
$m_{\tilde L}$ and $m_{\tilde R}$ are the average slepton
masses in the ${\tilde L}$ and ${\tilde R}$ slepton sectors,
respectively. 
The $f_{in}$'s and $f_{ic}$'s are loop functions from
neutralinos and charginos contributions, respectively, given by: 
\BEA 
f_{1n}(a)&=&\frac{-17a^3+9 a^2+9a-1+6a^2(a+3)\ln a}{24(1-a)^5}~, \nonumber \\
f_{2n}(a)&=&\frac{-5a^2+4a+1+2a(a+2) \ln a}{4(1-a)^4}~,\nonumber \\
f_{3n}(a)&=&\frac{1+2a \ln a -a^2}{2(1-a)^3}~,\nonumber \\
f_{1c}(a)&=&\frac{-a^3-9a^2+9a+1+6a(a+1) \ln a}{6(1-a)^5}~,\nonumber \\
f_{2c}(a)&=&\frac{-a^2-4a+5+2(2a+1) \ln a}{2(1-a)^4}~, \nonumber \\
f_{2n}(a,b)&=&f_{2n}(a)-f_{2n}(b)~, \non \\ 
f_{2c}(a,b)&=&f_{2c}(a)-f_{2c}(b)~.
\label{functions}
\EEA
It is also very illustrative to compare the forthcoming results with
those of the MIA for the case of equal mass scales, $m_{\tilde
  L}=m_{\tilde R}=\mu=M_2=M_1 \equiv m_S$. From the previous formulas we
get: 
\BEA
(A_{ij}^L)_{\rm MIA}&=&\frac{\alpha_2}{4 \pi} \delta^{LL}_{ij} \left[ \frac{1}{240} \frac{1}{m_S^2}+\tb \frac{1}{15} \frac{1}{m_S^2} \right] \nonumber \\ 
&+&\frac{\alpha_1}{4 \pi} \delta^{LL}_{ij} \left[ \frac{-1}{80}\frac{1}{m_S^2}+
\tb \frac{1}{12}\frac{1}{m_S^2} \right] \nonumber \\ 
&+&\frac{\alpha_1}{4 \pi} \delta^{LR}_{ij} \left[\frac{1}{m_S m_{l_j}} \right]
\label{MIA-L}
\EEA
and 
\BEA
(A_{ij}^R)_{\rm MIA}&=& 
 \frac{\alpha_1}{4 \pi} \delta^{RR}_{ij} \left[ \frac{-1}{20}\frac{1}{m_S^2}-
\tb \frac{1}{60}\frac{1}{m_S^2} \right] \nonumber \\ 
&+&\frac{\alpha_1}{4 \pi} \delta^{RL}_{ij} \left[\frac{1}{m_S m_{l_j}} \right]~.
\label{MIA-R}
\EEA
In all these basic MIA formulas (see also \cite{Hisano:2009ae}) one can see clearly the scaling of
the BRs with all the deltas, in the single mass insertion approximation, and
with the most relevant parameters for the present study, namely, the
common/average SUSY mass $m_{S}$,  and $\tb$. These formulas will be
used below in the interpretation of the full numerical results.


\subsection{Experimental bounds on LFV}
\label{sec:expbounds}

So far, LFV has not been observed. 
The best present (90\% CL) experimental bounds on the previously
selected LFV processes are summarized in the following: 
 \BEA
\br(\mu \to e \gamma) < 5.7 \times 10^{-13} &\mbox{\cite{Adam:2013mnn}} \nonumber \\
\br(\tau \to \mu \gamma) < 4.4 \times 10^{-8} &\mbox{\cite{Aubert:2009ag}}   \nonumber\\
\br(\tau \to e \gamma) < 3.3 \times 10^{-8} &\mbox{\cite{Aubert:2009ag}} \nonumber\\
\br(\mu \to eee) < 1.0 \times 10^{-12}& \mbox{\cite{Bellgardt:1987du}} \nonumber\\
\br(\tau \to \mu\mu\mu) < 2.1 \times 10^{-8}&\mbox{\cite{Hayasaka:2010np}}\nonumber\\
\br(\tau \to e e e)< 2.7 \times 10^{-8}&\mbox{\cite{Hayasaka:2010np}}\nonumber\\
\CR(\mu-e, {\rm Au}) < 7.0 \times 10^{-13}&\mbox{\cite{Bertl:2006up}}\nonumber\\
\br(\tau \to \mu \eta) < 2.3\times10^{-8}&\mbox{\cite{Hayasaka:2010et}}\nonumber\\
\br(\tau \to e \eta) < 4.4\times10^{-8}&\mbox{\cite{Hayasaka:2010et}}
\EEA
 At present, the most constraining bounds are from
$\br(\mu \to e \gamma)$, which has been just improved by the MEG
collaboration, and from $\CR(\mu-e, {\rm Au})$, both being at the 
\order{10^{-13}} level. Therefore, the 12 slepton mixings are by far the
most constrained ones. 
All these nine upper bounds above will be applied next to extract the
maximum allowed $|\delta^{AB}_{ij}|$ values.


\section{Results on LFV rates and constraints on slepton flavour mixings} 
\label{sec:results}

We have studied the previous observables in the pMSSM scenarios defined in Sections \ref{frameworkc} and \ref{sec:f2}. We work in a complete basis of $\delta^{AB}_{ij}$, that is we
take into account the full set of twelve parameters.  
For simplicity, we will assume real values for these flavour slepton
mixing parameters, therefore we will not have to be concerned with the
Lepton Electric Dipole Moments (EDM). Concretely, the scanned interval
in our estimates of LFV rates will be:  
\BE
-1 \le \delta^{AB}_{ij} \le +1 
\EE
For each explored non-vanishing single delta, $\delta^{AB}_{ij}$, or
pair of deltas, $(\delta^{AB}_{ij}, \delta^{CD}_{kl})$, the
corresponding slepton and sneutrino physical masses,  the slepton and
sneutrino rotation matrices, as well as the LFV rates will be numerically
computed with our private Fortran code. 

The results will be presented in the following two sections.

\subsection{Numerical results for specific pMSSM points}

\begin{figure}[ht!]
\begin{center}
\psfig{file=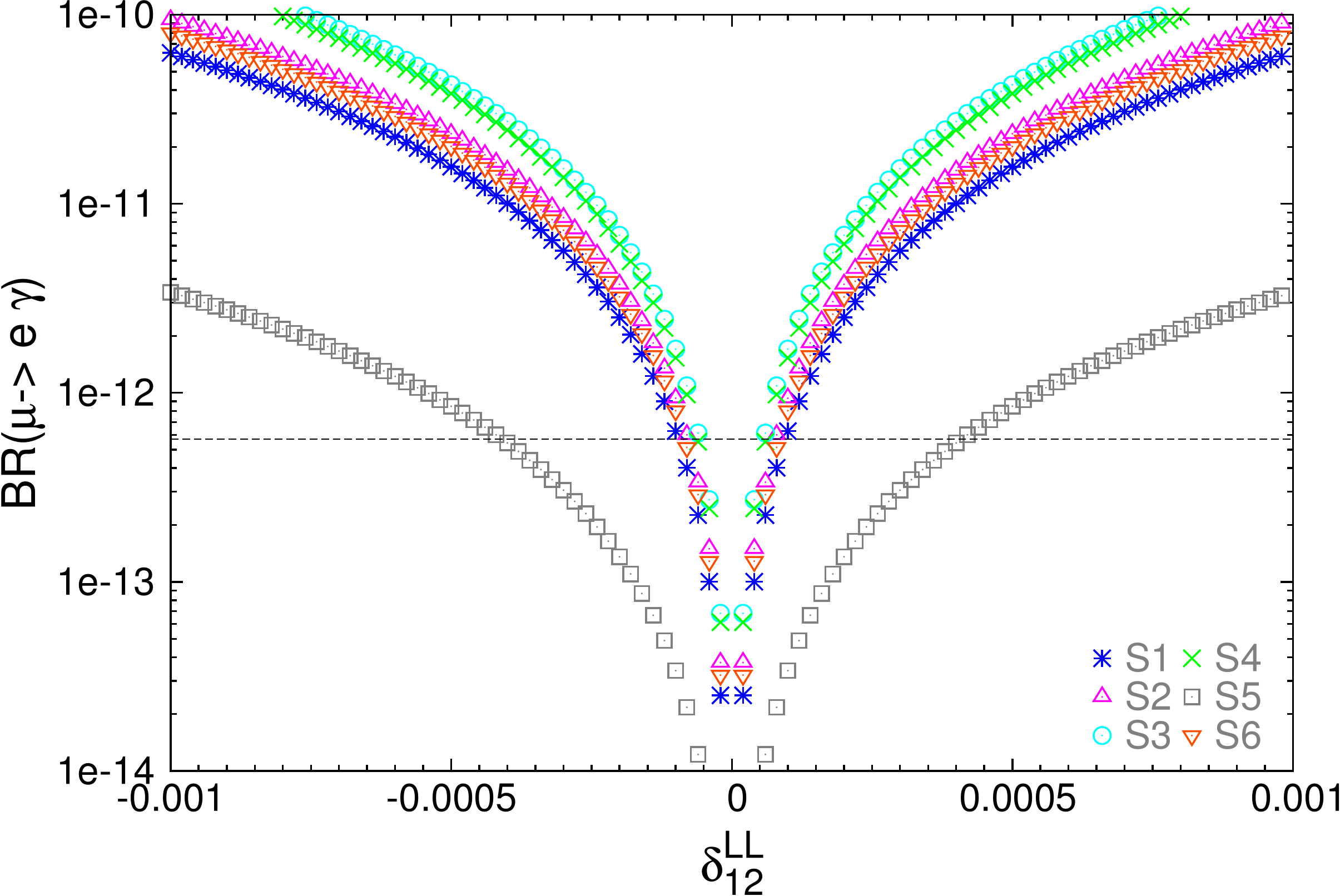   ,scale=0.45,angle=270,clip=}\\
\vspace{1em}
\psfig{file=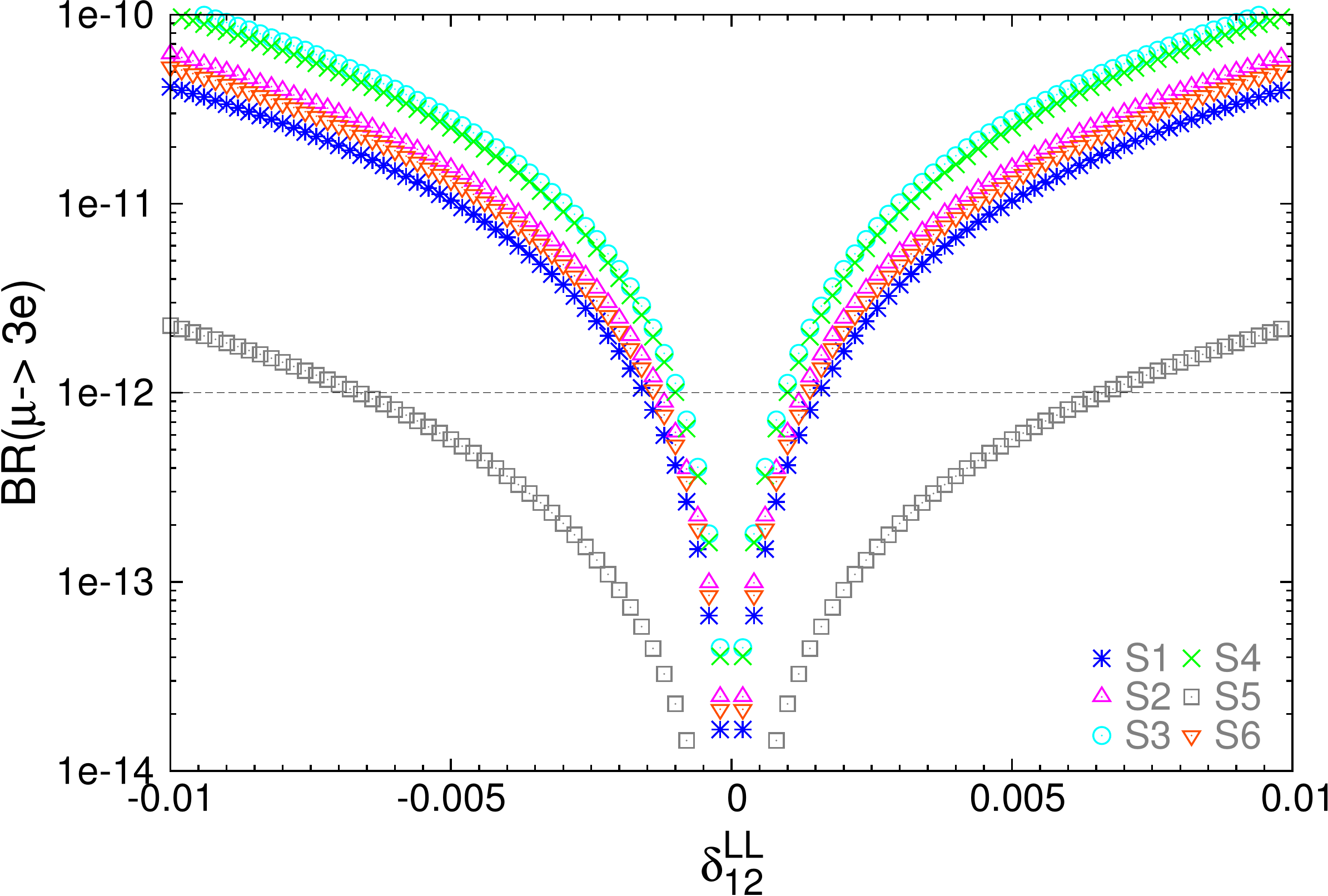     ,scale=0.30,angle=270,clip=}
\hspace{1em}
\psfig{file=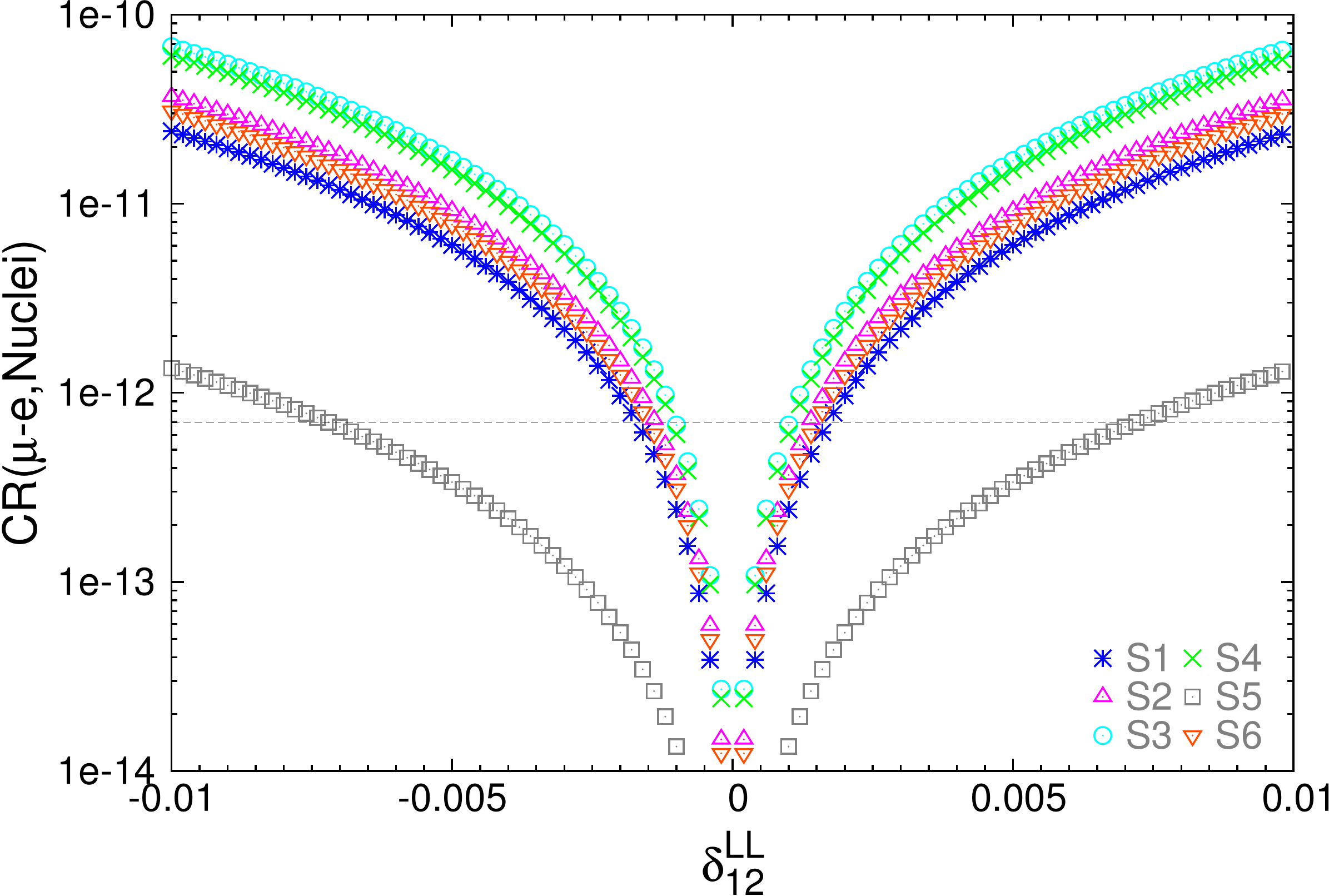    ,scale=0.30,angle=270,clip=}
\end{center}
\caption{LFV rates for $\mu-e$ transitions as a function of slepton
  mixing $\delta_{12}^{LL}$ for scenarios S1 to S6 defined in Section \ref{frameworkc}. The horizontal lines are the corresponding upper bounds collected in Section \ref{sec:expbounds}.}  
\label{mixing12LL}
\end{figure}

\begin{figure}[ht!]
\begin{center}
\psfig{file=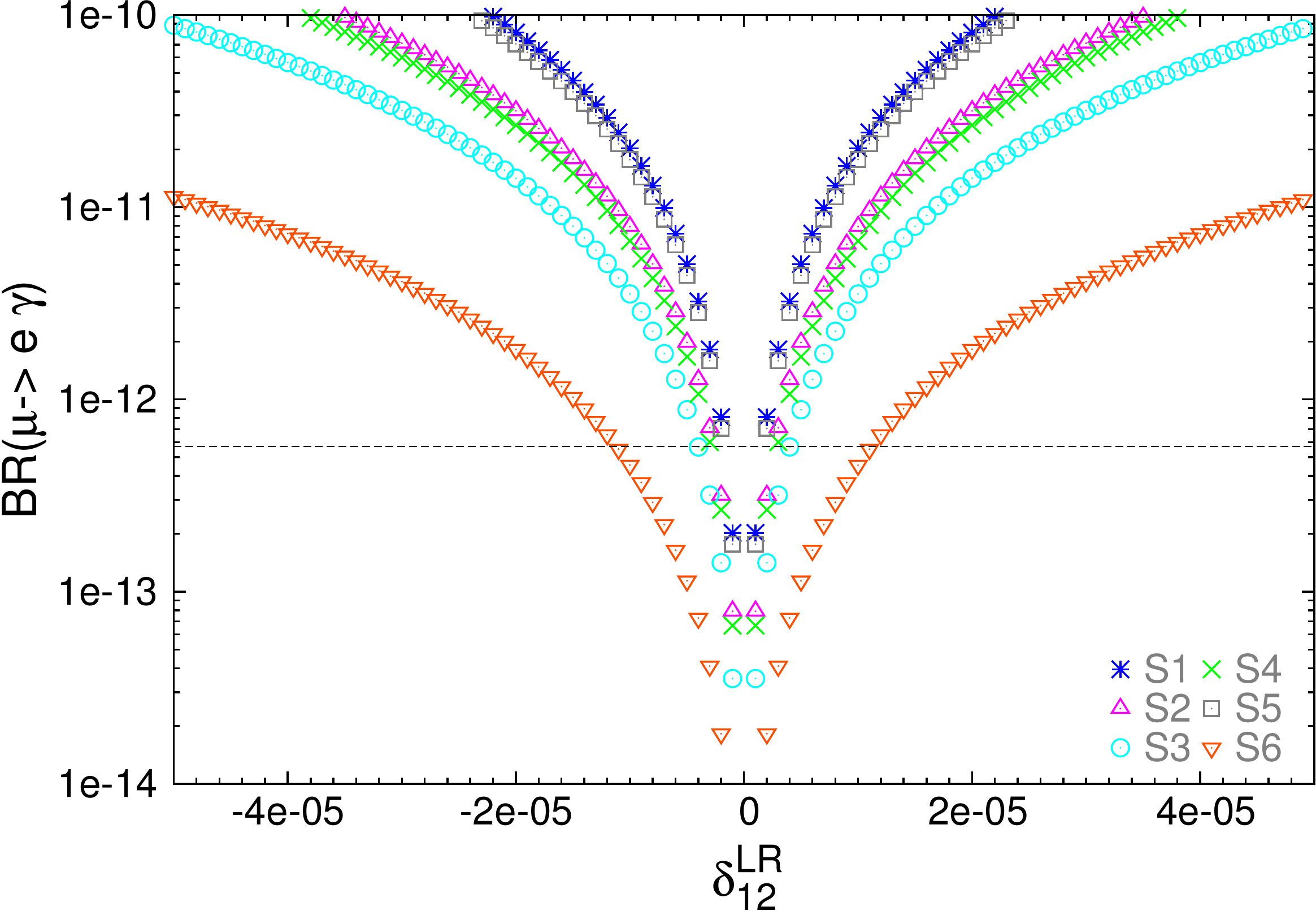   ,scale=0.45,angle=270,clip=}\\
\vspace{1em}
\psfig{file=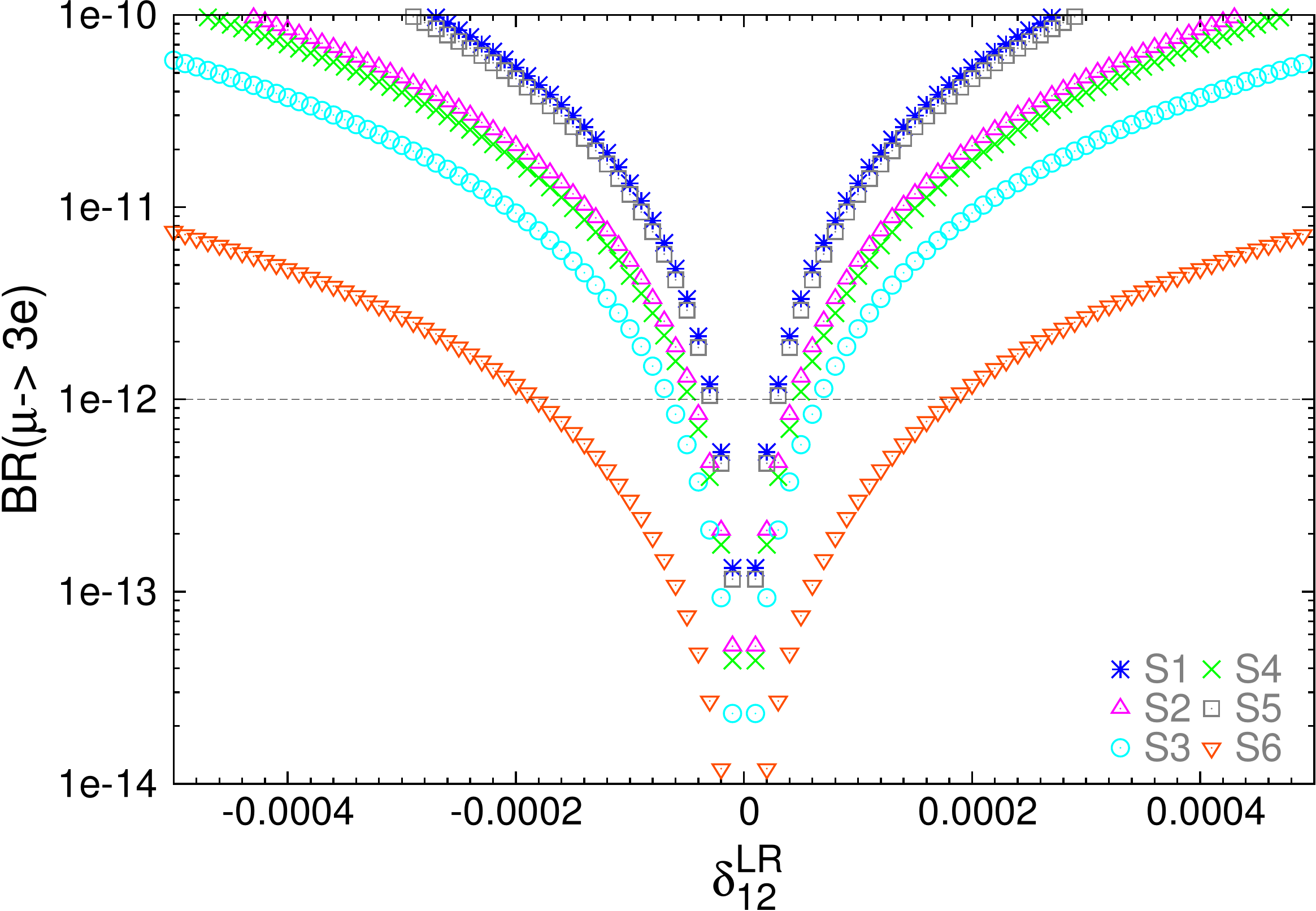     ,scale=0.30,angle=270,clip=}
\hspace{1em}
\psfig{file=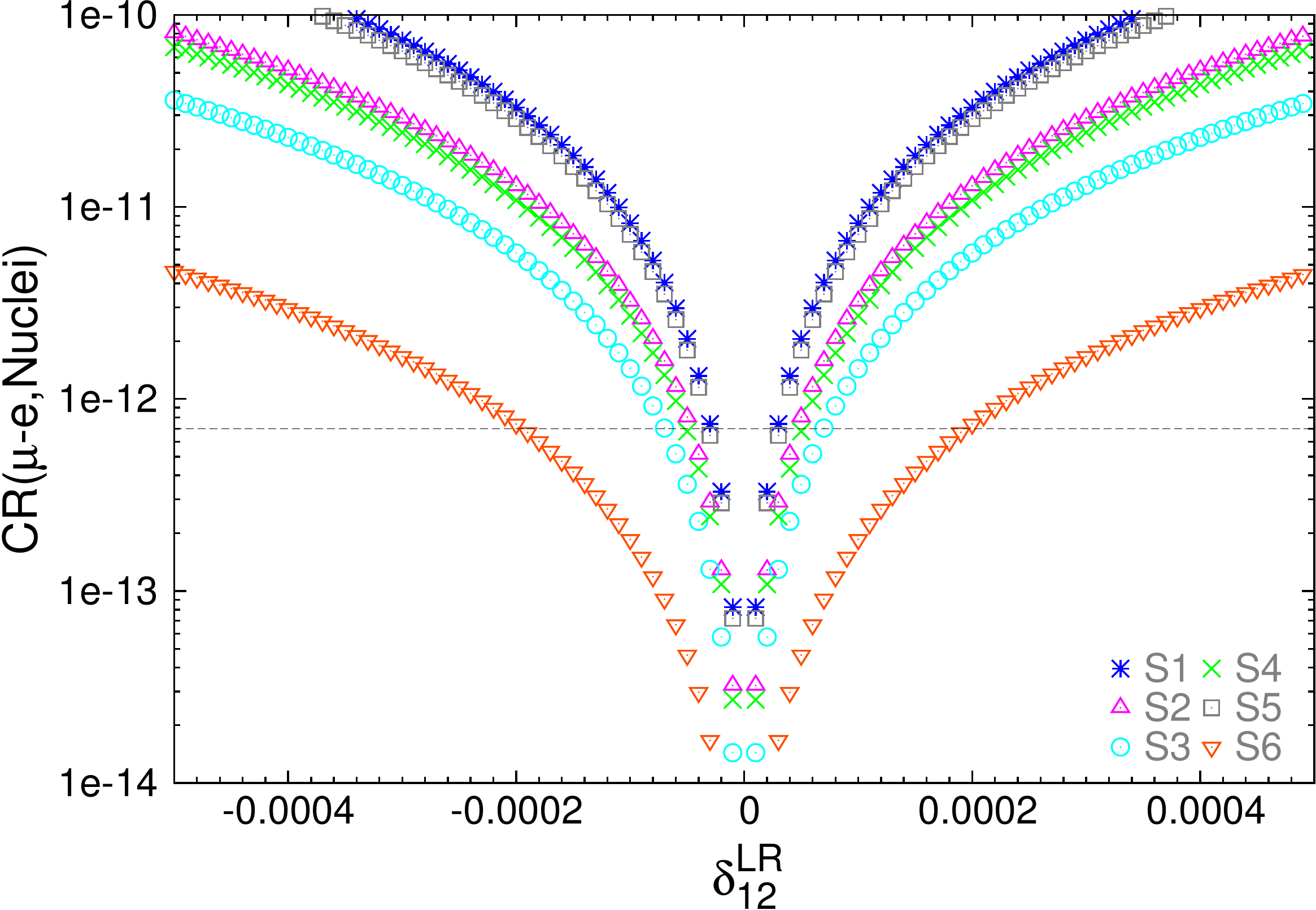    ,scale=0.30,angle=270,clip=}
\end{center}
\caption{LFV rates for $\mu-e$ transitions as a function of slepton
  mixing $\delta_{12}^{LR}$ for scenarios S1 to S6 defined in Section \ref{frameworkc}. The horizontal lines are the corresponding upper bounds collected in Section \ref{sec:expbounds}. The corresponding plots for
  $\delta_{12}^{RL}$, not shown here, are indistinguishable from these.}  
\label{mixing12LR}
\end{figure}

\begin{figure}[ht!]
\begin{center}
\psfig{file=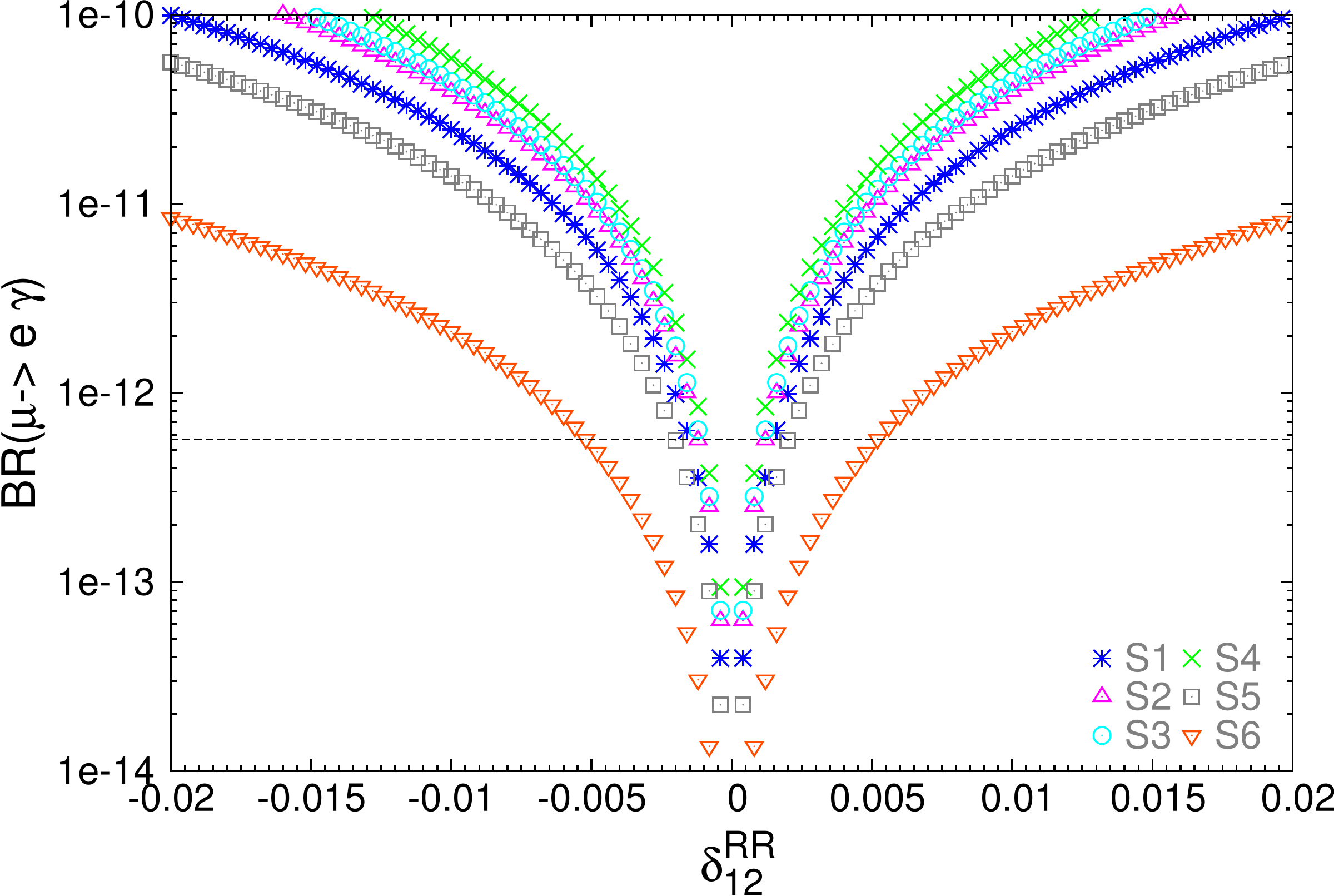   ,scale=0.45,angle=270,clip=}\\
\vspace{1em}
\psfig{file=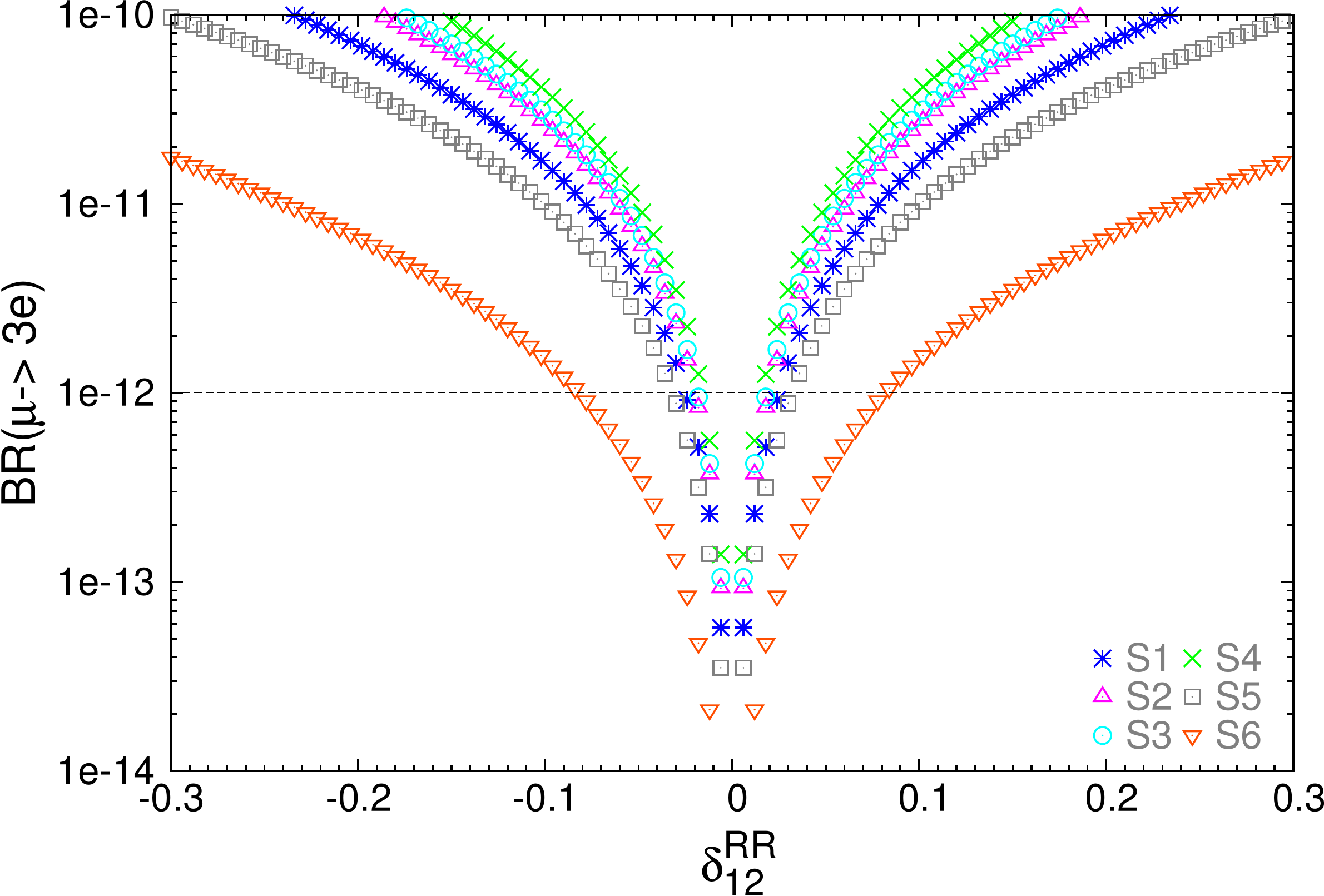     ,scale=0.30,angle=270,clip=}
\hspace{1em}
\psfig{file=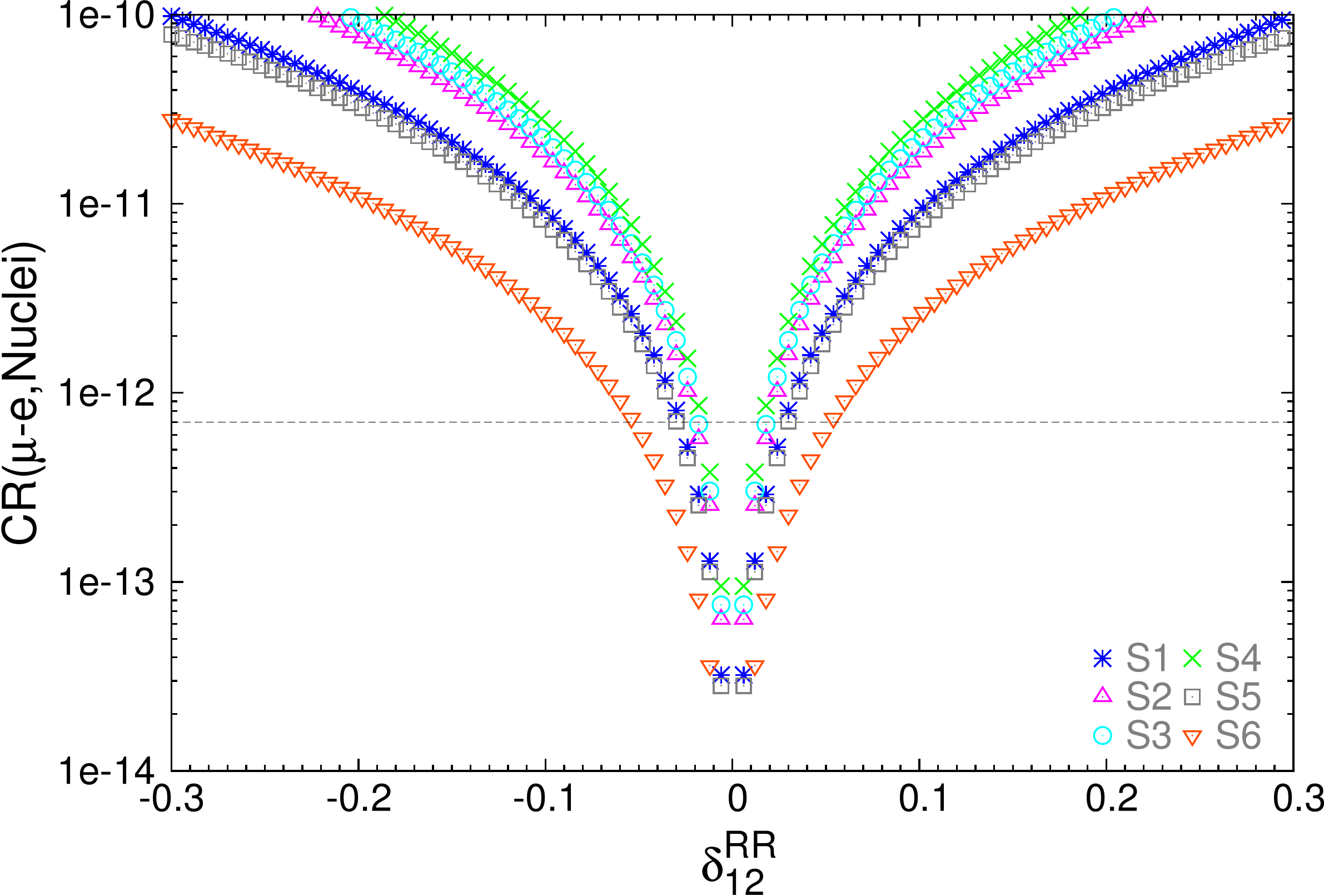    ,scale=0.30,angle=270,clip=}
\end{center}
\caption{LFV rates for $\mu-e$ transitions as a function of slepton
  mixing $\delta_{12}^{RR}$ for scenarios S1 to S6 defined in Section \ref{frameworkc}. The horizontal lines are the corresponding upper bounds collected in Section \ref{sec:expbounds}.}  
\label{mixing12RR}
\end{figure}

\begin{figure}[ht!]
\begin{center}
\psfig{file=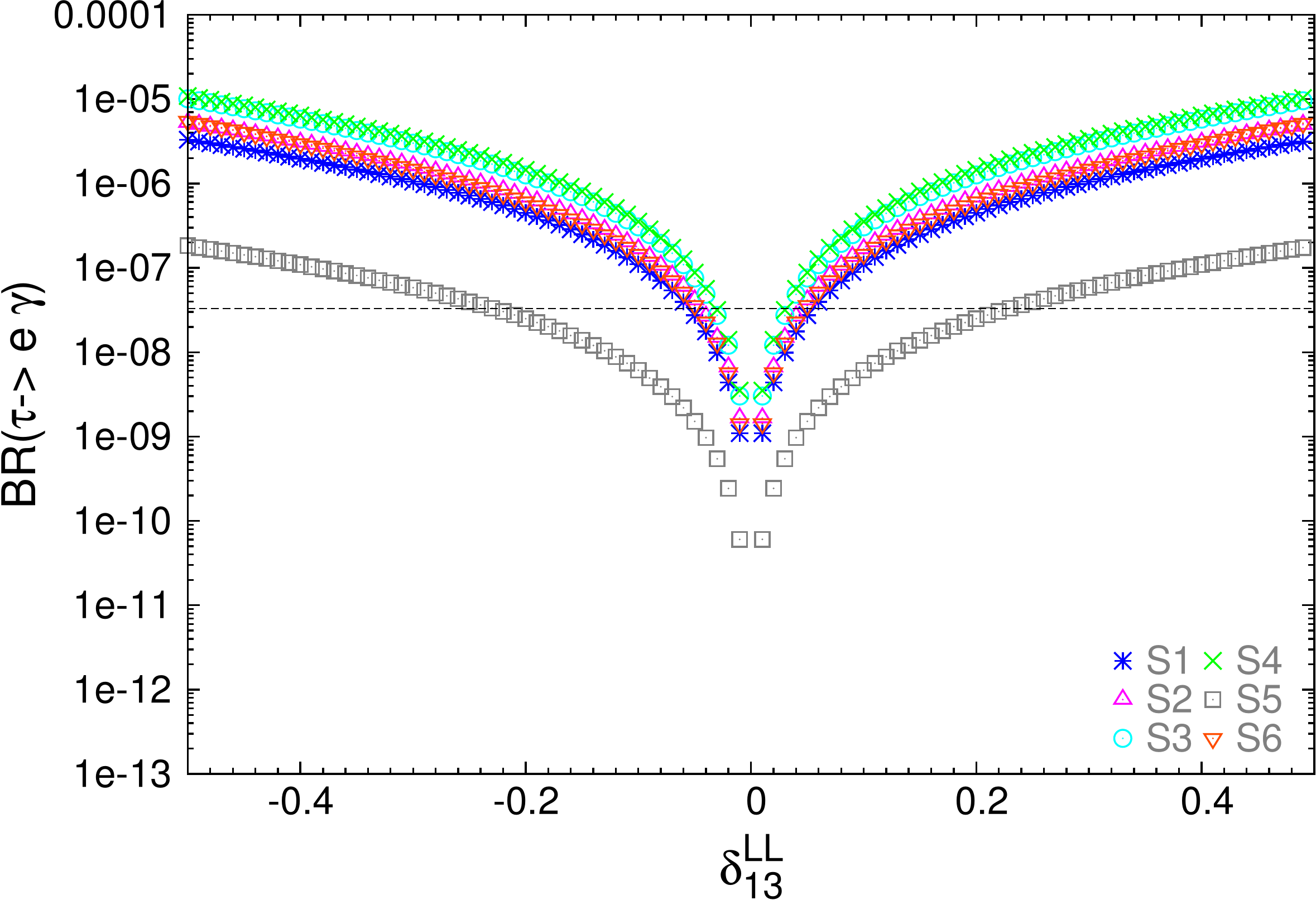  ,scale=0.45,angle=270,clip=}\\
\vspace{1em}
\psfig{file=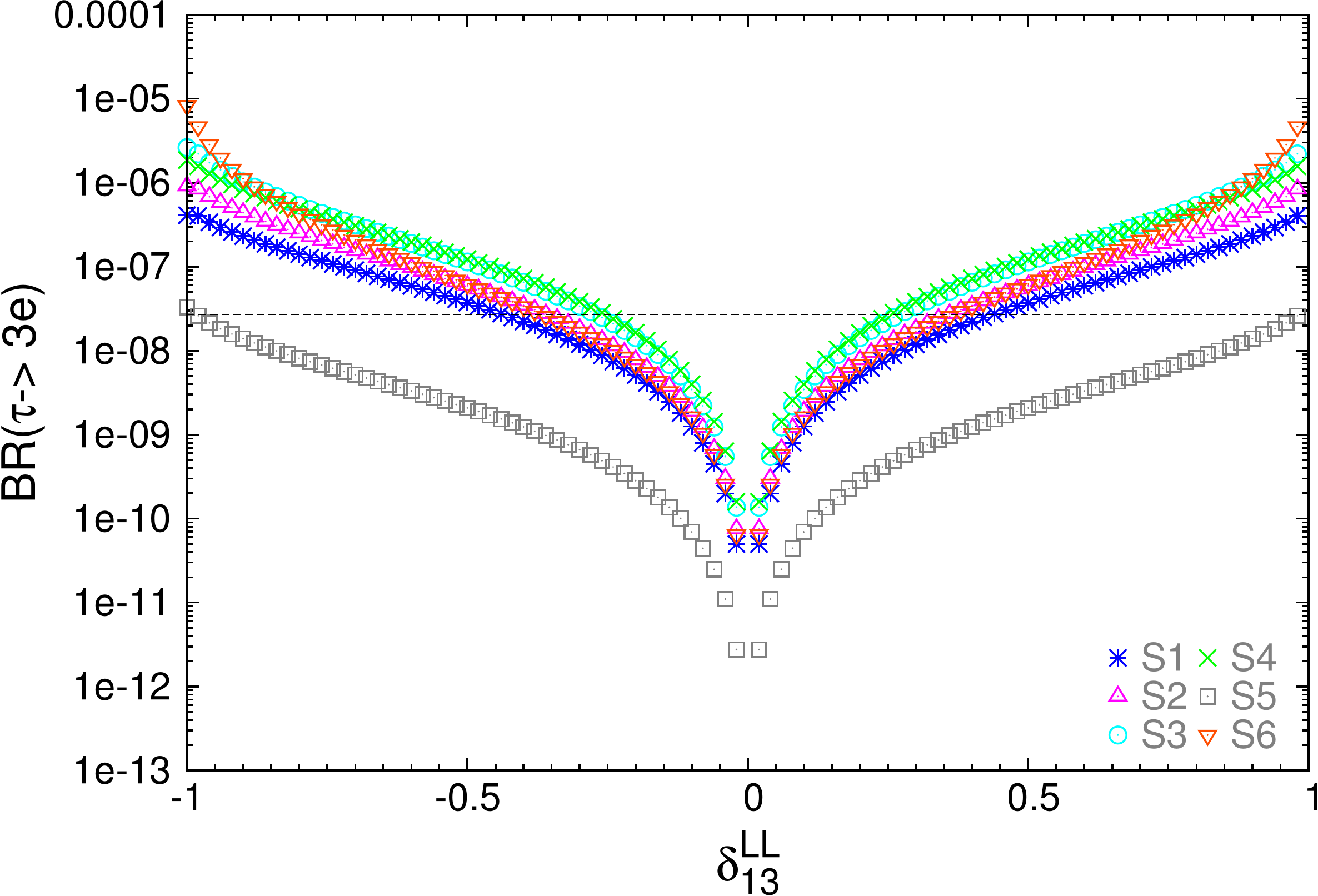    ,scale=0.30,angle=270,clip=}
\hspace{1em}
\psfig{file=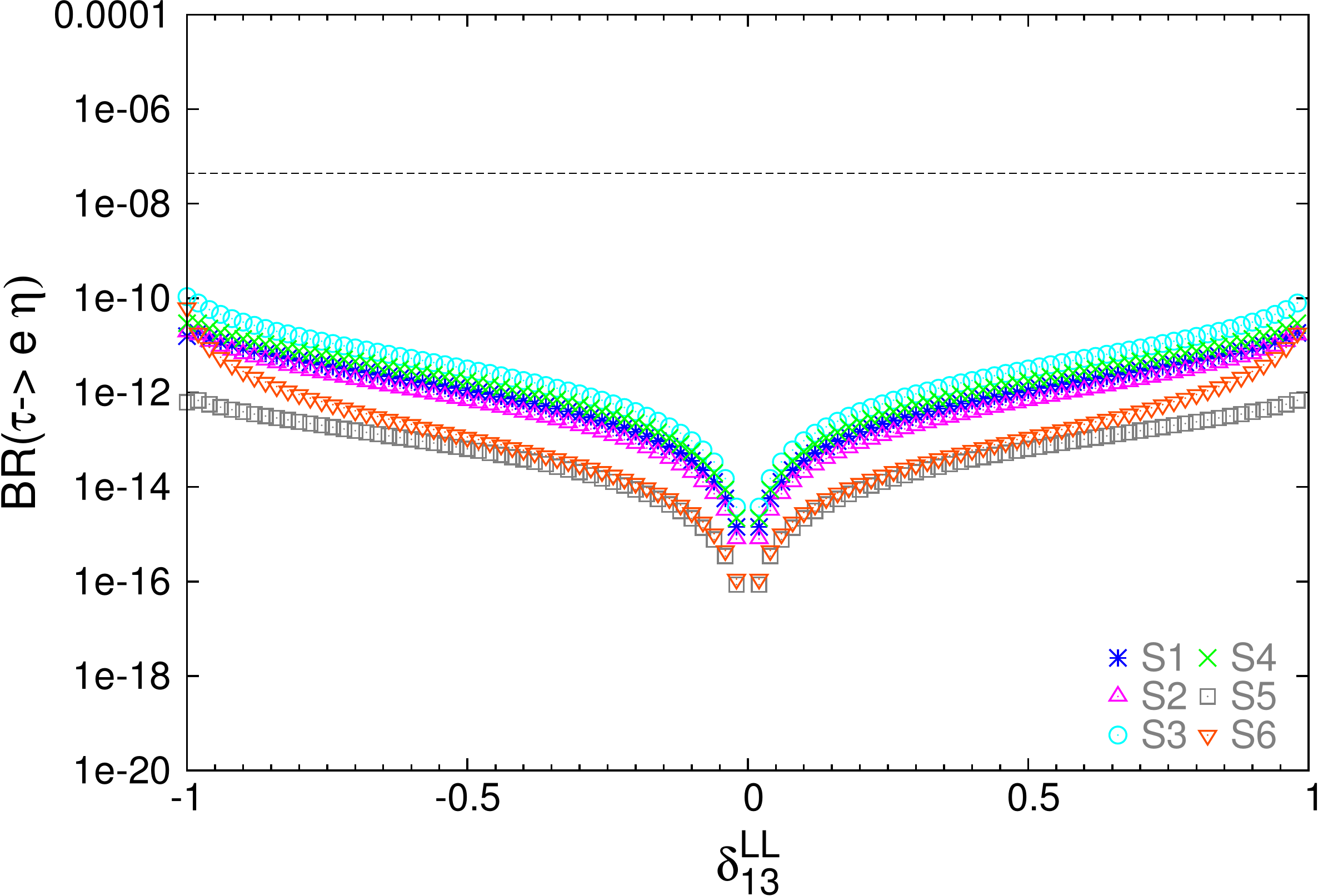   ,scale=0.30,angle=270,clip=}
\end{center}
\caption{LFV rates for $\tau-e$ transitions as a function of slepton 
mixing $\delta_{13}^{LL}$ for scenarios S1 to S6 defined in Section \ref{frameworkc}. The horizontal lines are the corresponding upper bounds collected in Section \ref{sec:expbounds}.} 
\label{mixing13LL}
\end{figure}

\begin{figure}[ht!]
\begin{center}
\psfig{file=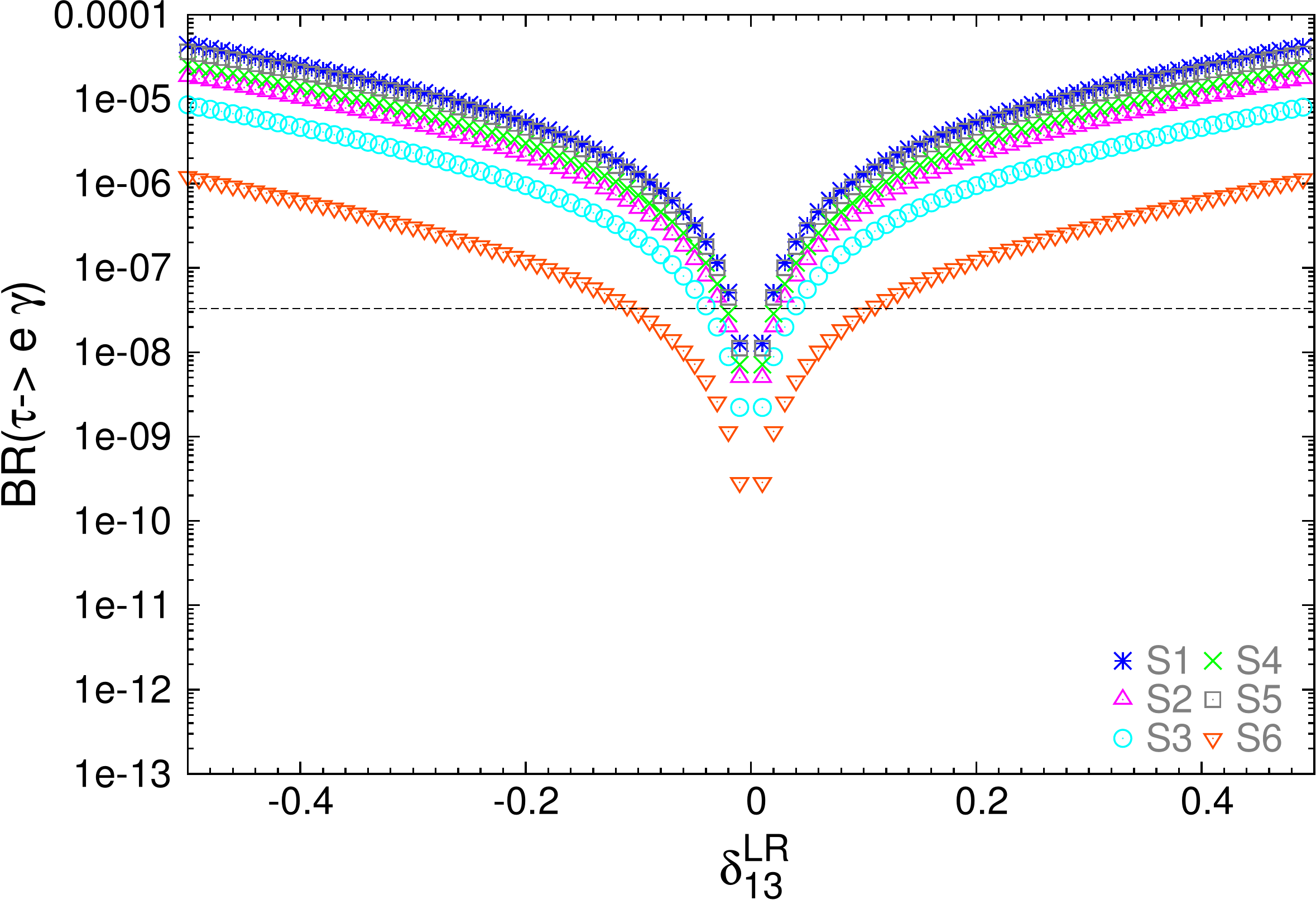  ,scale=0.45,angle=270,clip=}\\
\vspace{1em}
\psfig{file=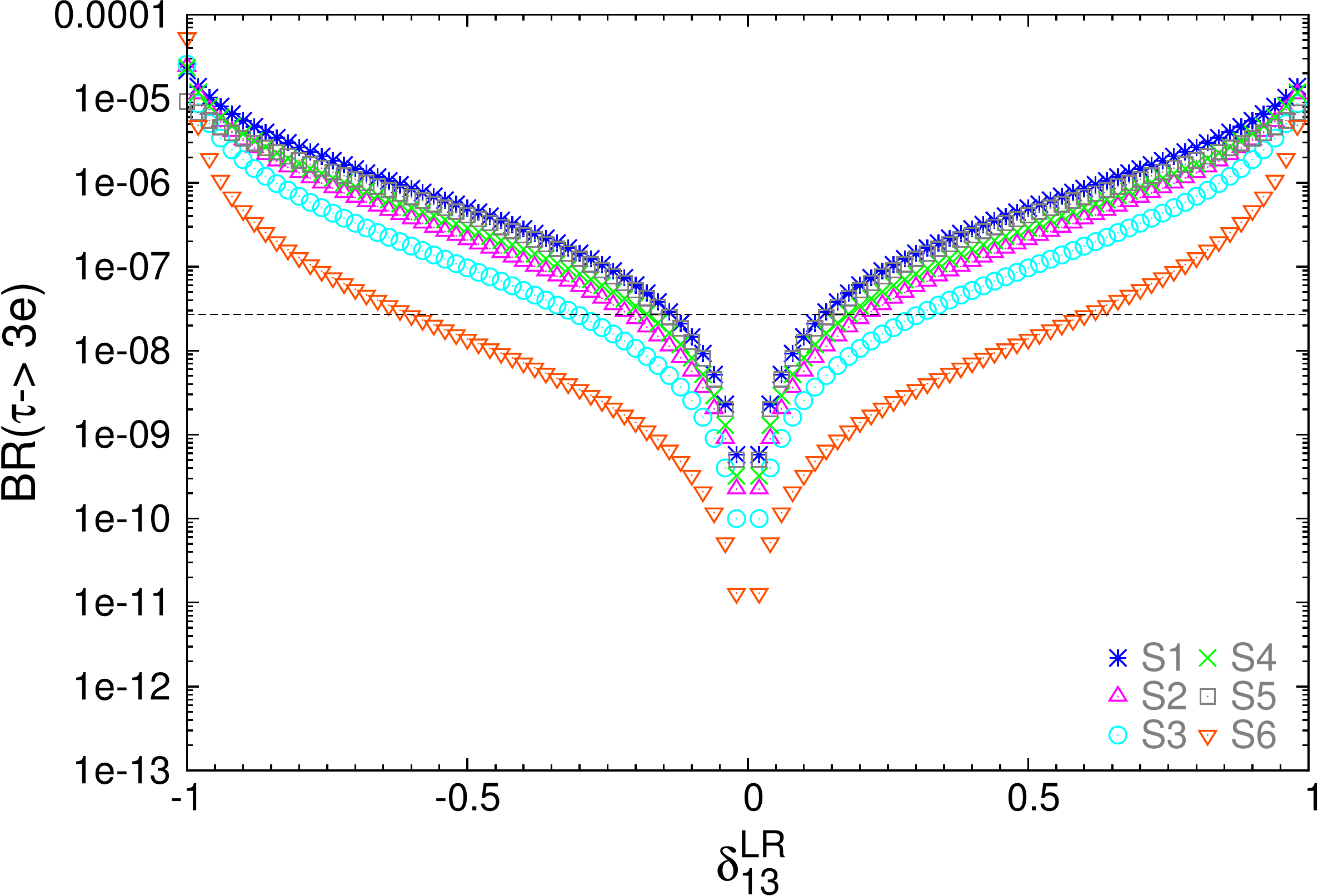    ,scale=0.30,angle=270,clip=}
\hspace{1em}
\psfig{file=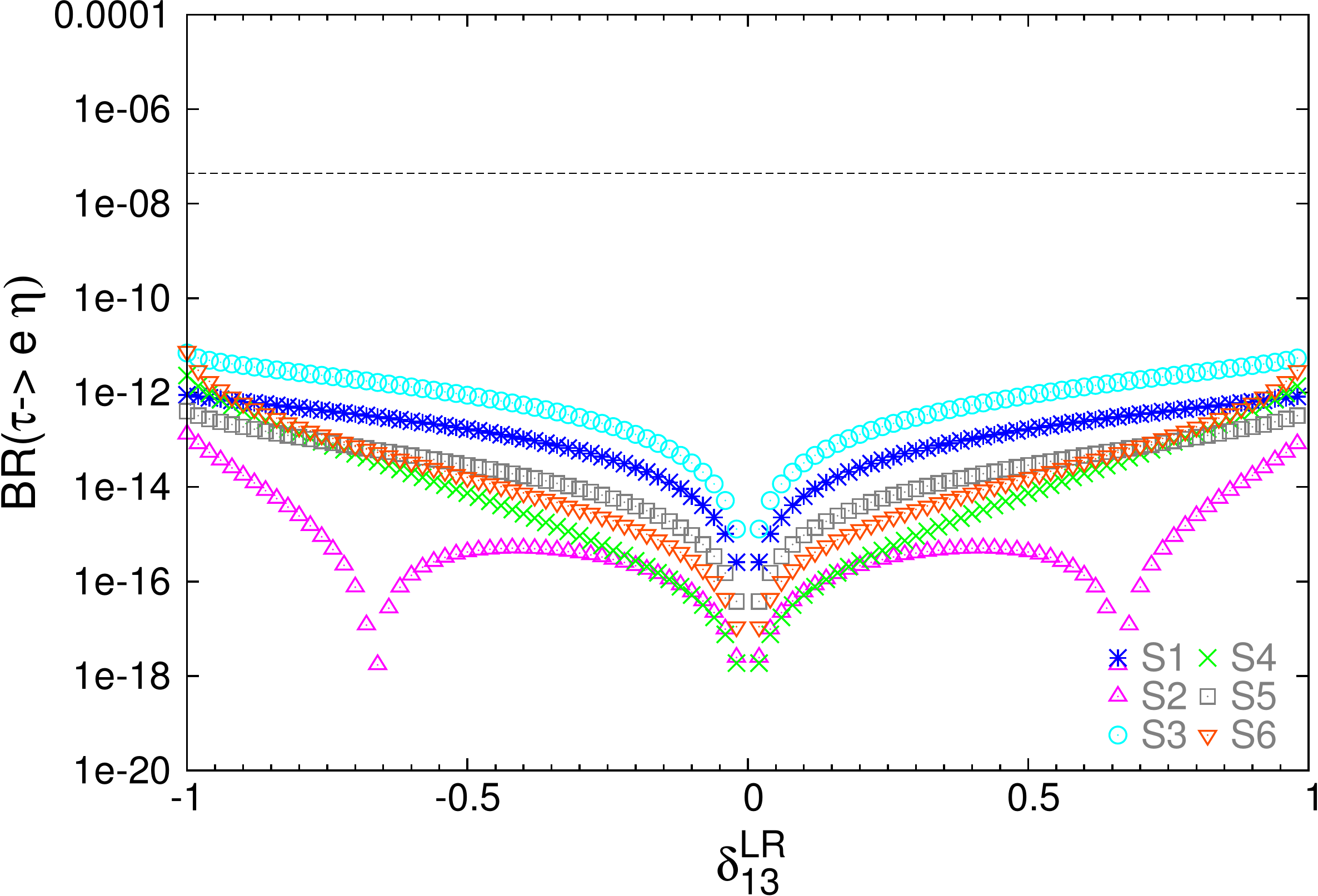   ,scale=0.30,angle=270,clip=}
\end{center}
\caption{LFV rates for $\tau-e$ transitions as a function of slepton
  mixing $\delta_{13}^{LR}$ for scenarios S1 to S6 defined in Section \ref{frameworkc}. The horizontal lines are the corresponding upper bounds collected in Section \ref{sec:expbounds}. The corresponding plots for
  $\delta_{13}^{RL}$, not shown here, are indistinguishable from these.}  
\label{mixing13LR}
\end{figure}

\begin{figure}[ht!]
\begin{center}
\psfig{file=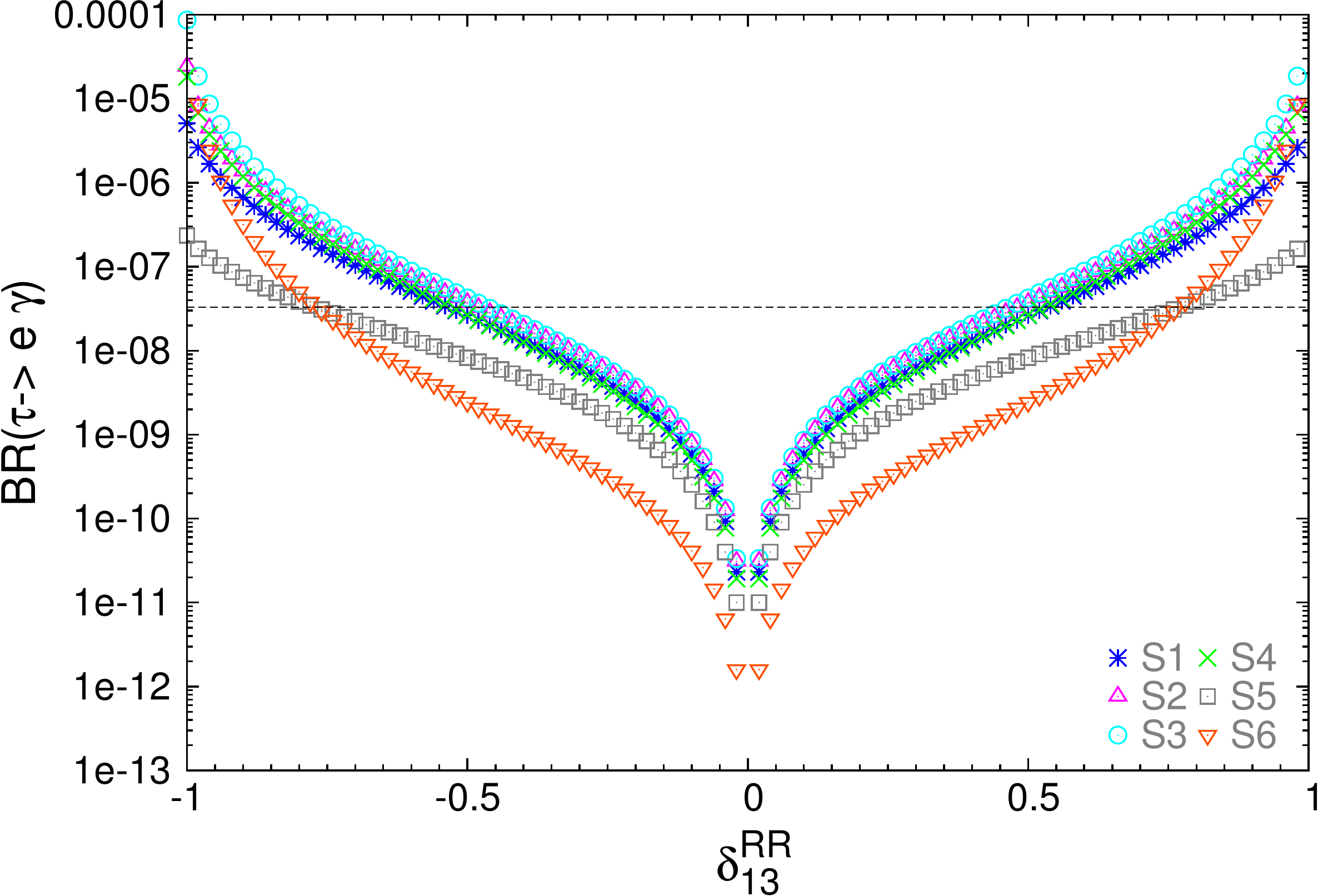  ,scale=0.45,angle=270,clip=}\\
\vspace{1em}
\psfig{file=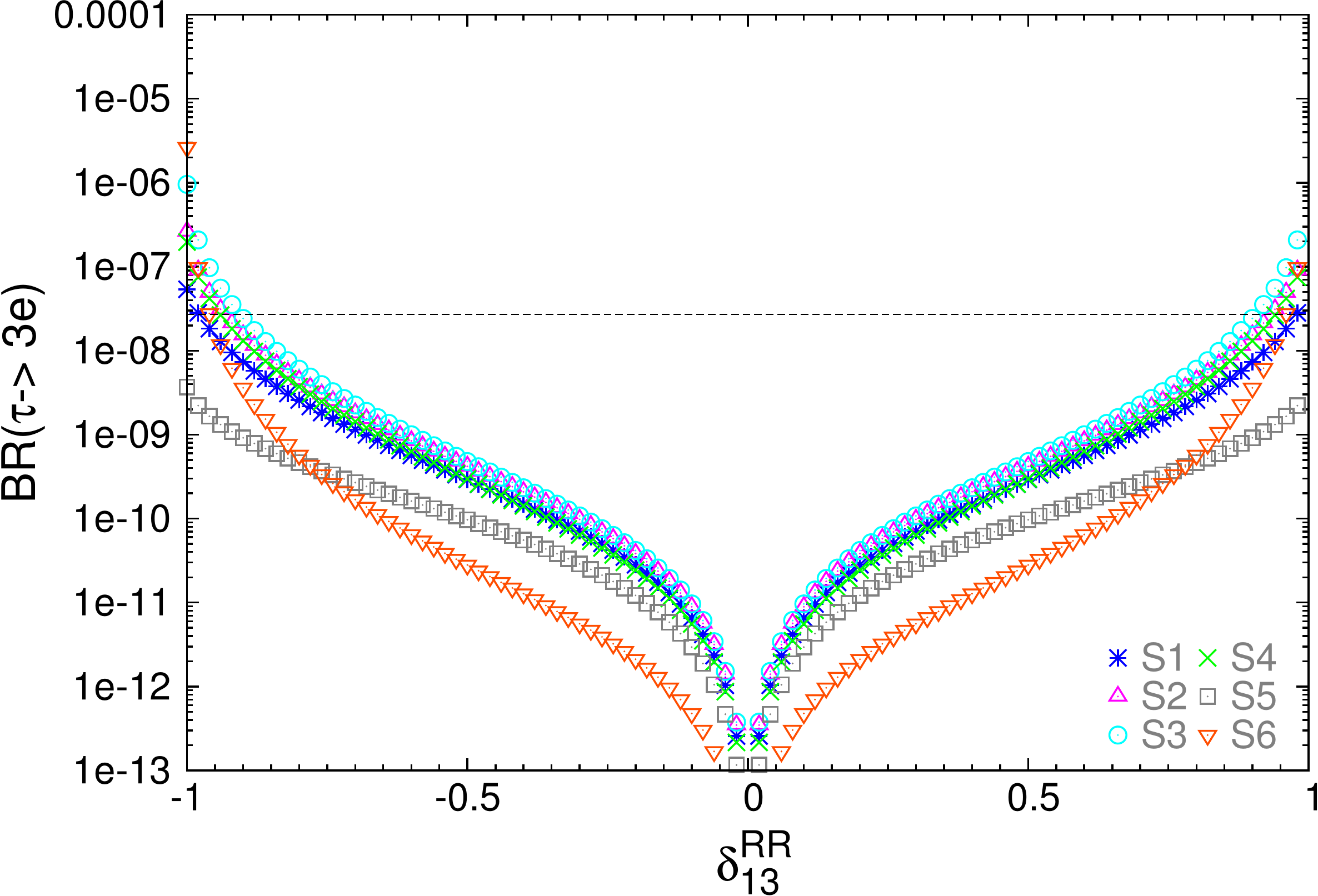    ,scale=0.30,angle=270,clip=}
\hspace{1em}
\psfig{file=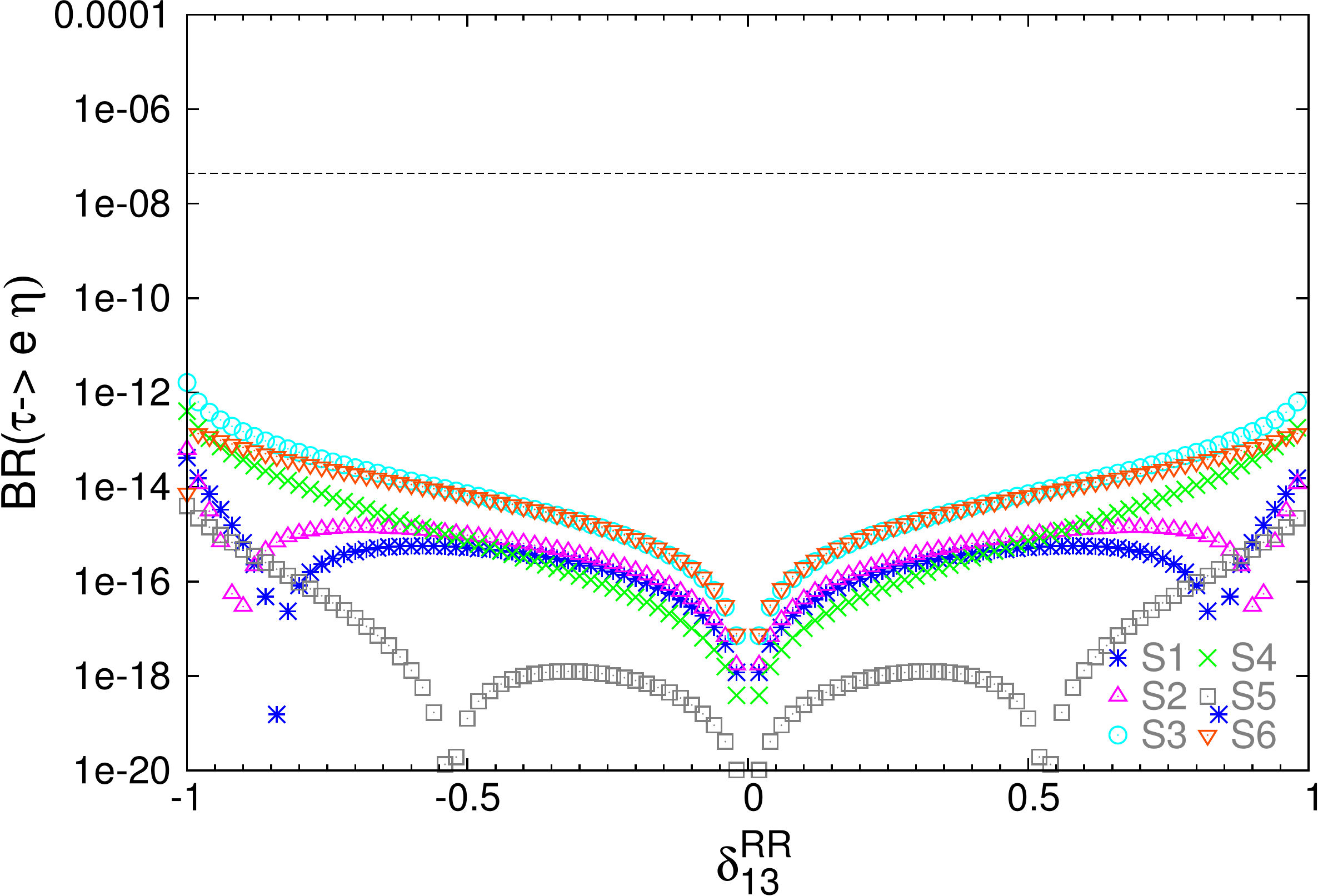   ,scale=0.30,angle=270,clip=}
\end{center}
\caption{LFV rates for $\tau-e$ transitions as a function of slepton
  mixing $\delta_{13}^{RR}$ for scenarios S1 to S6 defined in Section \ref{frameworkc}. The horizontal lines are the corresponding upper bounds collected in Section \ref{sec:expbounds}.}
\label{mixing13RR}
\end{figure}

\begin{figure}[ht!]
\begin{center}
\psfig{file=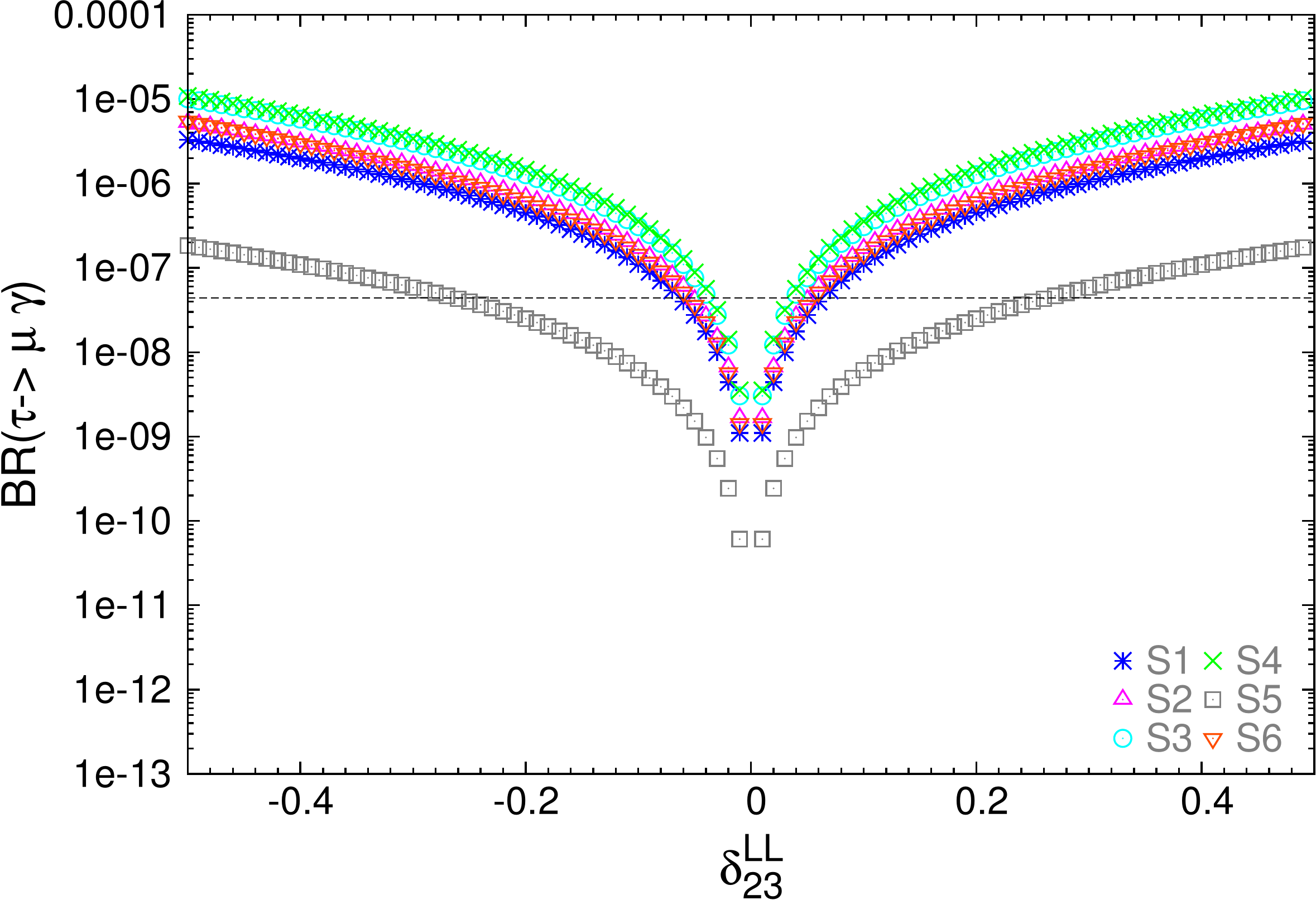  ,scale=0.45,angle=270,clip=}\\
\vspace{1em}
\psfig{file=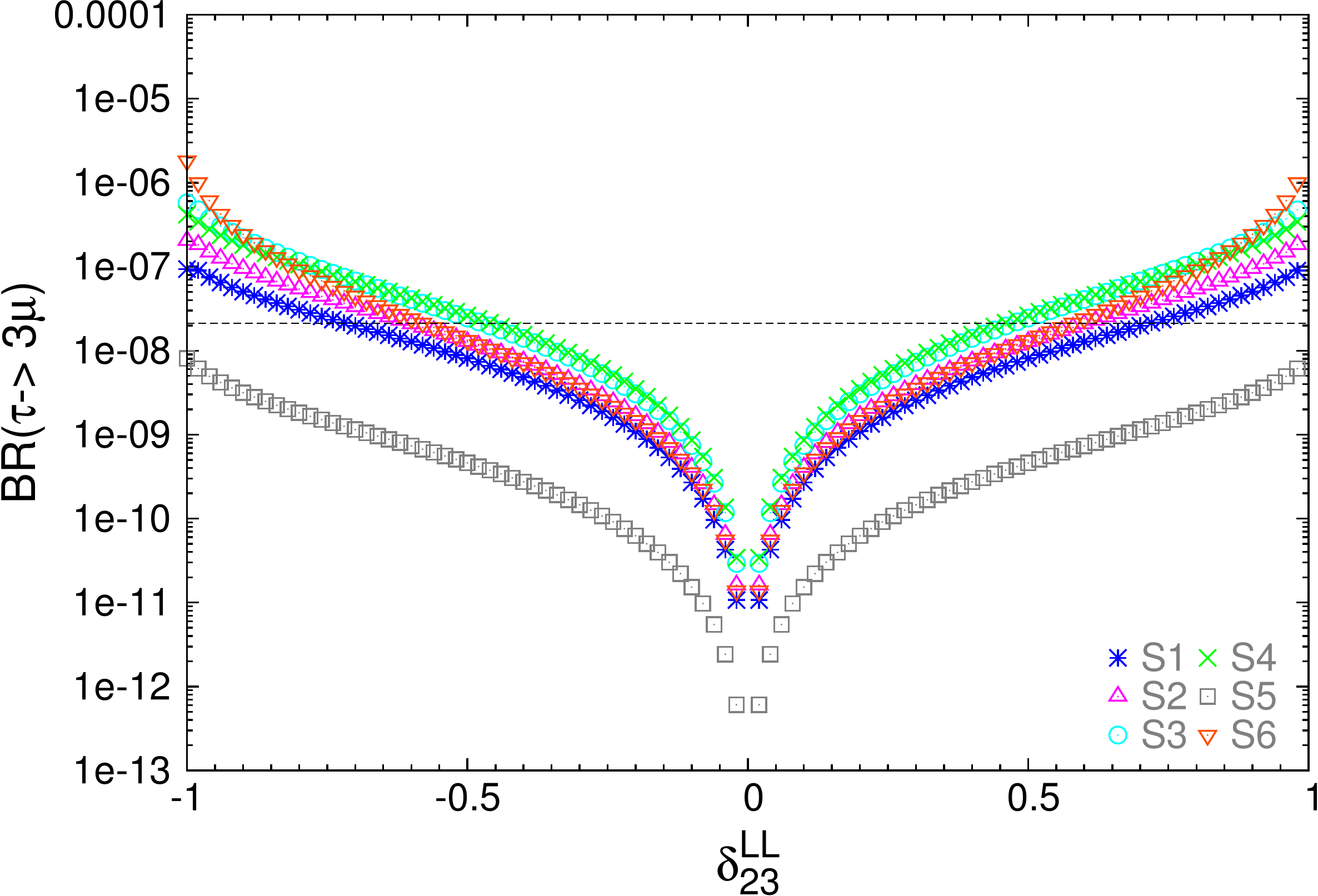    ,scale=0.30,angle=270,clip=}
\hspace{1em}
\psfig{file=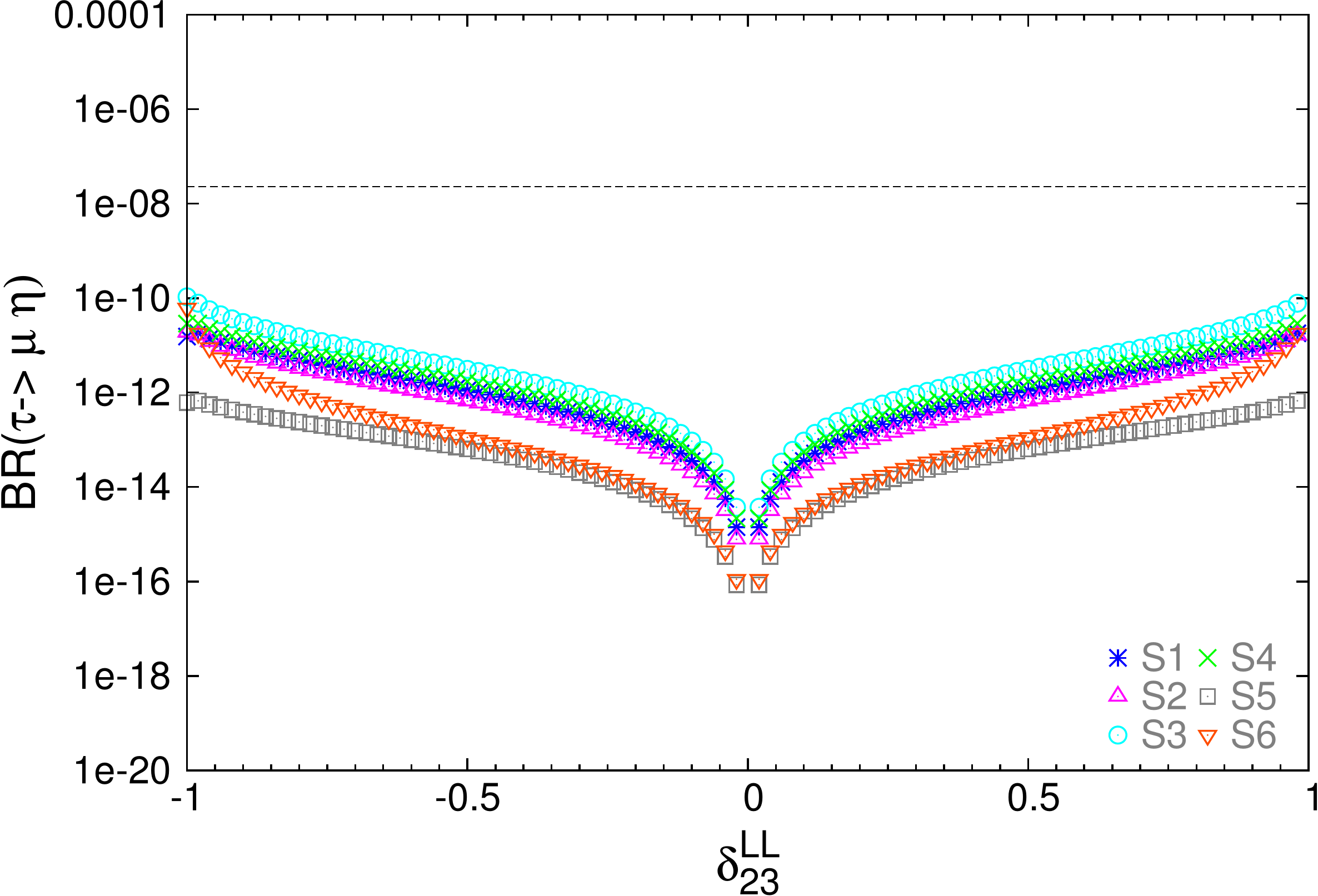   ,scale=0.30,angle=270,clip=}
\end{center}
\caption{LFV rates for $\tau-\mu$ transitions as a function of slepton mixing $\delta_{23}^{LL}$ for scenarios S1 to S6 defined in Section \ref{frameworkc}. The horizontal lines are the corresponding upper bounds collected in Section \ref{sec:expbounds}.} 
\label{mixing23LL}
\end{figure}

\begin{figure}[ht!]
\begin{center}
\psfig{file=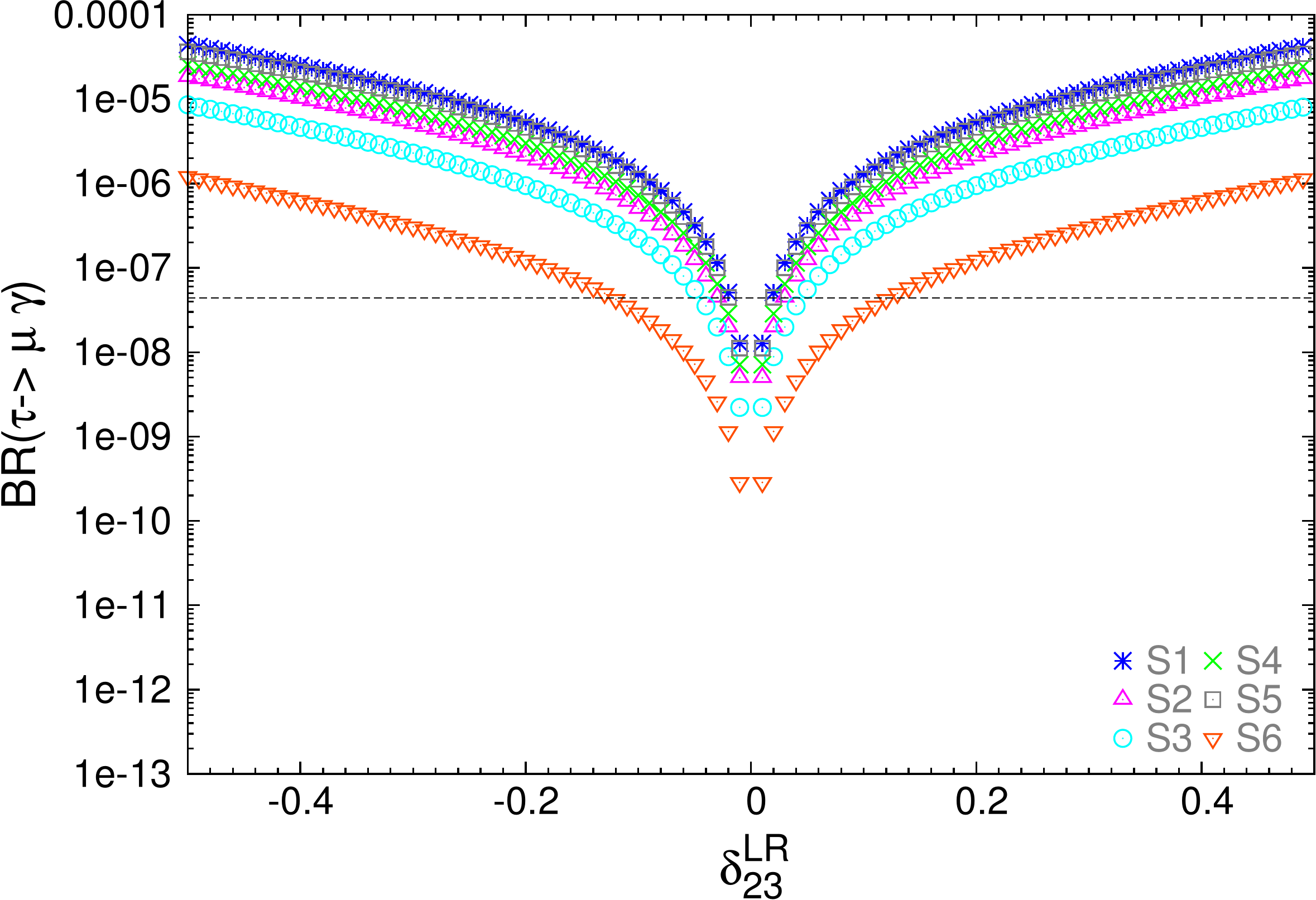  ,scale=0.45,angle=270,clip=}\\
\vspace{1em}
\psfig{file=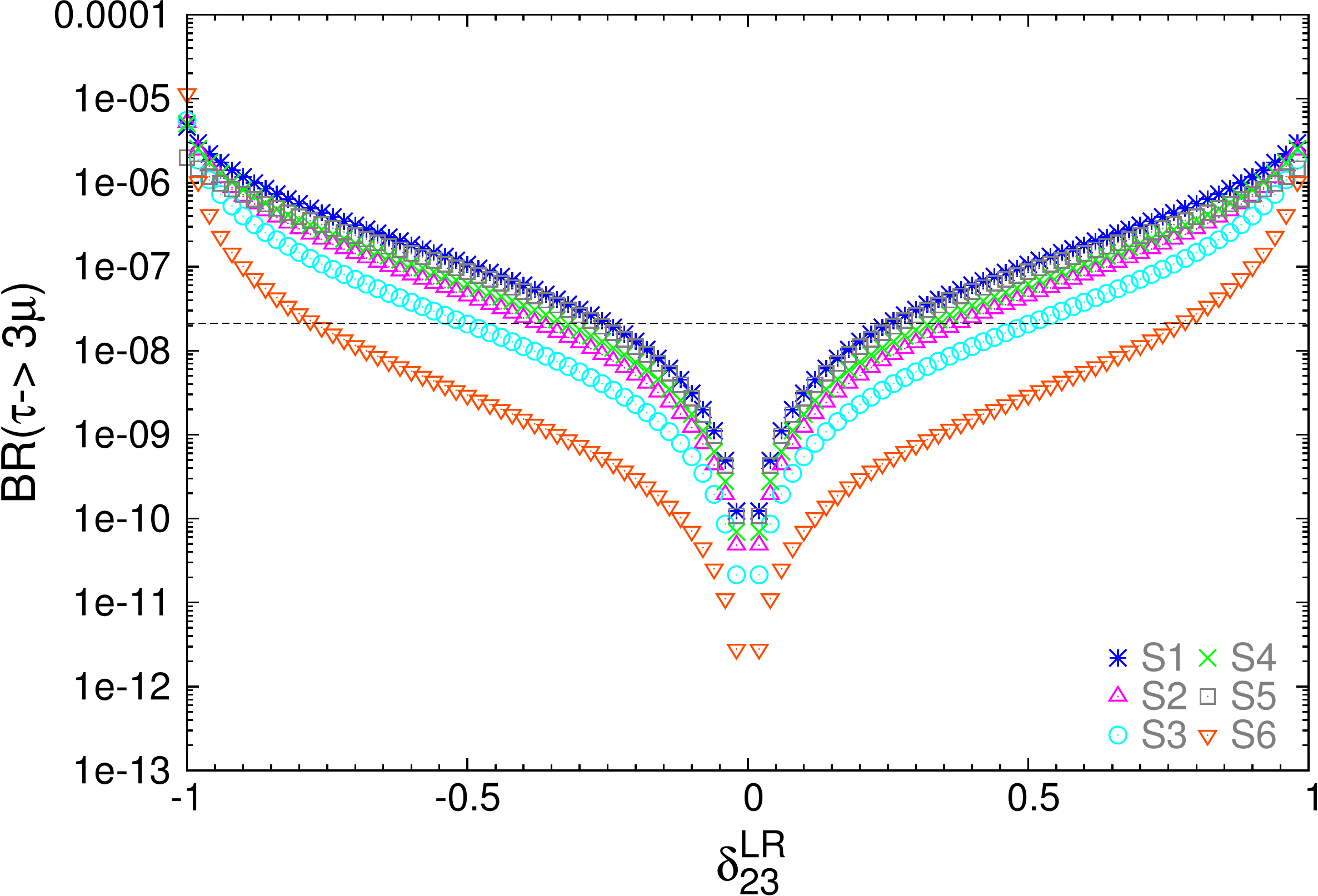    ,scale=0.30,angle=270,clip=}
\hspace{1em}
\psfig{file=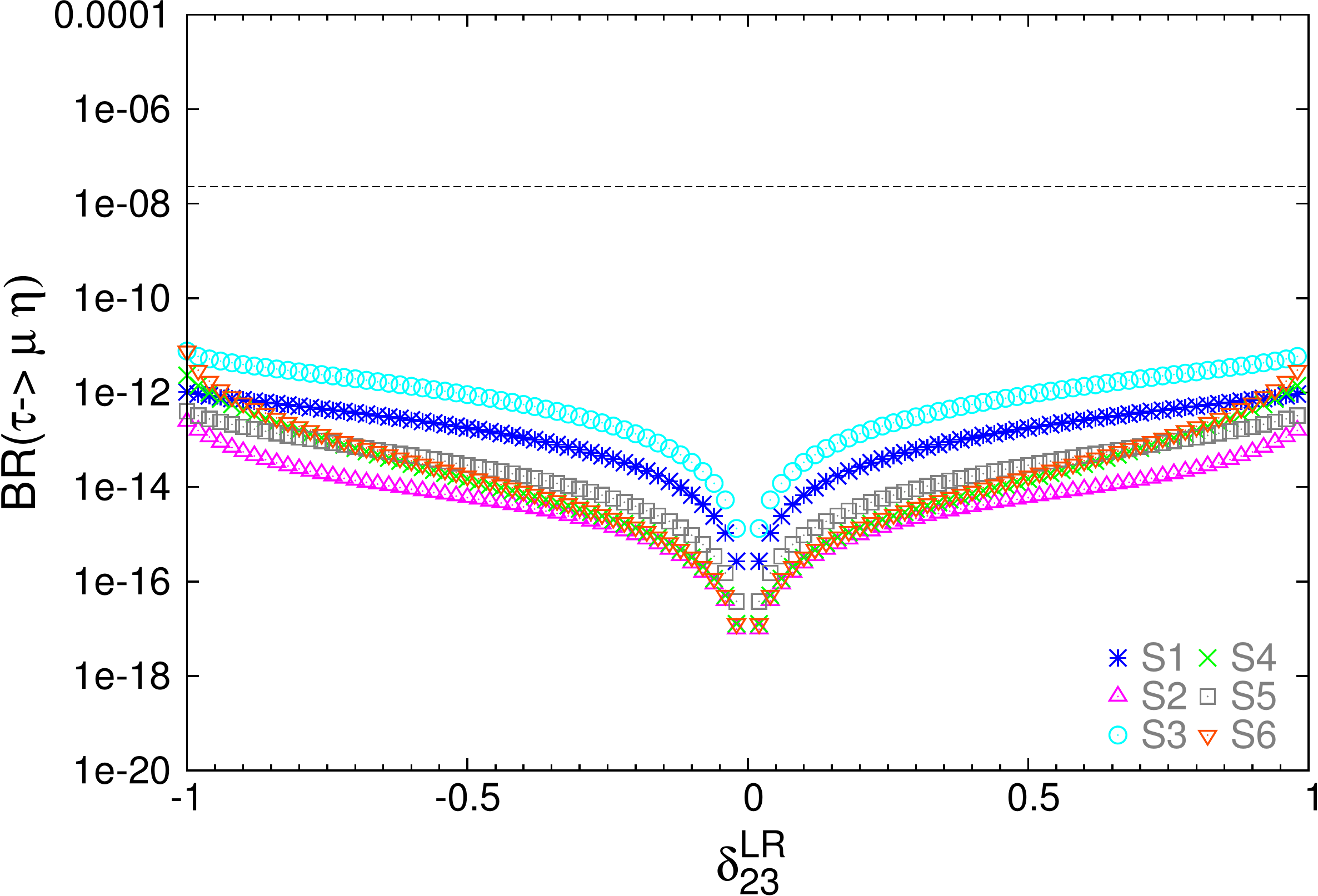   ,scale=0.30,angle=270,clip=}
\end{center}
\caption{LFV rates for $\tau-\mu$ transitions as a function of slepton
  mixing $\delta_{23}^{LR}$ for scenarios S1 to S6 defined in Section \ref{frameworkc}. The horizontal lines are the corresponding upper bounds collected in Section \ref{sec:expbounds}. The corresponding plots for
  $\delta_{23}^{RL}$, not shown here, are indistinguishable from these.}  
\label{mixing23LR}
\end{figure}

\begin{figure}[ht!]
\begin{center}
\psfig{file=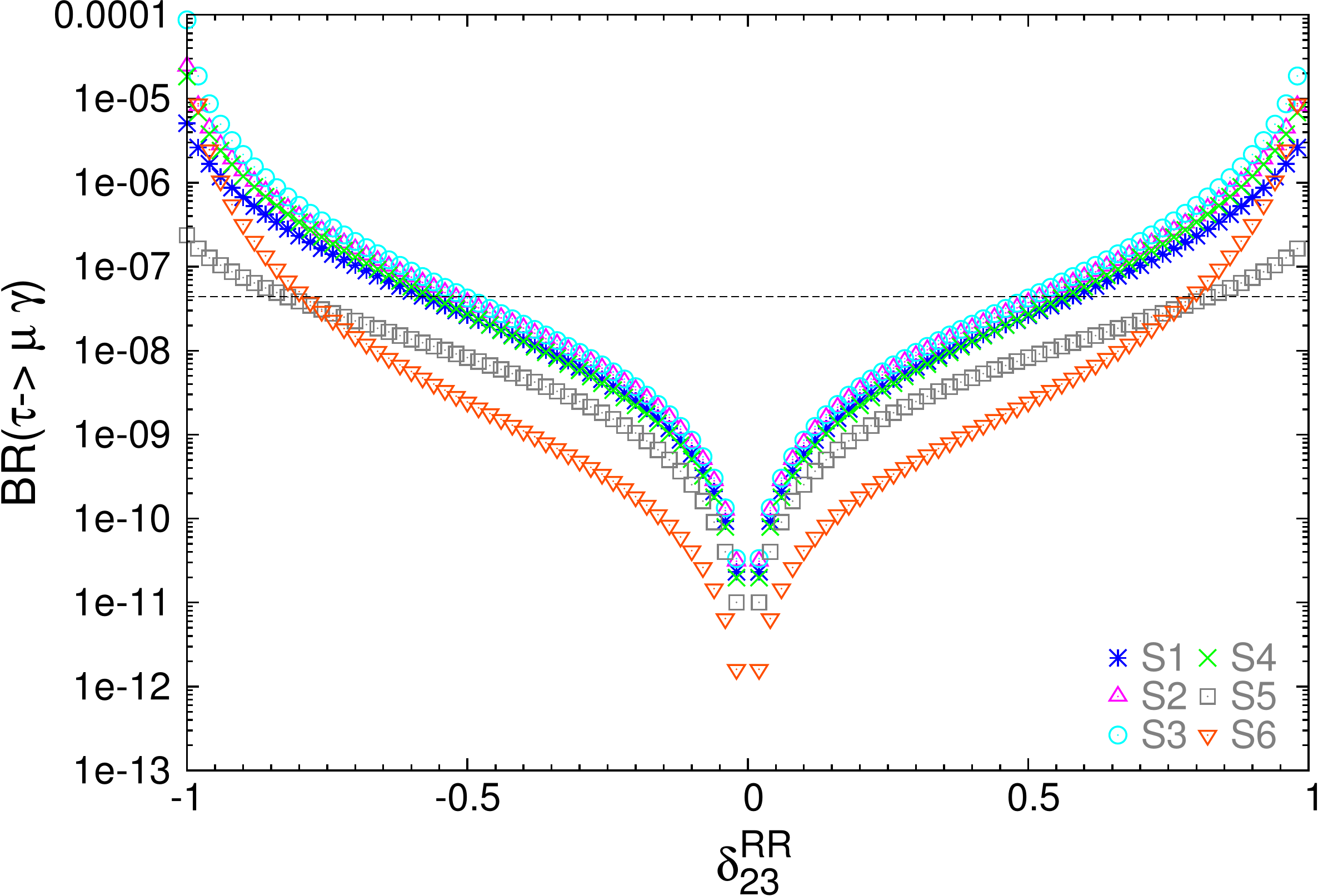  ,scale=0.45,angle=270,clip=}\\
\vspace{1em}
\psfig{file=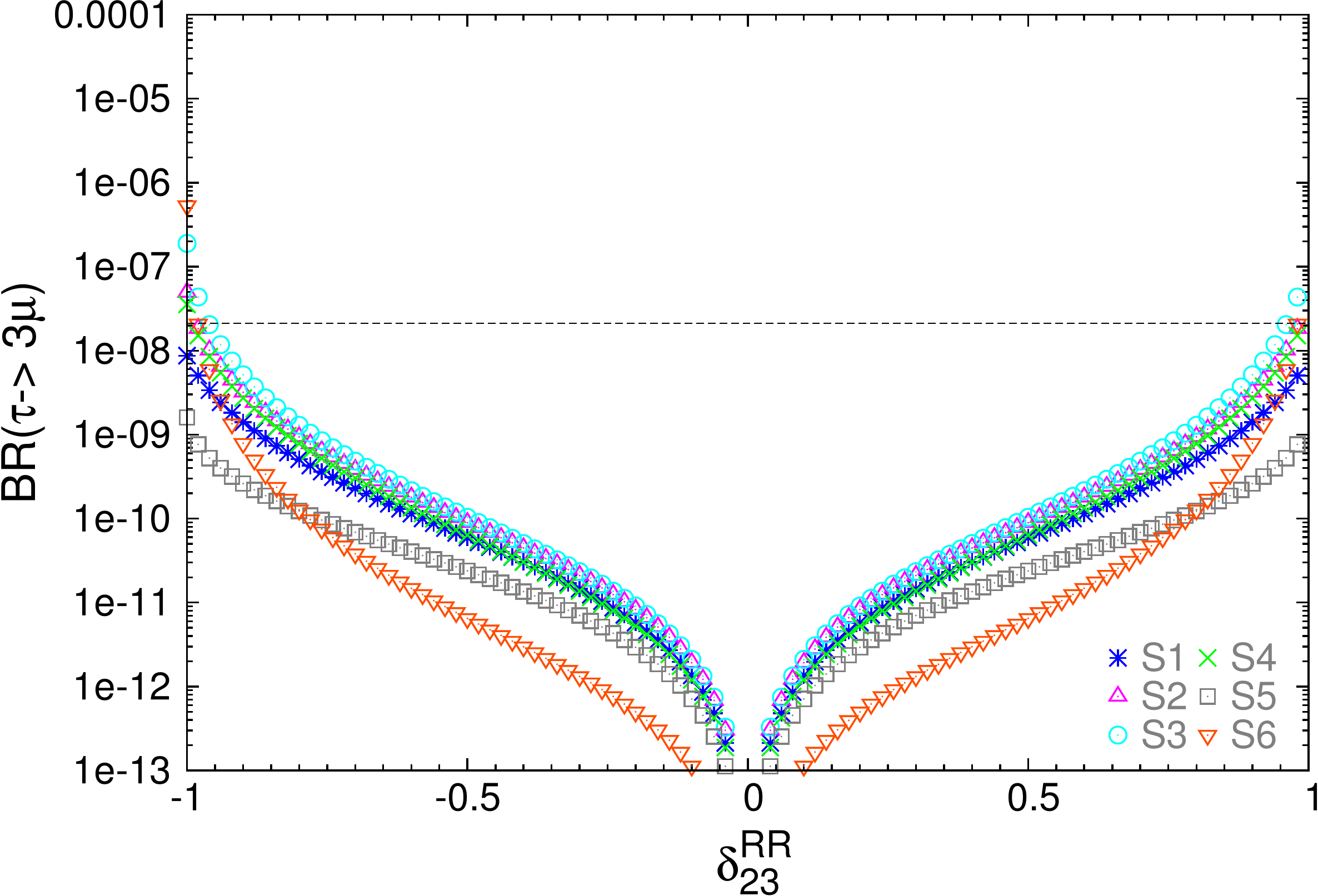    ,scale=0.30,angle=270,clip=}
\hspace{1em}
\psfig{file=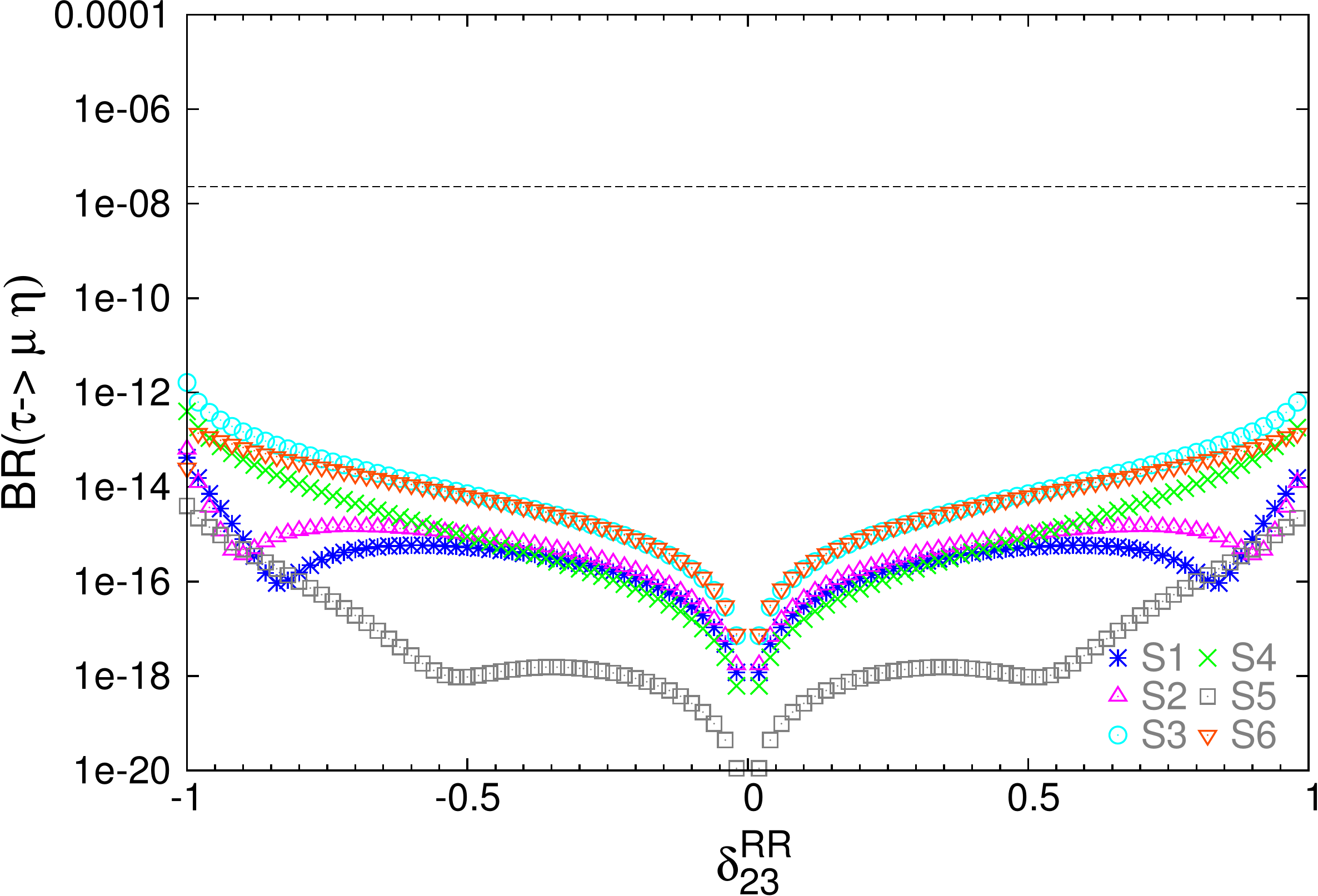   ,scale=0.30,angle=270,clip=}
\end{center}
\caption{LFV rates for $\tau-\mu$ transitions as a function of slepton
  mixing $\delta_{23}^{RR}$ for scenarios S1 to S6 defined in Section \ref{frameworkc}. The horizontal lines are the corresponding upper bounds collected in Section \ref{sec:expbounds}.}
\label{mixing23RR}
\end{figure}

The results of our numerical predictions of the branching ratios as
functions of the single deltas $\delta^{AB}_{ij}$, for the various
selected LFV processes and for the various scenarios S1 to S6 defined in
Section \ref{frameworkc}, are collected in figures \ref{mixing12LL} through
\ref{mixing23RR}, where a comparison with the corresponding present
upper experimental bound (the horizontal line in all these figures) is also included, given in \refse{sec:expbounds}. 
Figure \ref{mixing12LL}
summarizes the status of $\delta^{LL}_{12}$, \reffi{mixing12LR}
that of $\delta^{LR}_{12}$, \reffi{mixing12RR} that of
$\delta^{RR}_{12}$.
The analyzed experimental results are from $\br(\mu \to e \ga)$, 
$\br(\mu \to 3 e)$ and $\CR(\mu-e, \mbox{Nuclei})$.
Figure \ref{mixing13LL} depicts the results  of $\delta^{LL}_{13}$,  \reffi{mixing13LR} that of $\delta^{LR}_{13}$,
\reffi{mixing13RR} that of $\delta^{RR}_{13}$.
The analyzed experimental results are from $\br(\tau \to e \ga)$,
$\br(\tau \to 3 e)$ and $\br(\tau \to e \eta)$.
Figure \ref{mixing23LL} shows the results of 
$\delta^{LL}_{23}$, \reffi{mixing23LR} that of
$\delta^{LR}_{23}$, and \reffi{mixing23RR} that of
$\delta^{RR}_{23}$, where the experimental results are from 
$\br(\tau \to \mu \ga)$, $\br(\tau \to 3 \mu)$ and $\br(\tau \to \mu\eta)$.
The results for $\delta^{RL}_{ij}$ are indistinguishable from the corresponding ones for $\delta^{LR}_{ij}$, 
and consequently they have been omitted here.

 
A first look at these plots confirms the well
known result that the most stringent bounds are for the mixings between
the the first and the second slepton generations, 12. It is also evident
that the bounds for the mixings between the second and the third slepton
generations, 23,  are similar to the bounds for the mixings between the
first and the third generations, 13, and both are much weaker than the
bounds on the 12-mixings. 
As another general result one can observe that, 
whereas all the 12-mixings are constrained by
the three selected LFV processes,  
$\mu \to e \gamma$, $\mu \to 3 e$ and $\mu-e$ conversion in heavy (Au)
nuclei, the 23-mixings are not constrained, for the studied points, by
the semileptonic tau decay $\tau \to \mu \eta$. Similarly, the
13-mixings are not constrained either, by  $\tau \to e \eta$. The main
reason for this is that the studied points 
S1-S6 all have very heavy $A^0$ Higgs bosons, $M_{A}=500-1500\gev$ and
therefore the decay channel mediated by this $A^0$ is much
suppressed, even at large $\tb$, where the contribution from $A^0$ to 
$\br(\tau \to \mu \eta)$ and $\br(\tau \to e \eta)$, which is the dominant
one, grows as $(\tb)^6$ \cite{Arganda:2008jj,Herrero:2009tm}. It should also 
be noted the appearance of two symmetric minima in 
$\br(\tau \to \mu \eta)$ and $\br(\tau \to e \eta)$ of figs \ref{mixing23RR}
and \ref{mixing13RR} respectively, in the scenarios S5, S1 and S2. A similar 
feature can also be observed in
$\br(\tau \to e \eta)$ of fig \ref{mixing13LR} in scenario S2. For 
instance, in S5 these minima in $\br(\tau \to e \eta)$ appear at  
$\delta_{13}^{RR} \sim \pm 0.5$. We have checked that the origin of these minima
is due to the competing diagrams mediated by $A^0$ and $Z$ which give
contributions of similar size for  $\tb \lesssim 30$ but with opposite sign, and 
this produces strong cancellations in the total rates. Similar comments
apply to $\br(\tau \to \mu \eta)$.
 Another general result, which confirms some features known
in the literature for  particular models like SUSY-Seesaw models
\cite{Arganda:2005ji}, is the evident correlation between the $\br(l_j
\to 3 l_i)$ and  $\br(l_j \to l_i \gamma)$ rates. It should be
emphasized that we get these correlations in a model independent way and
without the use of any approximation, like the mass insertion
approximation or the large $\tb$ approximation. Since our
computation is full-one loop and has been performed in terms of
physical masses, our findings are valid for any  
value of $\tb$ and $\delta^{AB}_{ij}$'s. 
These correlations, confirmed in our plots, indicate that the general
prediction guided by the photon-dominance behaviour in $\br(l_j \to 3 l_i)$  
indeed works quite well for all the studied $\delta^{AB}_{ij}$'s and all
the studied  
S1-S6 points. This dominance of the $\gamma$-mediated channel in the
$l_j \to 3 l_i$ decays allows to derive the following simplified
relation:
\begin{equation}\label{BRlj3li_approx}
\frac{\br(l_j \to 3 l_i)}{\br(l_j \to l_i \gamma)} =
\frac{\alpha}{3\pi}\left(\log\frac{m_{l_j}^2}{m_{l_i}^2}-\frac{11}{4}\right) \,,
\end{equation}
which gives the approximate values of
$\frac{1}{440}$, $\frac{1}{94}$ and $\frac{1}{162}$ for 
$(l_jl_i)= (\tau \mu), (\tau e)$ and  $(\mu e)$, respectively.
The ${\cal O}(\alpha)$ suppression in the predicted rates of  
$\br(l_j \to 3 l_i)$ versus $\br(l_j \to l_i \gamma)$ yields, 
despite the fact that the experimental sensitivities to the leptonic decays $l_j
\to 3 l_i$ have been improved considerably in the last years, 
that the
radiative decays $l_j \to l_i \gamma$ are still the most efficient
decay channels in setting constraints to the slepton mixing
parameters. This holds for all the intergenerational mixings,
12, 13 and 23. 
As discussed in \cite{Arganda:2005ji}, in the context of SUSY, there
could be just a chance of departure from these ${\cal O}(\alpha)$
reduced ratios if the Higgs-mediated channels dominate the rates of the
leptonic decays, but this does not happen in our S1-S6 scenarios, with
rather heavy $H^0$ and $A^0$. We have checked that the
contribution from these Higgs channels are very small and can be safely
neglected, a scenario that is favoured by the recent results from
the heavy MSSM Higgs boson searches at the LHC~\cite{CMS-PAS-HIG-12-050}.

This same behaviour can be seen in the comparison between the $\br(\mu
\to e  \gamma)$ and $\CR(\mu-e, {\rm Nuclei})$ rates. Again there is an
obvious correlation in our plots for these two rates that can be
explained by the same argument as above, namely,  the photon-mediated
contribution in $\mu-e$ conversion dominates the other contributions,
for all the studied cases, and therefore the corresponding rates are
suppressed by a ${\cal O}(\alpha)$ factor respect to the radiative decay
rates. These correlations are clearly seen in all our plots for all the
studied $\delta^{AB}_{ij}$'s and in all S1-S6 scenarios. The
relevance of 
$\CR(\mu-e, {\rm Nuclei})$ as compared to $\br(\mu \to 3 e)$ is
given by the fact that not only the
present  experimental bound is slightly better, but also
that the future perspectives for the expected sensitivities are
clearly more promising in the $\mu-e$ conversion case (see below). 
In general, as
can be seen in our plots, the present bounds for  $\delta^{AB}_{ij}$'s
as obtained from $\CR(\mu-e, {\rm Nuclei})$ and  $\br(\mu \to 3 e)$ are
indeed very similar. 

In summary, the best bounds that one can infer from our results in
figures \ref{mixing12LL} through \ref{mixing23RR} come from the
radiative $l_j \to l_i \gamma$ decays and we get the maximal allowed
values for all $|\delta^{AB}_{ij}|$'s that are collected in
\refta{boundsSpoints} for each of the studied scenarios S1 to S6. 
They give an overall idea of the size of the bounds with respect to
the latest experimental data.
When comparing the results in
this table for the various scenarios, we see that scenario S3 gives the
most stringent constraints to the  $\delta^{LL}_{ij}$ and
$\delta^{RR}_{ij}$ mixings, in spite of having rather heavy sleptons
with masses close to 1 TeV. The reason is well understood from the $\tb$
dependence of the BRs which enhances the rates in the case of  $LL$
and/or $RR$ single deltas at large $\tb$, in agreement with the simple
results of the MIA formulas in \refeqs{MIA-L} and (\ref{MIA-R}). 
Here it should be noted that within S3 we have $\tb = 50$, which
is the largest considered value in these S1-S6 scenarios. Something
similar happens in S4 with $\tb =40$. In contrast, the most stringent
constraints on the $\delta^{LR}_{ij}$ mixings occur in scenarios S1 and
S5. Here it is important to note that there are not enhancing
$\tb$ factors in the $\delta^{LR}_{ij}$ case. In fact, the contributions
from the $\delta^{LR}_{ij}$'s  to the most constraining LFV radiative
decay rates are $\tb$ independent, in agreement again with the MIA simple
expectations (see \refse{sec:MIA}). Consequently, the stringent
constraints on $\de^{LR}_{i,j}$ in S1 and S5 arise due to the relatively
light sleptons in these scenarios.

\begin{table}[h!]
\centering
\begin{tabular}{|c|c|c|c|c|c|c|}
\hline
 &  S1 &  S2 &  S3 &  S4 &  S5 & S6  
 \\ \hline
 & & & & & & \\
$|\delta^{LL}_{12}|_{\rm max}$ & $10 \times 10^{-5}$ & $7.5\times 10^{-5}$ &   $5 \times 10^{-5}$& $6 \times 10^{-5}$ & $42\times 10^{-5}$  &  $8\times 10^{-5}$  \\ 
& & & & & & \\
\hline
& & & & & & \\
$|\delta^{LR}_{12}|_{\rm max}$ & $2\times 10^{-6}$ & $3\times 10^{-6}$ &
$4\times 10^{-6}$  & $3\times 10^{-6}$ & $2\times 10^{-6}$  & $1.2\times 10^{-5}$   \\ 
& & & & & & \\
\hline
& & & & & & \\
$|\delta^{RR}_{12}|_{\rm max}$ & $1.5 \times 10^{-3}$& $1.2 \times 10^{-3}$ & 
$1.1 \times 10^{-3}$ & $1 \times 10^{-3}$ & $2 \times 10^{-3}$ & $5.2 \times 10^{-3}$   \\ 
& & & & & & \\
\hline
& & & & & & \\
$|\delta^{LL}_{13}|_{\rm max} $ &  $5 \times 10^{-2}$ & $5 \times 10^{-2}$ & 
$3 \times 10^{-2}$ &  $3 \times 10^{-2}$& $23 \times 10^{-2}$ & $5 \times 10^{-2}$   \\ 
& & & & & & \\
\hline
& & & & & & \\
 $|\delta^{LR}_{13}|_{\rm max}$& $2\times 10^{-2}$  & $3\times 10^{-2}$ & $4\times 10^{-2}$ & $2.5\times 10^{-2}$ & $2\times 10^{-2}$ & $11\times 10^{-2}$   \\ 
& & & & & & \\
 \hline
 & & & & & & \\
$|\delta^{RR}_{13}|_{\rm max}$ & $5.4\times 10^{-1}$  & $5\times 10^{-1}$ & 
 $4.8\times 10^{-1}$ &$5.3\times 10^{-1}$  & $7.7\times 10^{-1}$ & $7.7\times 10^{-1}$ 
  \\ 
 & & & & & & \\ 
  \hline
 & & & & & & \\ 
$|\delta^{LL}_{23}|_{\rm max}$ & $6\times 10^{-2}$  & $6\times 10^{-2}$ & 
 $4\times 10^{-2}$& $4\times 10^{-2}$ & $27\times 10^{-2}$ & $6\times 10^{-2}$ 
  \\ 
 & & & & & & \\ 
  \hline
 & & & & & & \\ 
$|\delta^{LR}_{23}|_{\rm max}$ & $2\times 10^{-2}$   & $3\times 10^{-2}$ & 
$4\times 10^{-2}$ & $3\times 10^{-2}$ & $2\times 10^{-2}$ & $12\times 10^{-2}$ 
  \\ 
 & & & & & & \\ 
  \hline
 & & & & & & \\ 
$|\delta^{RR}_{23}|_{\rm max}$ & $5.7\times 10^{-1}$  & $5.2\times 10^{-1}$ & 
 $5\times 10^{-1}$& $5.6\times 10^{-1}$ & $8.3\times 10^{-1}$ & $8\times 10^{-1}$ 
  \\ 
 & & & & & & \\  
  \hline
\end{tabular}
\caption{ Present upper bounds on the slepton mixing parameters $|\delta^{AB}_{ij}|$ for the selected S1-S6 MSSM points defined in \refta{tab:spectra}. The bounds for $|\delta^{RL}_{ij}|$ are similar 
to those of $|\delta^{LR}_{ij}|$.}
\label{boundsSpoints}
\end{table}

\bigskip

So far, we have studied the case where just one mixing delta is allowed
to be non-vanishing. However, it is known in the
literature\cite{Masina:2002mv,Paradisi:2005fk} that one can get more
stringent or more loose bounds in some particular cases if,
instead, two (or even more) deltas are 
allowed to be non-vanishing. In order to study the implications of these
scenarios with two deltas, we have analyzed the improved bounds on pairs
of mixings of the 13 and 23 type which are at present the less
constrained as long each delta is analyzed singly.

\begin{figure}[ht!]
\begin{center}
\psfig{file=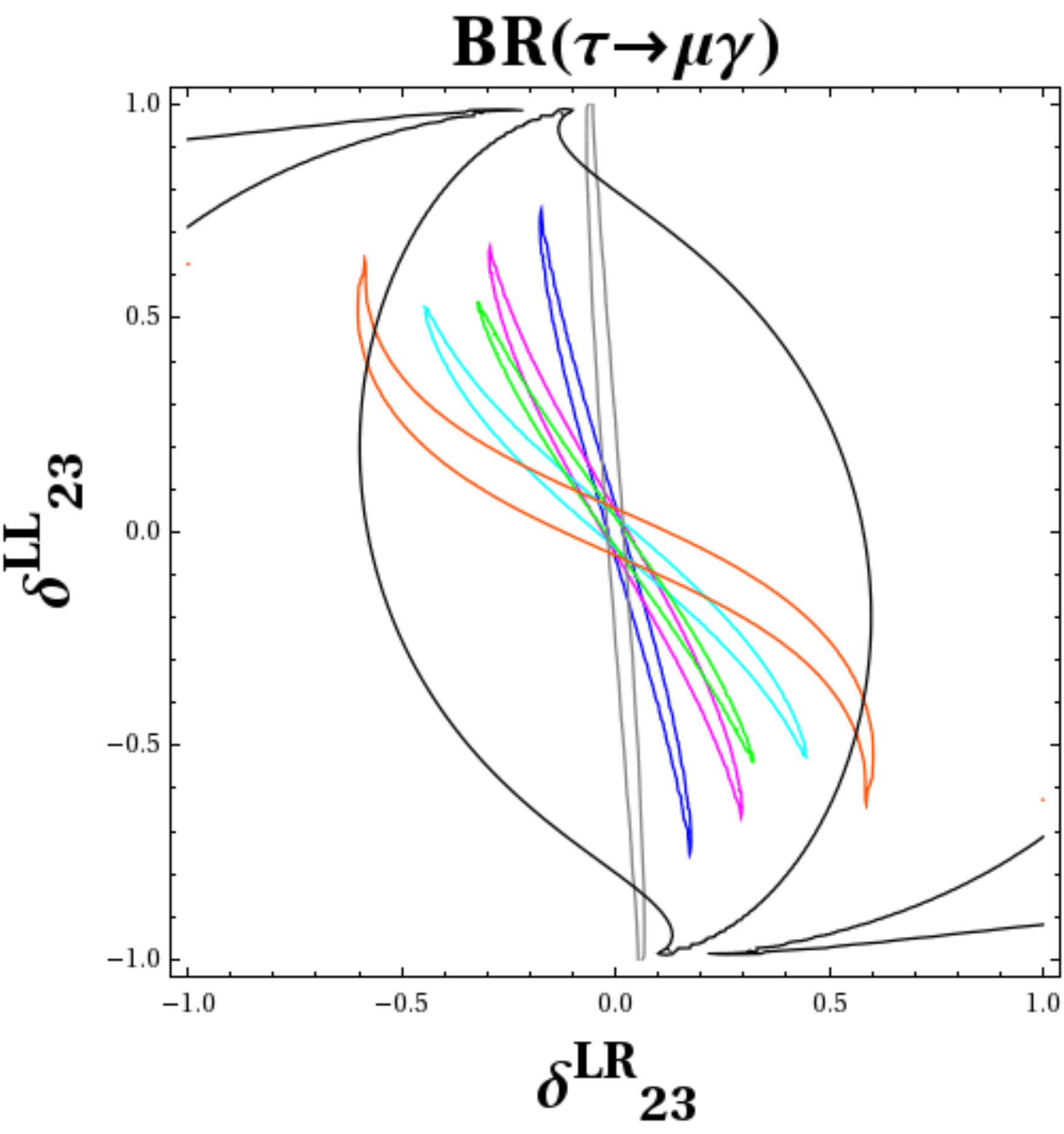,scale=0.70,clip=}
\end{center}
\caption{Maximum allowed values of  $(\delta^{LR}_ {23},\delta^{LL}_
  {23})$ in the scenarios S1 (dark blue), S2 (magenta), S3 (light blue),
  S4 (green), S5 (grey), S6 (orange) defined in Section \ref{frameworkc} and S7 (black) defined in Eq. \ref{s7def}. The contourlines
  shown correspond to the present experimental upper limit: $\br(\tau
  \to \mu \gamma)_{\rm max}=4.4\times 10^{-8}$. For each scenario the
  allowed deltas are those inside the corresponding contourline.}  
\label{23LR-23LL}
\end{figure}

First we have looked into the various delta pairings of 23 type,
$(\delta^{AB}_{23},\delta^{CD}_{23})$, and we have found that some of
them lead to interesting interferences in the $\br(\tau \to \mu
\gamma)$ rates that can be either constructive or destructive, depending
on the relative delta signs, therefore leading to either a reduction or
an enhancement, respectively, in the  maximum allowed delta values as
compared to the one single delta case. More specifically, we have found
interferences in $\br(\tau \to \mu \gamma)$ for the case of
non-vanishing  $(\delta^{LR}_ {23},\delta^{LL}_ {23})$ pairs that are
constructive if these deltas are of equal sign, and destructive if they
are of opposite sign. Similarly, we have also found interferences in
$\br(\tau \to \mu \gamma)$ for the case of non-vanishing  $(\delta^{RL}_
{23},\delta^{RR}_ {23})$ pairs that are constructive if they are of
equal sign, and destructive if they are of opposite sign. However, in
this latter case the size of the interference is very small and does not
lead to very relevant changes with respect to the single delta case. The
numerical results for the most interesting case of $(\delta^{LR}_
{23},\delta^{LL}_ {23})$ are shown in \reffi{23LR-23LL}. We have
analyzed the six previous points, S1 through S6 defined in Section \ref{frameworkc}, and a new point S7 with
extremely heavy sleptons and whose relevant parameters for this analysis
of the 23 delta bounds are as follows: 
\BEA
S7& : &  m_{\tilde L_{1,2,3}}=m_{\tilde E_{1,2,3}}=10000\,\, {\rm GeV} \nonumber \\
&& \mu = 2000\,\,{\rm GeV}; \tb= 60  \nonumber \\
&& M_2=2000 \,\,{\rm GeV}; M_1=1000\,\,{\rm GeV} \label{s7def}
\EEA  

This figure exemplifies in a clear way
that for some of the studied scenarios the destructive
interferences can be indeed quite relevant and produce new areas in the
$(\delta^{LR}_ {23},\delta^{LL}_ {23})$ plane with relatively large
allowed values of both $|\delta^{LR}_ {23}|$ and $|\delta^{LL}_ {23}|$
mixings. For instance, the orange contour which corresponds to the
maximum allowed values for scenario S6, leads to allowed mixings as
large as  $(\delta^{LR}_ {23},\delta^{LL}_ {23}) \sim (\pm 0.6, \mp
0.6)$. We also learn from this plot, that the relevance of this
$\delta^{LR}_ {23}-\delta^{LL}_ {23}$ interference grows in the
following order: Scenario S5 (grey contour) that has the smallest
interference effect, then S1, S2, S4, S3 and S6 that has the largest
interference effect. This growing interference effect is seen in the
plot as the contour being rotated anti-clockwise from
the most vertical one (S1) to the most inclined one
(S6). Furthermore, the size of the parameter space bounded by
these contours also grows, implying that 
``more'' parameter combinations are available for these two deltas.
It should be noted that, whereas the existence of the interference
effect can be already expected from the simple MIA formulas of
\refeqs{MIA-L} and (\ref{MIA-R}), the final found shape of these contours
in Fig.\ref{23LR-23LL} and their quantitative relevance cannot be
explained by these simple formulas. The separation from the MIA
expectations are even larger in the new studied scenario S7, as can be
clearly seen in this figure. The big black contour, centred at zero,
contains a rather large allowed area in the  $(\delta^{LR}_
{23},\delta^{LL}_ {23})$ plane, allowing  values, for instance, of
$(\delta^{LR}_ {23},\delta^{LL}_ {23}) \sim (\pm 0.5, \mp
0.5)$. Furthermore, in this S7 there appear new allowed regions at the
upper left and lower right corners of the plot with extreme allowed
values as large as  $(\pm 0.9, \mp 0.9)$. 
These ``extreme'' solutions are only captured by a full one-loop
calculation
and cannot be explained by the simple MIA formulas.

\begin{figure}[ht!]
\begin{center}
\psfig{file=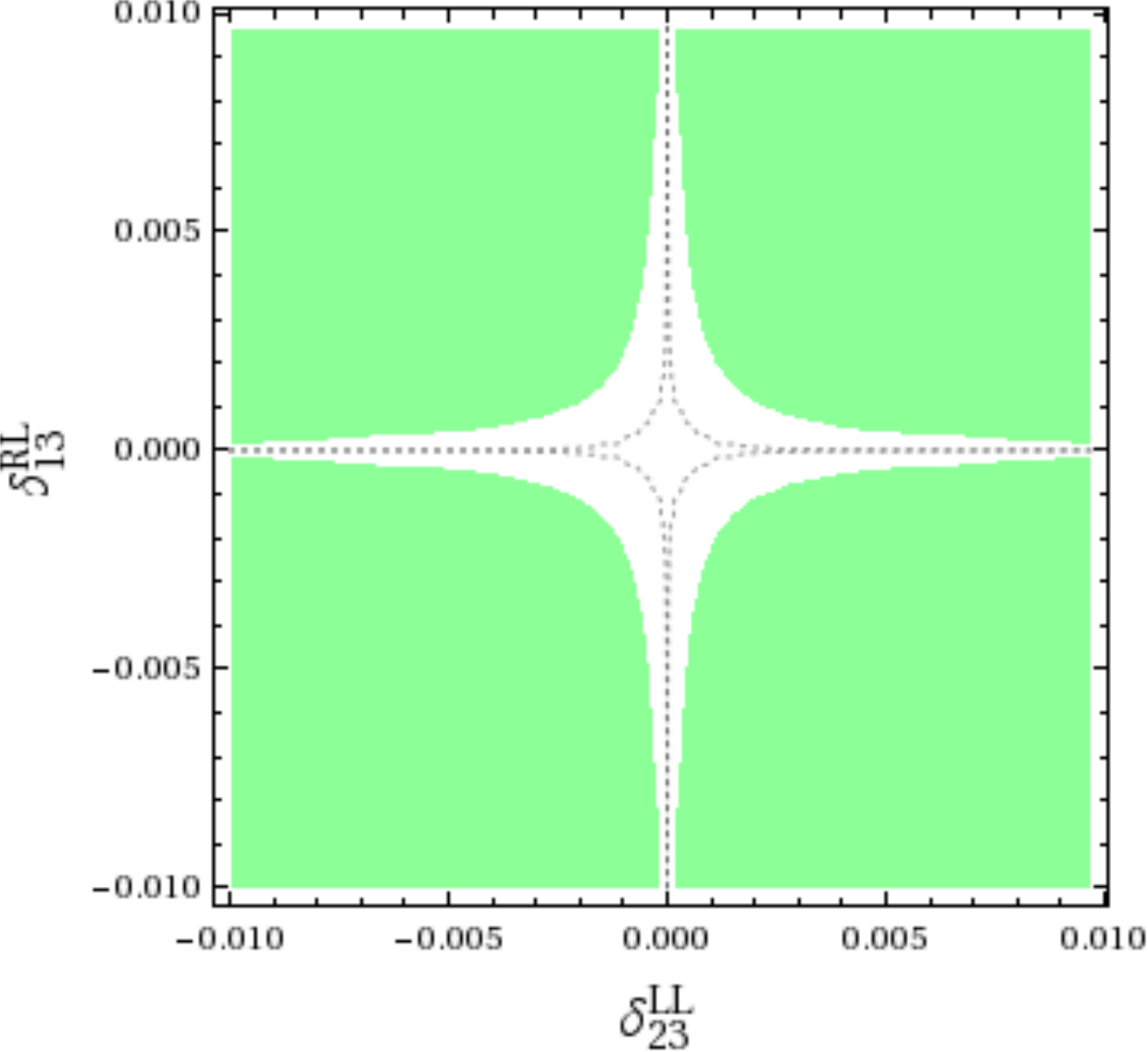     ,scale=0.63,clip=}
\psfig{file=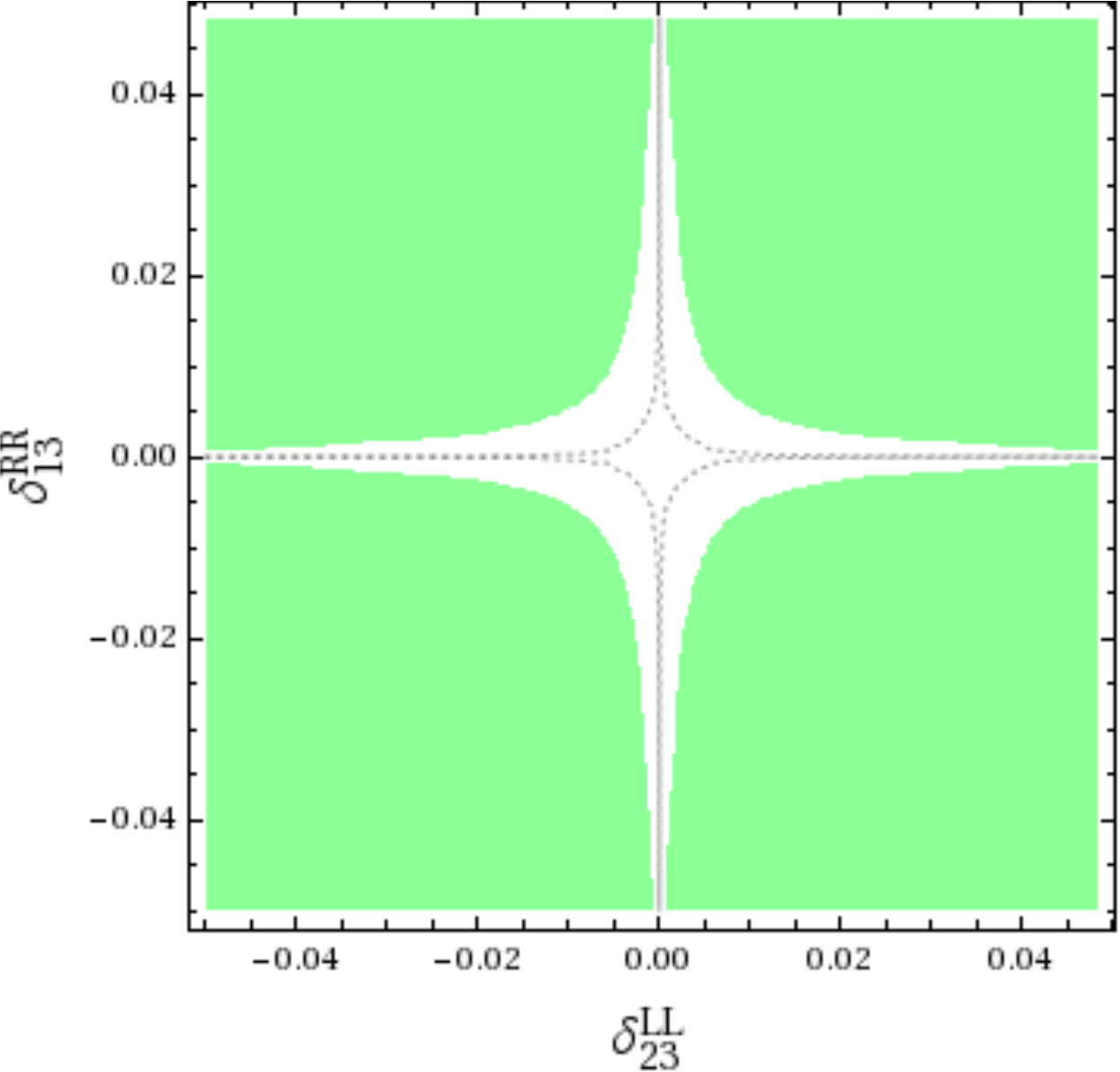   ,scale=0.595,clip=}\\
\vspace{1em}
\psfig{file=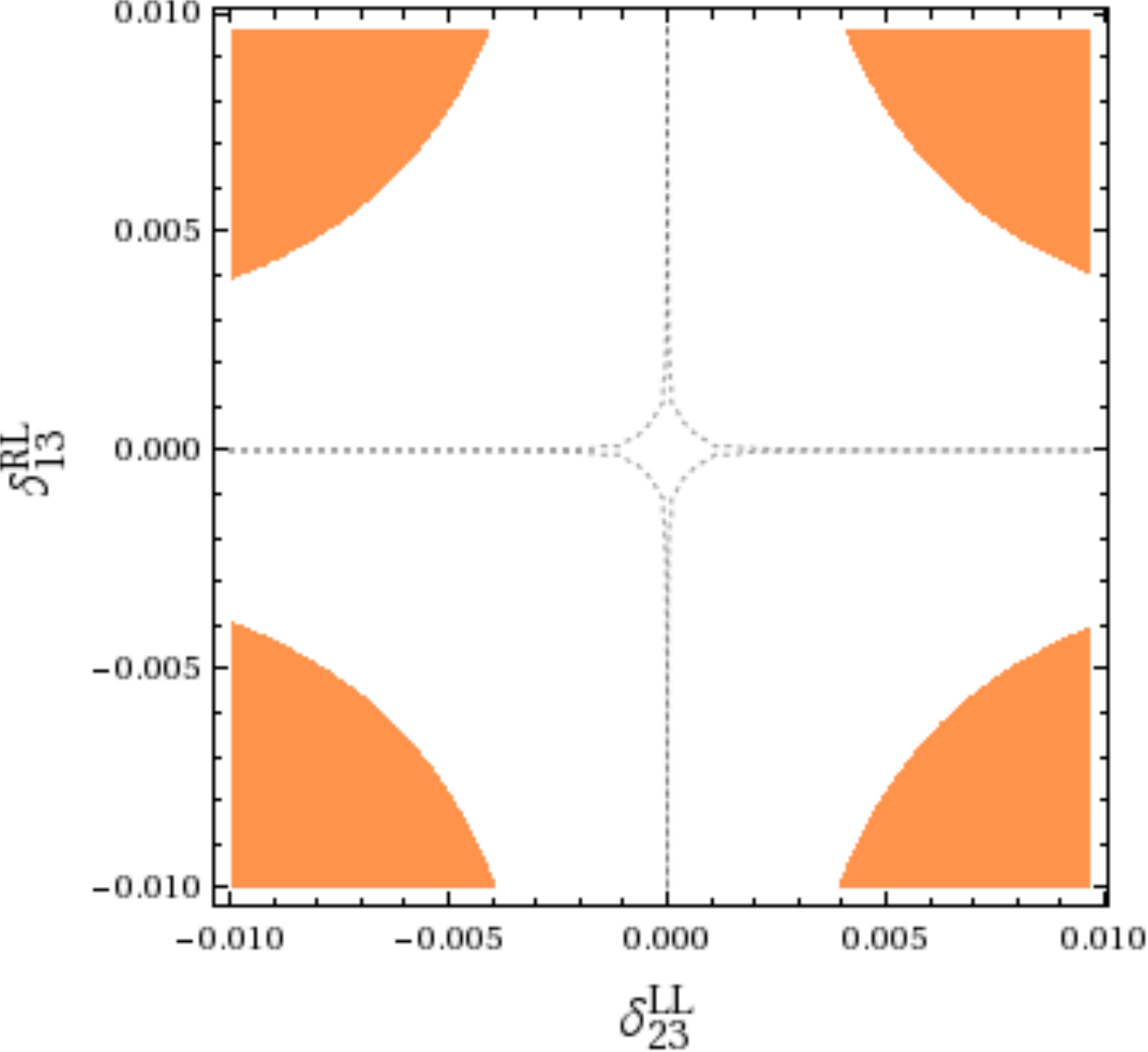    ,scale=0.63,clip=}
\psfig{file=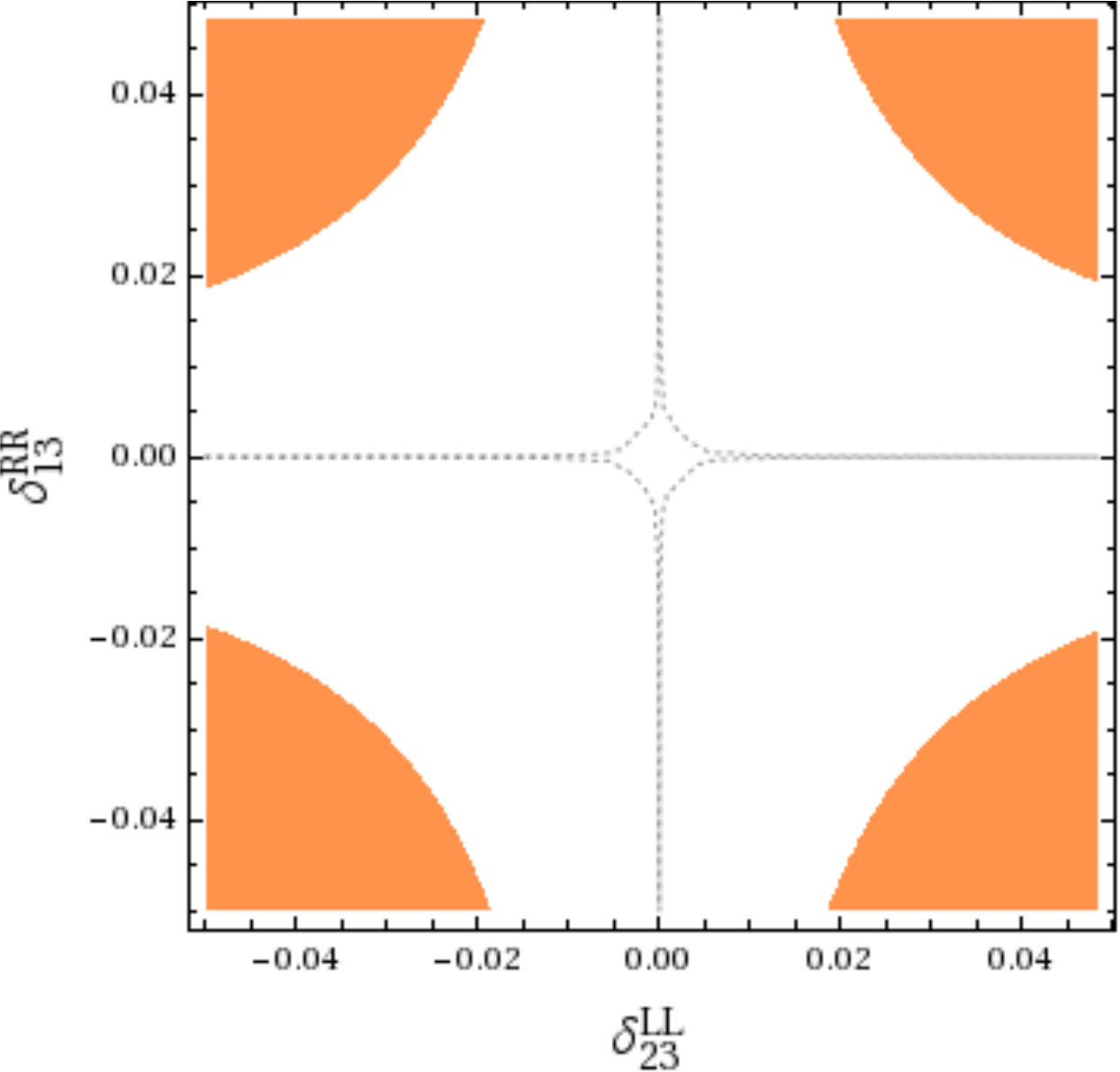    ,scale=0.595,clip=}
\end{center}
\caption{Bounds on pairs of slepton mixing parameters of (23,13) type
  for scenario S1: a) ($\delta_{23}^{LL}$,$\delta_{13}^{RL}$) in first
  column. Identical plots, not shown here, are found for:
  ($\delta_{23}^{RL}$,$\delta_{13}^{LL}$),
  ($\delta_{23}^{LR}$,$\delta_{13}^{RR}$), and
  ($\delta_{23}^{RR}$,$\delta_{13}^{LR}$); b)
  ($\delta_{23}^{LL}$,$\delta_{13}^{RR}$) in second column. Identical
  plots, not shown here, are found for:
  ($\delta_{23}^{RR}$,$\delta_{13}^{LL}$),
  ($\delta_{23}^{LR}$,$\delta_{13}^{RL}$), and
  ($\delta_{23}^{RL}$,$\delta_{13}^{LR}$). First row: Shaded regions (in
  green) are disallowed by the present upper experimental limit on 
  $\br(\mu \to e \gamma)$. 
  Second row: Shaded regions (in orange) are disallowed by the
  present upper experimental limit on $\CR(\mu - e, {\rm Nuclei})$. The
  allowed central areas in white will be shrinked by the future expected
  sensitivities in both $\mu \to e \gamma$ and $\mu-e$ conversion
  experimental searches (see text) to the small areas around the
  origin delimited 
  by the dotted lines.}  
\label{LL-RL}
\end{figure}

\begin{figure}[ht!]
\begin{center}
\psfig{file=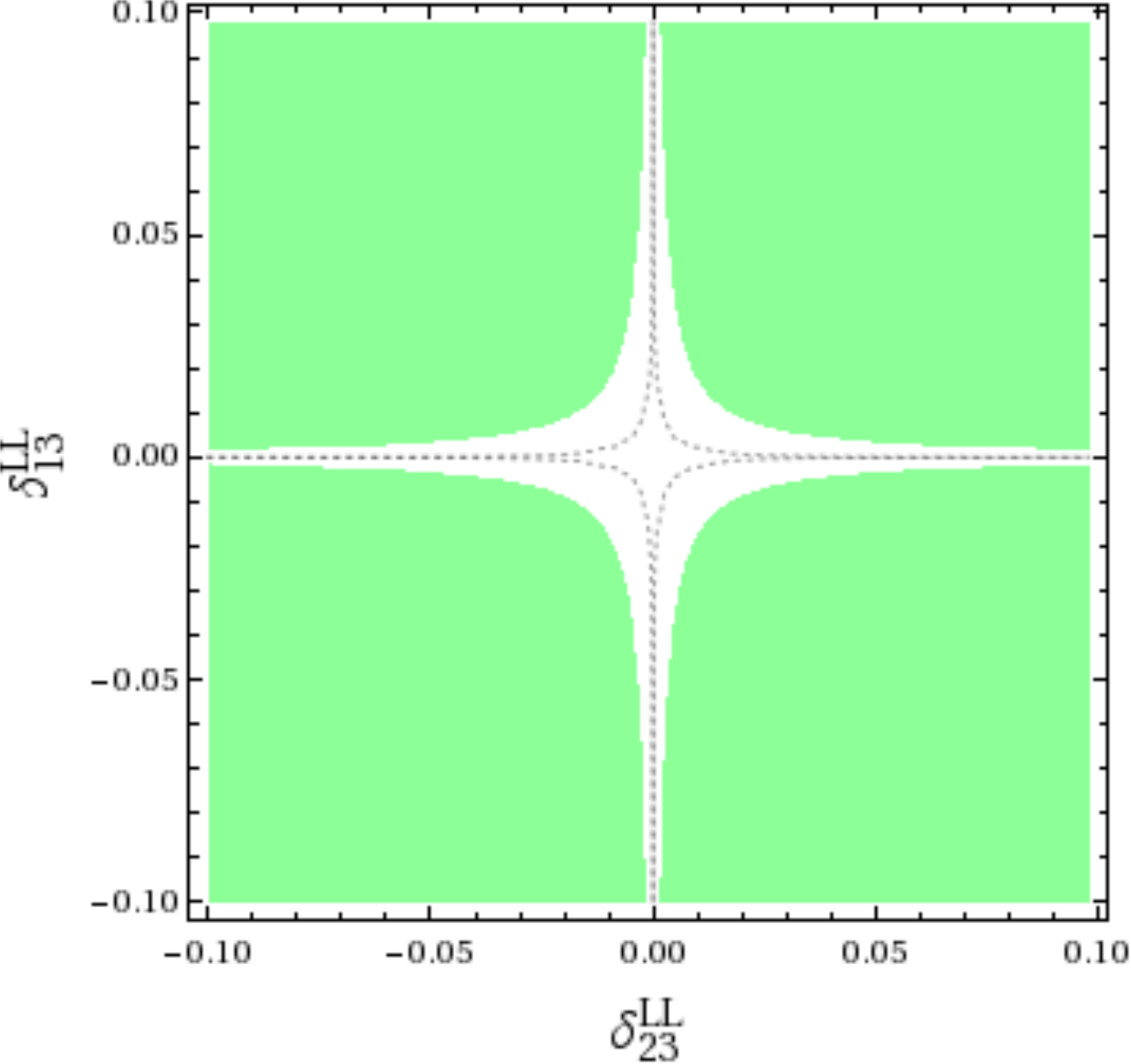     ,scale=0.625,clip=}
\psfig{file=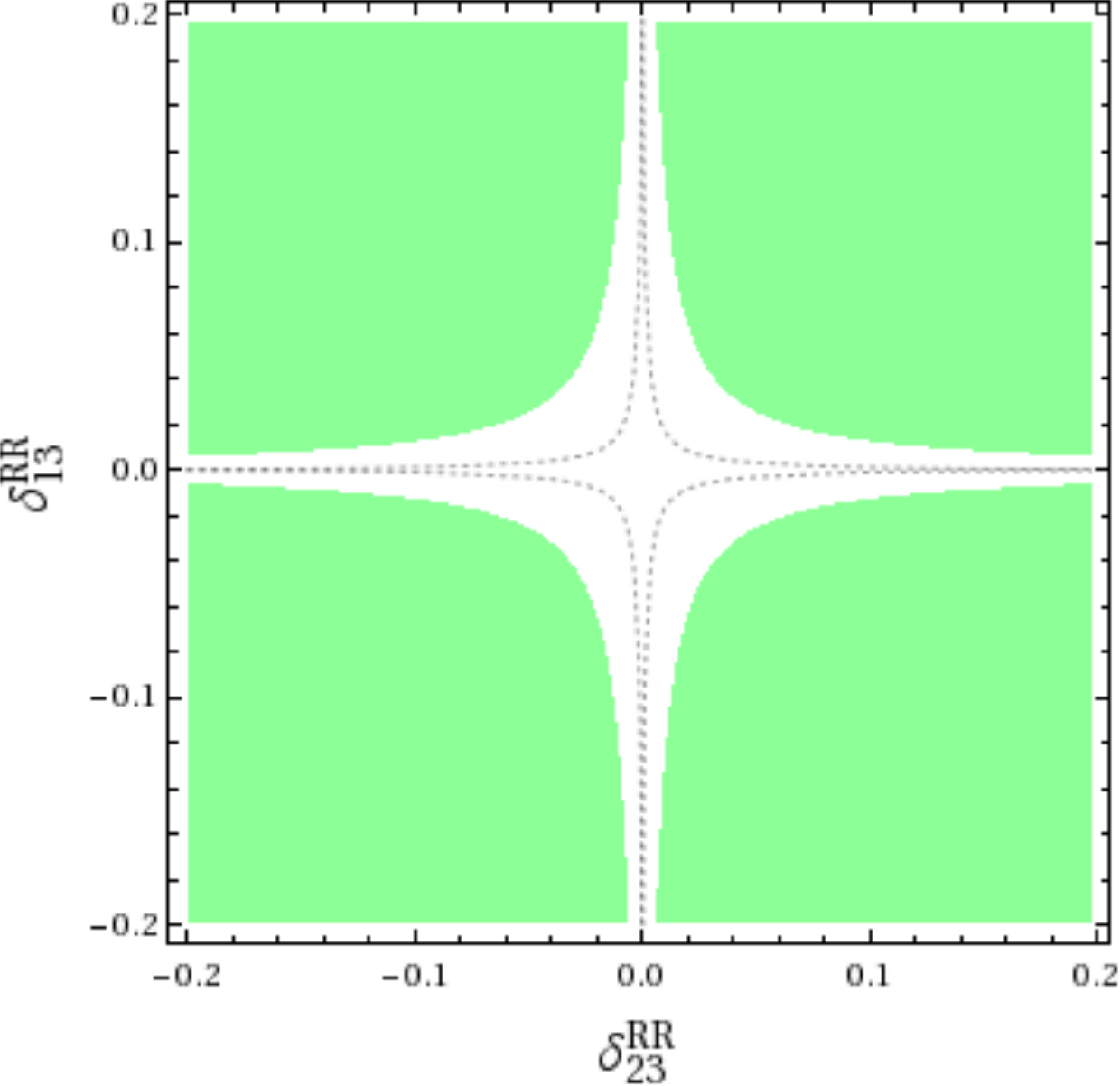   ,scale=0.60,clip=}\\
\vspace{1em}
\psfig{file=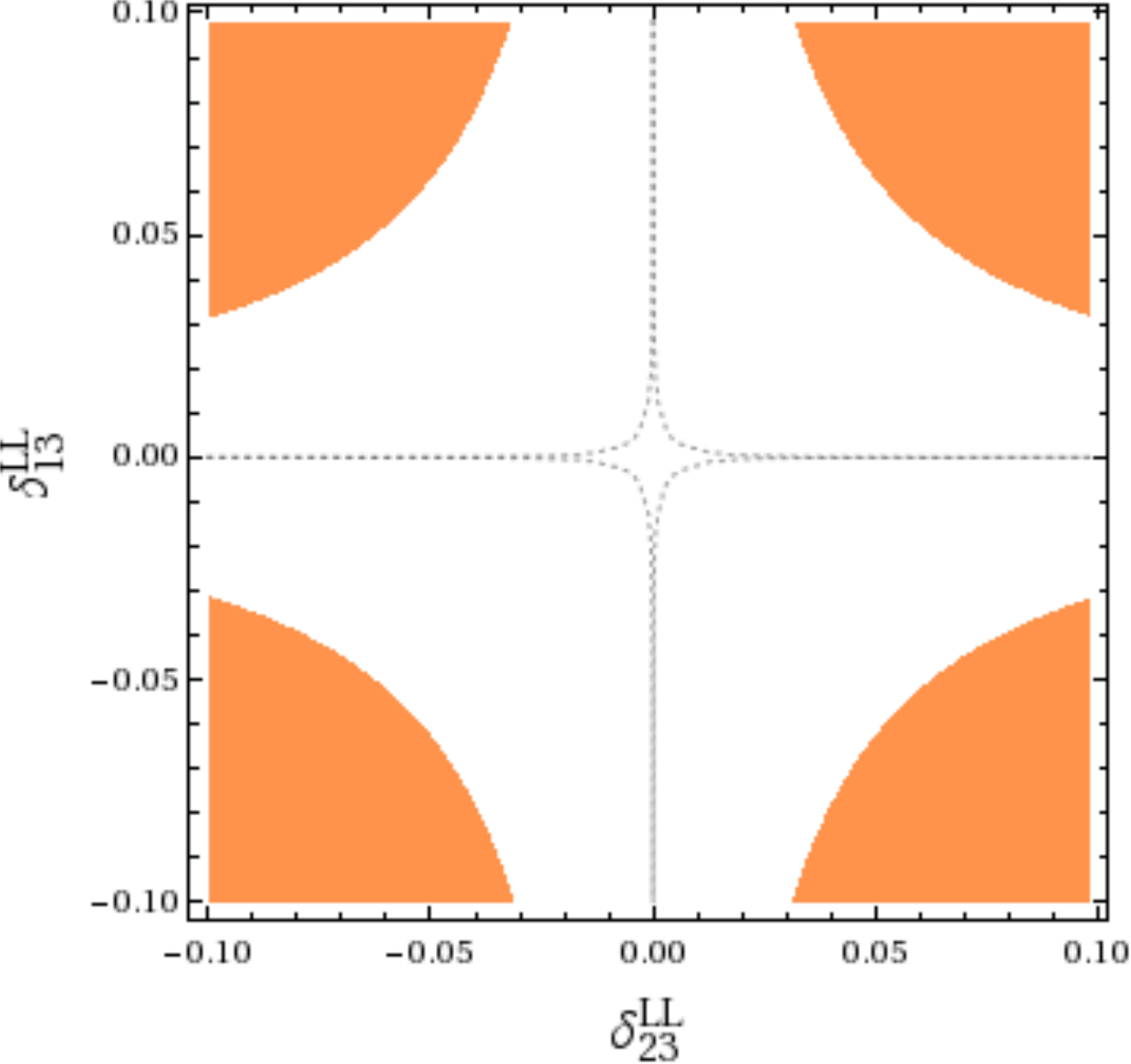     ,scale=0.625,clip=}
\psfig{file=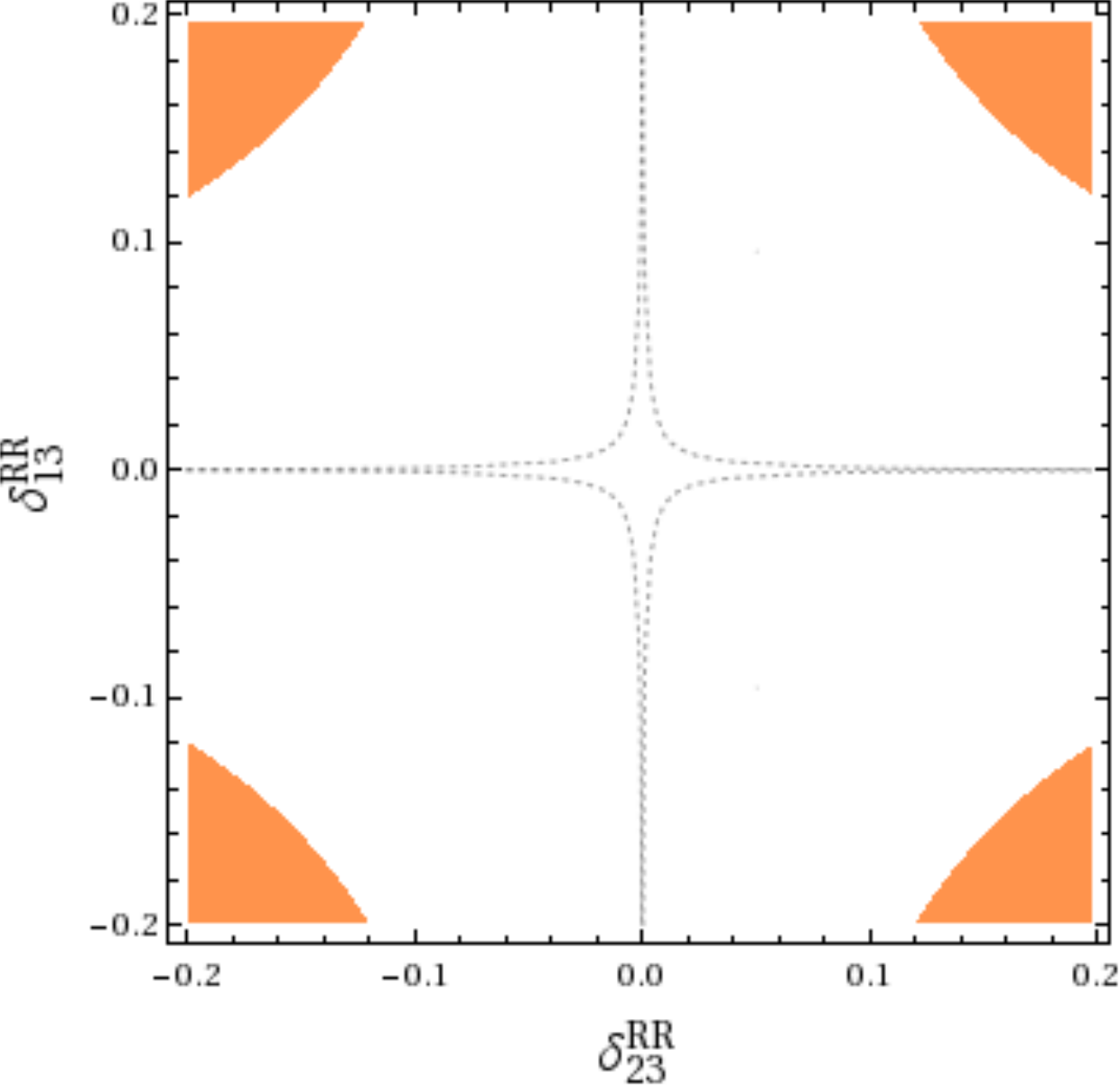    ,scale=0.60,clip=}
\end{center}
\caption{Bounds on pairs of slepton mixing parameters of (23,13) type
  for scenario S1: a)  ($\delta_{23}^{LL}$,$\delta_{13}^{LL}$) in first
  column; b)  ($\delta_{23}^{RR}$,$\delta_{13}^{RR}$) in second
  column. First row: Shaded regions (in green) are disallowed by 
  $\br(\mu \to e \gamma)$. 
  Second row: Shaded regions (in orange) are disallowed
  by $\CR(\mu - e, {\rm Nuclei})$. All inputs and explanations are
  as in \reffi{LL-RL}.}  
\label{LL-LLandRR-RR}
\vspace{-3em}
\end{figure}

\medskip
We now turn to examples in which more stringent bounds on 
  combinations of two deltas are derived. In particular,
 we have explored the restrictions that are obtained on the 
(13,23)  mixing pairs from the present bounds on $\br(\mu \to e \gamma)$
and $\CR(\mu-e, {\rm nuclei})$. In figures \ref{LL-RL} and
\ref{LL-LLandRR-RR} we show the results of this analysis for the S1
point. We have only selected  the pairs where we have found  improved
bounds respect to the previous single delta analysis. From \reffi{LL-RL}
we conclude that, for S1,  the maximal allowed values  
by present $\mu \to e \gamma$ (($ \mu-e$ conversion)) searches are
(given specifically here for equal input deltas):  
\BEA
(|\delta^{LL}_{23}|_{\rm max},|\delta^{RL}_{13}|_{\rm max})
&=&(0.0015,0.0015)\,\,((0.0062,0.0062))\,\,  
\label{double1}
\EEA
These numbers can be understood as follows: if, for instance, 
$\de^{RL}_{13} = 0.0015$ then $|\de^{LL}_{23}| < 0.0015$. If, on the
other hand, one delta goes to zero the bound on the other delta
disappears (from this particular observable).
We find equal bounds as in \refeq{double1} for: 
$(|\delta^{RL}_{23}|_{\rm max},|\delta^{LL}_{13}|_{\rm max})$, 
$(|\delta^{LR}_{23}|_{\rm max},|\delta^{RR}_{13}|_{\rm max})$ and   
$(|\delta^{RR}_{23}|_{\rm max},|\delta^{LR}_{13}|_{\rm max})$.  

Other pairings of deltas give less stringent bounds than
\refeq{double1} but still more stringent than the ones from the
single delta analysis. In particular, we get: 
\BEA
 (|\delta^{LL}_{23}|_{\rm max},|\delta^{RR}_{13}|_{\rm max}) 
&=&(0.0073,0.0073)\,\,((0.031,0.031))\,\,
\label{double2}
\EEA
And we get equal bounds as in \refeq{double2} for:
$(|\delta^{RR}_{23}|_{\rm max},|\delta^{LL}_{13}|_{\rm max})$, $(|\delta^{LR}_{23}|_{\rm max},|\delta^{RL}_{13}|_{\rm max})$ and $(|\delta^{RL}_{23}|_{\rm max},|\delta^{LR}_{13}|_{\rm max})$.

Finally, from  \reffi{LL-LLandRR-RR} we get:
\BEA
(|\delta^{LL}_{23}|_{\rm max},|\delta^{LL}_{13}|_{\rm max}) &=&(0.013,0.013)\,\,((0.056,0.056))\,\,
\label{double3}
\EEA
and 
\BEA
(|\delta^{RR}_{23}|_{\rm max},|\delta^{RR}_{13}|_{\rm max}) &=&(0.036,0.036)\,\,((0.16,0.16))\,\,
\label{double4}
\EEA
We have also studied the implications of the future expected
sensitivities in both 
$\br(\mu \to e \gamma) < 10^{-14}$~\cite{Mihara:2012zz} and 
$\CR(\mu - e, {\rm Nuclei}) < 2.6 \times 10^{-17}$~\cite{mueconvfuture},
 which are anticipated from future searches.
From our results in 
\reffis{LL-RL} and \ref{LL-LLandRR-RR} we conclude that the previous bounds
in \refeqs{double1}, (\ref{double2}), (\ref{double3}) and (\ref{double4})
will be improved (for both $\mu \to e
\gamma$ and  $\mu-e$ conversion)  to  
$(0.0005,0.0005)$, $(0.0025,0.0025)$, $(0.005,0.005)$ and $(0.01,0.01)$,
respectively. 

\clearpage
\newpage

\subsection{Numerical results for pMSSM-4 scenarios}

The main goal of this part is to investigate how the upper
bounds for the slepton mixing deltas that we have found previously could
change for different areas of the MSSM parameter space, 
other than the specific selected S1-S6 points. In order to explore the variation of these bounds for different choices
in the MSSM parameter space, we investigate the four qualitatively
different pMSSM-4 scenarios (a), (b), (c) and (d) defined in \refeqs{Sa},
(\ref{Sb}), (\ref{Sc}) and (\ref{Sd}), respectively, in Section \ref{sec:f2}. 
As explained before, the idea is to explore generic scenarios that are
compatible with present data, in particular with the measurement of
a Higgs boson mass, which we interpret as the mass of the light
$\cp$-even Higgs boson in the MSSM, and the present
experimental measurement of $(g-2)_\mu$. Taking these experimental
results into account, we have re-analyzed the full set of bounds for
the single deltas that are 
extracted from the most restrictive LFV processes as a function of the
two most relevant parameters in those scenarios: the generic
SUSY mass scale $m_{\rm SUSY}$ $(\equiv m_{\rm SUSY-EW})$ and
$\tb$. In order to find $\Mh$ around $\sim 125 - 126 \gev$ the
scale $m_{\rm SUSY-QCD}$ as well as the trilinear squark couplings have been
chosen to sufficiently high values, see \refse{sec:f2}.
For the analysis in those scenarios, we
use the bounds on the radiative decays, $l_j \to l_i \gamma$ which, as
we have already shown, are at present the most restrictive ones in the
case of one single non-vanishing delta. And to simplify the analysis in
this part of the work, we use the mass insertion approximation
formulas of \refeqs{BRs} through (\ref{functions}) to evaluate the
$\br(l_j \to l_i \gamma)$ rates. We have checked that these simple MIA
formulas provide a sufficiently accurate estimate of the
LFV rates for the present scenarios in the case of single 
deltas, in agreement with \citere{Paradisi:2005fk}. 
 
\begin{figure}[ht!]
\begin{center}
\psfig{file=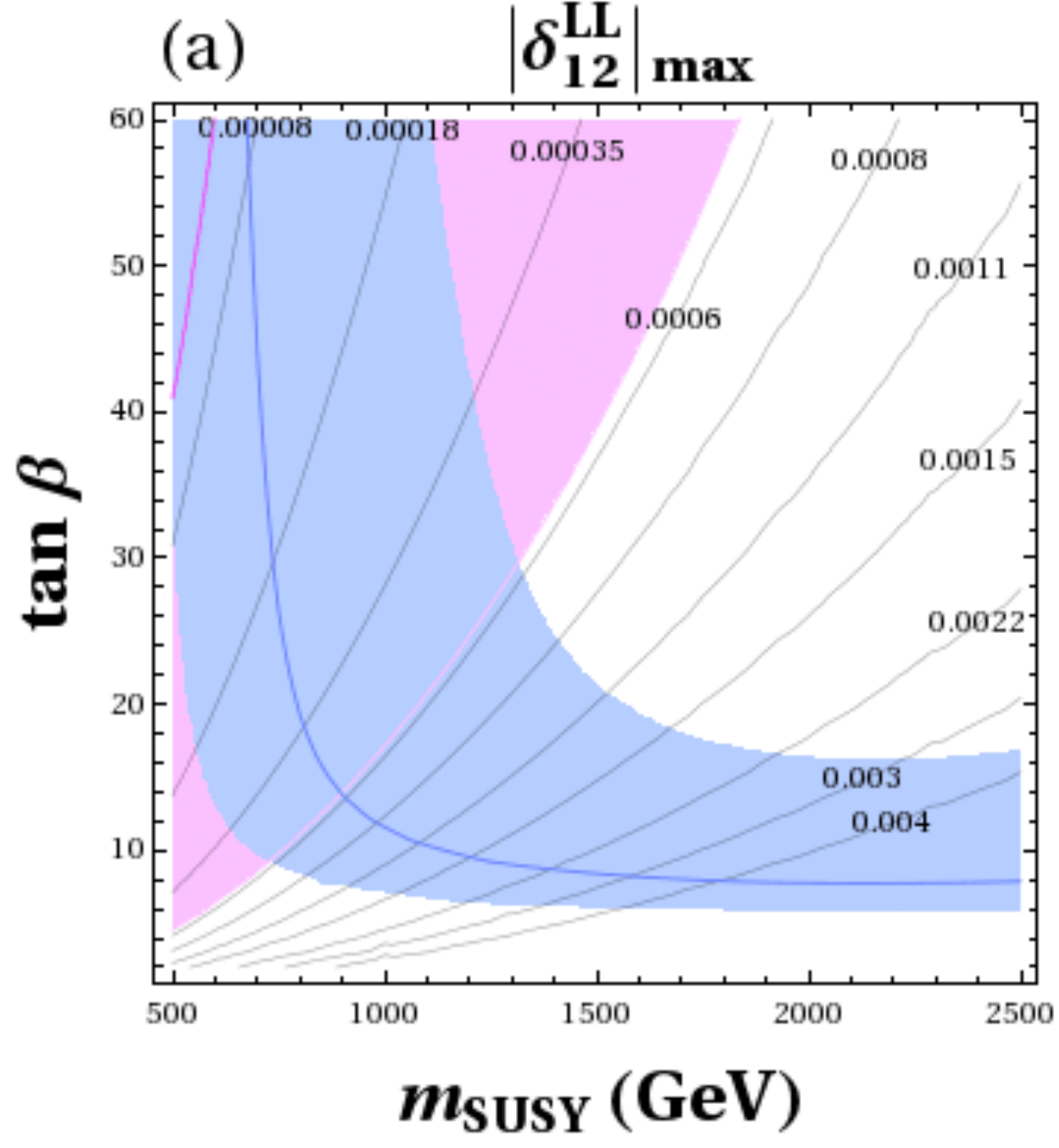,scale=0.60,clip=}
\psfig{file=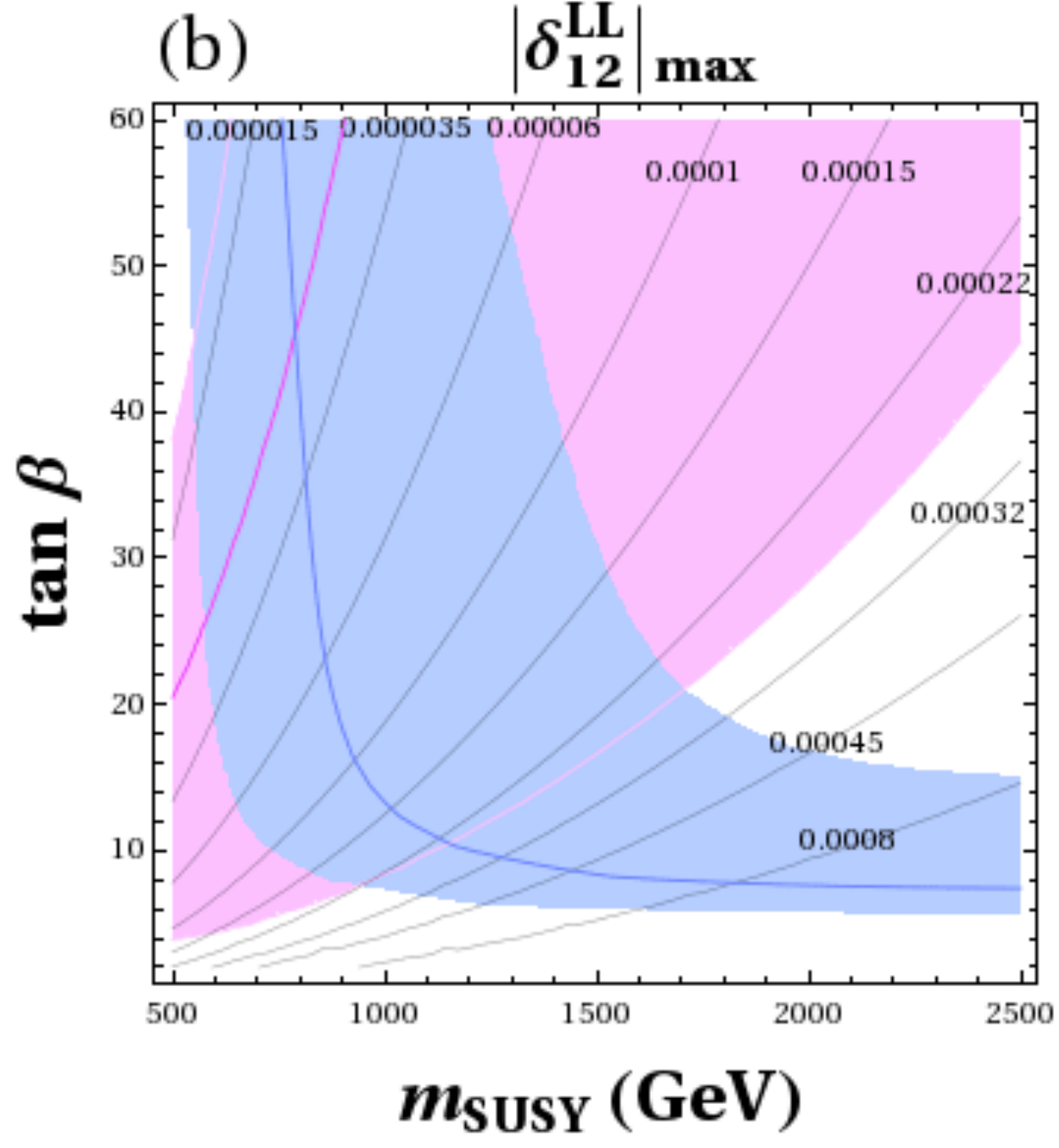,scale=0.60,clip=}\\
\psfig{file=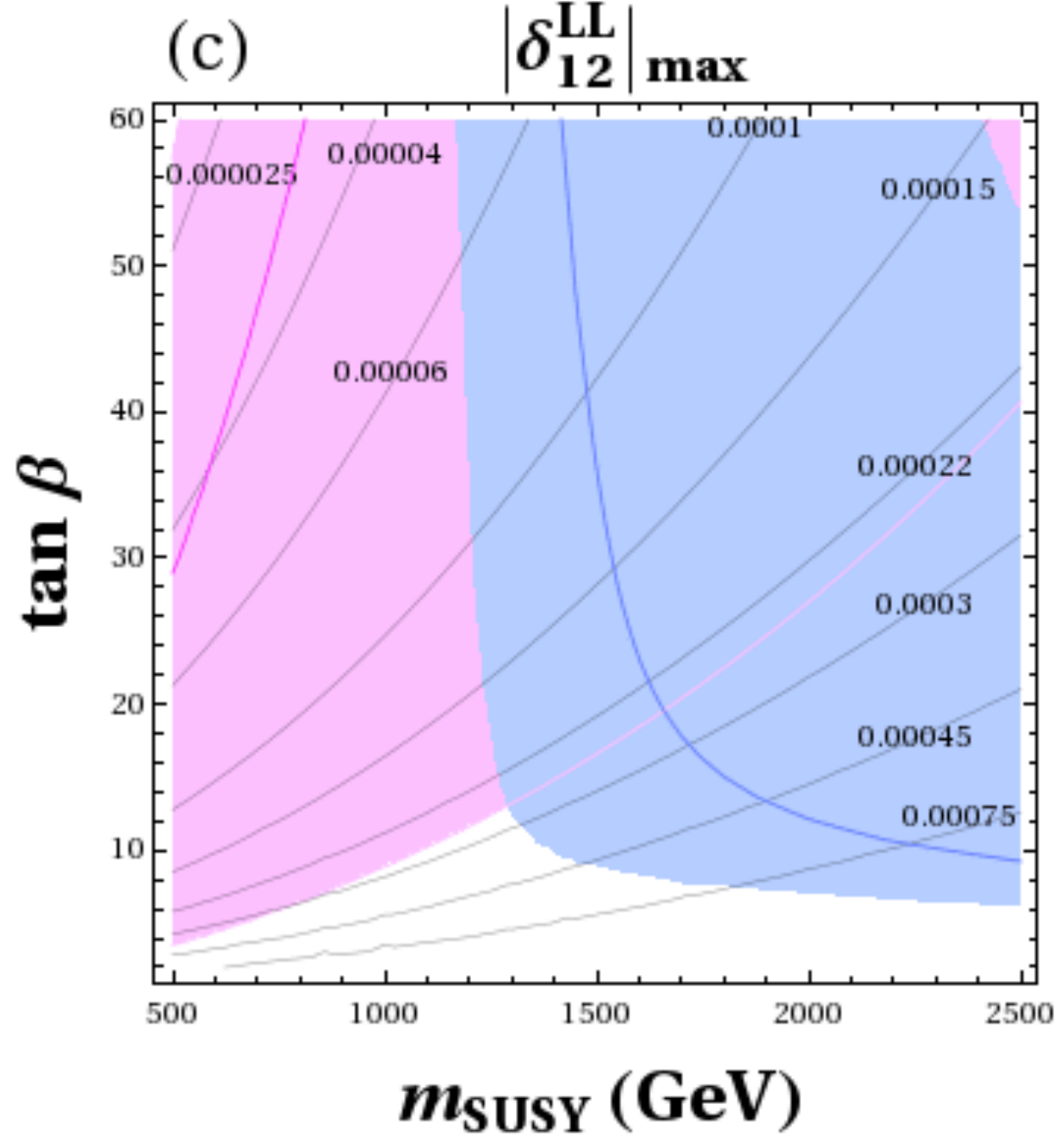,scale=0.60,clip=}
\psfig{file=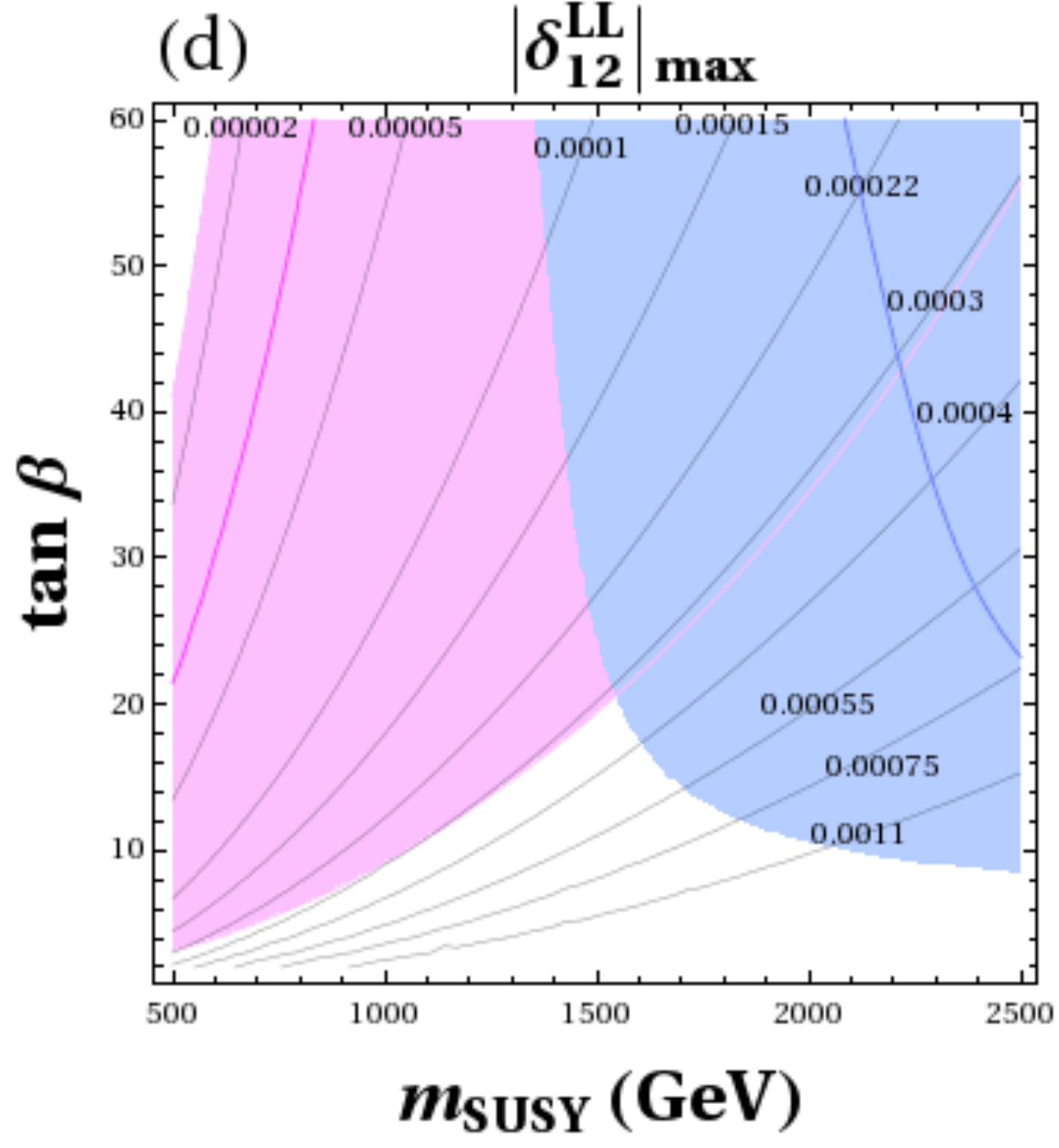,scale=0.60,clip=}
\end{center}
\caption{ Contourlines in the 
($m_{\rm SUSY}$, $\tb$) plane of maximum slepton mixing
  $|\delta_{12}^{LL}|_{\rm max}$ that are allowed by LFV searches in
  $\mu \to e \gamma$ for scenarios {\bf (a)}, {\bf (b)}, {\bf (c)} and
  {\bf (d)} of the pMSSM-4 type, defined in Section \ref{sec:f2}. The shaded
  areas in pink are the regions leading to a $(g-2)_\mu^{\rm SUSY}$
  prediction in the $(3.2,57.2) \times 10^{-10}$ interval. The interior
  pink contourline (without number) corresponds to setting $(g-2)_\mu^{\rm SUSY}$ exactly
  at the central value of the discrepancy $(g-2)_\mu^{\rm
    exp}-(g-2)_\mu^{\rm SM}=30.2 \times 10^{-10}$ . 
The shaded overimposed areas in blue are the regions leading to a
$\Mh$ prediction within the  $(123,127)\gev$ interval. The
interior blue contourline (without number) corresponds to the particular 
$\Mh=125\gev$ value.}
\label{msusytb-LL12}
\vspace{1em}
\end{figure}

\begin{figure}[ht!]
\vspace{4em}
\begin{center}
\psfig{file=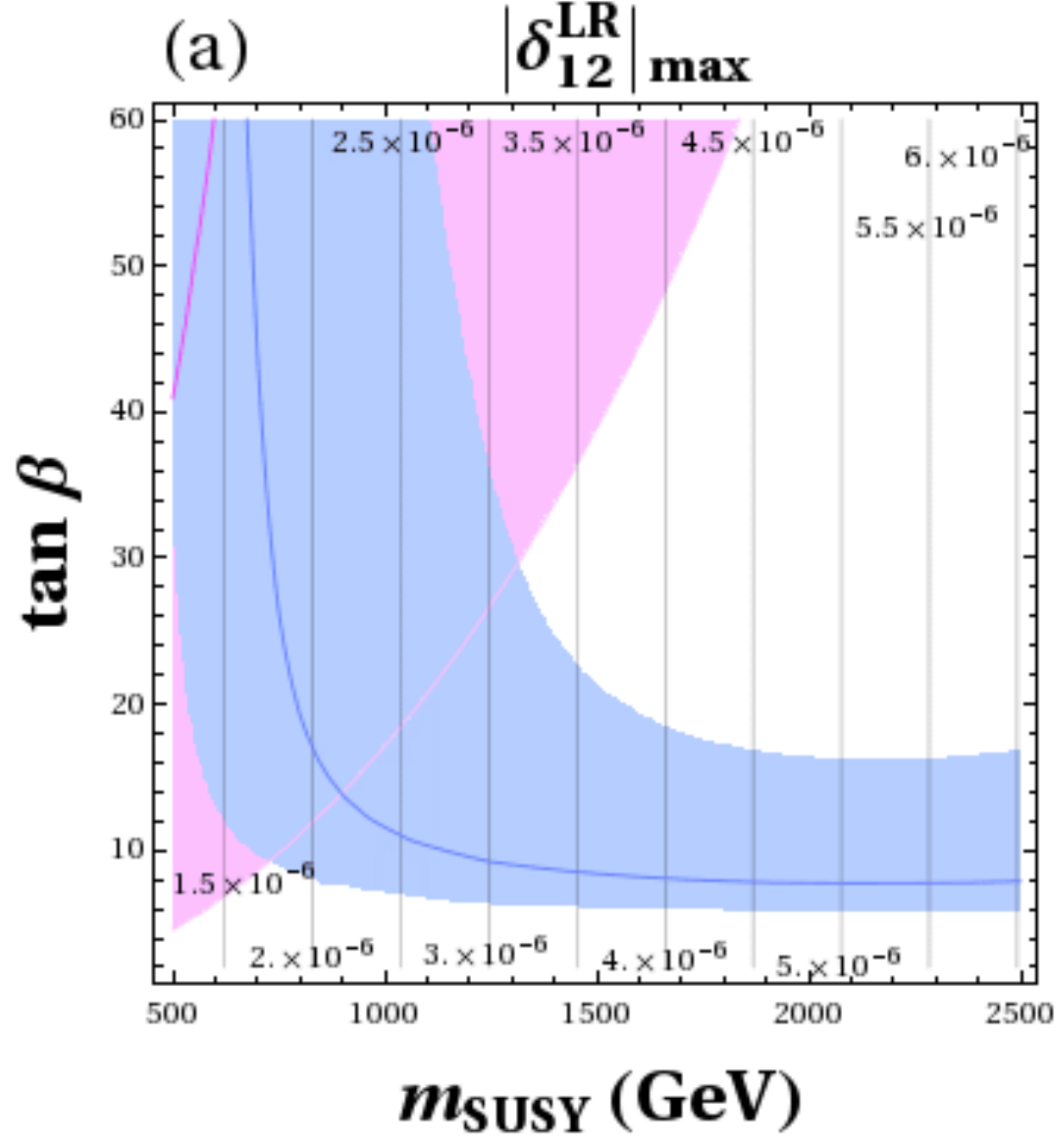,scale=0.60,clip=}
\psfig{file=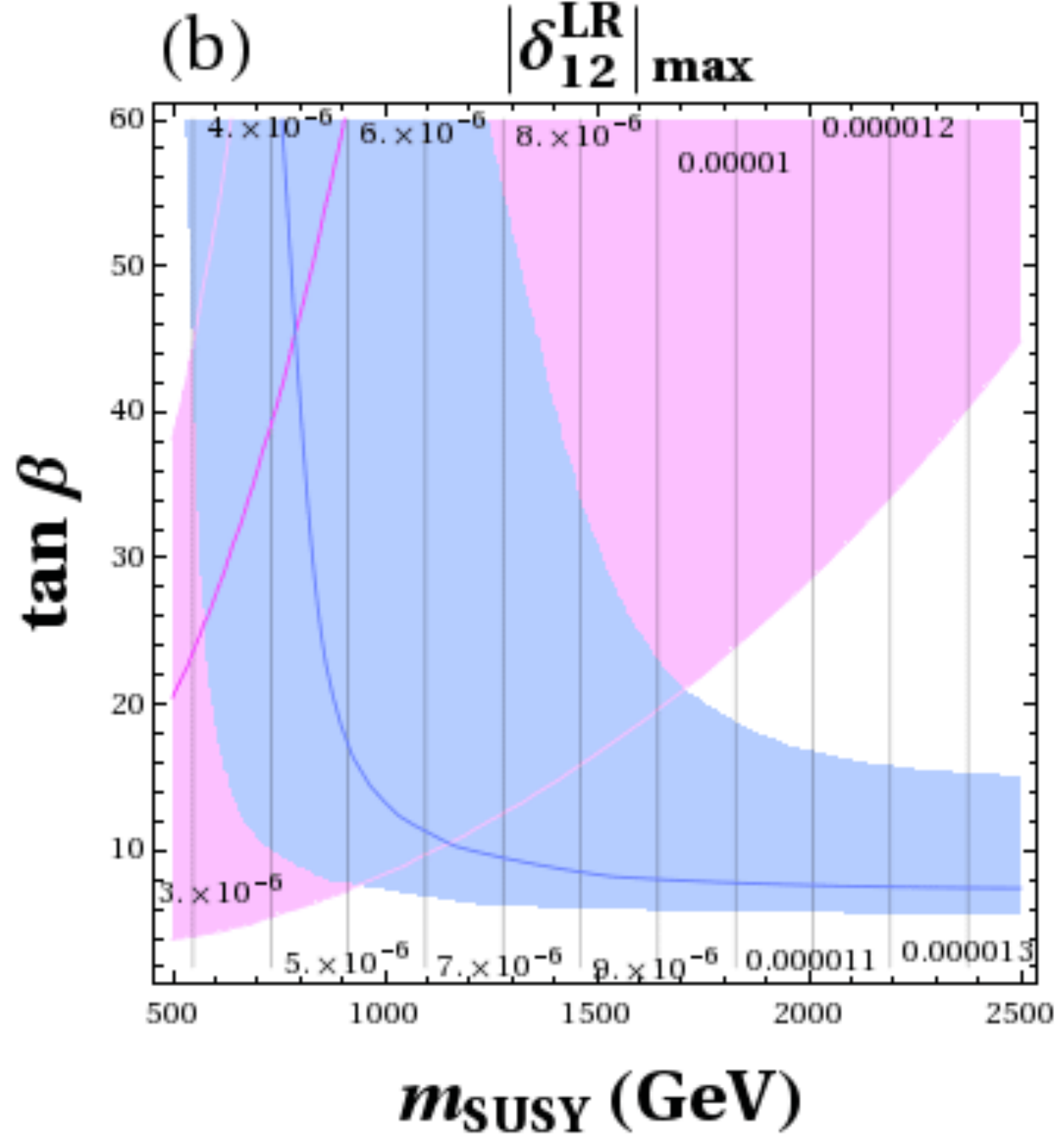,scale=0.60,clip=}\\
\psfig{file=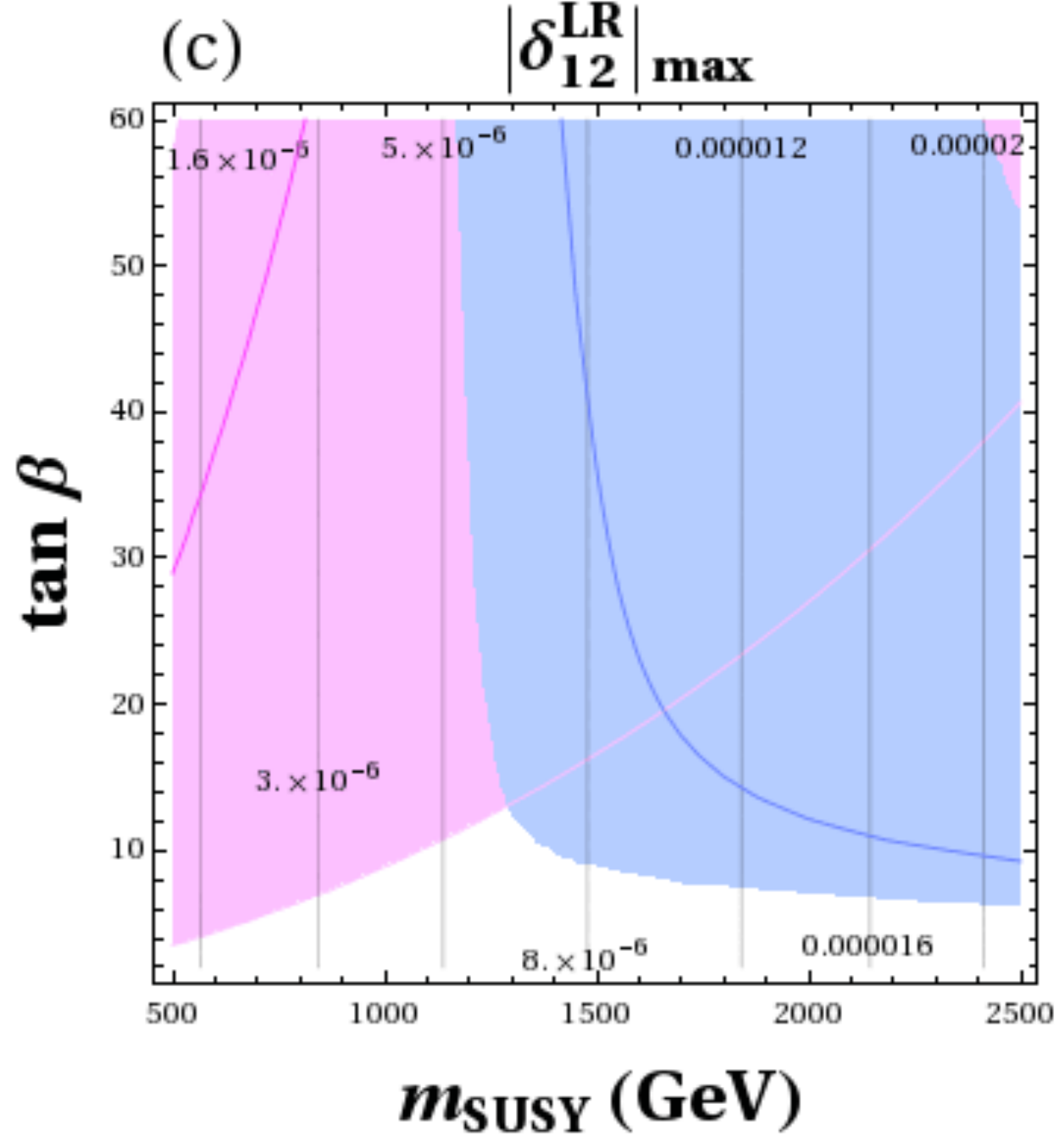,scale=0.60,clip=}
\psfig{file=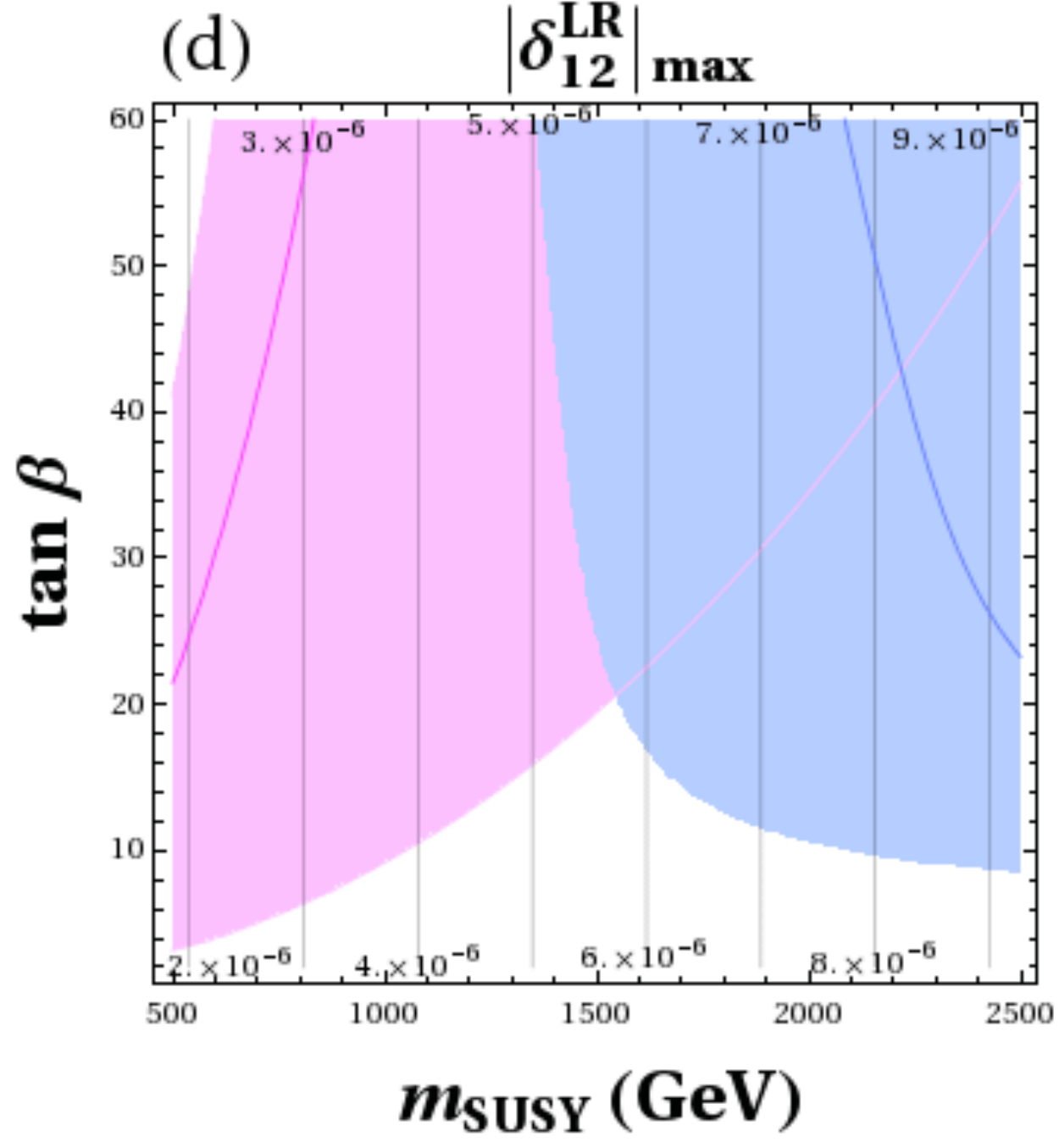,scale=0.60,clip=}
\end{center}
\caption{ Contourlines in the 
($m_{\rm SUSY}$, $\tb$) plane of maximum slepton mixing
  $|\delta_{12}^{LR}|_{\rm max}$ that are allowed by LFV searches in
  $\mu \to e \gamma$. All inputs and explanations are as in
  \reffi{msusytb-LL12}.} 
\label{msusytb-LR12}
\vspace{4em}
\end{figure}

\begin{figure}[ht!]
\vspace{4em}
\begin{center}
\psfig{file=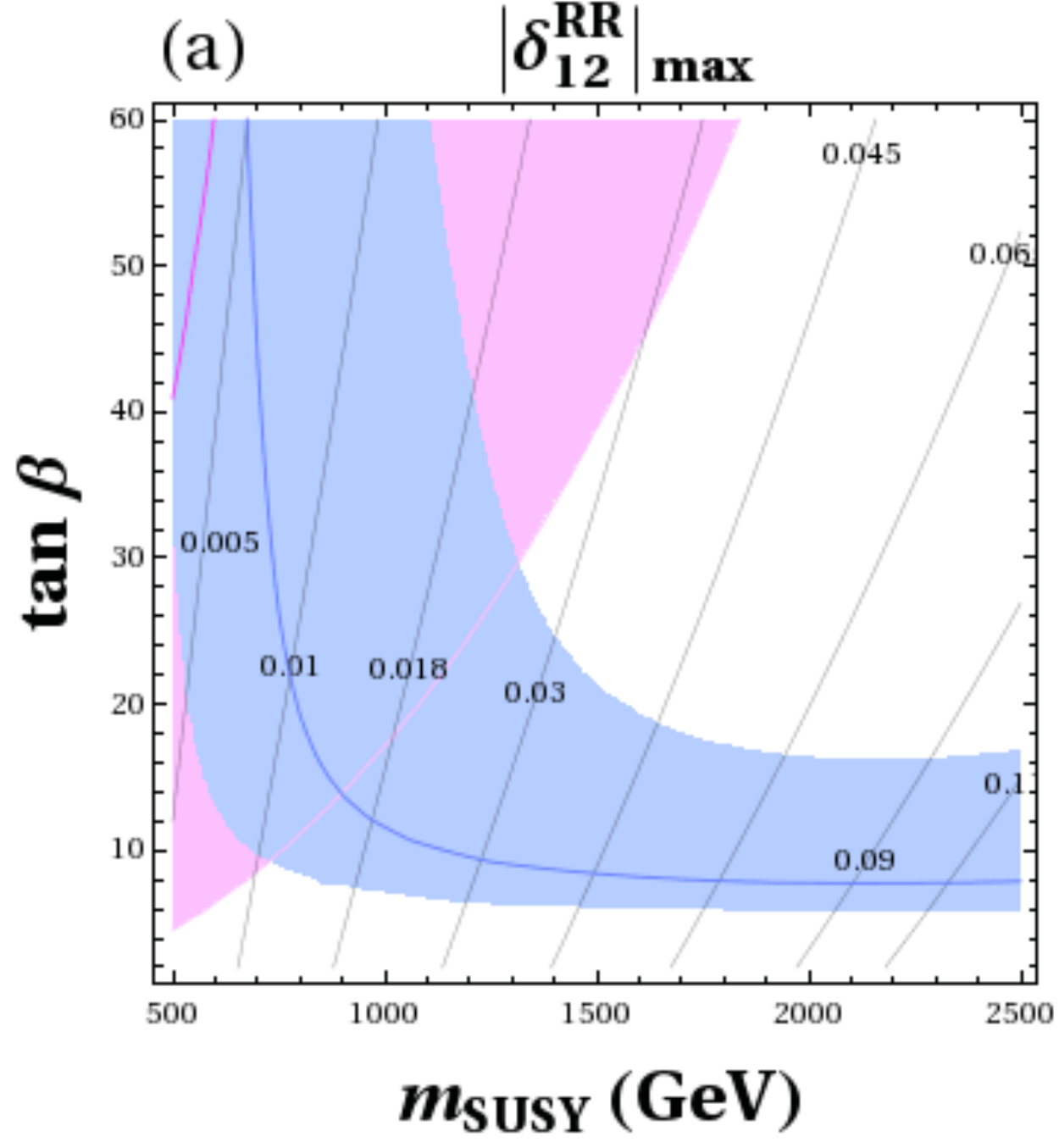,scale=0.60,clip=}
\psfig{file=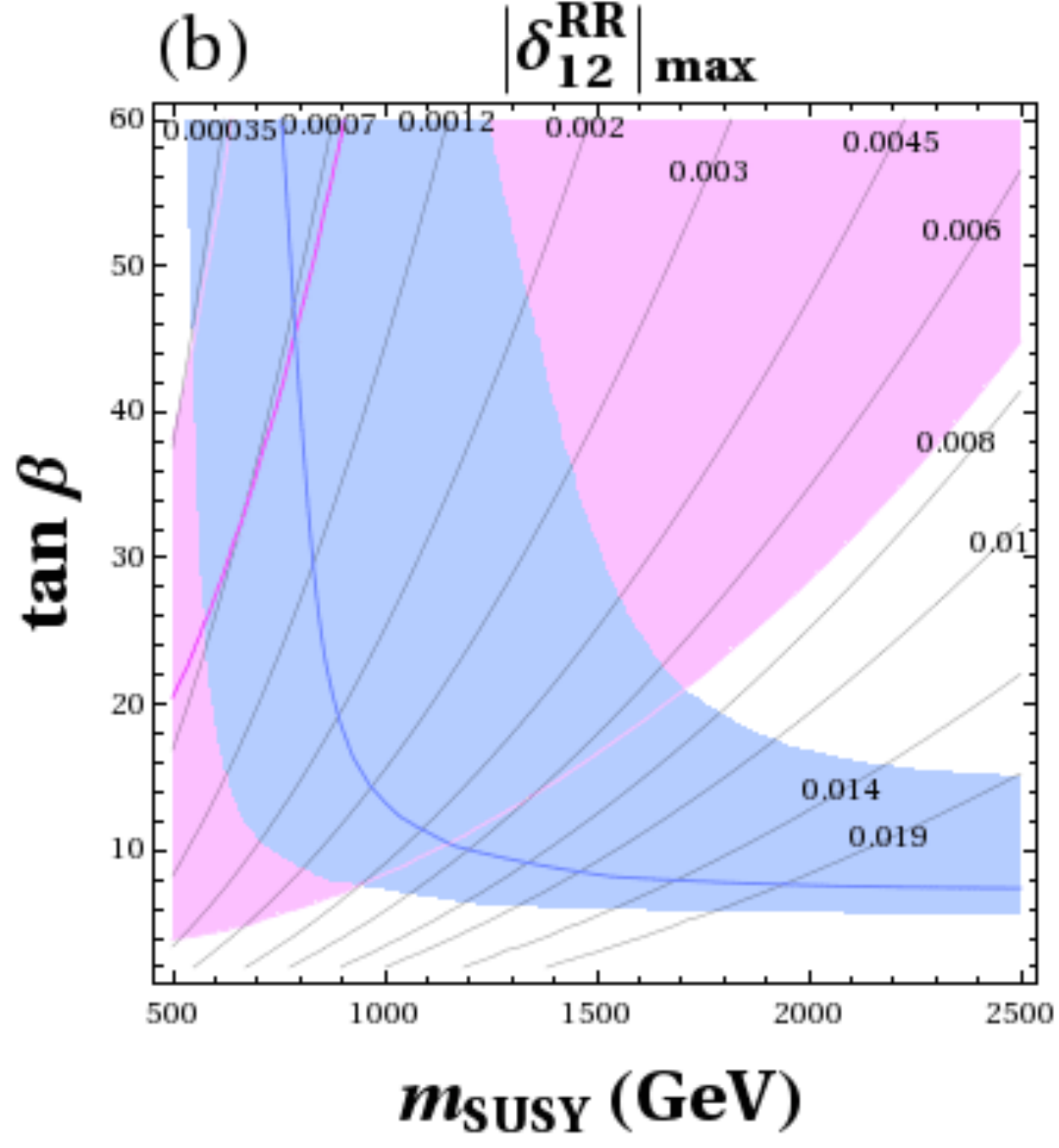,scale=0.60,clip=}\\
\psfig{file=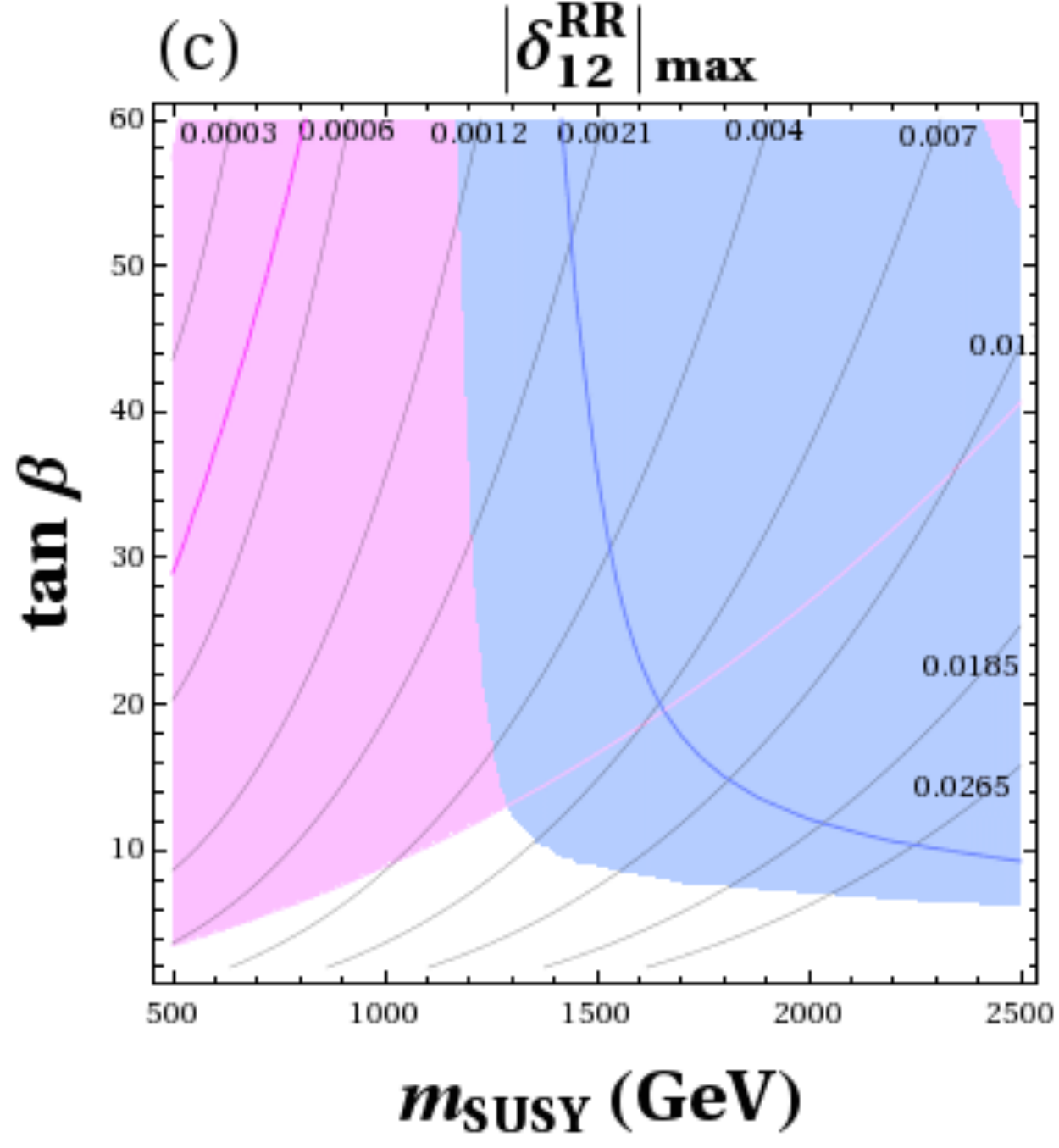,scale=0.60,clip=}
\psfig{file=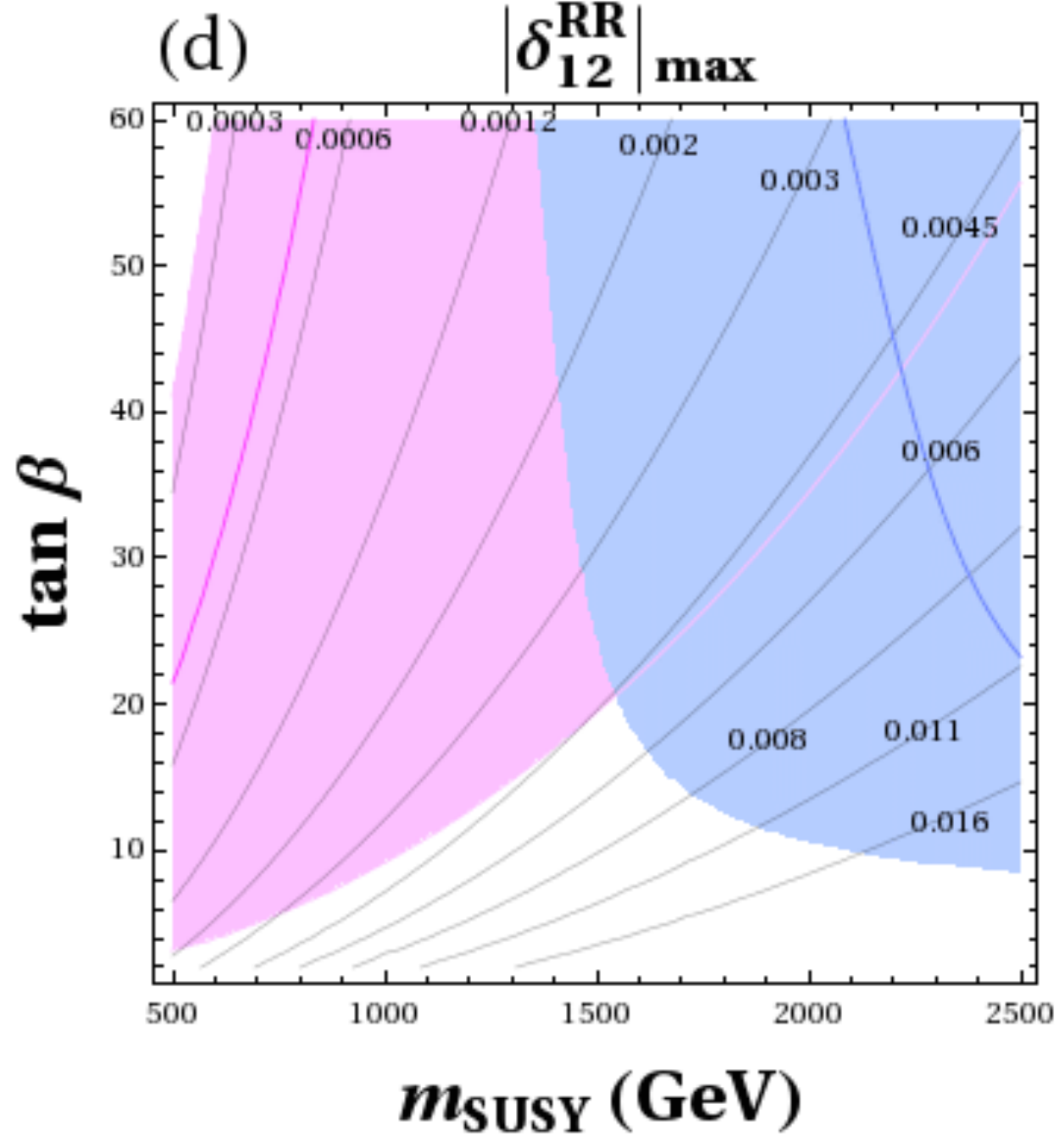 ,scale=0.60,clip=}
\end{center}
\caption{ Contourlines in the 
($m_{\rm SUSY}$, $\tb$) plane of maximum slepton mixing
  $|\delta_{12}^{RR}|_{\rm max}$ that are allowed by LFV searches in
  $\mu \to e \gamma$. All inputs and explanations are as in
  \reffi{msusytb-LL12}.} 
\label{msusytb-RR12}
\vspace{4em}
\end{figure}

\begin{figure}[ht!]
\vspace{3em}
\begin{center}
\psfig{file=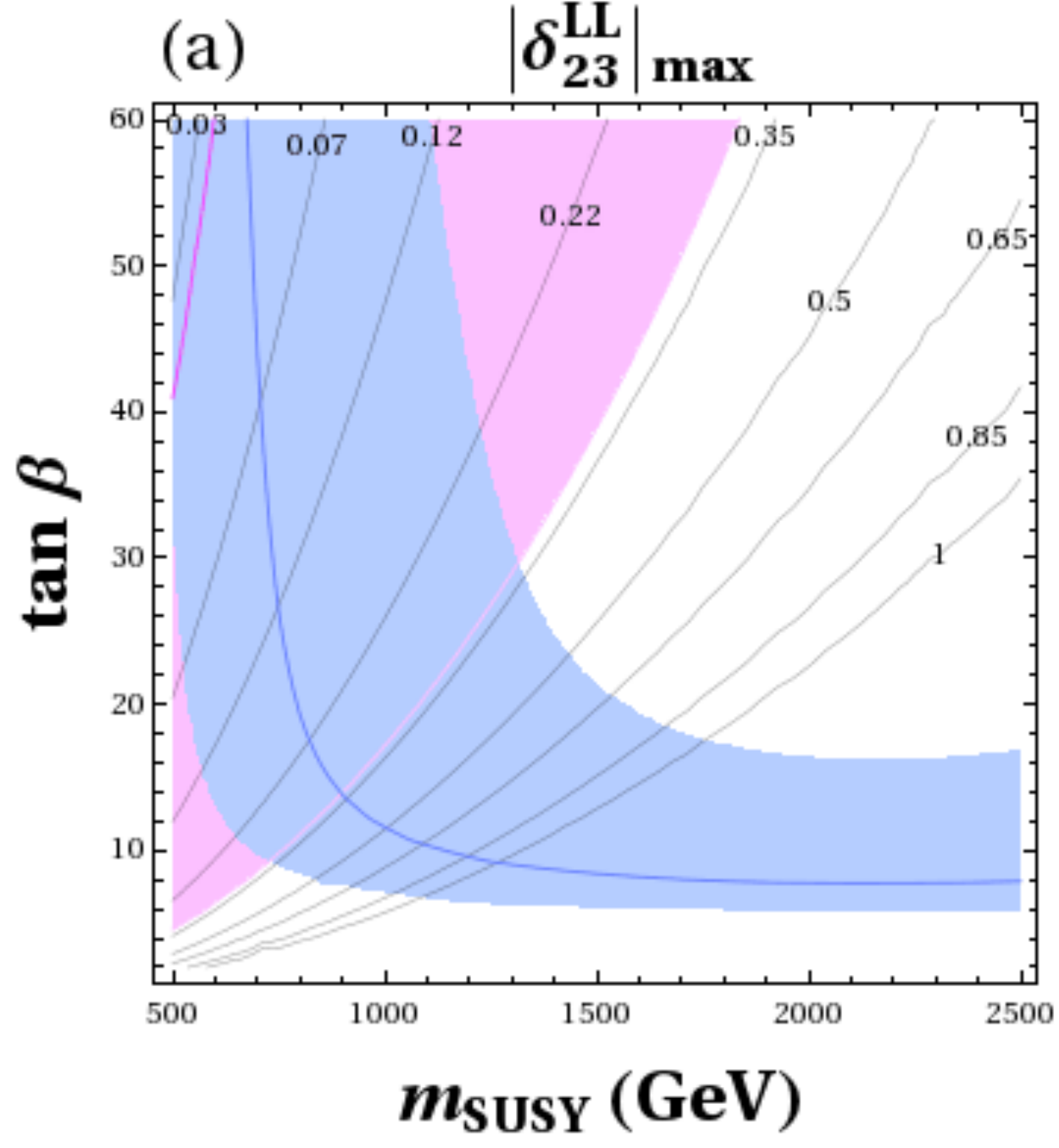,scale=0.60,clip=}
\psfig{file=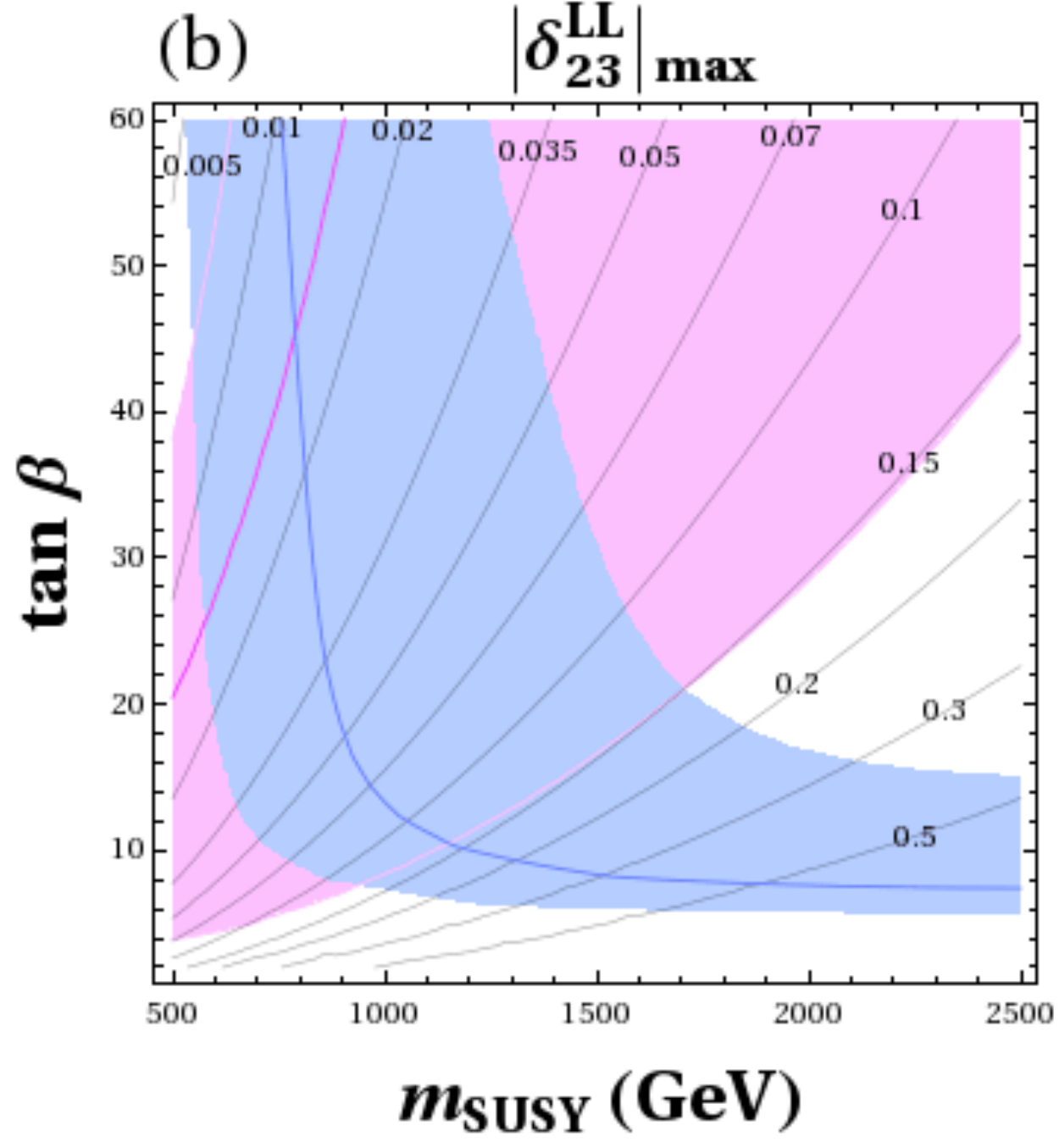,scale=0.60,clip=}\\
\psfig{file=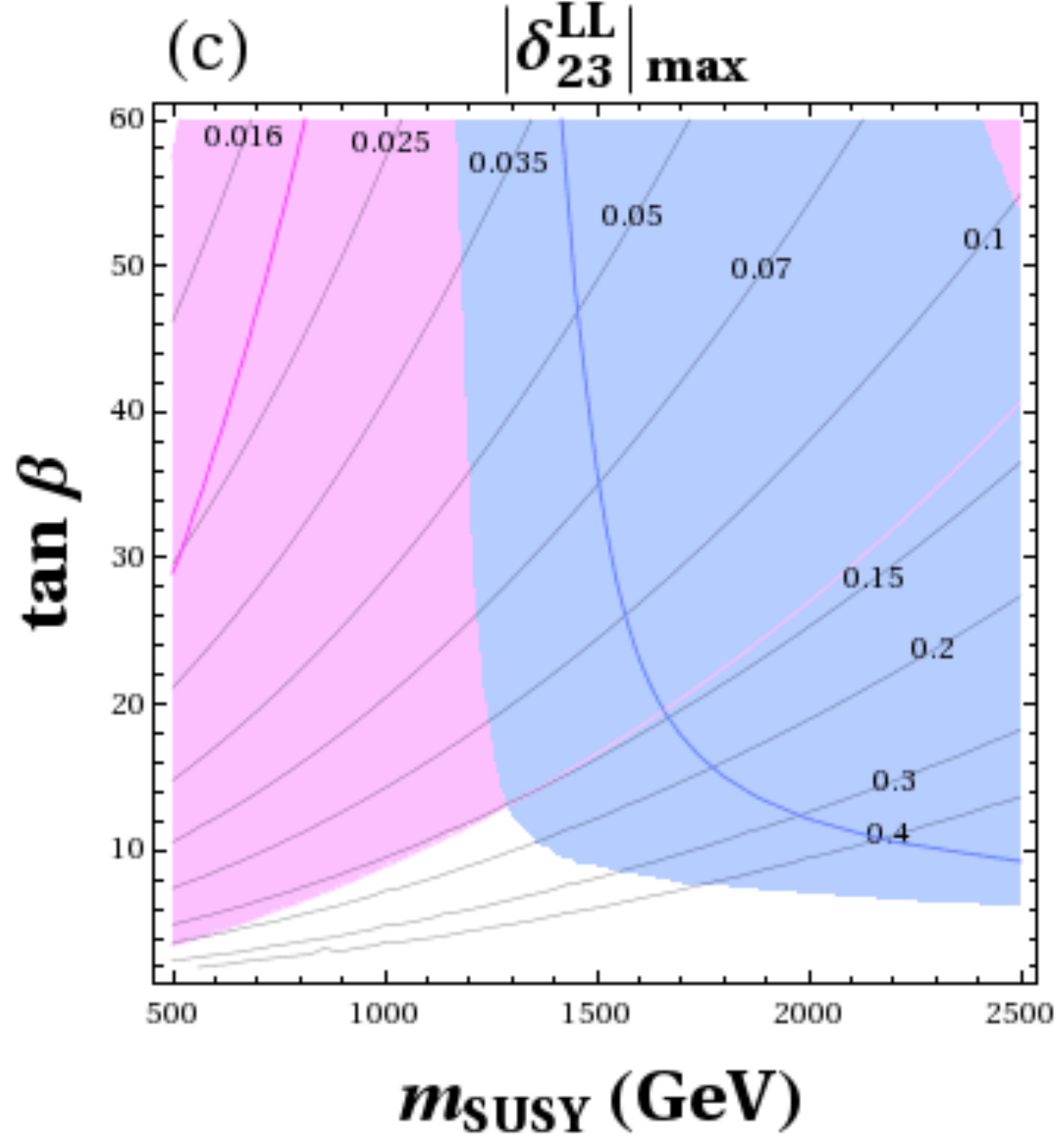,scale=0.60,clip=}
\psfig{file=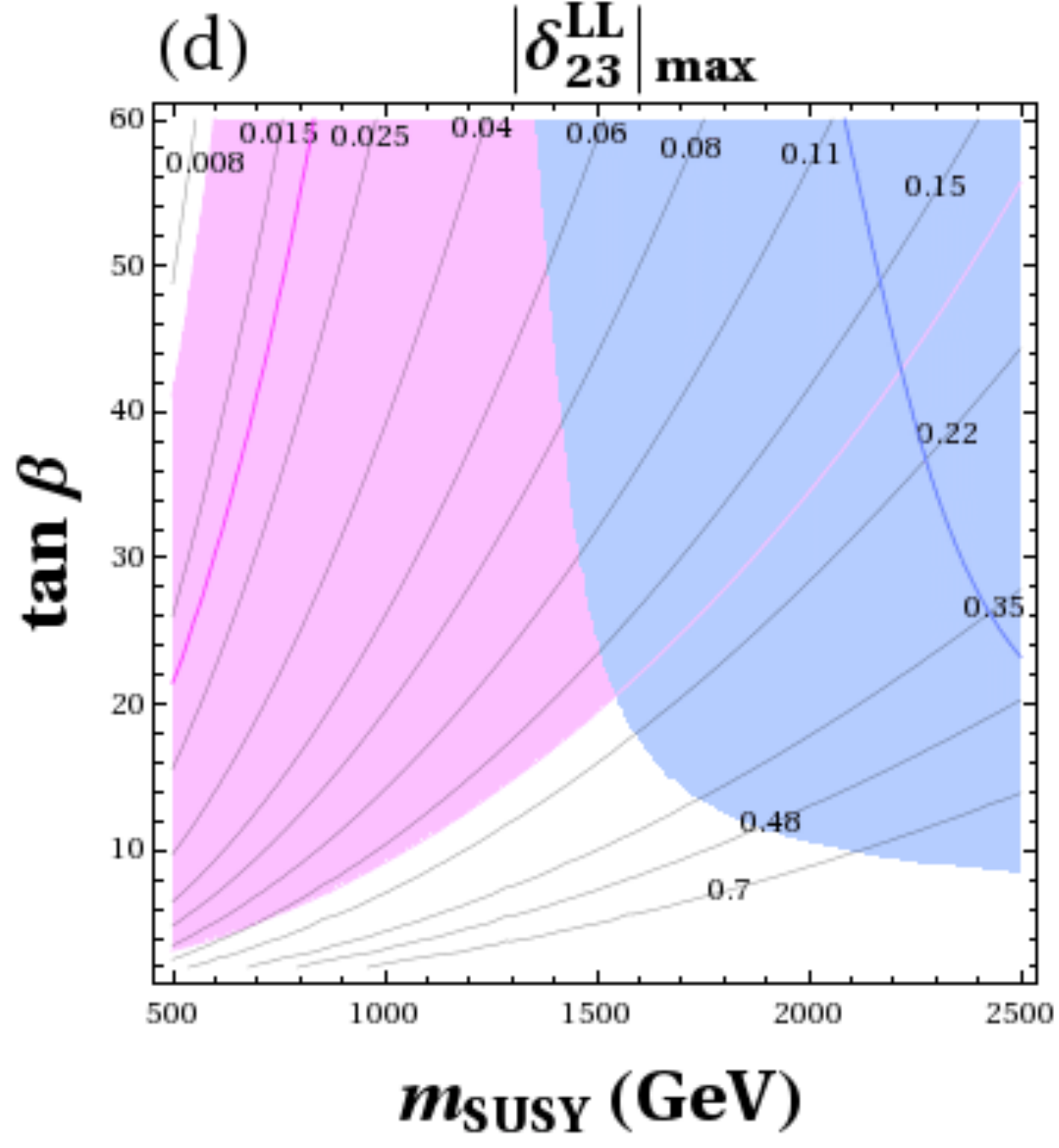 ,scale=0.60,clip=}
\end{center}
\caption{ Contourlines in the 
($m_{\rm SUSY}$, $\tb$) plane of maximum slepton mixing
  $|\delta_{23}^{LL}|_{\rm max}$ that are allowed by LFV searches in
  $\tau \to \mu \gamma$. All inputs and explanations are as in
  \reffi{msusytb-LL12}. Similar results/plots (not shown) are
    obtained for
  contourlines of maximum slepton mixing  $|\delta_{13}^{LL}|_{\rm max}$
  that are allowed by LFV searches in $\tau \to e \gamma$.} 
\label{msusytb-LL23}
\vspace{3em}
\end{figure}

\begin{figure}[ht!]
\vspace{3em}
\begin{center}
\psfig{file=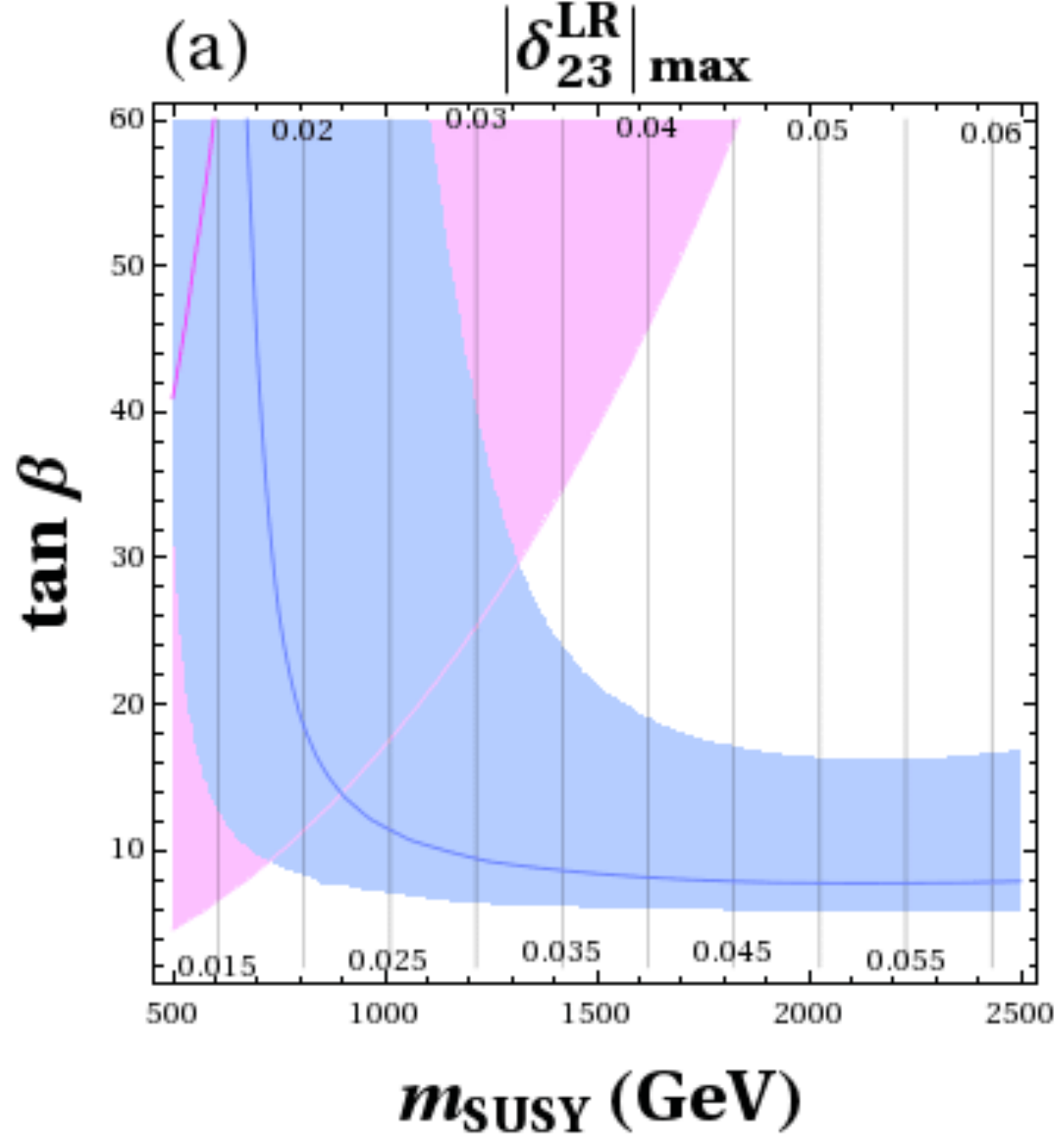,scale=0.60,clip=}
\psfig{file=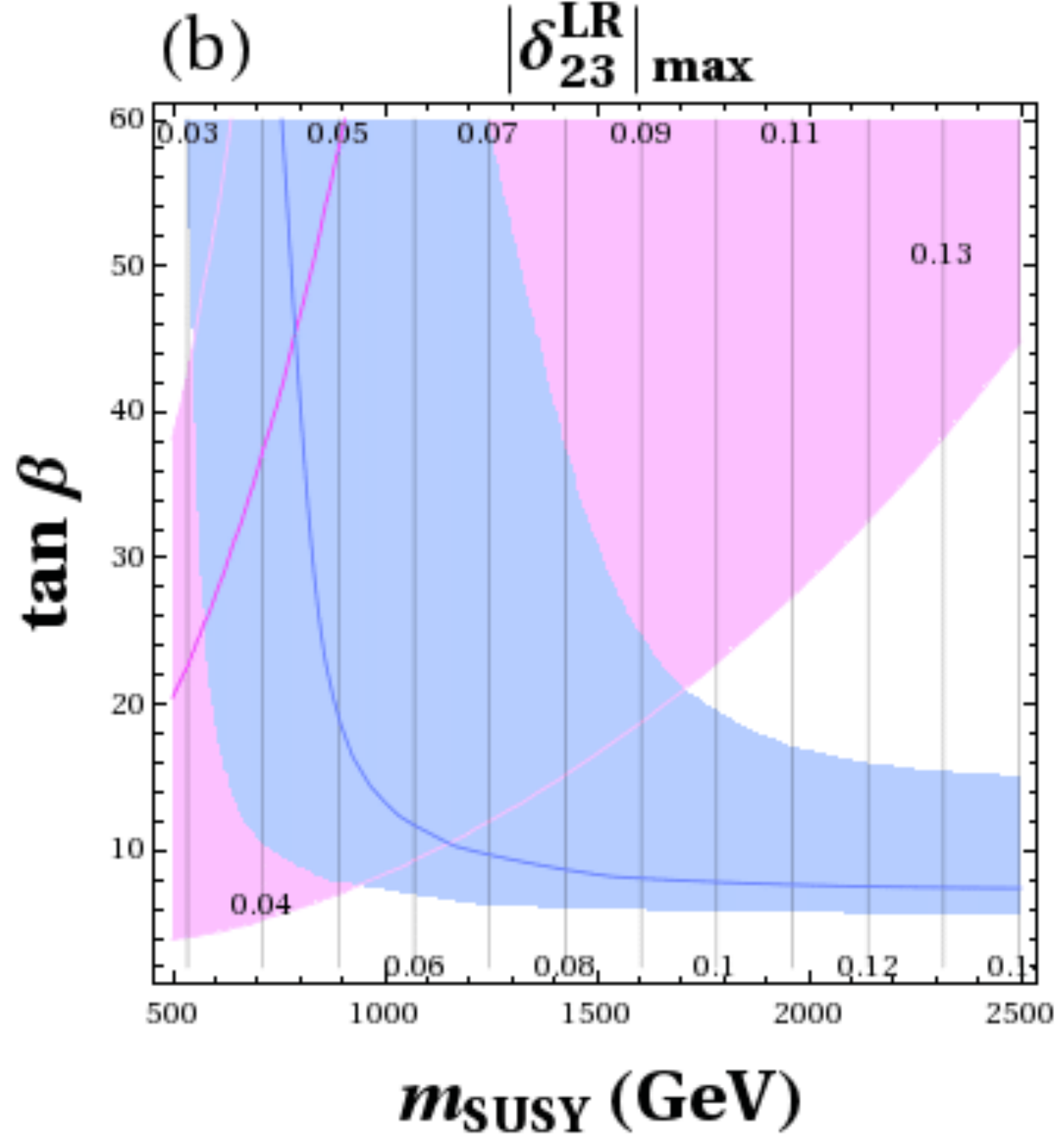,scale=0.60,clip=}\\
\psfig{file=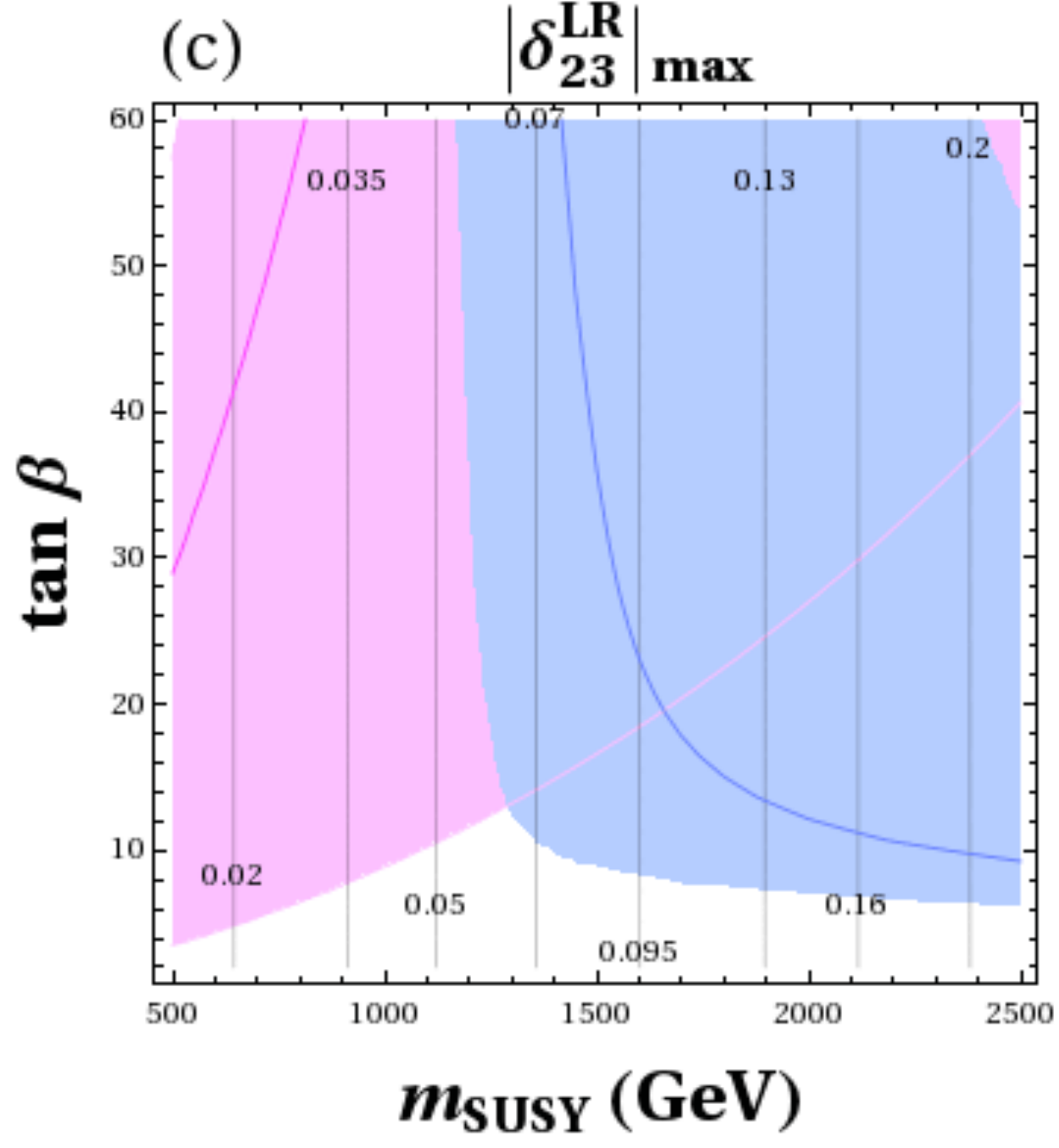,scale=0.60,clip=}
\psfig{file=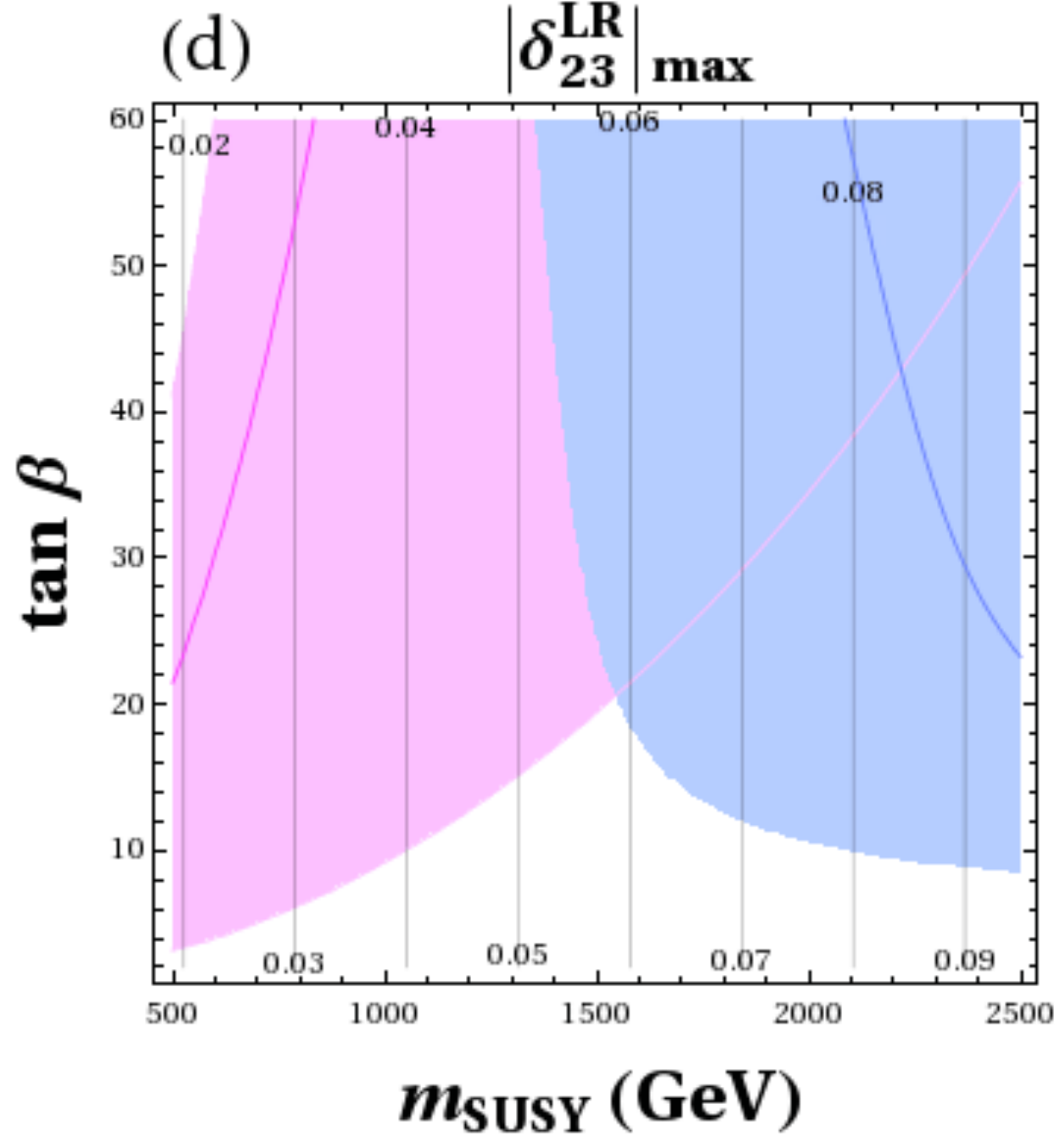,scale=0.60,clip=}
\end{center}
\caption{ Contourlines in the 
($m_{\rm SUSY}$, $\tb$) plane of maximum slepton mixing
  $|\delta_{23}^{LR}|_{\rm max}$ that are allowed by LFV searches in
  $\tau \to \mu \gamma$. All inputs and explanations are as in
  \reffi{msusytb-LL12}. Similar results/plots (not shown) are
    obtained for
  contourlines of maximum slepton mixing  $|\delta_{13}^{LR}|_{\rm max}$
  that are allowed by LFV searches in $\tau \to e \gamma$.} 
\label{msusytb-LR23}
\vspace{3em}
\end{figure}

\begin{figure}[ht!]
\vspace{3em}
\begin{center}
\psfig{file=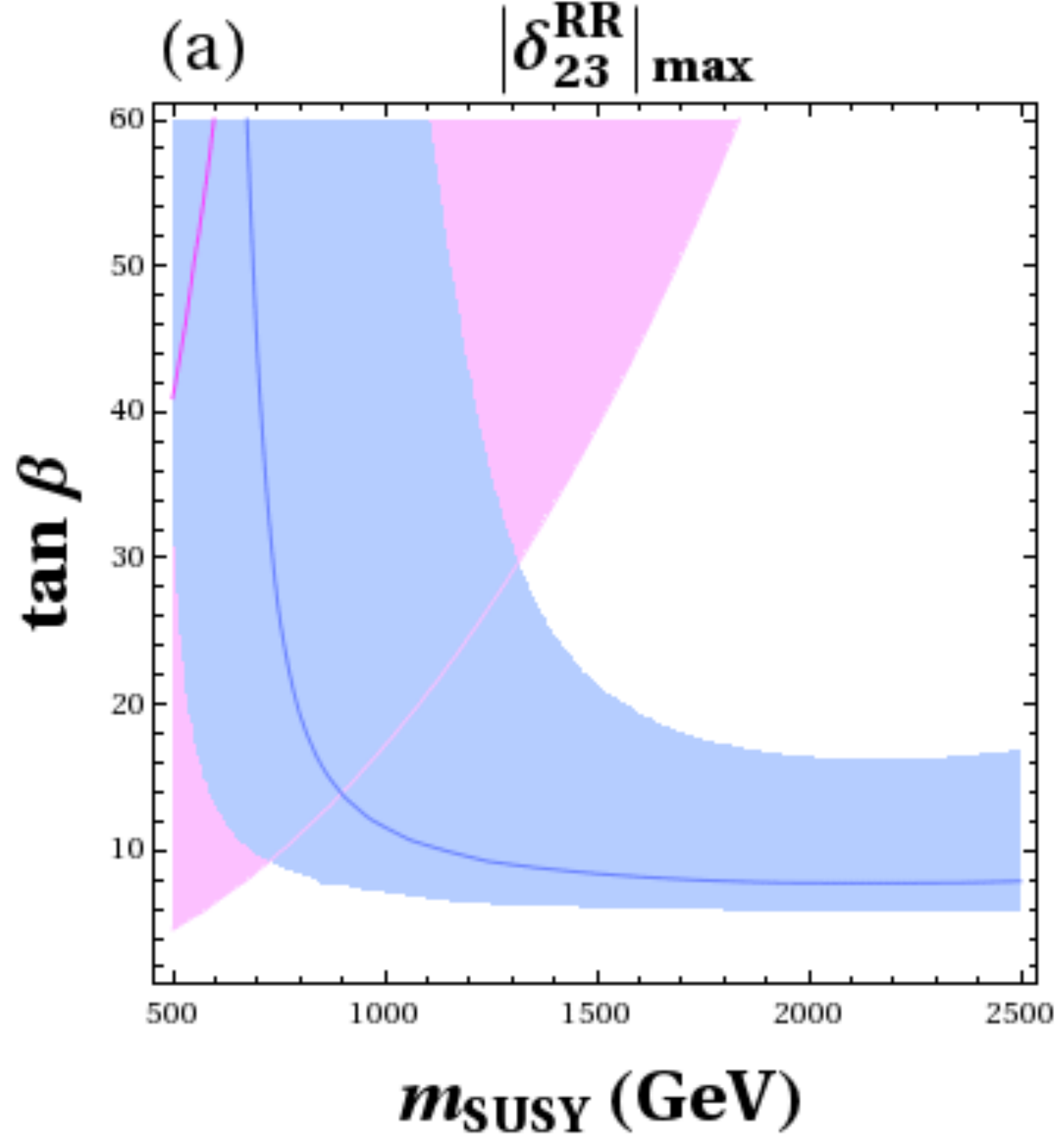,scale=0.60,clip=}
\psfig{file=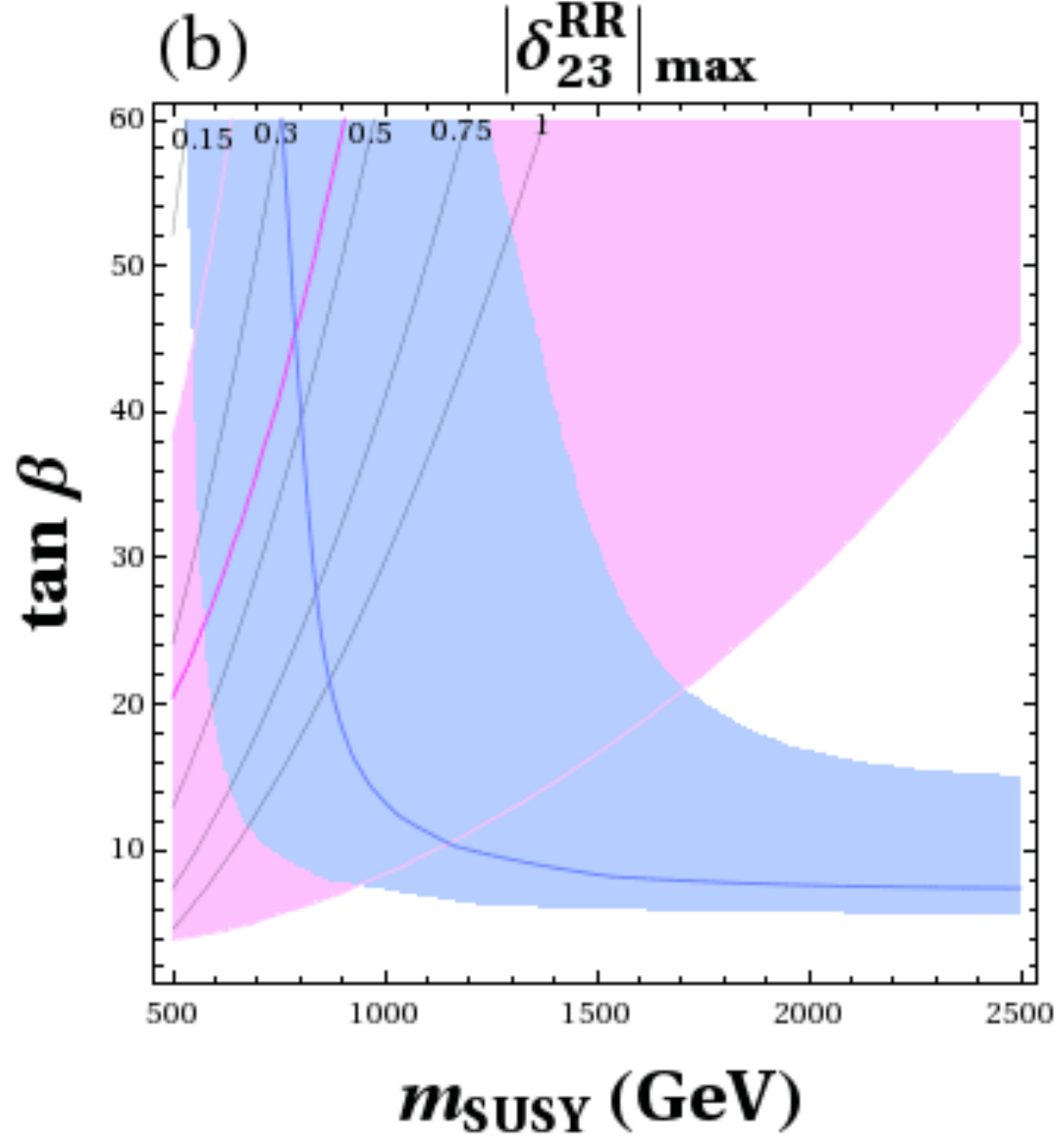,scale=0.60,clip=}\\
\psfig{file=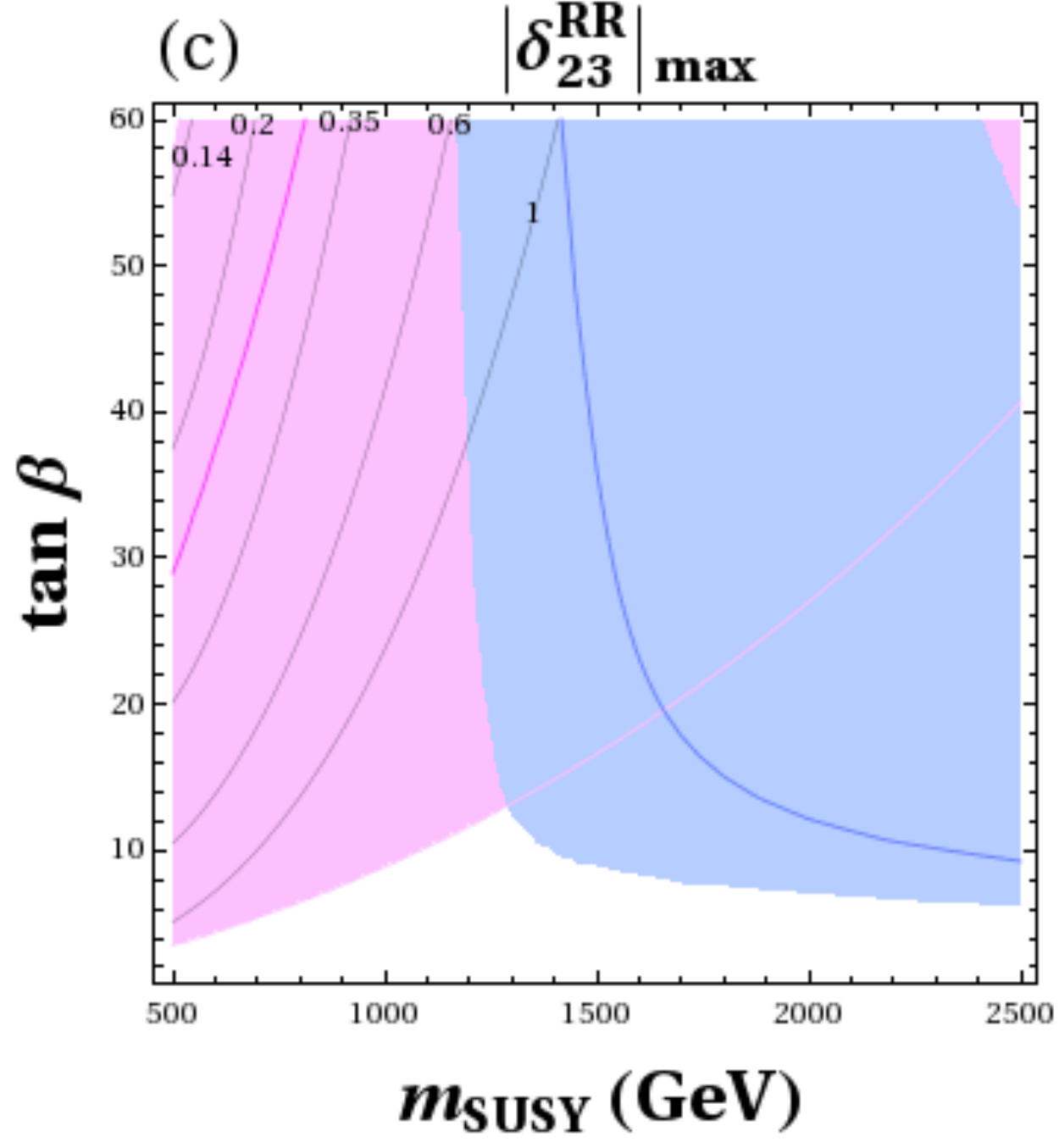,scale=0.60,clip=}
\psfig{file=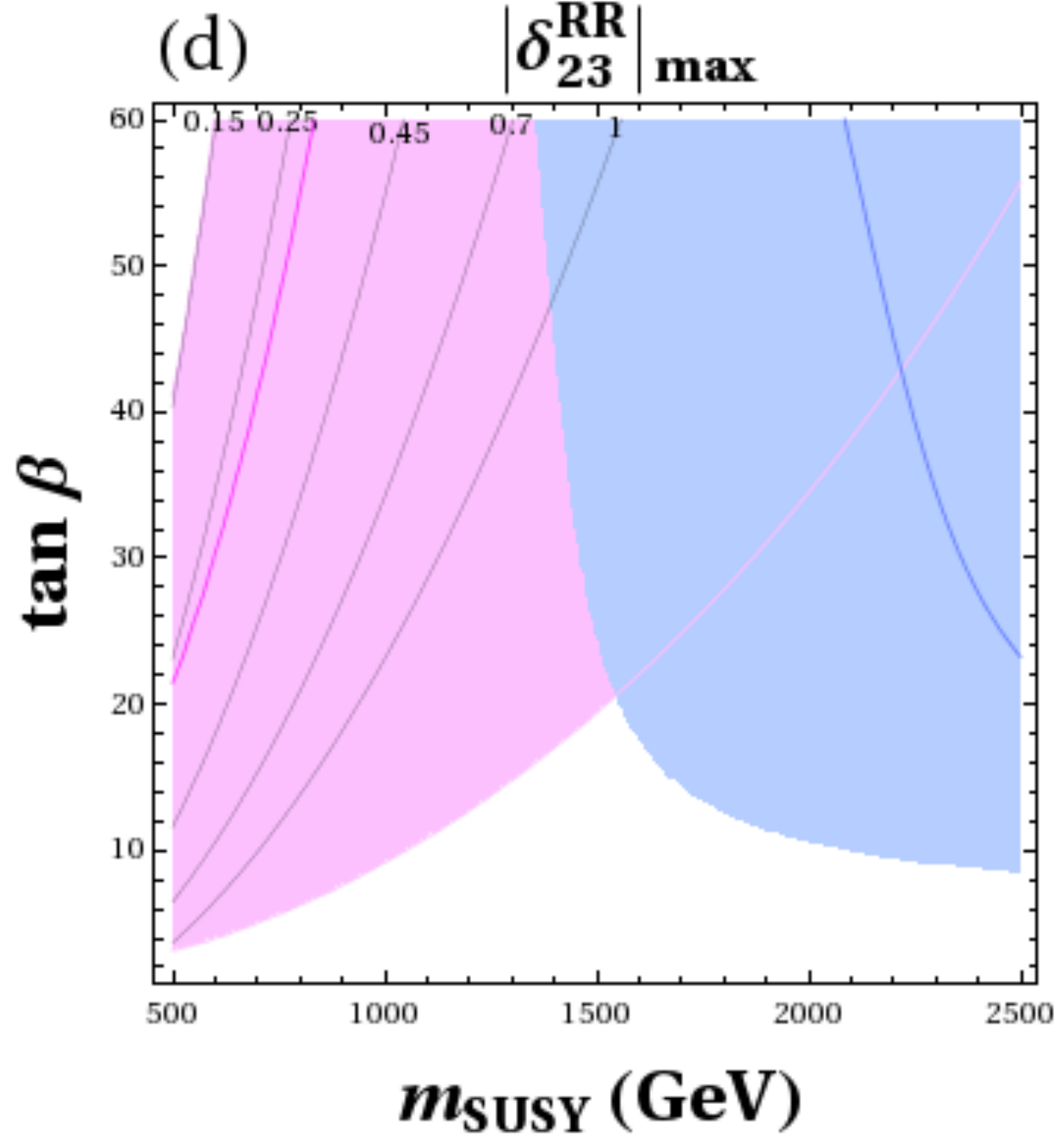,scale=0.60,clip=}
\end{center}
\caption{ Contourlines in the 
($m_{\rm SUSY}$, $\tb$) plane of maximum slepton mixing
  $|\delta_{23}^{RR}|_{\rm max}$ that are allowed by LFV searches in
  $\tau \to \mu \gamma$. All inputs and explanations are as in
  \reffi{msusytb-LL12}. Notice that only contourlines with $|\delta_{23}^{RR}|_{\rm max} \leq 1$ are included. In the scenario {\bf (a)} 
the countourlines with $|\delta_{23}^{RR}|_{\rm max} \leq 1$ are out of the region in the parameter space shown in this figure. Similar 
results/plots (not shown) are obtained for
  contourlines of maximum slepton mixing  $|\delta_{13}^{RR}|_{\rm max}$
  that are allowed by LFV searches in $\tau \to e \gamma$.} 
\label{msusytb-RR23}
\end{figure}

We present the numerical results of our analysis in the pMSSM-4 scenarios that are
shown in \reffis{msusytb-LL12} through 
\ref{msusytb-RR23}. Figures \ref{msusytb-LL12}, \ref{msusytb-LR12} and
\ref{msusytb-RR12} show the bounds for the slepton mixing of 12-type as
extracted from present $\mu \to e \gamma$ searches. Figures
\ref{msusytb-LL23}, \ref{msusytb-LR23} and \ref{msusytb-RR23} show the
bounds for the slepton mixing of 23-type as extracted from present $\tau
\to \mu \gamma$ searches. It should be noted that the bounds for
the slepton mixings of 13-type (not shown here) are equal (in the MIA)
to those of 23-type.  
In each plot we show the resulting contourlines in the 
($m_{\rm SUSY}$, $\tb$) plane of maximum allowed slepton mixing. 
In addition we also show in each plot the areas in the
pMSSM-4 parameter space for that particular scenario that lead 
to values of the lightest Higgs boson mass compatible with LHC data, and
at the same time to predictions of the muon anomalous magnetic
moment also compatible with 
data. As in the set of pMSSM points of the previous section, we use here again
\fh~\cite{feynhiggs,mhiggslong,mhiggsAEC,mhcMSSMlong} to evaluate 
$\Mh$ and {\it SPHENO}~\cite{SPheno} to evaluate  
$(g-2)_\mu$ (where \fh\ gives very similar results). 
The shaded areas in pink are the regions leading to a
$(g-2)_\mu^{\rm SUSY}$ prediction, from the SUSY one-loop contributions,
in the allowed by data $(3.2,57.2) \times 10^{-10}$ interval. The
interior pink contourline corresponds to setting $(g-2)_\mu^{\rm SUSY}$
exactly at the central value of the discrepancy $(g-2)_\mu^{\rm
  exp}-(g-2)_\mu^{\rm SM}=30.2 \times 10^{-10}$. 
The shaded overimposed areas in blue are the regions leading to a $\Mh$
prediction within the  $(123,127)\gev$ interval. The interior blue
contourline corresponds to the particular $\Mh=125\gev$ value.  

From these plots in the ($m_{\rm SUSY}$, $\tb$) plane one can draw
the following conclusions:
\begin{itemize}
\item[1.-]  For each scenario (a), (b), (c) and (d) one can derive the
  corresponding upper bound for each $|\delta^{AB}_{ij}|$ at a given
  ($m_{\rm SUSY}$, $\tb$) point in this plane.  
\item[2.-] The maximal allowed values of the $\delta^{LL}_{ij}$'s and
  $\delta^{RR}_{ij}$'s scale with $m_{\rm SUSY}$ and $\tb$ approximately
  as expected, growing with increasing $m_{\rm SUSY}$ as $\sim m_{\rm
    SUSY}^2$ and decreasing with increasing (large) $\tb$ as 
    $\sim 1/\tb$. The maximal allowed values of the $\delta^{LR}_{ij}$'s
  (and similarly $\delta^{RL}_{ij}$'s) are independent on $\tb$ and grow
  approximately as $\sim m_{\rm SUSY}$  with increasing $m_{\rm
    SUSY}$. This is in agreement with the qualitative behaviour 
found in the approximation formulas, 
  \refeqs{MIA-L} and (\ref{MIA-R}) of the MIA results in the simplest case
  of only one mass scale, $m_S$.  
\item[3.-] The intersections between the allowed areas by the required
  $(g-2)_\mu$ and $\Mh$ intervals move from the left side, $m_{\rm
  SUSY}\sim 500-1300 \gev$ to the right side of the plots,
  $m_{\rm SUSY}\sim 1300-2500 \gev$ from scenarios (a) through
  (d). This is clearly the consequence of the fact that
$(g-2)_\mu$ requires a rather light SUSY-EW sector, i.e.\
light charginos, neutralinos and sleptons, and a rather large $\tb$, and
that $\Mh$ requires a rather heavy SUSY squark sector.
Remember that here we are using a common reference SUSY scale  $m_{\rm SUSY}$,
relating all 
the SUSY sparticle masses, both in the SUSY-EW and SUSY-QCD sectors, 
leading to this ``tension''. (A more loose connection between these
two sectors would yield a more relaxed combination of the $(g-2)_\mu$
and $\Mh$ experimental results.)
In fact, in our plots one can observe that the particular
contourlines 
for the ``preferred'' values of $(g-2)_\mu$ and $\Mh$ by data (i.e.\ the
interior blue and pink contourlines) only cross in scenario (b) at
$m_{\rm SUSY}$ around $800\gev$ and $\tb \sim 45$ and get close,
although not crossing, in scenario (a) at  $m_{\rm SUSY}\sim 650\gev$
and very large $\tb\sim 60$. 
However, taking the uncertainties into account the overlap regions
  are quite substantial.
\item[4.-] By assuming a favoured region in the ($m_{\rm SUSY}$,$\tb$)
  parameter space given by the intersect of the two $(g-2)_\mu$ (in
  pink) and $\Mh$ (in blue) areas, one can extract improved bounds for
  the slepton mixing deltas valid in these intersects. 
Those bounds give a rough idea of which parameter regions in the
  pMSSM-4 are in better agreement with the experimental data on
  $(g-2)_\mu$ and $\Mh$.
The following
  intervals for the maximum allowed  $|\delta^{AB}_{ij}|$ values can be
  deduced from our plots in these intersecting regions:

Scenario (a): 

$|\delta^{LL}_{12}|_{\rm max} \sim (6,60) \times 10^{-5} $
 
$|\delta^{LR}_{12}|_{\rm max} \sim (1.2,3.2) \times 10^{-6} $

$|\delta^{RR}_{12}|_{\rm max} \sim  (3,25) \times 10^{-3} $

$|\delta^{LL}_{23}|_{\rm max} \sim  (3,35) \times 10^{-2}$

$|\delta^{LR}_{23}|_{\rm max} \sim (1,3.2) \times 10^{-2} $

$|\delta^{RR}_{23}|_{\rm max} \sim (10) \times 10^{-1} $

Scenario (b):

$|\delta^{LL}_{12}|_{\rm max} \sim (1.5,27) \times 10^{-5} $
 
$|\delta^{LR}_{12}|_{\rm max} \sim (3,9.2) \times 10^{-6} $

$|\delta^{RR}_{12}|_{\rm max} \sim  (0.35,7) \times 10^{-3} $

$|\delta^{LL}_{23}|_{\rm max} \sim  (0.7,15) \times 10^{-2}$

$|\delta^{LR}_{23}|_{\rm max} \sim (3,9.5) \times 10^{-2} $

$|\delta^{RR}_{23}|_{\rm max} \sim (2,10) \times 10^{-1} $

Scenario (c):

$|\delta^{LL}_{12}|_{\rm max} \sim (5,22) \times 10^{-5} $
 
$|\delta^{LR}_{12}|_{\rm max} \sim (5,22) \times 10^{-6} $

$|\delta^{RR}_{12}|_{\rm max} \sim  (1.2,10) \times 10^{-3} $

$|\delta^{LL}_{23}|_{\rm max} \sim  (3,15) \times 10^{-2}$

$|\delta^{LR}_{23}|_{\rm max} \sim (5,22) \times 10^{-2} $

$|\delta^{RR}_{23}|_{\rm max} \sim (6,10) \times 10^{-1} $

Scenario (d):

$|\delta^{LL}_{12}|_{\rm max} \sim (10,30) \times 10^{-5} $
 
$|\delta^{LR}_{12}|_{\rm max} \sim (5,9) \times 10^{-6} $

$|\delta^{RR}_{12}|_{\rm max} \sim  (1.2,4) \times 10^{-3} $

$|\delta^{LL}_{23}|_{\rm max} \sim  (5,20) \times 10^{-2}$

$|\delta^{LR}_{23}|_{\rm max} \sim (5,9.5) \times 10^{-2} $

$|\delta^{RR}_{23}|_{\rm max} \sim (7,10) \times 10^{-1} $

\end{itemize}

It should be noted that in the previous upper bounds, the
particular $10 \times 10^{-1}$ value appearing in
$|\delta^{RR}_{23}|_{\rm max}$ really means 1 or larger that 1, since we
have not explored out of the $-1 \leq \delta^{AB}_{ij} \leq 1$
intervals. Particularly, in scenario (a) which has the heaviest
gauginos, we find that all values in the $-1 \leq \delta^{RR}_{23} \leq
1$ interval are allowed by LFV data.  

Finally, one can shortly summarize the previous $|\delta^{AB}_{ij}|_{\rm
  max}$ intervals found from LFV searches, by just  signalling the
typical intervals for each delta, in the favoured by LHC and
$(g-2)_\mu$ data MSSM parameter space region, where the predictions 
in all scenarios lay at:   
$|\delta^{LL}_{12}|_{\rm max} \sim {\cal O} (10^{-5},10^{-4}) $, 
$|\delta^{LR}_{12}|_{\rm max} \sim {\cal O} (10^{-6},10^{-5}) $,
$|\delta^{RR}_{12}|_{\rm max} \sim {\cal O} (10^{-3},10^{-2}) $,
$|\delta^{LL}_{23}|_{\rm max} \sim {\cal O} (10^{-2},10^{-1}) $,
$|\delta^{LR}_{23}|_{\rm max} \sim {\cal O} (10^{-2},10^{-1}) $,
$|\delta^{RR}_{23}|_{\rm max} \sim {\cal O} (10^{-1},10^{0}) $. 
Very similar general bounds as for the 23 mixing are found for the
13 mixing.

\clearpage
\newpage
\chapter{LFV Higgs decays from flavour mixing in the slepton sector}
\label{lfvhiggsdecaysslepton}

In this chapter we complete our study of LFV processes with a research on some interesting LFV Higgs decays which will happen in presence of flavour mixing in the slepton sector. The detection of these decays in the LHC would be a clear signal of new physics beyond the SM, and as we will see through this chapter, these decays will provide a unique window to the MSSM even for a very heavy SUSY spectrum.

The results in this chapter have been published in \cite{Arana-Catania:2013xma}.

\section{Motivation for the study of LFV Higgs decays}
\label{motivlfvhiggs}

The absence of any experimental signal, so far, in the searches for supersymmetry at the LHC~\cite{SUSYsearches} and the discovery of a new Higgs-like particle by ATLAS~\cite{Aad:2012tfa} and CMS~\cite{Chatrchyan:2012ufa}, are pushing some SUSY mass parameters above the 1-TeV range. On one hand, the present lower mass bounds for squarks of the first and second generations and for gluinos are already above 1 TeV, and on the other hand, if the observed Higgs boson is identified with the lightest Higgs boson $h$ of the Minimal Supersymmetric Standard Model, a radiatively corrected mass $m_{h} \simeq$ 125-126 GeV also suggests that some third generation squarks could be in the TeV range. In principle, to place the SUSY masses at the multi-TeV range seems discouraging, both from an experimental point of view due to the inability to detect SUSY directly, and from a theoretical point of view, in regard to the naturalness of the theory, which contrarily suggests a soft SUSY-breaking scale, $m_{\rm SUSY}$, at or below the TeV scale. However, leaving the naturalness issue aside, the MSSM scenarios with very heavy SUSY masses can have other interesting aspects~\cite{heavySUSY}. In particular, this is the case of specific Higgs boson observables, like certain Higgs partial decay widths, which present a non-decoupling behaviour with $m_{\rm SUSY}$, as shown, for instance, in~\cite{Haber:2000kq,Dobado:2001mq,Curiel:2002pf,Curiel:2003uk,Arganda:2004bz}, opening a new window to the indirect detection of heavy SUSY. As it is well known, the decoupling of SUSY radiative corrections in the asymptotic large SUSY mass limit is valid for SUSY theories with an exact gauge symmetry, in agreement with the general decoupling behaviour of heavy states in Quantum Field Theory as stated in the famous Appelquist-Carazzone theorem~\cite{Appelquist:1974tg}. Nevertheless, it is also known that this theorem does not apply to theories with spontaneously broken gauge symmetries, nor with chiral fermions, which is the case of the MSSM. Furthermore, in order to have decoupling, the dimensionless couplings should not grow with the heavy masses. Otherwise, the mass suppression induced by the heavy-particle propagators can be compensated by the mass enhancement provided by the interaction vertices, with an overall non-decoupling effect, which is exactly what happens in some MSSM Higgs boson decays to fermions. For instance, it was studied in~\cite{Haber:2000kq} how non decoupling appears for large SUSY masses in the $h \to b \bar b$ decay through one-loop SUSY-QCD corrections, when the involved SUSY particle masses grow simultaneously with a generic soft SUSY-breaking scale $m_{\rm SUSY}$ (see also~\cite{Dobado:2001mq}). A similar non-decoupling behaviour was obtained for flavour changing neutral Higgs boson decays into quarks of the second and third generations through both SUSY-QCD corrections~\cite{Curiel:2002pf} and SUSY-EW corrections~\cite{Curiel:2003uk}. Other interesting non-decoupling SUSY-EW effects have also been seen in lepton flavour violating (LFV) Higgs boson decays within the context of the MSSM-seesaw model~\cite{Arganda:2004bz}; and in Higgs-mediated LFV processes like: $\tau \to 3\mu$ decays~\cite{Arganda:2005ji}, some semileptonic $\tau$ decays~\cite{Arganda:2008jj,Herrero:2009tm} and in $\mu-e$ conversion in heavy nuclei~\cite{Arganda:2007jw}; all of them within the MSSM-seesaw model too. Other non-decoupling effects from heavy SUSY particles have also been noticed within the context of the SUSY inverse-seesaw model~\cite{InverseSeesaw}. Some studies of SUSY non decoupling within the MSSM have also been performed in the effective field theory approach where the effective Higgs-fermion-fermion vertices are induced from one-loop corrections of heavy SUSY particles~\cite{Dobado:2001mq,Crivellin:2011jt,Crivellin:2010er} and in $b \bar b h$ production~\cite{Liu:2012qu}. 

In the present chapter, motivated by the recent discovery of the SM-like Higgs boson, we will focus on the study of the LFV Higgs boson decays at one-loop order with the hypothesis of general slepton flavour mixing. We will extend the study to the three neutral Higgs bosons of the MSSM $h$, $H$ and $A$, considering the new Higgs-like particle to be the lightest Higgs boson $h$. In particular, we will study the LFV Higgs decays $h \to \tau \mu$, $H \to \tau \mu$ and $A \to \tau \mu$. This kind of processes provide an unique window into new physics due to the strong suppression of flavour violation in the 
SM, where the flavour mixing is induced exclusively by 
the Yukawa couplings of the corresponding fermion sector. This is specially 
interesting in the lepton sector in which the flavour mixing will be hugely 
suppressed because of the smallness of the lepton Yukawa couplings. 
Therefore, the discovery of any process which violates the lepton flavour 
number would be a clear signal of physics beyond the SM. The specific case of LFV Higgs decays within supersymmetric 
models has deserved special attention in 
the literature~\cite{Arganda:2004bz,LFVHDworks}. Also encouraging results for
the reach of LFV Higgs decays at the LHC have been recently obtained in \cite{Blankenburg:2012ex} within the context of a general effective Lagrangian approach.               

Our purpose here is to take advantage of the mentioned non-decoupling behaviour with $m_\text{SUSY}$ in the LFV Higgs decay widths into charged leptons of different generations, $\Gamma (\phi \to l_il_j)$, with $i \neq j$ and $\phi=h,H,A$, in order to look for sizeable branching ratios which can give rise to detectable signals at the LHC. Here and from now on, $l_il_j$ with $i \neq j$ in the final state of the LFV decays refers to either $l_i {\bar l}_j$ or 
${\bar l}_i l_j$. In general, the radiatively corrected LFV Higgs couplings to leptons are proportional to the heaviest lepton mass involved, being this the reason why we select the $\tau \mu$ channel as the most promising one. In addition, the $\mu e$ channel is extremely restricted by the associated radiative decay, $\mu \to e \gamma$~\cite{Adam:2013mnn}, leaving us almost no room to move in the allowed parameter space of slepton flavour mixing, and driving us to extremely low rates, not measurable in any LHC detector.
The $\tau e$ channel, on the other hand, gives us very similar results to the $\tau \mu$ channel, and from an experimental point of view, the LHC sensitivity to the former should be equivalent to the latter~\cite{LHCsensitivity}. Therefore the results obtained along this chapter for the LFV Higgs decays into $\tau \mu$ are straightforwardly translated into the $\tau e$ channels.
For the present study we will focus then on the $\phi \to \tau \mu$ decays within the MSSM at one loop with general slepton flavour mixing between the second and the third generations, without assuming any particular source of flavour mixing. 

The general slepton mixing will be parametrized in a model-independent way as it was explained in Section \ref{sec:slepsector}. As we learned from Chapter \ref{phenoflavourslep}, when we studied the constraints to the flavour mixing parameters, the most important constraints to the four slepton mixings that we will study here come from the $\tau \to \mu \gamma$ searches~\cite{Aubert:2009ag}. The present bounds obtained there to BR$(\tau \to \mu \gamma)$ lead to constraints on the 23 mixings which for 500 GeV $\leq \, m_{\rm SUSY}\,\leq$ 1500 GeV and 5 $\leq\,\tan \beta \,\leq$ 60 are at $|\delta^{LL}_{23}|_{\rm max},|\delta^{LR}_{23}|_{\rm max},|\delta^{RL}_{23}|_{\rm max} \sim 
{\cal O}(10^{-2}-10^{-1})$, and $|\delta^{RR}_{23}|_{\rm max} \sim 
{\cal O}(10^{-1}-1)$, and for heavier SUSY masses lead to weaker bounds. We will check here, that by raising the SUSY mass scale into the multi-TeV range will turn into the needed relaxation of these bounds, since these LFV radiative decays present a decoupling behaviour with the SUSY scale. This will allow us to explore a high SUSY mass scale region, where we will find very promising values for the LFV Higgs rates. 

\section{Framework for the computation of the LFV decay rates BR$(\phi \to l_il_j)$}
\label{selectedlfvhiggs}

For the forthcoming estimates of the LFV Higgs decay rates, we use the complete one-loop formulas and the full set of diagrams contributing to the $\Gamma (\phi \to {\bar l_i} l_j)$ and $\Gamma (\phi \to l_i {\bar l_j})$ partial decay widths, with $i \neq j$, within the MSSM, which are written in terms of the mass eigenvalues for all the involved MSSM sparticles, including the physical slepton and sneutrino masses, $m_{\tilde l_i} \,\,(i=1,..,6)$ and $m_{\tilde \nu_i}\,\,(i=1,2,3)$, and the rotation matrices $R^{\tilde l}$ and $R^{\tilde \nu}$ of Eqs.~(\ref{sleptons}) and ~(\ref{sneutrinos}). We take these general one-loop formulas from \cite{Arganda:2004bz} and emphasize that they are valid for the general slepton mixing case considered here, with all the mixings effects from the $\delta^{AB}_{ij}$'s being transmitted to the LFV Higgs decay rates via the physical slepton and sneutrino masses and their corresponding rotations. The off-diagonal trilinear couplings in Eq.(\ref{v1Al}) also enter into this computation of the LFV Higgs decay rates. For a more detailed discussion about these analytical results we refer the reader to~\cite{Arganda:2004bz}. It should be also noted that, since we are assuming real $\delta^{AB}_{ij}$'s, the predictions for BR$(\phi \to l_i {\bar l_j})$ and the $\cp$-conjugate BR$(\phi \to {\bar l_i} l_j)$ are the same. Thus, we will perform our estimates for just one of them and will denote this rate generically by BR$(\phi \to l_il_j)$. Obviously, in the case that these two channels
cannot be differentiated experimentally one should then add the two contributions to the total final number of events. However, for the present computation we do not sum them and report instead the results for $\phi \to l_il_j$, meaning that they are valid for any of the two cases. 

As said above, we focus only on $h \to \tau \mu$, $H \to \tau \mu$ and $A \to \tau \mu$ decay channels and consider the constraints imposed over the parameter space by the current upper bound on the related LFV radiative decay $\tau \to \mu \gamma$~\cite{Aubert:2009ag}. The SUSY mass spectra are computed numerically with the code {\tt SPheno}~\cite{SPheno}. The slepton and sneutrino spectra are computed from an additional subroutine that we have implemented into {\tt SPheno} in order to include our parametrization of slepton mixing given by the 
$\delta^{AB}_{ij}$'s. 
The LFV decay rates are computed with our private FORTRAN code in which we have implemented the complete one-loop formulas for the LFV partial Higgs decay widths of~\cite{Arganda:2004bz} and the complete one-loop formulas for the LFV radiative $\tau$ decay widths which we take from~\cite{Arganda:2005ji}. Note that these latter formulas for the $\tau \to \mu \gamma$ decays are also written in terms of the physical sparticle eigenvalues and eigenstates and do not neglect any of the lepton masses. The mass spectrum of the MSSM Higgs sector, with two-loop corrections included, and their corresponding total widths are calculated by means of the code {\tt FeynHiggs}~\cite{feynhiggs} as in the previous chapters. 

In the numerical computations of this study of the Higgs decays we will work within the scenarios defined in Section \ref{scenarios} and we will restrict ourselves to the case where there are flavour mixings exclusively between the second and third generations of sleptons, thus we set all $\delta_{ij}^{AB}$'s to zero except for $ij=23$. On one hand, the LFV one-loop corrected Higgs couplings are proportional to the heaviest lepton mass involved~\cite{Arganda:2004bz} and, therefore, the Higgs decay rates into $\mu e$ are suppressed by a factor $m_\mu^2/m_\tau^2$ with respect to the $h, H, A \to \tau \mu, \tau e$ decay rates. On the other hand, the related LFV radiative decay $\mu \to e \gamma$ has a much more restrictive upper bound~\cite{Adam:2013mnn} than $\tau \to e \gamma$ and $\tau \to \mu \gamma$ decays~\cite{Aubert:2009ag}, and the present allowed values of the $\delta_{12}^{AB}$'s would not drive us to any measurable $\phi \to \mu e$ rates. 

\section{Numerical results for the branching ratios of the LFV decays}
\label{LFVrates}

\begin{figure}[t!]
\hspace{-0.5 cm}
\begin{tabular}{cc}
\includegraphics[width=80mm]{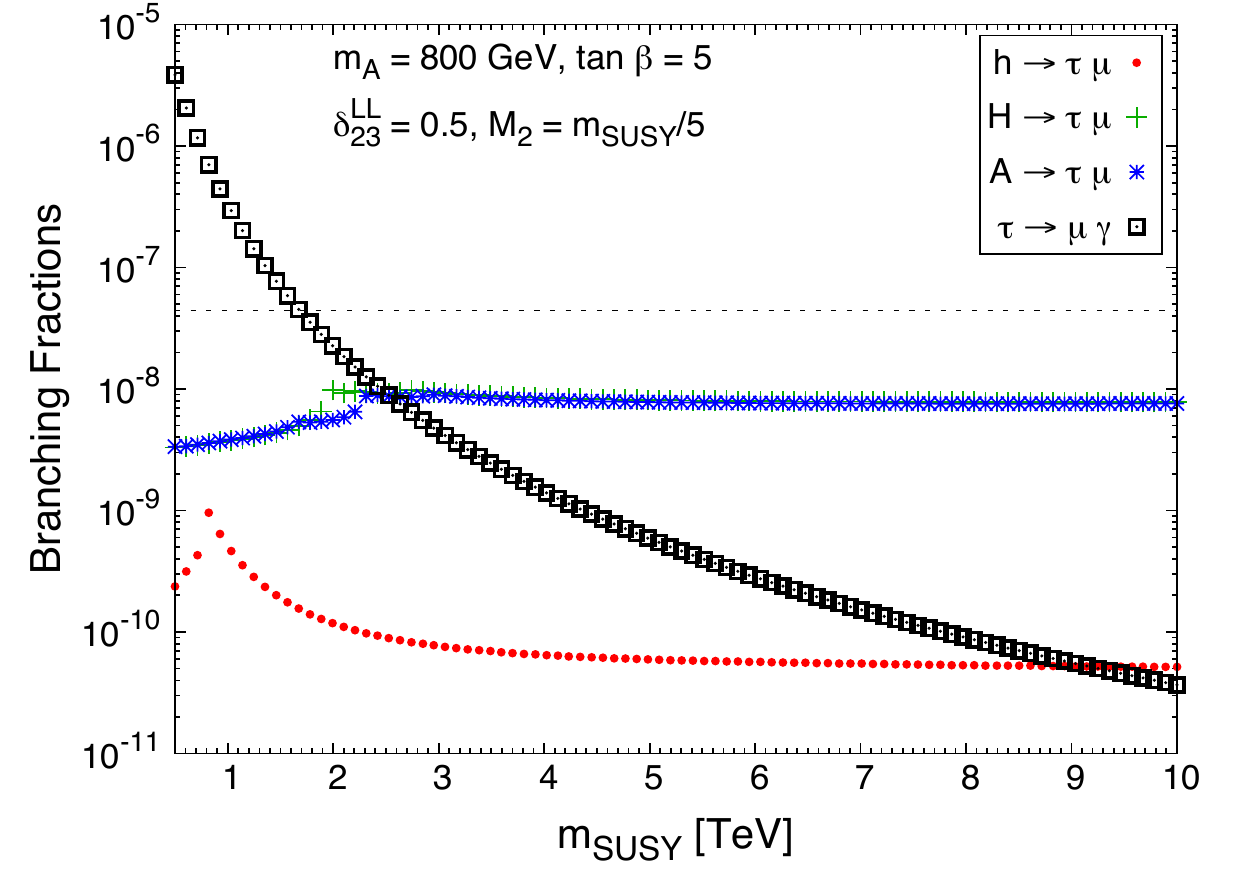} &
\includegraphics[width=80mm]{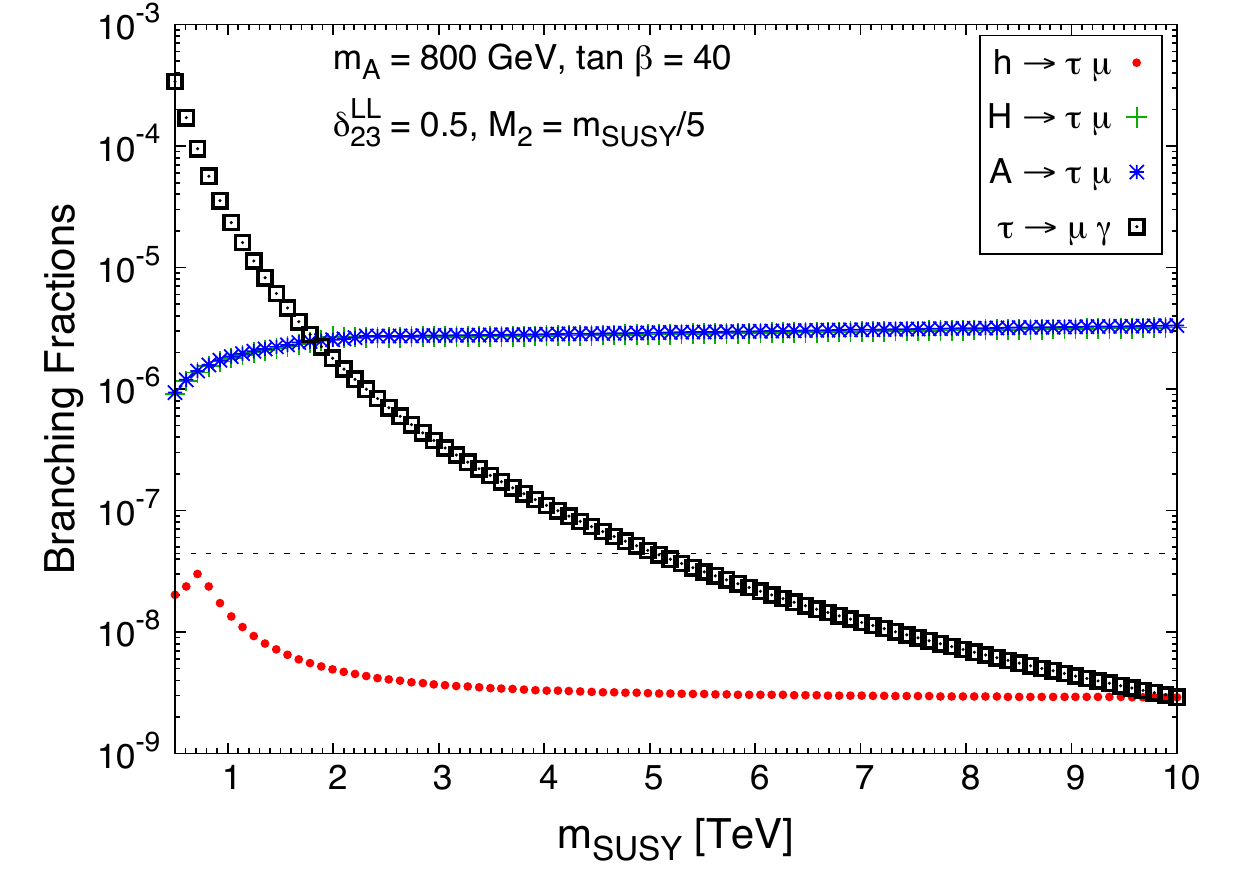} \\
\includegraphics[width=80mm]{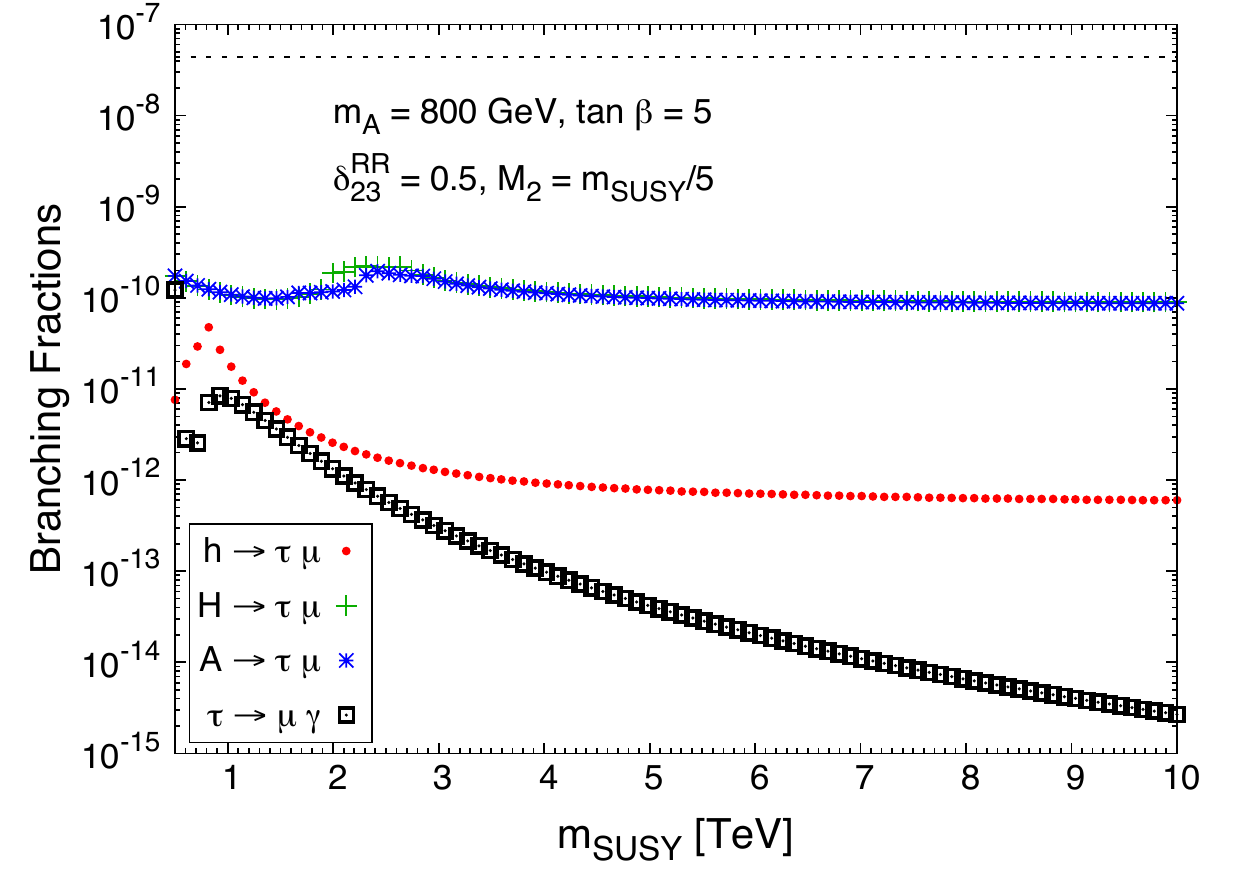} &
\includegraphics[width=80mm]{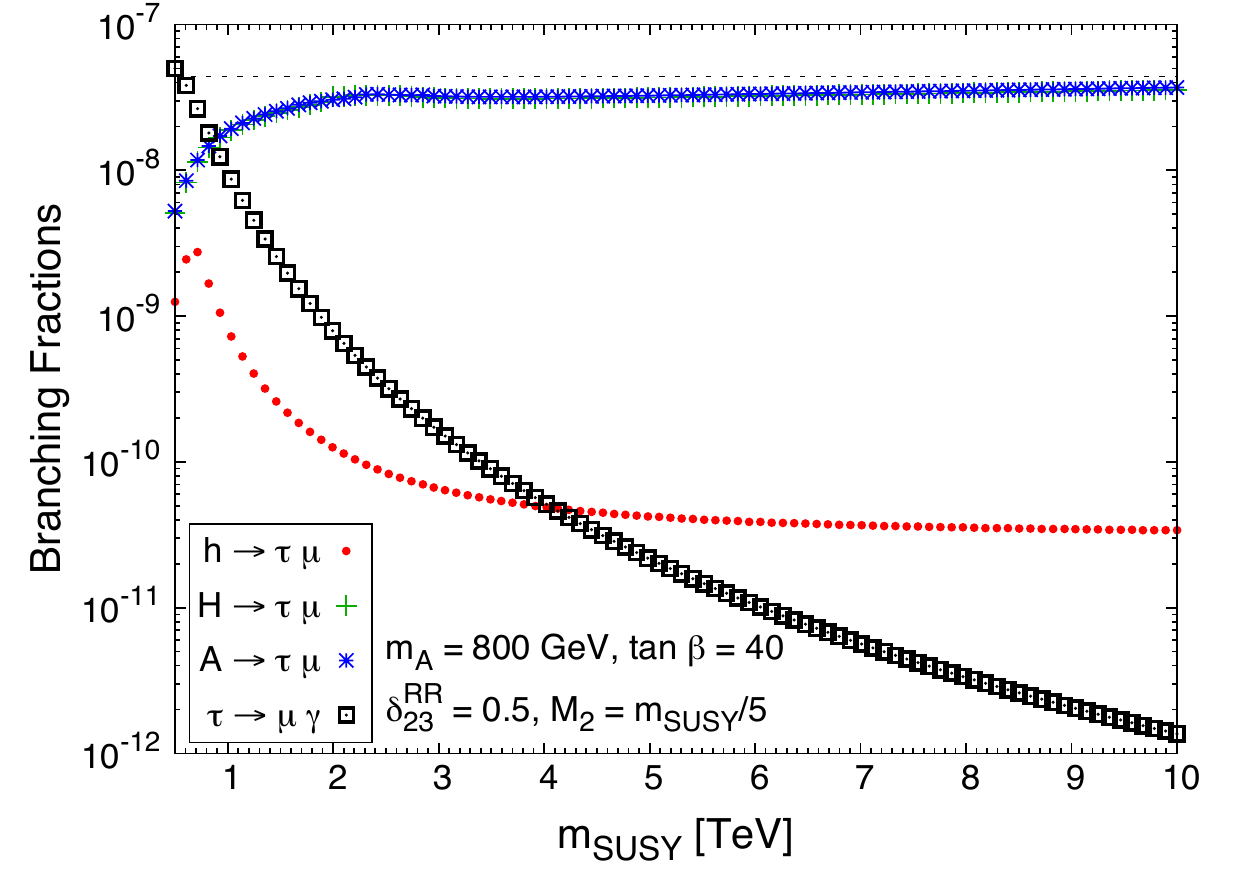} \\
\includegraphics[width=80mm]{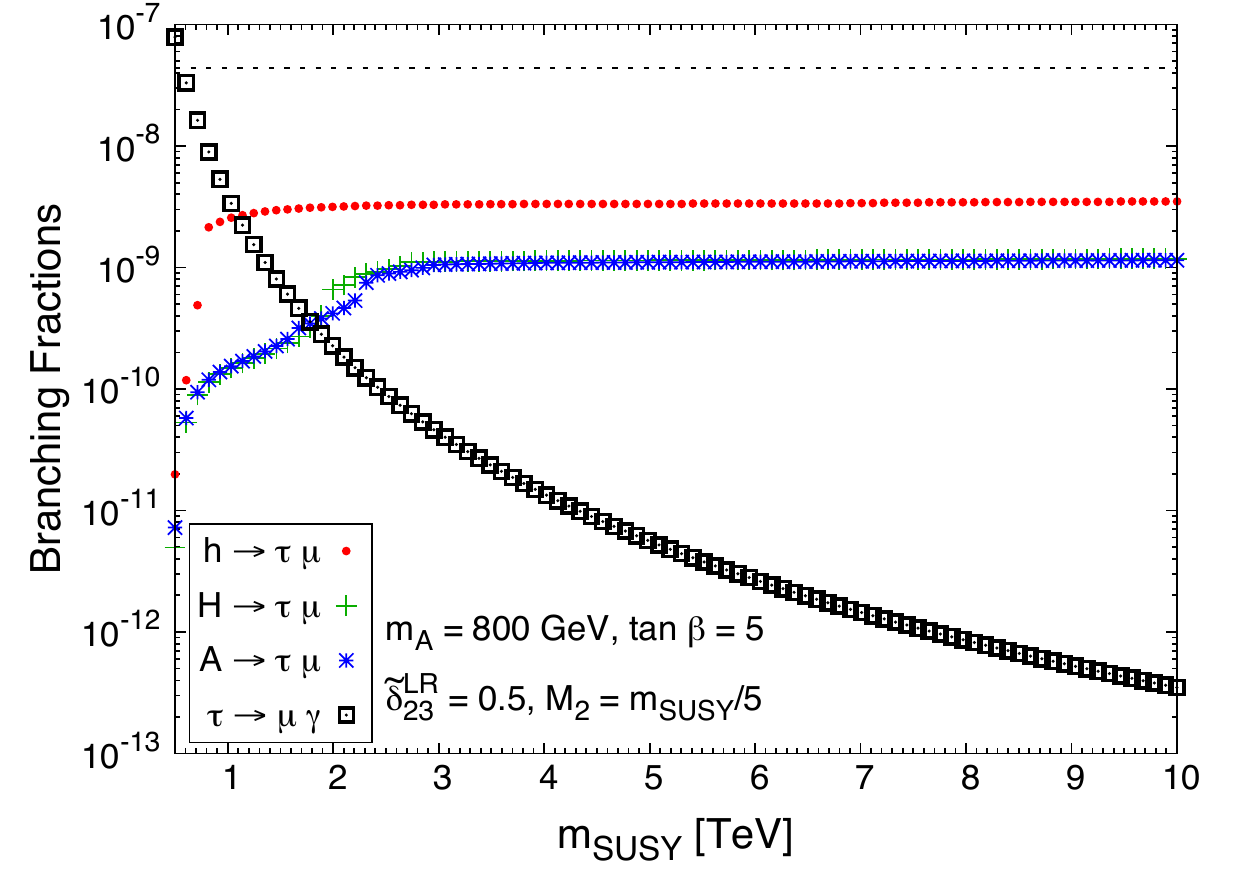} &
\includegraphics[width=80mm]{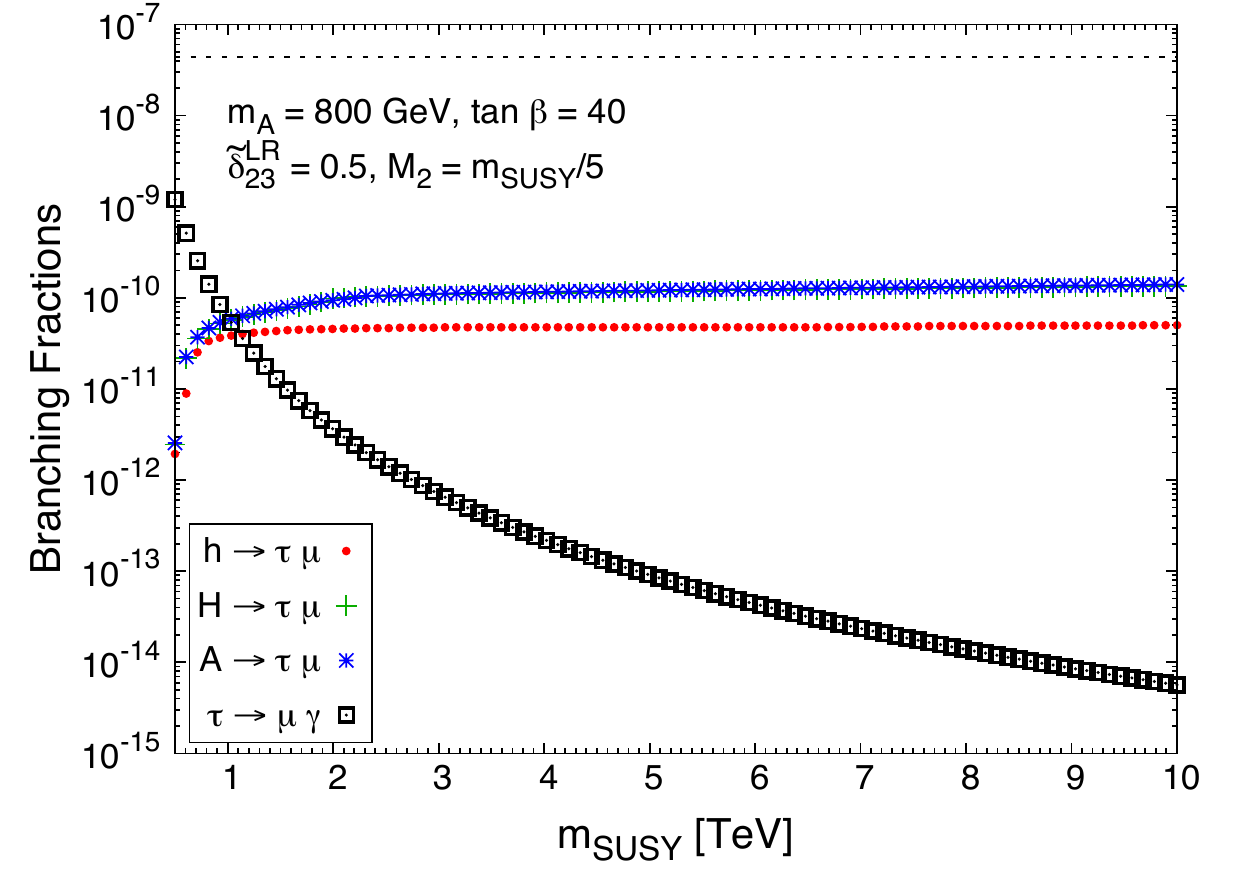}

\end{tabular}
\caption{Large $m_\text{SUSY}$ behaviour of the LFV decay rates for the scenarios defined in Section \ref{pmssmvheavysusy}: BR($h \to \tau \mu$), BR($H \to \tau \mu$), BR($A \to \tau \mu$) and 
BR($\tau \to \mu \gamma$) as functions of $m_\text{SUSY}$ for $\tan\beta = 5$ (left panels) and $\tan\beta = 40$ (right panels) with $\delta_{23}^{LL} = 0.5$ (upper panels), $\delta_{23}^{RR} = 0.5$ (middle panels) and $\tilde \delta_{23}^{LR} = 0.5$ (lower panels). The results for $\tilde \delta_{23}^{RL} = 0.5$ (not shown) are identical to those of $\tilde \delta_{23}^{LR} = 0.5$. In each case, the other flavour changing deltas are set to zero. In all panels, $m_A =$ 800 GeV and the other MSSM parameters are set to the values reported in the text, with $M_2 = \frac{1}{5} m_\text{SUSY}$. 
The horizontal dashed line denotes the current experimental upper bound for $\tau \to \mu \gamma$ channel, BR($\tau \to \mu \gamma$) $< 4.4 \times 10^{-8}$~\cite{Aubert:2009ag}.}\label{BRs-mSUSY}
\end{figure}

Next we present the numerical results for the branching ratios of the LFV decays. The scenarios that will be evaluated are described in Section \ref{pmssmvheavysusy}, with a detailed discussion of the possible values of the parameters.

We recall that in these scenarios the relevant mass parameters are:
\begin{eqnarray}
m_{\tilde L} &=& m_{\tilde E} = m_\text{SUSY} \,, \\ 
\mu &=& M_2 = a \,m_\text{SUSY} \,,
\end{eqnarray}
where $a$ is a constant coefficient that we will fix to different values, namely, $a = \frac{1}{5}$, $\frac{1}{3}$, 1.

We also set an approximate GUT inspired relation for the gaugino masses:
\begin{equation}
M_2 = 2 M_1 = M_3/4 \,.
\end{equation}

And the trilinear couplings in the slepton sector have been fixed to the generic SUSY mass scale, $A_\mu=A_e=A_\tau = m_\text{SUSY}$.

Regarding the non-diagonal trilinear couplings in the slepton sector we have also assumed a rather simple but realistic setting by relating them with the single soft SUSY-breaking mass scale, $m_\text{SUSY}$. Specifically, we assume the following linear relation:
\begin{eqnarray}
 {\cal A}^l_{23} &=& {\tilde \delta}^{LR}_{23} m_\text{SUSY} \,, \,\,\,\,\,  
 {\cal A}^l_{32} = {\tilde \delta}^{LR}_{32} m_\text{SUSY} \,.
\end{eqnarray} 

We show in Figure~\ref{BRs-mSUSY} the behaviour of the branching ratios for the two types of LFV decays, BR$(\phi \to \tau \mu)$ and BR$(\tau \to \mu \gamma$), as functions of $m_\text{SUSY}$, and we consider two different values of $\tan\beta$, namely, $\tan\beta = 5$ (left panels) and $\tan\beta = 40$ (right panels). In each case we set one single delta to be non vanishing with the particular values: $\delta_{23}^{LL} = 0.5$ (upper panels), $\delta_{23}^{RR} = 0.5$ (middle panels) and ${\tilde \delta}_{23}^{LR} = 0.5$ (lower panels). All the other flavour changing deltas are set to zero. We find identical results 
for ${\tilde \delta}_{23}^{RL} = 0.5$ as for ${\tilde \delta}_{23}^{LR} = 0.5$ and, for brevity, we have omitted the plots for ${\tilde \delta}_{23}^{RL}$ in Figure~\ref{BRs-mSUSY}.   

\begin{figure}[t!]
\hspace{-0.5 cm}
\begin{tabular}{cc}
\includegraphics[width=80mm]{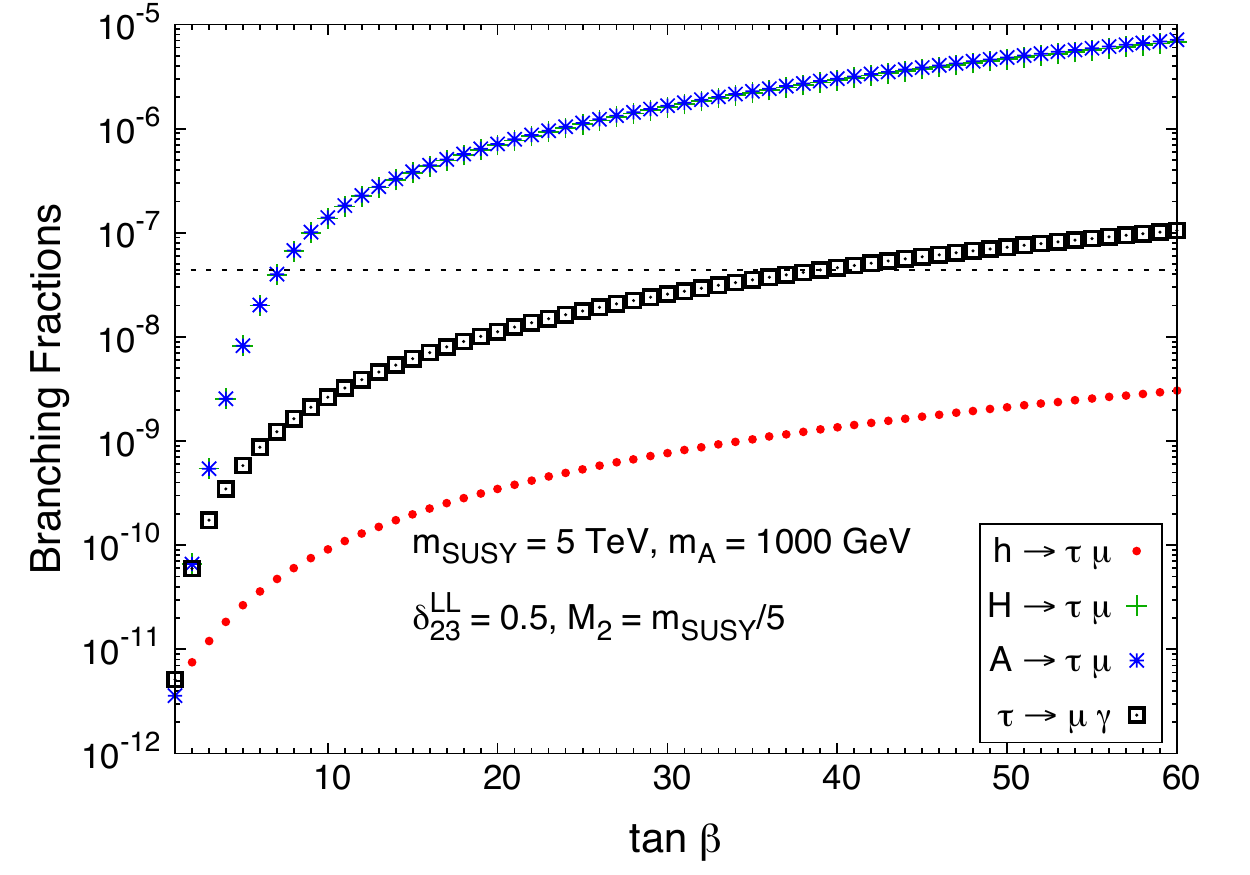} &
\includegraphics[width=80mm]{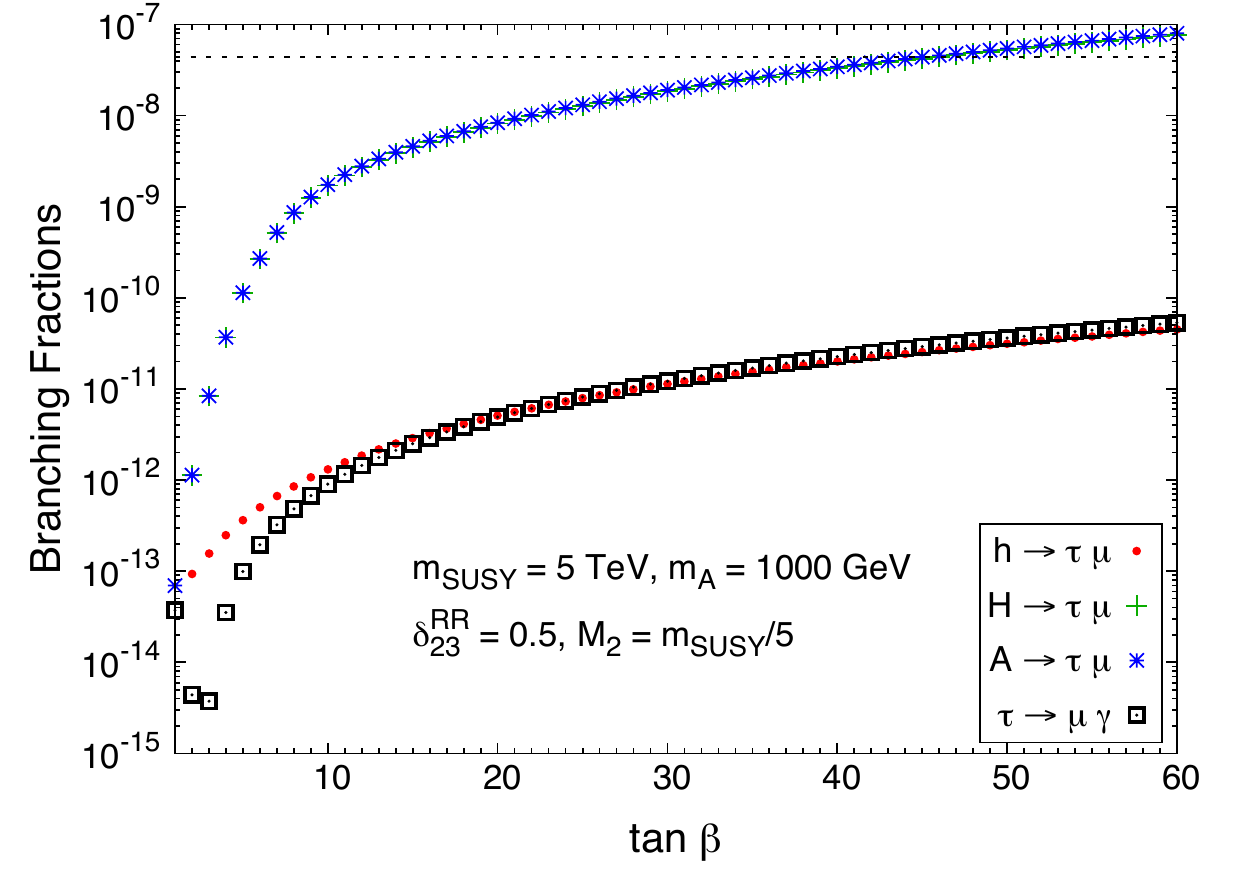} \\
\includegraphics[width=80mm]{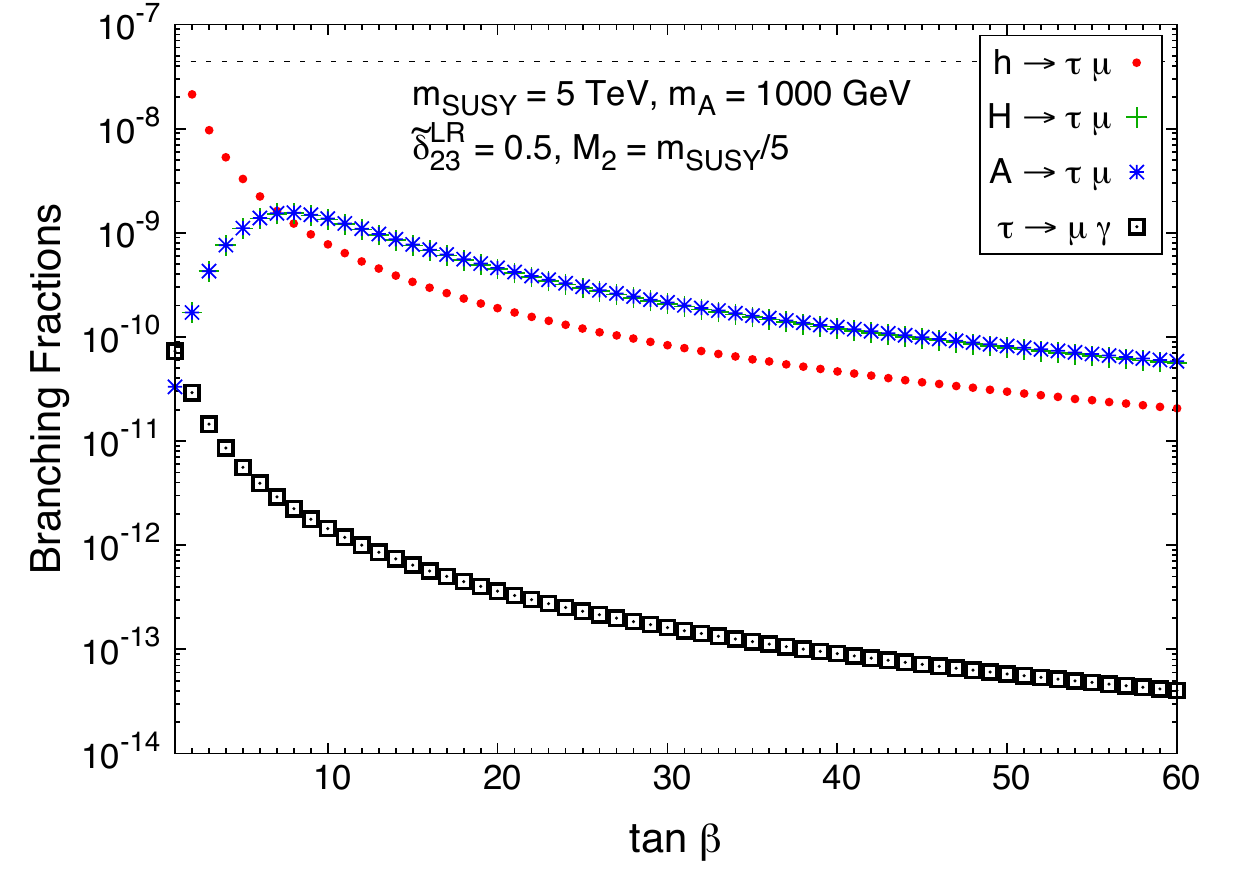} &
\includegraphics[width=80mm]{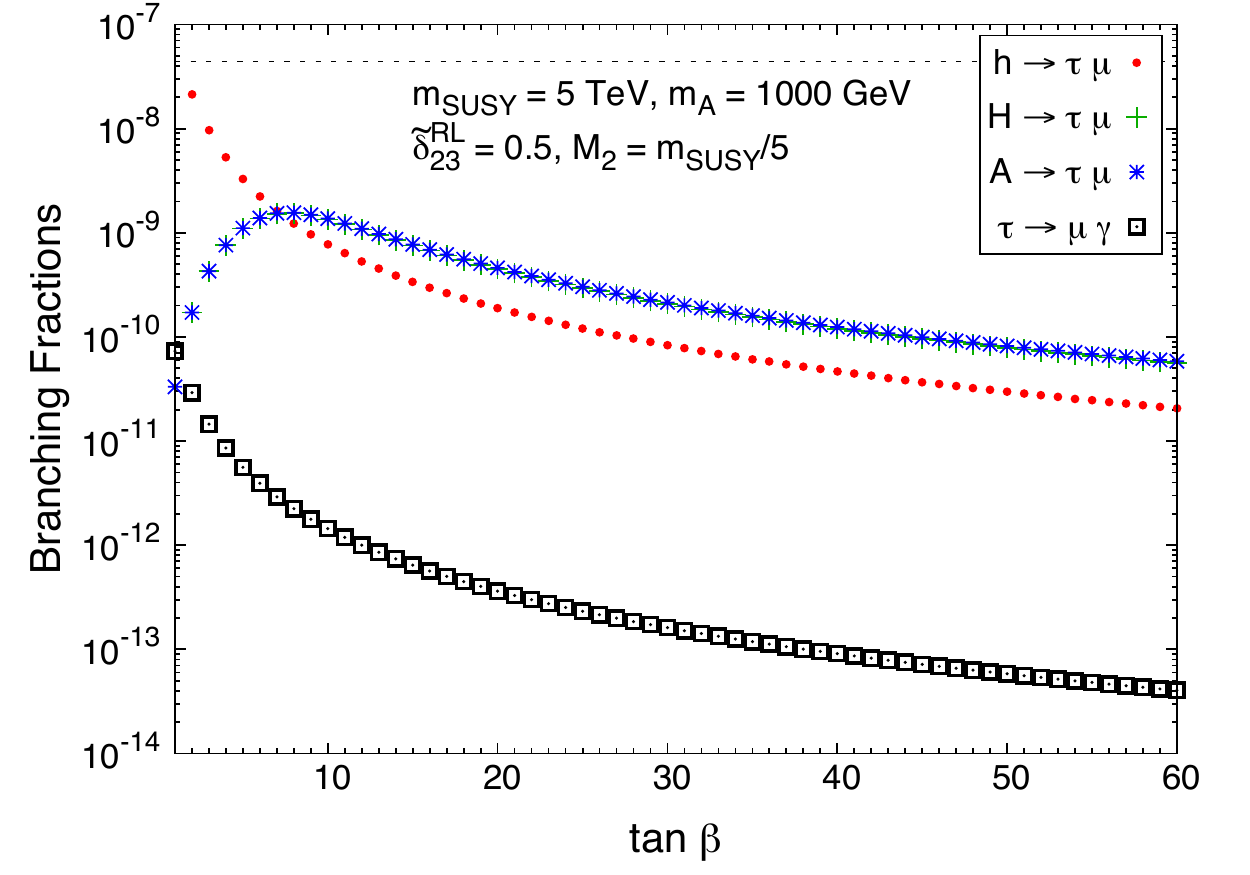}

\end{tabular}
\caption{BR($h \to \tau \mu$), BR($H \to \tau \mu$), BR($A \to \tau \mu$) and 
BR($\tau \to \mu \gamma$) for the scenarios defined in Section \ref{pmssmvheavysusy} as functions of $\tan\beta$ for $\delta_{23}^{LL} =$ 0.5 (upper left panel), $\delta_{23}^{RR} =$ 0.5 (upper right panel), $\tilde \delta_{23}^{LR} =$ 0.5 (lower left panel) and $\tilde \delta_{23}^{RL} =$ 0.5 (lower right panel). In each case, the other flavour changing deltas are set to zero. In all panels, $m_A =$ 1000 GeV, $m_\text{SUSY} =$ 5 TeV and the other MSSM parameters are set to the values reported in the text, with $M_2 = m_\text{SUSY}/5$. The horizontal dashed line denotes the current experimental upper bound for $\tau \to \mu \gamma$ channel, BR($\tau \to \mu \gamma$) $< 4.4 \times 10^{-8}$~\cite{Aubert:2009ag}. The green crosses are overimposed with the blue stars.}\label{BRs-tanb}
\end{figure}

On the upper panels of Figure~\ref{BRs-mSUSY}, when the responsible for the flavour mixing between the second and the third generations is $\delta_{23}^{LL}$, the branching ratios of the LFV Higgs decays show a clear non-decoupling behaviour with $m_\text{SUSY}$, which remain constant from $m_\text{SUSY} \simeq$ 1 TeV to $m_\text{SUSY} =$ 10 TeV, with values of BR$(h \to \tau \mu) \simeq 5 \times 10^{-11}$ and BR$(H, A \to \tau \mu) \simeq 8 \times 10^{-9}$ for $\tan\beta =$ 5, and BR$(h \to \tau \mu) \simeq 3 \times 10^{-9}$ and BR$(H, A \to \tau \mu) \simeq 3 \times 10^{-6}$ for $\tan\beta =$ 40. Another important feature that should be noted is the fast growing with $\tan\beta$ of these decays which increase the $H$ and $A$ LFV decay rates almost three orders of magnitude from $\tan\beta =$ 5 to $\tan\beta =$ 40. Furthermore, we have numerically checked that for large values of $\tan\beta \geq 10$ the partial decay widths $\Gamma(H, A \to \tau \mu)$ go approximately as $\sim (\tan\beta)^4$~\cite{Arganda:2004bz}, whereas $\Gamma(h \to \tau \mu)$ goes as $\sim (\tan\beta)^2$. This implies that the corresponding branching ratios go all at large $\tan\beta \geq 10$ as BR($h, H, A \to \tau \mu$) $\sim (\tan\beta)^2$ in this $LL$ case, since the total widths go as $\Gamma_\text{tot}(H, A) \sim (\tan\beta)^2$ and $\Gamma_\text{tot}(h)$ is approximately constant with $\tan\beta$. This behaviour of the BRs with $\tan\beta$ is confirmed numerically in our forthcoming Figure~\ref{BRs-tanb}. In contrast, the branching ratio of the $\tau \to \mu \gamma$ decay presents a decoupling behaviour with $m_\text{SUSY}$, decreasing as $\sim 1/m_\text{SUSY}^4$, and it is reduced around five orders of magnitude from $m_\text{SUSY} =$ 500 GeV to $m_\text{SUSY} =$ 10 TeV. In all these figures we have also included, for comparison, the experimental upper  bound for the $\tau \to \mu \gamma$ channel, whose present value is BR($\tau \to \mu \gamma$) $< 4.4 \times 10^{-8}$~\cite{Aubert:2009ag}. Thus, each $m_\text{SUSY}$ point which leads to a prediction of BR$(\tau \to \mu \gamma)$ above this line is excluded by data. Therefore, only values of $m_\text{SUSY} \gtrsim$ 2 TeV for $\tan\beta =$ 5 and $m_\text{SUSY} \gtrsim$ 5 TeV for $\tan\beta =$ 40, and their corresponding predictions for the LFV rates, are allowed for $\delta_{23}^{LL} = 0.5$.

On the middle panels of Figure~\ref{BRs-mSUSY} we have plotted the LFV Higgs and radiative decay rates as functions of $m_\text{SUSY}$, considering $\delta_{23}^{RR}$ as the responsible for $\tau-\mu$ mixing. A similar non-decoupling behaviour to the $\delta_{23}^{LL}$ case can be observed for $\delta_{23}^{RR}$, whose branching ratios stay again constant as $m_\text{SUSY}$ grows. However, the numerical contribution of $\delta_{23}^{RR}$ to the LFV processes is much less important than that of $\delta_{23}^{LL}$, and all the $RR$ rates are in comparison around two orders of magnitude smaller than the $LL$ rates. In the $RR$ case, another important feature is that all the predictions found of BR($\tau \to \mu \gamma$) for $5 \leq \tan \beta \leq 40$ and $m_\text{SUSY}$ values above 500 GeV are allowed by the present BR$(\tau \to \mu \gamma)$ experimental upper bound. 

The predictions of BR($h \to \tau \mu$), BR($H \to \tau \mu$), BR($A \to \tau
\mu$) and BR($\tau \to \mu \gamma$) as functions of $m_\text{SUSY}$, for the
case ${\tilde \delta}_{23}^{LR} =$ 0.5, are shown on the lower panels of
Figure~\ref{BRs-mSUSY}. We see clearly again a non-decoupling behaviour with
$m_\text{SUSY}$, since the branching ratios of the Higgs decays tend to a constant
value as $m_\text{SUSY}$ grows, in contrast to the BR$(\tau \to \mu \gamma)$
rates that display again a decoupling behaviour and decrease with
$m_\text{SUSY}$. For this particular choice of  ${\tilde \delta}_{23}^{LR} =$
0.5 we also see that the predicted branching ratios for the LFV $h$ decays at low $\tan
\beta= 5$ are  larger than in the previous $LL$ and $RR$ cases, whereas the branching ratios
for the LFV $H$ and $A$ decays are larger than those of the $RR$ case but
smaller than the $LL$ ones.  The lower right panel shows that for
$\tan\beta=40$ the branching ratios of the three Higgs bosons, $h$, $H$ and $A$, are comparatively smaller than for $\tan\beta=5$. This decreasing with $\tan\beta$ of the LFV decay rates for the $LR$ case with fixed value of ${\tilde \delta}_{23}^{LR}$ is confirmed in our forthcoming  Figure~\ref{BRs-tanb}. In consequence, the largest LFV Higgs decay rates that will be obtained in the $LR$ (and $RL$) case will be for low $\tan \beta$ values and this will be taken into account in our next studies in order to maximize the event rates from these decays at the LHC.  

To sum up the main results of Figure~\ref{BRs-mSUSY}, the most relevant 
$\delta_{23}^{AB}$ parameter at low $\tan \beta$ values for the lightest Higgs boson $h$ is ${\tilde \delta}_{23}^{LR}$ (and ${\tilde\delta}_{23}^{RL}$), 
which gives rise to larger LFV Higgs decay rates than $\delta_{23}^{LL}$ 
and $\delta_{23}^{RR}$, whereas for the $H$ and $A$ Higgs bosons the most relevant parameter is 
$\delta_{23}^{LL}$. At large $\tan \beta$ values, the most relevant parameter for all the three Higgs bosons is  $\delta_{23}^{LL}$. All these branching ratios, as we will see later, can be further enhanced by
considering two non-vanishing deltas at the same time, by exploring with larger size of these deltas, and also by considering different signs for the various deltas. Overall, the main conclusion from this Figure~\ref{BRs-mSUSY} is that  
if one wants to obtain sizeable and 
allowed by data branching ratios, one needs large values of $m_\text{SUSY}$, 
which plays a double role: on one hand, it keeps constant values of the 
LFV Higgs decay rates (due to the non-decoupling behaviour of these decays 
with $m_\text{SUSY}$) and, on the other hand, it brings down 
$\tau \to \mu \gamma$ below its experimental upper bound 
(because of the decoupling effect of LFV radiative decays with $m_\text{SUSY}$).

As we have said above, we show in Figure~\ref{BRs-tanb} the behaviour of LFV branching ratios as functions of $\tan\beta$ for $\delta_{23}^{LL} =$ 0.5 (upper left panel), $\delta_{23}^{RR} =$ 0.5 (upper right panel), $\tilde \delta_{23}^{LR} =$ 0.5 (lower left panel) and $\tilde \delta_{23}^{RL} =$ 0.5 (lower right panel) with $m_A =$ 1000 GeV, $m_\text{SUSY} =$ 5 TeV and $M_2 = m_\text{SUSY}/5$. All the LFV rates have a very similar behaviour with $\tan\beta$ for both $LL$ and $RR$ mixing cases and grow as $\sim (\tan\beta)^2$ for large values of $\tan\beta \gtrsim$ 10, as indicated in the previous paragraphs. In contrast, the LFV rates present a decreasing behaviour with $\tan\beta$ in the $LR$ and $RL$ cases, which are identical. BR$(\tau \to \mu \gamma)$ and BR$(h \to \tau \mu)$ go for all studied $\tan\beta$ values approximately as $\sim (\tan\beta)^{-2}$ while BR$(H \to \tau \mu)$ BR$(A \to \tau \mu)$ grow at low $\tan\beta$ around two orders of magnitude from $\tan\beta =$ 1 to $\tan\beta =$ 5, and from this value decrease in the same way as $\tau \to \mu \gamma$ and $h \to \tau \mu$. Therefore, within the large $\tan\beta$ regime ($\tan\beta \gtrsim$ 10), in the $LL$ and $RR$ mixing cases the LFV rates grow as $\sim (\tan\beta)^2$ whilst in the $LR$ and $LR$ ones these rates present the opposite behaviour and decrease as $\sim (\tan\beta)^{-2}$.

\begin{figure}[t!]
\hspace{-0.5 cm}
\begin{tabular}{cc}
\includegraphics[width=80mm]{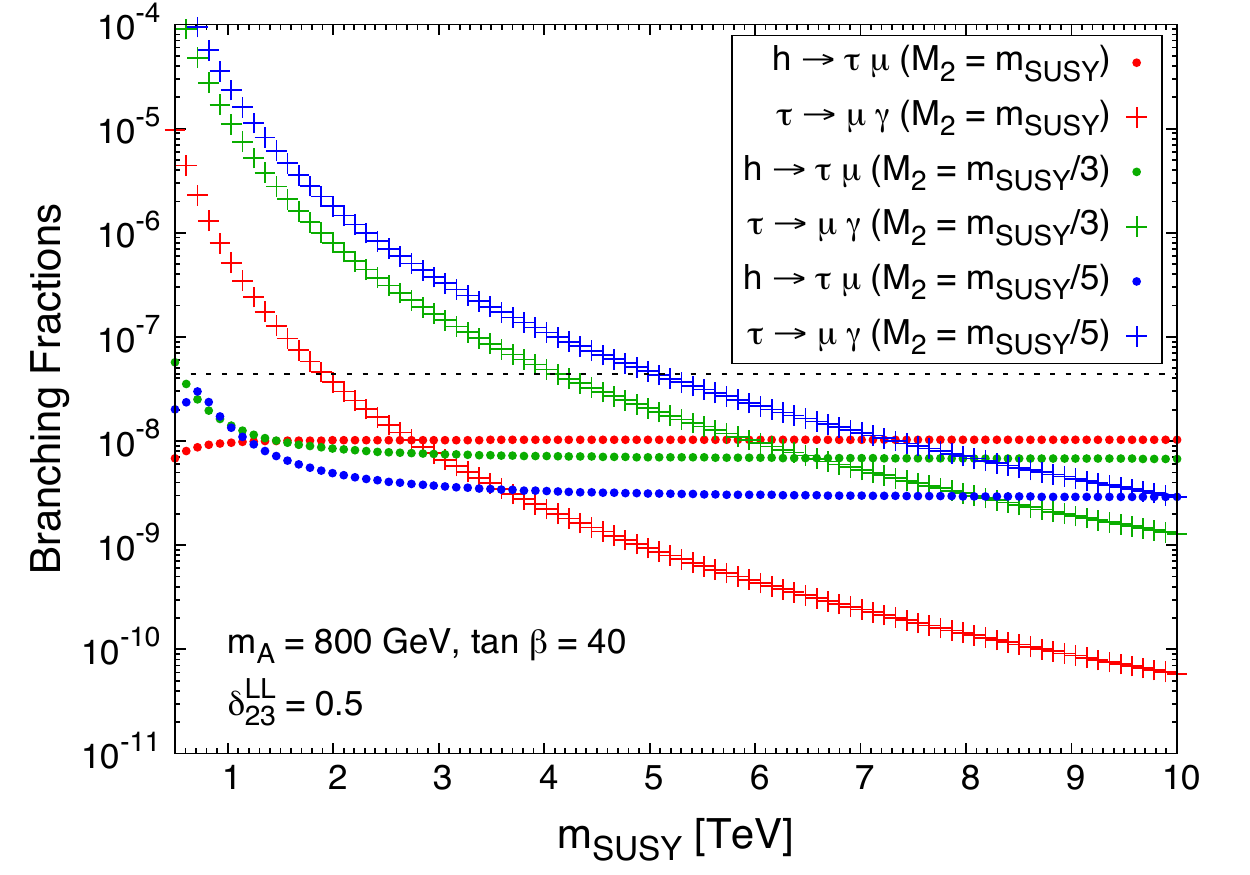} &
\includegraphics[width=80mm]{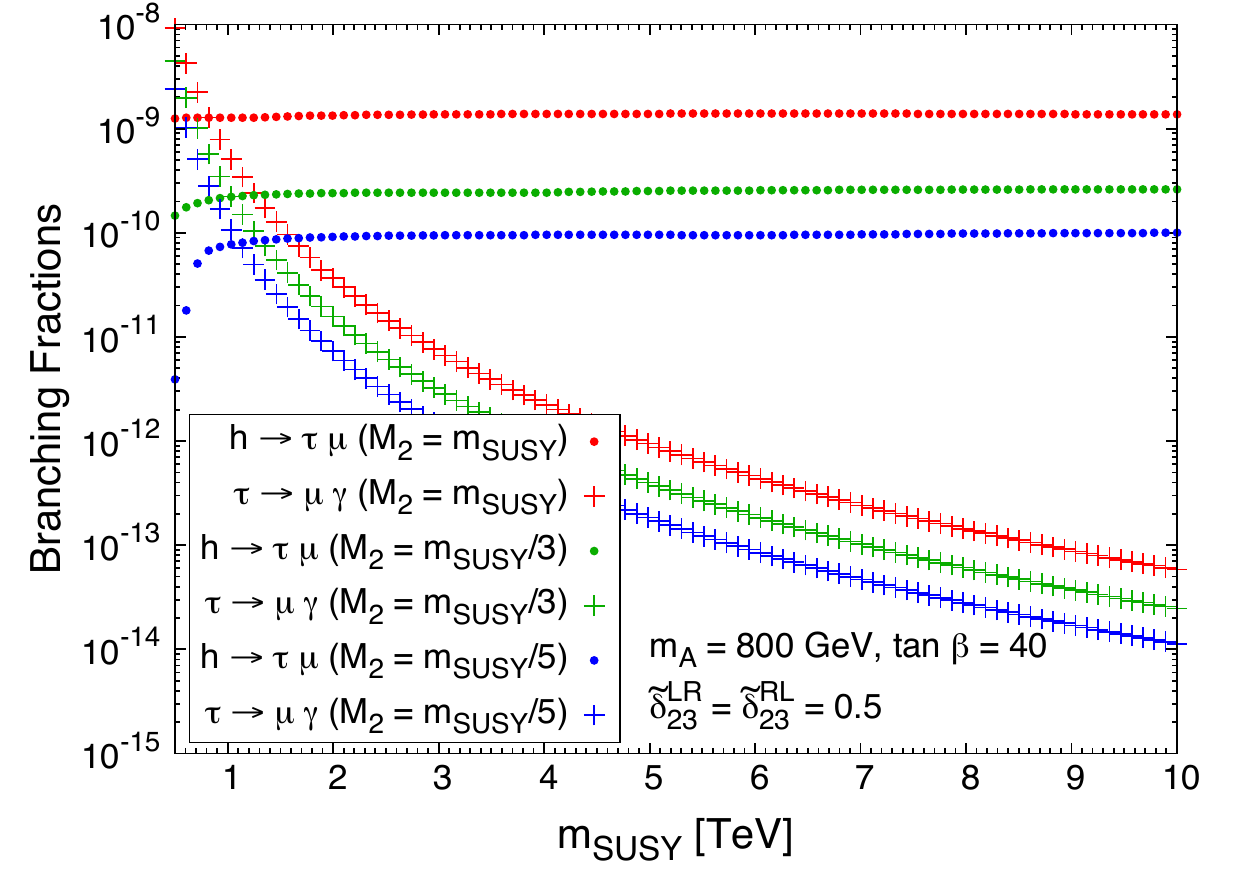} \\
\includegraphics[width=80mm]{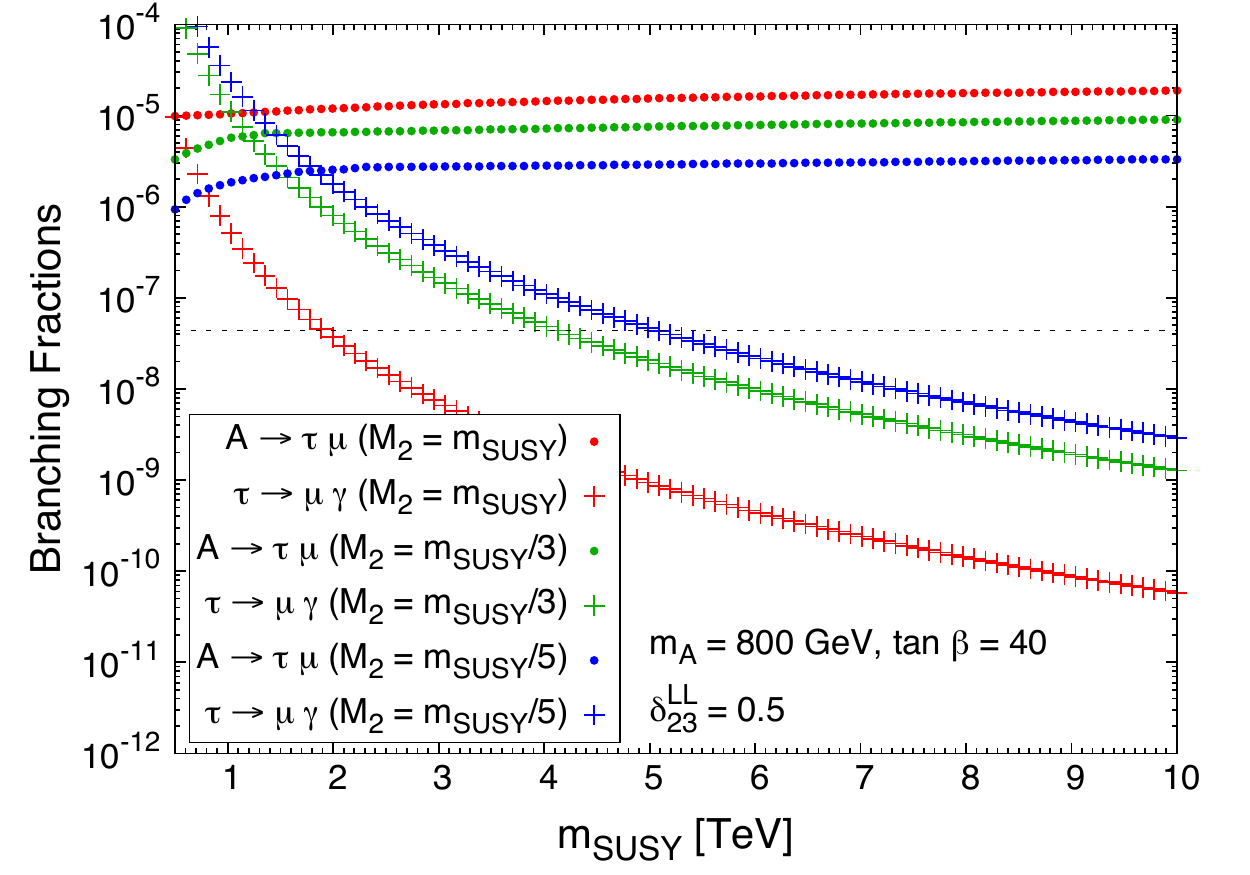} &
\includegraphics[width=80mm]{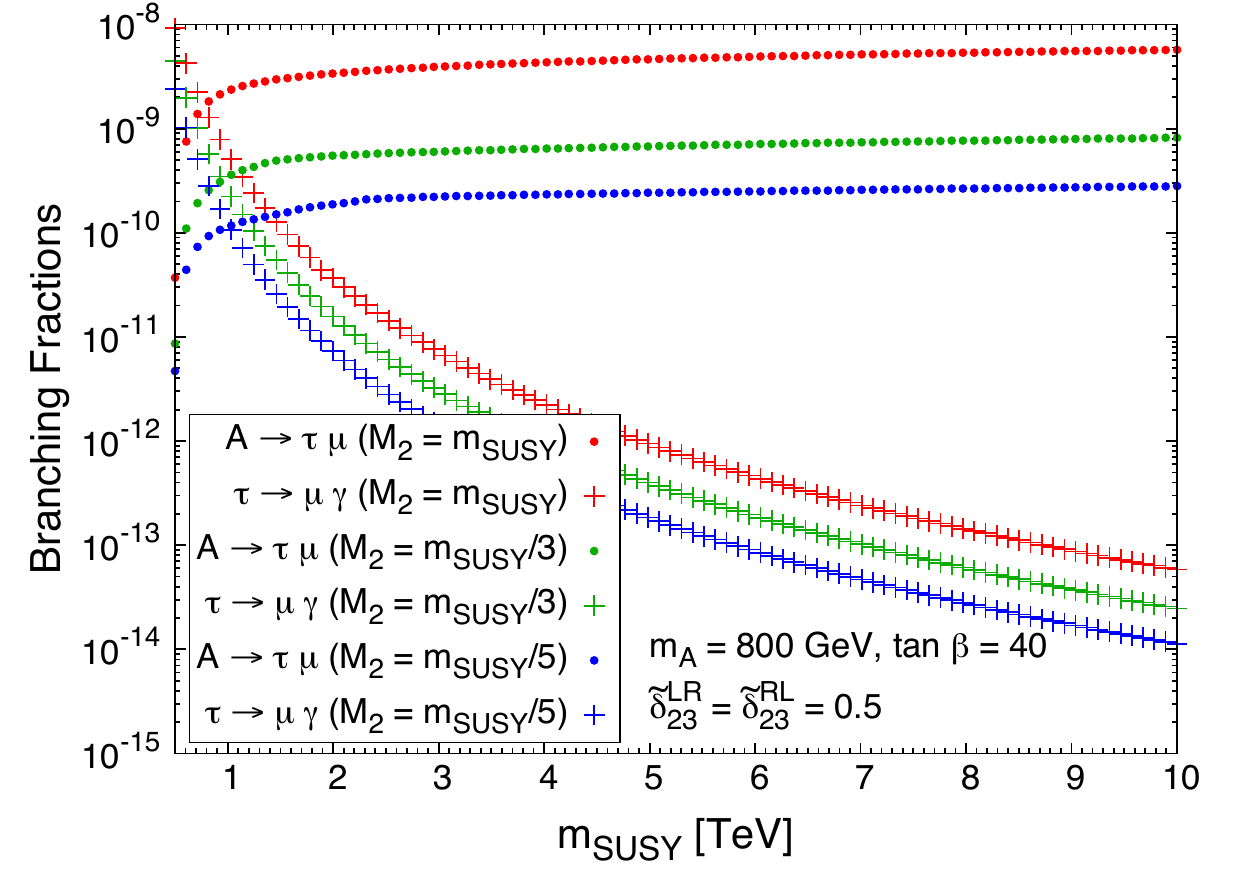} 
\end{tabular}
\caption{Sensitivity to $M_2$ for the scenarios defined in Section \ref{pmssmvheavysusy}: LFV Higgs decay rates (dots) and BR($\tau \to \mu \gamma$) (crosses) 
as functions of $m_\text{SUSY}$ with $\delta_{23}^{LL} = 0.5$ (left panels) 
and $\tilde \delta_{23}^{LR} = \tilde \delta_{23}^{RL} = 0.5$ (right panels) for different choices of 
$M_2$: $M_2 = m_\text{SUSY}$ (in red), $M_2 = \frac{1}{3} m_\text{SUSY}$ 
(in green) and $M_2 = \frac{1}{5} m_\text{SUSY}$ (in blue). The
results for $H$ (not shown) are nearly identical to those of $A$.
In each case, the other flavour changing deltas are set to zero. 
In all panels, $m_A =$ 800 GeV, $\tan\beta =$ 40 and the other MSSM 
parameters are set to the values reported in the text. The horizontal dashed 
line denotes the current experimental upper bound for $\tau \to \mu \gamma$ 
channel, BR($\tau \to \mu \gamma$) $< 4.4 \times 10^{-8}$~\cite{Aubert:2009ag}.}\label{BRs-M2}
\end{figure}

Now, we are interested in investigating if other choices of $M_2$ alter these 
previous results. For this purpose, we have plotted in Figure~\ref{BRs-M2} the 
predictions of BR($h \to \tau \mu$) (dots in upper panels), BR($A \to \tau \mu$) 
(dots in lower panels) and BR($\tau \to \mu \gamma$) (crosses in all panels) 
as functions of $m_\text{SUSY}$ for different values of $a$ 
(see Eq.~(\ref{M2mu})), $a = 1$ (in red), $a = \frac{1}{3}$ (in green) 
and $a = \frac{1}{5}$ (in blue), with $\delta_{23}^{LL} = 0.5$ (left panels) 
and $\tilde \delta_{23}^{LR} = \tilde \delta_{23}^{RL} = 0.5$ (right panels). The
results for the $H \to \tau \mu$ channel are nearly identical to those of $A \to \tau \mu$ and not shown here for shortness. In the case of $LL$ mixing,
 all the LFV Higgs rates, which present the same behaviour with $a$, increase 
 around a factor of 7 from $a = \frac{1}{5}$ to $a = 1$, while 
the $\tau \to \mu \gamma$ rates present the opposite behaviour with $a$, decreasing in a factor about 40 for the same values of $a$. Therefore, if $\delta_{23}^{LL}$ is the responsible for the slepton mixing, and for the explored interval $1/5 \le a \le 1$, the larger $M_2$ is (and consequently $M_1$ and $\mu$), the larger the LFV Higgs branching ratios are and the lower BR($\tau \to \mu \gamma$) is. 
In the 
$LR$-mixing case we see that again BR($h, H, A \to \tau \mu$) rise as $a$ 
grows and the enhancement is by a large factor of about 15 by 
changing $a = \frac{1}{5}$ to $a = 1$. In contrast to the $LL$ case, 
BR($\tau \to \mu \gamma$) also increases with $a$ for $LR$ mixing, 
although softer than the LFV Higgs rates. In summary, we learn from Figure~\ref{BRs-M2} 
that the best choice, for a fixed delta parameter, in order 
to obtain the largest LFV Higgs rates is $M_2 = m_\text{SUSY}$. However, 
we must be very careful, because it is possible that these large rates 
are excluded by the $\tau \to \mu \gamma$ upper bound, depending basically 
on the specific values of $\delta_{23}^{LL}$, $\tilde \delta_{23}^{LR}$ and 
$\tan \beta$.

\begin{figure}[t!]
\hspace{-0.5 cm}
\begin{tabular}{cc}
\includegraphics[width=80mm]{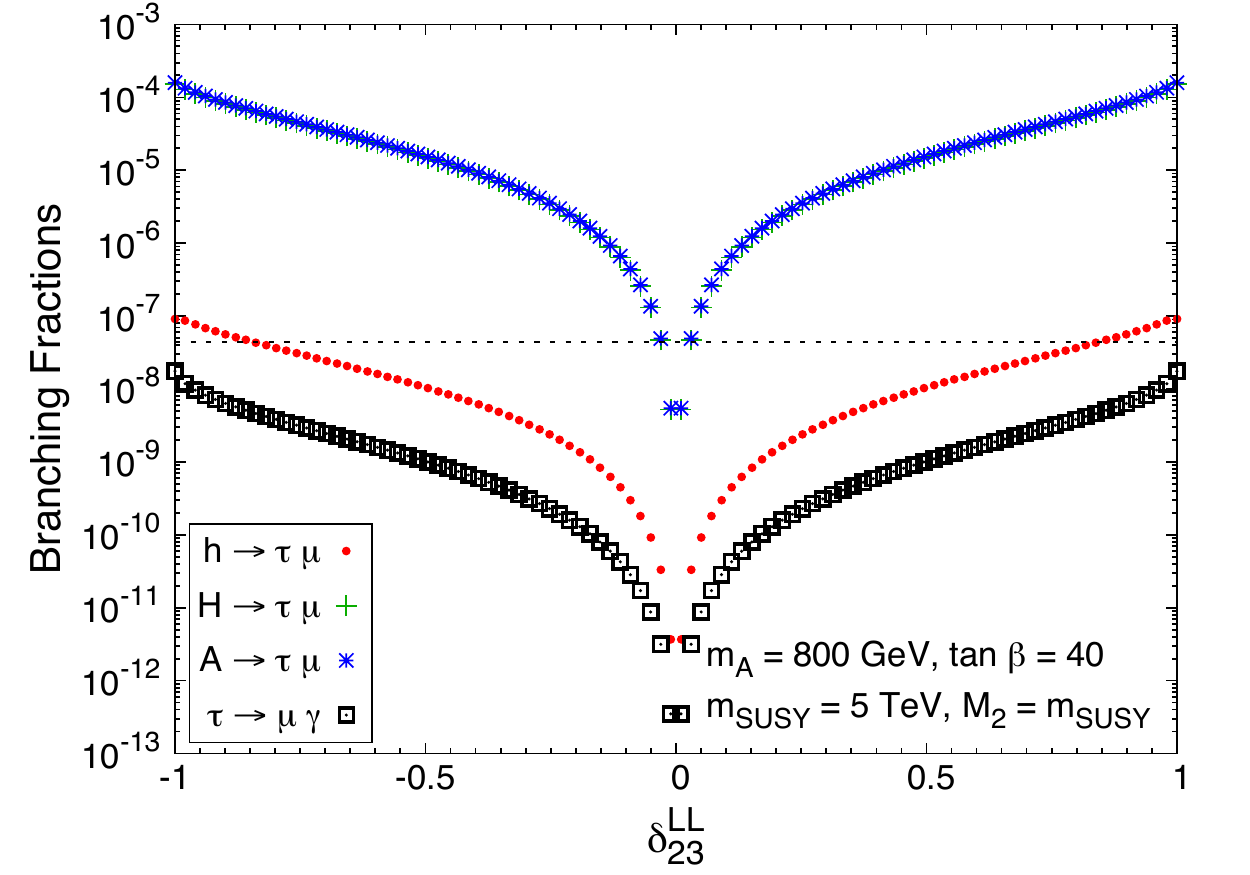} &
\includegraphics[width=80mm]{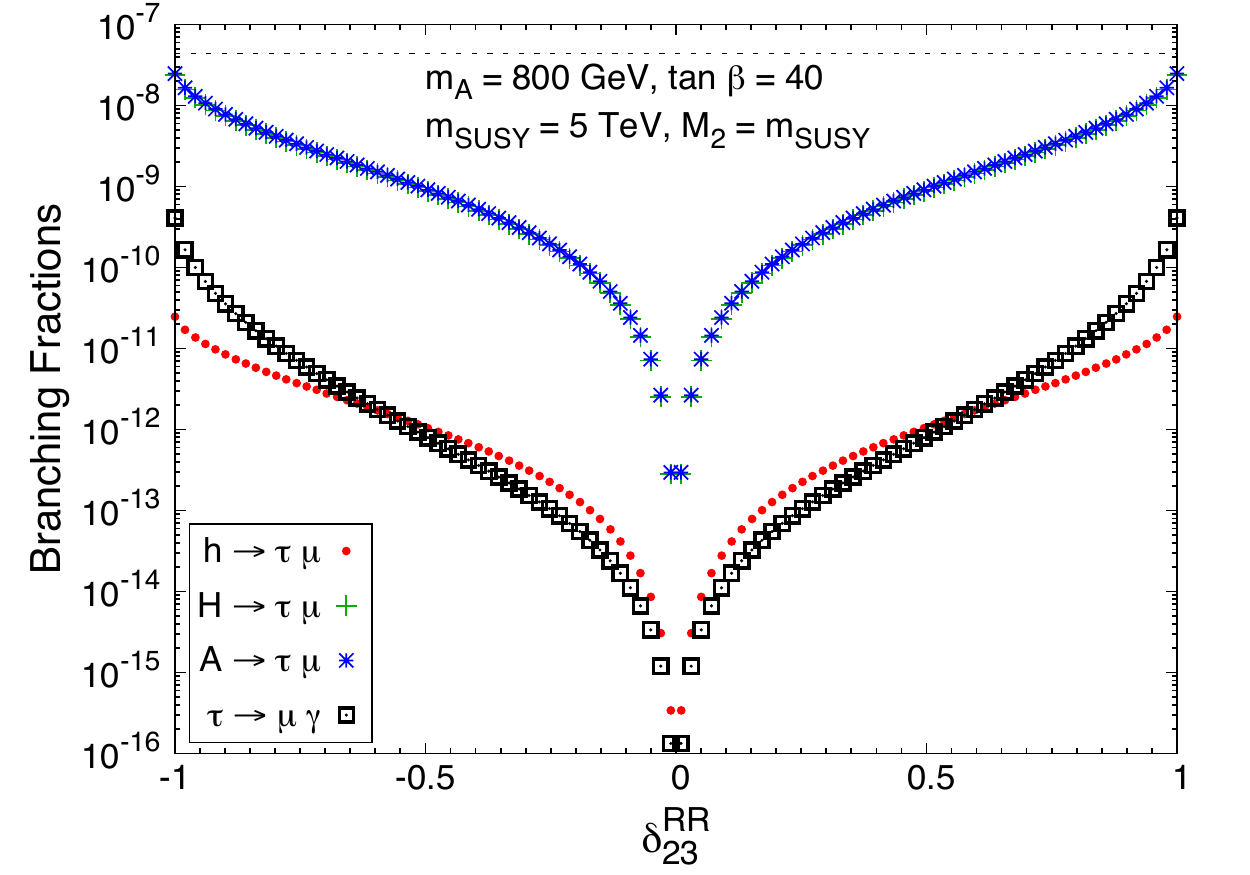} \\
\includegraphics[width=80mm]{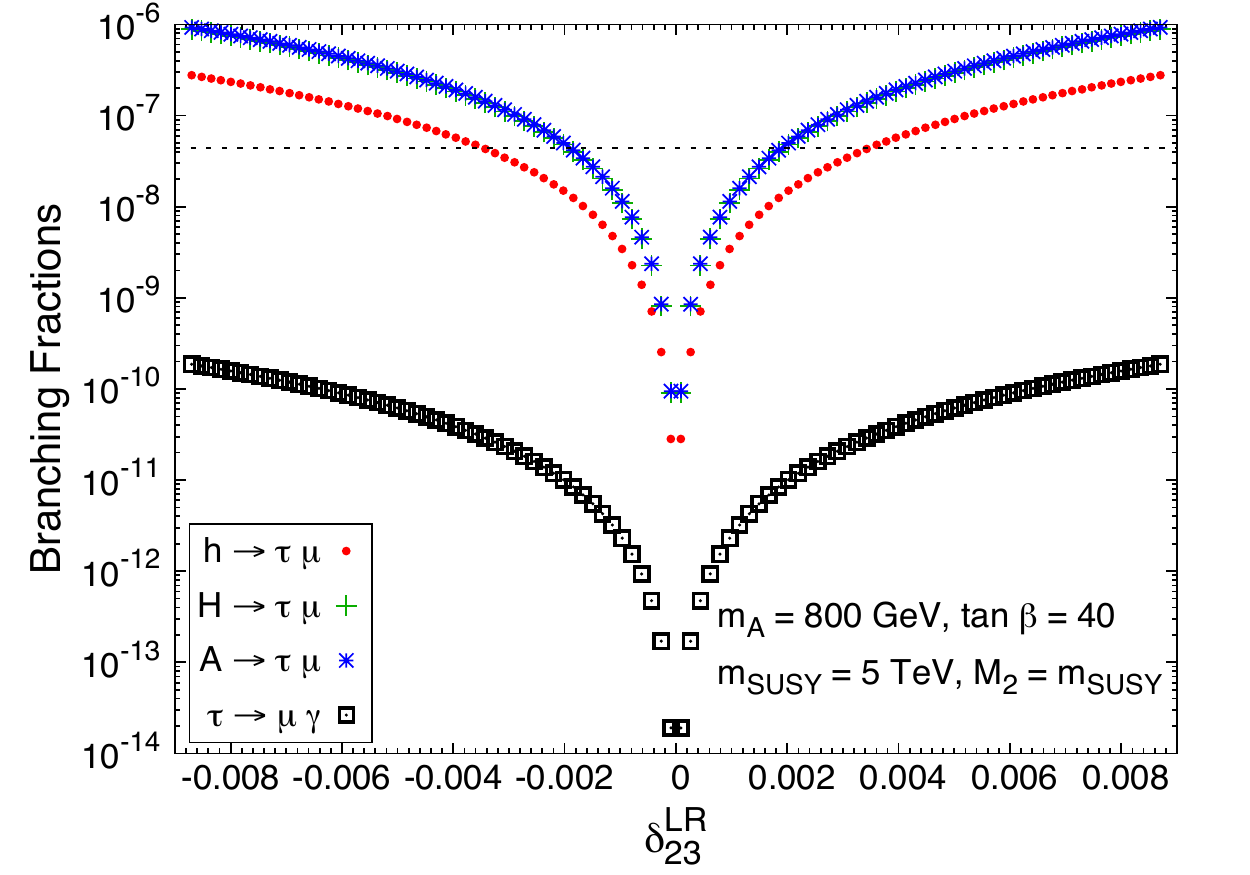} &
\includegraphics[width=80mm]{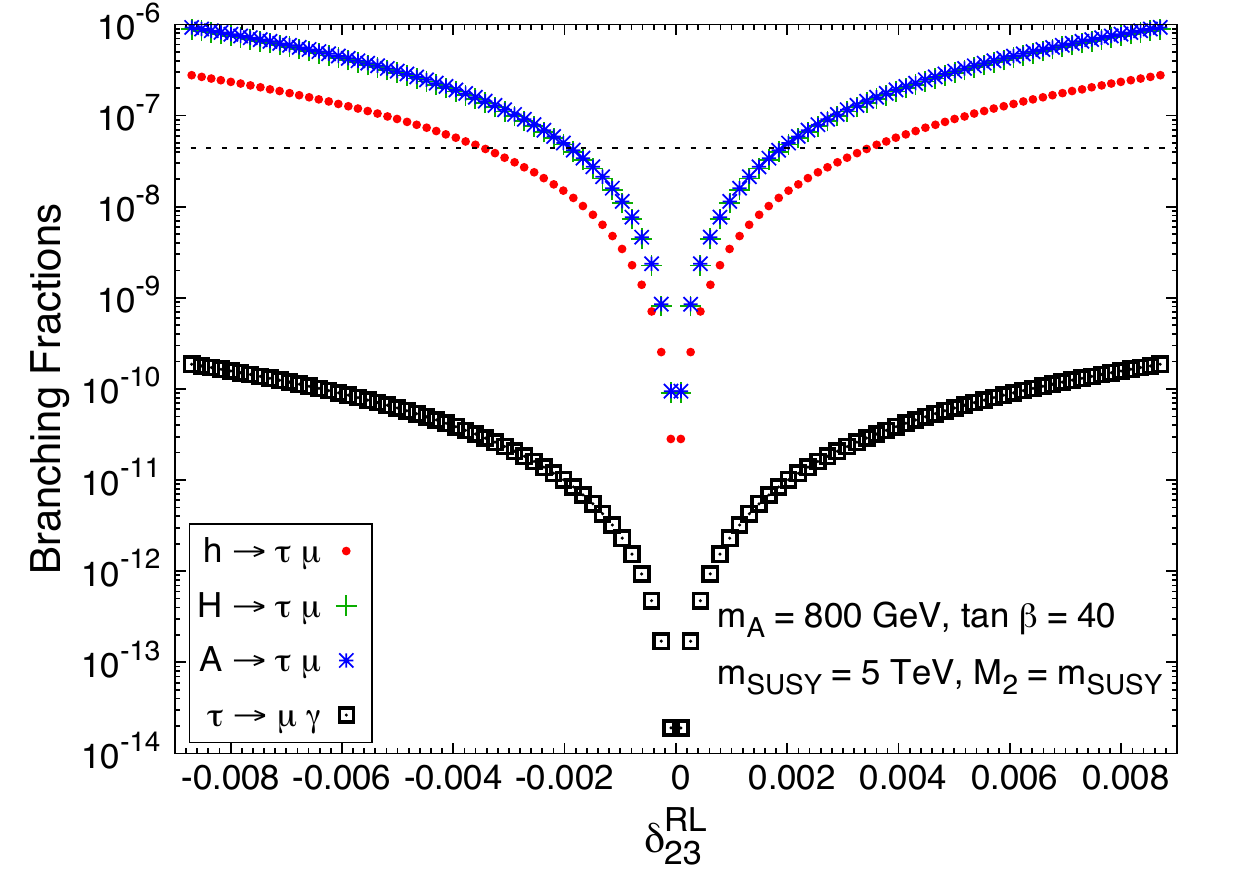}

\end{tabular}
\caption{BR($h \to \tau \mu$), BR($H \to \tau \mu$), BR($A \to \tau \mu$) and 
BR($\tau \to \mu \gamma$) for the scenarios defined in Section \ref{pmssmvheavysusy} as functions of $\delta_{23}^{LL}$ 
(upper left panel), $\delta_{23}^{RR}$ (upper right panel), $\delta_{23}^{LR}$ 
(lower left panel) and $\delta_{23}^{RL}$ (lower right panel). In each case, 
the other flavour changing deltas are set to zero. In all panels, $m_A =$ 800 GeV, $\tan\beta = 40$, $m_\text{SUSY} =$ 5 TeV and the other MSSM parameters are set to the values reported in the text, with $M_2 = m_\text{SUSY}$. The horizontal dashed line denotes the current experimental upper bound for $\tau \to \mu \gamma$ channel, BR($\tau \to \mu \gamma$) $< 4.4 \times 10^{-8}$~\cite{Aubert:2009ag}. The green crosses are overimposed with the blue stars.}\label{BRs-deltas}
\end{figure}

In order to look into the largest values of $\delta_{23}^{AB}$ allowed by data 
for the choice $M_2 = m_\text{SUSY}$, we show in Figure~\ref{BRs-deltas} the 
results of the branching fractions for the LFV Higgs decays into 
$\tau \mu$ and the related LFV radiative decay $\tau \to \mu \gamma$ as 
functions of the four deltas considered along this chapter. We have fixed in these
plots $\tan\beta=40$. For completeness, 
we have also presented here the results for $\delta_{23}^{RR}$, which are 
irrelevant for the present work since all the branching ratios obtained are 
extremely small to be detectable at the LHC, and the $\delta_{23}^{RL}$ 
results, which are identical to the $\delta_{23}^{LR}$ ones. The plots in 
Figure~\ref{BRs-deltas} show the expected growing of the LFV rates with the 
$|\delta_{23}^{AB}|$'s, and all of them are clearly symmetric 
$\delta_{23}^{AB} \to -\delta_{23}^{AB}$. On the upper left panel we have 
the results for the $LL$ case and it is clear that all the values of 
$\delta_{23}^{LL}$, from $-1$ to 1, are allowed by data, due to the large 
suppression that $\tau \to \mu \gamma$ suffers for $m_\text{SUSY} =$ 5 TeV. 
The predictions for $H \to \tau \mu$ (green crosses) are indistinguishable 
from $A \to \tau \mu$ ones (blue asterisks), which are superimposed in these 
plots. One can reach values of BR($h \to \tau \mu$) $\simeq 10^{-7}$ and 
BR($H, A \to \tau \mu$) $\simeq 2 \times 10^{-4}$ at the most for 
$\delta_{23}^{LL} = \pm 1$. The predictions for the LFV rates as functions 
of $\delta_{23}^{LR}$ are presented on the lower left panel of 
Figure~\ref{BRs-deltas}. In this case, all the values of 
$\left|\delta_{23}^{LR}\right|$ are allowed by the $\tau \to \mu \gamma$ 
upper bound and the largest value of $\left|\delta_{23}^{LR}\right| \simeq$ 
0.009 (which corresponds to $\tilde \delta_{23}^{LR} \simeq$ 10) gives rise 
to a branching fraction of $3 \times 10^{-7}$ for the $h \to \tau \mu$ 
channel, while BR($H, A \to \tau \mu$) reach values of $1 \times 10^{-6}$. The
low rates in the $h \to \tau \mu$ channel for this $LR$-mixing case can be 
notably increased, as we have
said previously, by assuming a lower $\tan\beta$ value closer to the 
low $\tan\beta$ region with $\tan\beta \lesssim$ 5. 

\begin{figure}[t!]
\hspace{-0.5 cm}
\begin{tabular}{cc}
\includegraphics[width=80mm]{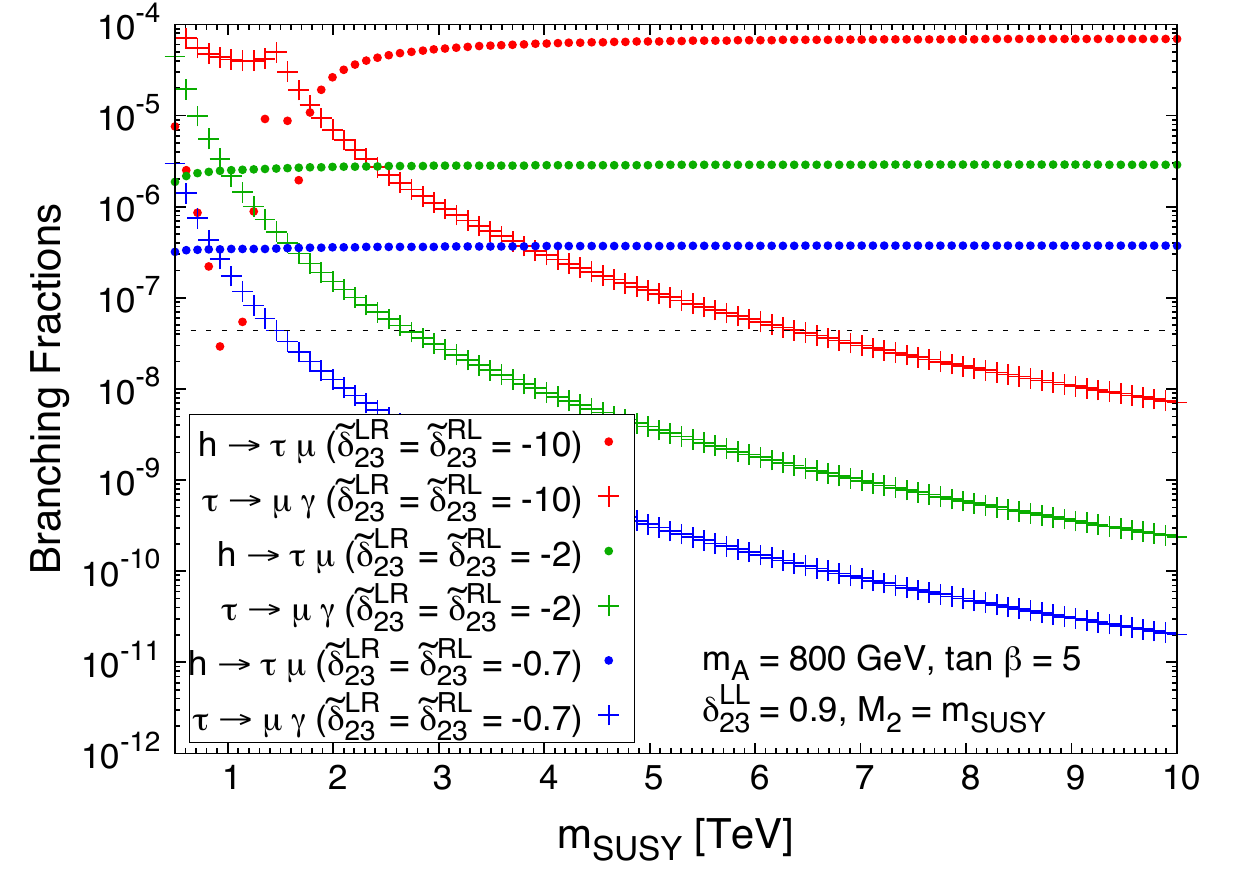} &
\includegraphics[width=80mm]{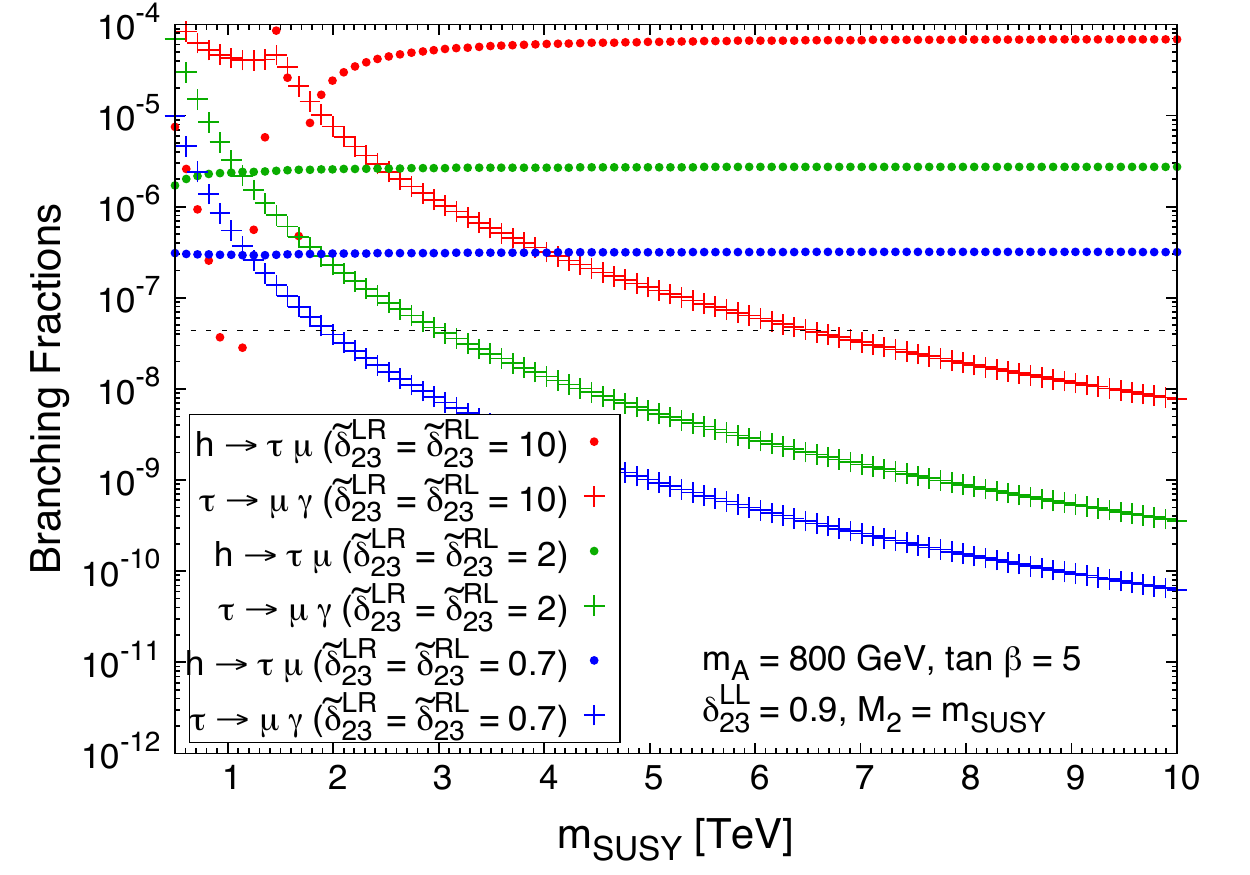} \\
\includegraphics[width=80mm]{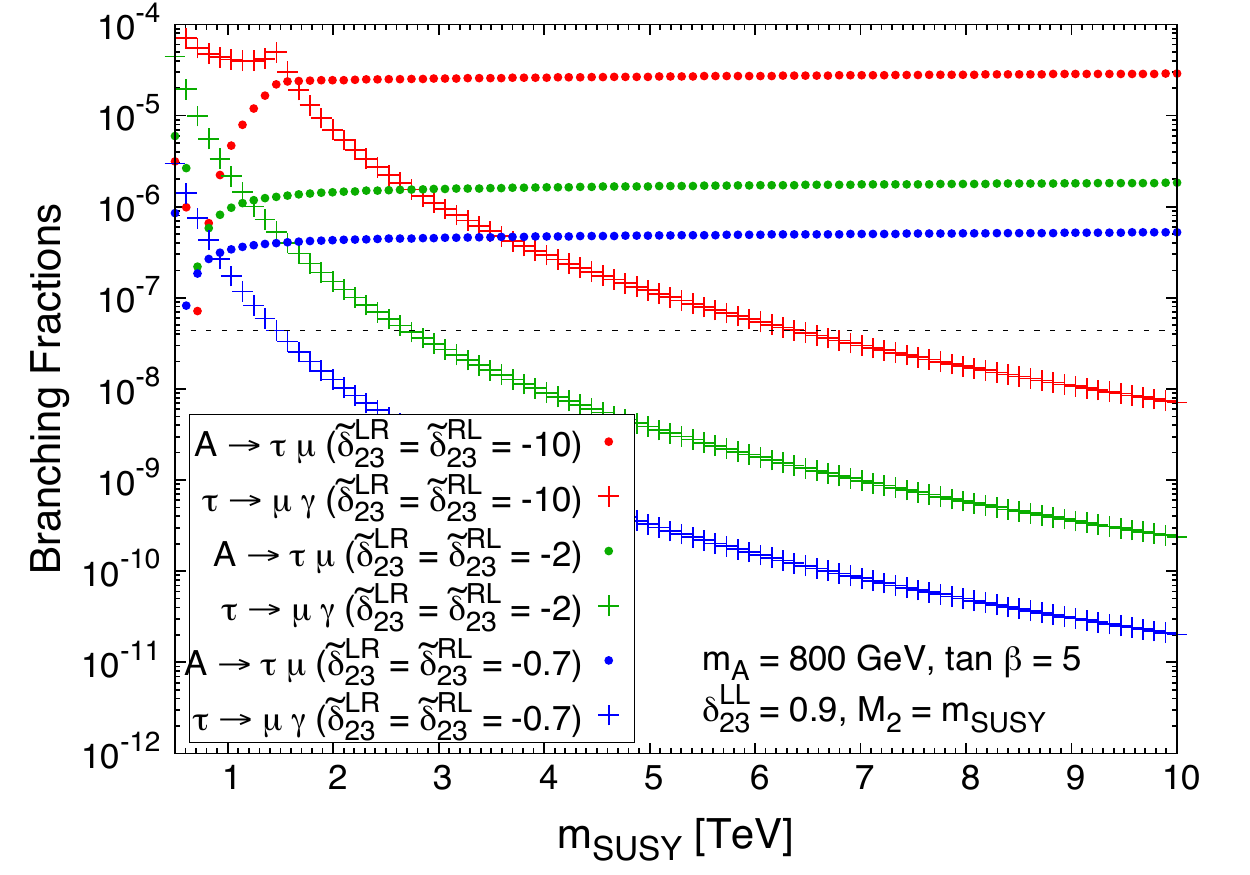} &
\includegraphics[width=80mm]{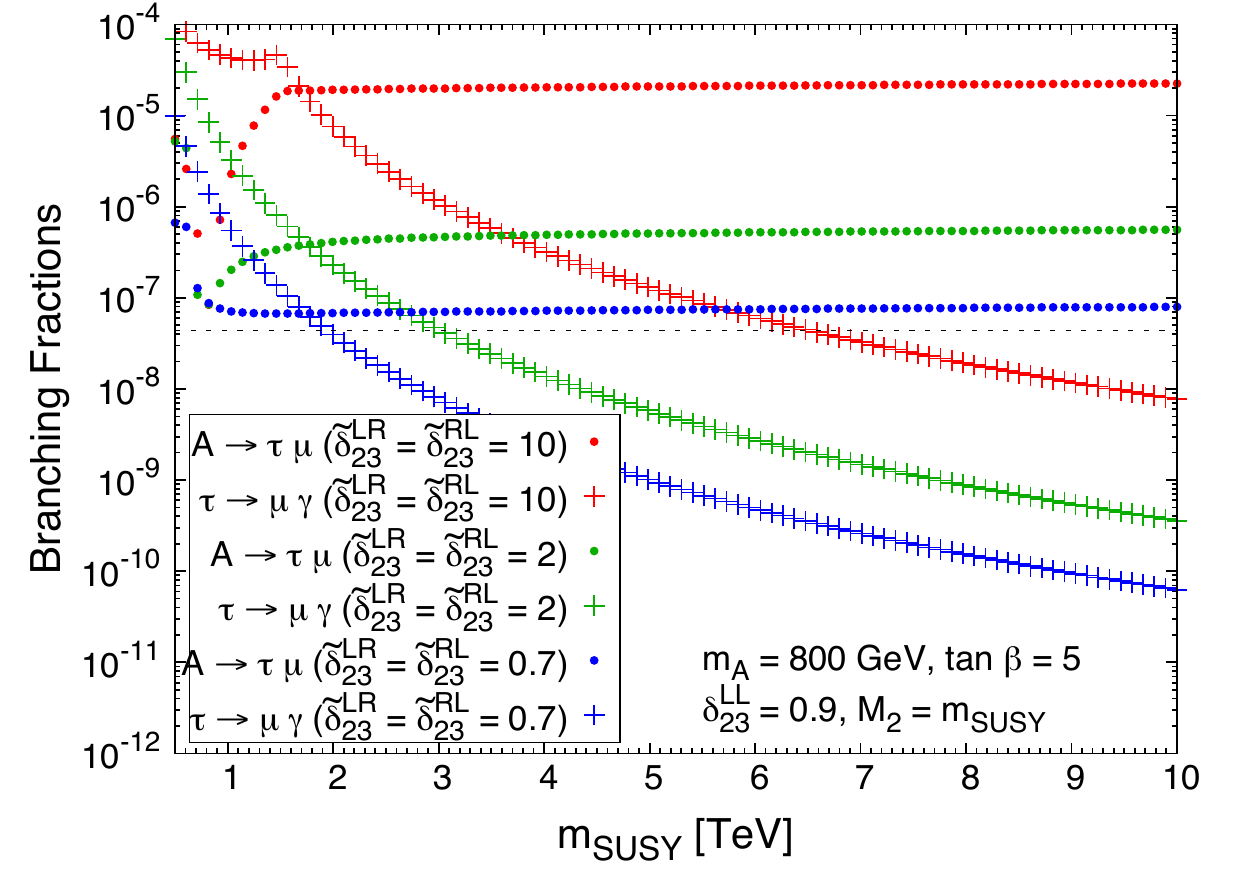} 
\end{tabular}
\caption{Sensitivity to double $LL$ and $LR$ mixing deltas for the scenarios defined in Section \ref{pmssmvheavysusy}: 
LFV Higgs decay rates (dots) and BR($\tau \to \mu \gamma$) (crosses) 
as functions of $m_\text{SUSY}$ with $\delta_{23}^{LL} = 0.9$ 
for different choices of
negative $LR$ mixings (left panels),
$\tilde \delta_{23}^{LR} = \tilde \delta_{23}^{RL}$: $-0.7$ (in blue), $-2$ 
(in green) and $-10$ (in red), and of positive $LR$ mixings (right panels), 
$\tilde \delta_{23}^{LR} = \tilde \delta_{23}^{RL}$: $+0.7$ (in blue), $+2$ 
(in green) and $+10$ (in red). The
results for $H$ (not shown) are nearly identical to those of $A$. 
In each case, the other flavour changing deltas are set to zero. 
In all panels, $m_A =$ 800 GeV, $\tan\beta =$ 5, $M_2 = m_\text{SUSY}$ and the other 
MSSM parameters are set to the values reported in the text. 
The horizontal dashed line denotes the current experimental 
upper bound for $\tau \to \mu \gamma$ channel, 
BR($\tau \to \mu \gamma$) $< 4.4 \times 10^{-8}$~\cite{Aubert:2009ag}.}
\label{BRs-deltatildeLRRL}
\end{figure}

Finally, we have studied the possibility of switching on several deltas 
simultaneously. Specifically, we have fixed $\delta_{23}^{LL} = 0.9$ and 
considered different choices of 
$\tilde \delta_{23}^{LR} = \tilde \delta_{23}^{RL}$
with either negative values:  
($-0.7$, $-2$ and $-10$), or positive values ($+0.7$,
$+2$ and $+10$). 
The results are depicted in Figure~\ref{BRs-deltatildeLRRL} for the case of low 
$\tan\beta =$ 5 that is the most interesting one since the $LL$ and $LR$
(and $RL$) contributions are of similar size and their interferences can be
relevant for some regions of the parameter space.    
As expected, the four LFV decay rates increase as 
$|\tilde \delta_{23}^{LR}| = |\tilde \delta_{23}^{RL}|$ grows, 
and they are slightly
higher than for single $LL$ or $LR$ mixings. The most 
important conclusion in this case is that we are able to obtain large 
branching ratios for all the three LFV Higgs decays, reaching values close 
to $10^{-4}$ for $h$ and about $3\times 10^{-5}$ for $A$ and $H$, if 
$\tilde \delta_{23}^{LR} = \tilde \delta_{23}^{RL} = \pm 10$. By comparing the
results for negative versus positive $LR$ mixings, we also learn from this
figure that there are not relevant differences. The LFV Higgs decay rates for 
negative mixings
are slightly higher than the corresponding rates for positive mixings, and this
difference is more visible in the $A$ and $H$ LFV decays than in the $h$ LFV 
decay. It should also be noted that the rates for $\tau \to \mu \gamma$ decays
go the other way around, namely, they are slightly larger for positive $LR$
mixings than for negative $LR$ mixings, indicating that the interference
between the $LL$ and $LR$ contributions must be of opposite sign in the LFV 
Higgs decays versus the $\tau \to \mu \gamma$ decays.

To close this subsection, we can conclude from Figures~\ref{BRs-mSUSY}, 
\ref{BRs-tanb}, \ref{BRs-M2}, \ref{BRs-deltas} and~\ref{BRs-deltatildeLRRL} 
that, for the explored intervals of the parameter space, the largest LFV 
Higgs rates that are allowed by the $\tau \to \mu \gamma$ upper bound are 
obtained for the following values of the model parameters: large 
$m_\text{SUSY} \gtrsim$ 5 TeV, $M_2$ close to $m_\text{SUSY}$ and
$|\delta_{23}^{LL}|$ and $|\tilde\delta_{23}^{LR}|$ (and/or
$|\tilde\delta_{23}^{RL}|$) close to their 
maximum explored values of 1 and 10 respectively. According to these previous findings, 
in the forthcoming computation of cross sections and
event rates at the LHC, whenever we have to fix them, we will set the following 
particular reference 
model parameters: 
$m_\text{SUSY} =$ 5 TeV, $M_2 = m_\text{SUSY}$, $\delta_{23}^{LL} = 0.9$ and $\tilde \delta_{23}^{LR} = \tilde \delta_{23}^{RL} = \pm 5$ , which are approximately the largest allowed values by the metastability bounds (see Section \ref{pmssmvheavysusy}). 
The corresponding predictions for other choices of $\delta_{23}^{LL}$, 
$\tilde \delta_{23}^{LR}$, $\tilde \delta_{23}^{RL}$, $M_2$, $m_\text{SUSY}$ 
and $\tan\beta$ can be easily derived from these commented five figures, from \ref{BRs-mSUSY} to \ref{BRs-deltatildeLRRL}.

\section{Results for the LFV cross sections and event rates at the LHC}
\label{LFVxsections}

In this section we present the results of the LFV cross sections and event 
rates at the LHC which are mediated by the production of neutral MSSM Higgs 
bosons and their subsequent LFV decays into $\tau \mu$. The scenarios evaluated are the same as in the previous section.
 The production cross 
sections of the neutral Higgs bosons are calculated here by means of the 
code {\tt FeynHiggs}~\cite{feynhiggs}. For low values of $\tan\beta$, 
the production cross sections of the three neutral MSSM Higgs bosons are 
dominated by gluon fusion. By contrast, for moderate and large values of 
$\tan\beta$ ($\gtrsim 10$), the production cross sections of $H$ and $A$ Higgs 
bosons 
via bottom-antibottom quark annihilation become the dominant ones, 
while the $h$ production cross section is still dominated by gluon fusion. 
In the following, we consider centre-of-mass energies at the LHC 
of $\sqrt{s} =$ 8 TeV (current phase) and $\sqrt{s} =$ 14 TeV (future phase) and focus in the two cases 
with the largest LFV Higgs decay rates, with either $LL$ or $LR$ or both 
slepton 
$\tau-\mu$ mixings. Although we do not expect any competitive background 
to these singular LFV signals at the LHC, a more realistic and devoted 
study of the potential backgrounds should be done, but this is beyond the 
scope of this work. 

\subsection{Predictions of LFV cross sections for single \texorpdfstring{$LL$}{LL} mixing}
\label{LLmixing}

\begin{figure}[t!]
\hspace{-0.5 cm}
\begin{tabular}{cc}
\includegraphics[width=80mm]{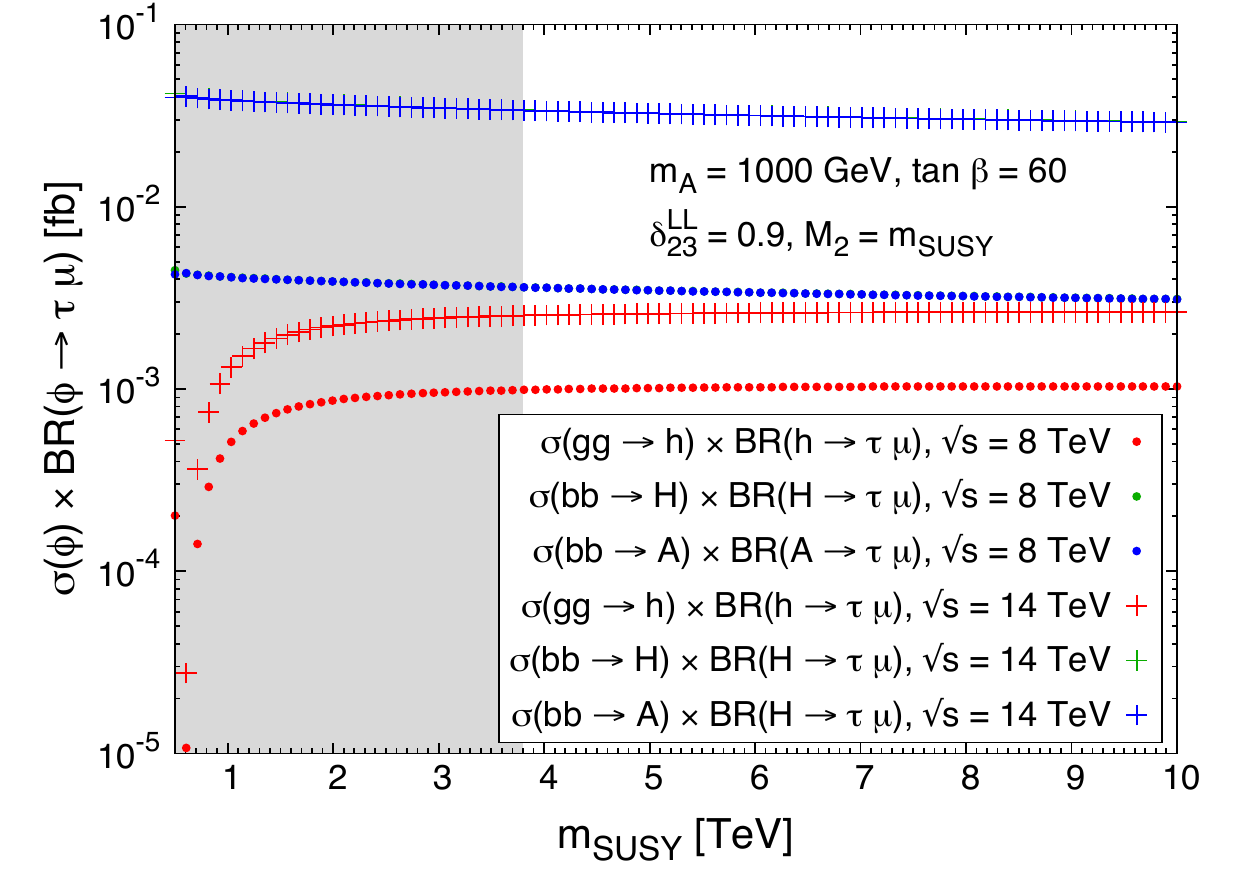} &
\includegraphics[width=80mm]{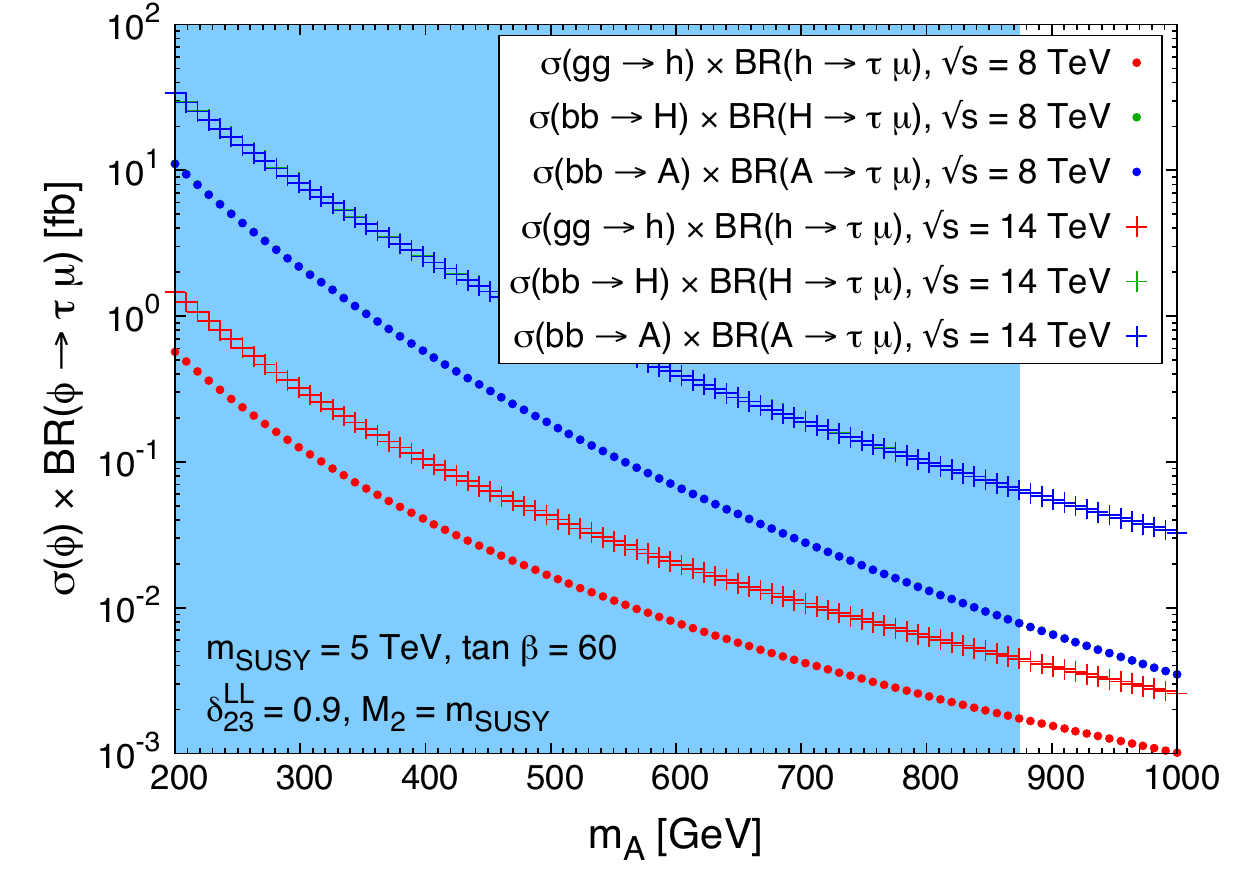}
\end{tabular}
\caption{LFV cross sections at the LHC for single $LL$ mixing for the scenarios defined in Section \ref{pmssmvheavysusy}. Left panel: $\sigma(\phi)$ $\times$ BR($\phi \to \tau \mu$) as a 
function of $m_\text{SUSY}$, where $\phi = h, H, A$, with $m_A =$ 1000 GeV and $\tan\beta =$ 60, 
for $\sqrt{s} =$ 8 TeV (dots) and $\sqrt{s} =$ 14 TeV (crosses). The shaded 
grey area is excluded by the $\tau \to \mu \gamma$ upper bound.
Right panel: $\sigma(\phi)$ $\times$ BR($\phi \to \tau \mu$) as a function 
of $m_A$, where $\phi = h, H, A$, with $\tan\beta =$ 60 and $m_\text{SUSY} =$ 5 TeV for 
$\sqrt{s} =$ 8 TeV (dots) and $\sqrt{s} =$ 14 TeV (crosses). The shaded blue 
area is excluded by CMS searches for MSSM neutral Higgs bosons 
decaying to $\tau {\bar \tau}$ pairs ~\cite{CMS-PAS-HIG-12-050} in the $m_A-\tan\beta$ parameter space 
for the MSSM $m_h^\text{max}$ scenario \cite{Carena:2002qg}\cite{Carena:2013qia}. 
The production cross section of the light Higgs boson $h$ is calculated via 
gluon fusion and the $H$ and $A$ production cross sections are calculated 
via bottom-antibottom quark annihilation. In both panels $\delta_{23}^{LL} =$ 0.9 and the other MSSM 
parameters are set to the values reported in the text, with $M_2 = m_\text{SUSY}$. The green dots and crosses are overimposed with the blue ones.}\label{delta23LL0.9}
\end{figure}

We first consider the case of single $LL$ mixing which is the delta parameter 
best motivated from theory. In fact, most of the top-down approaches to 
LFV physics, where the mixing parameters are generated through RGE running, 
as for instance in SUSY-seesaw models, predict that the largest mixing comes 
from $\delta_{23}^{LL}$, with the following hierarchy: 
$\delta_{23}^{LL}$ $\gg$ $\delta_{23}^{LR}$ $\gg$ $\delta_{23}^{RR}$. This hierarchy can be understood by the solution to the RGE-induced intergenerational mixing, by using the one-loop leading-log approximation where the approximate solution for the off-diagonal terms can be written as \cite{Hisano:1995cp}:

\begin{align}
(\Delta m_{\tilde{L}}^2)_{ij}&\,=\,
-\frac{1}{8\, \pi^2}\, (3\, M_0^2+ A_0^2)\, (Y_{\nu}^\dagger\, 
L\, Y_{\nu})_{ij} 
\,,\nonumber \\
(\Delta {\cal A}^l)_{ij}&\,=\,
- \frac{3}{16 \,\pi^2}\, A_0\, y_{l}\, (Y_{\nu}^\dagger\, L\, Y_{\nu})_{ij}
\,,\nonumber \\
(\Delta m_{\tilde{E}}^2)_{ij}&\,=\,
0\,\,;\, L_{kl}\, \equiv \,\log \left( \frac{M_X}{m_{M_k}}\right) \,
\delta_{kl}.
\end{align}
where $i,j,k=1...3$, $m_{M_k}$ are the Majorana Masses, $M_X$ the GUT scale, $y_{l}$ the Yukawa of the lepton, and $A_0$ and $M_0$ the universal scalar soft mass and trilinear coupling at the GUT scale. This motivates us to investigate further this case and to look for potential 
sizeable branching ratios and measurable cross sections at the LHC with 
just this single $\delta_{23}^{LL}$. In this case, we focus on the large $\tan
\beta$ region where, as we have seen previously, the LFV rates are maximized.  
The most optimistic numbers which we have been able 
to achieve for $\delta_{23}^{LL}$ are displayed in Figure~\ref{delta23LL0.9}, 
which correspond to the following choice of parameters: 
$M_2 = m_\text{SUSY}$, $\delta_{23}^{LL} =$ 0.9, 
$\tan\beta = 60$ and $m_A =$ 1000 GeV (in order to be in accord with CMS searches~\cite{CMS-PAS-HIG-12-050} for MSSM neutral Higgs bosons 
decaying to $\tau {\bar \tau}$ pairs). On the left panel of Figure~\ref{delta23LL0.9} one can observe the cross sections for all the $h, H, A \to \tau \mu$ channels, for the current LHC phase (dots) and for the future one (crosses), where the shaded grey area is the excluded region by the $\tau \to \mu \gamma$ upper bound. As we can clearly see in this plot, 
the moderate cross sections predicted in the allowed region of 
$m_\text{SUSY}$ indicate that total integrated luminosities of 
${\cal L} \gtrsim 300$ fb$^{-1}$ and ${\cal L} \gtrsim 30$ fb$^{-1}$ 
would be required at least for the $H, A \to \tau \mu$ channels, 
in the current LHC phase and in the future one, respectively, in order to 
obtain events if the unique source of lepton flavour violation is 
$\delta_{23}^{LL}$. For the $h \to \tau \mu$ channel it would be needed 
even larger values of ${\cal L}$, namely, $1 \times 10^3$ fb$^{-1}$ in the present stage of the LHC and 400 fb$^{-1}$ in the future one.

The low rates obtained for the $H, A \to \tau \mu$ channels are due 
to the large value of $m_A$ considered up to now, which reduces the 
production cross sections of these two heavy Higgs bosons. Now we are 
going to scan over a wider region of $m_A$ values, from 200 GeV to 1000 GeV, 
fixing $m_\text{SUSY}$ to our reference  value of 5 TeV. The
right panel of Figure~\ref{delta23LL0.9} shows the results of 
$\sigma(\phi)$ $\times$ BR($\phi \to \tau \mu$) as a function of 
$m_A$ with a LHC centre-of-mass energy of $\sqrt{s} =$ 8 TeV (dots) 
and $\sqrt{s} =$ 14 TeV (crosses). The cross sections of LFV Higgs decays 
are strongly dependent on $m_A$ 
and fall down more than three orders of magnitude from $m_A =$ 200 GeV to 
$m_A =$ 1000 GeV. On one hand, the LFV $H$ and $A$ channels decrease 
with $m_A$, as expected, because this parameter gives precisely the mass 
of the Higgs particles being produced. On the other hand, the cross section for
the
$h \to \tau \mu$ process diminishes with this parameter since our settings here lay in the 
so-called decoupling limit, meaning that for large $m_A$ values 
the lightest Higgs boson $h$ behaves as a SM-like Higgs boson, and in the SM the coupling between the Higgs boson and a $\tau$ and a $\mu$ is zero, therefore leading
to extremely suppressed 
LFV couplings. For low values of $m_A \sim $ 200 GeV, we obtain large cross sections, 
$\sigma(gg \to h) \times$ BR$(h \to \tau \mu) \simeq 0.6 \textrm{fb}$ and 
$\sigma(b \bar b \to H, A) \times$ BR$(H, A \to \tau \mu) \simeq 10 \textrm{fb}$ 
in the present LHC phase, and $\sigma(gg \to h) \times$ 
BR$(h \to \tau \mu) \simeq 2 \textrm{fb}$ and 
$\sigma(b \bar b \to H, A) \times$ BR$(H, A \to \tau \mu) \simeq 30 \textrm{fb}$ in the 
future one. However, these low values of $m_A$ are not allowed by present LHC
data. Indeed, we can see in the shaded area (in blue) of this figure 
that values of 
$m_A \lesssim$ 875 GeV for 
$\tan\beta =$ 60 are already excluded by CMS searches for MSSM neutral Higgs bosons 
decaying to $\tau {\bar \tau}$ pairs ~\cite{CMS-PAS-HIG-12-050} in the $m_A-\tan\beta$ parameter space 
for the MSSM $m_h^\text{max}$ scenario \cite{Carena:2002qg}\cite{Carena:2013qia}. 

The overall conclusion for the $LL$ case is that, in order to get some Higgs-mediated LFV event at the LHC for $m_A$ and $\tan \beta$ input values within the CMS allowed region, total integrated luminosities of ${\cal L} \gtrsim 100$ fb$^{-1}$ and ${\cal L} \gtrsim 20$ fb$^{-1}$ would be required for the $H, A \to \tau \mu$ channels, in the current LHC phase and in the future one, respectively. For the $h \to \tau \mu$ channel it would be needed larger values of ${\cal L}$: 500 fb$^{-1}$ in the present LHC phase and 200 fb$^{-1}$ in the future one.

\subsection{Predictions of LFV cross sections for single \texorpdfstring{$LR$}{LR} mixing}

\label{LRmixing}

\begin{figure}[t!]
\hspace{-0.5 cm}
\begin{tabular}{cc}
\includegraphics[width=80mm]{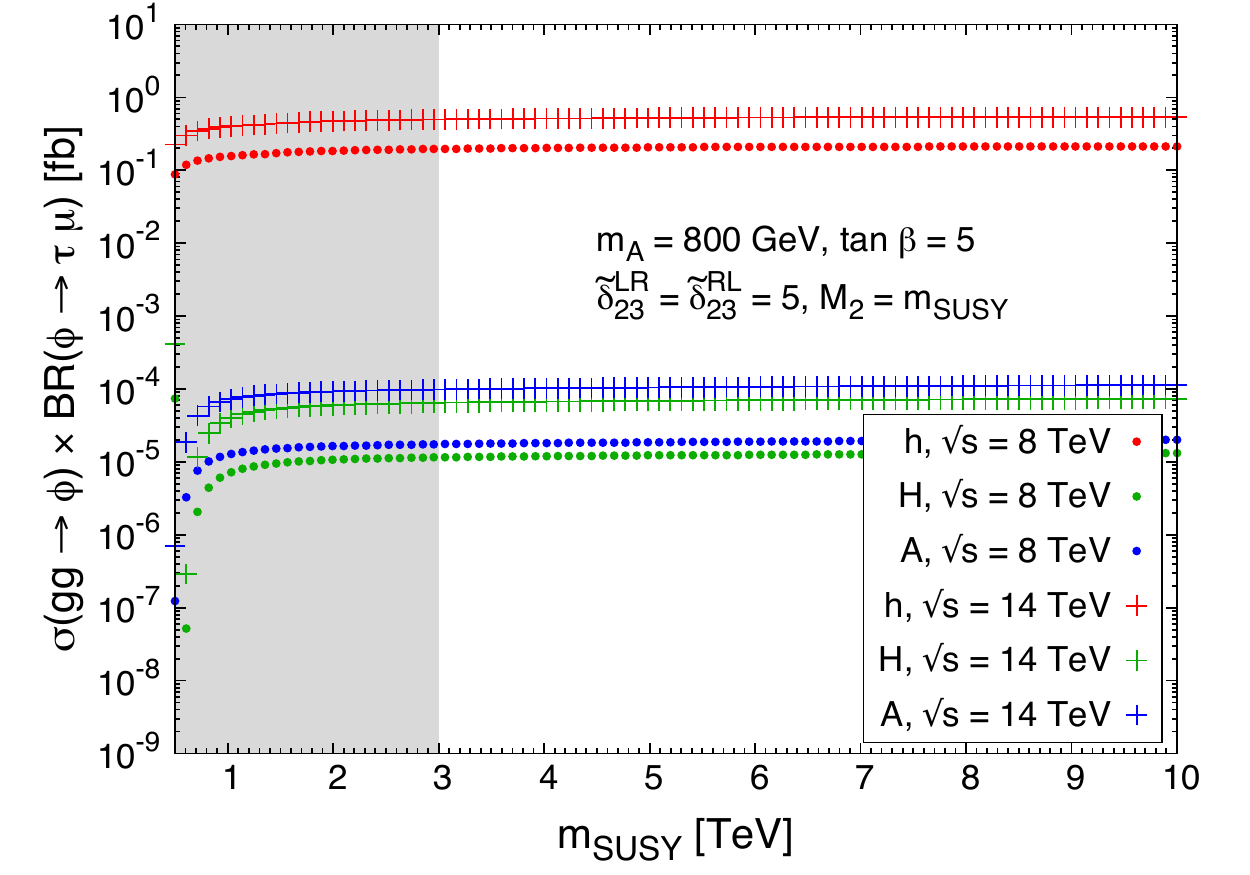} &
\includegraphics[width=80mm]{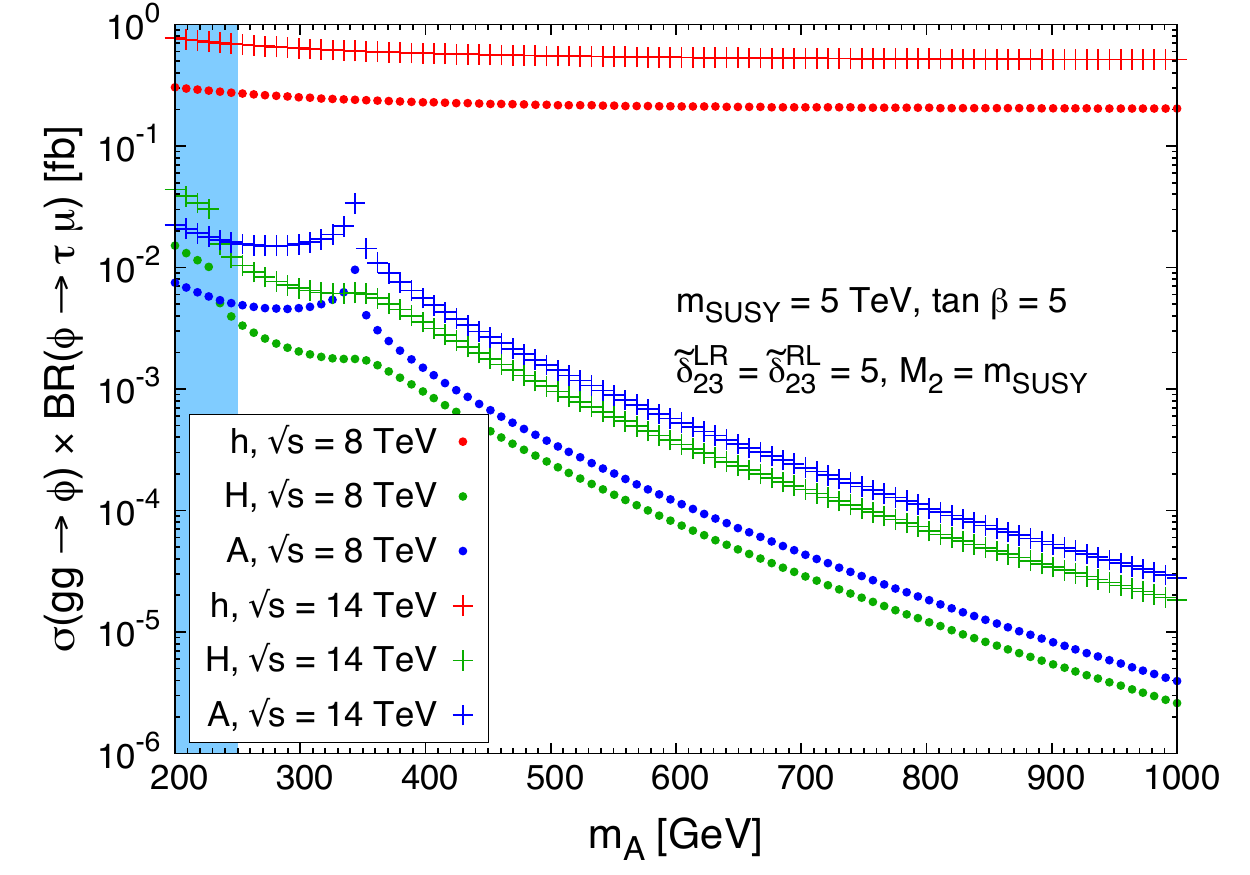}
\end{tabular}
\caption{LFV cross sections at the LHC for single $LR$ mixing for the scenarios defined in Section \ref{pmssmvheavysusy}. Left panel: $\sigma(\phi)$ $\times$ BR($\phi \to \tau \mu$) 
as a function of $m_\text{SUSY}$, where $\phi = h, H, A$, with $m_A =$ 800 GeV and $\tan\beta =$ 5,
for $\sqrt{s} =$ 8 TeV (dots) and $\sqrt{s} =$ 14 TeV (crosses). 
The shaded grey area is excluded by the $\tau \to \mu \gamma$ upper bound.
Right panel: $\sigma(\phi)$ $\times$ BR($\phi \to \tau \mu$) as a 
function of $m_A$, where $\phi = h, H, A$, with  $\tan\beta =$ 5 and $m_\text{SUSY} =$ 5 TeV 
for $\sqrt{s} =$ 8 TeV (dots) and $\sqrt{s} =$ 14 TeV (crosses). 
The shaded blue area is excluded by CMS searches for MSSM neutral 
Higgs bosons. The Higgs production cross sections are calculated via gluon 
fusion. In both panels $\tilde \delta_{23}^{LR} = \tilde \delta_{23}^{RL} =$ 5 and the other MSSM parameters are set to the values reported in the text, with $M_2 = m_\text{SUSY}$.}\label{deltahat23LRRL5}
\end{figure}

Next we consider the case of single $LR$ mixing. For the numerical 
estimates here we fix 
$\tilde \delta_{23}^{LR} = \tilde \delta_{23}^{RL} =$ 5 and focus in the low
$\tan \beta$ region where, as we have seen previously, the LFV rates are
maximized. 
The predictions of the cross sections of LFV Higgs processes at 
the LHC are exhibited in Figure~\ref{deltahat23LRRL5} for 
$\sqrt{s} =$ 8 TeV (dots) and $\sqrt{s} =$ 14 TeV (crosses), with 
$\tan\beta =$ 5 and $M_2 = m_\text{SUSY}$. On the left panel we show 
the predicted LFV cross sections as functions of $m_\text{SUSY}$, 
with $m_A =$ 800 GeV. For the $m_\text{SUSY}$ region allowed by the 
$\tau \to \mu \gamma$ upper bound ($m_\text{SUSY} \gtrsim$ 3 TeV), 
$\sigma(gg \to h)$ $\times$ BR($h \to \tau \mu$) can reach values 
around 0.2 fb in the present LHC phase and up to 0.5 fb in the future one. 
The predictions for $\sigma(gg \to H, A)$ $\times$ BR($H, A \to \tau \mu$) 
are much smaller than the LFV $h$ channel, since the value of $m_A$ 
considered here is rather heavy (800 GeV) and consequently leads to 
small production cross sections for both $H$ and $A$ Higgs bosons, 
and reach maximum values of about $10^{-4}$ fb for the future LHC phase, 
meaning that a total integrated luminosity of at least $10^4$ fb$^{-1}$ would be 
required in order 
to get at least one event.

The results for $\sigma(\phi)$ $\times$ BR($\phi \to \tau \mu$) 
(with $\phi = h, H, A$) as a 
function of $m_A$  are displayed on the right panel of 
Figure~\ref{deltahat23LRRL5} 
for $m_\text{SUSY} =$ 5 TeV. The largest cross section is, by far, 
for the $h \to \tau \mu$ channel again, even for low values of $m_A$. The cusp appearing around $m_A \simeq$ 350 GeV for the $A \to \tau \mu$ channel is due to the well known threshold effect when crossing by two on-shell top quarks\footnote{This threshold effect does not appear on the right panel of Figure~\ref{delta23LL0.9} because at large $\tan\beta$ the gluon fusion production is dominated by the one-loop diagrams with 
bottom-quark exchange.}. This effect also occurs for $H$ production via gluon fusion, but it is softer, resulting in a plateau instead of a cusp. For the LFV $A$ channel, we reach cross sections of 0.01 fb and 0.04 fb at the most in the present and the future LHC phases, respectively, around the top-quark threshold effect, while the largest LFV $H$ cross sections, for values of $m_A \lesssim$ 250 GeV, are excluded by CMS searches for MSSM neutral Higgs bosons decaying to $\tau {\bar \tau}$ pairs ~\cite{CMS-PAS-HIG-12-050} in the $m_A-\tan\beta$ parameter space for the MSSM $m_h^\text{max}$ scenario \cite{Carena:2002qg}\cite{Carena:2013qia} (shaded blue area). The largest LFV $H$ cross section allowed by CMS searches is $4 \times 10^{-3}$ fb ($1 \times 10^{-2}$ fb) in the present (future) phase of the LHC. These results show us that, in the future LHC phase, a total integrated luminosity of ${\cal L} =$ 100 fb$^{-1}$ could be enough to produce some LFV event coming from the decays of the heavy neutral Higgs bosons. In the case of $h \to \tau \mu$ channel, this same luminosity could give rise to tens of events. 

\subsection{Predictions of LFV cross sections for double $LL$ and $LR$ mixings}
\label{LLLRmixing}

\begin{figure}[t!]
\hspace{-0.5 cm}
\begin{tabular}{cc}
\includegraphics[width=80mm]{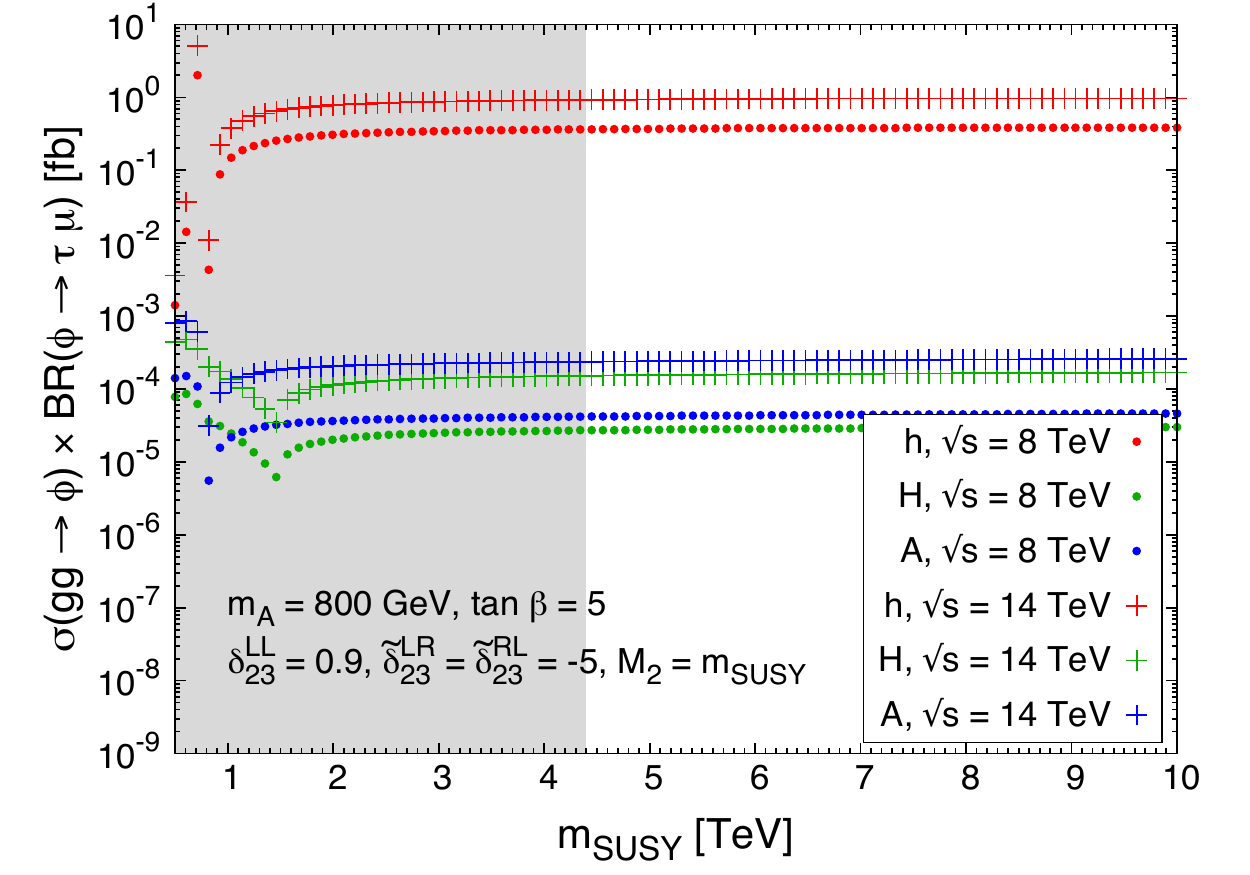} &
\includegraphics[width=80mm]{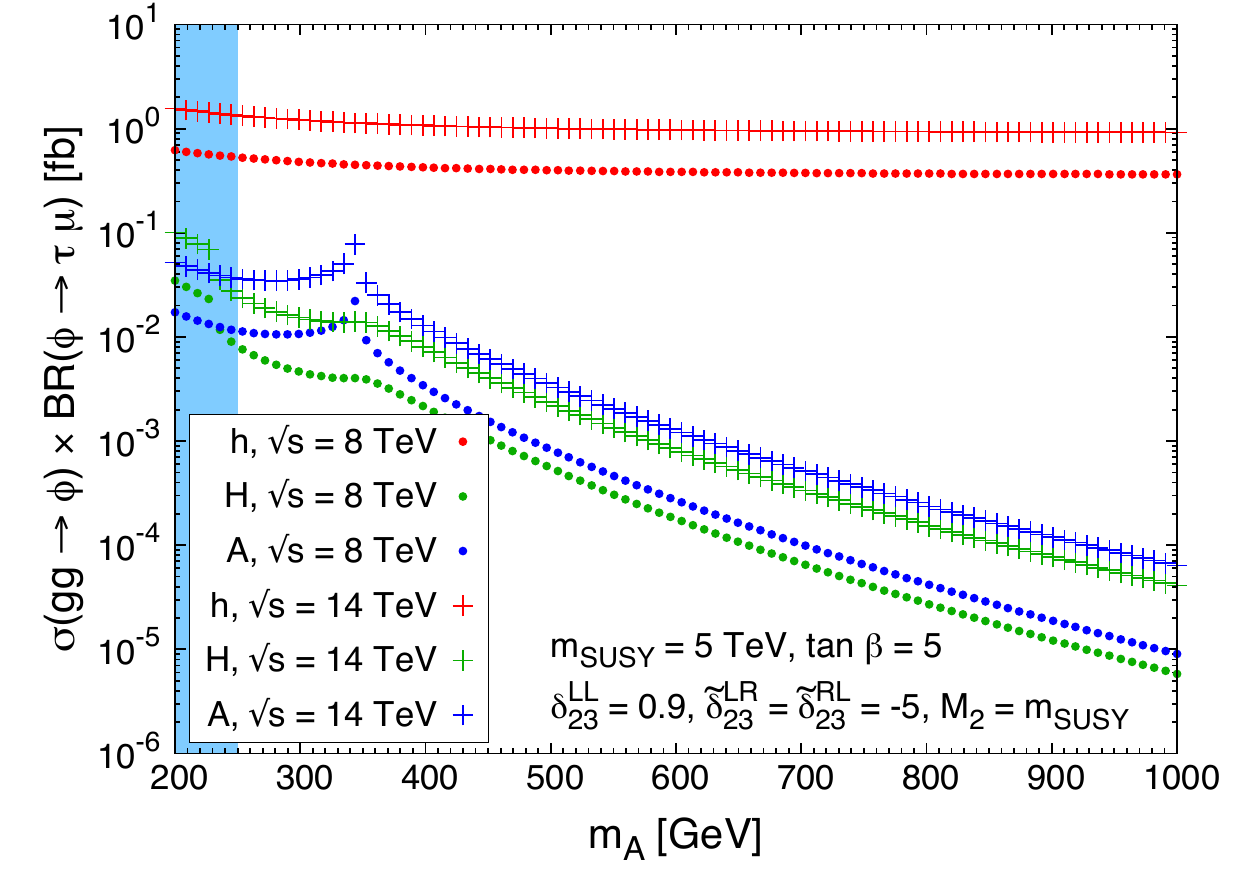}
\end{tabular}
\caption{LFV cross sections at the LHC for double $LL$ and $LR$ mixings and low
$\tan\beta$ for the scenarios defined in Section \ref{pmssmvheavysusy}. Left panel: $\sigma(\phi)$ $\times$ BR($\phi \to \tau \mu$) as a function of $m_\text{SUSY}$, where $\phi = h, H, A$, with $m_A =$ 800 GeV and $\tan\beta =$ 5 for $\sqrt{s} =$ 8 TeV (dots) and $\sqrt{s} =$ 14 TeV (crosses). The shaded grey area is excluded by the $\tau \to \mu \gamma$ upper bound.
Right panel: $\sigma(\phi)$ $\times$ BR($\phi \to \tau \mu$) as a function of $m_A$, where $\phi = h, H, A$, with $\tan\beta =$ 5 and $m_\text{SUSY} =$ 7 TeV for $\sqrt{s} =$ 8 TeV (dots) and $\sqrt{s} =$ 14 TeV (crosses). The shaded blue areas are excluded by CMS searches for MSSM neutral Higgs bosons. The Higgs production cross sections are calculated via gluon fusion. In both panels $\delta_{23}^{LL} =$ 0.9, $\tilde \delta_{23}^{LR} = \tilde \delta_{23}^{RL} =-5$ and the other MSSM parameters are set to the values reported in the text, with $M_2 = m_\text{SUSY}$.}\label{delta2309-LRRL10tanb5}
\end{figure}

In this section we present the cross sections of LFV Higgs processes at the 
LHC in the case of switching on several deltas at the same time. Specifically, 
we fix $\delta_{23}^{LL} =$ 0.9 and 
$\tilde \delta_{23}^{LR} = \tilde \delta_{23}^{RL} =-5$ and focus on both cases
with either low or large $\tan\beta$ values. We next present the predictions
of the LFV cross sections as functions of $m_\text{SUSY}$ and $m_A$. 
Figure~\ref{delta2309-LRRL10tanb5} is devoted to the low $\tan\beta$ case 
with $\tan\beta =$ 5. Both panels show very similar results to 
those of Figure~\ref{deltahat23LRRL5}, but it is interesting to notice a 
slight increase in the LFV cross sections when compared to the previous single 
$LR$ and
$LL$ cases that allow us to require smaller total 
integrated luminosities in order to 
get events at the LHC. For example, in the future phase of the LHC, 
around the top-quark pair production threshold, 
${\cal L} \lesssim$ 10 fb$^{-1}$ would be enough to obtain one event in the 
LFV pseudoscalar channel. 
The $h \to \tau \mu$ channel is even more promising and, 
for the same LHC phase, with a total integrated luminosity of just 1 fb$^{-1}$ 
we would get already more than one LFV event. As stated in the previous
sections, all these rates can be further enhanced if larger values of the
$LL$ and/or 
$LR$ mixings above those here assumed were taken.   

\begin{figure}[t!]
\hspace{-0.5 cm}
\begin{tabular}{cc}
\includegraphics[width=80mm]{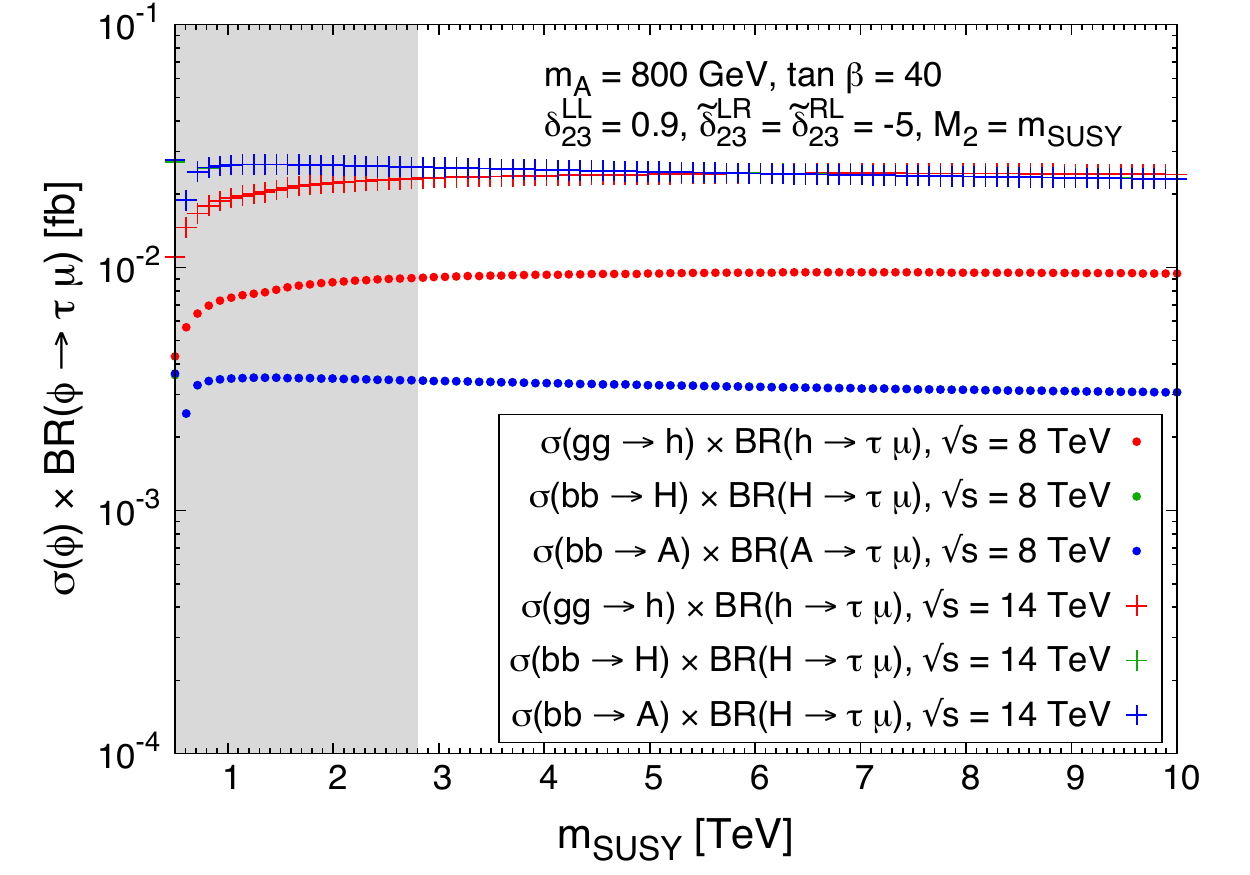} &
\includegraphics[width=80mm]{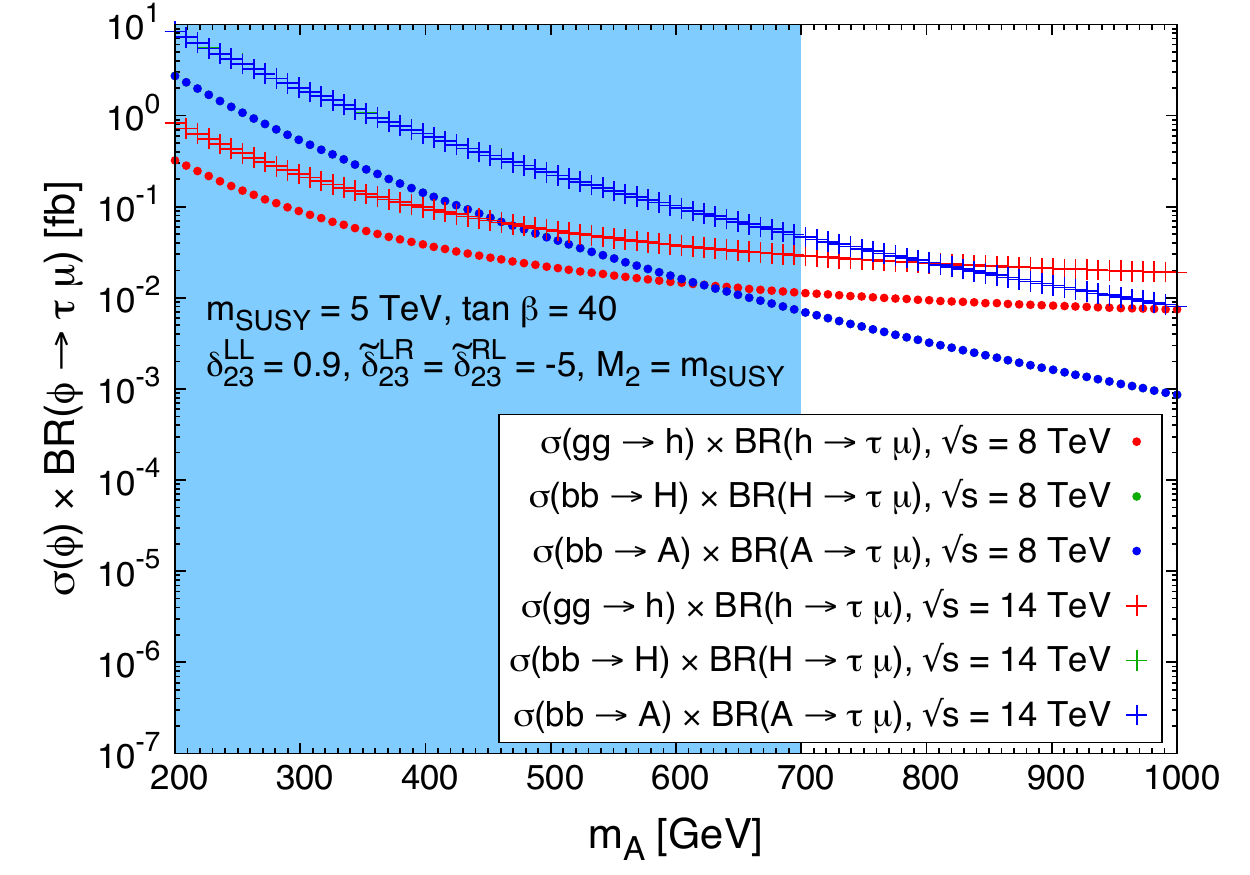}
\end{tabular}
\caption{LFV cross sections at the LHC for double $LL$ and $LR$ mixings and large
$\tan\beta$ for the scenarios defined in Section \ref{pmssmvheavysusy}. Left panel: $\sigma(\phi)$ $\times$ BR($\phi \to \tau \mu$) as a function of $m_\text{SUSY}$, where $\phi = h, H, A$, with $m_A =$ 800 GeV and $\tan\beta =$ 40 for $\sqrt{s} =$ 8 TeV (dots) and $\sqrt{s} =$ 14 TeV (crosses). The shaded grey area is excluded by the $\tau \to \mu \gamma$ upper bound.
Right panel: $\sigma(\phi)$ $\times$ BR($\phi \to \tau \mu$) as a function of
$m_A$, where $\phi = h, H, A$, with $\tan\beta =$ 40 and $m_\text{SUSY} =$ 5 TeV for $\sqrt{s} =$ 8
TeV (dots) and $\sqrt{s} =$ 14 TeV (crosses). The shaded blue area is 
excluded by 
CMS searches for MSSM neutral Higgs bosons. The production cross section of the light Higgs boson $h$ is calculated via gluon fusion and the $H$ and $A$ production cross sections are calculated 
via bottom-antibottom quark annihilation. In both panels, $\delta_{23}^{LL} =$ 0.9, $\tilde \delta_{23}^{LR} = \tilde \delta_{23}^{RL} =-5$ and the other MSSM parameters are set to the values reported in the text, with $M_2 = m_\text{SUSY}$.}\label{delta2309-LRRL10tanb40}
\end{figure}

The results for the LFV cross sections in the large $\tan\beta$ region are 
shown in Figure~\ref{delta2309-LRRL10tanb40}, for $\tan\beta =$ 40, 
as functions of $m_\text{SUSY}$ (left panel) and as functions of $m_A$ 
(right panel). These results are qualitatively similar to those of 
Figure~\ref{delta23LL0.9}, but the main difference in this case is that 
all the three LFV cross sections, for $h$, $H$ and $A$, 
are numerically of the same order of magnitude, due to the enhancement from 
both $LR/RL$ and
$LL$ mixing acting simultaneously and compensating each other in the two
studied regions of low and large $\tan\beta$ values. This is particularly
interesting for the future LHC phase where we find that a total integrated 
luminosity of just 50 fb$^{-1}$ would be needed to get a few LFV events 
in either of the three LFV Higgs channels.

\subsection{Predictions of LFV event rates at the LHC in the \texorpdfstring{$(m_A,\tan\beta)$}{mA-tanb} plane}
\label{NeventsLHC}

Once we have set up the most relevant parameters for the present 
study of LFV at the LHC, which are the two flavour mixing 
deltas $\delta_{23}^{LL}$ and $\delta_{23}^{LR}$ 
(and correspondingly $\delta_{23}^{RL}$), $m_\text{SUSY}$ and $m_A$ masses 
and $\tan\beta$, we will present next the results for the final rates 
at the LHC, both in the present and future phases, in the most convenient 
way for comparison with future experimental analysis, namely, in the 
($m_A,\tan\beta$) plane. Since the results
for the $H \to \tau \mu$ event rates turn out to be nearly equal 
to those of the 
$A \to \tau \mu$ ones will not exhibited them here for shortness. Thus we will focus
in the LFV rates of $h$ and $A$ decays. In the following plots we have
also specified the areas of the $(m_A,\tan\beta)$ plane (blue areas) that  
are excluded by 
the recent CMS searches for MSSM neutral Higgs bosons decaying to 
$\tau {\bar \tau}$ pairs ~\cite{CMS-PAS-HIG-12-050} in the so-called $m_h^\text{max}$ 
scenario \cite{Carena:2002qg}\cite{Carena:2013qia} (variations are possible due to deviations
from that scenario, see \cite{Carena:2013qia}).  All the predictions shown next are allowed
by the present upper $\tau \to \mu \gamma$ bound. 

\begin{figure}[t!]
\hspace{-0.5 cm}
\begin{tabular}{cc}
\includegraphics[width=80mm]{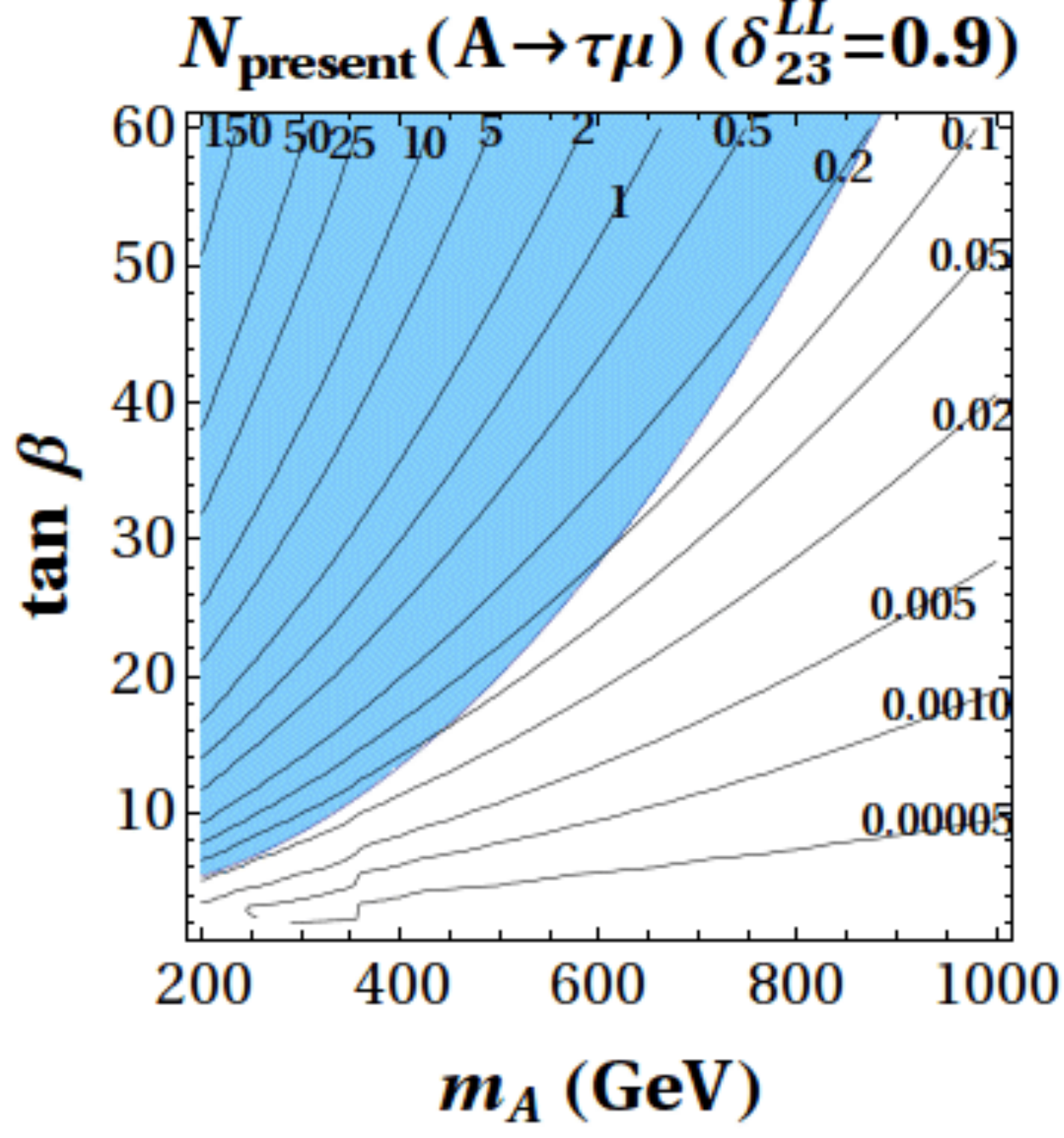} &
\includegraphics[width=80mm]{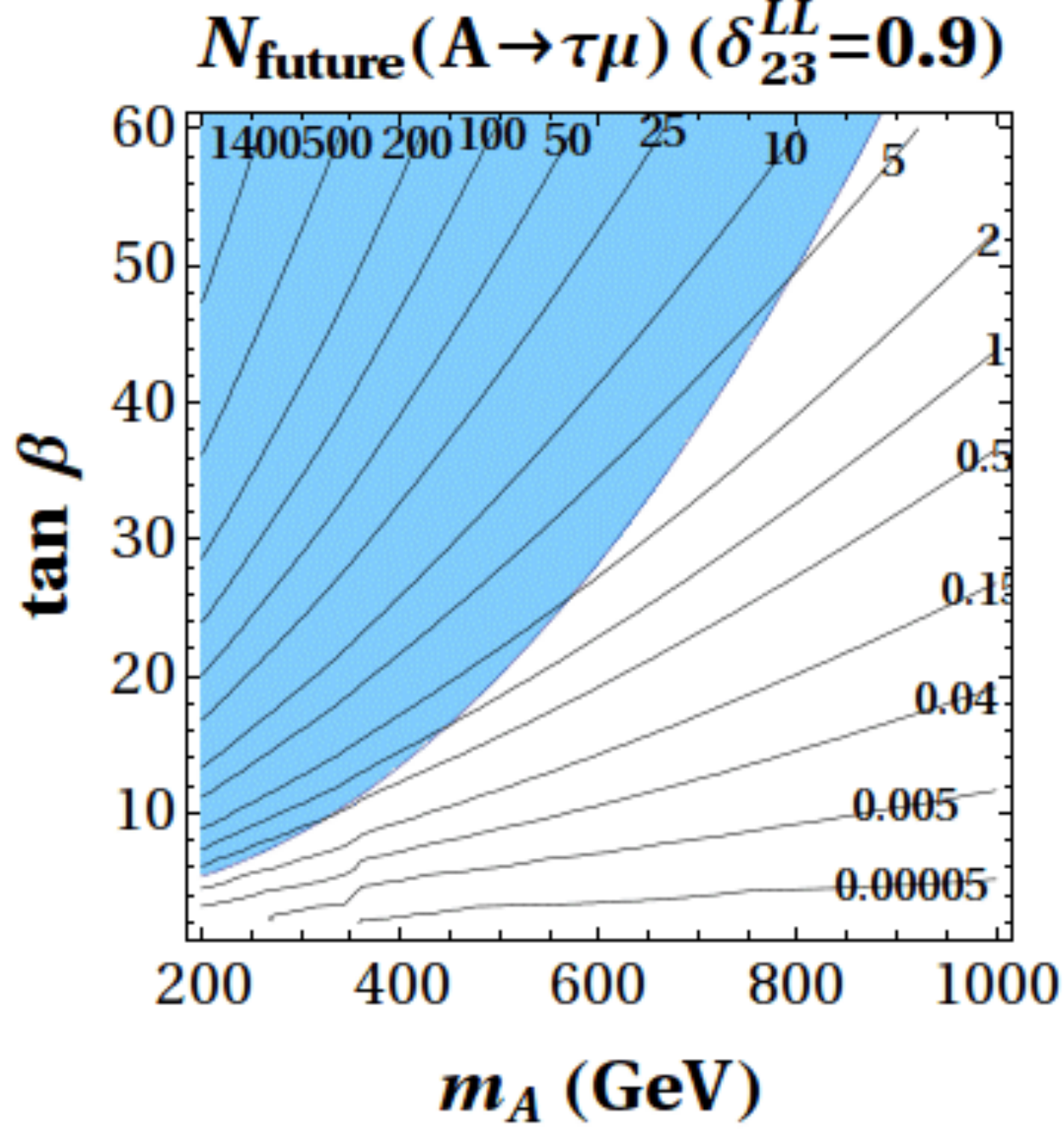}
\end{tabular}
\caption{Number of expected LFV events in the ($m_A,\tan\beta$) plane 
from $A \to \tau \mu$ for $\delta_{23}^{LL} =$ 0.9 and 
$m_\text{SUSY} =$ 5 TeV for the scenarios defined in Section \ref{pmssmvheavysusy}. Left panel: present phase of the 
LHC with $\sqrt{s} =$ 8 TeV and ${\cal L} =$ 25 fb$^{-1}$. 
Right panel: future phase of the LHC with $\sqrt{s} =$ 14 TeV and 
${\cal L} =$ 100 fb$^{-1}$. In both panels the other MSSM parameters 
are set to the values reported in the text, with $M_2 = m_\text{SUSY}$. 
The shaded blue areas are excluded by CMS 
searches~\cite{CMS-PAS-HIG-12-050}  (see variations from these limits for different scenarios in \cite{Carena:2013qia}). The results for $H$ (not shown)
are nearly equal to these ones for $A$.}\label{NeventsA-mAtanb}
\end{figure}

We start this analysis with the $LL$ case and plot in 
Figure~\ref{NeventsA-mAtanb} the number of events expected in the 
($m_A,\tan\beta$) plane for the $A \to \tau \mu$ channel with 
$\delta_{23}^{LL} =$ 0.9 and $m_\text{SUSY} =$ 5 TeV, considering both 
the present and future LHC phases (left and right panels, respectively). 
 On the other hand, the $h \to \tau \mu$ channel (not shown) cannot supply any 
 significant signal 
 at the LHC in this $LL$ case, as one can infer from 
 Figure~\ref{delta23LL0.9}, unless extremely large total 
 integrated luminosities were considered (larger than 500 fb$^{-1}$). 
 Due to the CMS exclusion region in the ($m_A,\tan\beta$) 
 plane~\cite{CMS-PAS-HIG-12-050} for searches of MSSM neutral Higgs bosons 
decaying to $\tau {\bar \tau}$ pairs,  
 it is evident that we cannot expect any LFV event from neither 
 the $h \to \tau \mu$
 channel nor the $A,H\to \tau \mu$ channels in the present phase 
 of the LHC if the unique responsible for $\tau-\mu$ mixing is the 
 $\delta_{23}^{LL}$ parameter and $|\delta_{23}^{LL}| <$ 1. 
 The event rates from $A, H \to \tau \mu$ for the future phase of the LHC 
 are more promising, as shown 
 on the right panel of Figure~\ref{NeventsA-mAtanb}. For instance, 
 for $m_A \simeq$ 450 GeV and $\tan\beta \simeq$ 15, we could expect
 at least 1 event, and up to 5 for larger values of $m_A$ and 
 values of $\tan\beta \gtrsim$ 50.

\begin{figure}[t!]
\hspace{-0.5 cm}
\begin{tabular}{cc}
\includegraphics[width=80mm]{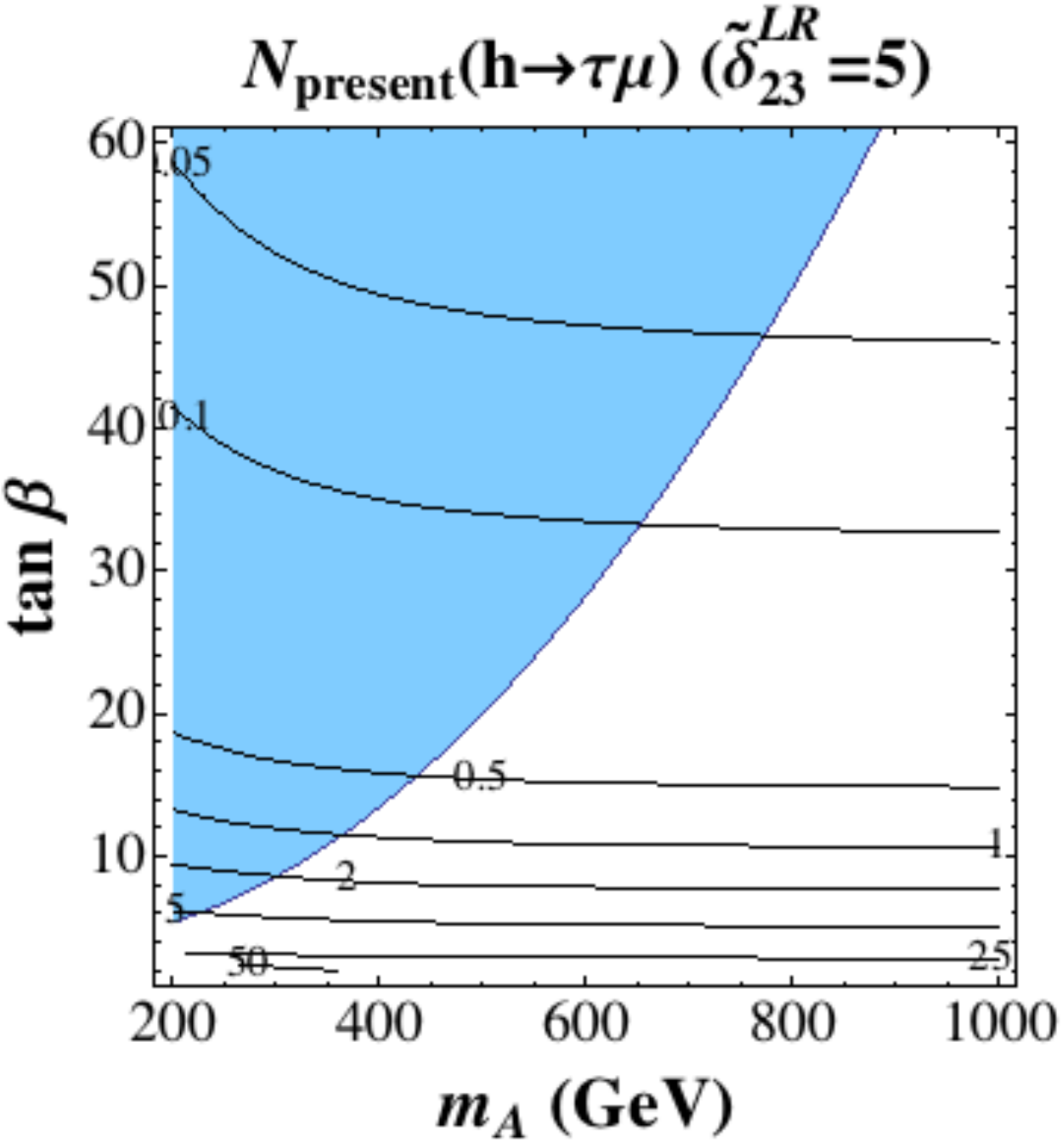} &
\includegraphics[width=80mm]{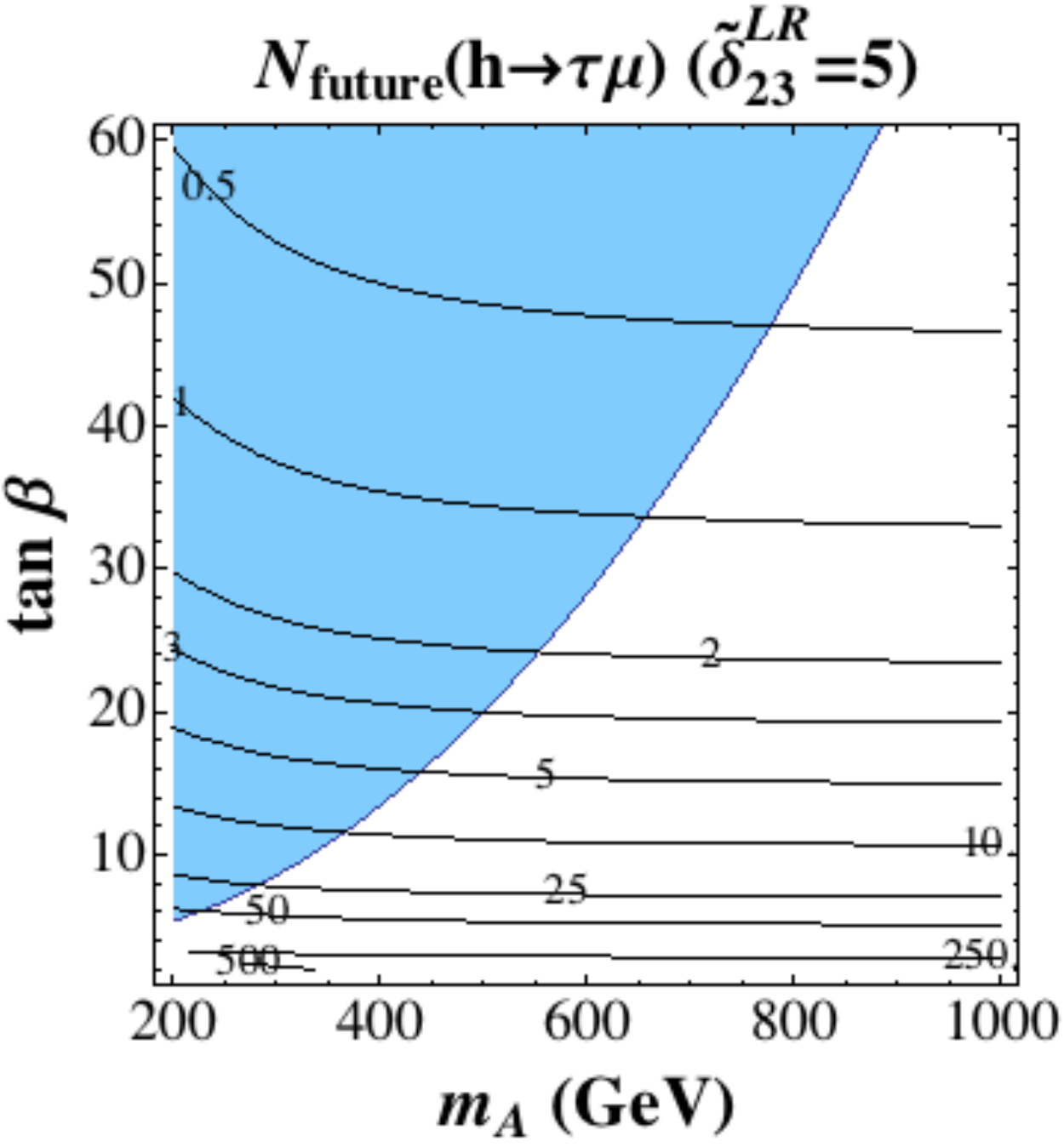}
\end{tabular}
\caption{Number of expected LFV events in the ($m_A,\tan\beta$) plane from 
$h \to \tau \mu$ for $\tilde \delta_{23}^{LR} = \tilde \delta_{23}^{RL} =$ 5 
and $m_\text{SUSY} =$ 5 TeV for the scenarios defined in Section \ref{pmssmvheavysusy}. Left panel: present phase of the LHC 
with $\sqrt{s} =$ 8 TeV and ${\cal L} =$ 25 fb$^{-1}$. Right panel: future phase of the LHC with $\sqrt{s} =$ 14 TeV and ${\cal L} =$ 100 fb$^{-1}$. In both panels the other MSSM parameters are set to the values reported in the text, with $M_2 = m_\text{SUSY}$. The shaded blue areas are excluded by CMS searches~\cite{CMS-PAS-HIG-12-050} (see variations from these limits for different scenarios in \cite{Carena:2013qia}).}\label{Neventsh-mAtanb-deltaLR}
\end{figure}

\begin{figure}[t!]
\hspace{-0.5 cm}
\begin{tabular}{cc}
\includegraphics[width=80mm]{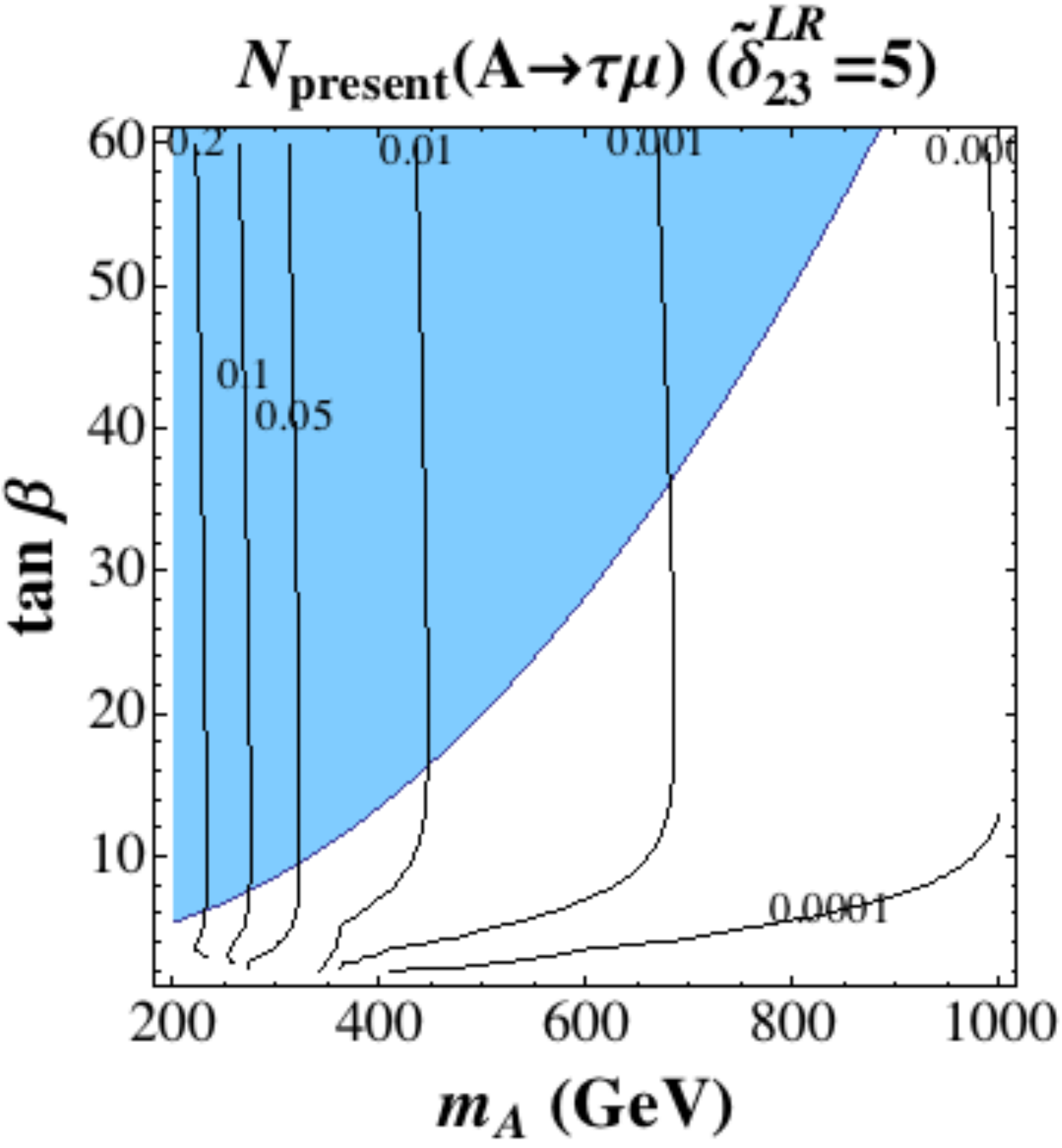} &
\includegraphics[width=80mm]{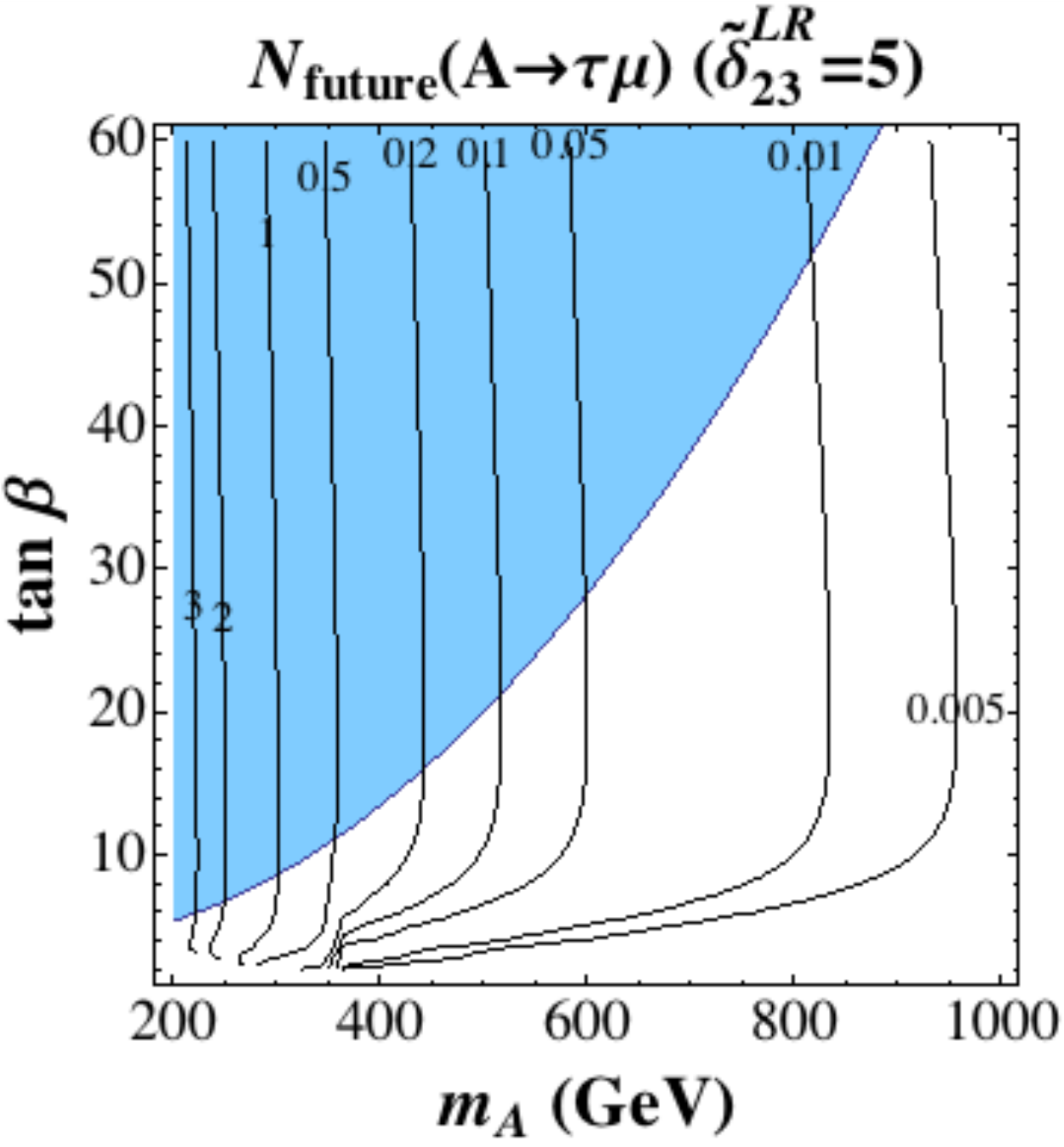}
\end{tabular}
\caption{Number of expected LFV events in the ($m_A,\tan\beta$) 
plane from $A \to \tau \mu$ for 
$\tilde \delta_{23}^{LR} = \tilde \delta_{23}^{RL} =$ 5 and 
$m_\text{SUSY} =$ 5 TeV for the scenarios defined in Section \ref{pmssmvheavysusy}. 
Left panel: present phase of the LHC with $\sqrt{s} =$ 8 TeV 
and ${\cal L} =$ 25 fb$^{-1}$. Right panel: future phase of 
the LHC with $\sqrt{s} =$ 14 TeV and ${\cal L} =$ 100 fb$^{-1}$. 
In both panels the other MSSM parameters are set to the values reported 
in the text, with $M_2 = m_\text{SUSY}$. The shaded blue areas 
are excluded by CMS searches~\cite{CMS-PAS-HIG-12-050} (see variations from these limits for different scenarios in \cite{Carena:2013qia}). 
The results for $H$ (not shown)
are nearly equal to these ones for $A$.}\label{NeventsA-mAtanb-deltaLR}
\end{figure}

Next we analyze the results for the case of $LR$ and $RL$ mixings 
in the 
($m_A,\tan\beta$) plane. Figures~\ref{Neventsh-mAtanb-deltaLR} 
and~\ref{NeventsA-mAtanb-deltaLR} summarize the results for the 
$h \to \tau \mu$ and $A \to \tau \mu$ channels, respectively, 
in the present and future LHC stages. 
 
On the left panel of Figure~\ref{Neventsh-mAtanb-deltaLR}, 
where the number of expected events from the $h \to \tau \mu$ channel 
in the present phase of the LHC are shown as a function of 
$m_A$ and $\tan\beta$, for $m_\text{SUSY} =$ 5 TeV and 
$\tilde \delta_{23}^{LR} = \tilde \delta_{23}^{RL} =$ 5, we see again
that the maximum allowed number of events are obtained in the low 
$\tan\beta$ region. Tens of events are expected, up to 50 for 
$\tan\beta \lesssim$ 3, in all the studied $m_A$ interval. 
In any case, in all the allowed region the number of predicted 
events are softly dependent on $m_A$ and at least 
one event is obtained, even for large values of $m_A$ and 
$\tan\beta \lesssim$ 10. On the right panel 
of Figure~\ref{Neventsh-mAtanb-deltaLR}, the predictions for the 
$h \to \tau \mu$ channel, in the future LHC phase with a 
centre-of-mass energy of $\sqrt{s} =$ 14 TeV and a total integrated 
luminosity of ${\cal L} =$ 100 fb$^{-1}$, show the same behaviour 
with respect to the two pair of parameters as on the left panel but 
with an increase in the number of events of around one 
order of magnitude. Again the maximum amount 
of events are for the lowest $\tan\beta$ values, being these 
nearly independent on $m_A$, and the rates decrease as we raise 
$\tan\beta$, showing a small variation with respect to 
$m_A$ for the allowed region by data (in white), 
as in the previous mentioned plot. Specifically, we 
obtain up to 500 events for $\tan\beta \simeq$ 2, and between 250 and 1 events 
for the region between $\tan\beta =$ 2 and $\tan\beta =$ 35.

The corresponding results for the $A \to \tau \mu$ channel, displayed 
in Figure~\ref{NeventsA-mAtanb-deltaLR}, show a very different behaviour with 
$m_A$ and $\tan\beta$ than the previous $h$ case. The number of expected 
LFV events at the LHC via $A \to \tau \mu$ decays 
diminish as $m_A$ increases, due to the suppression in the production 
cross section of a heavy pseudoscalar Higgs boson, and stay constant 
with $\tan\beta$, due mainly to the compensation between the 
growing of the $A$ production cross section via 
bottom-antibottom quark annihilation and the reduction 
of BR$(A \to \tau \mu)$ with this parameter, as previously illustrated 
in Figure~\ref{BRs-tanb}. In the present phase of the LHC we cannot 
expect any event, as shown on the left panel of 
Figure~\ref{NeventsA-mAtanb-deltaLR}. The right panel of 
Figure~\ref{NeventsA-mAtanb-deltaLR}, containing the predictions 
for the $A \to \tau \mu$ channel in the future LHC phase, shows an 
analogous behaviour to that of the left panel. 
The number of expected events increase around one order of magnitude 
respect the present LHC phase, and for values of $m_A$ below 300 GeV, 
one could expect between 1 and 3 events independently on the value 
of $\tan\beta$.

\begin{figure}[t!]
\hspace{-0.5 cm}
\begin{tabular}{cc}
\includegraphics[width=80mm]{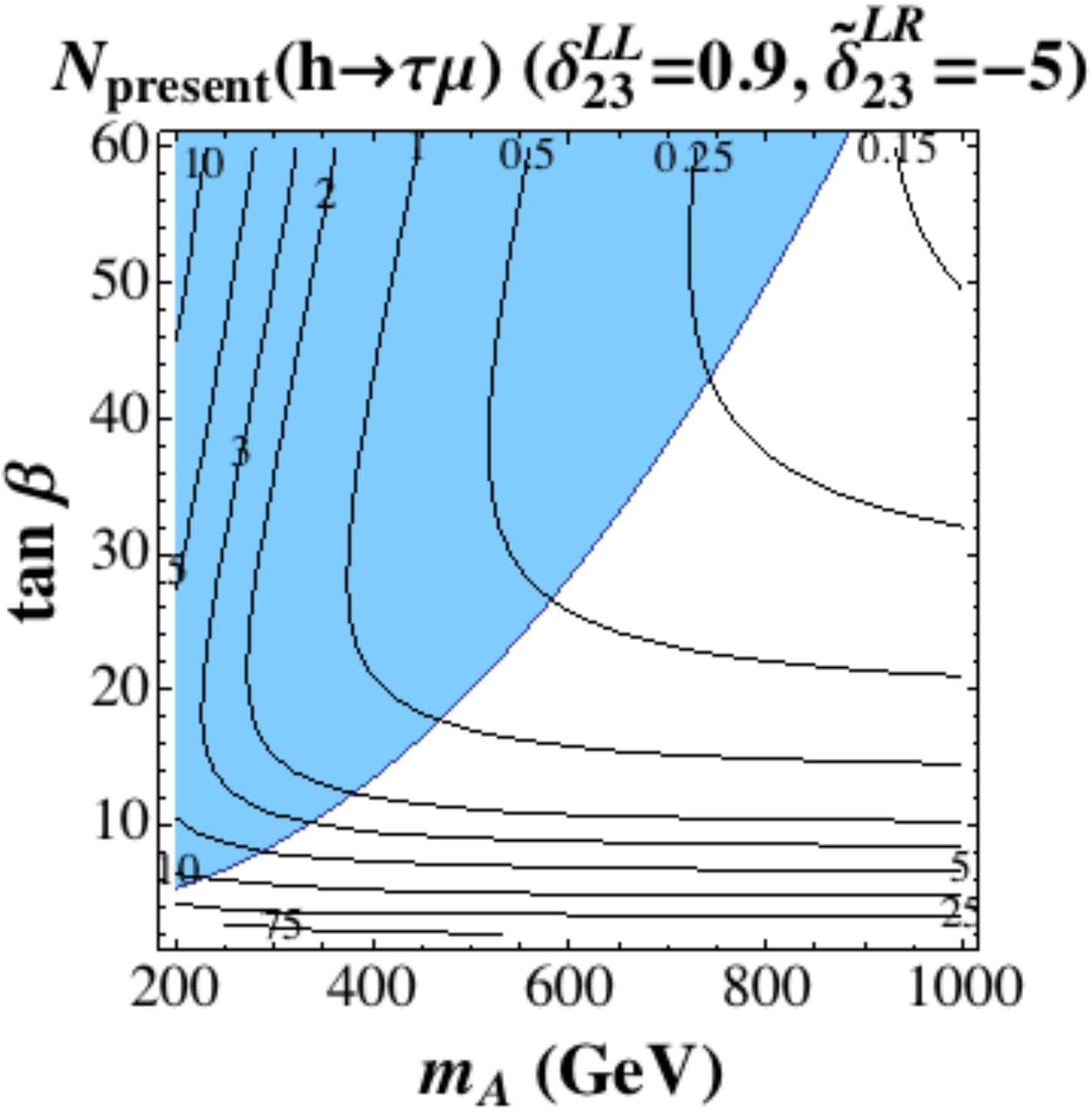} &
\includegraphics[width=80mm]{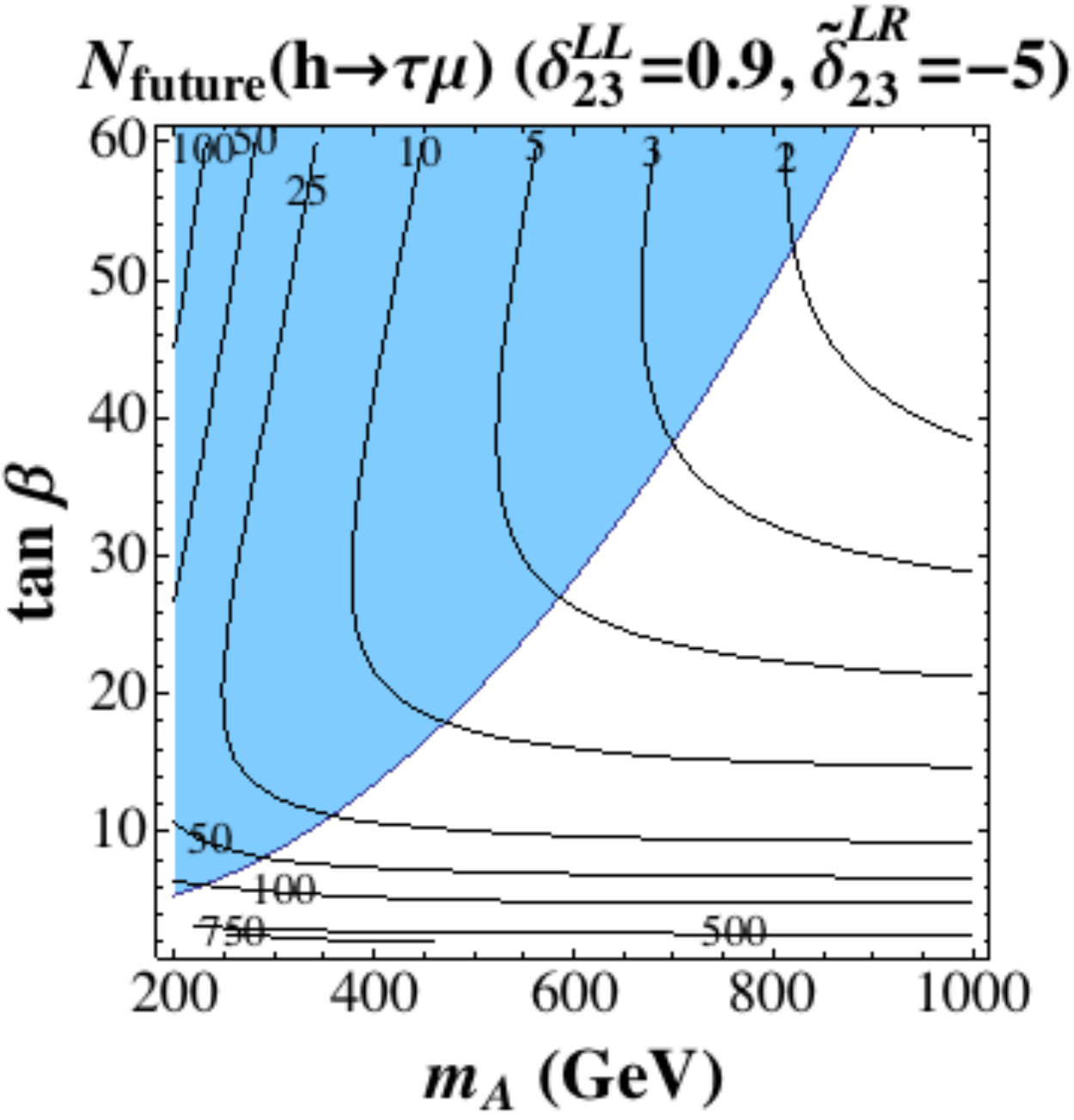}\\
\includegraphics[width=75mm]{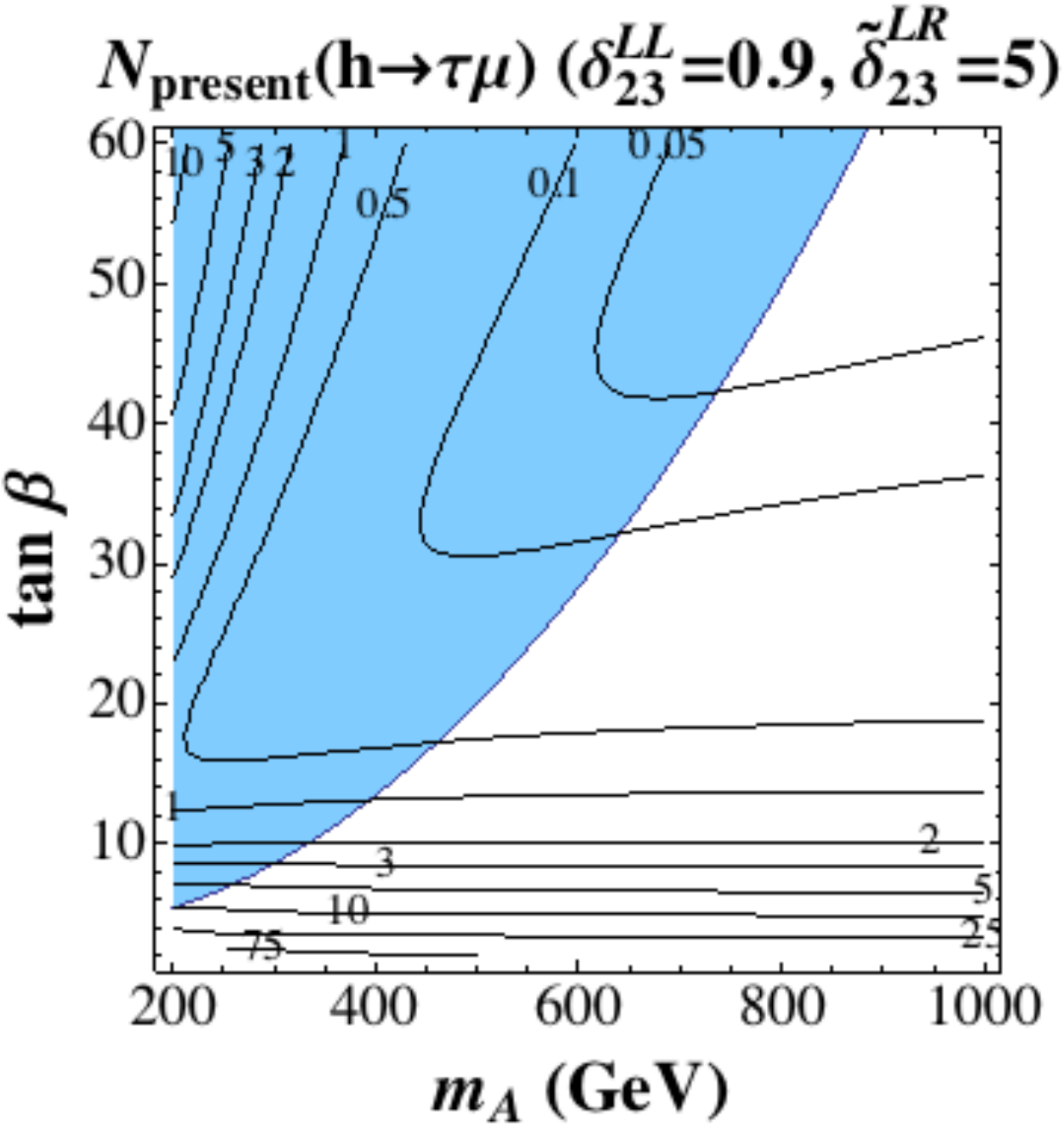} &
\includegraphics[width=75mm]{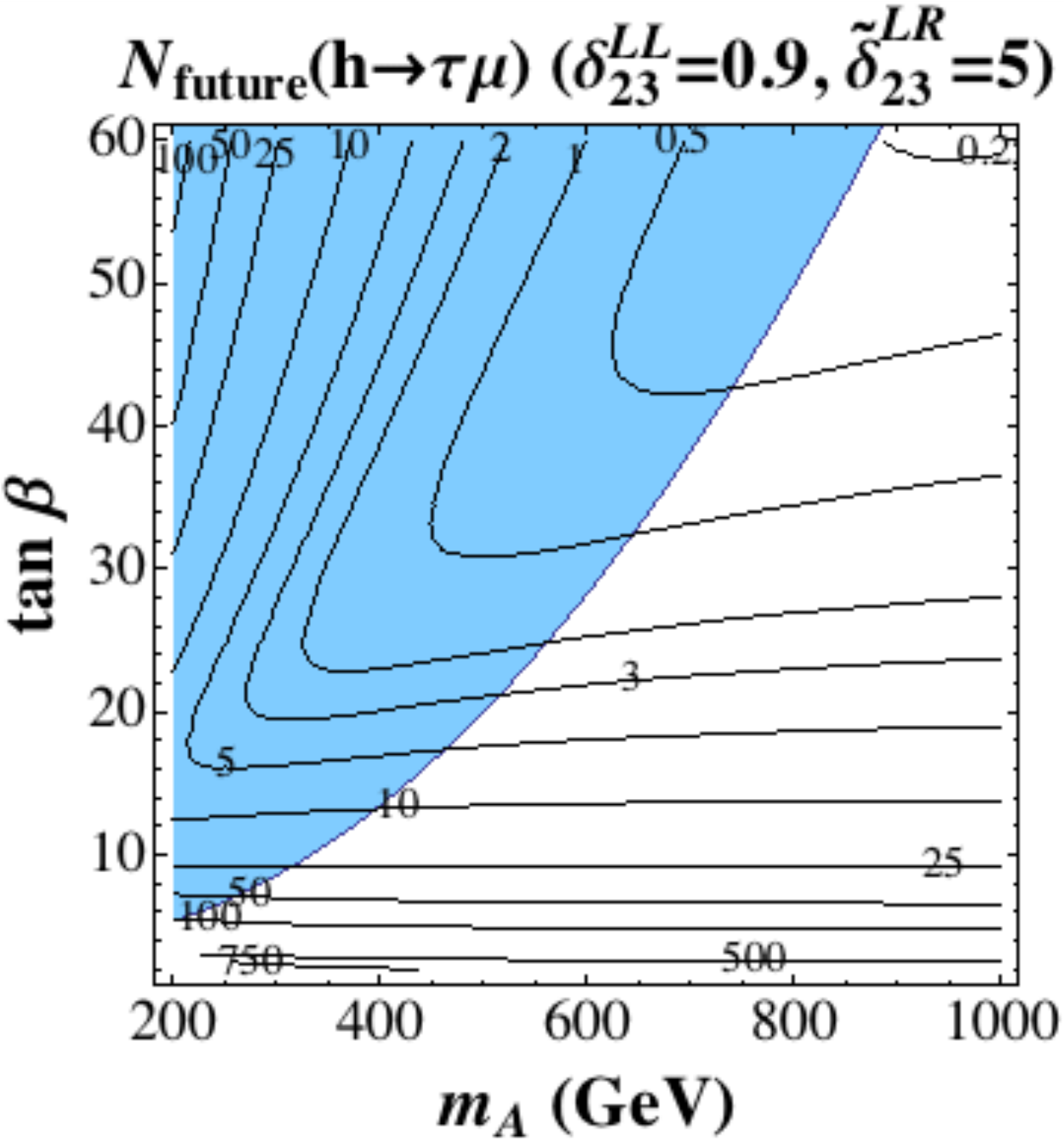}

\end{tabular}
\caption{Number of expected LFV events in the ($m_A,\tan\beta$) plane from 
$h \to \tau \mu$ for $m_\text{SUSY} =$ 5 TeV, $\delta_{23}^{LL} =$ 0.9 and 
$\tilde \delta_{23}^{LR} = \tilde \delta_{23}^{RL} =$ -5 (upper panels) or 
$\tilde \delta_{23}^{LR} = \tilde \delta_{23}^{RL} =$ +5 (lower panels) for the scenarios defined in Section \ref{pmssmvheavysusy}. 
Left panels: present phase of the LHC with 
$\sqrt{s} =$ 8 TeV and ${\cal L} =$ 25 fb$^{-1}$. Right panels: future phase 
of the LHC with $\sqrt{s} =$ 14 TeV and ${\cal L} =$ 100 fb$^{-1}$. 
In all panels the other MSSM parameters are set to the values reported 
in the text, with $M_2 = m_\text{SUSY}$. The shaded blue areas are 
excluded by CMS searches~\cite{CMS-PAS-HIG-12-050} (see variations from these limits for different scenarios in \cite{Carena:2013qia}).}
\label{Neventsh-mAtanb-deltaLLdeltaLR}
\end{figure}

\begin{figure}[t!]
\hspace{-0.5 cm}
\begin{tabular}{cc}
\includegraphics[width=80mm]{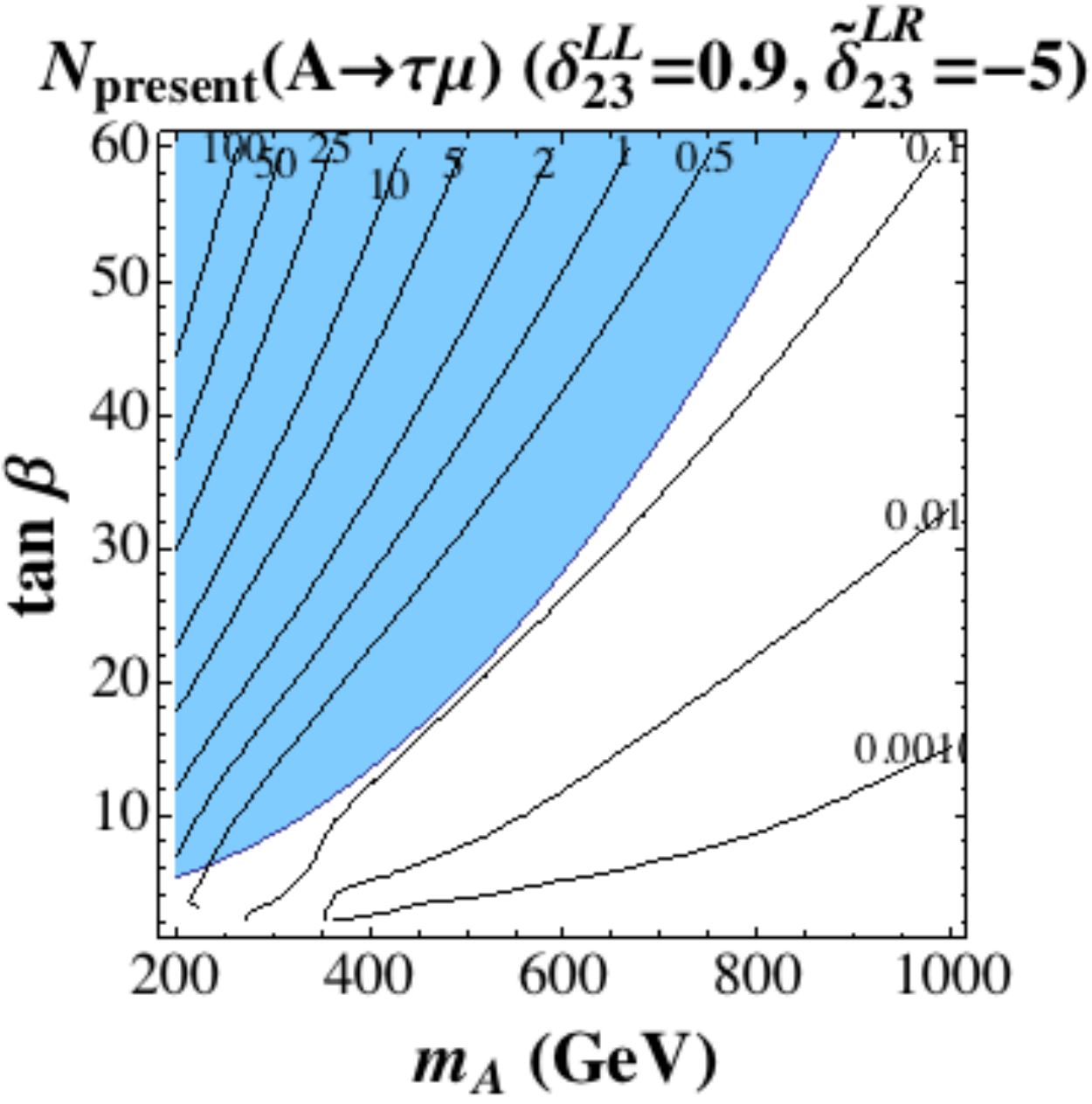} &
\includegraphics[width=80mm]{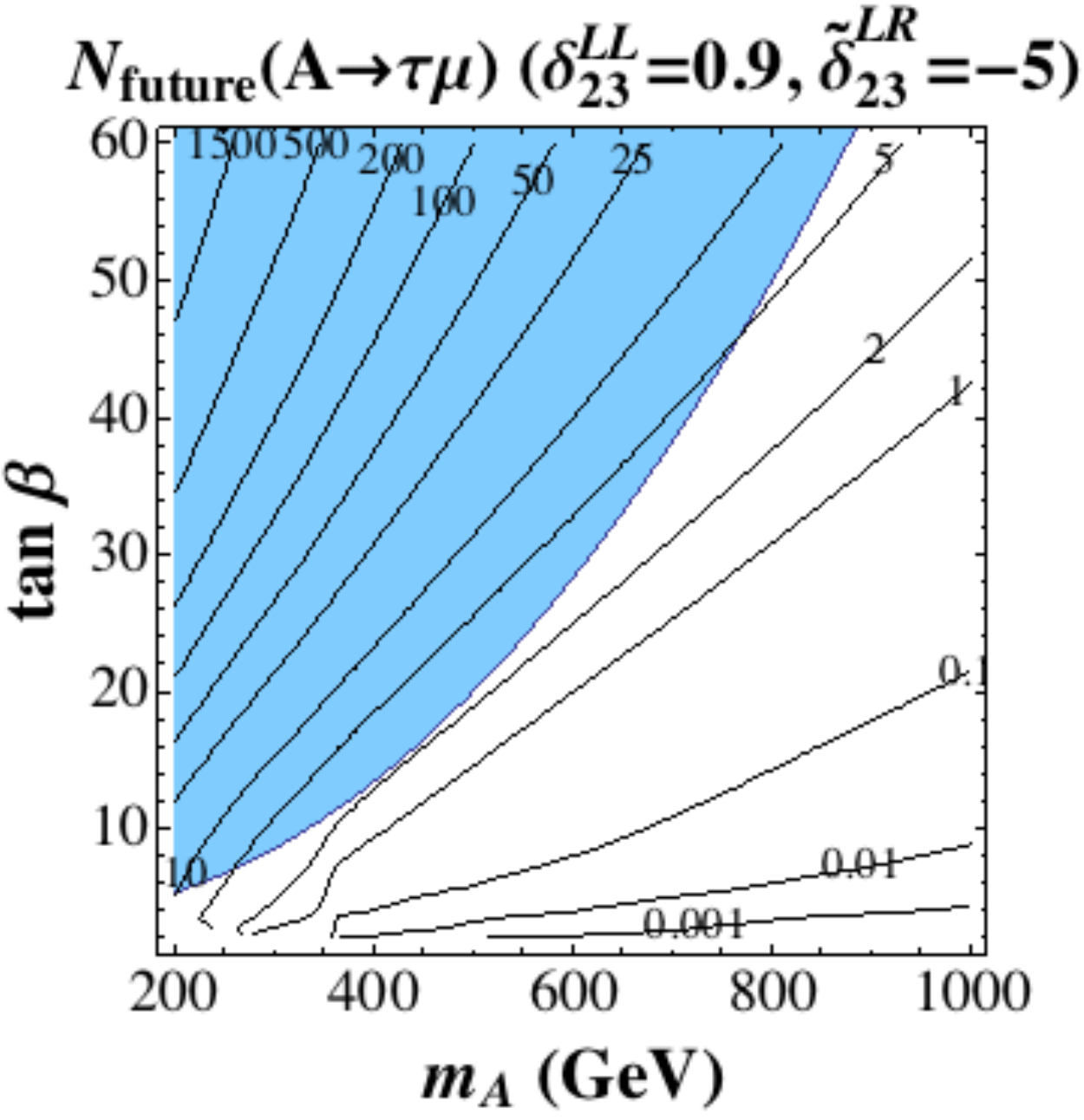}
\end{tabular}
\caption{Number of expected LFV events in the ($m_A,\tan\beta$) plane 
from $A \to \tau \mu$ for $\delta_{23}^{LL} =$ 0.9, 
$\tilde \delta_{23}^{LR} = \tilde \delta_{23}^{RL} =$ -5 and 
$m_\text{SUSY} =$ 5 TeV for the scenarios defined in Section \ref{pmssmvheavysusy}. Left panel: 
present phase of the LHC with $\sqrt{s} =$ 8 TeV and ${\cal L} =$ 
25 fb$^{-1}$. Right panel: future phase of the LHC with $\sqrt{s} =$ 14 TeV 
and ${\cal L} =$ 100 fb$^{-1}$. In both panels the other MSSM parameters 
are set to the values reported in the text, with $M_2 = m_\text{SUSY}$. 
The shaded blue areas are excluded by CMS searches~\cite{CMS-PAS-HIG-12-050} (see variations from these limits for different scenarios in \cite{Carena:2013qia}).
The results for $H$ (not shown)
are nearly equal to these ones for $A$.}
\label{NeventsA-mAtanb-deltaLLdeltaLR}
\end{figure}

Finally, the results for the $h \to \tau \mu$ and $A \to \tau \mu$ channels 
in the double $LL$ and $LR$ mixings case are summarized in 
Figures~\ref{Neventsh-mAtanb-deltaLLdeltaLR} 
and~\ref{NeventsA-mAtanb-deltaLLdeltaLR}, respectively. 
 
The predictions for the contour lines of $h \to \tau \mu$ event rates  
(Figure~\ref{Neventsh-mAtanb-deltaLLdeltaLR}) show a clear different 
pattern than in the previous cases of single deltas. In both LHC phases, 
we achieve 
an increase in the number of events respect to the single $LL$ and 
$LR$ (and/or
$RL$) mixing cases. The most interesting numbers are at the lower part 
of these plots with $\tan \beta < 10$. We find as large as 75 LFV events 
in the present phase 
of the 
LHC for very low values of $\tan\beta\simeq $ 2 and 
$m_A \lesssim$ 600 GeV.  
In the future LHC phase we predict up 
to 750 events, 
for $\tan\beta \simeq$ 2 and $m_A \lesssim$ 450 GeV. It should be also noted that 
these conclusions apply to
both choices for the 
$LR/RL$ mixings of -5
and +5, as can be seen in Figure~\ref{Neventsh-mAtanb-deltaLLdeltaLR}. 
The only notable 
differences that we find between the results of these two cases are in the
slight different patterns of the contour lines, signalling a small different
sensitivity to $m_A$ and/or $\tan\beta$. Also one can appreciate in these plots 
that one gets a bit 
lower rates for +5 than for -5, in agreement with our previous findings 
reported in Figure~\ref{BRs-deltatildeLRRL}  

The number of events for the $A \to \tau \mu$ channel in the double $LL$ and 
$LR$ mixings case are displayed in 
Figure~\ref{NeventsA-mAtanb-deltaLLdeltaLR}. The $LL$ mixing is set here 
to 0.9 and the $LR=RL$ mixings are fixed to -5. 
As expected, the predicted event rates increase as $\tan\beta$ grows and are 
reduced as $m_A$ gets bigger, due to the suppression in the production cross 
section of a heavy pseudoscalar Higgs boson. However, 
for the present LHC phase one cannot say much about this channel, 
since all the $m_A-\tan\beta$ regions which could 
produce 
a relevant number of 
LFV 
events are excluded at present by CMS searches. We find at the most one event 
in the tiny low left corner of the allowed region in this $(m_A,\tan\beta)$ 
plot. In contrast,
the predictions in the future phase of the LHC are more promising. 
We predict up to about 5 LFV events for large values of $m_A$ and 
$\tan\beta$ and up to 10 in the low $\tan\beta$ region with 
$m_A \simeq$ 200 GeV. Similar conclusions are found for the case 
of positive $LR$=+5 mixing (not shown). The shape of the contour lines 
in this case are slightly
modified at low $\tan\beta$ but with no relevant implications in terms of event
rates.

\bigskip

\clearpage
\newpage

\chapter{Conclusions}
\label{conclusions}

The recent discovery of the Higgs boson, if it is confirmed that is indeed the SM Higgs boson, would close the particle content of the SM, and would establish the Brout-Englert-Higgs mechanism as the correct way of breaking the $SU(2)_L\times U(1)_Y$ electroweak symmetry and giving mass to the elementary particles. However the SM is not the final model of particle physics since it is believed to be an effective theory of a wider and more complete theory, with new features as for example: including gravity in its framework, explaining the more fundamental dynamical origin for the value of the masses and couplings of the particles and the cause of dark matter, probably unifying the forces, and giving an answer to other unknown phenomena. The absence of any direct signal of such a wider theory in the experiments so far, has lead us in the present work to study observables where it could be possible to observe indirect effects of this new physics before it is directly discovered. 

Flavour physics has been during the history of Particle Physics a unique place from where to extract clues and information about non yet discovered particles, as it happened before the discovery of the top quark or the masses of the neutrinos, and thus flavour has been our choice through this work where to look for these new physics effects. This choice has been done in combination with the new observation subject that is the Higgs boson, whose properties as mass, couplings or decay rates, still unknown or measured with low precision could reveal us surprises, departing from the SM-like behaviour that until the moment seems to exhibit the present data.

The model of new physics studied through this work has been the MSSM. Within this model we have carried out a research on the most relevant phenomenology implied by the hypothesis of NMFV from general flavour mixing in the sfermion sector both for squarks and sleptons. In order to do this, in Chapter \ref{paramnmfvscen} we have introduced a general parametrization of the flavour mixing elements between different sfermion generations in terms of a series of generic and dimensionless parameters $\deXYij$ ($X,Y=L,R$; $i$ and $j$ are the generations involved in the mixing), and then we have explored through this thesis the consequences of these parameters on various observables. 

The first part of our research has been focused on the squark sector. Chapter \ref{sec:Bphysics} was devoted to the study of the flavour mixing between the second and the third generation squarks, through the following $B$ physics observables: \bsg, ${\rm BR}(B_s \to \mu^+ \mu^-)$ and $\Delta M_{B_s}$. We have studied the phenomenology of these observables with respect to the MSSM and the $\deXYij$ parameters, and constrained these latter with the last experimental observations. It has also been compared the situation before and after the LHC. The numerical evaluation of these observables has been done with the BPHYSICS subroutine of the SuFla code, to which we have added the one-loop gluino boxes contribution in the case of the $\Delta M_{B_s}$, that was previously not available and it is known to be relevant in the context of NMFV scenarios. We have estimated the size of these corrections and compared them with the other relevant contributions from
chargino boxes and double Higgs penguins for all values of $\tan \beta$ for the first time. We have concluded that gluino boxes dominate for moderate and low $\tan \beta \leq 20$, while at large $\tan \beta$ the double $H$-penguins dominate. For the other observables, the large $\tan \beta$ effects could also very relevant specially because of the effects of the gluino and chargino loops in the case of \bsg, and of the Higgs penguin diagrams for ${\rm BR}(B_s \to \mu^+ \mu^-)$. The constraints for the studied scenarios from each of the $B$ physics observables to the $\deXYij$ before and after the LHC data can be found in tables \ref{tableintervals} and \ref{tableintervals2} respectively. In tables \ref{tabdeltasummary} and \ref{tabdeltasummary2} one can find the combined constraints of the three observables for the pre-LHC and post-LHC situations respectively. From this study, we have also concluded that \bsg\ and $\Delta M_{B_s}$ are the most restrictive observables. 

The absence of SUSY in the LHC so far has moved the allowed values of the masses of the sparticles to higher energy regions. The scenarios studied for the pre-LHC situation have lighter squarks than are not allowed anymore in the post-LHC situation, and thus the corrections to the $B$ physics observables are smaller for the latter, and consequently the allowed regions for the deltas are at present larger than they were before the LHC started. 

For the pre-LHC situation, the NMFV deltas in the top-sector could be sizeable $|\de_{ct}^{XY}|$  larger than ${\cal O}(0.1)$ and still compatible with $B$ data. In particular, $\delta^{RL}_{ct}$ and $\delta^{RR}_{ct}$ were the less constrained parameters, and to a lesser extent also $\delta^{LR}_{ct}$. The parameters on the bottom-sector were, in contrast, quite constrained. The most tightly constrained were clearly  $\delta^{LR}_{sb}$ and $\delta^{RL}_{sb}$, specially the first one with just some singular allowed values: either positive and of the order of  $3-5 \times 10^{-2}$, or negative and with a small size of the order of $-7 \times 10^{-3}$; for the second the limits were around  $2 \times 10^{-2}$ for both positive and negative values. $\delta^{RR}_{sb}$ was the less constrained parameter in the bottom sector, with larger allowed intervals of  $|\delta^{RR}_{sb}| \lsim 0.4-0.9$ depending on the scenario. For the current post-LHC situation, the less restricted deltas are $\delta^{RR}_{ct}$ and $\delta^{RR}_{sb}$ than in general could reach a size of ${\cal O}(1)$, then $\delta^{LL}_{23}$, $\delta^{LR}_{ct}$ and $\delta^{RL}_{ct}$ with bounds one order of magnitude smaller, with the first one being more restricted than the last two, and finally the most restricted deltas are $\delta^{LR}_{sb}$ and $\delta^{RL}_{sb}$ that can reach $|\deXYij|$ at most of ${\cal O}(0.01)$. 

In general terms the sensitivity of \bsg\ follows the behaviour we just described for the different deltas. For $\Delta M_{B_s}$ it is possible to find more or less sensitivity to each different delta depending on the point of study in the MSSM parameter space, due mainly to the different contributions as the gluino boxes and the double Higgs penguins, with very different behaviour with $\tan \beta$, as explained before. Although we remark again than in general large sensitivity is found. In the case of ${\rm BR}(B_s \to \mu^+ \mu^-)$, the main sensitivity is to $\delta^{LL}_{23}$, and then depending on the MSSM point also to the other deltas. This behaviour with $\delta^{LL}_{23}$ depends to a great extent of the enhancement at large $\tan \beta$, but the present experimental constraints force $M_A$ to be large in this last case, producing a decrease on the sensitivity to the delta that can counteract the former effect.

From the numerical evaluations it could also be observed than some excluded MSSM points with MFV (i.e. for vanishing deltas) by $B$ physics observables could be recovered as allowed scenarios after setting some of the deltas to non-zero values.

After having a general picture of the restrictions of the different deltas by the $B$ physics observables, and the behaviour of these observables, we have considered in Chapter \ref{higgsmasssquark} the corrections to the Higgs boson mass generated from squark flavour mixing from these same deltas. In first place, we performed an analytical study on these corrections, deriving the one-loop formulas for them with \fa\ and \fc. The new Feynman rules included in the model file are listed in the Appendix A, and all the results for the unrenormalized self-energies and tadpoles are collected in Appendix B. In the MFV case, up to three loop corrections to the neutral Higgs bosons masses are available, but the study of the NMFV case has been restricted in the past to the study of some particular flavour mixing elements, as the one-loop calculation taking into account the $LL$-mixing between the third and second generation of scalar up-type quarks. Our one-loop study in this thesis has been done in full generality, including all generations and all flavour mixing terms for the up- and down-type of squarks. We have also checked with this generality the finiteness of the analytical results for these renormalized Higgs self-energies. These corrections presented in the Appendix B have been compared with the code included in \fh, and in the process of checking the finiteness of the formulas we have found discrepancies with the charged Higgs part of \fh, leading to an updated version of the code.

The next step has been a numerical evaluation of the Higgs bosons mass corrections from squark flavour mixing between the second and the third generation, taking into account the restrictions from the $B$ physics observables of the previous chapter. The numerical evaluation has been performed for the cases of having one or two $\deXYij$ with a non-zero value, comparing the situation previous and following to the LHC data. With respect to the one delta studies, it can be seen from our plots (see for instance figs.  \ref{figdeltamh0}, \ref{figdeltamH0} and
\ref{figdeltamHp}) how the corrections to all masses are symmetric in the sign of the deltas, as a difference of what happened with the $B$ physics observables, and that can be positive or negative depending on the scenario and the value of the delta. In particular for the light Higgs boson, the corrections tend to have an ``M form'', being positive and small for small deltas, and then negative and larger in size as the deltas grow. For the pre-LHC situation, we have found a very strong sensitivity to $\delta^{LR}_{sb}$ and $\delta^{RL}_{sb}$ for all the Higgs bosons, that are also the most constrained deltas from $B$ physics. For the light Higgs boson we found also an important sensitivity to $\delta^{LR}_{ct}$ and $\delta^{RL}_{ct}$, reaching values of up to tens of GeV for the low $\tan \beta$ case where the $B$ physics observables are less restrictive. For the post-LHC situation, the general behaviour of the corrections with the various deltas remains approximately the same. The values of the corrections are very large as compared with the precision of the LHC, and should be taken into account into any Higgs boson analysis in the NMFV framework. In fact, we find that the LHC is able to put constraints on some of these delta parameters like $\delta^{LR}_{ct}$ and $\delta^{RL}_{ct}$ that are allowed by other observables, as it is the case for the $B$ physics observables.

The numerical analysis has been extended for the light Higgs boson corrections to the case of having simultaneously two deltas with non-zero values, in the pre-LHC situation. The largest corrections allowed by $B$ data have been found for the  $(\de^{LL}_{23},\de^{LR}_{ct} )$ and  $(\de^{LL}_{23},\de^{RL}_{ct} )$ cases. These corrections have reached several tens of GeV in the pre-LHC situation. The largest allowed region by $B$ data and largest corrections have corresponded to scenarios with low $\tan \beta$ and heavy SUSY spectra. Also important corrections have been found in the $(\de^{LL}_{23},\de^{RR}_{ct} )$ plane up to tens of GeV for some specific points. The corrections found to the heavy Higgs bosons in the double delta plane have been much smaller than the previous ones.

The next part of our research, has been the study of the slepton flavour mixing. In Chapter \ref{phenoflavourslep} we have developed a study of the phenomenology on different LFV observables and concluded with the constraints that they set on the flavour mixing parameters  $\deABij$ of the slepton sector. The LFV processes whose phenomenology have been studied in this chapter have been: radiative decays  $\mu \to e \gamma$, $\tau \to e \gamma$ and $\tau \to \mu \gamma$ ($l_j \to l_i \gamma$), leptonic decays $\mu \to 3 e$, $\tau \to 3 e$ and $\tau \to 3 \mu$ ($l_j \to 3 l_i$), semileptonic tau decays $\tau \to \mu \eta$ and $\tau \to e \eta$ and conversion of $\mu$ into $e$ in heavy nuclei. The radiative decays are usually the most restrictive ones, but the leptonic and semileptonic decays are included because they give access to the Higgs sector parameters, since they can be mediated by Higgs bosons. Our numerical study has been done for scenarios that are allowed by the last data from the LHC, including a mass for the light Higgs boson around 125 - 126 GeV, and also the current value for the muon anomalous magnetic moment, $(g-2)_\mu$. 

Our study has shown that the radiative decays are still the most restrictive observables to the slepton flavour mixing parameters. Comparing the flavour mixings between different pairs of generations, the experimental bounds for the observables related with the 12-mixings are around four orders of magnitude smaller than the others in all observables, and thus these mixings are much more constrained. The bounds for 23-mixings are similar to the ones for 13-mixings. It was also shown that all the observables behave symmetrically respect to the sign of the deltas. The constraints for the different $\deABij$ for the studied points are collected in \refta{boundsSpoints}. The most restrictive constraints to $LL$ and $RR$ mixings have been found to happen for the large $\tan \beta$ case, while the $LR$ mixings are independent of $\tan \beta$ and have stronger constraints for the cases with lightest sleptons.

We also found that the effect of the pseudoscalar Higgs boson mediation in the semileptonic tau decays was not very important since the present exclusion limits for the mass of this Higgs boson lead us to assume a quite heavy $A$ boson, whose effects are then reduced. These observables set no constraints to the 23- and 13-mixings. With respect to the photon mediation in the leptonic decays and in the conversion of $\mu$ into $e$ in heavy nuclei, we confirmed that this photon channel dominates over the other contributions, and thus there was a correlation between the predictions for the leptonic decay rates, the $\mu - e$ conversion rates and the radiative decay rates. 

After this, our study was extended to the case of having simultaneously two deltas different from zero. First we studied the case $(\delta^{AB}_{23},\delta^{CD}_{23})$ with $AB\neq CD$ and found constructive or destructive interferences on $\br(\tau \to \mu \gamma)$ depending on the signs of the deltas. The most relevant case was $(\delta^{LR}_{23},\delta^{LL}_ {23})$ where important destructive interferences were found for the case of opposite signs of the deltas, that made larger the maximum allowed value for each delta as compared with having just that delta. The shape and size of the allowed regions with extreme solutions as  $(\pm 0.9, \mp 0.9)$ for the deltas came from our full mass matrix diagonalization computation, and thus they can not be described with the simplest MIA formulas that are usually used in the literature. As a second study on the two-delta plane, we studied the constraints in the (13,23) mixings coming from $\mu \to e \gamma$ and $\mu-e$ conversion as compared with the bounds coming from the observables related to each single delta, and found surprisingly better constraints from this double delta situation. 

The study on the bounds on slepton flavour mixings from the radiative decays was completed with a study on the dependence of the bounds on the deltas with respect to the most relevant MSSM parameters in this LFV context: $\tan \beta$ and the mass scale of the sleptons $m_{\rm SUSY}$. The study of the bounds was done comparing it with the allowed regions by the measurements of the Higgs boson mass and the value of $(g-2)_\mu$. We observed how the maximal allowed values on the $\delta^{LL}_{ij}$'s and $\delta^{RR}_{ij}$'s scale with $m_{\rm SUSY}$ and $\tb$ approximately growing with increasing $m_{\rm SUSY}$ as $\sim m_{\rm SUSY}^2$ and decreasing with increasing (large) $\tb$ as $\sim 1/\tb$. The maximal allowed values of the $\delta^{LR}_{ij}$'s (and similarly $\delta^{RL}_{ij}$'s) are independent on $\tb$ and grow approximately as $\sim m_{\rm SUSY}$  with increasing $m_{\rm SUSY}$. It was also observed how $(g-2)_\mu$ requires a rather light SUSY-EW sector, i.e.\ light charginos, neutralinos and sleptons, and a rather large $\tb$, and $\Mh$ requires a rather heavy SUSY squark sector, creating some tension between the two observables when the slepton and the squark sector were related, as it happened in our scenarios.

The maximum values for the deltas found for our scenarios in the favoured by LHC and
$(g-2)_\mu$ data MSSM parameter space region, lay at the following intervals:   
$|\delta^{LL}_{12}|_{\rm max} \sim {\cal O} (10^{-5},10^{-4}) $, 
$|\delta^{LR}_{12}|_{\rm max} \sim {\cal O} (10^{-6},10^{-5}) $,
$|\delta^{RR}_{12}|_{\rm max} \sim {\cal O} (10^{-3},10^{-2}) $,
$|\delta^{LL}_{23}|_{\rm max} \sim {\cal O} (10^{-2},10^{-1}) $,
$|\delta^{LR}_{23}|_{\rm max} \sim {\cal O} (10^{-2},10^{-1}) $,
$|\delta^{RR}_{23}|_{\rm max} \sim {\cal O} (10^{-1},10^{0}) $. Very similar general bounds as for the 23 mixing are found for the 13 mixing. 

Our work was concluded with Chapter \ref{lfvhiggsdecaysslepton}, where we studied the effects on LFV Higgs decays from slepton flavour mixing. This research was done focusing on the possibility of a very heavy SUSY spectra above the TeV scale and we were mainly motivated in looking for SUSY non-decoupling effects in the LFV Higgs decays being governed by the general slepton flavour mixing. 

The observables chosen for this last study in this thesis were the LFV Higgs decays $\phi \to \tau \mu$, with $\phi=h,H,A$, and the LFV radiative decay $\tau \to \mu \gamma$. Similar results and conclusions can be extrapolated for the $\phi \to \tau e$ and $\tau \to e \gamma$ decays. As we saw in the previous chapter, the  $\tau \to \mu \gamma$ radiative decay was the most constraining one of our studied observables to the slepton flavour mixing between the second and the third generations. The LFV Higgs couplings to leptons are proportional to the heaviest lepton mass involved, and thus we were more interested on the decays to $\tau$ than to other leptons. Besides we saw in the previous chapter that the constraints from $\mu \to e \gamma$ on the slepton flavour mixings are very tight and therefore do not lead to measurable $\phi \to \mu e$ rates. The predictions for the $\phi \to \tau e$ decay are very similar to the ones for the $\phi \to \tau \mu$ decays, and from the experimental point of view the sensitivity is also very similar, so our results are easily translated into this other channel.

The numerical evaluation of our study has been done with {\it SPHENO} and a private FORTRAN code, where we have implemented the full one-loop formulas for the LFV Higgs and radiative decays. The spectrum of the Higgs sector and the total widths of the Higgs bosons have been calculated as before with \fh\ at two-loop order.

We have shown that the three LFV Higgs decays $\phi \to \tau \mu$, with $\phi=h,H,A$, present a non-decoupling behaviour with $m_\text{SUSY}$, remaining constant at large $m_\text{SUSY}>2$ TeV. This is complemented with the radiative decays behaviour showing on the contrary a fast decoupling with $m_\text{SUSY}$. Thus we have an optimal situation for detecting indirectly the effects of SUSY even with a very heavy SUSY spectra scale. With respect to other parameters, the branching ratios grow at large $\tan \beta$ as BR($h, H, A \to \tau \mu$) $\sim (\tan\beta)^2$ for the $LL$ and $RR$ case, and decrease as $(\tan\beta)^{-2}$ in the $\tilde \delta_{23}^{LR}$ case. The most relevant $\delta_{23}^{AB}$ parameter at low $\tan \beta$ values for the lightest Higgs boson $h$ is ${\tilde \delta}_{23}^{LR}$, which gives rise to larger LFV Higgs decay rates than $\delta_{23}^{LL}$ and $\delta_{23}^{RR}$, whereas for the $H$ and $A$ Higgs bosons the most relevant parameter is $\delta_{23}^{LL}$. At large $\tan \beta$ values, the most relevant parameter for all the three Higgs bosons is $\delta_{23}^{LL}$. It was also shown that by taking large $M_2=\mu$ the branching ratios were also enhanced. With respect to $M_A$ it would be desirable to have a low $M_A$ in the case of large $\tan \beta$ to enhance the branching ratio of the heavy Higgs bosons, but the exclusion limits in the $M_A-\tan\beta$ plane make it impossible.

The results were shown to be symmetric respect to the sign of the $\delta_{23}^{AB}$, and equal for ${\tilde \delta}_{23}^{LR}$ and ${\tilde\delta}_{23}^{RL}$. And we also saw that considering two $\delta_{23}^{AB}$ different from zero produced interferences that could enhance or reduce branching ratios.

This chapter was concluded with a computation of the LFV event rates that will be produced by these $\phi \to \tau \mu$ decays at the LHC in the $M_A-\tan\beta$ plane with large $m_\text{SUSY} \geq 5$ TeV, considering the present phase with a centre-of-mass energy of $\sqrt{s} =$ 8 TeV and the future phase with $\sqrt{s} =$ 14 TeV. For the single mixing cases, we concluded that the $RR$ mixing will not be able to be measured at the LHC. The $LL$ case would produce up to 5 events for the heavy Higgs bosons decays in the future phase of the LHC at large $\tan \beta$, and none for the light Higgs boson. And finally $LR$ is found to be the most relevant parameter, producing up to 50 events for $m_\text{SUSY} \geq 5$ TeV for low $\tan \beta$ for the light Higgs boson in the present phase, and hundreds of events in the future phase. In the case of the heavy Higgs bosons we obtained a few events only in the future phase. All these results are of course allowed by the relevant radiative decay $\tau \to \mu \gamma$. 

In the double-mixing case, the best situation was found for the case of $LL$ and $LR$ mixings. In this case the sensitivity with respect to $\tan \beta$ was improved, and we obtained for instance, a few events for the light Higgs boson at the present stage of the LHC even for moderate $\tan\beta \sim 15$ and large $m_A \geq $ 500 GeV. In the future LHC phase the reach to larger $\tan \beta$ values increased and we got some event even at very large $\tan\beta\sim 40$ and $m_A \geq$ 800 GeV. The largest rates found in this double-mixing situation were in any case for $h \to \tau \mu$ and were clearly localised at the low $\tan\beta$ region where we predicted for the present LHC phase up to about 75 events, and  up to about 750 LFV events for the future LHC phase. In the future LHC phase, we got about 10 events at the most for the $H, A \to \tau \mu$ channels in the low $\tan \beta$ region and 5 events at the most in the high $\tan \beta$ region. 

As a summary of our work, we have presented an exhaustive study of the phenomenological implications from general sfermion flavour mixing in a NMFV-MSSM context on $B$ physics observables, Higgs bosons masses corrections, LFV in leptonic observables and LFV Higgs decays. Through our study, we have set constraints on the sfermion flavour mixing parameters from the last experimental measurements. With respect to the Higgs bosons, it has been shown the important effects of the squark flavour mixing in the value of the Higgs bosons masses, and the possibility of using the value of the masses to set constraints on the squark flavour mixing. We have also shown how LFV Higgs decays are unique observables to indirectly test SUSY even in the experimentally discouraging situation of having a very heavy supersymmetry not directly reachable at the colliders. 

Supersymmetry is one of the most interesting proposals for physics beyond the SM, and we hope that our work made clear the importance of flavour mixing not only in its characterization, but in the new windows it opens to the indirect detection of SUSY. Now let the experiments speak.


\chapter{Conclusiones}
\label{conclusiones}

El reciente descubrimiento del bos\'on de Higgs, si se confirma que es el bos\'on de Higgs del SM, cerrar\'\i{}a el contenido de part\'\i{}culas del SM, y establecer\'\i{}a el mecanismo de Brout-Englert-Higgs como la forma correcta de romper la simetr\'\i{}a electrod\'ebil  $SU(2)_L\times U(1)_Y$ y dar masa a las part\'\i{}culas elementales. Sin embargo el SM no es un modelo final de f\'\i{}sica de part\'\i{}culas dado que se cree que es una teor\'\i{}a efectiva de una teor\'\i{}a m\'as amplia y completa, con nuevas caracter\'\i{}sticas como por ejemplo: incluir gravedad en su mismo marco, explicar el origen din\'amico m\'as fundamental de los valores de las masas y los acoplamientos de las part\'\i{}culas y la causa de la materia oscura, probablemente unificar las fuerzas, y dar respuesta a otros fen\'omenos desconocidos. La ausencia hasta el momento en los experimentos de se\~nales directas de dicha teor\'\i{}a m\'as amplia, nos ha llevado a estudiar en el presente trabajo observables donde sea posible observar efectos indirectos de esta nueva f\'\i{}sica antes de que sea directamente descubierta.

La f\'\i{}sica del sabor ha sido durante la historia de la f\'\i{}sica de part\'\i{}culas un lugar \'unico del que extraer pistas e informaci\'on sobre part\'\i{}culas todav\'\i{}a no descubiertas, como ocurri\'o antes del descubrimiento del quark top o las masas de los neutrinos, y por lo tanto el sabor ha sido nuestra elecci\'on a lo largo de este trabajo donde buscar estos efectos de nueva f\'\i{}sica. Esta elecci\'on ha sido hecha en combinaci\'on con el nuevo sujeto de observaci\'on que es el bos\'on de Higgs, cuyas propiedades como la masa, los acoplamientos o las tasas de desintegraci\'on, todav\'\i{}a desconocidas o medidas con poca precisi\'on pueden revelarnos sorpresas, alej\'andose del comportamiento tipo SM que hasta el momento exhiben los datos.

El modelo de nueva f\'\i{}sica estudiado a lo largo de este trabajo ha sido el MSSM. Usando este modelo hemos llevado a cabo una investigaci\'on sobre la fenomenolog\'\i{}a m\'as relevante inducida por la hip\'otesis de NMFV de mezcla general de sabor en el sector sfermi\'onico tanto para squarks como para sleptones. Para llevar a cabo esto, en el Cap\'\i{}tulo \ref{paramnmfvscen} hemos introducido una parametrizaci\'on general de los elementos de mezcla de sabor entre las diferentes generaciones de sfermiones en t\'erminos de una serie de par\'ametros gen\'ericos sin dimensiones $\deXYij$ ($X,Y=L,R$; $i$ y $j$ son las generaciones involucradas en la mezcla), y despu\'es hemos explorado a lo largo de la tesis las consecuencias de estos par\'ametros en varios observables.

La primera parte de nuestra investigaci\'on se ha centrado en el sector de los squarks. El Cap\'\i{}tulo \ref{sec:Bphysics} se ha dedicado al estudio de la mezcla de sabor entre la segunda y la tercera generaciones de squarks, a trav\'es de los siguientes observables de $B$: \bsg, ${\rm BR}(B_s \to \mu^+ \mu^-)$ y $\Delta M_{B_s}$. Hemos estudiado la fenomenolog\'\i{}a de estos observables con respecto a los par\'ametros del MSSM y los $\deXYij$, y restringido estos \'ultimos con las \'ultimas observaciones experimentales. Tambi\'en se ha comparado la situaci\'on anterior y posterior al LHC. La evaluaci\'on num\'erica de estos observables ha sido hecha con la subrutina BPHYSICS del c\'odigo SuFla, a la cual hemos a\~nadido la contribuci\'on de las cajas de gluinos a un loop en el caso de  $\Delta M_{B_s}$, que previamente no estaban disponibles y es sabido que son relevantes en el contexto de escenarios NMFV. Hemos estimado el tama\~no de estas correcciones y las hemos comparado por primera vez con las otras contribuciones relevantes de cajas de charginos y ping\"uinos dobles de Higgs para todos los valores de $\tan \beta$. Hemos concluido que las cajas de gluinos dominan para $\tan \beta \leq 20$ bajo y moderado, mientras que a gran $\tan \beta$ dominan los dobles ping\"uinos de $H$. Para los dem\'as observables, los efectos de gran $\tan \beta$ pueden ser tambi\'en muy relevantes especialmente debido a los efectos de loops de charginos y gluinos en el caso de \bsg, y de los diagramas de ping\"uinos de Higgs para ${\rm BR}(B_s \to \mu^+ \mu^-)$. Las restricciones para los escenarios estudiados para cada uno de los observables de f\'\i{}sica de mesones $B$ a los $\deXYij$ antes y despu\'es de los datos del LHC pueden encontrarse en las tablas \ref{tableintervals} y \ref{tableintervals2} respectivamente. En las tablas  \ref{tabdeltasummary} y \ref{tabdeltasummary2} se pueden encontrar las restricciones combinadas de los tres observables para las situaciones pre-LHC y post-LHC respectivamente. De este estudio tambi\'en hemos concluido que \bsg\ y $\Delta M_{B_s}$ son los observables m\'as restrictivos.

La ausencia por el momento de SUSY en el LHC ha movido los valores permitidos de las masas de las spart\'\i{}culas a regiones de energ\'\i{}a m\'as alta. Los escenarios estudiados para la situaci\'on pre-LHC tienen squarks m\'as ligeros que no est\'an permitidos en la situaci\'on post-LHC, y por lo tanto las correcciones a los observables de f\'\i{}sica de mesones $B$ son m\'as peque\~nas para los \'ultimos, y consecuentemente las regiones permitidas para los deltas son actualmente mayores de lo que eran antes de que el LHC empezara a funcionar.

Para la situaci\'on pre-LHC, las deltas NMFV en el sector top pod\'\i{}an ser grandes con  $|\de_{ct}^{XY}|$ mayores que ${\cal O}(0.1)$ y todav\'\i{}a compatibles con los datos de f\'\i{}sica de mesones $B$. En particular, $\delta^{RL}_{ct}$ y $\delta^{RR}_{ct}$ eran los par\'ametros menos restringidos, y en menor extensi\'on tambi\'en $\delta^{LR}_{ct}$. Los par\'ametros en el sector bottom, por el contrario, estaban muy restringidos. Los m\'as restringidos eran claramente  $\delta^{LR}_{sb}$ y $\delta^{RL}_{sb}$, especialmente el primero con s\'olo algunos valores singulares permitidos: positivos y del orden de  $3-5 \times 10^{-2}$, o bien negativos y con un tama\~no del orden de  $-7 \times 10^{-3}$; para el segundo los l\'\i{}mites se situaban alrededor de  $2 \times 10^{-2}$, tanto para valores positivos como negativos. $\delta^{RR}_{sb}$ era el par\'ametro menos restringido en el sector bottom, con intervalos permitidos de $|\delta^{RR}_{sb}| \lsim 0.4-0.9$, dependiendo del escenario. Para la actual situaci\'on post-LHC, los deltas menos restringidos son  $\delta^{RR}_{ct}$ y $\delta^{RR}_{sb}$, que en general pueden alcanzar un tama\~no de  ${\cal O}(1)$, despu\'es  $\delta^{LL}_{23}$, $\delta^{LR}_{ct}$ y $\delta^{RL}_{ct}$, con l\'\i{}mites un orden de magnitud m\'as peque\~nos,   estando el primero m\'as restringido que los otros dos, y finalmente los deltas m\'as restringidos son $\delta^{LR}_{sb}$ y $\delta^{RL}_{sb}$, que pueden alcanzar $|\deXYij|$ como m\'aximo de ${\cal O}(0.01)$. 

En t\'erminos generales la sensibilidad de  \bsg\ sigue el comportamiento que hemos descrito para los diferentes deltas. Para  $\Delta M_{B_s}$ es posible encontrar mayor o menor sensibilidad para cada diferente delta dependiendo del punto a estudiar en el espacio de par\'ametros del MSSM, debido principalmente a las diferentes contribuciones como las cajas de gluinos y los ping\"uinos dobles de Higgs, con muy diferentes comportamientos con  $\tan \beta$, como hemos explicado antes. Aunque se\~nalamos de nuevo que en general se encuentra una gran sensibilidad. En el caso de ${\rm BR}(B_s \to \mu^+ \mu^-)$, la mayor sensibilidad es para $\delta^{LL}_{23}$, y despu\'es dependiendo del punto del espacio de par\'ametros del MSSM para los otros deltas. Este comportamiento con $\delta^{LL}_{23}$ depende en gran medida del aumento a gran $\tan \beta$, pero las actuales cotas experimentales fuerzan a $M_A$ a ser grande en este \'ultimo caso, produciendo una disminuci\'on en la sensibilidad al delta que contrarresta el anterior efecto.

De las evaluaciones num\'ericas se puede observar tambi\'en que algunos puntos del MSSM excluidos con MFV (es decir, para deltas nulos) por observables de f\'\i{}sica de mesones $B$ pueden ser recuperados como escenarios permitidos despu\'es de fijar alguno de los deltas a un valor no nulo.

Despu\'es de tener una imagen general de las restricciones de los diferentes deltas por los observables de f\'\i{}sica de mesones $B$, y del comportamiento de estos observables, hemos considerado en el Cap\'\i{}tulo  \ref{higgsmasssquark} las correcciones a la masa del bos\'on de Higgs producidas por la mezcla de sabor de los squarks a partir de estos mismos deltas. En primer lugar hemos realizado un estudio anal\'\i{}tico de estas correcciones, derivando sus f\'ormulas a un loop mediante \fa\ and \fc. Las nuevas reglas de Feynman inclu\'\i{}das en el archivo del modelo est\'an listadas en el Ap\'endice A, y todos los resultados para las auto-energ\'\i{}as no renormalizadas y los tadpoles est\'an recogidos en el Ap\'endice B. En el caso de MFV, est\'an disponibles hasta correcciones a tres loops para las masas de los bosones de Higgs neutros, pero el caso de NMFV se ha limitado en el pasado al estudio de algunos elementos de mezcla de sabor en particular, como los c\'alculos a un loop teniendo en cuenta la mezcla $LL$ entre la tercera y la segunda generaci\'on de quarks escalares de tipo up. Nuestro estudio a un loop en esta tesis ha sido realizado con total generalidad, incluyendo todas las generaciones y todos los t\'erminos de mezcla de sabor para los squarks de tipo up y down. Tambi\'en hemos comprobado con esta generalidad la finitud de los resultados anal\'\i{}ticos para estas auto-energ\'\i{}as renormalizadas del Higgs. Estas correcciones presentadas en el Ap\'endice B han sido comparadas con el c\'odigo incluido en \fh, y en el proceso de comprobaci\'on de la finitud de las f\'ormulas hemos encontrado discrepancias entre la parte de los Higgs cargados de \fh, conduciendo a una versi\'on actualizada del c\'odigo.

El siguiente paso ha sido una evaluaci\'on num\'erica de las correcciones a la masa de los bosones de Higgs de mezcla de sabores de squarks entre la segunda y la tercera generaci\'on, teniendo en cuenta las restricciones de los observables de f\'\i{}sica de mesones $B$ de los cap\'\i{}tulos previos. La evaluaci\'on num\'erica ha sido realizada para los casos con uno o dos  $\deXYij$ distintos de cero, comparando las situaciones anterior y posterior a los datos del LHC. Con respecto a los estudios con un delta, se puede ver de nuestros gr\'aficos (por ejemplo figs.  \ref{figdeltamh0}, \ref{figdeltamH0} y
\ref{figdeltamHp}) c\'omo las correcciones a todas las masas son sim\'etricas en el signo de los deltas, a diferencia de lo que ocurr\'\i{}a con los observables de la f\'\i{}sica de mesones $B$, y que pueden ser positivas o negativas dependiendo del escenario y del valor de delta. En particular para el bos\'on de Higgs ligero, las correcciones tienden a adoptar una ``forma de M'', siendo positivas y peque\~nas para deltas peque\~nos, y despu\'es negativas y grandes seg\'un crecen los deltas. Para la situaci\'on pre-LHC, hemos encontrado una sensibilidad muy fuerte a  $\delta^{LR}_{sb}$ y $\delta^{RL}_{sb}$ para todos los bosones de Higgs, que son tambi\'en los deltas m\'as restringidos de f\'\i{}sica de mesones $B$. Para el bos\'on de Higgs ligero hemos encontrado una importante sensibilidad a  $\delta^{LR}_{ct}$ y $\delta^{RL}_{ct}$, alcanzando valores de hasta decenas de GeV para el caso de bajo $\tan \beta$ donde los observables de f\'\i{}sica de mesones $B$ son menos restrictivos. Para la situaci\'on post-LHC, el comportamiento general de las correcciones con los diferentes deltas se mantiene aproximadamente igual. Los valores de las correcciones son muy grandes comparados con la precisi\'on del LHC, y deben ser tenidos en cuenta en cualquier an\'alisis del bos\'on de Higgs en el marco de NMFV. De hecho, encontramos que el LHC es capaz de poner l\'\i{}mites a algunos de estos par\'ametros delta como $\delta^{LR}_{ct}$ y $\delta^{RL}_{ct}$ que son permitidos por otros observables, como es el caso de los observables de f\'\i{}sica de mesones $B$.

El an\'alisis num\'erico ha sido extendido para las correcciones al bos\'on de Higgs ligero al caso de tener simult\'aneamente dos deltas con valores no nulos, en la situaci\'on pre-LHC. Las correcciones m\'as grandes permitidas por los datos de f\'\i{}sica de mesones $B$ se ha obtenido para los casos de $(\de^{LL}_{23},\de^{LR}_{ct} )$ y $(\de^{LL}_{23},\de^{RL}_{ct} )$. Estas correcciones alcanzan varias decenas de GeV en la situaci\'on pre-LHC. Las regiones m\'as amplias permitidas por los observables de f\'\i{}sica de mesones $B$ y las mayores correcciones corresponden a escenarios con bajo $\tan \beta$  y espectro de SUSY pesado. Tambi\'en se han obtenido importantes correcciones en el plano $(\de^{LL}_{23},\de^{RR}_{ct} )$ de hasta decenas de GeV para algunos puntos espec\'\i{}ficos. Las correcciones encontradas para los bosones de Higgs pesados en el plano con dos deltas han sido mucho menores que las previas.

La siguiente parte de nuestro investigaci\'on ha tratado el estudio de la mezcla de sabor de los sleptones. En el Cap\'\i{}tulo \ref{phenoflavourslep} hemos desarrollado un estudio de la fenomenolog\'\i{}a de diferentes observables LFV que hemos concluido con las restricciones que imponen en los par\'ametros de mezcla de sabor  $\deABij$ del sector de los sleptones. Los procesos LFV cuya fenomenolog\'\i{}a ha sido estudiada en este cap\'\i{}tulo han sido desintegraciones radiativas $\mu \to e \gamma$, $\tau \to e \gamma$ y $\tau \to \mu \gamma$ ($l_j \to l_i \gamma$), desintegraciones lept\'onicas  $\mu \to 3 e$, $\tau \to 3 e$ y $\tau \to 3 \mu$, desintegraciones semilept\'onicas $\tau \to \mu \eta$ y $\tau \to e \eta$ y conversi\'on de $\mu$ en $e$ en n\'ucleos pesados. Las desintegraciones radiativas son generalmente las m\'as restrictivas, pero se han incluido las desintegraciones lept\'onicas y semilept\'onicas porque dan acceso a los par\'ametros del sector de Higgs, dado que pueden estar mediadas por bosones de Higgs. Nuestro estudio num\'erico ha sido realizado para escenarios permitidos por los \'ultimos datos del LHC, incluyendo una masa para el bos\'on de Higgs ligero alrededor de 125 - 126 GeV, y tambi\'en para el valor actual del momento magn\'etico an\'omalo del mu\'on,  $(g-2)_\mu$.

Nuestro estudio ha mostrado que las desintegraciones radiativas son todav\'\i{}a los observables m\'as restrictivos a los par\'ametros de mezcla de sabor de los sleptones. Comparando las mezclas de sabor entre diferentes pares de generaciones, los l\'\i{}mites experimentales para los observables relacionados con las mezclas 12 son alrededor de cuatro \'ordenes de magnitud m\'as peque\~nos que los otros en todos los observables, y por tanto estas mezclas est\'an mucho m\'as restringidas. Los l\'\i{}mites para las mezclas 23 son similares a los de las mezclas 13. Tambi\'en se ha mostrado que todos los observables se comportan sim\'etricamente respecto a los signos de los deltas. Los l\'\i{}mites para los diferentes $\deABij$ para los puntos estudiados est\'an recogidos en la Tabla \ref{boundsSpoints}. Los l\'\i{}mites m\'as restrictivos a las mezclas $LL$ y $RR$ se han encontrado para el caso de gran  $\tan \beta$, mientras que las mezclas $LR$ son independientes de  $\tan \beta$ y tienen mayores restricciones para los casos con sleptones ligeros.

Tambi\'en hemos encontrado que el efecto de la mediaci\'on por parte del bos\'on de Higgs pseudoescalar en las desintegraciones semilept\'onicas del tau no era muy importante dado que los l\'\i{}mites actuales de exclusi\'on para la masa de este bos\'on de Higgs nos llevan a asumir un bos\'on $A$ bastante pesado, cuyos efectos son por lo tanto reducidos. Estos observables no imponen restricciones a las mezclas 23 y 13. Con respecto a la mediaci\'on por parte del fot\'on en las desintegraciones lept\'onicas y en la conversi\'on de $\mu$ en $e$ en n\'ucleos pesados, confirmamos que este canal fot\'onico domina sobre las otras contribuciones, y por tanto hay una correlaci\'on entre las predicciones de las tasas de desintegraci\'on lept\'onicas, las tasas de conversi\'on  $\mu - e$ y las tasas de desintegraciones radiativas.

Despu\'es de esto, nuestro estudio se ha extendido al caso de tener simult\'aneamente dos deltas diferentes de cero. Primero hemos estudiado el caso $(\delta^{AB}_{23},\delta^{CD}_{23})$ con $AB\neq CD$ donde hemos encontrado interferencias constructivas o destructivas en $\br(\tau \to \mu \gamma)$ dependiendo de los signos de los deltas. El caso m\'as relevante ha sido $(\delta^{LR}_{23},\delta^{LL}_ {23})$ donde se han encontrado importantes interferencias destructivas para el caso de deltas con signos opuestos, que hac\'\i{}an mayor el valor m\'aximo permitido para cada delta en comparaci\'on con el caso de tener s\'olo ese delta. La forma y tama\~no de las regiones permitidas con soluciones extremas para los deltas como $(\pm 0.9, \mp 0.9)$ provienen de nuestro c\'alculo con diagonalizaci\'on de la matriz de masa completa, y por tanto no pueden ser descritas con las f\'ormulas m\'as simples de MIA que se usan frecuentemente en la literatura. Como segundo estudio en el plano con dos deltas, hemos estudiado las restricciones en las mezclas (13,23) provenientes de  $\mu \to e \gamma$ y de la conversi\'on $\mu-e$ en comparaci\'on con los l\'\i{}mites provenientes de los observables relacionados con cada delta individual, y hemos descubierto restricciones sorprendentemente mejores provenientes de esta situaci\'on de doble delta.

El estudio de los l\'\i{}mites para la mezcla de sabor de los sleptones de las desintegraciones radiativas ha sido completado con un estudio sobre la dependencia de los l\'\i{}mites de los deltas con respecto a los par\'ametros m\'as relevantes del MSSM en el siguiente contexto de LFV: $\tan \beta$ y la escala de masa de los sleptones $m_{\rm SUSY}$. El estudio de los l\'\i{}mites ha sido realizado compar\'andolo con la regiones permitidas por las medidas de la masa del bos\'on de Higgs y el valor de  $(g-2)_\mu$. Observamos c\'omo los valores m\'aximos permitidos para los deltas $\delta^{LL}_{ij}$ y $\delta^{RR}_{ij}$ escalan con $m_{\rm SUSY}$ y $\tb$ creciendo aproximadamente con $m_{\rm SUSY}$ creciente como $\sim m_{\rm SUSY}^2$ y decreciendo con (gran) $\tb$ decreciente como $\sim 1/\tb$. Los valores m\'aximos permitidos de los deltas $\delta^{LR}_{ij}$ (y similarmente $\delta^{RL}_{ij}$) son independientes de $\tb$ y crecen aproximadamente como  $\sim m_{\rm SUSY}$  con $m_{\rm SUSY}$ creciente. Tambi\'en se ha observado c\'omo $(g-2)_\mu$ requiere de un sector SUSY-EW, es decir charginos, neutralinos y sleptones, bastante ligero y una $\tb$ bastante grande, y $\Mh$ requiere un sector de los squarks bastante pesado, creando cierta tensi\'on entre estos dos observables cuando los sectores de los squarks y de los sleptones est\'an relacionados, como ocurr\'\i{}a en nuestros escenarios.

Los valores m\'aximos para los deltas encontrados en nuestros escenarios en la regi\'on del espacio de par\'ametros del MSSM favorecida por los datos del LHC y de $(g-2)_\mu$ se encuentran en los siguientes intervalos:
$|\delta^{LL}_{12}|_{\rm max} \sim {\cal O} (10^{-5},10^{-4}) $, 
$|\delta^{LR}_{12}|_{\rm max} \sim {\cal O} (10^{-6},10^{-5}) $,
$|\delta^{RR}_{12}|_{\rm max} \sim {\cal O} (10^{-3},10^{-2}) $,
$|\delta^{LL}_{23}|_{\rm max} \sim {\cal O} (10^{-2},10^{-1}) $,
$|\delta^{LR}_{23}|_{\rm max} \sim {\cal O} (10^{-2},10^{-1}) $,
$|\delta^{RR}_{23}|_{\rm max} \sim {\cal O} (10^{-1},10^{0}) $. L\'\i{}mites generales muy similares a los de las mezclas 23 se han encontrado para las mezclas 13.

Nuestro trabajo ha sido concluido con el Cap\'\i{}tulo  \ref{lfvhiggsdecaysslepton}, donde hemos estudiado los efectos en las desintegraciones LFV del Higgs de mezcla de sabor slept\'onico. Esta investigaci\'on ha sido realizada centr\'andonos en la posibilidad de un espectro de SUSY muy pesado por encima de la escala del TeV y nuestra motivaci\'on principal ha sido encontrar efectos de SUSY no desacoplantes en las desintegraciones LFV del Higgs dominadas por la mezcla general de sabor slept\'onico.

Los observables elegidos para este \'ultimo estudio en esta tesis han sido las desintegraciones LFV del Higgs $\phi \to \tau \mu$, con $\phi=h,H,A$, y la desintegraci\'on radiativa LFV  $\tau \to \mu \gamma$. Resultados y conclusiones similares pueden ser extrapolados para las desintegraciones  $\phi \to \tau e$ y $\tau \to e \gamma$. Como vimos en los cap\'\i{}tulos previos, la desintegraci\'on radiativa $\tau \to \mu \gamma$  es la m\'as restrictiva de los observables estudiados para la mezcla de sabor de sleptones entre las generaciones segunda y tercera. Los acoplamientos LFV del Higgs a los leptones son proporcionales a las masas de los leptones m\'as pesadas involucradas, y por tanto hemos estado m\'as interesados en las desintegraciones a $\tau$ que a otros leptones. Adem\'as vimos en los cap\'\i{}tulos anteriores que las restricciones de $\mu \to e \gamma$ en la mezcla de sabor lept\'onico est\'an muy ajustadas y por lo tanto no llevan a tasas de $\phi \to \mu e$ medibles. Las predicciones para la desintegraci\'on  $\phi \to \tau e$ son muy similares a las de la desintegraci\'on $\phi \to \tau \mu$, y desde el punto de vista experimental la sensibilidad es tambi\'en muy parecida, as\'\i{} que nuestros resultados son f\'acilmente traducibles a este otro canal.

La evaluaci\'on num\'erica de nuestro estudio ha sido realizada con {\it SPHENO} y un c\'odigo privado de FORTRAN, donde hemos implementado las f\'ormulas completas a un loop para las desintegraciones LFV del Higgs y las radiativas. El espectro del sector de Higgs y las anchuras totales de los bosones de Higgs han sido calculadas como antes con \fh\ a dos loops.

Hemos mostrado que las tres desintegraciones LFV del Higgs $\phi \to \tau \mu$, con $\phi=h,H,A$, presentan un comportamiento no desacoplante con $m_\text{SUSY}$, manteni\'endose constantes a gran  $m_\text{SUSY}>2$ TeV. Esto se complementa con el comportamiento de las desintegraciones radiativas que muestran por el contrario un desacoplamiento r\'apido con  $m_\text{SUSY}$. De esta forma tenemos una situaci\'on \'optima para detectar indirectamente los efectos de SUSY incluso con un espectro de SUSY de escala muy pesado. Con respecto a los otros par\'ametros, los cocientes de ramificaci\'on crecen a gran  $\tan \beta$ como  BR($h, H, A \to \tau \mu$) $\sim (\tan\beta)^2$ para los casos $LL$ y $RR$, y decrecen como  $(\tan\beta)^{-2}$ en el caso $\tilde \delta_{23}^{LR}$. El par\'ametro $\delta_{23}^{AB}$ m\'as relevante a bajo $\tan \beta$ para el bos\'on de Higgs ligero $h$ es  ${\tilde \delta}_{23}^{LR}$, que da lugar a mayores tasas de desintegraci\'on LFV del Higgs que  $\delta_{23}^{LL}$ y $\delta_{23}^{RR}$, mientras que para los bosones de Higgs $H$ y $A$ el par\'ametro m\'as relevante es $\delta_{23}^{LL}$. Para valores grandes de $\tan \beta$, el par\'ametro m\'as relevante para los tres bosones de Higgs es  $\delta_{23}^{LL}$. Tambi\'en se mostr\'o que tomando $M_2=\mu$ grande los cocientes de ramificaci\'on tambi\'en aumentaban. Con respecto a $M_A$ ser\'\i{}a deseable tener un caso con bajo $M_A$ y gran $\tan \beta$ para aumentar el cociente de ramificaci\'on de los bosones de Higgs pesados, pero los l\'\i{}mites de exclusi\'on en el plano $M_A-\tan\beta$  hacen que esto sea imposible.

Se ha mostrado c\'omo los resultados son sim\'etricos respecto al signo de los  $\delta_{23}^{AB}$, e iguales para  ${\tilde \delta}_{23}^{LR}$ y ${\tilde\delta}_{23}^{RL}$. Y tambi\'en hemos visto que considerar dos deltas $\delta_{23}^{AB}$ diferentes de cero produce interferencias que pueden aumentar o reducir los cocientes de ramificaci\'on.

Este cap\'\i{}tulo se ha concluido con el c\'alculo de las tasas de eventos LFV que ser\'an producidas por estas desintegraciones $\phi \to \tau \mu$ en el LHC en el plano $M_A-\tan\beta$ con gran $m_\text{SUSY} \geq 5$ TeV, considerando la fase actual con una energ\'\i{}a en el centro de masas de $\sqrt{s} =$ 8 TeV y en la fase futura con $\sqrt{s} =$ 14 TeV. Para el caso de una \'unica mezcla, hemos concluido que para la mezcla $RR$ no ser\'a posible que \'esta sea medida en el LHC. El caso $LL$ producir\'\i{}a hasta 5 eventos para las desintegraciones de los bosones de Higgs pesados en la fase futura del LHC a gran $\tan \beta$, y ninguno para el bos\'on de Higgs ligero. Y finalmente se ha encontrado que $LR$ es el par\'ametro m\'as relevante, produciendo hasta 50 eventos para  $m_\text{SUSY} \geq 5$ TeV con baja  $\tan \beta$ para el bos\'on de Higgs ligero en la fase actual, y cientos de eventos en la fase futura. En el caso de los bosones de Higgs pesados hemos obtenido unos pocos eventos s\'olo en la fase futura. Todos estos resultados son por supuesto permitidos por la desintegraci\'on radiativa relevante  $\tau \to \mu \gamma$. 

En el caso de la mezcla doble, la mejor situaci\'on se ha encontrado para el caso de las mezclas $LL$ y $LR$. En este caso la sensibilidad con respecto a $\tan \beta$ mejor\'o, y hemos obtenido por ejemplo, unos pocos eventos para el bos\'on de Higgs ligero en la fase actual del LHC incluso para $\tan\beta \sim 15$ moderada y gran $m_A \geq $ 500 GeV. En la futura fase del LHC ha aumentado el alcance para valores mayores de $\tan \beta$ y hemos obtenido algunos eventos para $\tan\beta\sim 40$ muy grande y $m_A \geq$ 800 GeV. Las mayores tasas encontradas en esta situaci\'on de mezcla doble han sido en cualquier caso para $h \to \tau \mu$ y est\'an claramente localizadas en la regi\'on de bajo $\tan\beta$ donde hemos predicho para la fase actual del LHC hasta aproximadamente 75 eventos, y hasta 750 eventos para la fase futura del LHC. En la fase futura hemos obtenido alrededor de 10 eventos como m\'aximo para los canales  $H, A \to \tau \mu$ en la regi\'on de baja $\tan \beta$ y 5 eventos como m\'aximo en la regi\'on de gran $\tan \beta$.

Como resumen de nuestro trabajo, hemos presentado un estudio exhaustivo de las implicaciones fenomenol\'ogicas de la mezcla general de sabor sfermi\'onico en un contexto de NMFV-MSSM en observables de f\'\i{}sica de mesones $B$, correcciones a las masas de los bosones de Higgs, observables LFV lept\'onicos y desintegraciones LFV del Higgs. A lo largo de nuestro estudio, hemos impuesto restricciones a los par\'ametros de mezcla de sabor sfermi\'onicos a partir de las \'ultimas medidas experimentales. Con respecto a los bosones de Higgs, se han mostrado los importantes efectos de la mezcla de sabor de squarks en el valor de las masas de los bosones de Higgs, y la posibilidad de usar el valor de las masas para imponer restricciones en la mezcla de sabor de los squarks. Tambi\'en hemos mostrado c\'omo las desintegraciones LFV del Higgs son observables \'unicos para probar SUSY incluso en la desalentadora situaci\'on experimental de tener una supersimetr\'\i{}a muy pesada que no fuera alcanzable directamente en los colisionadores.

Supersimetr\'\i{}a es una de las propuestas m\'as interesantes para la f\'\i{}sica m\'as all\'a del SM, y esperamos que nuestro trabajo haya dejado claro la importancia de la mezcla de sabor no s\'olo en su caracterizaci\'on, sino en las nuevas ventanas que abre a la detecci\'on indirecta de SUSY. Ahora dejemos a los experimentos hablar.

\bigskip

\clearpage
\newpage
\begin{appendix}

\chapter{Relevant Feynman rules of the NMFV scenarios} 

We list the new Feynman rules of the NMFV scenarios that are 
involved in the present computation of Chapter \ref{higgsmasssquark}.
The corresponding couplings to the Higgs boson $H$ are obtained
from the ones listed here for the lightest Higgs boson $h$ by replacing 

\begin{equation}
c_{\alpha}\rightarrow s_{\alpha}\quad;\quad s_{\alpha}\rightarrow-c_{\alpha}\quad;\quad s_{\alpha+\beta}\rightarrow-c_{\alpha+\beta}\quad;\quad c_{2\alpha}\rightarrow-c_{2\alpha}\label{eq:cambio}
\end{equation}

The abbreviated notation used in the following formulas for some quantities is the following:
$s_{x}=\sin x;\, c_{x}=\cos x;\, \sw=\sin\theta_{W};\, \cw=\cos\theta_{W}=\frac{\MW}{\MZ};\, t_{\beta}=\tan\beta$. For the definitions of the other appearing quantities in this appendix see Chapters \ref{supersymmetry} and \ref{paramnmfvscen}.

{\it{1. Couplings of two squarks and one/two Higgs bosons}}\\

\begin{table}[H]
\begin{tabular}{ll}
\parbox[c]{1em}{\includegraphics{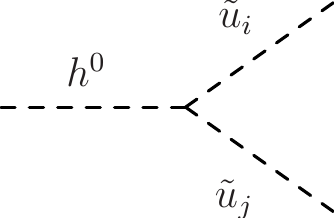}}
 &\hspace*{3.0cm} -$\sum_{k,l=1}^{3} \frac{ie}{6\MW\cw\sw s_{\beta}}
\left\{ R^{\tilde{u}\,*}_{i,k}\left\{ \delta_{kl} R^{\tilde{u}}_{j,l} 
\left(6\ca\cw {m_{{u}_{k}}^2}-\MW\MZ s_{\alpha+\beta} s_{\beta} 
(3-4\sw^2)\right)\right.
\right.$\vspace*{-0.6cm}\\
&\hspace*{6.0cm}$\left.+3\cw R^{\tilde{u}}_{j,3+l}(A^{u}_{k,l}\ca\, 
m_{{u}_{k}}+
\delta_{kl}\,m_{{u}_{k}}\,\mu^{*}\sa)
\right\}$\\
&\hspace*{6.0cm}$+R^{\tilde{u}\,*}_{i,3+k}\left\{
\delta_{kl} R^{\tilde{u}}_{j,3+l} 
\left(6\ca\cw {m_{{u}_{k}}^2}-4\MW\MZ s_{\alpha+\beta} s_{\beta}\sw^2
\right)\right.$\\
&\hspace*{6.0cm}$\left.\left.+
3\cw R^{\tilde{u}}_{j,l}(A^{u*}_{l,k}\ca\, m_{{u}_{l}}+
\delta_{kl}\,m_{{u}_{k}}\,\mu\sa)
\right\}
\right\}$\\
&\\
\parbox[c]{1em}{\includegraphics{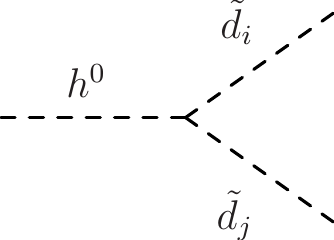}}
&\hspace*{3.0cm} $\sum_{k,l=1}^{3} \frac{ie}{6\MW\cw\sw c_{\beta}}
\left\{ R^{\tilde{d}\,*}_{i,k}\left\{ \delta_{kl} R^{\tilde{d}}_{j,l}
\left(6\sa\cw {m_{{d}_{k}}^2}-\MW\MZ s_{\alpha+\beta} c_{\beta} 
(3-2\sw^2)\right)\right.
\right.$\vspace*{-0.6cm}\\
&\hspace*{6.0cm}$\left.+3\cw R^{\tilde{d}}_{j,3+l}(A^{d}_{k,l}\sa\, 
m_{{d}_{k}}+
\delta_{kl}\,m_{{d}_{k}}\,\mu^{*}\ca)
\right\}$\\
&\hspace*{6.0cm}$+R^{\tilde{d}\,*}_{i,3+k}\left\{
\delta_{kl} R^{\tilde{d}}_{j,3+l} 
\left(6\sa\cw {m_{{d}_{k}}^2}-2\MW\MZ s_{\alpha+\beta} c_{\beta}\sw^2
\right)\right.$\\
&\hspace*{6.0cm}$\left.\left.+
3\cw R^{\tilde{d}}_{j,l}(A^{d*}_{l,k}\sa\, m_{{d}_{l}}+
\delta_{kl}\,m_{{d}_{k}}\,\mu\ca)
\right\}
\right\}$\\
&\\
\end{tabular}
\end{table}
\begin{table}[H]
\begin{tabular}{ll}
\parbox[c]{1em}{\includegraphics{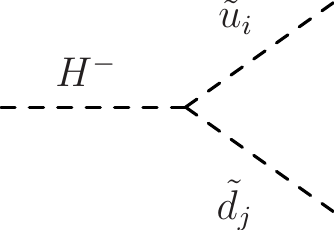}}
&$\sum_{k,l=1}^{3} \frac{ie}{\sqrt{2}\MW\sw t_{\beta}}
\left\{ R^{\tilde{u}\,*}_{i,3+k}\left\{ R^{\tilde{d}}_{j,l} 
\left(\sum_{n=1}^{3} A^{u*}_{n,k} \, m_{{u}_{n}} {V_{\rm{CKM}}^{*\,nl}}+
 m_{{u}_{k}}\,\mu {V_{\rm{CKM}}^{*\,kl}} t_{\beta}\right)\right.\right.$
\vspace*{-0.6cm}\\
&\hspace*{3.0cm}$\left.+ m_{{d}_{l}} m_{{u}_{k}} {V_{\rm{CKM}}^{*\,kl}}
R^{\tilde{d}}_{j,3+l}(1+{t_{\beta}}^{2})\right\}$\\
&\hspace*{3.0cm}$+R^{\tilde{u}\,*}_{i,k}\left\{ R^{\tilde{d}}_{j,3+l}t_{\beta} 
\left(\sum_{n=1}^{3} A^{d}_{n,l} \, m_{{d}_{n}} {V_{\rm{CKM}}^{*\,kn}} 
t_{\beta}+
 m_{{d}_{l}}\,\mu^{*} {V_{\rm{CKM}}^{*\,kl}}\right)\right.$\\
&\hspace*{3.0cm}$\left.\left.+ {V_{\rm{CKM}}^{*\,kl}}\, R^{\tilde{d}}_{j,l}
\left(m^{2}_{{u}_{k}}-t_{\beta}(\MW^2 s_{2\beta}-m^{2}_{{d}_{l}}t_{\beta})
\right)\right\}\right\}$\\
&\\
{\includegraphics{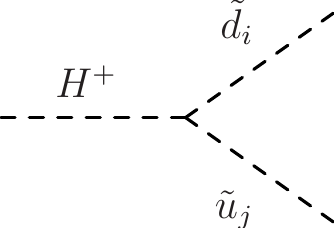}}
 &\put(10,30){$\sum_{k,l=1}^{3} \frac{ie}{\sqrt{2}\MW\sw t_{\beta}}
\left\{ R^{\tilde{d}\,*}_{i,3+l}\left\{ R^{\tilde{u}}_{j,k} t_{\beta}
\left(\sum_{n=1}^{3} A^{d*}_{n,l} \, m_{{d}_{n}} {V^{kn}_{\rm{CKM}}}t_{\beta}+
 m_{{d}_{l}}\,\mu {V^{kl}_{\rm{CKM}}}\right)\right.\right.$}
\vspace*{-0.6cm}\\
&\hspace*{3.0cm}$\left.+ m_{{d}_{l}} m_{{u}_{k}} {V^{kl}_{\rm{CKM}}}
R^{\tilde{u}}_{j,3+k}(1+{t_{\beta}}^{2})\right\}$\\
&\hspace*{3.0cm}$+R^{\tilde{d}\,*}_{i,l}\left\{ R^{\tilde{u}}_{j,3+k} 
\left(\sum_{n=1}^{3} A^{u}_{n,k} \, m_{{u}_{n}} {V^{nl}_{\rm{CKM}}}+
 m_{{u}_{k}}\,\mu^{*} {V^{kl}_{\rm{CKM}}}t_{\beta}\right)\right.$\\
&\hspace*{3.0cm}$\left.\left.+ {V^{kl}_{\rm{CKM}}}\, R^{\tilde{u}}_{j,k}
\left(m^{2}_{{u}_{k}}-t_{\beta}(\MW^2 s_{2\beta}-m^{2}_{{d}_{l}}t_{\beta})
\right)\right\}\right\}$\\
\parbox[c]{1em}{\includegraphics{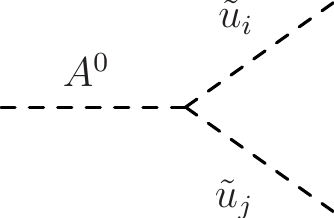}}
& -$\sum_{k,l=1}^{3} \frac{e}{2\MW\sw t_{\beta}}
\left\{ R^{\tilde{u}\,*}_{i,3+k} R^{\tilde{u}}_{j,l}
\left(A^{u\,*}_{l,k}\,m_{{u}_{l}}+\delta_{kl}\,m_{{u}_{k}}\,\mu\, t_{\beta}
\right)\right.$\vspace*{-0.6cm}\\
&\hspace*{3.0cm}$\left.-
R^{\tilde{u}\,*}_{i,k} R^{\tilde{u}}_{j,3+l}
\left(A^{u}_{k,l}{m_{u}}_{k}+\delta_{kl}\,m_{{u}_{k}}\,\mu^{*}\,t_{\beta}
\right)\right\}$\\
&\\
&\\
\parbox[c]{1em}{\includegraphics{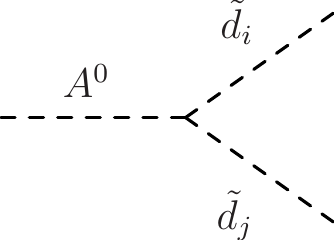}}
& -$\sum_{k,l=1}^{3} \frac{e}{2\MW\sw}
\left\{ R^{\tilde{d}\,*}_{i,3+k} R^{\tilde{d}}_{j,l}
\left(A^{d\,*}_{l,k}{m_{d}}_{l}t_{\beta}+\delta_{kl}\,m_{{d}_{k}}\mu
\right)\right.$\vspace*{-0.6cm}\\
&\hspace*{3.0cm}$\left.-
R^{\tilde{d}\,*}_{i,k} R^{\tilde{d}}_{j,3+l}
\left(A^{d}_{k,l}\,m_{{d}_{k}}\,t_{\beta}+
\delta_{kl}\,m_{{d}_{k}}\,\mu^{*}
\right)\right\}$\\
&\\
&\\
\parbox[c]{1em}{\includegraphics{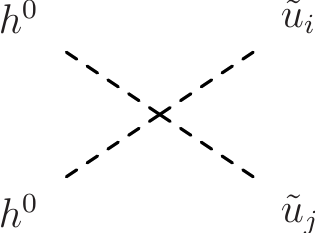}}
 & -$\sum_{k=1}^{3} \frac{ie^2}{12\MW^2\cw^2\sw^2 s^{2}_{\beta}}
\left\{ R^{\tilde{u}}_{i,k} R^{\tilde{u}\,*}_{j,k}
\left(6\ca^2\cw^2 {m_{{u}_{k}}^2}-c_{2\alpha}\MW^2 s^{2}_{\beta} 
(3-4\sw^2)\right)
\right.$\vspace*{-0.6cm}\\
&\hspace*{3.0cm}$\left.+2 R^{\tilde{u}}_{i,3+k} R^{\tilde{u}\,*}_{j,3+k}
\left(3\ca^2\cw^2 {m_{{u}_{k}}^2}-2c_{2\alpha}\MW^2 s^{2}_{\beta}\sw^2
\right)\right\}$\\
&\\
\parbox[c]{1em}{\includegraphics{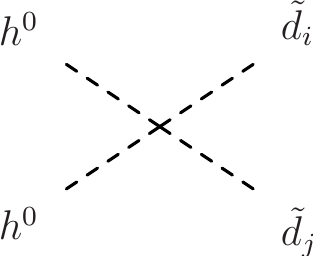}}
 &$-\sum_{k=1}^{3} \frac{ie^2}{12\MW^2\cw^2\sw^2 c^{2}_{\beta}}
\left\{ R^{\tilde{d}}_{i,k} R^{\tilde{d}\,*}_{j,k}
\left(6\sa^2\cw^2 {m_{{d}_{k}}^2}+c_{2\alpha}\MW^2 c^{2}_{\beta} 
(3-2\sw^2)\right)
\right.$\vspace*{-0.6cm}\\
&\hspace*{3.0cm}$\left.+2 R^{\tilde{d}}_{i,3+k} R^{\tilde{d}\,*}_{j,3+k}
\left(3\sa^2\cw^2 {m_{{d}_{k}}^2}+c_{2\alpha}\MW^2 c^{2}_{\beta}\sw^2
\right)\right\}$\\
&\\
\end{tabular}
\end{table}

\begin{table}[H]
\begin{tabular}{ll}
\parbox[c]{1em}{\includegraphics{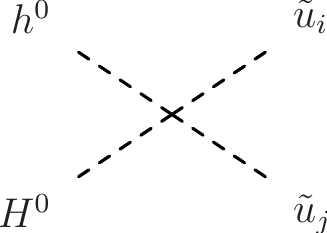}}
 & -$\sum_{k=1}^{3} \frac{ie^2 s_{2\alpha}}{12\MW^2\cw^2\sw^2 s^{2}_{\beta}}
\left\{ R^{\tilde{u}}_{i,k} R^{\tilde{u}\,*}_{j,k}
\left(3\cw^2 {m_{{u}_{k}}^2}-\MW^2 s^{2}_{\beta} 
(3-4\sw^2)\right)
\right.$\vspace*{-0.6cm}\\
&\hspace*{3.0cm}$\left.+R^{\tilde{u}}_{i,3+k} R^{\tilde{u}\,*}_{j,3+k}
\left(3\cw^2 {m_{{u}_{k}}^2}-4\MW^2 s^{2}_{\beta}\sw^2
\right)\right\}$\\
&\\
\parbox[c]{1em}{\includegraphics{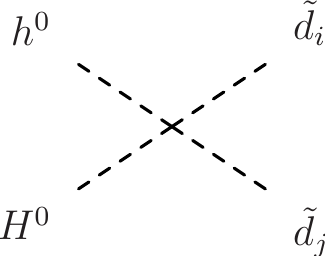}}
 &$\sum_{k=1}^{3} \frac{ie^2 s_{2\alpha}}{12\MW^2\cw^2\sw^2 c^{2}_{\beta}}
\left\{ R^{\tilde{d}}_{i,k} R^{\tilde{d}\,*}_{j,k}
\left(3\cw^2 {m_{{d}_{k}}^2}-\MW^2 c^{2}_{\beta} 
(3-2\sw^2)\right)
\right.$\vspace*{-0.6cm}\\
&\hspace*{3.0cm}$\left.+ R^{\tilde{d}}_{i,3+k} R^{\tilde{d}\,*}_{j,3+k}
\left(3\cw^2 {m_{{d}_{k}}^2}-2\MW^2 c^{2}_{\beta}\sw^2
\right)\right\}$\\
&\\
{\includegraphics{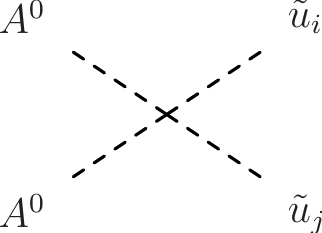}}
 &\put(10,30){$-\sum_{k=1}^{3} \frac{ie^2}{12\MW^2\cw^2\sw^2 t^{2}_{\beta}}
\left\{ R^{\tilde{u}}_{i,k} R^{\tilde{u}\,*}_{j,k}
\left(6\cw^2 {m_{{u}_{k}}^2}-c_{2\beta}\MW^2 t^{2}_{\beta} 
(3-4\sw^2)\right)
\right.$}\vspace*{-0.6cm}\\
&\hspace*{3.0cm}$\left.+2 R^{\tilde{u}}_{i,3+k} R^{\tilde{u}\,*}_{j,3+k}
\left(3\cw^2 {m_{{u}_{k}}^2}-2c_{2\beta}\MW^2 t^{2}_{\beta}\sw^2
\right)\right\}$\\
&\\
&\\
\parbox[c]{1em}{\includegraphics{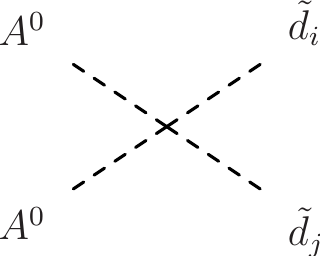}}
 &$-\sum_{k=1}^{3} \frac{ie^2}{12\MW^2\cw^2\sw^2}
\left\{ R^{\tilde{d}}_{i,k} R^{\tilde{d}\,*}_{j,k}
\left(6\cw^2 {m_{{d}_{k}}^2}t^{2}_{\beta}+c_{2\beta}\MW^2
(3-2\sw^2)\right)
\right.$\vspace*{-0.6cm}\\
&\hspace*{3.0cm}$\left.+2 R^{\tilde{d}}_{i,3+k} R^{\tilde{d}\,*}_{j,3+k}
\left(3\cw^2 {m_{{d}_{k}}^2}t^{2}_{\beta}+c_{2\beta}\MW^2 \sw^2
\right)\right\}$\\
&\\
\parbox[c]{1em}{\includegraphics{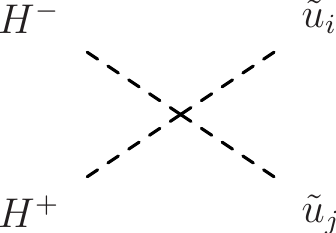}}
 &$-\sum_{k,l=1}^{3} \frac{ie^2}{12\MW^2\sw^2\cw^2 t^{2}_{\beta}}
\left\{ R^{\tilde{u}\,*}_{i,k}R^{\tilde{u}}_{j,l}t^{2}_{\beta}
\left(6\sum_{n=1}^{3} m^{2}_{{d}_{n}} {V_{\rm{CKM}}^{*\,kn}} {V^{ln}_{\rm{CKM}}}
\cw^2t^{2}_{\beta}\right.\right.$
\vspace*{-0.6cm}\\
&\hspace*{3.0cm}$\left.+c_{2\beta}\delta_{kl}\MW^2(1+2\cw^2)\right)$\\
&\hspace*{3.0cm}$\left.+2\delta_{kl}
R^{\tilde{u}\,*}_{i,3+k}R^{\tilde{u}}_{j,3+l}
\left(3\cw^2m^{2}_{{u}_{k}}-2c_{2\beta}\MW^2\sw^2t^{2}_{\beta}\right)
\right\}$\\
&\\
{\includegraphics{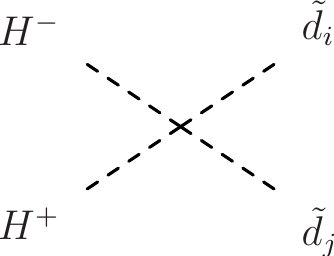}}
 &\put(10,30){$-\sum_{k,l=1}^{3} \frac{ie^2}{12\MW^2\sw^2\cw^2 t^{2}_{\beta}}
\left\{ R^{\tilde{d}\,*}_{i,k}R^{\tilde{d}}_{j,l}
\left(6\sum_{n=1}^{3} m^{2}_{{u}_{n}} {V^{nk}_{\rm{CKM}}} {V_{\rm{CKM}}^{*\,nl}}
\cw^2
\right.\right.$}
\vspace*{-0.6cm}\\
&\hspace*{3.0cm}$\left.+c_{2\beta}\delta_{kl}\MW^2\,t^{2}_{\beta}
(1-4\cw^2)\right)$\\
&\hspace*{3.0cm}$\left.+2\delta_{kl}
R^{\tilde{d}\,*}_{i,3+k}R^{\tilde{d}}_{j,3+l} t^{2}_{\beta}
\left(3\cw^2 t^{2}_{\beta} m^{2}_{{d}_{k}}+c_{2\beta}\MW^2\sw^2\right)
\right\}$\\
&\\
&\\
\end{tabular}
\end{table}

{\it{3. Couplings of two squarks and one/two gauge bosons}}\\
\vspace{-1.0cm}
\begin{table}[H]
\begin{tabular}{ll}
\parbox[c]{1em}{\includegraphics{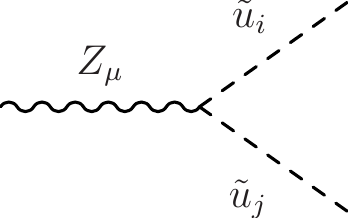}}
 & $\sum_{k=1}^{3} \frac{ie}{6\cw\sw}
\left( 4 R^{\tilde{u}\,*}_{i,3+k} R^{\tilde{u}}_{j,3+k}\sw^2-
R^{\tilde{u}\,*}_{i,k} R^{\tilde{u}}_{j,k} (3-4\sw^2)
\right)(p+p^{'})_{\mu}$\\
&\\
\parbox[c]{1em}{\includegraphics{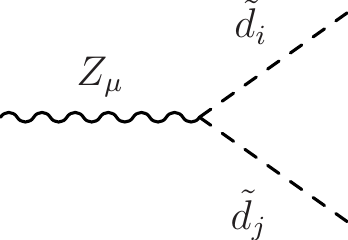}}
& $-\sum_{k=1}^{3} \frac{ie}{6\cw\sw}
\left( 2 R^{\tilde{d}\,*}_{i,3+k} R^{\tilde{d}}_{j,3+k}\sw^2-
R^{\tilde{d}\,*}_{i,k} R^{\tilde{d}}_{j,k} (3-2\sw^2)
\right)(p+p^{'})_{\mu}$\\
&\\
{\includegraphics{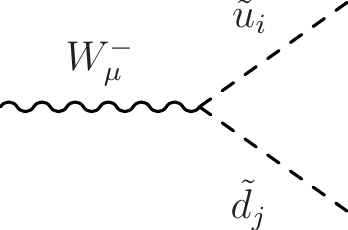}}
 &\put(10,30){$-\sum_{k,l=1}^{3} \frac{ie}{\sqrt{2}\sw}
 {V_{\rm{CKM}}^{*\,kl}}R^{\tilde{u}\,*}_{i,k} R^{\tilde{d}}_{j,l}
\,(p+p^{'})_{\mu}$}\\
&\\
\parbox[c]{1em}{\includegraphics{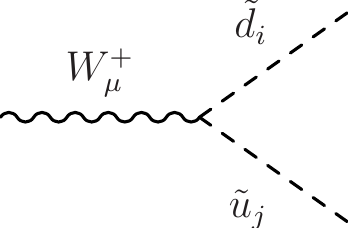}}
 & $-\sum_{k,l=1}^{3} \frac{ie}{\sqrt{2}\sw}
 {V_{\rm{CKM}}^{kl}}R^{\tilde{u}}_{j,k} R^{\tilde{d\,*}}_{i,l}\,(p+p^{'})_{\mu}$\\
&\\
\parbox[c]{1em}{\includegraphics{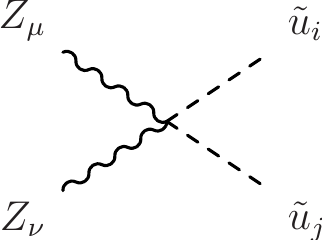}}
& $\sum_{k=1}^{3} \frac{i e^2}{18\cw^2\sw^2}
\left( R^{\tilde{u}\,*}_{i,k} R^{\tilde{u}}_{j,k} (3-4\sw^2)^2+
16 R^{\tilde{u}\,*}_{i,3+k} R^{\tilde{u}}_{j,3+k} \sw^4
\right)g_{\mu \nu}$\\
&\\
\parbox[c]{1em}{\includegraphics{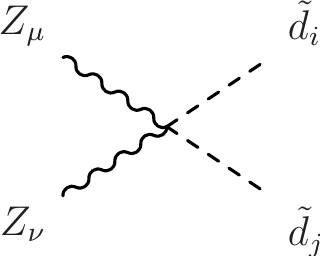}}
& {$\sum_{k=1}^{3} \frac{i e^2}{18\cw^2\sw^2}
\left( R^{\tilde{d}\,*}_{i,k} R^{\tilde{d}}_{j,k} (3-2\sw^2)^2+
4 R^{\tilde{d}\,*}_{i,3+k} R^{\tilde{d}}_{j,3+k} \sw^4
\right)g_{\mu \nu}$}\\
&\\
\parbox[c]{1em}{\includegraphics{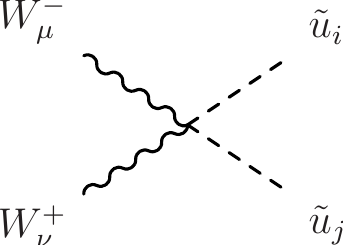}}
 & $\sum_{k=1}^{3} \frac{ie^2}{2\sw^2}
 R^{\tilde{u}\,*}_{i,k} R^{\tilde{u}}_{j,k}\,g_{\mu \nu}$\\
&\\
\end{tabular}
\end{table}

\begin{table}[H]
\begin{tabular}{ll}
{\includegraphics{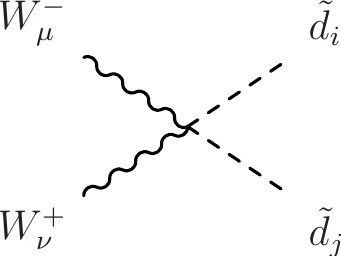}}
 &\put(10,30){$\sum_{k=1}^{3} \frac{ie^2}{2\sw^2}
 R^{\tilde{d\,*}}_{i,k} R^{\tilde{d}}_{j,k}\,g_{\mu \nu}$}\\
&\\
&\\
\end{tabular}
\end{table}

\clearpage
\newpage
\chapter{One-loop diagrams for the Higgs boson self-energies}

In this appendix we collect the one-loop Feynman diagrams and analytical results contributing to the Higgs boson self-energies, Higgs boson tadpoles and gauge boson self-energies within our NMFV framework, that enter into the computation of the radiative corrections to the MSSM Higgs boson masses presented in Chapter \ref{higgsmasssquark}.

The results in this chapter have been published in \cite{AranaCatania:2011ak}.

All the following Feynman diagrams have been calculated using FeynArts
3.5~\cite{feynarts} and FormCalc 6.0~\cite{formcalc}.
The notation used here is the same as in Appendix A. Furthermore we use the functions~\cite{a0b0c0}

\begin{equation}
\frac{i}{16\pi}A_{0}\left[m^{2}\right]\equiv\int\frac{\mu^{4-D}d^{D}k}{\left(2\pi\right)^{D}}\frac{1}{k^{2}-m^{2}}\end{equation}

\begin{equation}
\frac{i}{16\pi}B_{0}\left[p^{2},m_{1}^{2},m_{2}^{2}\right]\equiv\int\frac{\mu^{4-D}d^{D}k}{\left(2\pi\right)^{D}}\frac{1}{\left[k^{2}-m_{1}^{2}\right]\left[\left(k+p\right)^{2}-m_{2}^{2}\right]}\end{equation}

\begin{equation}
\frac{i}{16\pi}p^{2}B_{1}\left[p^{2},m_{1}^{2},m_{2}^{2}\right]\equiv\int\frac{\mu^{4-D}d^{D}k}{\left(2\pi\right)^{D}}\frac{pk}{\left[k^{2}-m_{1}^{2}\right]\left[\left(k+p\right)^{2}-m_{2}^{2}\right]}\end{equation}

The generic diagrams from the quark and squark sectors have been ordered according to its topologies, and the
particles involved in the internal loops (quarks $q$ or squarks $\tilde{q}$), and are collected in fig \ref{figfdall} of this appendix. The bare functions can then be expressed as a sum of several parts corresponding to the different contributions according to fig \ref{figfdall}:
\begin{equation}
\Sigma_{\phi\phi^{\prime}}=\Sigma_{\phi\phi^{\prime}}^{2q}+\Sigma_{\phi\phi^{\prime}}^{2\tilde{q}}+\Sigma_{\phi\phi^{\prime}}^{1\tilde{q}}\qquad\Sigma_{VV}=\Sigma_{VV}^{2q}+\Sigma_{VV}^{2\tilde{q}}+\Sigma_{VV}^{1\tilde{q}}\qquad T_{\phi}=T_{\phi}^{q}+T_{\phi}^{\tilde{q}}\label{selfpartssum}\end{equation}
where $\phi,\,\phi^{\prime}=h,\, H,\, A,\, H^{\pm}$ and
$V=W,\, Z$. All the self-energies $\Sigma$ correspond to 
$\Sigma\left(p^{2}\right)$. The self-energies for $H$ are 
obtained by the replacements of Eq.~\ref{eq:cambio} of Appendix A on the results of $h$. We summarize in the following the analytical results for the various parts in Eq. \ref{selfpartssum}:

\begin{figure}[H]
\centering
\begin{tabular}{ccc}
\includegraphics[scale=0.33]{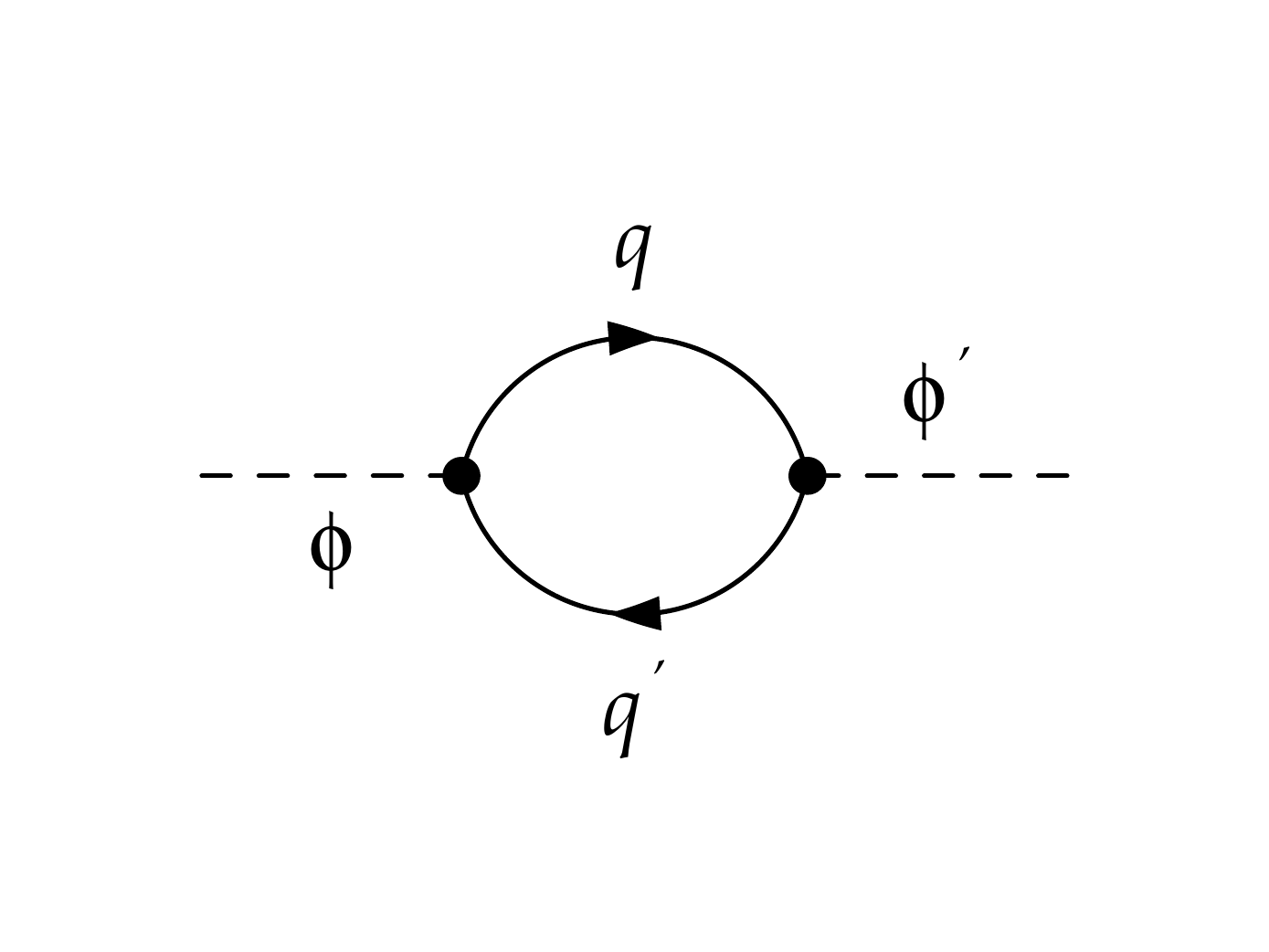}&
\includegraphics[scale=0.33]{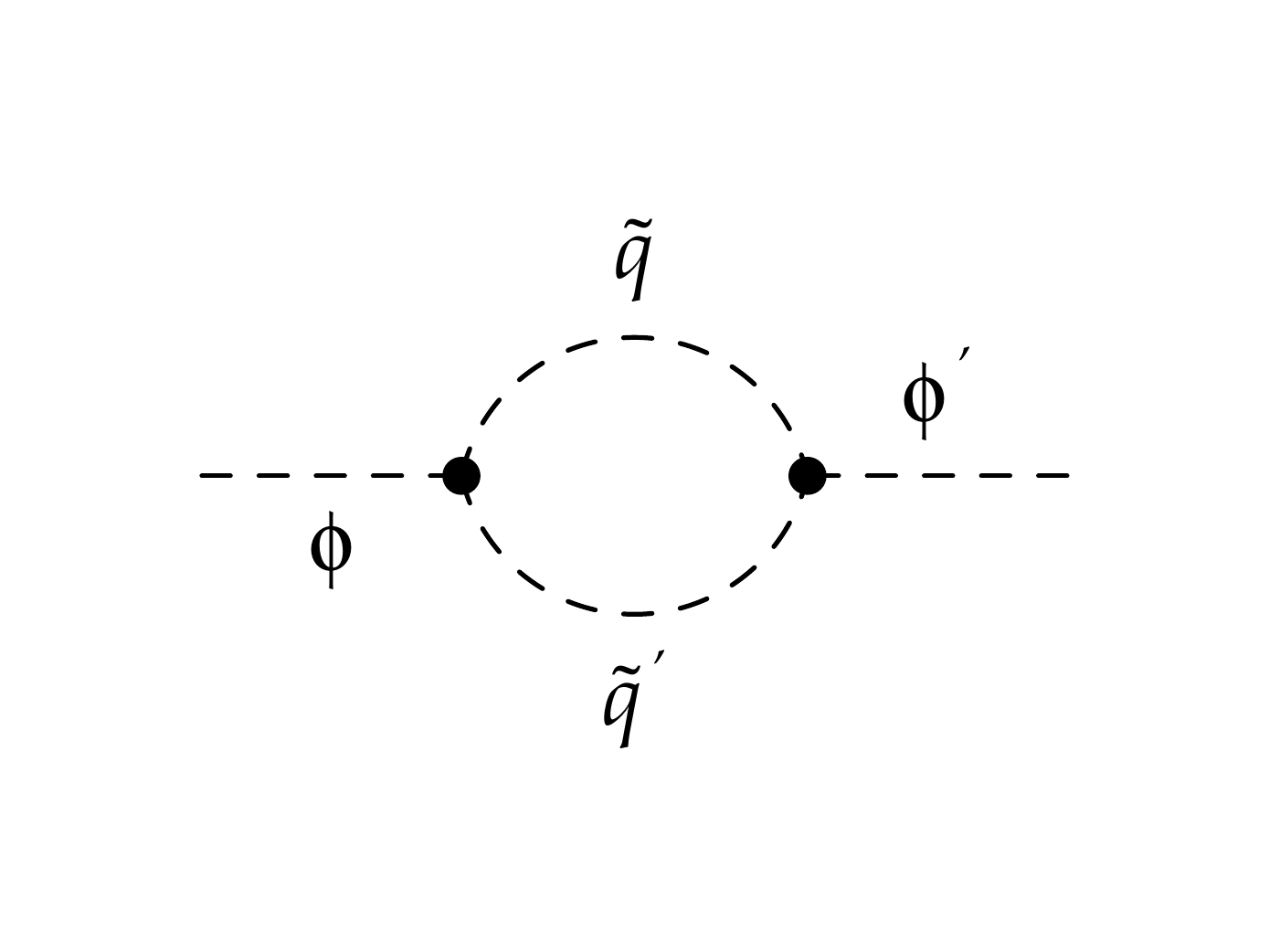}&
\includegraphics[scale=0.33]{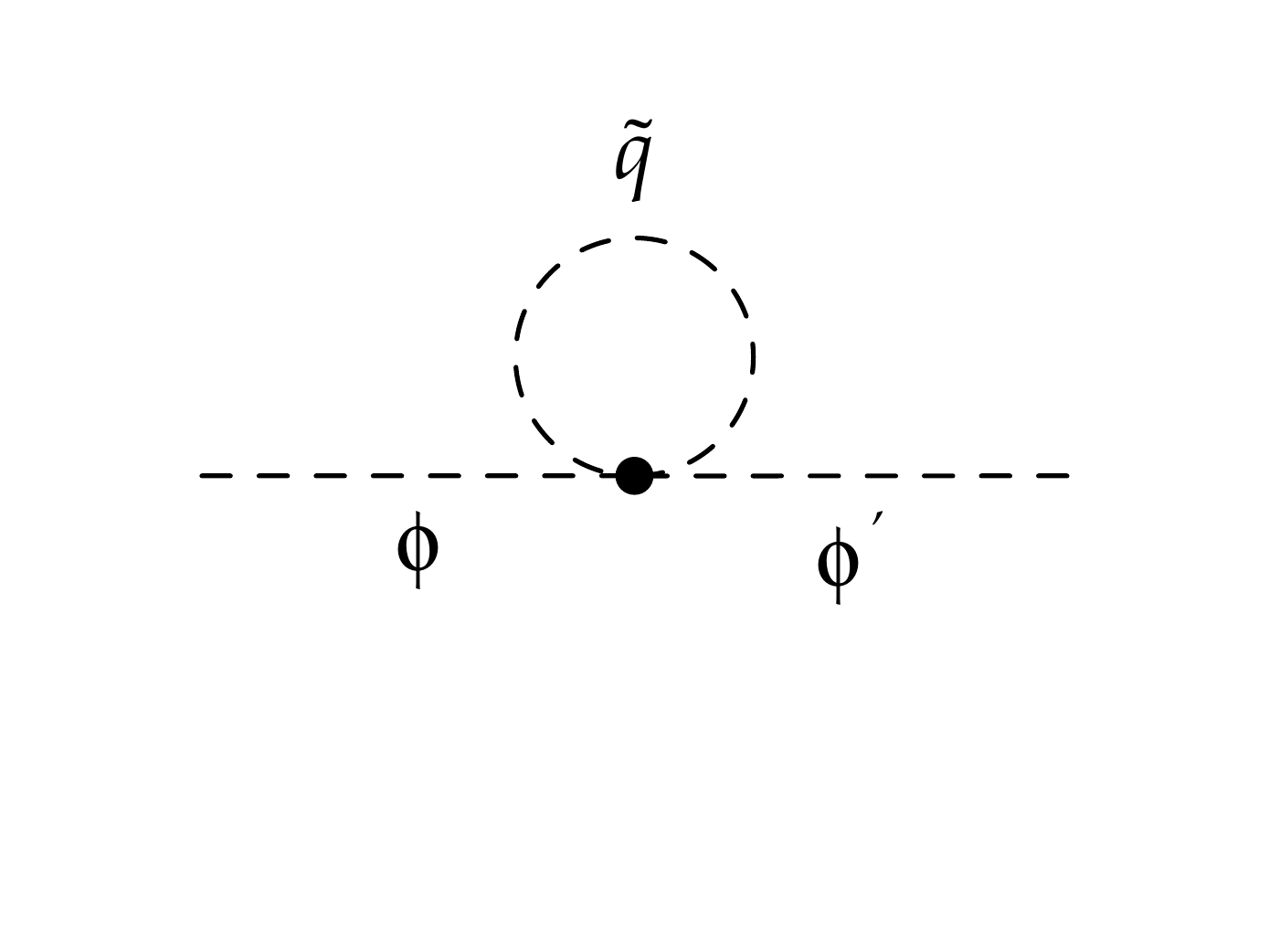}\\
\includegraphics[scale=0.33]{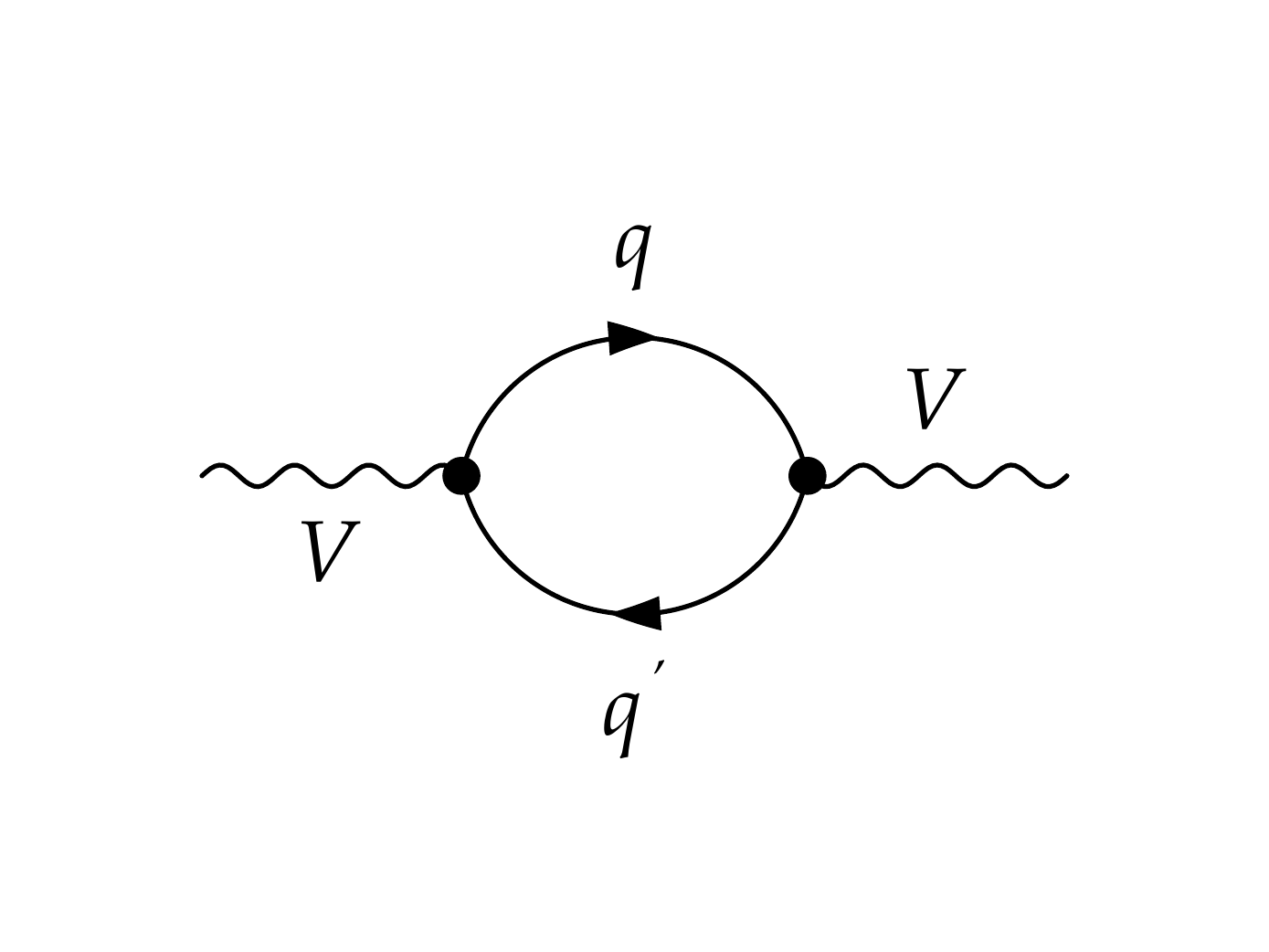}&
\includegraphics[scale=0.33]{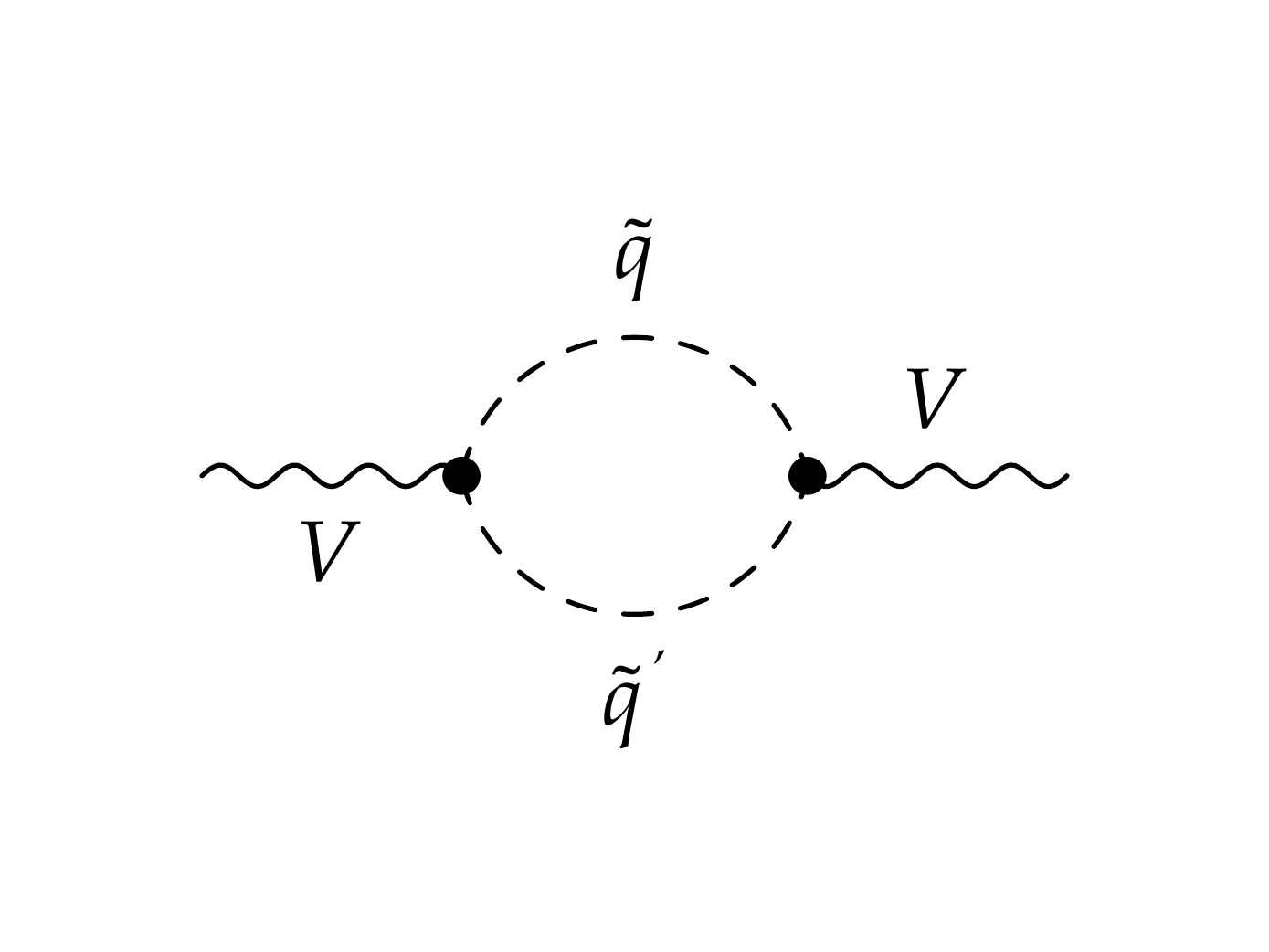}&
\includegraphics[scale=0.33]{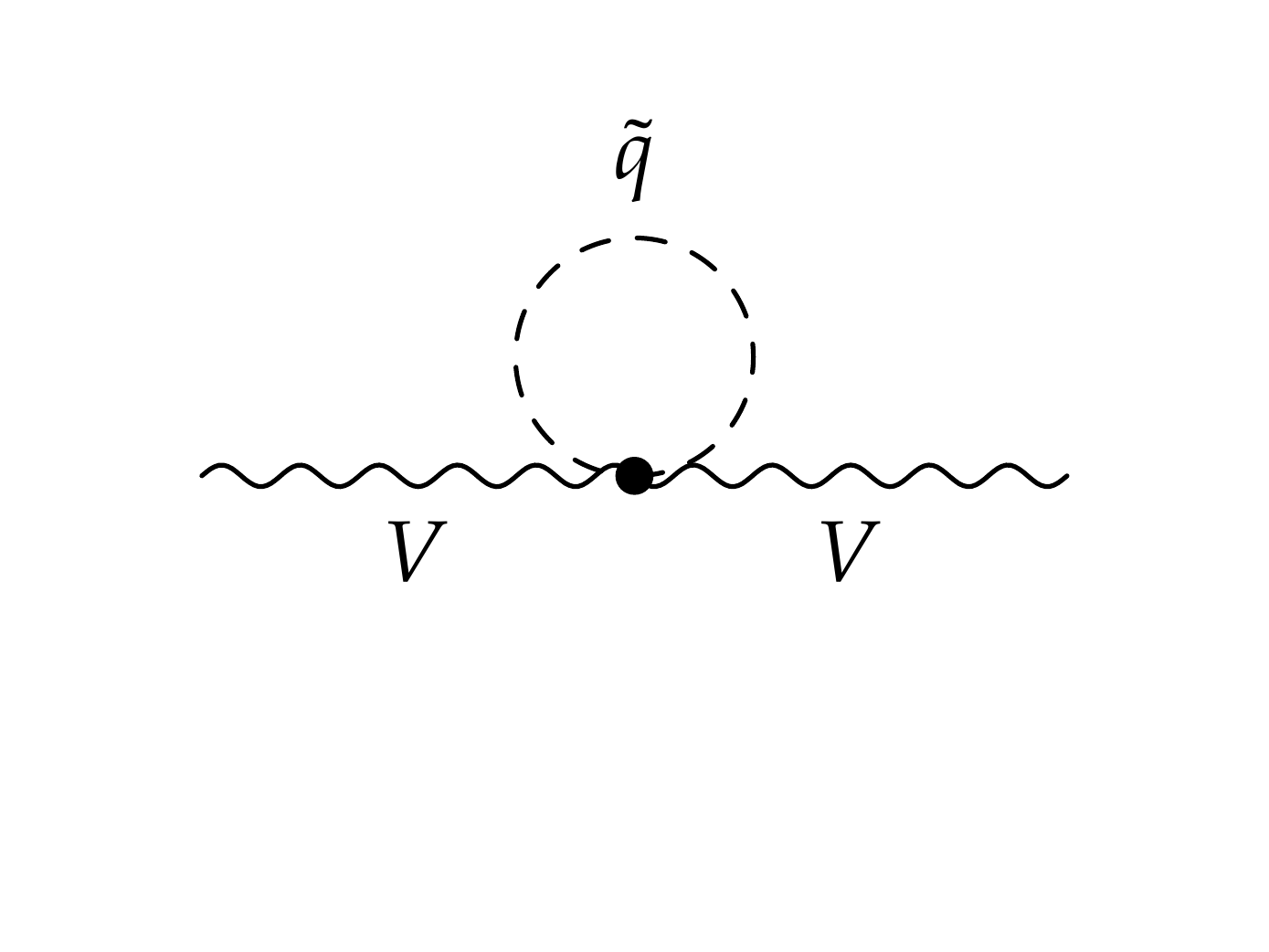}\\
\includegraphics[scale=0.33]{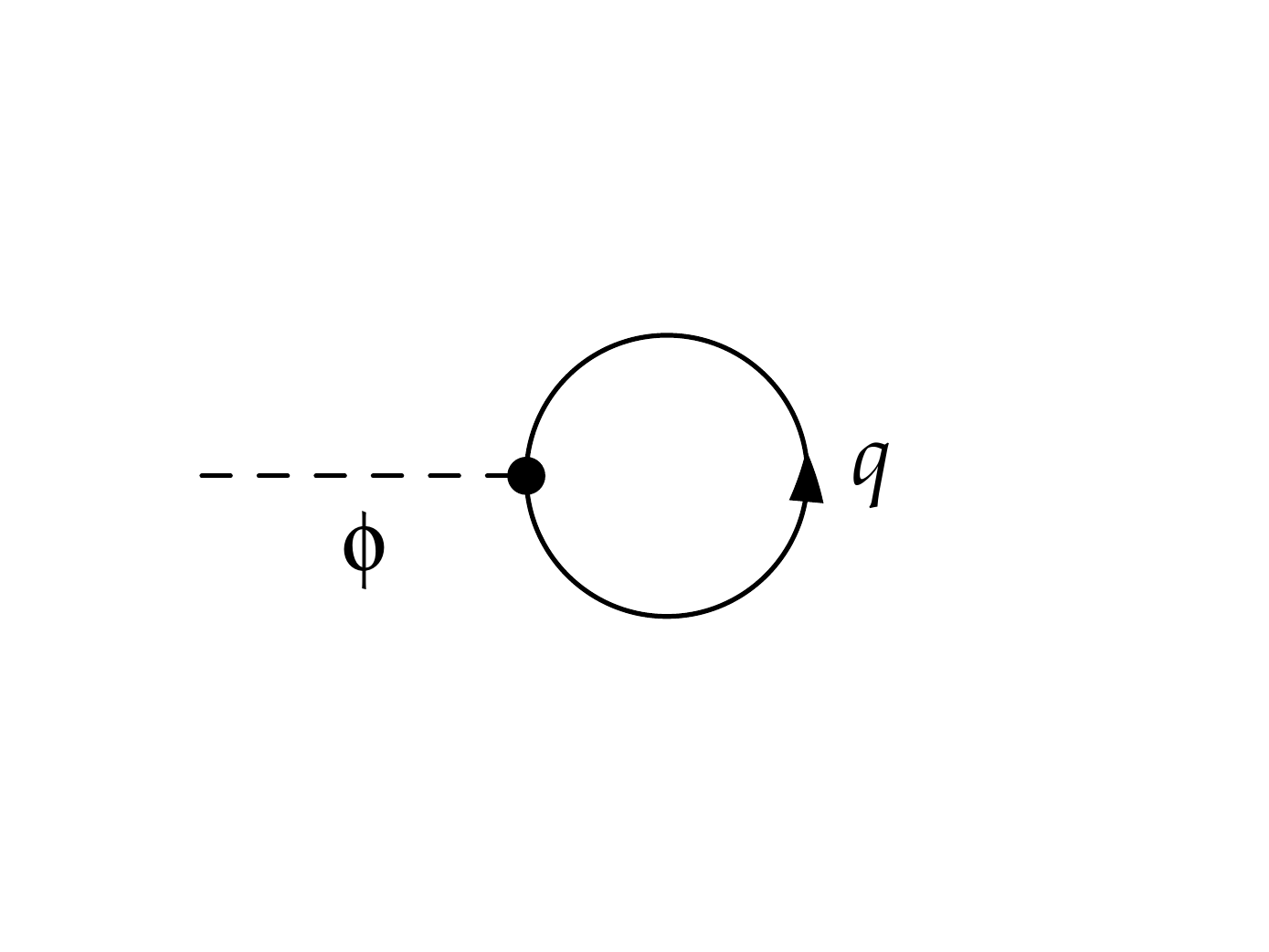}&
\includegraphics[scale=0.33]{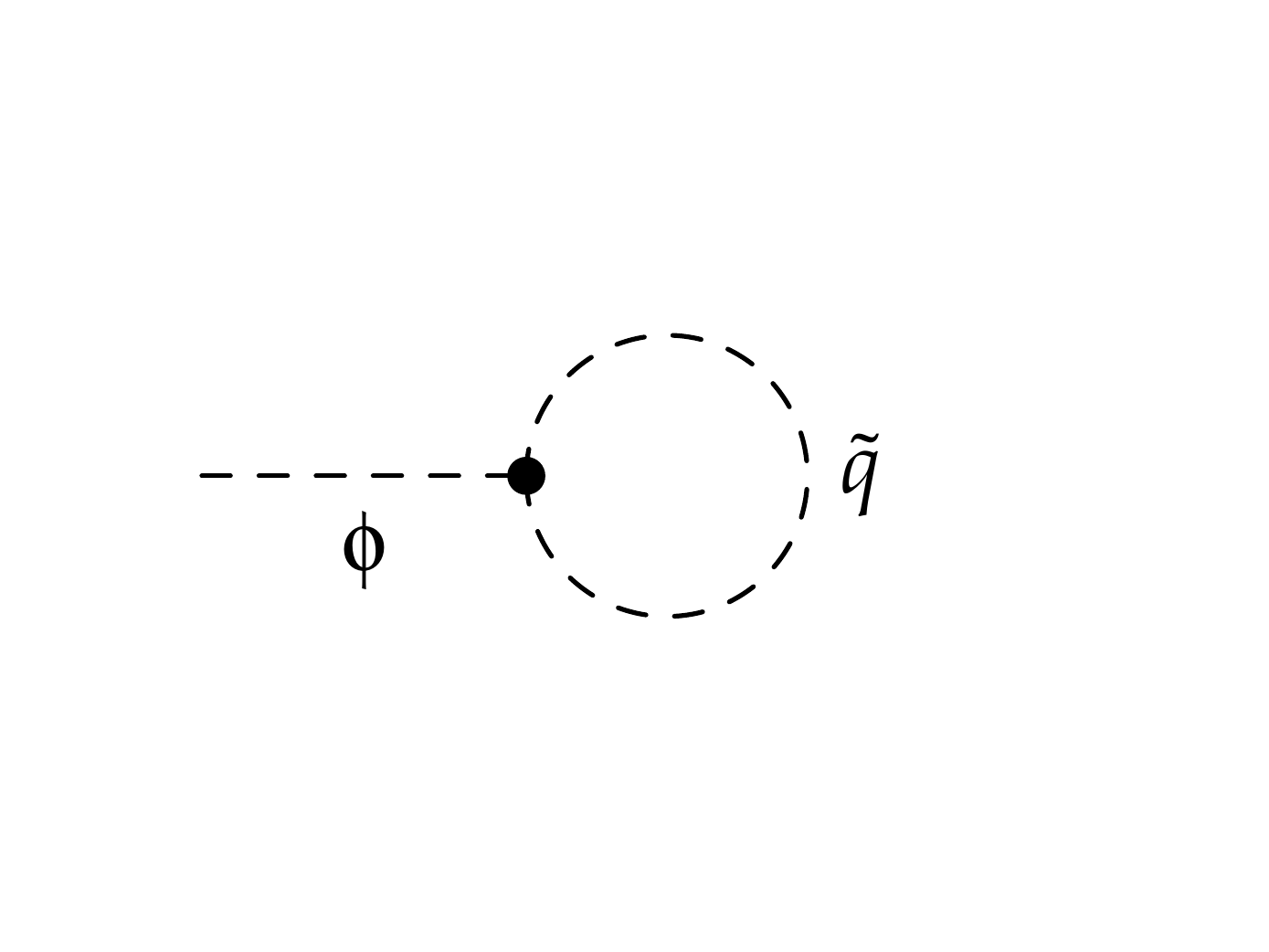}&
\end{tabular}
\caption{Different topologies for 
$\Sigma_{\phi\phi^{\prime}},\,\Sigma_{VV},\, T_{\phi}$}
\label{figfdall}
\end{figure}

\begin{itemize}
\item $h$
\end{itemize}
\begin{eqnarray}
\Sigma_{hh}^{2q} & = & -\sum_{i=1}^{3}\frac{3\alpha c_{\alpha}^{2}m_{u_{i}}^{2}}{4\pi M_{W}^{2}s_{\beta}^{2}s_{W}^{2}}\left\{ A_{0}\left[m_{u_{i}}^{2}\right]+p^{2}B_{1}\left[p^{2},m_{u_{i}}^{2},m_{u_{i}}^{2}\right]+2m_{u_{i}}^{2}B_{0}\left[p^{2},m_{u_{i}}^{2},m_{u_{i}}^{2}\right]\right\} \nonumber \\
 && -\sum_{i=1}^{3}\frac{3\alpha s_{\alpha}^{2}m_{d_{i}}^{2}}{4\pi M_{W}^{2}c_{\beta}^{2}s_{W}^{2}}\left\{ A_{0}\left[m_{d_{i}}^{2}\right]+p^{2}B_{1}\left[p^{2},m_{d_{i}}^{2},m_{d_{i}}^{2}\right]+2m_{d_{i}}^{2}B_{0}\left[p^{2},m_{d_{i}}^{2},m_{d_{i}}^{2}\right]\right\}\\
&&\nonumber \\
\Sigma_{hh}^{2\tilde{q}}&= & \sum_{m,n}^{6}\sum_{i,j,k,l}^{3}\frac{1}{48c_{W}^{2}M_{W}^{2}\pi s_{\beta}^{2}s_{W}^{2}}\alpha B_{0}\left[p^{2},m_{\tilde{u}_{m}}^{2},m_{\tilde{u}_{n}}^{2}\vphantom{m_{\tilde{d}_{l}}^{2}}\right]\nonumber \\
 && \times\left\{ \vphantom{\vphantom{\sum_{k=1}^{3}}}\delta_{i,j}\left(M_{W}m_{Z}s_{\alpha+\beta}s_{\beta}\left(-3+4s_{W}^{2}\right)+6c_{\alpha}c_{W}m_{u_{i}}^{2}\right)R_{n,j}^{\tilde{u}}R_{m,i}^{\tilde{u}*}\right.\nonumber \\
 && +3c_{W}\left(c_{\alpha}A_{i,j}^{u}+\mu^{*}s_{\alpha}\delta_{i,j}\right)m_{u_{i}}R_{n,3+j}^{\tilde{u}}R_{m,i}^{\tilde{u}*}\nonumber \\
&&+\left(3c_{\alpha}c_{W}A_{j,i}^{u*}m_{u_{j}}R_{n,j}^{\tilde{u}}+3c_{W}\mu s_{\alpha}\delta_{i,j}m_{u_{i}}R_{n,j}^{\tilde{u}}\right)R_{m,3+i}^{\tilde{u}*}\nonumber \\
 && \left.+2\delta_{i,j}\left(-2M_{W}m_{Z}s_{\alpha+\beta}s_{\beta}s_{W}^{2}+3c_{\alpha}c_{W}m_{u_{i}}^{2}\right)R_{n,3+j}^{\tilde{u}}R_{m,3+i}^{\tilde{u}*}\vphantom{\sum_{k=1}^{3}}\right\} \nonumber \\
 && \times\left\{ \vphantom{\vphantom{\sum_{k=1}^{3}}}\delta_{k,l}\left(M_{W}m_{Z}s_{\alpha+\beta}s_{\beta}\left(-3+4s_{W}^{2}\right)+6c_{\alpha}c_{W}m_{u_{k}}^{2}\right)R_{m,l}^{\tilde{u}}R_{n,k}^{\tilde{u}*}\right.\nonumber \\
 && +3c_{W}\left(c_{\alpha}A_{k,l}^{u}+\mu^{*}s_{\alpha}\delta_{k,l}\right)m_{u_{k}}R_{m,3+l}^{\tilde{u}}R_{n,k}^{\tilde{u}*}\nonumber \\
&&+\left(3c_{\alpha}c_{W}A_{l,k}^{u*}m_{u_{l}}R_{m,l}^{\tilde{u}}+3c_{W}\mu s_{\alpha}\delta_{k,l}m_{u_{k}}R_{m,l}^{\tilde{u}}\right)R_{n,3+k}^{\tilde{u}*}\nonumber \\
 && \left.+2\delta_{k,l}\left(-2M_{W}m_{Z}s_{\alpha+\beta}s_{\beta}s_{W}^{2}+3c_{\alpha}c_{W}m_{u_{k}}^{2}\right)R_{m,3+l}^{\tilde{u}}R_{n,3+k}^{\tilde{u}*}\vphantom{\vphantom{\sum_{k=1}^{3}}}\right\} \nonumber \\
 && +\sum_{m,n}^{6}\sum_{i,j,k,l}^{3}\frac{1}{48c_{W}^{2}M_{W}^{2}\pi c_{\beta}^{2}s_{W}^{2}}\alpha B_{0}\left[p^{2},m_{\tilde{d}_{m}}^{2},m_{\tilde{d}_{n}}^{2}\right]\nonumber \\
 && \times\left\{ \vphantom{\vphantom{\sum_{k=1}^{3}}}\delta_{i,j}\left(M_{W}m_{Z}s_{\alpha+\beta}c_{\beta}\left(-3+2s_{W}^{2}\right)+6s_{\alpha}c_{W}m_{d_{i}}^{2}\right)R_{n,j}^{\tilde{d}}R_{m,i}^{\tilde{d}*}\right.\nonumber \\
 && +3c_{W}\left(s_{\alpha}A_{i,j}^{d}+\mu^{*}c_{\alpha}\delta_{i,j}\right)m_{d_{i}}R_{n,3+j}^{\tilde{d}}R_{m,i}^{\tilde{d}*}\nonumber \\
&&+\left(3s_{\alpha}c_{W}A_{j,i}^{d*}m_{d_{j}}R_{n,j}^{\tilde{d}}+3c_{W}\mu c_{\alpha}\delta_{i,j}m_{d_{i}}R_{n,j}^{\tilde{d}}\right)R_{m,3+i}^{\tilde{d}*}\nonumber \\
 && \left.-2\delta_{i,j}\left(M_{W}m_{Z}s_{\alpha+\beta}c_{\beta}s_{W}^{2}-3s_{\alpha}c_{W}m_{d_{i}}^{2}\right)R_{n,3+j}^{\tilde{d}}R_{m,3+i}^{\tilde{d}*}\vphantom{\sum_{k=1}^{3}}\right\} \nonumber \\
 && \times\left\{ \vphantom{\sum_{k=1}^{3}}\left(\delta_{k,l}\left(M_{W}m_{Z}s_{\alpha+\beta}c_{\beta}\left(-3+2s_{W}^{2}\right)+6s_{\alpha}c_{W}m_{d_{k}}^{2}\right)R_{m,l}^{\tilde{d}}R_{n,k}^{\tilde{d}*}\right)\right.\nonumber \\
 && +3c_{W}\left(s_{\alpha}A_{k,l}^{d}+\mu^{*}c_{\alpha}\delta_{k,l}\right)m_{d_{k}}R_{m,3+l}^{\tilde{d}}R_{n,k}^{\tilde{d}*}\nonumber \\
&&+\left(3s_{\alpha}c_{W}A_{l,k}^{d*}m_{d_{l}}R_{m,l}^{\tilde{d}}+3c_{W}\mu c_{\alpha}\delta_{k,l}m_{d_{k}}R_{m,l}^{\tilde{d}}\right)R_{n,3+k}^{\tilde{d}*}\nonumber \\
 && \left.-2\delta_{k,l}\left(M_{W}m_{Z}s_{\alpha+\beta}c_{\beta}s_{W}^{2}-3s_{\alpha}c_{W}m_{d_{k}}^{2}\right)R_{m,3+l}^{\tilde{d}}R_{n,3+k}^{\tilde{d}*}\vphantom{\sum_{k=1}^{3}}\right\}\\
&&\nonumber \\
\Sigma_{hh}^{1\tilde{q}}&= & \sum_{l=1}^{6}\sum_{i=1}^{3}\frac{1}{16c_{W}^{2}M_{W}^{2}\pi s_{\beta}^{2}s_{W}^{2}}\alpha A_{0}\left[m_{\tilde{u}_{l}}^{2}\vphantom{m_{\tilde{d}_{l}}^{2}}\right]\left\{ \vphantom{R_{l,i}^{\tilde{d}}}R_{l,i}^{\tilde{u}}R_{l,i}^{\tilde{u}*}\left(c_{2\alpha}m_{w}^{2}s_{\beta}^{2}\left(-3+4s_{W}^{2}\right)+6c_{\alpha}^{2}c_{W}^{2}m_{u_{i}}^{2}\right)\right.\nonumber \\
 && \left.+2R_{l,3+i}^{\tilde{u}}R_{l,3+i}^{\tilde{u}*}\left(-2c_{2\alpha}M_{W}^{2}s_{\beta}^{2}s_{W}^{2}+3c_{\alpha}^{2}c_{W}^{2}m_{u_{i}}^{2}\right)\vphantom{R_{l,i}^{\tilde{d}}}\right\} \nonumber \\
 && -\sum_{l=1}^{6}\sum_{i=1}^{3}\frac{1}{16c_{W}^{2}M_{W}^{2}\pi c_{\beta}^{2}s_{W}^{2}}\alpha A_{0}\left[m_{\tilde{d}_{l}}^{2}\right]\left\{ R_{l,i}^{\tilde{d}}R_{l,i}^{\tilde{d}*}\left(c_{2\alpha}m_{w}^{2}c_{\beta}^{2}\left(-3+2s_{W}^{2}\right)-6s_{\alpha}^{2}c_{W}^{2}m_{d_{i}}^{2}\right)\right.\nonumber \\
 && \left.-2R_{l,3+i}^{\tilde{d}}R_{l,3+i}^{\tilde{d}*}\left(c_{2\alpha}M_{W}^{2}c_{\beta}^{2}s_{W}^{2}+3s_{\alpha}^{2}c_{W}^{2}m_{d_{i}}^{2}\right)\vphantom{R_{l,i}^{\tilde{d}}}\right\} 
\end{eqnarray}

\begin{itemize}
\item $hH$
\end{itemize}
\begin{eqnarray}
\Sigma_{hH}^{2q} & = & -\sum_{i=1}^{3}\frac{3\alpha c_{\alpha}s_{\alpha}m_{u_{i}}^{2}}{4\pi M_{W}^{2}s_{\beta}^{2}s_{W}^{2}}\left\{ A_{0}\left[m_{u_{i}}^{2}\right]+p^{2}B_{1}\left[p^{2},m_{u_{i}}^{2},m_{u_{i}}^{2}\right]+2m_{u_{i}}^{2}B_{0}\left[p^{2},m_{u_{i}}^{2},m_{u_{i}}^{2}\right]\right\} \nonumber \\
 &  & +\sum_{i=1}^{3}\frac{3\alpha c_{\alpha}s_{\alpha}m_{d_{i}}^{2}}{4\pi M_{W}^{2}c_{\beta}^{2}s_{W}^{2}}\left\{ A_{0}\left[m_{d_{i}}^{2}\right]+p^{2}B_{1}\left[p^{2},m_{d_{i}}^{2},m_{d_{i}}^{2}\right]+2m_{d_{i}}^{2}B_{0}\left[p^{2},m_{d_{i}}^{2},m_{d_{i}}^{2}\right]\right\}\,\,\\
\Sigma_{hH}^{2\tilde{q}}&= & \sum_{m,n}^{6}\sum_{i,j,k,l}^{3}\frac{1}{48c_{W}^{2}M_{W}^{2}\pi s_{\beta}^{2}s_{W}^{2}}\alpha B_{0}\left[p^{2},m_{\tilde{u}_{m}}^{2},m_{\tilde{u}_{n}}^{2}\vphantom{m_{\tilde{d}_{l}}^{2}}\right]\nonumber \\
 && \times\left\{ \vphantom{\vphantom{\sum_{k=1}^{3}}}\delta_{i,j}\left(M_{W}m_{Z}s_{\alpha+\beta}s_{\beta}\left(-3+4s_{W}^{2}\right)+6c_{\alpha}c_{W}m_{u_{i}}^{2}\right)R_{n,j}^{\tilde{u}}R_{m,i}^{\tilde{u}*}\right.\nonumber \\
 && +3c_{W}\left(c_{\alpha}A_{i,j}^{u}+\mu^{*}s_{\alpha}\delta_{i,j}\right)m_{u_{i}}R_{n,3+j}^{\tilde{u}}R_{m,i}^{\tilde{u}*}\nonumber \\
&&+\left(3c_{\alpha}c_{W}A_{j,i}^{u*}m_{u_{j}}R_{n,j}^{\tilde{u}}+3c_{W}\mu s_{\alpha}\delta_{i,j}m_{u_{i}}R_{n,j}^{\tilde{u}}\right)R_{m,3+i}^{\tilde{u}*}\nonumber \\
 && \left.+2\delta_{i,j}\left(-2M_{W}m_{Z}s_{\alpha+\beta}s_{\beta}s_{W}^{2}+3c_{\alpha}c_{W}m_{u_{i}}^{2}\right)R_{n,3+j}^{\tilde{u}}R_{m,3+i}^{\tilde{u}*}\vphantom{\vphantom{\sum_{k=1}^{3}}}\right\} \nonumber \\
 && \times\left\{ \vphantom{\vphantom{\sum_{k=1}^{3}}}\delta_{k,l}\left(M_{W}m_{Z}c_{\alpha+\beta}s_{\beta}\left(3-4s_{W}^{2}\right)+6s_{\alpha}c_{W}m_{u_{k}}^{2}\right)R_{m,l}^{\tilde{u}}R_{n,k}^{\tilde{u}*}\right.\nonumber \\
 && +3c_{W}\left(s_{\alpha}A_{k,l}^{u}-\mu^{*}c_{\alpha}\delta_{k,l}\right)m_{u_{k}}R_{m,3+l}^{\tilde{u}}R_{n,k}^{\tilde{u}*}\nonumber \\
&&+\left(3s_{\alpha}c_{W}A_{l,k}^{u*}m_{u_{l}}R_{m,l}^{\tilde{u}}-3c_{W}\mu c_{\alpha}\delta_{k,l}m_{u_{k}}R_{m,l}^{\tilde{u}}\right)R_{n,3+k}^{\tilde{u}*}\nonumber \\
 & &\left.+2\delta_{k,l}\left(2M_{W}m_{Z}c_{\alpha+\beta}s_{\beta}s_{W}^{2}+3s_{\alpha}c_{W}m_{u_{k}}^{2}\right)R_{m,3+l}^{\tilde{u}}R_{n,3+k}^{\tilde{u}*}\vphantom{\vphantom{\sum_{k=1}^{3}}}\right\} \nonumber \\
 && -\sum_{m,n}^{6}\sum_{i,j,k,l}^{3}\frac{1}{48c_{W}^{2}M_{W}^{2}\pi c_{\beta}^{2}s_{W}^{2}}\alpha B_{0}\left[p^{2},m_{\tilde{d}_{m}}^{2},m_{\tilde{d}_{n}}^{2}\right]\nonumber \\
 && \times\left\{ \vphantom{\vphantom{\sum_{k=1}^{3}}}\delta_{i,j}\left(M_{W}m_{Z}s_{\alpha+\beta}c_{\beta}\left(-3+2s_{W}^{2}\right)+6s_{\alpha}c_{W}m_{d_{i}}^{2}\right)R_{n,j}^{\tilde{d}}R_{m,i}^{\tilde{d}*}\right.\nonumber \\
 && +3c_{W}\left(s_{\alpha}A_{i,j}^{d}+\mu^{*}c_{\alpha}\delta_{i,j}\right)m_{d_{i}}R_{n,3+j}^{\tilde{d}}R_{m,i}^{\tilde{d}*}\nonumber \\
&&+\left(3s_{\alpha}c_{W}A_{j,i}^{d*}m_{d_{j}}R_{n,j}^{\tilde{d}}+3c_{W}\mu c_{\alpha}\delta_{i,j}m_{d_{i}}R_{n,j}^{\tilde{d}}\right)R_{m,3+i}^{\tilde{d}*}\nonumber \\
 && \left.-2\delta_{i,j}\left(M_{W}m_{Z}s_{\alpha+\beta}c_{\beta}s_{W}^{2}-3s_{\alpha}c_{W}m_{d_{i}}^{2}\right)R_{n,3+j}^{\tilde{d}}R_{m,3+i}^{\tilde{d}*}\vphantom{\vphantom{\sum_{k=1}^{3}}}\right\} \nonumber \\
 && \times\left\{ \vphantom{\vphantom{\sum_{k=1}^{3}}}\delta_{k,l}\left(M_{W}m_{Z}c_{\alpha+\beta}c_{\beta}\left(-3+2s_{W}^{2}\right)+6c_{\alpha}c_{W}m_{d_{k}}^{2}\right)R_{m,l}^{\tilde{d}}R_{n,k}^{\tilde{d}*}\right.\nonumber \\
 && +3c_{W}\left(c_{\alpha}A_{k,l}^{d}-\mu^{*}s_{\alpha}\delta_{k,l}\right)m_{d_{k}}R_{m,3+l}^{\tilde{d}}R_{n,k}^{\tilde{d}*}\nonumber \\
&&+\left(3c_{\alpha}c_{W}A_{l,k}^{d*}m_{d_{l}}R_{m,l}^{\tilde{d}}-3c_{W}\mu s_{\alpha}\delta_{k,l}m_{d_{k}}R_{m,l}^{\tilde{d}}\right)R_{n,3+k}^{\tilde{d}*}\nonumber \\
 && \left.-2\delta_{k,l}\left(2M_{W}m_{Z}c_{\alpha+\beta}c_{\beta}s_{W}^{2}-3c_{\alpha}c_{W}m_{d_{k}}^{2}\right)R_{m,3+l}^{\tilde{d}}R_{n,3+k}^{\tilde{d}*}\vphantom{\vphantom{\sum_{k=1}^{3}}}\right\}\\
\Sigma_{hH}^{1\tilde{q}}&= & \sum_{l=1}^{6}\sum_{i=1}^{3}\frac{1}{16c_{W}^{2}M_{W}^{2}\pi s_{\beta}^{2}s_{W}^{2}}\alpha A_{0}\left[m_{\tilde{u}_{l}}^{2}\vphantom{m_{\tilde{d}_{l}}^{2}}\right]\left\{ \vphantom{R_{l,i}^{\tilde{d}}}R_{l,i}^{\tilde{u}}R_{l,i}^{\tilde{u}*}\left(s_{2\alpha}m_{w}^{2}s_{\beta}^{2}\left(-3+4s_{W}^{2}\right)+3s_{2\alpha}c_{W}^{2}m_{u_{i}}^{2}\right)\right.\nonumber \\
 && \left.+R_{l,3+i}^{\tilde{u}}R_{l,3+i}^{\tilde{u}*}\left(-4s_{2\alpha}M_{W}^{2}s_{\beta}^{2}s_{W}^{2}+3s_{2\alpha}c_{W}^{2}m_{u_{i}}^{2}\right)\vphantom{R_{l,i}^{\tilde{d}}}\right\} \nonumber \\
 && -\sum_{l=1}^{6}\sum_{i=1}^{3}\frac{1}{16c_{W}^{2}M_{W}^{2}\pi c_{\beta}^{2}s_{W}^{2}}\alpha A_{0}\left[m_{\tilde{d}_{l}}^{2}\right]\left\{ R_{l,i}^{\tilde{d}}R_{l,i}^{\tilde{d}*}\left(s_{2\alpha}m_{w}^{2}c_{\beta}^{2}\left(-3+4s_{W}^{2}\right)+3s_{2\alpha}c_{W}^{2}m_{d_{i}}^{2}\right)\right.\nonumber \\
 && \left.+R_{l,3+i}^{\tilde{d}}R_{l,3+i}^{\tilde{d}*}\left(-2s_{2\alpha}M_{W}^{2}c_{\beta}^{2}s_{W}^{2}+3s_{2\alpha}c_{W}^{2}m_{d_{i}}^{2}\right)\vphantom{R_{l,i}^{\tilde{d}}}\right\}
\end{eqnarray}

\newpage
\begin{itemize}
\item $A$
\end{itemize}
\begin{eqnarray}
\Sigma_{AA}^{2q} & = & -\sum_{i=1}^{3}\frac{3\alpha m_{u_{i}}^{2}}{4\pi M_{W}^{2}t_{\beta}^{2}s_{W}^{2}}\left\{ A_{0}\left[m_{u_{i}}^{2}\right]+p^{2}B_{1}\left[p^{2},m_{u_{i}}^{2},m_{u_{i}}^{2}\right]\right\} \nonumber \\
 &  & -\sum_{i=1}^{3}\frac{3\alpha t_{\beta}^{2}m_{d_{i}}^{2}}{4\pi M_{W}^{2}s_{W}^{2}}\left\{ A_{0}\left[m_{d_{i}}^{2}\right]+p^{2}B_{1}\left[p^{2},m_{d_{i}}^{2},m_{d_{i}}^{2}\right]\right\} \\
&&\nonumber \\
\Sigma_{AA}^{2\tilde{q}}&= & -\sum_{m,n}^{6}\sum_{i,j,k,l}^{3}\frac{3}{16M_{W}^{2}\pi t_{\beta}^{2}s_{W}^{2}}\alpha B_{0}\left[p^{2},m_{\tilde{u}_{m}}^{2},m_{\tilde{u}_{n}}^{2}\vphantom{m_{\tilde{d}_{l}}^{2}}\right]\nonumber \\
 && \times\left\{ \left(-A_{i,j}^{u}-\mu^{*}t_{\beta}\delta_{i,j}\right)m_{u_{i}}R_{n,3+j}^{\tilde{u}}R_{m,i}^{\tilde{u}*}+\left(A_{j,i}^{u*}m_{u_{j}}+\mu t_{\beta}\delta_{i,j}m_{u_{i}}\right)R_{n,j}^{\tilde{u}}R_{m,3+i}^{\tilde{u}*}\vphantom{R_{m,i}^{\tilde{d}*}}\right\} \nonumber \\
 & &\times\left\{ \left(-A_{k,l}^{u}-\mu^{*}t_{\beta}\delta_{k,l}\right)m_{u_{k}}R_{m,3+l}^{\tilde{u}}R_{n,k}^{\tilde{u}*}+\left(A_{l,k}^{u*}m_{u_{l}}+\mu t_{\beta}\delta_{k,l}m_{u_{k}}\right)R_{m,l}^{\tilde{u}}R_{n,3+k}^{\tilde{u}*}\vphantom{R_{m,i}^{\tilde{d}*}}\right\} \nonumber \\
 && -\sum_{m,n}^{6}\sum_{i,j,k,l}^{3}\frac{3}{16M_{W}^{2}\pi s_{W}^{2}}\alpha B_{0}\left[p^{2},m_{\tilde{d}_{m}}^{2},m_{\tilde{d}_{n}}^{2}\right]\nonumber \\
 && \times\left\{ \left(-t_{\beta}A_{i,j}^{d}-\mu^{*}\delta_{i,j}\right)m_{d_{i}}R_{n,3+j}^{\tilde{d}}R_{m,i}^{\tilde{d}*}+\left(t_{\beta}A_{j,i}^{d*}m_{d_{j}}+\mu\delta_{i,j}m_{d_{i}}\right)R_{n,j}^{\tilde{d}}R_{m,3+i}^{\tilde{d}*}\right\} \nonumber \\
 && \times\left\{ \left(-t_{\beta}A_{k,l}^{d}-\mu^{*}\delta_{k,l}\right)m_{d_{k}}R_{m,3+l}^{\tilde{d}}R_{n,k}^{\tilde{d}*}+\left(t_{\beta}A_{l,k}^{d*}m_{d_{l}}+\mu\delta_{k,l}m_{d_{k}}\right)R_{m,l}^{\tilde{d}}R_{n,3+k}^{\tilde{d}*}\right\}\\
&&\nonumber \\
\Sigma_{AA}^{1\tilde{q}}&= & \sum_{l=1}^{6}\sum_{i=1}^{3}\frac{1}{16c_{W}^{2}M_{W}^{2}\pi t_{\beta}^{2}s_{W}^{2}}\alpha A_{0}\left[m_{\tilde{u}_{l}}\vphantom{m_{\tilde{d}_{l}}^{2}}\right]\left\{ \vphantom{R_{l,i}^{\tilde{d}}}R_{l,i}^{\tilde{u}}R_{l,i}^{\tilde{u}*}\left(c_{2\beta}m_{w}^{2}t_{\beta}^{2}\left(-3+4s_{W}^{2}\right)+6c_{W}^{2}m_{u_{i}}^{2}\right)\right.\nonumber \\
 && \left.+2R_{l,3+i}^{\tilde{u}}R_{l,3+i}^{\tilde{u}*}\left(-2c_{2\beta}M_{W}^{2}t_{\beta}^{2}s_{W}^{2}+3c_{W}^{2}m_{u_{i}}^{2}\right)\vphantom{R_{l,i}^{\tilde{d}}}\right\} \nonumber \\
 && +\sum_{l=1}^{6}\sum_{i=1}^{3}\frac{1}{16c_{W}^{2}M_{W}^{2}\pi s_{W}^{2}}\alpha A_{0}\left[m_{\tilde{d}_{l}}\vphantom{m_{\tilde{d}_{l}}^{2}}\right]\left\{ R_{l,i}^{\tilde{d}}R_{l,i}^{\tilde{d}*}\left(c_{2\beta}m_{w}^{2}\left(3-2s_{W}^{2}\right)+6c_{W}^{2}t_{\beta}^{2}m_{d_{i}}^{2}\right)\right.\nonumber \\
 && \left.+2R_{l,3+i}^{\tilde{d}}R_{l,3+i}^{\tilde{d}*}\left(c_{2\beta}M_{W}^{2}s_{W}^{2}+3c_{W}^{2}t_{\beta}^{2}m_{d_{i}}^{2}\right)\vphantom{R_{l,i}^{\tilde{d}}}\right\} \end{eqnarray}

\begin{itemize}
\item $H^{\pm}$
\end{itemize}
\begin{align}
\Sigma_{H^{-}H^{+}}^{2q}=- & \sum_{i=1}^{3}\sum_{j=1}^{3}\frac{3\alpha}{4\pi M_{W}^{2}s_{W}^{2}}\left\{ \vphantom{\vphantom{\sum_{k=1}^{3}}}m_{u_{i}}^{2}\left(2m_{d_{j}}^{2}+m_{u_{i}}^{2}/t_{\beta}^{2}+m_{d_{j}}^{2}t_{\beta}^{2}\right)\VCKM^{i,j}\VCKM^{i,j*}B_{0}\left[p^{2},m_{d_{j}}^{2},m_{u_{i}}^{2}\right]\right.\nonumber \\
 & +\left(m_{u_{i}}^{2}/t_{\beta}^{2}+m_{d_{j}}^{2}t_{\beta}^{2}\right)\VCKM^{i,j}\VCKM^{i,j*}p^{2}B_{1}\left[p^{2},m_{u_{i}}^{2},m_{d_{j}}^{2}\right]\nonumber \\
 & \left.+A_{0}\left[m_{d_{j}}^{2}\right]\left(m_{u_{i}}^{2}\VCKM^{i,j}\VCKM^{i,j*}/t_{\beta}^{2}+m_{d_{j}}^{2}\VCKM^{i,j}\VCKM^{i,j*}t_{\beta}^{2}\right)\vphantom{\vphantom{\sum_{k=1}^{3}}}\right\} \end{align}

\begin{align}
\Sigma_{H^{-}H^{+}}^{2\tilde{q}}= & -\sum_{m,n}^{6}\sum_{i,j,k,l}^{3}\frac{3}{8M_{W}^{2}\pi t_{\beta}^{2}s_{W}^{2}}\alpha B_{0}\left[p^{2},m_{\tilde{u}_{m}}^{2},m_{\tilde{u}_{n}}^{2}\right]\left\{ \vphantom{\vphantom{\sum_{k=1}^{3}}}\right.\nonumber \\
 & \sum_{p,q}^{3}\left[\vphantom{\sum_{k=1}^{3}}\left(t_{\beta}^{2}A_{p,i}^{d}\VCKM^{k,p*}m_{d_{p}}R_{n,3+i}^{\tilde{d}}R_{m,k}^{\tilde{u}*}+A_{p,k}^{u*}\VCKM^{p,i*}m_{u_{p}}R_{n,i}^{\tilde{d}}R_{m,3+k}^{\tilde{u}*}\right)\right.\nonumber \\
 & \left.\times\left(t_{\beta}^{2}A_{q,j}^{d*}\VCKM^{l,q}m_{d_{q}}R_{m,l}^{\tilde{u}}R_{n,3+j}^{\tilde{d}*}+A_{q,l}^{u}\VCKM^{q,j}m_{u_{q}}R_{m,3+l}^{\tilde{u}}R_{n,j}^{\tilde{d}*}\right)\vphantom{\sum_{k=1}^{3}}\right]\nonumber \\
 & +\sum_{p}^{3}\left[\vphantom{\sum_{k=1}^{3}}\VCKM^{k,i*}\left\{ R_{n,i}^{\tilde{d}}\left(\left(-M_{W}^{2}s_{2\beta}t_{\beta}+m_{u_{k}}^{2}+t_{\beta}^{2}m_{d_{i}}^{2}\vphantom{R_{n,j}^{\tilde{d}*}}\right)R_{m,k}^{\tilde{u}*}+\mu t_{\beta}m_{u_{k}}R_{m,3+k}^{\tilde{u}*}\right)\right.\right.\nonumber \\
 & \left.+m_{d_{i}}R_{n,3+i}^{\tilde{d}}\left(\mu^{*}t_{\beta}R_{m,k}^{\tilde{u}*}+\left(1+t_{\beta}^{2}\right)m_{u_{k}}R_{m,3+k}^{\tilde{u}*}\right)\right\} \nonumber \\
 & \times\left\{ A_{p,l}^{u}\VCKM^{p,j}m_{u_{p}}R_{m,3+l}^{\tilde{u}}R_{n,j}^{\tilde{d}*}+t_{\beta}^{2}A_{p,j}^{d*}\VCKM^{l,p}m_{d_{p}}R_{m,l}^{\tilde{u}}R_{n,3+j}^{\tilde{d}*}\right\} \nonumber \\
 & +\VCKM^{l,j}\left\{ R_{m,l}^{\tilde{u}}\left(\left(-M_{W}^{2}s_{2\beta}t_{\beta}+m_{u_{l}}^{2}+t_{\beta}^{2}m_{d_{j}}^{2}\vphantom{R_{n,j}^{\tilde{d}*}}\right)R_{n,j}^{\tilde{d}*}+\mu t_{\beta}m_{d_{j}}R_{n,3+j}^{\tilde{d}*}\right)\right.\nonumber \\
 & \left.+m_{u_{l}}R_{m,3+l}^{\tilde{u}}\left(\mu^{*}t_{\beta}R_{n,j}^{\tilde{d}*}+\left(1+t_{\beta}^{2}\right)m_{d_{j}}R_{n,3+j}^{\tilde{d}*}\right)\right\} \nonumber \\
 & \left.\times\left\{ A_{p,k}^{u*}\VCKM^{p,i}m_{u_{p}}R_{n,i}^{\tilde{d}}R_{m,3+k}^{\tilde{u}*}+t_{\beta}^{2}A_{p,i}^{d}\VCKM^{k,p*}m_{d_{p}}R_{n,3+i}^{\tilde{d}}R_{m,k}^{\tilde{u}*}\right\} \vphantom{\sum_{k=1}^{3}}\right]\nonumber \\
 & +\left[\vphantom{\sum_{k=1}^{3}}\VCKM^{l,j}\VCKM^{k,i*}\left\{ R_{n,i}^{\tilde{d}}\left(\left(-M_{W}^{2}s_{2\beta}t_{\beta}+m_{u_{k}}^{2}+t_{\beta}^{2}m_{d_{i}}^{2}\right)R_{m,k}^{\tilde{u}*}+\mu t_{\beta}m_{u_{k}}R_{m,3+k}^{\tilde{u}*}\right)\right.\right.\nonumber \\
 & \left.+m_{d_{i}}R_{n,3+i}^{\tilde{d}}\left(\mu^{*}t_{\beta}R_{m,k}^{\tilde{u}*}+\left(1+t_{\beta}^{2}\right)m_{u_{k}}R_{m,3+k}^{\tilde{u}*}\right)\right\} \nonumber \\
 & \times\left\{ R_{m,l}^{\tilde{u}}\left(\left(-M_{W}^{2}s_{2\beta}t_{\beta}+m_{u_{l}}^{2}+t_{\beta}^{2}m_{d_{j}}^{2}\right)R_{n,j}^{\tilde{d}*}+\mu t_{\beta}m_{d{}_{j}}R_{n,3+j}^{\tilde{d}*}\right)\right.\nonumber \\
 & \left.\left.\left.+m_{u_{l}}R_{m,3+l}^{\tilde{u}}\left(\mu^{*}t_{\beta}R_{n,j}^{\tilde{d}*}+\left(1+t_{\beta}^{2}\right)m_{d_{j}}R_{n,3+j}^{\tilde{d}*}\right)\right\} \vphantom{\sum_{k=1}^{3}}\right]\vphantom{\vphantom{\sum_{k=1}^{3}}}\right\} \end{align}

\begin{align}
\Sigma_{H^{-}H^{+}}^{1\tilde{q}}= & \sum_{l=1}^{6}\sum_{i=1}^{3}\frac{1}{16c_{W}^{2}M_{W}^{2}\pi t_{\beta}^{2}s_{W}^{2}}\alpha A_{0}\left[m_{\tilde{u}_{l}}^{2}\vphantom{m_{\tilde{d}_{l}}^{2}}\right]\left\{ 2R_{l,3+i}^{\tilde{u}}R_{l,3+i}^{\tilde{u}*}\left(-2c_{2\beta}M_{W}^{2}t_{\beta}^{2}s_{W}^{2}+3c_{W}^{2}m_{u_{i}}^{2}\right)\vphantom{\vphantom{\sum_{k=1}^{3}}}\right.\nonumber \\
 & \left.+t_{\beta}^{2}R_{l,i}^{\tilde{u}}\left(R_{l,i}^{\tilde{u}*}c_{2\beta}m_{w}^{2}\left(1+2c_{W}^{2}\right)+\sum_{j=1}^{3}\sum_{k=1}^{3}6R_{l,j}^{\tilde{u}*}c_{W}^{2}t_{\beta}^{2}\VCKM^{i,k}\VCKM^{j,k*}m_{d_{k}}^{2}\right)\right\} \nonumber \\
 & +\sum_{l=1}^{6}\sum_{i=1}^{3}\frac{1}{16c_{W}^{2}M_{W}^{2}\pi t_{\beta}^{2}s_{W}^{2}}\alpha A_{0}\left[m_{\tilde{d}_{l}}^{2}\right]\left\{ \vphantom{\vphantom{\sum_{k=1}^{3}}}2R_{l,3+i}^{\tilde{d}}R_{l,3+i}^{\tilde{d}*}\left(c_{2\beta}M_{W}^{2}t_{\beta}^{2}s_{W}^{2}+3c_{W}^{2}t_{\beta}^{4}m_{d_{i}}^{2}\right)\right.\nonumber \\
 & \left.+R_{l,i}^{\tilde{d}}\left(R_{l,i}^{\tilde{d}*}c_{2\beta}m_{w}^{2}t_{\beta}^{2}\left(1-4c_{W}^{2}\right)+\sum_{j=1}^{3}\sum_{k=1}^{3}6R_{l,j}^{\tilde{d}*}c_{W}^{2}\VCKM^{k,j}\VCKM^{k,i*}m_{u_{k}}^{2}\right)\right\} \end{align}

\begin{itemize}
\item $Z$
\end{itemize}
\begin{eqnarray}
\Sigma_{ZZ}^{2q} & =- & \sum_{i=1}^{3}\frac{\left(9-24s_{W}^{2}+32s_{W}^{4}\right)\alpha}{36c_{W}^{2}\pi s_{W}^{2}}\left\{ A_{0}\left[m_{u_{i}}^{2}\right]+p^{2}B_{1}\left[p^{2},m_{u_{i}}^{2},m_{u_{i}}^{2}\right]\right\} \nonumber \\
 &  & -\sum_{i=1}^{3}\frac{\left(9+48s_{W}^{2}-64s_{W}^{4}\right)\alpha}{72c_{W}^{2}\pi s_{W}^{2}}\left\{ m_{u_{i}}^{2}B_{0}\left[p^{2},m_{u_{i}}^{2},m_{u_{i}}^{2}\right]\right\} \nonumber \\
 &  & -\sum_{i=1}^{3}\frac{\left(9-12s_{W}^{2}+8s_{W}^{4}\right)\alpha}{36c_{W}^{2}\pi s_{W}^{2}}\left\{ A_{0}\left[m_{d_{i}}^{2}\right]+p^{2}B_{1}\left[p^{2},m_{d_{i}}^{2},m_{d_{i}}^{2}\right]\right\} \nonumber \\
 &  & -\sum_{i=1}^{3}\frac{\left(9+24s_{W}^{2}-16s_{W}^{4}\right)\alpha}{72c_{W}^{2}\pi s_{W}^{2}}\left\{ m_{d_{i}}^{2}B_{0}\left[p^{2},m_{d_{i}}^{2},m_{d_{i}}^{2}\right]\right\}\\
&&\nonumber \\
\Sigma_{ZZ}^{2\tilde{q}}&=&- \sum_{m,n}^{6}\sum_{i,j}^{3}\frac{\alpha}{72c_{W}^{2}\pi s_{W}^{2}}\left\{ A_{0}\left[m_{\tilde{u}_{n}}^{2}\vphantom{m_{\tilde{d}_{l}}^{2}}\right]+2m_{\tilde{u}_{m}}^{2}B_{0}\left[p^{2},m_{\tilde{u}_{m}}^{2},m_{\tilde{u}_{n}}^{2}\vphantom{m_{\tilde{d}_{l}}^{2}}\right]\right.\nonumber \\
 && \left.+\left(p^{2}+m_{\tilde{u}_{m}}^{2}-m_{\tilde{u}_{n}}^{2}\right)B_{1}\left[p^{2},m_{\tilde{u}_{m}}^{2},m_{\tilde{u}_{n}}^{2}\vphantom{m_{\tilde{d}_{l}}^{2}}\right]-\frac{p^{2}}{3}+m_{\tilde{u}_{m}}^{2}-m_{\tilde{u}_{n}}^{2}\right\} \nonumber \\
 && \times\left\{ \left(-3+4s_{W}^{2}\right)R_{n,i}^{\tilde{u}}R_{m,i}^{\tilde{u}*}+4s_{W}^{2}R_{n,3+i}^{\tilde{u}}R_{m,3+i}^{\tilde{u}*}\right\}\nonumber \\ 
&& \,\,\,\left\{ \left(-3+4s_{W}^{2}\right)R_{m,j}^{\tilde{u}}R_{n,j}^{\tilde{u}*}+4s_{W}^{2}R_{m,3+j}^{\tilde{u}}R_{n,3+j}^{\tilde{u}*}\right\} \nonumber \\
 && -\sum_{m,n}^{6}\sum_{i,j}^{3}\frac{\alpha}{72c_{W}^{2}\pi s_{W}^{2}}\left\{ A_{0}\left[m_{\tilde{d}_{n}}^{2}\vphantom{m_{\tilde{d}_{l}}^{2}}\right]+2m_{\tilde{d}_{m}}^{2}B_{0}\left[p^{2},m_{\tilde{d}_{m}}^{2},m_{\tilde{d}_{n}}^{2}\vphantom{m_{\tilde{d}_{l}}^{2}}\right]\right.\nonumber \\
 && \left.+\left(p^{2}+m_{\tilde{d}_{m}}^{2}-m_{\tilde{d}_{n}}^{2}\right)B_{1}\left[p^{2},m_{\tilde{d}_{m}}^{2},m_{\tilde{d}_{n}}^{2}\vphantom{m_{\tilde{d}_{l}}^{2}}\right]-\frac{p^{2}}{3}+m_{\tilde{d}_{m}}^{2}-m_{\tilde{d}_{n}}^{2}\right\} \nonumber \\
 && \times\left\{ \left(-3+2s_{W}^{2}\right)R_{n,i}^{\tilde{d}}R_{m,i}^{\tilde{d}*}+2s_{W}^{2}R_{n,3+i}^{\tilde{d}}R_{m,3+i}^{\tilde{d}*}\right\}\nonumber \\ 
&&\,\,\,\left\{ \left(-3+2s_{W}^{2}\right)R_{m,j}^{\tilde{d}}R_{n,j}^{\tilde{d}*}+2s_{W}^{2}R_{m,3+j}^{\tilde{d}}R_{n,3+j}^{\tilde{d}*}\right\}\\
&&\nonumber \\
\Sigma_{ZZ}^{1\tilde{q}}&= & \sum_{l=1}^{6}\sum_{i=1}^{3}\frac{1}{24c_{W}^{2}\pi s_{W}^{2}}\alpha A_{0}\left[m_{\tilde{u}_{l}}^{2}\vphantom{m_{\tilde{d}_{l}}^{2}}\right]\left(\vphantom{R_{l,i}^{\tilde{d}}}R_{l,i}^{\tilde{u}}R_{l,i}^{\tilde{u}*}\left(-3+4s_{W}^{2}\right)^{2}+16s_{W}^{4}R_{l,3+i}^{\tilde{u}}R_{l,3+i}^{\tilde{u}*}\right)\nonumber \\
 && +\sum_{l=1}^{6}\sum_{i=1}^{3}\frac{1}{24c_{W}^{2}\pi s_{W}^{2}}\alpha A_{0}\left[m_{\tilde{d}_{l}}^{2}\right]\left(R_{l,i}^{\tilde{d}}R_{l,i}^{\tilde{d}*}\left(3-2s_{W}^{2}\right)^{2}+4s_{W}^{4}R_{l,3+i}^{\tilde{d}}R_{l,3+i}^{\tilde{d}*}\right)
\end{eqnarray}

\begin{itemize}
\item $W$
\end{itemize}
\begin{eqnarray}
\Sigma_{WW}^{2q} & =- & \sum_{i,l}^{3}\frac{\alpha}{4\pi s_{W}^{2}}\VCKM^{l,i}\VCKM^{l,i*}\left\{ 2A_{0}\left[m_{d_{i}}^{2}\right]+m_{u_{l}}^{2}B_{0}\left[p^{2},m_{d_{i}}^{2},m_{u_{l}}^{2}\right]\right.\nonumber \\
 &  & \left.+\left(m_{d_{l}}^{2}-m_{u_{l}}^{2}+2p^{2}\right)B_{1}\left[p^{2},m_{u_{i}}^{2},m_{d_{l}}^{2}\right]\right\}\\
&&\nonumber \\
\Sigma_{WW}^{2\tilde{q}}&= & -\sum_{m,n}^{6}\sum_{i,j,k,l}^{3}\frac{3\alpha}{12\pi s_{W}^{2}}\VCKM^{k,i}\VCKM^{l,j*}R_{m,k}^{\tilde{u}}R_{m,l}^{\tilde{u}*}R_{n,j}^{\tilde{d}}R_{n,i}^{\tilde{d}*}\left\{ A_{0}\left[m_{\tilde{d}_{n}}^{2}\vphantom{m_{\tilde{d}_{l}}^{2}}\right]\right.\nonumber \\
&&+2m_{\tilde{u}_{m}}^{2}B_{0}\left[p^{2},m_{\tilde{u}_{m}}^{2},m_{\tilde{d}_{n}}^{2}\vphantom{m_{\tilde{d}_{l}}^{2}}\right]\nonumber \\
 && \left.+\left(p^{2}+m_{\tilde{u}_{m}}^{2}-m_{\tilde{d}_{n}}^{2}\right)B_{1}\left[p^{2},m_{\tilde{u}_{m}}^{2},m_{\tilde{d}_{n}}^{2}\vphantom{m_{\tilde{d}_{l}}^{2}}\right]-\frac{p^{2}}{3}+m_{\tilde{u}_{m}}^{2}+m_{\tilde{d}_{n}}^{2}\right\}\\
&&\nonumber \\
\Sigma_{WW}^{1\tilde{q}}&= & \sum_{l=1}^{6}\sum_{i=1}^{3}\frac{3}{8\pi s_{W}^{2}}\alpha A_{0}\left[m_{\tilde{u}_{l}}^{2}\vphantom{m_{\tilde{d}_{l}}^{2}}\right]R_{l,i}^{\tilde{u}}R_{l,i}^{\tilde{u}*}+\sum_{l=1}^{6}\sum_{i=1}^{3}\frac{3}{8\pi s_{W}^{2}}\alpha A_{0}\left[m_{\tilde{d}_{l}}^{2}\vphantom{m_{\tilde{d}_{l}}^{2}}\right]R_{l,i}^{\tilde{d}}R_{l,i}^{\tilde{d}*}
\end{eqnarray}

\newpage
\begin{itemize}
\item Tadpoles
\end{itemize}
\begin{eqnarray}
T_{h}^{q} & = & -\sum_{i}^{3}\frac{3}{8\pi^{2}M_{W}s_{\beta}s_{W}}c_{\alpha}m_{u_{i}}^{2}eA_{0}\left[m_{u_{i}}^{2}\right]\\
T_{h}^{\tilde{q}} & = & \sum_{m}^{6}\sum_{i,j}^{3}\frac{1}{32\pi^{2}c_{W}M_{W}s_{\beta}s_{W}}eA_{0}\left[m_{\tilde{u}_{m}}^{2}\right]\nonumber \\
 &  & \times\left[\vphantom{\sum_{k=1}^{3}}\left\{ \vphantom{\sum_{k=1}^{3}}\delta_{i,j}\left(M_{W}m_{Z}s_{\alpha+\beta}s_{\beta}\left(-3+4s_{W}^{2}\right)+6c_{\alpha}c_{W}m_{u_{i}}^{2}\right)R_{m,j}^{\tilde{u}}\right.\right.\nonumber \\
 &  & \left.+3c_{W}\left(c_{\alpha}A_{i,j}^{u}+\mu^{*}s_{\alpha}\delta_{i,j}\right)m_{u_{i}}R_{m,3+j}^{\tilde{u}}\vphantom{\sum_{k=1}^{3}}\right\} R_{m,i}^{\tilde{u}*}
\nonumber \\
&&+\left\{ \vphantom{\sum_{k=1}^{3}}3c_{\alpha}c_{W}A_{j,i}^{u*}m_{u_{j}}R_{m,j}^{\tilde{u}}+3c_{W}\mu s_{\alpha}\delta_{i,j}m_{u_{i}}R_{m,j}^{\tilde{u}}\right.\nonumber \\
 &  & \left.\left.+2\delta_{i,j}\left(-2M_{W}m_{Z}s_{\alpha+\beta}s_{\beta}s_{W}^{2}+3c_{\alpha}c_{W}m_{u_{i}}^{2}\right)R_{m,3+j}^{\tilde{u}}\vphantom{\sum_{k=1}^{3}}\right\} R_{m,3+i}^{\tilde{u}*}\vphantom{\sum_{k=1}^{3}}\right]\nonumber \\
 &  & +\sum_{i}^{3}\frac{3}{8\pi^{2}M_{W}c_{\beta}s_{W}}s_{\alpha}m_{d_{i}}^{2}eA_{0}\left[m_{d_{i}}^{2}\right]-\sum_{m}^{6}\sum_{i,j}^{3}\frac{1}{32\pi^{2}c_{W}M_{W}c_{\beta}s_{W}}eA_{0}\left[m_{\tilde{d}_{m}}^{2}\right]\nonumber \\
 &  & \times\left[\vphantom{\sum_{k=1}^{3}}\left\{ \vphantom{\sum_{k=1}^{3}}\delta_{i,j}\left(M_{W}m_{Z}s_{\alpha+\beta}c_{\beta}\left(-3+2s_{W}^{2}\right)+6s_{\alpha}c_{W}m_{d_{i}}^{2}\right)R_{m,j}^{\tilde{d}}\right.\right.\nonumber \\
 &  & \left.+3c_{W}\left(s_{\alpha}A_{i,j}^{d}+\mu^{*}c_{\alpha}\delta_{i,j}\right)m_{d_{i}}R_{m,3+j}^{\tilde{d}}\vphantom{\sum_{k=1}^{3}}\right\} R_{m,i}^{\tilde{d}*}
\nonumber \\
&&+\left\{ \vphantom{\sum_{k=1}^{3}}3s_{\alpha}c_{W}A_{j,i}^{d*}m_{d_{j}}R_{m,j}^{\tilde{d}}+3c_{W}\mu c_{\alpha}\delta_{i,j}m_{d_{i}}R_{m,j}^{\tilde{d}}\right.\nonumber \\
 &  & \left.\left.-2\delta_{i,j}\left(M_{W}m_{Z}s_{\alpha+\beta}c_{\beta}s_{W}^{2}-3s_{\alpha}c_{W}m_{d_{i}}^{2}\right)R_{m,3+j}^{\tilde{d}}\vphantom{\sum_{k=1}^{3}}\right\} R_{m,3+i}^{\tilde{d}*}\vphantom{\sum_{k=1}^{3}}\right]\end{eqnarray}

The expressions for $T_{H}$ are obtained using the replacements
of Eq.~\ref{eq:cambio} of Appendix A on the results of~$T_{h}$.

\end{appendix}


\end{document}